\documentclass[psfig,graphicx,useAMS,10pt]{iopart}
\usepackage{txfonts,color,iopams,graphicx}
\usepackage[landscape]{geometry}

\begin{document}

\title[D.E.U.S. XI]{D.E.U.S.
\\ (Dimension Embedded in Unified Symmetry) \\XI: SU(2) and SU(3) Groups}

\author[A.S. Popescu]{Adrian Sabin Popescu\dag\
\footnote[3]{sabinp@aira.astro.ro}}
\address{\dag\ Astronomical Institute of Romanian Academy, Str. Cutitul de Argint 5, RO-040557 
Bucharest, Romania}

\begin{abstract}
From DEUS model results that the SU(2) and the SU(3) groups, describing the weak, respectively, the strong interaction, have different generators than the classical ones.
\end{abstract}


\section{Introduction}

Without giving to many details, we can say that the electromagnetic interactions are mathematically described through the invariance under the U(1) group. This group is having only one parameter, the existence of only one field quanta (the photon) being enough to characterize these interactions. For the case of weak interactions we have a good description through the SU(2) group, having three independent parameters. Here we need three field quanta: $W^{+}$, $W^{-}$ and $Z^{0}$. In classical particle theory, the strong interactions are described by the "color" SU(3) coresponding to eight gluons.

From the local charts of the hyper-surfaces crossed by an event that travels from the external observer's spacetime toward the interior of the DEUS object \cite{deus1}, and following the external observer's perception of the event, we will derive the infinitesimal element for the DEUS object symmetry, which will prove to be the SU(2) infinitesimal element for a DEUS object with all its self-similar content evaporated, or the SU(3) infinitesimal element for a DEUS object containing another "visible" self-similar DEUS level. The path followed by this event is:

\begin{equation}
\begin{array}{ccccc}
\textnormal{{\bf 1.}}\;\; x_{pure\; helicoidal\; hyper-surface}^{\nu} & \longleftarrow x_{helicoid\; C=1;\; \theta\neq arctg (t/\phi)}^{\mu} & \longleftarrow x_{ergosphere's\; helicoid}^{\sigma} & &  \\
 & & \uparrow & \nwarrow \textnormal{{\scriptsize {\bf D}}} \leftarrow \textnormal{{\scriptsize {\bf A}}} & \\
 & & \frac{t}{\phi}=-\frac{\phi_{k}}{t_{k}} & & \!\!\!\!\!\!\!\!\!\!\!\!\!\!\!\!\!\circledS\;\; x_{BHIG-FRW}^{\alpha} \\
 & & \downarrow & \swarrow \textnormal{{\scriptsize {\bf D}}} \leftarrow \textnormal{{\scriptsize {\bf D}}} & \\
\textnormal{{\bf 2.}}\;\; x_{pure\; catenoidal\; hyper-surface}^{\nu} & \longleftarrow x_{catenoid\; C=1;\; \theta\neq arctg (t/\phi)}^{\mu} & \longleftarrow x_{ergosphere's\; catenoid}^{\sigma} & & 
\end{array}
\label{eq_1}
\end{equation}

\section{SU(2) Group}

For the first path ({\bf 1.}), the transformation which relates each point with coordinates $x_{pure\; helicoidal\; hyper-surface}^{\nu}$ to $x_{helicoid\; C=1;\; \theta\neq arctg (t/\phi)}^{\mu}$ is, in the {\bf D} case \cite{deus1}, $A=\left\{a_{\mu}^{\nu} \right\}$ where:

\begin{equation}
x_{pure\; helicoidal\; hyper-surface}^{\nu}=a_{\mu}^{\nu}\; x_{helicoid\; C=1;\; \theta\neq arctg (t/\phi)}^{\mu}\; ,
\label{eq_2}
\end{equation}
\noindent
having:
{\tiny
\begin{equation}
\!\!\!\!\!\!\!\!\!\!\!\!\!\!\!\!\!\!\!\!\!\!\!\!\!\!\!\!\!\!\!\!\!\!\!\!\!\!\!\!\!\!\!\!\!\!\!\!\!\!\!\!\!\!\!\!\!\!\!\!\!\!\!\!\!\!\!\!\!\!\!\!\!\!\!\!\!\!\!\!\!\!\!\!\!\!\!\!A=\left(
\begin{array}{ccccc}
1 & \displaystyle \frac{t_{k}}{\phi_{k}} & \displaystyle \sinh \theta\; \frac{\displaystyle \sin\left[\textnormal{arctg}\left(\frac{t_{k}}{\phi_{k}}\right) \right]}{\textnormal{arctg}\displaystyle \left(\frac{t_{k}}{\phi_{k}}\right)} & \begin{array}{cc}(\mp ) \\ (\pm ) \end{array} \displaystyle \frac{i}{\sqrt{2}}\; \sinh \theta\; \frac{\sin\left[\textnormal{arctg}\displaystyle \left(\frac{t_{k}}{\phi_{k}}\right) \right]}{\textnormal{arctg} \displaystyle \left(\frac{t_{k}}{\phi_{k}}\right)} & \begin{array}{cc} (\mp ) \\ (\pm ) \end{array} \displaystyle i\; \sin\left[\textnormal{arctg} \left(\displaystyle \frac{t_{k}}{\phi_{k}}\right) \right] \\
\mp \displaystyle i\; \frac{\phi_{k}}{t_{k}}\sqrt{\displaystyle \frac{\phi_{k}}{t_{k}\; \textnormal{tg}\; \theta}} & \mp \displaystyle i\; \sqrt{\displaystyle \frac{\phi_{k}}{t_{k}\; \textnormal{tg}\; \theta}} & \mp \displaystyle i\; \sinh \theta\; \frac{\displaystyle \cos\left[\textnormal{arctg}\left(\frac{t_{k}}{\phi_{k}}\right) \right]}{\textnormal{arctg}\displaystyle \left(\frac{t_{k}}{\phi_{k}}\right)} \sqrt{\displaystyle \frac{\phi_{k}}{t_{k}\; \textnormal{tg}\; \theta}} & \mp \displaystyle \frac{1}{\sqrt{2}} \sinh \theta \; \frac{\displaystyle \cos\left[\textnormal{arctg}\left(\frac{t_{k}}{\phi_{k}}\right) \right]}{\textnormal{arctg}\displaystyle \left(\frac{t_{k}}{\phi_{k}}\right)} \sqrt{\displaystyle \frac{\phi_{k}}{t_{k}\; \textnormal{tg}\; \theta}} & \mp \displaystyle \cos\left[\textnormal{arctg}\left(\frac{t_{k}}{\phi_{k}}\right) \right] \sqrt{\displaystyle \frac{\phi_{k}}{t_{k}\; \textnormal{tg}\; \theta}} \\
\displaystyle \frac{1}{\sinh \theta}\; \frac{\textnormal{arctg}\left(\displaystyle \frac{t_{k}}{\phi_{k}}\right)}{\displaystyle \sin\left[\textnormal{arctg}\left(\frac{t_{k}}{\phi_{k}}\right) \right]} & \displaystyle \frac{1}{\sinh \theta}\; \frac{\textnormal{arctg}\left(\displaystyle \frac{t_{k}}{\phi_{k}}\right)}{\displaystyle \cos\left[\textnormal{arctg}\left(\frac{t_{k}}{\phi_{k}}\right) \right]} & 1 & \begin{array}{cc} (\mp ) \\ (\pm ) \end{array} \displaystyle \frac{i}{\sqrt{2}} & \begin{array}{cc} (\mp ) \\ (\pm ) \end{array} \displaystyle i\; \frac{1}{\sinh \theta}\; \textnormal{arctg}\left(\frac{t_{k}}{\phi_{k}}\right) \\
\displaystyle \pm i\; \frac{1}{\sinh \theta}\; \frac{\textnormal{arctg}\left(\displaystyle \frac{t_{k}}{\phi_{k}}\right)}{\displaystyle \sin\left[\textnormal{arctg}\left(\frac{t_{k}}{\phi_{k}}\right) \right]} & \displaystyle \pm i\; \frac{1}{\sinh \theta}\; \frac{\textnormal{arctg}\left(\displaystyle \frac{t_{k}}{\phi_{k}}\right)}{\displaystyle \cos\left[\textnormal{arctg}\left(\frac{t_{k}}{\phi_{k}}\right) \right]} & \pm i & \pm \displaystyle \frac{1}{\sqrt{2}} & \pm \displaystyle \frac{1}{\sinh \theta}\; \textnormal{arctg}\left(\frac{t_{k}}{\phi_{k}}\right) \\
0 & 0 & 0 & 0 & 0
\end{array}
\right)
\label{eq_3}
\end{equation}}
\noindent
where, for few elements (for example, for the $a^{1}_{4}$ element), the transformation is permissive in having either $\mp$, either $\pm$, but keeping the first or the second choice between two elements having the same "problem". 

Now, the transformation from $x_{helicoid\; C=1;\; \theta\neq arctg (t/\phi)}^{\mu}$ to $x_{ergosphere's\; helicoid}^{\sigma}$ coordinates is $A'=\left\{(a')_{\sigma}^{\mu}\right\}$ where:

\begin{equation}
x_{helicoid\; C=1;\; \theta\neq arctg (t/\phi)}^{\mu}=(a')_{\sigma}^{\mu}\; x_{ergosphere's\; helicoid}^{\sigma} \; ,
\label{eq_4}
\end{equation}
\noindent
with:

{\tiny \begin{equation}
\!\!\!\!\!\!\!\!\!\!\!\!\!\!\!\!\!\!\!\!\!\!\!\!\!\!\!\!\!\!\!\!\!\!\!\!\!\!\!\!\!\!\!\!\!\!\!\!\!\!\!\!A'=\left(
\begin{array}{ccccc}
1 & -\displaystyle \frac{t_{k}}{\phi_{k}} & \displaystyle \sinh \theta\; \frac{\displaystyle \sin\left[\textnormal{arctg}\left(\frac{t_{k}}{\phi_{k}}\right) \right]}{\textnormal{arctg}\displaystyle \left(\frac{t_{k}}{\phi_{k}}\right)} & 0 & \mp \displaystyle i\; \sin\left[\textnormal{arctg} \displaystyle \left(\frac{t_{k}}{\phi_{k}}\right) \right] \\
\displaystyle \frac{\phi_{k}}{t_{k}} & -1 & \displaystyle \sinh \theta \frac{\displaystyle \cos\left[\textnormal{arctg}\left(\frac{t_{k}}{\phi_{k}}\right) \right]}{\textnormal{arctg}\displaystyle \left(\frac{t_{k}}{\phi_{k}}\right)} & 0 & \mp \displaystyle i\; \cos\left[\textnormal{arctg} \displaystyle \left(\frac{t_{k}}{\phi_{k}}\right) \right] \\
\displaystyle \frac{1}{\sinh \theta}\; \frac{\textnormal{arctg}\displaystyle \left(\frac{t_{k}}{\phi_{k}}\right)}{\displaystyle \sin\left[\textnormal{arctg}\left(\frac{t_{k}}{\phi_{k}}\right) \right]} & - \displaystyle \frac{1}{\sinh \theta}\; \frac{\textnormal{arctg}\displaystyle \left(\frac{t_{k}}{\phi_{k}}\right)}{\displaystyle \cos\left[\textnormal{arctg}\left(\frac{t_{k}}{\phi_{k}}\right) \right]} & 1 & 0 & \mp \displaystyle i\; \frac{1}{\sinh \theta}\; \textnormal{arctg}\left(\frac{t_{k}}{\phi_{k}}\right) \\
\begin{array}{cc} (\pm ) \\ (\mp ) \end{array} \displaystyle i\; \sqrt{2}\; \frac{1}{\sinh \theta}\; \frac{\textnormal{arctg}\displaystyle \left(\frac{t_{k}}{\phi_{k}}\right)}{\displaystyle \sin\left[\textnormal{arctg}\left(\frac{t_{k}}{\phi_{k}}\right) \right]} & \begin{array}{cc} (\mp ) \\ (\pm ) \end{array} \displaystyle i\; \sqrt{2}\; \frac{1}{\sinh \theta}\; \frac{\textnormal{arctg}\displaystyle \left(\frac{t_{k}}{\phi_{k}}\right)}{\displaystyle \cos\left[\textnormal{arctg}\left(\frac{t_{k}}{\phi_{k}}\right) \right]} & \begin{array}{cc} (\pm ) \\ (\mp ) \end{array} \displaystyle i\; \sqrt{2} & 0 & \pm \displaystyle \sqrt{2}\; \frac{1}{\sinh \theta}\; \textnormal{arctg} \displaystyle \left(\frac{t_{k}}{\phi_{k}}\right) \\
\begin{array}{cc} (\pm ) \\ (\mp ) \end{array} \displaystyle i\; \frac{1}{\sin\left[\textnormal{arctg} \displaystyle \left(\frac{t_{k}}{\phi_{k}}\right) \right]} & \begin{array}{cc} (\mp ) \\ (\pm ) \end{array} \displaystyle i\; \frac{1}{\cos\left[\textnormal{arctg} \displaystyle \left(\frac{t_{k}}{\phi_{k}}\right) \right]} & \begin{array}{cc} (\pm ) \\ (\mp ) \end{array} \displaystyle i\; \sinh \theta\; \frac{1}{\textnormal{arctg} \displaystyle \left(\frac{t_{k}}{\phi_{k}}\right)} & 0 & \pm 1
\end{array}
\right)
\label{eq_5}
\end{equation}}

In conclusion, the transformation from $x_{pure\; helicoidal\; hyper-surface}^{\nu}$ to $x_{ergosphere's\; helicoid}^{\sigma}$ reads:

{\tiny \begin{equation}
\!\!\!\!\!\!\!\!\!\!\!\!\!\!\!\!\!\!\!\!\!\!\!\!\!\!\!\!\!\!\!\!\!\!\!\!\!\!\!\!\!\!\!\!\!\!\!\!\!\!\!\!AA'=\lambda\cdot \left(
\begin{array}{ccccc}
1 & -\displaystyle \frac{t_{k}}{\phi_{k}} & \displaystyle \sinh \theta\; \frac{\displaystyle \sin\left[\textnormal{arctg}\left(\frac{t_{k}}{\phi_{k}}\right) \right]}{\textnormal{arctg}\displaystyle \left(\frac{t_{k}}{\phi_{k}}\right)} & 0 & \mp \displaystyle i\; \sin\left[\textnormal{arctg} \displaystyle \left(\frac{t_{k}}{\phi_{k}}\right) \right] \\
\mp \displaystyle i\; \frac{\phi_{k}}{t_{k}}\sqrt{\displaystyle \frac{\phi_{k}}{t_{k}\; \textnormal{tg}\; \theta}} & \pm \displaystyle i\; \sqrt{\displaystyle \frac{\phi_{k}}{t_{k}\; \textnormal{tg}\; \theta}} & \mp \displaystyle i\; \sinh \theta \frac{\displaystyle \cos\left[\textnormal{arctg}\left(\frac{t_{k}}{\phi_{k}}\right) \right]}{\textnormal{arctg}\displaystyle \left(\frac{t_{k}}{\phi_{k}}\right)} \sqrt{\displaystyle \frac{\phi_{k}}{t_{k}\; \textnormal{tg}\; \theta}} & 0 & - \displaystyle \cos\left[\textnormal{arctg}\left(\frac{t_{k}}{\phi_{k}}\right) \right] \sqrt{\displaystyle \frac{\phi_{k}}{t_{k}\; \textnormal{tg}\; \theta}} \\
\displaystyle \frac{1}{\sinh \theta}\; \frac{\textnormal{arctg}\displaystyle \left(\frac{t_{k}}{\phi_{k}}\right)}{\displaystyle \sin\left[\textnormal{arctg}\left(\frac{t_{k}}{\phi_{k}}\right) \right]} & - \displaystyle \frac{1}{\sinh \theta}\; \frac{\textnormal{arctg}\displaystyle \left(\frac{t_{k}}{\phi_{k}}\right)}{\displaystyle \cos\left[\textnormal{arctg}\left(\frac{t_{k}}{\phi_{k}}\right) \right]} & 1 & 0 & \mp \displaystyle i\; \frac{1}{\sinh \theta}\; \textnormal{arctg}\left(\frac{t_{k}}{\phi_{k}}\right) \\
\displaystyle \pm i\; \frac{1}{\sinh \theta}\; \frac{\textnormal{arctg}\displaystyle \left(\frac{t_{k}}{\phi_{k}}\right)}{\displaystyle \sin\left[\textnormal{arctg}\left(\frac{t_{k}}{\phi_{k}}\right) \right]} & \displaystyle \mp i\; \frac{1}{\sinh \theta}\; \frac{\textnormal{arctg}\left(\displaystyle \frac{t_{k}}{\phi_{k}}\right)}{\displaystyle \cos\left[\textnormal{arctg}\left(\frac{t_{k}}{\phi_{k}}\right) \right]} & \pm i & 0 & \displaystyle \frac{1}{\sinh \theta}\; \textnormal{arctg}\left(\frac{t_{k}}{\phi_{k}}\right) \\
0 & 0 & 0 & 0 & 0
\end{array}
\right)
\label{eq_6}
\end{equation}}
\noindent
where the product result is the same when working with the {\bf A} case (real helicoid and imaginary catenoid) instead of {\bf D}, and $\lambda$ is 5 for the {\bf D} case and 4 for the {\bf A} case \cite{deus1}, and indicates the number of spacetime dimensions in which the event must be seen by the external observer. 

Following the other possible path ({\bf 2.}) from (\ref{eq_1}) (the catenoid is the five-dimensional timelike equivalent of the five-dimensional spacelike helicoid), the transformation from $x_{pure\; catenoidal\; hyper-surface}^{\nu}$ to $x_{catenoid\; C=1;\; \theta\neq arctg (t/\phi)}^{\mu}$, in the {\bf D} case \cite{deus1}, is:

\begin{equation}
x_{pure\; catenoidal\; hyper-surface}^{\nu}=b_{\mu}^{\nu}\; x_{catenoid\; C=1;\; \theta\neq arctg (t/\phi)}^{\mu} \; ,
\label{eq_7}
\end{equation}
\noindent
with $B=\left\{b_{\mu}^{\nu} \right\}$:

{\tiny \begin{equation}
\!\!\!\!\!\!\!\!\!\!\!\!\!\!\!\!\!\!\!\!\!\!\!\!\!\!\!\!\!\!\!\!\!\!\!\!\!\!\!\!\!\!\!\!\!\!\!\!\!\!\!\!\!\!\!\!\!\!\!\!\!\!\!\!\!\!\!\!\!\!\!\!\!\!\!\!\!\!\!\!\!\!\!\!\!\!\!B=\left(
\begin{array}{ccccc}
1 & \displaystyle \frac{1}{\textnormal{tg}\; \theta} & \displaystyle \cos \theta\; \frac{\sqrt{t^{2}+\phi^{2}}}{A}\; \cosh \left[\frac{\sqrt{t^{2}+\phi^{2}}}{A} \right] & \displaystyle \mp i\; \frac{\cos \theta}{\theta} & 0 \\
\pm \displaystyle \textnormal{tg}\; \theta\; \sqrt{\displaystyle \frac{t}{\phi\; \textnormal{tg}\; \theta}} & \pm \displaystyle \sqrt{\displaystyle \frac{t}{\phi\; \textnormal{tg}\; \theta}} & \pm \displaystyle \sin \theta\; \frac{\sqrt{t^{2}+\phi^{2}}}{A}\; \cosh \left[\frac{\sqrt{t^{2}+\phi^{2}}}{A} \right]\; \sqrt{\displaystyle \frac{t}{\phi\; \textnormal{tg}\; \theta}} & - \displaystyle i\; \frac{\sin \theta}{\theta}\; \sqrt{\displaystyle \frac{t}{\phi\; \textnormal{tg}\; \theta}} & 0 \\
\displaystyle \frac{1}{\cos \theta}\; \frac{A}{\sqrt{t^{2}+\phi^{2}}}\; \cosh^{-1}\left[\frac{\sqrt{t^{2}+\phi^{2}}}{A} \right] & \displaystyle \frac{1}{\sin \theta}\; \frac{A}{\sqrt{t^{2}+\phi^{2}}}\; \cosh^{-1}\left[\frac{\sqrt{t^{2}+\phi^{2}}}{A} \right] & 1 & \displaystyle \mp i\; \frac{1}{\theta}\; \frac{A}{\sqrt{t^{2}+\phi^{2}}}\; \cosh^{-1}\left[\frac{\sqrt{t^{2}+\phi^{2}}}{A} \right] & 0 \\
\pm \displaystyle i\; \frac{1}{\cos \theta}\; \frac{A}{\sqrt{t^{2}+\phi^{2}}}\; \cosh^{-1}\left[\frac{\sqrt{t^{2}+\phi^{2}}}{A} \right] & \pm \displaystyle i\; \frac{1}{\sin \theta}\; \frac{A}{\sqrt{t^{2}+\phi^{2}}}\; \cosh^{-1}\left[\frac{\sqrt{t^{2}+\phi^{2}}}{A} \right] & \pm i & \displaystyle \frac{1}{\theta}\; \frac{A}{\sqrt{t^{2}+\phi^{2}}}\; \cosh^{-1}\left[\frac{\sqrt{t^{2}+\phi^{2}}}{A} \right] & 0 \\
0 & 0 & 0 & 0 & 0
\end{array}
\right)
\label{eq_8}
\end{equation}}

From $x_{catenoid\; C=1;\; \theta\neq arctg (t/\phi)}^{\mu}$ to $x_{ergosphere's\; catenoid}^{\sigma}$ we have:

\begin{equation}
x_{catenoid\; C=1;\; \theta\neq arctg (t/\phi)}^{\mu}=(b')_{\sigma}^{\mu}\; x_{ergosphere's\; catenoid}^{\sigma} \; ,
\label{eq_9}
\end{equation}
\noindent
where $B'=\left\{(b')_{\sigma}^{\mu}\right\}$:

{\tiny \begin{equation}
\!\!\!\!\!\!\!\!\!\!\!\!\!\!\!\!\!\!\!\!\!\!\!\!\!\!\!\!\!\!\!\!\!\!\!\!\!\!\!\!\!\!\!\!\!\!\!\!\!\!\!\!\!\!\!\!\!\!\!\!\!\!B'=\left(
\begin{array}{ccccc}
1 & \displaystyle \frac{1}{\textnormal{tg}\; \theta} & \displaystyle \cos \theta\; \frac{\sqrt{t^{2}+\phi^{2}}}{A}\; \cosh \left[\frac{\sqrt{t^{2}+\phi^{2}}}{A} \right] & 0 & \displaystyle \mp i\; \cos \theta \\
\displaystyle \textnormal{tg}\; \theta & 1 & \displaystyle \sin \theta\; \frac{\sqrt{t^{2}+\phi^{2}}}{A}\; \cosh \left[\frac{\sqrt{t^{2}+\phi^{2}}}{A} \right] & 0 & \displaystyle \mp i\; \sin \theta \\
\displaystyle \frac{1}{\cos \theta}\; \frac{A}{\sqrt{t^{2}+\phi^{2}}}\; \cosh^{-1}\left[\frac{\sqrt{t^{2}+\phi^{2}}}{A} \right] & \displaystyle \frac{1}{\sin \theta}\; \frac{A}{\sqrt{t^{2}+\phi^{2}}}\; \cosh^{-1}\left[\frac{\sqrt{t^{2}+\phi^{2}}}{A} \right] & 1 & 0 & \displaystyle \mp i\; \frac{A}{\sqrt{t^{2}+\phi^{2}}}\; \cosh^{-1}\left[\frac{\sqrt{t^{2}+\phi^{2}}}{A} \right] \\
\pm \displaystyle i\; \frac{\theta}{\cos \theta} & \pm \displaystyle i\; \frac{\theta}{\sin \theta} & \pm \displaystyle i\; \theta\; \frac{\sqrt{t^{2}+\phi^{2}}}{A}\; \cosh \left[\frac{\sqrt{t^{2}+\phi^{2}}}{A} \right] & 0 & \theta \\
0 & 0 & 0 & 0 & 0
\end{array}
\right)
\label{eq_10}
\end{equation}}

In conclusion, for $x_{ergosphere's\; catenoid}^{\sigma} \longrightarrow x_{catenoid\; C=1;\; \theta\neq arctg (t/\phi)}^{\mu}\longrightarrow x_{pure\; catenoidal\; hyper-surface}^{\nu}$ we have:

{\tiny \begin{equation}
\!\!\!\!\!\!\!\!\!\!\!\!\!\!\!\!\!\!\!\!\!\!\!\!\!\!\!\!\!\!\!\!\!\!\!\!\!\!\!\!\!\!\!\!\!\!\!\!\!\!\!\!\!\!\!\!\!\!\!\!\!\!\!\!\!\!\!\!\!\!\!\!\!\!\!\!\!\!\!\!\!\!\!\!\!\!\!BB'=\lambda\cdot \left(
\begin{array}{ccccc}
1 & \displaystyle \frac{1}{\textnormal{tg}\; \theta} & \displaystyle \cos \theta\; \frac{\sqrt{t^{2}+\phi^{2}}}{A}\; \cosh \left[\frac{\sqrt{t^{2}+\phi^{2}}}{A} \right] & 0 & \displaystyle \mp i\; \cos \theta \\
\pm \displaystyle \textnormal{tg}\; \theta\; \sqrt{\displaystyle \frac{t}{\phi\; \textnormal{tg}\; \theta}} & \pm \displaystyle \sqrt{\displaystyle \frac{t}{\phi\; \textnormal{tg}\; \theta}} & \pm \displaystyle \sin \theta\; \frac{\sqrt{t^{2}+\phi^{2}}}{A}\; \cosh \left[\frac{\sqrt{t^{2}+\phi^{2}}}{A} \right]\; \sqrt{\displaystyle \frac{t}{\phi\; \textnormal{tg}\; \theta}} & 0 & \displaystyle - i\; \sin \theta\; \sqrt{\displaystyle \frac{t}{\phi\; \textnormal{tg}\; \theta}} \\
\displaystyle \frac{1}{\cos \theta}\; \frac{A}{\sqrt{t^{2}+\phi^{2}}}\; \cosh^{-1}\left[\frac{\sqrt{t^{2}+\phi^{2}}}{A} \right] & \displaystyle \frac{1}{\sin \theta}\; \frac{A}{\sqrt{t^{2}+\phi^{2}}}\; \cosh^{-1}\left[\frac{\sqrt{t^{2}+\phi^{2}}}{A} \right] & 1 & 0 & \displaystyle \mp i\; \frac{A}{\sqrt{t^{2}+\phi^{2}}}\; \cosh^{-1}\left[\frac{\sqrt{t^{2}+\phi^{2}}}{A} \right] \\
\pm \displaystyle i\; \frac{1}{\cos \theta}\; \frac{A}{\sqrt{t^{2}+\phi^{2}}}\; \cosh^{-1}\left[\frac{\sqrt{t^{2}+\phi^{2}}}{A} \right] & \pm \displaystyle i\; \frac{1}{\sin \theta}\; \frac{A}{\sqrt{t^{2}+\phi^{2}}}\; \cosh^{-1}\left[\frac{\sqrt{t^{2}+\phi^{2}}}{A} \right] & \pm \displaystyle i & 0 & \displaystyle \frac{A}{\sqrt{t^{2}+\phi^{2}}}\; \cosh^{-1}\left[\frac{\sqrt{t^{2}+\phi^{2}}}{A} \right] \\
0 & 0 & 0 & 0 & 0
\end{array}
\right)
\label{eq_11}
\end{equation}}
\noindent
where the product result is the same when working with the {\bf A} case (real helicoid and imaginary catenoid) instead of {\bf D}, where $\lambda$ is equal with 4 for the {\bf D} case, or equal with 5 for the {\bf A} case \cite{deus1}.

With the $\displaystyle \frac{t}{\phi}=-\frac{\phi_{k}}{t_{k}}$ correlation between catenoidal and helicoidal coordinates and with the transformations $\displaystyle t_{FRW}=\textnormal{arctg} \left(\frac{t_{k}}{\phi_{k}}\right)$ and $\displaystyle r=\sqrt{t^{2}+\phi^{2}}$ to FRW coordinates, we can write:

{\tiny \begin{equation}
\!\!\!\!\!\!\!\!\!\!\!\!\!\!\!\!\!\!\!\!\!\!\!\!\!AA'=\lambda\cdot \left(
\begin{array}{ccccc}
1 & -\displaystyle \textnormal{tg}\; t_{FRW} & \displaystyle \sinh \theta\; \frac{\displaystyle \sin t_{FRW}}{t_{FRW}} & 0 & \mp \displaystyle i\; \sin t_{FRW} \\
\mp \displaystyle i\; \frac{1}{\textnormal{tg}\; t_{FRW}}\sqrt{\displaystyle \frac{1}{\textnormal{tg}\; t_{FRW}\; \textnormal{tg}\; \theta}} & \pm \displaystyle i\; \sqrt{\displaystyle \frac{1}{\textnormal{tg}\; t_{FRW}\; \textnormal{tg}\; \theta}} & \mp \displaystyle i\; \sinh \theta\; \frac{\displaystyle \cos t_{FRW}}{t_{FRW}} & 0 & - \displaystyle \cos t_{FRW} \sqrt{\displaystyle \frac{1}{\textnormal{tg}\; t_{FRW}\; \textnormal{tg}\; \theta}} \\
\displaystyle \frac{1}{\sinh \theta}\; \frac{t_{FRW}}{\displaystyle \sin t_{FRW}} & - \displaystyle \frac{1}{\sinh \theta}\; \frac{t_{FRW}}{\displaystyle \cos t_{FRW}} & 1 & 0 & \mp \displaystyle i\; \frac{1}{\sinh \theta}\; t_{FRW} \\
\displaystyle \pm i\; \displaystyle \frac{1}{\sinh \theta}\; \frac{t_{FRW}}{\displaystyle \sin t_{FRW}} & \displaystyle \mp i\; \displaystyle \frac{1}{\sinh \theta}\; \frac{t_{FRW}}{\displaystyle \cos t_{FRW}} & \pm i & 0 & \displaystyle \frac{1}{\sinh \theta}\; t_{FRW} \\
0 & 0 & 0 & 0 & 0
\end{array}
\right)
\label{eq_12}
\end{equation}}
\noindent
and:

{\tiny \begin{equation}
\!\!\!\!\!\!\!\!\!\!\!\!\!\!\!\!\!\!\!\!\!\!\!\!\!\!\!\!\!\!\!\!\!\!\!\!\!\!\!\!\!\!\!\!\!\!\!\!\!\!\!\!\!\!\!\!\!\!\!\!\!\!\!\!\!\!\!\!\!\!\!BB'=\lambda\cdot \left(
\begin{array}{ccccc}
1 & \displaystyle \frac{1}{\textnormal{tg}\; \theta} & \displaystyle \cos \theta\; \left( \frac{r}{A}\right)\; \cosh \left(\frac{r}{A}\right) & 0 & \mp \displaystyle i\; \cos \theta \\
\pm \displaystyle i\; \textnormal{tg}\; \theta\; \sqrt{\displaystyle \frac{1}{\textnormal{tg}\; t_{FRW}\; \textnormal{tg}\; \theta}} & \pm \displaystyle i\; \sqrt{\displaystyle \frac{1}{\textnormal{tg}\; t_{FRW}\; \textnormal{tg}\; \theta}} & \pm \displaystyle i\; \sin \theta\; \left(\frac{r}{A}\right)\; \cosh \left(\frac{r}{A}\right)\; \sqrt{\displaystyle \frac{1}{\textnormal{tg}\; t_{FRW}\; \textnormal{tg}\; \theta}} & 0 & \displaystyle \sin \theta\; \sqrt{\displaystyle \frac{1}{\textnormal{tg}\; t_{FRW}\; \textnormal{tg}\; \theta}} \\
\displaystyle \frac{1}{\cos \theta}\; \left(\frac{A}{r}\right)\; \cosh^{-1} \left(\frac{r}{A}\right) & \displaystyle \frac{1}{\sin \theta}\; \left(\frac{A}{r}\right)\; \cosh^{-1} \left(\frac{r}{A}\right) & 1 & 0 & \mp \displaystyle i\; \left(\frac{A}{r}\right) \cosh^{-1} \left(\frac{r}{A}\right) \\
\displaystyle \pm i\; \frac{1}{\cos \theta}\; \left(\frac{A}{r}\right)\; \cosh^{-1} \left(\frac{r}{A}\right) & \pm i\; \displaystyle \frac{1}{\sin \theta}\; \left(\frac{A}{r}\right)\; \cosh^{-1} \left(\frac{r}{A}\right) & \pm i & 0 & \displaystyle \left(\frac{A}{r}\right)\; \cosh^{-1} \left(\frac{r}{A}\right) \\
0 & 0 & 0 & 0 & 0
\end{array}
\right)
\label{eq_13}
\end{equation}}

For {\bf D} situation, at the point $\circledS$ we must have equivalence between $AA'$ and $BB'$. From this equality results:

\begin{equation}
\begin{array}{cccc}
\displaystyle \textnormal{tg}\; \theta=-\frac{1}{\textnormal{tg}\; t_{FRW}} \\
\displaystyle \cos \theta=\sin t_{FRW} \\
\displaystyle \sin \theta=-\cos t_{FRW} \\
\displaystyle \sinh \theta=\frac{r\; t_{FRW}}{A}\; \cosh \left(\frac{r}{A}\right)
\end{array}
\label{eq_14}
\end{equation}

Apart from that, we saw that in the ergosphere it is valid the relation (in {\it r} and $t_{FRW}$ coordinates):

\begin{equation}
A^{2}\cosh^{-2}\left(\frac{r}{A} \right)+t_{FRW}^{2}=0 \; ,
\label{eq_15}
\end{equation}
\noindent
which can be written as:

\begin{equation}
\cosh^{-1} \left(\frac{r}{A} \right)=i\; \frac{t_{FRW}}{A} \; .
\label{eq_16}
\end{equation}

Also, for having observable four-dimensional (into BHIG-FRW spacetime) effects:

\begin{equation}
r=i\; t_{FRW} \; .
\label{eq_17}
\end{equation}

With the (\ref{eq_14}), (\ref{eq_16}) and (\ref{eq_17}) substitutions in $C\stackrel{not}{=}AA'=BB'$ we get:

\begin{equation}
C=\lambda\cdot \left(
\begin{array}{ccccc}
1 & i\; \tanh r & \displaystyle -i\; \sinh r & 0 & \mp \displaystyle \sinh r \\
\pm i\; \displaystyle \frac{1}{\tanh r} & \mp 1 & \pm \displaystyle \cosh r & 0 & - \displaystyle i\; \cosh r \\
i\; \displaystyle \frac{1}{\sinh r} & - \displaystyle \frac{1}{\cosh r} & 1 & 0 & \mp \displaystyle i \\
\mp \displaystyle \frac{1}{\sinh r} & \mp i\; \displaystyle \frac{1}{\cosh r} & \pm i & 0 & 1 \\
0 & 0 & 0 & 0 & 0
\end{array}
\right)
\label{eq_18}
\end{equation}
where $c^{\alpha}_{\beta}\; c^{\beta}_{\alpha}=\pm\lambda$, ($\lambda=$ 4 or 5).

We note now with $C'$ the matrix {\it C} without the zero rows and columns and without the number of spacetime dimensions $\lambda$:

\begin{equation}
C'=\left(
\begin{array}{cccc}
1 & i\; \tanh r & \displaystyle -i\; \sinh r & \mp \displaystyle \sinh r \\
\pm i\; \displaystyle \frac{1}{\tanh r} & \mp 1 & \pm \displaystyle \cosh r & - \displaystyle i\; \cosh r \\
i\; \displaystyle \frac{1}{\sinh r} & - \displaystyle \frac{1}{\cosh r} & 1 & \mp \displaystyle i \\
\mp \displaystyle \frac{1}{\sinh r} & \mp i\; \displaystyle \frac{1}{\cosh r} & \pm i & 1 \\
\end{array}
\right)
\label{eq_19}
\end{equation}
which, in the case of:

\begin{equation}
\begin{array}{cc}
\sinh^{-1}r \rightarrow \cosh r \\
\cosh^{-1}r \rightarrow \sinh r
\end{array}
\label{eq_20}
\end{equation}
\noindent
and, after that, $r=0$ (pointlike particle), becomes:

\begin{equation}
C'=\left(
\begin{array}{cccc}
1 & i & 0 & 0 \\
\pm i & \mp 1 & \pm 1 & - i \\
i & 0 & 1 & \mp i \\
\mp 1 & 0 & \pm i & 1 \\
\end{array}
\right)
\label{eq_21}
\end{equation}

Until now we neglected the rotation between the fifth and the third coordinates, from DEUS system to the external's observer system of coordinates. In (\ref{eq_21}) this rotation translates in an interchanging between the elements $(c')^{1}_{3}\leftrightarrow (c')^{2}_{4}$, $(c')^{1}_{4}\leftrightarrow (c')^{2}_{3}$, $(c')^{3}_{3}\leftrightarrow (c')^{4}_{4}$ and $(c')^{3}_{4}\leftrightarrow (c')^{4}_{3}$ (the fourth column was excluded previously). Keeping only the down signs in (\ref{eq_21}), the new matrix will be:

\begin{equation}
C''=\left(
\begin{array}{cccc}
1 & i & -i & -1 \\
-i & 1 & 0 & 0 \\
i & 0 & 1 & -i \\
1 & 0 & i & 1 \\
\end{array}
\right)
\label{eq_22}
\end{equation}
\noindent
where $(c'')^{\alpha}_{\beta}$ form a group.

For $x_{BHIG-FRW}^{\alpha}\rightarrow x_{DEUS}^{\sigma}$ we have the transformation $x_{DEUS}^{\sigma}=C'' x_{BHIG-FRW}^{\alpha}$. We define an infinitesimal transformation $x_{DEUS}^{\sigma}=\left(I+dC'' \right)x_{BHIG-FRW}^{\alpha}$. Then:

\begin{equation}
dC''=\left(
\begin{array}{cccc}
0 & i & -i & -1 \\
-i & 0 & 0 & 0 \\
i & 0 & 0 & -i \\
1 & 0 & i & 0 \\
\end{array}
\right)
\label{eq_23}
\end{equation}
\noindent
where $det(I+dC'')=1$.

If the above infinitesimal element is the infinitesimal element for the SU(2) group then we should be able to write:

\begin{equation}
dC''=\left(
\begin{array}{cc}
i\; c_{1} & c_{2}+i\; c_{3} \\
-c_{2}+i\; c_{3} & -i\; c_{1}
\end{array}
\right)
\label{eq_24}
\end{equation}

This proves to be right for:

\begin{equation}
c_{1}=\left( \begin{array}{cc}
0 & 1 \\
-1 & 0 \\
\end{array} \right) \;\;\; ; \;\;\;
c_{2}=\left( \begin{array}{cc}
-i & -\displaystyle \frac{1}{2} \\
-\displaystyle \frac{1}{2} & 0 \\
\end{array} \right) \;\;\; ; \;\;\;
c_{3}=\displaystyle \frac{1}{2} \left( \begin{array}{cc}
0 & i \\
-i & 0 \\
\end{array} \right)
\label{eq_25}
\end{equation}

In this case, the group generators are:

\begin{equation}
\begin{array}{ccc}
\displaystyle X^{1}=\left(
\begin{array}{cc}
0 & 1 \\
-1 & 0 \\
\end{array} \right) i\; {\cal X}^{1}\frac{\partial}{\partial {\cal X}^{1}}-
\left(
\begin{array}{cc}
0 & 1 \\
-1 & 0 \\
\end{array} \right) i\; {\cal X}^{2}\frac{\partial}{\partial {\cal X}^{2}} \\
\displaystyle X^{2}=\left(
\begin{array}{cc}
-i & -\displaystyle \frac{1}{2} \\
-\displaystyle \frac{1}{2} & 0 \\
\end{array} \right) {\cal X}^{2}\frac{\partial}{\partial {\cal X}^{1}}-
\left(
\begin{array}{cc}
-i & -\displaystyle \frac{1}{2} \\
-\displaystyle \frac{1}{2} & 0 \\
\end{array} \right) {\cal X}^{1}\frac{\partial}{\partial {\cal X}^{2}} \\
\displaystyle X^{3}=\frac{1}{2} \left( \begin{array}{cc}
0 & i \\
-i & 0 \\
\end{array} \right) i\; {\cal X}^{2}\frac{\partial}{\partial {\cal X}^{1}}+
\frac{1}{2} \left( \begin{array}{cc}
0 & i \\
-i & 0 \\
\end{array} \right) i\; {\cal X}^{1}\frac{\partial}{\partial {\cal X}^{2}}
\label{eq_26}
\end{array}
\end{equation}

If we write:

\begin{equation}
X^{1}=\left( \begin{array}{cc}
X_{a}^{1} & X_{b}^{1} \\
X_{c}^{1} & X_{d}^{1} \\
\end{array} \right) \;\;\; ; \;\;\;
X^{2}=\left( \begin{array}{cc}
X_{a}^{2} & X_{b}^{2} \\
X_{c}^{2} & X_{d}^{2} \\
\end{array} \right) \;\;\; ; \;\;\;
X^{3}=\left( \begin{array}{cc}
X_{a}^{3} & X_{b}^{3} \\
X_{c}^{3} & X_{d}^{3} \\
\end{array} \right)
\label{eq_27}
\end{equation}
\noindent
then the generators satisfy the commutation relations:

\begin{equation}
 \left[X_{\nu}^{i}\; , \; X_{\mu}^{j} \right]=2\; i\; X_{\sigma}^{k} \; ,
\label{eq_28}
\end{equation}
\noindent
where {\it i}, {\it j}, {\it k} = 1, 2, or 3 and $\nu$, $\mu$, $\sigma$ = {\it a}, {\it b}, {\it c}, or {\it d}.
 
\section{SU(3) Group}

We will begin with the matrix:

\begin{equation}
dM=\left(
\begin{array}{cccccccccc}
\textcolor{red}{0} & \textcolor{red}{0} & \textcolor{red}{0} & \textcolor{red}{0} & \textcolor{red}{0} & \textcolor{red}{0} & \textcolor{red}{0} & \textcolor{red}{0} & \textcolor{blue}{0} & \textcolor{blue}{0} \\
\textcolor{red}{0} & 0 & \textcolor{red}{0} & i & \textcolor{red}{-i} & -i & \textcolor{red}{0} & -1 & \textcolor{blue}{-1} & \textcolor{blue}{0} \\
\textcolor{red}{0} & \textcolor{red}{0} & \textcolor{red}{0} & \textcolor{red}{0} & \textcolor{red}{0} & \textcolor{red}{0} & \textcolor{red}{0} & \textcolor{red}{0} & \textcolor{blue}{0} & \textcolor{blue}{0} \\
\textcolor{red}{0} & -i & \textcolor{red}{0} & 0 & \textcolor{red}{0} & 0 & \textcolor{red}{0} & 0 & \textcolor{blue}{0} & \textcolor{blue}{0} \\
\textcolor{red}{0} & \textcolor{red}{i} & \textcolor{red}{0} & \textcolor{red}{0} & \textcolor{red}{0} & \textcolor{red}{0} & \textcolor{red}{0} & \textcolor{red}{-i} & \textcolor{blue}{-i} & \textcolor{blue}{0} \\
\textcolor{red}{0} & i & \textcolor{red}{0} & 0 & \textcolor{red}{0} & 0 & \textcolor{red}{0} & -i & \textcolor{blue}{-i} & \textcolor{blue}{0} \\
\textcolor{red}{0} & \textcolor{red}{0} & \textcolor{red}{0} & \textcolor{red}{0} & \textcolor{red}{0} & \textcolor{red}{0} & \textcolor{red}{0} & \textcolor{red}{0} & \textcolor{blue}{0} & \textcolor{blue}{0} \\
\textcolor{red}{0} & 1 & \textcolor{red}{0} & 0 & \textcolor{red}{i} & i & \textcolor{red}{0} & 0 & \textcolor{blue}{0} & \textcolor{blue}{0} \\
\textcolor{blue}{0} & \textcolor{blue}{1} & \textcolor{blue}{0} & \textcolor{blue}{0} & \textcolor{blue}{i} & \textcolor{blue}{i} & \textcolor{blue}{0} & \textcolor{blue}{0} & \textcolor{blue}{0} & \textcolor{blue}{0} \\
\textcolor{blue}{0} & \textcolor{blue}{0} & \textcolor{blue}{0} & \textcolor{blue}{0} & \textcolor{blue}{0} & \textcolor{blue}{0} & \textcolor{blue}{0} & \textcolor{blue}{0} & \textcolor{blue}{0} & \textcolor{blue}{0} \\
\end{array}
\right)
\label{eq_29}
\end{equation}

The infinitesimal element of SU(3) is constructed as a mixing of three self-similarity levels, where the basic structure for each level is the one of the infinitesimal element $dC''$ (the SU(2) representation). In (\ref{eq_29}), the black elements are from level three of self-similarity, the red elements from level four of self-similarity and the blue elements from the fifth coordinate of two of the self-similarity levels.

Because from the FRW Universe we are able to see {\bf only} to a complete level four (particle and wave, or helicoid and catenoid), level five of self-similarity appears, in the infinitesimal element for SU(3), only in the last two rows and columns, which means that we are able to observe only the quark wave representation (catenoids). The quark particle representation (quark-helicoids contained in the catenoid of level four quark wave representation) and the sub-quark BHIG representation of the fifth self-similarity level are not observable.

The first row and column (excepting the blue elements) are coming from the $x^{1}$ coordinate of level four of self-similarity; The second row and column (excepting the red elements from level four, and the blue level five's elements) are coming from the $x^{1}$ coordinate of level three of self-similarity; The third row and column (excepting the blue level five's elements) are coming from the $x^{2}$ coordinate of level four of self-similarity; The fourth row and column (excepting the red elements from level four, and the blue level five's elements) are coming from the $x^{2}$ coordinate of three of self-similarity; The fifth row and column (excepting the blue level five's elements) are coming from the $x^{3}$ coordinate of level four of self-similarity; The sixth row and column (excepting the red elements from level four, and the blue level five's elements) are coming from the $x^{3}$ coordinate of three of self-similarity; The seventh row and column (excepting the blue le!
 vel five's elements) are coming from the $x^{4}$ coordinate of level four of self-similarity; The eighth row and column (excepting the red elements from level four, and the blue level five's elements) are coming from the $x^{4}$ coordinate of three of self-similarity; The ninth row and column are coming from the $(x^{5})_{level\; four}\rightarrow (x^{4})_{level\; five}$ coordinate; The tenth row and column are coming from the $(x^{5})_{level\; three}\rightarrow (x^{4})_{level\; two}$ coordinate.

For keeping into discussion only the self-similarity levels at which the effects take place, we will have to exclude from the (\ref{eq_29}) matrix the last row and column, the superior self-similarity level not affecting the effect perception, as is already contained in the space of the observer. This is also the reason for which we excluded the fifth line and column in the SU(2) previous study.

The new matrix for the four-dimensional SU(3) infinitesimal element is:

\begin{equation}
dM'=\left(
\begin{array}{ccccccccc}
0 & 0 & 0 & 0 & 0 & 0 & 0 & 0 & 0 \\
0 & 0 & 0 & i & -i & -i & 0 & -1 & -1 \\
0 & 0 & 0 & 0 & 0 & 0 & 0 & 0 & 0 \\
0 & -i & 0 & 0 & 0 & 0 & 0 & 0 & 0 \\
0 & i & 0 & 0 & 0 & 0 & 0 & -i & -i \\
0 & i & 0 & 0 & 0 & 0 & 0 & -i & -i \\
0 & 0 & 0 & 0 & 0 & 0 & 0 & 0 & 0 \\
0 & 1 & 0 & 0 & i & i & 0 & 0 & 0 \\
0 & 1 & 0 & 0 & i & i & 0 & 0 & 0 \\
\end{array}
\right)
\label{eq_30}
\end{equation}

For SU(3), the infinitesimal element must be possible to be written as:

\begin{equation}
dM'=\left(
\begin{array}{cccc}
i\; \lambda_{1} & \lambda_{2}+i\; \lambda_{3} & \lambda_{4}+i\; \lambda_{5} \\
-\lambda_{2}+i\; \lambda_{3} & i\; \lambda_{6} & \lambda_{7}+i\; \lambda_{8} \\
-\lambda_{4}+i\; \lambda_{5} & -\lambda_{7}+\lambda_{8} & -i\; \lambda_{1}-i\; \lambda_{6} \\
\end{array}
\right)
\label{eq_31}
\end{equation}

The equivalence between (\ref{eq_30}) and (\ref{eq_31}) occurs when:

\begin{equation}
\begin{array}{cc}
\lambda_{1}=\lambda_{6}=\lambda_{8}=\left( \begin{array}{ccc}
0 & 0 & 0 \\
0 & 0 & 0 \\
0 & 0 & 0 \\
\end{array} \right) \; ; \;
\lambda_{7}=\left( \begin{array}{ccc}
0 & 0 & 0 \\
0 & -i & -i \\
0 & -i & -i \\
\end{array} \right) \; ; \;
\lambda_{5}=\displaystyle \frac{1}{2}\left( \begin{array}{ccc}
0 & 0 & 0 \\
0 & 0 & i \\
0 & -i & 0 \\
\end{array} \right) \\
\lambda_{4}=\left( \begin{array}{ccc}
0 & 0 & 0 \\
0 & -1 & \displaystyle -\frac{1}{2} \\
0 & \displaystyle -\frac{1}{2} & 0 \\
\end{array} \right) \; ; \;
\lambda_{3}=\left( \begin{array}{ccc}
0 & \displaystyle -\frac{1}{2} & 0 \\
\displaystyle \frac{1}{2} & 0 & \displaystyle -\frac{1}{2} \\
0 & \displaystyle \frac{1}{2} & 0 \\
\end{array} \right) \; ; \;
\lambda_{2}=\left( \begin{array}{ccc}
0 & \displaystyle \frac{i}{2} & 0 \\
\displaystyle \frac{i}{2} & -i & \displaystyle -\frac{i}{2} \\
0 & \displaystyle -\frac{i}{2} & 0 \\
\end{array} \right)
\end{array}
\label{eq_32}
\end{equation}
\noindent
which are the SU(3) generators as they result from the DEUS model.

\section{Conclusions}

From the local charts of the hyper-surfaces crossed by an event that travels from the external observer's spacetime toward the interior of the DEUS object we derived the infinitesimal element for the DEUS object symmetry, which proved to be the SU(2) infinitesimal element for a DEUS object with all its self-similar content evaporated, or the SU(3) infinitesimal element for a DEUS object containing another "visible" self-similar DEUS level.

In a future paper we will show that the "flavor" and the "color" SU(3) symmetries are "unified" into one SU(3) symmetry having the (\ref{eq_32}) generators, where $\lambda_{1}$, $\lambda_{6}$ and $\lambda_{8}$ describe gluon interactions and some of the other $\lambda$'s quark interactions.

At level three of self-similarity (and also in SU(2) describing it) we see, from the external FRW spacetime, only what remains after the evaporation of its level four (described by SU(3)) DEUS content. A consequence of having level four of self-similarity contained in level three and, also, the SU(2) representation as a simplified SU(3) representation, is the impossibility of having SU(2) $\otimes$ SU(3) or U(1) $\otimes$ SU(2)$\otimes$ SU(3), this requiring either, for the external observer, a same spatial-temporal scale of the two levels under discussion, the model and the observed interaction range interdicting that, or, placing ourselves at that level (being {\bf five-dimensional} and {\bf included} in the catenoidal hyper-surface of self-similarity level three containing the level four DEUS objects).

The unification with the gravitation is also possible, but as a local unification, at each DEUS level separately taken. We can have, for example, U(1) $\otimes$ SU(2) $\otimes$ {\it Gravitation} at the scale at which the corresponding graviton acts \cite{deus9} ($10^{-22}-10^{-23}$ meters and inside the non-evaporated level three DEUS object), the scale at which U(1) $\otimes$ SU(3) $\otimes$ {\it Gravitation} occurs being much lower than that. Again, the evaporated pre-Big Bang DEUS object (level one) is having a different scale than the other contained levels, making possible the global unification only at the beginning or at the end of the FRW Universe.

In FRW spacetime, the massless fields resulting from level one (pre-Big Bang DEUS object) interact with the other levels of self-similarity expelled at the collapse of level one (in a comparable way as for the previously described \cite{deus3,deus4,deus5,deus6,deus7} level's two black hole-electromagnetic radiation interaction), but non-evaporated, being possible to talk about U(1) $\otimes$ SU(2) or U(1) $\otimes$ SU(3).

\ack
Manuscript registered at the Astronomical Institute, Romanian Academy, record No. 259 from 04/19/07.

\end{document}


\title[D.E.U.S.]{I: Five-Dimensional Manifolds and World-Lines}

\begin{abstract}
The goal of this paper is to give a possible theoretical approach for a five-dimensional black hole internal geometry by using self-similar minimal surfaces for the representation of the three-dimensional projections of the considered manifolds. Knowing that for each minimal surface it must exist a conjugate surface and an associate family capable to describe both these surfaces, we construct the supplementary dimensions for which the partial derivatives of our conjugate manifolds (catenoid and helicoid) satisfy the Cauchy-Riemann equations. After constructing the five-dimensional coordinate system and {\it ansatz} for each of the resulting particular manifold we conclude by giving the physical meanings hidden behind the spacetime geometry presented in this paper.
\end{abstract}


\section{Introduction}

In a classic approach, it is well known that the Einstein equations are the mathematical description of the way in which the mass-energy generates curvature and so the geometry of the spacetime. Geometric and topological methods in which the geometry precedes physics and are {\it a posteriori} linked to the physical reality are not something new in theory. On one side we have the study of Calabi-Yau manifolds, derived from physics at energies or lengths not reached yet, which describe the string compactification with unbroken supersymmetry \cite{green}. On the other side we have the Topological Quantum Field Theories as Seiberg-Witten theory derived from the Donaldson's work on smooth manifolds \cite{iga,donald} or Chern-Simons gauge theory in which the observables are knot and link invariants \cite{labastida}. The way in which the DEUS model is constructed is by having, instead of a stress-energy tensor or an action, an {\it ab initio} fundamental geometry of spacetime and a unique transformation of coordinates from the catenoid-helicoid coordinates to a system of coordinates that proves to be orthogonal. This is possible if we consider that this geometry contains (and is contained by) the geometry of the external observer and then proving that the cosmological parameters of this external observer are the ones of a FRW spacetime.
In $(t,\phi)$ and $(t_{k},\phi_{k})$ the DEUS object exhibits, for an external observer in $(r,t_{FRW})$ coordinates, Black Hole properties as known from literature \cite{chandra,novi,wald,weinberg}, while in $(r,t_{FRW})$ the DEUS object's helicoid {\bf is} the FRW (globally) or the Minkowski (locally) spacetime.
This paper will set the fundament of a five-dimensional structure (e.g., Black Hole) departing from the trivial three-dimensional minimal surface representation of the catenoid and its conjugate surface, the helicoid. The second section is dedicated to the construction of a five-dimensional coordinate system that satisfies the Cauchy-Riemann equations, while the last two sections will concentrate on resulted spacetime geometry and its composing manifolds (ansatz, properties and correlations). 

\section{Catenary and Catenoid: A Short History}

Galileo was the first to investigate the catenary which he mistook for parabola. Galileo's suggestion that a heavy rope would hang in the shape of a parabola was disproved by Jungius in 1669, but the true shape of the "chain-curve", the catenary, was not found until 1690-1691, when Huygens, Leibniz and John Bernoulli replied to a challenge by James Bernoulli. The name was first used by Huygens in a letter to Leibniz in 1690. In 1691 James Bernoulli obtained its true form and gave some of its properties. David Gregory, wrote a comprehensive treatise on the "catenarian" in 1697.

Minimal surfaces are surfaces that connect given curves in space and have the least area. This is one simple way to define minimal surfaces. The minimal surface called catenoid, first discovered by Jean Baptiste Meusnier in 1776, is the surface of the revolution of a catenary. In nature it can be seen as soap films, by bending wires into two parallel circles and dipping them into soap solution. The soap film between the two circles is the catenoid.

A very interesting and actual application for the catenoid is the Penning and Paul trap used for the storage of charged particles nearly at rest in space, in the same way in which a storage ring is used for the storage of relativistic, highly charged ions.

\begin{figure}
\centering
\leavevmode
\includegraphics[width=7.0cm,height=7.0cm]{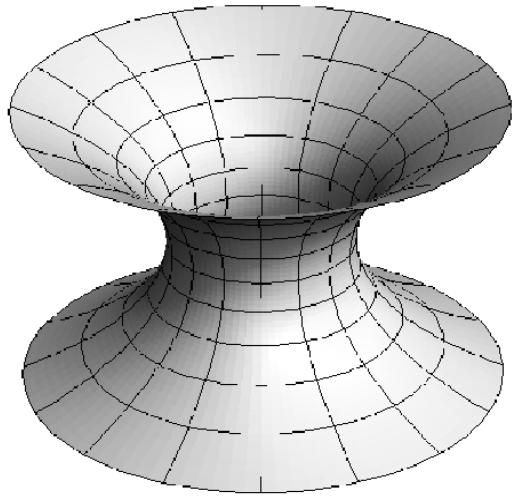}
\caption{Catenoid}
\label{catenoid}
\end{figure}
\begin{figure}
\centering
\leavevmode
\includegraphics[width=6.0cm,height=9.0cm]{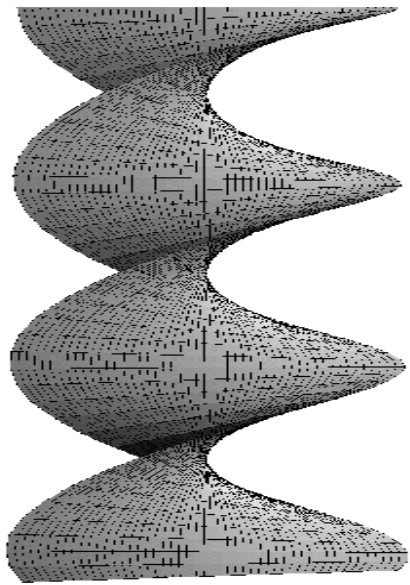}
\caption{Helicoid}
\label{helicoid}
\end{figure}

\section{Gravitational Instantons from Minimal Surfaces}

Gravitational instantons which are given by hyper-K\"{a}hler metrics were studied intensively in the framework of supergravity and M-theory as well as Seiberg-Witten theory \cite{ketov,gib,seiberg}. Generally speaking, the gravitational instantons are given by regular complete metrics with Euclidian signature and self-dual curvature which implies that they satisfy the vacuum Einstein field equations \cite{hawking,gib1,gib2,yau,ati}. The simplest gravitational instantons are obtained from the Schwarzschild-Kerr and Taub-NUT solutions by analytically continuing them to the Euclidian sector \cite{aliev}.

It was observed \cite{nut} that minimal surfaces in Euclidian space can be used in the construction of instanton solutions, even for the case of Yang-Mills instantons \cite{com}. For every minimal surface in three-dimensional Euclidian space there exists a gravitational instanton which is an exact solution of the Einstein field equations with Euclidian signature and self-dual curvature. If the surface is defined by the Monge ansatz $\phi=\phi(x,t)$, then the metric establishing this correspondence is given by \cite{aliev}:

\begin{equation}
\begin{array}{ll}
ds^{2} =& \frac{\displaystyle 1}{\displaystyle \sqrt{1+\phi_{t}^{2}+\phi_{x}^{2}}}[(1+\phi_{t}^{2}) (dt^{2}+dy^{2}) + (1+\phi_{x}^{2}) (dx^{2}+dz^{2}) +\\ &+ 2\phi_{t}\phi_{x}(dt\; dx+dy\; dz)]\; .
\end{array}
\label{eq_e}
\end{equation}

The Einstein field equations reduce to the classical equation \cite{one}:

\begin{equation}
\label{eq_f}
\left(1+\phi_{x}^{2}\right) \phi_{tt}-2\phi_{t}\phi_{x}\phi_{tx}+\left(1+\phi_{t}^{2}\right) \phi_{xx}=0
\end{equation}
\noindent
governing minimal surfaces in $R^{3}$. This solution seems to generate many gravitational instantons, like the {\bf helicoid} (Fig. \ref{helicoid}) defined by $\phi(x,t)$ = arctg$(x/t)$, {\bf catenoid} (Fig. \ref{catenoid}) defined by $\phi(x,t)$ = cosh$^{-1}\left(\sqrt{x^{2}+t^{2}}\right)$, or {\bf Scherk surface} given as $\phi(x,t)$ = log$\left({\rm cos}\; x/{\rm cos}\; t\right)$.  Gravitational instantons that follow from this construction will admit at least two commuting Killing vectors $\partial_{y}$, $\partial_{z}$, which implies that they may be the complete non-compact Ricci-flat K\"{a}hler metrics that have been considered in the context of cosmic strings \cite{gre,gib3}. The general gravitational instanton metric that results from Weierstrass' general local solution for minimal surfaces has been constructed in \cite{aliev1} using the correspondence (\ref{eq_e}) between minimal surfaces and gravitational instantons. But an important theorem by Bernstein states that {\bf there are no non-trivial solutions of (\ref{eq_f}) which are defined on the whole ({\it x}, {\it t}) plane} \cite{osse}. From the above examples it is also clear that the solutions are singular at some points. These singularities are not easy to remove and, therefore, one should be careful in calling these solutions gravitational instantons, which are supposed to be complete non-singular solutions.

Within the framework of Dirichlet {\it p}-branes Gibbons \cite{gib4}, considering the Born-Infeld theory BIon particles as ends of strings intersecting the brane and ignoring gravity effects, obtains topologically non-trivial electrically neutral catenoidal solutions looking like two {\it p}-branes joined by a throat. His general solution is a non-singular deformation of the catenoid if the charge is not too large and a singular deformation of the BIon solution for charges above that limit. Performing a duality rotation Gibbons \cite{gib4} obtained a monopole solution, the BPS limit being a solution of the abelian Bogolmol'nyi equations. This situation resembles that of sub and super extreme black-brane solutions of the supergravity theories. He also shows that some specific Lagrangians submanifolds may be regarded as supersymmetric configurations consisting of {\it p}-branes at angles joined by throats which are the sources of global monopoles. These catenoids are strikingly similar to the Einstein-Rosen bridges (surfaces of constant time) that one encounters in classical super-gravity solutions representing Black Holes or black {\it p}-branes. 

\section{Associate Family of Catenoid and Helicoid}

Among the fundamental observations in the minimal surface theory is that every minimal surface comes in a family of minimal surfaces, the so-called associate family or Bonnet family \cite{kar}. 
Until now there are known just few associate families of minimal surfaces \cite{http}:
\begin{itemize}
\item the catenoid-helicoid associate family containing the catenoid and the helicoid, classic embedded minimal surface without periodicities. 
\item the P-G-D associate family  which contains the Schwarz P, the G (Gyroid) and and the D (Diamond) triply periodic minimal surfaces (periodic in three directions).
\item the Scherk associate family containing the Scherk's first (classic doubly periodic) and second (classic singly-periodic) minimal surfaces.
\item Scherk Torii associate family which contains the Schwarz first and second Torii minimal surfaces.
\end{itemize}
All the above families, excepting the catenoid-helicoid family, contain periodic surfaces (for example, a Schwarz surface is connected at three of its ends with similar Schwarz surfaces, only the resulting surface being associate with the G or P minimal surface of the family), all the composing surfaces coexisting in the considered Euclidian space or in the general $R^{n}$. So, from logical point of view is more natural to consider that, if we intend to construct a geometry of a spacetime based on minimal surfaces our first choice must be the simplest possible, or, in other words, to take the catenoid and helicoid as the minimal surfaces of interest. This choice is favorable also for the simplicity of the mathematical description of the catenoid and the helicoid.
With the catenoid given by:

\begin{equation}
C(u,v) = \begin{pmatrix}
{\cos v\; \cosh u \cr
\sin v\; \cosh u \cr
u}
\end{pmatrix}
\label{eq_g}
\end{equation}
\noindent
and the helicoid by:

\begin{equation}
H(u,v) = \begin{pmatrix}
{\sin v\; \sinh u \cr
-\; \cos v\; \sinh u \cr
v}
\end{pmatrix}
\label{eq_h}
\end{equation}
\noindent
we can obtain the associate family $F^{\varphi}(u,v)$ of both minimal surfaces:

\begin{equation}
F^{\varphi}(u,v)={\rm cos}\; \varphi\cdot C(u,v)+{\rm sin}\; \varphi\cdot H(u,v)
\label{eq_i}
\end{equation}
\noindent

The minimal surfaces represented graphs of the functions $u: \Omega\in R^{n}\rightarrow R$ and $v: \Omega\in  R^{n}\rightarrow R$ \cite{dorf}.

The parameter $\varphi \in [0,2\pi]$ is the family parameter. For $\varphi=\pi/2$ the surface is called the {\it conjugate} of the surface with $\varphi=0$, and $\varphi=\pi$ leads to a point mirror image. The helicoid is called the {\it conjugate surface} of the catenoid and, in general, each pair of surfaces  $F^{\varphi}$ and $F^{\varphi+\frac{\pi}{2}}$ are conjugate to each other.

In (\ref{eq_g}) we may have either $u=A\; {\rm cosh}^{-1}\left(\displaystyle \frac{\sqrt{t^{2}+\phi^{2}}}{A} \right)$, either $u=\theta$ with $\theta\in [0, 2\pi]$. In the same way, in (\ref{eq_h}) we may have either $v={\rm arctg}\left(\displaystyle \frac{t}{\phi}\right)$, either $v=\theta$ where $\theta$ is the same as before.

The conjugate surfaces and the associate family display the following properties \cite{kar}: 

\begin{enumerate}
\item The surface normals at points corresponding to an arbitrary point ($u_{0}, v_{0}$) in the domain are identical, $N_{F^{\varphi}}(u_{0}, v_{0})=N_{C}(u_{0}, v_{0})=N_{H}(u_{0}, v_{0})$;
\item The partial derivatives fulfill the following correspondences:

\begin{equation}
F_{u}^{\varphi}(u_{0}, v_{0})={\rm cos}\; \varphi\cdot C_{u}(u_{0}, v_{0})-{\rm sin}\; \varphi\cdot C_{v}(u_{0}, v_{0})\nonumber
\end{equation}

\begin{equation}
F_{v}^{\varphi}(u_{0}, v_{0})={\rm sin}\; \varphi\cdot C_{u}(u_{0}, v_{0})+{\rm cos}\; \varphi\cdot C_{v}(u_{0}, v_{0})
\label{eq_j}
\end{equation}
\noindent
in particular, the partials of catenoid and helicoid satisfy the Cauchy-Riemann equations:

\begin{eqnarray}
C_{u}(u_{0}, v_{0})=H_{v}(u_{0}, v_{0})\nonumber\\
C_{v}(u_{0}, v_{0})=-\; H_{u}(u_{0}, v_{0})\; ;
\label{eq_k}
\end{eqnarray}
\item If a minimal patch is bounded by a straight line, then its conjugate patch is bounded by a planar symmetry line and vice versa. This can be seen in the catenoid-helicoid case, where the planar meridians of the catenoid correspond to the straight lines of the helicoid;
\item Since at every point the length and the angle between the partial derivatives are identical for the surface and its conjugate (both surfaces are isomeric) we have as a result, that the angles at corresponding boundary vertices of surface and conjugate surface are identical.
\end{enumerate}

In a complex ($C^{3}$ or $C^{n}$) analytical description of minimal surfaces the associate family is an easy concept. It is a basic fact in complex analysis that every harmonic map is the real part of a complex holomorphic map. The three coordinate functions of a minimal surface in Euclidian space $R^{3}$ are harmonic maps. Therefore, there exist three other harmonic maps which define together with the original coordinate functions three holomorphic functions, or {\bf a single complex vector-valued function}. The real part of this function is the original minimal surface, the imaginary part is the conjugate surface, and the projections in-between define surfaces of the associate family.

Let now rewrite the catenoid as:

\begin{equation}
C(u_{1},v_{1}) = \begin{pmatrix}
{\cos v_{1}\; \cosh u_{1} \cr
\sin v_{1}\; \sinh u_{1}}
\end{pmatrix}
\label{eq_l}
\end{equation}
\noindent
and the helicoid as:

\begin{equation}
H(u_{2},v_{2}) = \begin{pmatrix}
{\sin v_{2}\; \sinh u \cr
-\; \cos v_{2}\; \sinh u}
\end{pmatrix}
\label{eq_m}
\end{equation}
\noindent

The correspondence between (\ref{eq_g}) and (\ref{eq_l}) may be given trough $u=A\; {\rm cosh}^{-1}\left(\displaystyle \frac{u_{1}}{A} \right)$, function which can generate the catenoid, where $u_{1}=\sqrt{t^{2}+\phi^{2}}$ and $v_{1}=\theta$ with $\theta \in [0, 2\pi]$. Here $\phi$ is a spatial coordinate and {\it t} a temporal coordinate.
In the same way, the correspondence between (\ref{eq_h}) and (\ref{eq_m}) is given through $v=v_{2}={\rm arctg}\left(\displaystyle \frac{t}{\phi}\right)$, the function which generates the helicoid, and $u_{2}={\rm sinh}\; u$.

With $u_{1}=\sqrt{t^{2}+\phi^{2}}$, $v_{1}=\theta$, $u_{2}={\rm sinh}\; \theta$, $v_{2}={\rm arctg}\left(\displaystyle \frac{t}{\phi} \right)$ in (\ref{eq_g}) and (\ref{eq_h}):

\begin{equation}
C(u_{1},v_{1}) = \begin{pmatrix}
{u_{1}\; \cos v_{1} \cr
u_{1}\; \sin v_{1} \cr
A\; \cosh^{-1}\left(\displaystyle \frac{u_{1}}{A} \right)}
\end{pmatrix}
\label{eq_n}
\end{equation}
\begin{equation}
H(u_{2},v_{2}) = \begin{pmatrix}
{u_{2}\; \sin v_{2} \cr
-\; u_{2}\; \cos v_{2} \cr
v_{2}}
\end{pmatrix}
\label{eq_o}
\end{equation}

In this way it will be possible to construct two distinct surfaces, the bridge between them being the associate family $F^{\varphi}(u,v)$. We will have the $x^{\mu}$ coordinates of the associate family as function of $u_{1}$, $v_{1}$, $u_{2}$ and $v_{2}$ using (\ref{eq_n}) and (\ref{eq_o}) in (\ref{eq_i}) and an angle $\chi$ (for an object moving inside the associate family surface from the catenoid to the helicoid and back, but never getting out from it; {\it C} is the radius of this motion):

\begin{equation}
\left\{ \begin{array}{lll}
x^{1}=\left(u_{1}\; \textnormal{cos}\; v_{1}\; \textnormal{cos}\; \varphi + u_{2}\; \textnormal{sin}\; v_{2}\; \textnormal{sin}\; \varphi \right) \cdot \textnormal{cos}\; \chi \\
x^{2}=\left(u_{1}\; \textnormal{sin}\; v_{1}\; \textnormal{cos}\; \varphi - u_{2}\; \textnormal{cos}\; v_{2}\; \textnormal{sin}\; \varphi \right)\cdot C \\
x^{3}=\left[ A\; \textnormal{cosh}^{-1}\left(\displaystyle \frac{u_{1}}{A} \right)\; \textnormal{cos}\; \varphi + v_{2}\; \textnormal{sin}\; \varphi \right]\cdot \textnormal{sin}\; \chi
\end{array} \right.
\label{eq_p}
\end{equation}

\section{Cauchy-Riemann Equations}

We will derive a fourth and fifth coordinate for the associate family in the way in which to have the Cauchy-Riemann equations satisfied for the five-dimensional tensor $\hat{x}$:

\begin{equation}
\frac{\partial \hat{x}}{\partial u} 
\left| {\tiny \begin{array}{llll} \\ \\ \\
\; \\
\varphi_{1}=0\; \textnormal{or}\; 2\pi \\
\chi_{1}=0\; \textnormal{or}\; 2\pi \\
C=1
\end{array}} \right.
=\frac{\partial \hat{x}}{\partial v} 
\left| {\tiny \begin{array}{llll} \\ \\ \\
\; \\
\varphi_{2}=-\; \frac{\pi}{2}\; \textnormal{or}\; \frac{\pi}{2} \\
\chi_{2}=-\; \frac{\pi}{2}\; \textnormal{or}\; \frac{\pi}{2} \\
C=1
\end{array}} \right.
\label{eq_r}
\end{equation}
\begin{equation}
\frac{\partial \hat{x}}{\partial v} 
\left| {\tiny \begin{array}{llll} \\ \\ \\
\; \\
\varphi_{1}=0\; \textnormal{or}\; 2\pi \\
\chi_{1}=0\; \textnormal{or}\; 2\pi \\
C=1
\end{array}} \right.
=-\; \frac{\partial \hat{x}}{\partial u} 
\left| {\tiny \begin{array}{llll} \\ \\ \\
\; \\
\varphi_{2}=-\; \frac{\pi}{2}\; \textnormal{or}\; \frac{\pi}{2} \\
\chi_{2}=-\; \frac{\pi}{2}\; \textnormal{or}\; \frac{\pi}{2} \\
C=1
\end{array}} \right.
\label{eq_s}
\end{equation}
\noindent
where $\hat{x} 
\left| {\tiny \begin{array}{llll} \\ \\ \\
\; \\
\varphi_{1}=0\; \textnormal{or}\; 2\pi \\
\chi_{1}=0\; \textnormal{or}\; 2\pi \\
C=1
\end{array}} \right.$ is for the five-dimensional catenoid and $\hat{x} 
\left| {\tiny \begin{array}{llll} \\ \\ \\
\; \\
\varphi_{2}=-\; \frac{\pi}{2}\; \textnormal{or}\; \frac{\pi}{2} \\
\chi_{2}=-\; \frac{\pi}{2}\; \textnormal{or}\; \frac{\pi}{2} \\ 
C=1
\end{array}} \right.$ is for the five-dimensional helicoid. ({\it u}, {\it v}) pair can be any combination from $u=A\; {\rm cosh}^{-1}\left(\displaystyle \frac{\sqrt{t^{2}+\phi^{2}}}{A} \right)$, $u=\theta$, $v={\rm arctg}\left(\displaystyle \frac{t}{\phi}\right)$, $v=\theta$. In order to satisfy the properties of the given conjugate surfaces and of their associate family, the catenoid and helicoid must exist in different spaces or at different times. In the associate family, if the five-dimensional catenoid exists spacelike the five-dimensional helicoid must be timelike and the opposite. One can chose $x^{4}$ and $x^{5}$ as:

{\bf A}. \begin{equation}
\left\{ \begin{array}{ll}
x^{4}=D\cdot u_{2}\; \textnormal{sin}\; \varphi + i\; E\cdot u_{1}\; \textnormal{cos}\; \varphi \\
x^{5}=F\cdot v_{1}\; \textnormal{cos}\; \varphi + i\; G\cdot v_{2}\; \textnormal{sin}\; \varphi
\end{array} \right.
\label{eq_t}
\end{equation}

{\bf B}. \begin{equation}
\left\{ \begin{array}{ll}
x^{4}=D\cdot v_{2}\; \textnormal{sin}\; \varphi + i\; E\cdot u_{1}\; \textnormal{cos}\; \varphi \\
x^{5}=F\cdot v_{1}\; \textnormal{cos}\; \varphi + i\; G\cdot u_{2}\; \textnormal{sin}\; \varphi
\end{array} \right.
\label{eq_u}
\end{equation}

{\bf C}. \begin{equation}
\left\{ \begin{array}{ll}
x^{4}=D\cdot u_{2}\; \textnormal{sin}\; \varphi + i\; E\cdot v_{1}\; \textnormal{cos}\; \varphi \\
x^{5}=F\cdot u_{1}\; \textnormal{cos}\; \varphi + i\; G\cdot v_{2}\; \textnormal{sin}\; \varphi
\end{array} \right.
\label{eq_v}
\end{equation}

{\bf D}. \begin{equation}
\left\{ \begin{array}{ll}
x^{4}=D\cdot v_{2}\; \textnormal{sin}\; \varphi + i\; E\cdot v_{1}\; \textnormal{cos}\; \varphi \\
x^{5}=F\cdot u_{1}\; \textnormal{cos}\; \varphi + i\; G\cdot u_{2}\; \textnormal{sin}\; \varphi
\end{array} \right.
\label{eq_w}
\end{equation}
\noindent
where {\it D}, {\it E}, {\it F} and {\it G} are some functions that must be determined.
Cauchy-Riemann conditions (\ref{eq_r}) and (\ref{eq_s}) are satisfied only for {\bf A} and {\bf D} cases (equations (\ref{eq_t}) and (\ref{eq_w})). Let analyze in detail these cases.

\subsection{Case {\bf A}}

From (\ref{eq_r}) and (\ref{eq_s}), in the case with $u=A\; {\rm cosh}^{-1}\left(\displaystyle \frac{\sqrt{t^{2}+\phi^{2}}}{A} \right)$ and $v={\rm arctg}\left(\displaystyle \frac{t}{\phi}\right)$ ({\bf catenoid-helicoid associate hyper-surface}), results that {\it E} = {\it G} = 0. This means that (\ref{eq_t}) is reduced to:

\begin{equation}
\left\{ \begin{array}{ll}
x^{4}=D\cdot u_{2}\; \textnormal{sin}\; \varphi = D\cdot \textnormal{sinh}\; \theta\; \textnormal{sin}\; \varphi \\
x^{5}=F\cdot v_{1}\; \textnormal{cos}\; \varphi = F\cdot \theta\; \textnormal{cos}\; \varphi
\end{array} \right.
\label{eq_x}
\end{equation}

When $u=A\; {\rm cosh}^{-1}\left(\displaystyle \frac{\sqrt{t^{2}+\phi^{2}}}{A} \right)$ and $v=\theta$ and $\textnormal{cos}\; \varphi =\pm 1$, $\textnormal{sin}\; \varphi =0$ we will be in the situation when we will have only the five-dimensional {\bf catenoid}. Because from the Cauchy-Riemann conditions (\ref{eq_r}) and (\ref{eq_s}) results that:

\begin{equation}
-\; 2i\; E = D\cdot \textnormal{cosh}\; \theta
\label{eq_y}
\end{equation}
\begin{equation}
F = i\; G\cdot \frac{1}{\sqrt{t^{2}+\phi^{2}}}\; \frac{\textnormal{cosh}^2\left(\displaystyle \frac{\sqrt{t^{2}+\phi^{2}}}{A}\right)} {\textnormal{sinh}\left(\displaystyle \frac{\sqrt{t^{2}+\phi^{2}}}{A}\right)}\; \left(\frac{\phi}{t}-\frac{t}{\phi} \right)
\label{eq_ys}
\end{equation}
\noindent
the supplementary coordinates will be:
\begin{equation}
\left\{ \begin{array}{ll}
x^{4}=\pm\; i\; E\cdot u_{1} = \pm\; i\; E\cdot \sqrt{t^{2}+\phi^{2}} \\
x^{5}=\pm\; F\cdot v_{1} = \pm\; F\cdot \theta
\end{array} \right.
\label{eq_z}
\end{equation}
\noindent
or:

\begin{equation}
\left\{ \begin{array}{ll}
x^{4}=\mp\; \displaystyle \frac{1}{2}\; D\cdot \textnormal{cosh}\; \theta\; u_{1} = \mp\; \displaystyle \frac{1}{2}\; D\cdot \textnormal{cosh}\; \theta\; \sqrt{t^{2}+\phi^{2}} \\
x^{5}=\pm\; \left[i\; G\cdot \displaystyle \frac{1}{\sqrt{t^{2}+\phi^{2}}}\; \frac{\textnormal{cosh}^2\left(\displaystyle \frac{\sqrt{t^{2}+\phi^{2}}}{A}\right)} {\textnormal{sinh}\left(\displaystyle \frac{\sqrt{t^{2}+\phi^{2}}}{A}\right)}\; \left(\frac{\phi}{t}-\frac{t}{\phi} \right)\right]\; \theta
\end{array} \right.
\label{eqa_a}
\end{equation}
\noindent
or:

\begin{equation}
\left\{ \begin{array}{ll}
x^{4}=\pm\; i\; E\cdot u_{1} = \pm\; i\; E\cdot \sqrt{t^{2}+\phi^{2}} \\
x^{5}=\pm\; \left[i\; G\cdot \displaystyle \frac{1}{\sqrt{t^{2}+\phi^{2}}}\; \frac{\textnormal{cosh}^2\left(\displaystyle \frac{\sqrt{t^{2}+\phi^{2}}}{A}\right)} {\textnormal{sinh}\left(\displaystyle \frac{\sqrt{t^{2}+\phi^{2}}}{A}\right)}\; \left(\frac{\phi}{t}-\frac{t}{\phi} \right)\right]\; \theta
\end{array} \right.
\label{eqa_b}
\end{equation}
\noindent
or:

\begin{equation}
\left\{ \begin{array}{ll}
x^{4}= \mp\; \displaystyle \frac{1}{2}\; D\cdot \textnormal{cosh}\; \theta\; \sqrt{t^{2}+\phi^{2}} \\
x^{5}=\pm\; F\cdot \theta
\end{array} \right.
\label{eqa_c}
\end{equation}

When $\textnormal{cos}\; \varphi =0$ and $\textnormal{sin}\; \varphi =\pm\; 1$, for $u=\theta$ and $v={\rm arctg}\left(\displaystyle \frac{t}{\phi}\right)$ we will have a only the five-dimensional {\bf helicoid}. From (\ref{eq_r}) and (\ref{eq_s}) results that:

\begin{equation}
F = 2i\; G
\label{eqa_d}
\end{equation}
\begin{equation}
i\; E\cdot \sqrt{t^{2}+\phi^{2}}\; \left(\frac{t}{\phi}-\frac{\phi}{t} \right)= -\; D\cdot \textnormal{cosh}\; \theta\; ,
\label{eqa_e}
\end{equation}
\noindent
which means that:

\begin{equation}
\left\{ \begin{array}{lll}
x^{4}= \mp\; \left[ i\; \displaystyle \frac{E}{\textnormal{cosh}\; \theta}\; \sqrt{t^{2}+\phi^{2}}\; \left(\displaystyle \frac{t}{\phi}-\frac{\phi}{t} \right)\right]\; u_{2} = \\
\; \; \; \; = \mp\; \left[ i\; \displaystyle \frac{E}{\textnormal{cosh}\; \theta}\; \sqrt{t^{2}+\phi^{2}}\; \left(\displaystyle \frac{t}{\phi}-\frac{\phi}{t} \right)\right]\; \textnormal{sinh}\; \theta  \\
x^{5}=\pm\; i\; G\cdot v_{2} = \pm\; i\; G\cdot \textnormal{arctg}\left(\displaystyle \frac{t}{\phi} \right)
\end{array} \right.
\label{eqa_f}
\end{equation}
\noindent
or:

\begin{equation}
\left\{ \begin{array}{lll}
x^{4}= \pm\; D\cdot u_{2} = \pm\; D\cdot \textnormal{sinh}\; \theta \\
x^{5}=\pm\; \displaystyle \frac{1}{2}\; F\cdot v_{2}  = \pm\; \displaystyle \frac{1}{2}\; F\cdot \textnormal{arctg}\left(\displaystyle \frac{t}{\phi} \right)
\end{array} \right.
\label{eqa_g}
\end{equation}
\noindent
or:

\begin{equation}
\left\{ \begin{array}{lll}
x^{4}= \mp\; \left[i\; \displaystyle \frac{E}{\textnormal{cosh}\; \theta}\; \sqrt{t^{2}+\phi^{2}}\; \left(\displaystyle \frac{t}{\phi}-\frac{\phi}{t} \right)\right]\; u_{2} = \\
\; \; \; \; = \mp\; \left[i\; \displaystyle \frac{E}{\textnormal{cosh}\; \theta}\; \sqrt{t^{2}+\phi^{2}}\; \left(\displaystyle \frac{t}{\phi}-\frac{\phi}{t} \right)\right]\; \textnormal{sinh}\; \theta  \\
x^{5}=\pm\; \displaystyle \frac{1}{2}\; F\cdot v_{2} = \pm\; \displaystyle \frac{1}{2}\; F\cdot \textnormal{arctg}\left(\displaystyle \frac{t}{\phi} \right)
\end{array} \right.
\label{eqa_h}
\end{equation}
\noindent
or:

\begin{equation}
\left\{ \begin{array}{ll}
x^{4}= \pm\; D\cdot u_{2} = \pm\; D\cdot \textnormal{sinh}\; \theta \\
x^{5}=\pm\; i\; G\cdot v_{2} = \pm\; i\; G\cdot \textnormal{arctg}\left(\displaystyle \frac{t}{\phi} \right)
\end{array} \right.
\label{eqa_i}
\end{equation}
\noindent

\subsection{Case {\bf D}}

When $\textnormal{cos}\; \varphi =\pm 1$ and $\textnormal{sin}\; \varphi =0$, for $u=A\; {\rm cosh}^{-1}\left(\displaystyle \frac{\sqrt{t^{2}+\phi^{2}}}{A} \right)$ and $v=\theta$ we will have only the five-dimensional {\bf catenoid}. From (\ref{eq_r}), (\ref{eq_s}) and (\ref{eq_w}) results that:

\begin{equation}
\left\{ \begin{array}{lll}
x^{4}= \pm\; \displaystyle \frac{D}{\sqrt{t^{2}+\phi^{2}}}\; \frac{\textnormal{cosh}^2\left(\displaystyle \frac{\sqrt{t^{2}+\phi^{2}}}{A}\right)} {\textnormal{sinh}\left(\displaystyle \frac{\sqrt{t^{2}+\phi^{2}}}{A}\right)}\; \left(\displaystyle \frac{\phi}{t}-\frac{t}{\phi} \right)\; v_{1} = \\
\; \; \; \; = \pm\; \displaystyle \frac{D}{\sqrt{t^{2}+\phi^{2}}}\; \frac{\textnormal{cosh}^2\left(\displaystyle \frac{\sqrt{t^{2}+\phi^{2}}}{A}\right)}{\textnormal{sinh}\left(\displaystyle \frac{\sqrt{t^{2}+\phi^{2}}}{A}\right)}\; \left(\displaystyle \frac{\phi}{t}-\frac{t}{\phi} \right)\; \theta \\
x^{5}=\pm\; F\cdot u_{1} = \pm\; F\cdot \sqrt{t^{2}+\phi^{2}}
\end{array} \right.
\label{eqa_j}
\end{equation}
\noindent
or:

\begin{equation}
\left\{ \begin{array}{lll}
x^{4}= \pm\; i\; E\cdot v_{1} = \pm\; iE\cdot \theta \\
\displaystyle x^{5}=\mp\; i\; G\cdot \frac{1}{2}\; \textnormal{cosh}\; \theta\; \frac{\textnormal{sinh}\left(\displaystyle \frac{\sqrt{t^{2}+\phi^{2}}}{A}\right)} {\textnormal{cosh}^{2}\left(\displaystyle \frac{\sqrt{t^{2}+\phi^{2}}}{A}\right)}\; u_{1} = \\
\displaystyle \; \; \; \; = \mp\; i\; G\cdot \frac{1}{2}\; \textnormal{cosh}\; \theta\; \frac{\textnormal{sinh}\left(\displaystyle \frac{\sqrt{t^{2}+\phi^{2}}}{A}\right)}{\textnormal{cosh}^{2}\left(\displaystyle \frac{\sqrt{t^{2}+\phi^{2}}}{A}\right)}\; \sqrt{t^{2}+\phi^{2}}
\end{array} \right.
\label{eqa_k}
\end{equation}
\noindent
or:

\begin{equation}
\left\{ \begin{array}{ll}
x^{4}= \pm\; i\; E\cdot v_{1} = \pm\; i\; E\cdot \theta \\
x^{5}=\pm\; F\cdot u_{1} = \pm\; F\cdot \sqrt{t^{2}+\phi^{2}}
\end{array} \right.
\label{eqa_l}
\end{equation}
\noindent
or:

\begin{equation}
\left\{ \begin{array}{llll}
x^{4}= \pm\; \displaystyle \frac{D}{\sqrt{t^{2}+\phi^{2}}}\; \frac{\textnormal{cosh}^2\left(\displaystyle \frac{\sqrt{t^{2}+\phi^{2}}}{A}\right)}{\textnormal{sinh}\left(\displaystyle \frac{\sqrt{t^{2}+\phi^{2}}}{A}\right)}\; \left(\displaystyle \frac{\phi}{t}-\frac{t}{\phi} \right)\; v_{1} = \\
\; \; \; \; = \pm\; \displaystyle \frac{D}{\sqrt{t^{2}+\phi^{2}}}\; \frac{\textnormal{cosh}^2\left(\displaystyle \frac{\sqrt{t^{2}+\phi^{2}}}{A}\right)}{\textnormal{sinh}\left(\displaystyle \frac{\sqrt{t^{2}+\phi^{2}}}{A}\right)}\; \left(\displaystyle \frac{\phi}{t}-\frac{t}{\phi} \right)\; \theta \\
x^{5}=\mp\; i\; G\cdot \displaystyle \frac{1}{2}\; \textnormal{cosh}\; \theta\; \frac{\textnormal{sinh}\left(\displaystyle \frac{\sqrt{t^{2}+\phi^{2}}}{A}\right)}{\textnormal{cosh}^{2}\left(\displaystyle \frac{\sqrt{t^{2}+\phi^{2}}}{A}\right)}\; u_{1} = \\
\; \; \; \; = \mp\; i\; G\cdot \displaystyle \frac{1}{2}\; \textnormal{cosh}\; \theta\; \frac{\textnormal{sinh}\left(\displaystyle \frac{\sqrt{t^{2}+\phi^{2}}}{A}\right)}{\textnormal{cosh}^{2}\left(\displaystyle \frac{\sqrt{t^{2}+\phi^{2}}}{A}\right)}\; \sqrt{t^{2}+\phi^{2}}
\end{array} \right.
\label{eqa_m}
\end{equation}
\noindent

When $\textnormal{cos}\; \varphi =0$ and $\textnormal{sin}\; \varphi =\pm\; 1$, for $u=\theta$ and $v={\rm arctg}\left(\displaystyle \frac{t}{\phi}\right)$ we will have only the five-dimensional {\bf helicoid}. From (\ref{eq_r}) and (\ref{eq_s}) results that:

\begin{equation}
i\; E = 2D
\label{eqa_n}
\end{equation}
\begin{equation}
F\cdot \sqrt{t^{2}+\phi^{2}}\; \left(\frac{t}{\phi}-\frac{\phi}{t} \right)= -\; i\; G\cdot \textnormal{cosh}\; \theta\; ,
\label{eqa_o}
\end{equation}
\noindent
which means that:

\begin{equation}
\left\{ \begin{array}{lll}
x^{4}= \pm\; D\cdot v_{2} = \pm\; D\cdot \textnormal{arctg}\left(\displaystyle \frac{t}{\phi} \right) \\
x^{5}=\mp\; \displaystyle \frac{1}{\textnormal{cosh}\; \theta}\; F\cdot \sqrt{t^{2}+\phi^{2}} \left(\displaystyle \frac{t}{\phi}-\frac{\phi}{t} \right)\; u_{2} = \\
\; \; \; \; = \mp\; \displaystyle \frac{1}{\textnormal{cosh}\; \theta}\; F\cdot \sqrt{t^{2}+\phi^{2}} \left(\displaystyle \frac{t}{\phi}-\frac{\phi}{t} \right)\; \textnormal{sinh}\; \theta
\end{array} \right.
\label{eqa_p}
\end{equation}
\noindent
or:

\begin{equation}
\left\{ \begin{array}{ll}
x^{4}= \pm\; i\; E\; \displaystyle \frac{1}{2}\cdot v_{2} = \pm\; i\; E\; \displaystyle \frac{1}{2}\cdot \textnormal{arctg}\left(\displaystyle \frac{t}{\phi} \right) \\
x^{5}=\pm\; i\; G\cdot u_{2} = \pm\; i\; G\cdot \textnormal{sinh}\; \theta
\end{array} \right.
\label{eqa_r}
\end{equation}
\noindent
or:

\begin{equation}
\left\{ \begin{array}{ll}
x^{4}= \pm\; D\cdot v_{2} = \pm\; D\cdot \textnormal{arctg}\left(\displaystyle \frac{t}{\phi} \right) \\
x^{5}=\pm\; i\; G\cdot u_{2} = \pm\; i\; G\cdot \textnormal{sinh}\; \theta
\end{array} \right.
\label{eqa_s}
\end{equation}
\noindent
or:

\begin{equation}
\left\{ \begin{array}{lll}
x^{4}= \pm\; i\; E\; \displaystyle \frac{1}{2}\cdot v_{2} = \pm\; i\; E\; \displaystyle \frac{1}{2}\cdot \textnormal{arctg}\left(\displaystyle \frac{t}{\phi} \right) \\
x^{5}=\mp\; \displaystyle \frac{1}{\textnormal{cosh}\; \theta}\; F\cdot \sqrt{t^{2}+\phi^{2}} \left(\displaystyle \frac{t}{\phi}-\frac{\phi}{t} \right)\; u_{2} = \\
\; \; \; \; = \mp\; \displaystyle \frac{1}{\textnormal{cosh}\; \theta}\; F\cdot \sqrt{t^{2}+\phi^{2}} \left(\displaystyle \frac{t}{\phi}-\frac{\phi}{t} \right)\; \textnormal{sinh}\; \theta
\end{array} \right.
\label{eqa_t}
\end{equation}
\noindent

The catenoid-helicoid associate hyper-surface ($u=A\; {\rm cosh}^{-1}\left(\displaystyle \frac{\sqrt{t^{2}+\phi^{2}}}{A} \right)$ and $v={\rm arctg}\left(\displaystyle \frac{t}{\phi}\right)$) fulfill the Cauchy-Riemann conditions only for $D=F=0$ which means that (see equation (\ref{eq_w})):

\begin{equation}
\left\{ \begin{array}{lll}
x^{4}= i\; E\cdot v_{1}\; \textnormal{cos}\; \varphi = i\; E\cdot \theta\; \textnormal{cos}\; \varphi \\
x^{5}= i\; G\cdot u_{2}\; \textnormal{sin}\; \varphi = i\; G\cdot \textnormal{sinh}\; \theta\; \textnormal{sin}\; \varphi
\end{array} \right.
\label{eqa_u}
\end{equation}
\noindent

\section{World-Lines}

\subsection{Ergosphere}

For a better image and clarity, even that only the three dimensional projections of our five-dimensional manifolds can be called "catenoids" or "helicoids", we will permit ourselves to use these name conventions also for our multi-dimensional manifolds. Also, in order to differentiate the helicoid coordinates from the catenoids coordinates we will change the notation for the helicoid coordinates from {\it t} to $t_{k}$ and from $\phi$ to $\phi_{k}$ keeping {\it t} and $\phi$ just for the catenoid.
Having (\ref{eq_p}) and (\ref{eq_x}) (or (\ref{eqa_u})) coordinates for the associate hyper-surface we may derive all the unknown quantities with the help the first fundamental form:

$$
g_{\mu \nu}\equiv \frac{\partial \hat{x}}{\partial x^{\mu}}\; \frac{\partial \hat{x}}{\partial x^{\nu}}
$$
\noindent
by solving the system of equations formed with the help of the metric tensor components. Here it is important to say that no matter which choice of pairs for our unknown quantities {\it D}, {\it E}, {\it F}, {\it G}, (e.g., D and G etc.) we are making in our fourth and fifth coordinates, the $g_{\mu \nu}$ expressions must be the same for all four possible choices, this being translated to only one ansatz. From this resulted that all the metric components, excepting $g_{t\phi}$ (or  $g_{t_{k}\phi_{k}}$), $g_{tt}$ (or $g_{t_{k}t_{k}}$) and $g_{\phi \phi}$ (or $g_{\phi_{k} \phi_{k}}$), must be 0. This applies not only for the associate hyper-surface but also for the catenoid and for the helicoid.
Solving the system for the associate hyper-surface ($u=A\; {\rm cosh}^{-1}\left(\displaystyle \frac{\sqrt{t^{2}+\phi^{2}}}{A} \right)$ and $v={\rm arctg}\left(\displaystyle \frac{t}{\phi}\right)$) without any constraints on {\it C} we got some "strong" and some "weak" constraints between $t$, $t_{k}$, $\phi$, $\phi_{k}$, $\theta$ and $\varphi$ coordinates. The strong constraints are:

\begin{equation}
\begin{array}{lll}
\displaystyle \frac{t}{\phi}=-\; \frac{\phi_{k}}{t_{k}} \\
A^{2}\; \textnormal{cosh}^{-\; 2} \left(\displaystyle \frac{\sqrt{t^{2}+\phi^{2}}}{A} \right)+\textnormal{arctg}^{2}\left(\displaystyle \frac{t_{k}}{\phi_{k}} \right)=0 \\
\theta=\textnormal{arctg}\left(\displaystyle \frac{t}{\phi} \right)
\end{array}
\label{eqa_s1}
\end{equation}
\noindent
while the weak ones are:

\begin{equation}
\begin{array}{lll}
\sqrt{t^{2}+\phi^{2}}\; \textnormal{cos}\; \theta\; \textnormal{cos}\; \varphi + \textnormal{sinh}\; \theta\; \textnormal{sin} \left[\textnormal{arctg}\left(\displaystyle \frac{t_{k}}{\phi_{k}} \right) \right]\; \textnormal{sin}\; \varphi=0 \\
\theta\; \textnormal{cosh}\; \theta\; \textnormal{sin}^{2}\; \varphi + \textnormal{sinh}\; \theta\; \textnormal{cos}^{2}\; \varphi =0 \\
\textnormal{cos}\; \varphi =\pm\; \displaystyle \frac{\phi^{2}}{t^{2}}
\end{array}
\label{eqa_t1}
\end{equation}

For the same hyper-surface:

\begin{itemize}
\item In {\bf D} case:
\begin{equation}
\left\{ \begin{array}{lllll}
x^{1}=\left\{\sqrt{t^{2}+\phi^{2}}\; \textnormal{cos}\; \theta\; \textnormal{cos}\; \varphi + \textnormal{sinh}\; \theta\; \textnormal{sin} \left[\textnormal{arctg}\left(\displaystyle \frac{t_{k}}{\phi_{k}} \right) \right]\; \textnormal{sin}\; \varphi \right\} \cdot \textnormal{cos}\; \chi \\
x^{2}=\left\{\sqrt{t^{2}+\phi^{2}}\; \textnormal{sin}\; \theta\; \textnormal{cos}\; \varphi - \textnormal{sinh}\; \theta\; \textnormal{cos} \left[\textnormal{arctg}\left(\displaystyle \frac{t_{k}}{\phi_{k}} \right) \right]\; \textnormal{sin}\; \varphi \right\}\; C_{ergos} \\
x^{3}=\left\{\displaystyle \textnormal{arctg}\left(\frac{t_{k}}{\phi_{k}} \right)\; \textnormal{sin}\; \varphi\; \pm\; i\; \textnormal{arctg}\left(\displaystyle \frac{t_{k}}{\phi_{k}} \right)\; \textnormal{cos}\; \varphi \right\}\cdot \textnormal{sin}\; \chi \\
x^{4}=\pm\; i \sqrt{t^{2}+\phi^{2}}\cdot \theta\; \textnormal{cos}\; \varphi \\
x^{5}=\pm\; i\; \textnormal{sinh}\; \theta\; \textnormal{sin}\; \varphi
\end{array} \right.
\label{eqa_v}
\end{equation}
\noindent
with $E_{ergos}=\pm\; \sqrt{t^{2}+\phi^{2}}$ and $G_{ergos}=\pm\; 1$. With $E_{ergos}$ and $G_{ergos}$ determined it resulted that $C_{ergos}=1$.
\item for {\bf A} case:
\begin{equation}
\left\{ \begin{array}{lllll}
x^{1}=\left\{\sqrt{t^{2}+\phi^{2}}\; \textnormal{cos}\; \theta\; \textnormal{cos}\; \varphi + \textnormal{sinh}\; \theta\; \textnormal{sin} \left[\textnormal{arctg}\left(\displaystyle \frac{t_{k}}{\phi_{k}} \right) \right]\; \textnormal{sin}\; \varphi \right\} \cdot \textnormal{cos}\; \chi \\
x^{2}=\left\{\sqrt{t^{2}+\phi^{2}}\; \textnormal{sin}\; \theta\; \textnormal{cos}\; \varphi - \textnormal{sinh}\; \theta\; \textnormal{cos} \left[\textnormal{arctg}\left(\displaystyle \frac{t_{k}}{\phi_{k}} \right) \right]\; \textnormal{sin}\; \varphi \right\}\; C_{ergos} \\
x^{3}=\left\{\textnormal{arctg}\left(\displaystyle \frac{t_{k}}{\phi_{k}} \right)\; \textnormal{sin}\; \varphi\; \pm\; i\; \textnormal{arctg}\left(\displaystyle \frac{t_{k}}{\phi_{k}} \right)\; \textnormal{cos}\; \varphi \right\}\cdot \textnormal{sin}\; \chi \\
x^{4}=\pm\; i\; \textnormal{sinh}\; \theta\; \textnormal{sin}\; \varphi \\
x^{5}=\pm\; i \sqrt{t^{2}+\phi^{2}}\cdot \theta\; \textnormal{cos}\; \varphi
\end{array} \right.
\label{eqa_va}
\end{equation}
\noindent
with $F_{ergos}=\pm\; i \sqrt{t^{2}+\phi^{2}}$ and $D_{ergos}=\pm\; i$. With $F_{ergos}$ and $D_{ergos}$ determined it resulted that $C_{ergos}=1$
\end{itemize}

By using the corresponding $\varphi$ values for the catenoid and helicoid in the associate hyper-surface (and also the above "strong" and "weak" constraints) we obtained (independently from the $g_{\mu \nu}=0$ equations which gave {\it D}, {\it E}, {\it F} and {\it G}) two possible results for {\it C}, the same for {\bf A} and {\bf D} cases:

\begin{equation}
\begin{array}{lll}
C^{2}_{cat}=1 \; \; \textnormal{and} \; \; C^{2}_{cat}=\frac{\textnormal{sin}\left[\textnormal{arctg}\; \left(\displaystyle \frac{t}{\phi} \right) \right]\; \textnormal{cos}\; \displaystyle \theta}{\textnormal{cos}\left[\textnormal{arctg}\; \left(\displaystyle \frac{t}{\phi} \right) \right]\; \textnormal{sin}\; \displaystyle \theta} \\
C^{2}_{hel}=1 \; \; \textnormal{and} \; \; C^{2}_{hel}=-\; \frac{\textnormal{cos}\left[\textnormal{arctg}\; \left(\displaystyle \frac{t_{k}}{\phi_{k}} \right) \right]\; \displaystyle \textnormal{cos}\; \theta}{\textnormal{sin}\left[\textnormal{arctg}\; \left(\displaystyle \frac{t_{t}}{\phi_{k}} \right) \right]\; \textnormal{sin}\; \displaystyle \theta} \end{array}
\label{eqb_a}
\end{equation}

With all the said above, the ansatz for the associate hyper-surface proved to be:

\begin{equation}
ds^{2}_{ergos}=Z\cdot \frac{\phi_{k}^{2}}{t_{k}^{2}+\phi_{k}^{2}}\; dt_{k}^{2}+Z\cdot \frac{t_{k}^{2}}{t_{k}^{2}+\phi_{k}^{2}}\; d\phi_{k}^{2}-Z\cdot \frac{t_{k}\; \phi_{k}}{t_{k}^{2}+\phi_{k}^{2}}\; dt_{k}\; d\phi_{k}
\label{eqb_b}
\end{equation}
\noindent
where:

\begin{equation}
\begin{array}{lll}
Z = \textnormal{sinh}^{2}\; \theta\; \textnormal{cos}^{2}\left[\textnormal{arctg}\left(\displaystyle \frac{t_{k}}{\phi_{k}} \right) \right]\; \displaystyle \frac{1}{t_{k}^{2}+\phi_{k}^{2}}\; \textnormal{sin}^{2}\; \varphi\; \textnormal{cos}^{2}\; \chi + \\ + \textnormal{sinh}^{2}\; \theta\; \textnormal{sin}^{2}\left[\textnormal{arctg}\left(\displaystyle \frac{t_{k}}{\phi_{k}} \right) \right]\; \displaystyle \frac{1}{t_{k}^{2}+\phi_{k}^{2}}\; \textnormal{sin}^{2}\; \varphi\cdot C^{2} + \\
+ \frac{\textnormal{sinh}^{2}\left(\displaystyle \frac{\sqrt{t^{2}+\phi^{2}}}{A} \right)}{\textnormal{cosh}^{4}\left(\displaystyle \frac{\sqrt{t^{2}+\phi^{2}}}{A} \right)}\; \textnormal{cos}^{2}\; \varphi\; \textnormal{sin}^{2}\; \chi + \displaystyle \frac{1}{t_{k}^{2}+\phi_{k}^{2}}\; \textnormal{sin}^{2}\; \varphi\; \textnormal{sin}^{2}\; \chi
\end{array}
\label{eqb_c}
\end{equation}

When $\theta = 0$ in (\ref{eqb_c}):

\begin{equation}
Z_{eff}\equiv Z_{\theta =0} = \frac{\textnormal{sinh}^{2}\left(\displaystyle \frac{\sqrt{t^{2}+\phi^{2}}}{A} \right)}{\textnormal{cosh}^{4}\left(\displaystyle \frac{\sqrt{t^{2}+\phi^{2}}}{A} \right)}\; \textnormal{cos}^{2}\; \varphi\; \textnormal{sin}^{2}\; \chi + \frac{1}{t_{k}^{2}+\phi_{k}^{2}}\; \textnormal{sin}^{2}\; \varphi\; \textnormal{sin}^{2}\; \chi
\label{eqb_d}
\end{equation}

From now on we will entitle the associate hyper-surface of the catenoid and helicoid with $C=C_{ergos}=C_{cat}=C_{hel}=1$ with the name {\bf ergosphere}. The reasons for choosing this name will be seen from the physical interpretations of the model equations. 

The strong constraint $\displaystyle \frac{t}{\phi}=-\; \frac{\phi_{k}}{t_{k}}$ is having as consequence not only the fact that passing from catenoid to helicoid the spatial coordinate transforms in a temporal coordinate and the temporal in spatial, but also that the ergosphere acts like a "mirror" for the objects passing through it (seen by an external observer situated, as we will see, on a warped external space, conformal with a Friedman-Robertson-Walker space). If the object falling through the ergosphere is seen as continuing its path to the interior of the central object, the external observer will see its fall reversed in time. So, in this case, the falling object will be seen like not falling at all but coming through the external observer (time reversal). Actually, because the external time becomes internal space the external observer will never see the object crossing the ergosphere but "frozen" on it.
The other strong constraint $\theta=\textnormal{arctg}\left(\displaystyle \frac{t}{\phi} \right)$ proved to be valid {\bf only} for $C=1$ ($u=u_{cat}=u_{hel}=v=v_{cat}=v_{hel}$). When $C\neq 1$ it won't be possible to speak anymore of existence of a combined hyper-surface of catenoid and helicoid ($u\neq v$), they existing without their conjugate hyper-surface. Also, the hyper-surfaces for which $C_{cat}$ and $C_{hel}$ are different from 1 will not satisfy the strong condition $\theta=\textnormal{arctg}\left(\displaystyle \frac{t}{\phi} \right)$.

Let us consider now an uni-dimensional, rigid string-like object which rotates in time in $\theta$ plane. Its rotation will generate a helicoidal spacetime sheet. Let denote with $t_{k}$ the temporal coordinate and with $\phi_{k}$ its length. Its five-dimensional coordinates (satisfying the Cauchy-Riemann conditions and following the same rules for their determination) will be:

\begin{itemize}
\item {\bf A} case
\begin{equation}
\left\{ \begin{array}{ll}
x^{1}=x^{2}=x^{3}=x^{4}=0 \\
x^{5}=\pm\; \textnormal{arctg}\left(\displaystyle \frac{t_{k}}{\phi_{k}} \right)
\end{array} \right.
\label{eqb_e}
\end{equation}
\item {\bf D} case
\begin{equation}
\left\{ \begin{array}{ll}
x^{1}=x^{2}=x^{3}=x^{5}=0 \\
x^{4}=\pm\; i\; \textnormal{arctg}\left(\displaystyle \frac{t_{k}}{\phi_{k}} \right)
\end{array} \right.
\label{eqb_f}
\end{equation}
\end{itemize}

The metric for this helicoidal spacetime sheet will be:

\begin{equation}
ds^{2}=-\; \frac{\phi_{k}^{2}}{\left(t_{k}^{2}+\phi_{k}^{2} \right)^{2}}\; dt_{k}^{2}-\frac{t_{k}^{2}}{\left(t_{k}^{2}+\phi_{k}^{2} \right)^{2}}\; d\phi_{k}^{2}+\frac{t_{k}\; \phi_{k}}{\left(t_{k}^{2}+\phi_{k}^{2} \right)^{2}}\; dt_{k}\; d\phi_{k}
\label{eqb_g}
\end{equation}

The first observation is that the {\bf A} case (\ref{eqb_e}) is spacelike, while {\bf D} case (\ref{eqb_f}) is timelike. The second remarque is that the ansatz (\ref{eqb_g}) is similar to the (\ref{eqb_b}) ansatz, with $Z=1$, and with reversed sign. This similarity will be useful when we will describe how our Black Hole will look like and how an observer will follow different hyper-surfaces in its way to the interior. Anticipating a little bit we will name the spacetime generated by this uni-dimensional object "Black Hole Internal Geometry" (BHIG).

Further on we will concentrate on the five-dimensional catenoid ($\textnormal{cos}\; \varphi =\pm\;1$ ; $\textnormal{sin}\; \varphi =0$; $u=A\; {\rm cosh}^{-1}\left(\displaystyle \frac{\sqrt{t^{2}+\phi^{2}}}{A} \right)$ ; $v=\theta$) and the helicoid ($\textnormal{cos}\; \varphi =0$ ; $\textnormal{sin}\; \varphi =\pm\; 1$ ; $u=\theta$ ; $v={\rm arctg}\left(\displaystyle \frac{t}{\phi}\right)$) from the ergosphere ($C=1$), with their corresponding $x^{4}$ and $x^{5}$ for {\bf A} and {\bf D} situations (equations (\ref{eq_z}), (\ref{eqa_a}), (\ref{eqa_b}), (\ref{eqa_c}) for the catenoid and equations (\ref{eqa_f}), (\ref{eqa_g}), (\ref{eqa_h}), (\ref{eqa_i}) for the helicoid in {\bf A} case; equations (\ref{eqa_j}), (\ref{eqa_k}), (\ref{eqa_l}), (\ref{eqa_m}) for the catenoid and equations (\ref{eqa_p}), (\ref{eqa_r}), (\ref{eqa_s}), (\ref{eqa_t}) for the helicoid in the {\bf D} case). For these we derived all the relations between {\it D}, {\it E}, {\it F} and {\it G}. In the {\bf D} case, we obtained for the {\bf catenoid}:

\begin{equation}
\begin{array}{ll}
F_{cat}=\pm\; i \\
D_{cat}=E_{cat}=0
\end{array}
\label{con2}
\end{equation}
\noindent
while for the {\bf helicoid}:

\begin{equation}
\begin{array}{lll}
E_{hel}=\pm\; 2i\; D_{hel} =0 \\
G_{hel}=\pm 1 \\
F_{hel}=0
\end{array}
\label{con1}
\end{equation}

Because the catenoid is not in the same spacetime as the helicoid it is understandable why the catenoid is not consistent anymore with $D=F=0$ of the {\bf D} case ergosphere. However it is consistent with the {\bf A} case ergosphere for which $E=G=0$.
Analogous, in the {\bf A} case we will have:

\begin{equation}
\begin{array}{ll}
E_{cat}=\pm\; 1 \\
F_{cat}=G_{cat}=0
\end{array}
\label{con3}
\end{equation}
\noindent
and:

\begin{equation}
\begin{array}{ll}
F_{hel}=\pm\; 2i\; G_{hel} \\
F_{hel}=G_{hel}=0 \\
D_{hel}=\pm\; i
\end{array}
\label{con4}
\end{equation}

So, for an {\bf A} case ergosphere we will have a {\bf A} case helicoid and a {\bf D} case catenoid and for a {\bf D} case ergosphere we will have a {\bf D} case helicoid and a {\bf A} case catenoid.

The large number of hyper-surfaces which are composing our five dimensional object and their consequences for the ensemble determines us to do the following development.

\subsection{{\bf D} Case}

\subsubsection{Ergosphere's Helicoid ($C=1$ ; $\theta=\textnormal{arctg}\left(\displaystyle \frac{t}{\phi} \right)$)}\ 

\noindent
Parameters:

\begin{equation}
\begin{array}{lll}
E=\pm\; 2i\; D =0 \\
G=\pm 1 \\
F=0
\end{array}
\label{eqb_h}
\end{equation}
\noindent
Coordinates:

\begin{equation}
\left\{ \begin{array}{lllll}
x^{1}=\textnormal{sinh}\; \theta\; \textnormal{sin} \left[\textnormal{arctg}\left(\displaystyle \frac{t_{k}}{\phi_{k}} \right) \right] \\
x^{2}=-\; \textnormal{sinh}\; \theta\; \textnormal{cos} \left[\textnormal{arctg}\left(\displaystyle \frac{t_{k}}{\phi_{k}} \right) \right] \\
x^{3}=\textnormal{arctg}\left(\displaystyle \frac{t_{k}}{\phi_{k}} \right) \\
x^{4}=0\equiv \sqrt{\left(x^{1} \right)^{2}+\left(x^{2} \right)^{2}+\left(x^{5} \right)^{2}} \\
x^{5}=\pm\; i\; \textnormal{sinh}\; \theta
\end{array} \right.
\label{eqb_i}
\end{equation}
\noindent
Ansatz:

\begin{equation}
ds^{2}=Z\cdot \frac{\phi_{k}^{2}}{\left(t_{k}^{2}+\phi_{k}^{2} \right)^{2}}\; dt_{k}^{2}+Z\cdot \frac{t_{k}^{2}}{\left(t_{k}^{2}+\phi_{k}^{2} \right)^{2}}\; d\phi_{k}^{2}-Z\cdot \frac{t_{k}\; \phi_{k}}{\left(t_{k}^{2}+\phi_{k}^{2} \right)^{2}}\; dt_{k}\; d\phi_{k}
\label{eqb_j}
\end{equation}
\noindent
where:
\begin{equation}
Z=\textnormal{cosh}^{2}\theta-\frac{E^{2}}{4}\equiv \textnormal{cosh}^{2}\theta + D^{2}\equiv \textnormal{cosh}^{2}\theta
\label{eqb_k}
\end{equation}

If we make $\theta=0$ in (\ref{eqb_j}) ansatz we see that $ds^{2}_{\theta=0}=-\; ds^{2}_{BHIG}$ meaning that, in the ergosphere, not only that the space and time coordinates switch their roles, but also that the angle $\theta$ "freezes" to 0. It also means that the object that crosses the ergosphere heading to BHIG doesn't violate the causality for the external observer which sees it, as exiting from the ergosphere. We will see in the further development that, if the object is to enter into the BHIG spacetime, it must do it from the helicoid and not from the catenoid.

Another important observation which arrises from the coordinates system is that, from the four-dimensional external observer's perception, $x^{4}$ is a "hidden" spacetime dimension, to this contributing the $x^{1}$, $x^{2}$ and $x^{5}$ dimensions.

\subsubsection{Ergosphere's Catenoid ($C=1$ ; $\theta=\textnormal{arctg}\left(\displaystyle \frac{t}{\phi} \right)$)}\ 

\noindent
Parameters:

\begin{equation}
\begin{array}{ll}
F=\pm\; i \\
D=E=0
\end{array}
\label{eqb_l}
\end{equation}
\noindent
Coordinates:

\begin{equation}
\left\{ \begin{array}{lllll}
x^{1}=\sqrt{t^{2}+\phi^{2}}\; \textnormal{cos}\; \theta \\
x^{2}=\sqrt{t^{2}+\phi^{2}}\; \textnormal{sin}\; \theta \\
x^{3}=A\; \textnormal{cosh}^{-\; 1}\left(\displaystyle \frac{\sqrt{t^{2}+\phi^{2}}}{A} \right) \\
x^{4}=0\equiv \sqrt{\left(x^{1} \right)^{2}+\left(x^{2} \right)^{2}+\left(x^{5} \right)^{2}} \\
x^{5}=\pm\; i\; \sqrt{t^{2}+\phi^{2}}
\end{array} \right.
\label{eqb_m}
\end{equation}
\noindent
Ansatz:

\begin{equation}
ds^{2}=Z\cdot \frac{t^{2}}{t^{2}+\phi^{2}}\; dt^{2}+Z\cdot \frac{\phi^{2}}{t^{2}+\phi^{2}}\; d\phi^{2}+Z\cdot \frac{t\; \phi}{t^{2}+\phi^{2}}\; dt\; d\phi
\label{eqb_n}
\end{equation}
\noindent
where:
\begin{equation}
Z =\frac{\textnormal{sinh}^{2}\left(\displaystyle \frac{\sqrt{t^{2}+\phi^{2}}}{A}\right)}{\textnormal{cosh}^{4}\left(\displaystyle \frac{\sqrt{t^{2}+\phi^{2}}}{A} \right)}+1+F^{2}\equiv \frac{\textnormal{sinh}^{2}\left(\displaystyle \frac{\sqrt{t^{2}+\phi^{2}}}{A}\right)}{\textnormal{cosh}^{4}\left(\displaystyle \frac{\sqrt{t^{2}+\phi^{2}}}{A} \right)}
\label{eqb_n1}
\end{equation}

By comparing the $Z_{eff}$ factor from (\ref{eqb_d}) with {\it Z} from (\ref{eqb_k}) or {\it Z} from (\ref{eqb_n1}) we see that, for a $\chi=\pm\; \pi /2$ and $\varphi$ for catenoid or helicoid, they are equal, meaning again that, in the ergosphere $\theta=0$.
Like in the previous case we observe again that here $x^{4}$ hidden, this happening because of $x^{1}$, $x^{2}$ and $x^{5}$.

\subsubsection{Catenoid with $C=1$ and $\theta\neq \textnormal{arctg}\left(\displaystyle \frac{t}{\phi} \right)$}\ 

\noindent
Parameters:

\begin{equation}
\begin{array}{lll}
F=G=0 \\
E=\pm\; \sqrt{t^{2}+\phi^{2}} \\
D=\pm\; i\; \left(t^{2}+\phi^{2} \right)\; \frac{\displaystyle \textnormal{sinh} \left(\displaystyle \frac{\sqrt{t^{2}+\phi^{2}}}{A} \right)}{\displaystyle \textnormal{cosh}^{2} \left(\displaystyle \frac{\sqrt{t^{2}+\phi^{2}}}{A} \right)}\; \frac{\displaystyle 1}{\displaystyle \frac{\phi}{t}-\frac{t}{\phi}}
\end{array}
\label{eqb_o}
\end{equation}
\noindent
Coordinates:

\begin{equation}
\left\{ \begin{array}{lllll}
x^{1}=\sqrt{t^{2}+\phi^{2}}\; \textnormal{cos}\; \theta \\
x^{2}=\sqrt{t^{2}+\phi^{2}}\; \textnormal{sin}\; \theta \\
x^{3}=A\; \textnormal{cosh}^{-\; 1}\left(\displaystyle \frac{\sqrt{t^{2}+\phi^{2}}}{A} \right) \\
x^{4}=\pm\; i\; \theta \sqrt{t^{2}+\phi^{2}} \\
x^{5}=0
\end{array} \right.
\label{eqb_p}
\end{equation}
\noindent
Ansatz:

\begin{equation}
ds^{2}=Z\cdot \frac{t^{2}}{t^{2}+\phi^{2}}\; dt^{2}+Z\cdot \frac{\phi^{2}}{t^{2}+\phi^{2}}\; d\phi^{2}+Z\cdot \frac{t\; \phi}{t^{2}+\phi^{2}}\; dt\; d\phi
\label{eqb_r}
\end{equation}
\noindent
where:
\begin{equation}
\begin{array}{lll}
Z &=1 + \frac{\displaystyle \textnormal{sinh}^{2} \left(\displaystyle \frac{\sqrt{t^{2}+\phi^{2}}}{A} \right)}{\displaystyle \textnormal{cosh}^{4} \left(\displaystyle \frac{\sqrt{t^{2}+\phi^{2}}}{A} \right)} + F^{2} \equiv \\ &\equiv 1 + \displaystyle \frac{\textnormal{sinh}^{2} \left(\displaystyle \frac{\sqrt{t^{2}+\phi^{2}}}{A} \right)}{\textnormal{cosh}^{4} \left(\displaystyle \frac{\sqrt{t^{2}+\phi^{2}}}{A} \right)} - \displaystyle \frac{1}{4}\; G^{2}\; \textnormal{cosh}^{2}\theta \equiv \\ &\equiv 1 + \displaystyle \frac{\textnormal{sinh}^{2} \left(\displaystyle \frac{\sqrt{t^{2}+\phi^{2}}}{A} \right)}{\textnormal{cosh}^{4} \left(\displaystyle \frac{\sqrt{t^{2}+\phi^{2}}}{A} \right)}
\end{array}
\label{eqb_s}
\end{equation}

Here by contrast with previous cases $x^{5}$ is the one which is hidden.

\subsubsection{Helicoid with $C=1$ and $\theta\neq \textnormal{arctg}\left(\displaystyle \frac{t}{\phi} \right)$}\ 

\noindent
Parameters:

\begin{equation}
\begin{array}{llll}
E=\pm\; 2i\; D=\pm\; 2\sqrt{2}\;i \\
D=\pm\; \sqrt{2} \\
G=\pm 1 \\
F=\pm\; i\; \textnormal{cosh}^{3}\theta\; \displaystyle \frac{1}{\sqrt{t_{k}^{2}+\phi_{k}^{2}}}\; \displaystyle \frac{1}{\displaystyle \frac{t_{k}}{\phi_{k}}-\frac{\phi_{k}}{t_{k}}}
\end{array}
\label{eqb_t}
\end{equation}
\noindent
Coordinates:

\begin{equation}
\left\{ \begin{array}{lllll}
x^{1}=\textnormal{sinh}\; \theta\; \textnormal{sin} \left[\textnormal{arctg}\left(\displaystyle \frac{t_{k}}{\phi_{k}} \right) \right] \\
x^{2}=\textnormal{sinh}\; \theta\; \textnormal{cos} \left[\textnormal{arctg}\left(\displaystyle \frac{t_{k}}{\phi_{k}} \right) \right] \\
x^{3}=\textnormal{arctg}\left(\displaystyle \frac{t_{k}}{\phi_{k}} \right) \\
x^{4}=\pm\; i\; \sqrt{2}\; \textnormal{arctg}\left(\displaystyle \frac{t_{k}}{\phi_{k}} \right)\; \; \textnormal{and}\; \; x^{4}=\mp\; i\; \sqrt{2}\; \textnormal{arctg}\left(\displaystyle \frac{t_{k}}{\phi_{k}} \right) \\
x^{5}=\pm\; i\; \textnormal{sinh}\; \theta\; \; \textnormal{and}\; \; x^{5}=\mp\; i\; \textnormal{sinh}\; \theta
\end{array} \right.
\label{eqb_u}
\end{equation}
\noindent
Ansatz:

\begin{equation}
ds^{2}=Z\cdot \frac{\phi_{k}^{2}}{\left(t_{k}^{2}+\phi_{k}^{2} \right)^{2}}\; dt_{k}^{2}+Z\cdot \frac{t_{k}^{2}}{\left(t_{k}^{2}+\phi_{k}^{2} \right)^{2}}\; d\phi_{k}^{2}-Z\cdot \frac{t_{k}\; \phi_{k}}{\left(t_{k}^{2}+\phi_{k}^{2} \right)^{2}}\; dt_{k}\; d\phi_{k}
\label{eqb_v}
\end{equation}
\noindent
where:
\begin{equation}
Z =\textnormal{cosh}^{2}\theta + D^{2}\equiv \textnormal{cosh}^{2}\theta - \frac{1}{4}\; E^{2}\equiv \textnormal{cosh}^{2}\theta + 2
\label{eqb_w}
\end{equation}

We see that the two different forms for $x^{4}$ and $x^{5}$ permit the existence of four spaces all with the same metric. These are, until now, the first real five-dimensional hyper-surfaces. Actually these hyper-surfaces are not allowing to the external observer to see the BHIG hyper-surface ("no hair" theorem equivalent).

\subsubsection{{\bf Pure Catenoidal Hyper-Surface} ($\theta\neq \textnormal{arctg}\left(\displaystyle \frac{t}{\phi} \right)$)}\ 

\noindent
Parameters:

\begin{equation}
\begin{array}{lll}
C=\pm\; \sqrt{\frac{\textnormal{sin}\left[\textnormal{arctg}\left(\displaystyle \frac{t}{\phi} \right) \right]\; \textnormal{cos}\; \displaystyle \theta}{\textnormal{cos}\left[\textnormal{arctg}\left(\displaystyle \frac{t}{\phi} \right) \right]\; \textnormal{sin}\; \displaystyle \theta}} \\
F=G=0 \\
E=\pm\; \frac{\displaystyle A\; \textnormal{cosh}^{-\; 1}\left(\frac{\sqrt{t^{2}+\phi^{2}}}{A} \right)}{\displaystyle \theta} 
\end{array}
\label{eqb_x}
\end{equation}
\noindent
Coordinates:

\begin{equation}
\left\{ \begin{array}{lllll}
x^{1}=\sqrt{t^{2}+\phi^{2}}\; \textnormal{cos}\; \theta \\
x^{2}=\pm\; \sqrt{t^{2}+\phi^{2}}\; \textnormal{sin}\; \theta\; \sqrt{\frac{\textnormal{sin}\left[\textnormal{arctg}\left(\displaystyle \frac{t}{\phi} \right) \right]\; \textnormal{cos}\; \displaystyle \theta}{\textnormal{cos}\left[\textnormal{arctg}\left(\displaystyle \frac{t}{\phi} \right) \right]\; \textnormal{sin}\; \displaystyle \theta}} \\
x^{3}=A\; \textnormal{cosh}^{-\; 1}\left(\displaystyle \frac{\sqrt{t^{2}+\phi^{2}}}{A} \right) \\
x^{4}=\pm\; i\; A\; \textnormal{cosh}^{-\; 1}\left(\displaystyle \frac{\sqrt{t^{2}+\phi^{2}}}{A} \right) \\
x^{5}=0\equiv \sqrt{\left(x^{3} \right)^{2} + \left(x^{4} \right)^{2}}
\end{array} \right.
\label{eqb_y}
\end{equation}
\noindent
Ansatz:

\begin{equation}
ds^{2}=Z\cdot \frac{t^{2}}{t^{2}+\phi^{2}}\; dt^{2}+Z\cdot \frac{\phi^{2}}{t^{2}+\phi^{2}}\; d\phi^{2}+Z\cdot \frac{t\; \phi}{t^{2}+\phi^{2}}\; dt\; d\phi
\label{eqb_z}
\end{equation}
\noindent
where:
\begin{equation}
\begin{array}{ll}
Z &=\textnormal{cos}^{2}\theta - \frac{\displaystyle \textnormal{sin}^{2} \displaystyle \theta\; \displaystyle \textnormal{cos}^{2} \displaystyle \theta}{\displaystyle \textnormal{sin}^{2} \displaystyle \theta - \displaystyle \frac{A^{2}\; \textnormal{cosh}^{-\; 2}\left(\displaystyle \frac{\sqrt{t^{2}+\phi^{2}}}{A} \right)}{\displaystyle \theta^{2}\left(\displaystyle t^{2}+\phi^{2} \right)}} + \frac{\displaystyle \textnormal{sinh}^{2} \left(\displaystyle \frac{\sqrt{t^{2}+\phi^{2}}}{A} \right)}{\displaystyle \textnormal{cosh}^{4} \left(\displaystyle \frac{\sqrt{t^{2}+\phi^{2}}}{A} \right)} + F^{2} \equiv \\ &\equiv \textnormal{cos}^{2}\theta - \frac{\textnormal{sin}^{2} \displaystyle \theta\; \textnormal{cos}^{2} \displaystyle \theta}{\textnormal{sin}^{2} \displaystyle \theta - \displaystyle \frac{A^{2}\; \textnormal{cosh}^{-\; 2}\left(\displaystyle \frac{\sqrt{t^{2}+\phi^{2}}}{A} \right)}{\displaystyle \theta^{2}\left(\displaystyle t^{2}+\phi^{2} \right)}} + \frac{\displaystyle \textnormal{sinh}^{2} \left(\displaystyle \frac{\sqrt{t^{2}+\phi^{2}}}{A} \right)}{\displaystyle \textnormal{cosh}^{4} \left(\displaystyle \frac{\sqrt{t^{2}+\phi^{2}}}{A} \right)}
\end{array}
\label{eqc_a}
\end{equation}

The hidden dimension for this case is $x^{5}$.

\subsubsection{{\bf Pure Helicoidal Hyper-Surface} ($\theta\neq \textnormal{arctg}\left(\displaystyle \frac{t}{\phi} \right)$)}\ 

\noindent
Parameters:

\begin{equation}
\begin{array}{llll}
C=\pm\; i\; \sqrt{\frac{\textnormal{cos}\left[\textnormal{arctg}\left(\displaystyle \frac{t_{k}}{\phi_{k}} \right) \right]\; \textnormal{cos}\; \displaystyle \theta}{\textnormal{sin}\left[\textnormal{arctg}\left(\displaystyle \frac{t_{k}}{\phi_{k}} \right) \right]\; \textnormal{sin}\; \displaystyle \theta}} \\
D=\pm\; i \\
E=\mp\; 2 \\
F=G=0
\end{array}
\label{eqc_b}
\end{equation}
\noindent
Coordinates:

\begin{equation}
\left\{ \begin{array}{lllll}
x^{1}=\textnormal{sinh}\; \theta\; \textnormal{sin} \left[\textnormal{arctg}\left(\displaystyle \frac{t_{k}}{\phi_{k}} \right) \right] \\
x^{2}=\mp\; i\; \textnormal{sinh}\; \theta\; \textnormal{cos} \left[\textnormal{arctg}\left(\displaystyle \frac{t_{k}}{\phi_{k}} \right) \right]\; \sqrt{\frac{\textnormal{cos}\left[\textnormal{arctg}\left(\displaystyle \frac{t_{k}}{\phi_{k}} \right) \right]\; \textnormal{cos}\; \displaystyle \theta}{\textnormal{sin}\left[\textnormal{arctg}\left(\displaystyle \frac{t_{k}}{\phi_{k}} \right) \right]\; \textnormal{sin}\; \displaystyle \theta}} \\
x^{3}=\textnormal{arctg}\left(\displaystyle \frac{t_{k}}{\phi_{k}} \right) \\
x^{4}=\pm\; i\; \textnormal{arctg}\left(\displaystyle \frac{t_{k}}{\phi_{k}}\right) \\
x^{5}=0\equiv \sqrt{\left(x^{3} \right)^{2} + \left(x^{4} \right)^{2}}
\end{array} \right.
\label{eqc_c}
\end{equation}
\noindent
Ansatz:

\begin{equation}
ds^{2}=Z\cdot \frac{\phi_{k}^{2}}{\left(t_{k}^{2}+\phi_{k}^{2} \right)^{2}}\; dt_{k}^{2}+Z\cdot \frac{t_{k}^{2}}{\left(t_{k}^{2}+\phi_{k}^{2} \right)^{2}}\; d\phi_{k}^{2}-Z\cdot \frac{t_{k}\; \phi_{k}}{\left(t_{k}^{2}+\phi_{k}^{2} \right)^{2}}\; dt_{k}\; d\phi_{k}
\label{eqc_d}
\end{equation}
\noindent
where:
\begin{equation}
\!\!\!\!\!\!\!\!\!\!\!\!\!\!\!\!\!\!\!\!\!\!\!\!\!\!\!\!\!\!\!\!\!\!\!\!\!\!\!\!\!\!\!\!\!\begin{array}{lll}
Z &=\displaystyle \textnormal{sinh}^{2}\theta\; \left\{\textnormal{cos}^{2}\left[\textnormal{arctg}\left(\displaystyle \frac{t_{k}}{\phi_{k}} \right) \right] - \frac{\textnormal{cos}\; \displaystyle \theta}{\textnormal{sin}\; \displaystyle \theta}\; \frac{\textnormal{cos}\left[\textnormal{arctg}\left(\displaystyle \frac{t_{k}}{\phi_{k}} \right) \right]}{\textnormal{sin}^{3}\left[\textnormal{arctg}\left(\displaystyle \frac{t_{k}}{\phi_{k}} \right) \right]}\; \left\{\textnormal{sin}^{2}\left[\textnormal{arctg}\left(\displaystyle \frac{t_{k}}{\phi_{k}} \right) \right] + \displaystyle \frac{1}{2} \right\}^{2} + 1 - \displaystyle \frac{1}{4}\; E^{2} \right\}\equiv \\
&\displaystyle \equiv \textnormal{sinh}^{2}\theta\; \left\{\textnormal{cos}^{2}\left[\textnormal{arctg}\left(\displaystyle \frac{t_{k}}{\phi_{k}} \right) \right] - \frac{\textnormal{cos}\; \displaystyle \theta}{\textnormal{sin}\; \displaystyle \theta}\; \frac{\textnormal{cos}\left[\textnormal{arctg}\left(\displaystyle \frac{t_{k}}{\phi_{k}} \right) \right]}{\textnormal{sin}^{3}\left[\textnormal{arctg}\left(\displaystyle \frac{t_{k}}{\phi_{k}} \right) \right]}\; \left\{\textnormal{sin}^{2}\left[\textnormal{arctg}\left(\displaystyle \frac{t_{k}}{\phi_{k}} \right) \right] + \displaystyle \frac{1}{2} \right\}^{2} + 1 + D^{2} \right\}\equiv \\
&\displaystyle \equiv \textnormal{sinh}^{2}\theta\; \left\{\textnormal{cos}^{2}\left[\textnormal{arctg}\left(\displaystyle \frac{t_{k}}{\phi_{k}} \right) \right] - \frac{\textnormal{cos}\; \displaystyle \theta}{\textnormal{sin}\; \displaystyle \theta}\; \frac{\textnormal{cos}\left[\textnormal{arctg}\left(\displaystyle \frac{t_{k}}{\phi_{k}} \right) \right]}{\textnormal{sin}^{3}\left[\textnormal{arctg}\left(\displaystyle \frac{t_{k}}{\phi_{k}} \right) \right]}\; \left\{\textnormal{sin}^{2}\left[\textnormal{arctg}\left(\displaystyle \frac{t_{k}}{\phi_{k}} \right) \right] + \displaystyle \frac{1}{2} \right\}^{2} \right\}
\end{array}
\label{eqc_e}
\end{equation}

In this particular case, because of $x^{3}$ and $x^{4}$, the hidden dimension for this case is $x^{5}$.

\subsection{{\bf A} Case}

\subsubsection{Ergosphere's Helicoid ($C=1$ ; $\theta=\textnormal{arctg}\left(\displaystyle \frac{t}{\phi} \right)$)}\ 

\noindent
Parameters:

\begin{equation}
\begin{array}{lll}
F=\pm\; 2i\; G \\
F=G=0 \\
D=\pm\; i
\end{array}
\label{eqc_f}
\end{equation}
\noindent
Coordinates:

\begin{equation}
\left\{ \begin{array}{lllll}
x^{1}=\textnormal{sinh}\; \theta\; \textnormal{sin} \left[\textnormal{arctg}\left(\displaystyle \frac{t_{k}}{\phi_{k}} \right) \right] \\
x^{2}=-\; \textnormal{sinh}\; \theta\; \textnormal{cos} \left[\textnormal{arctg}\left(\displaystyle \frac{t_{k}}{\phi_{k}} \right) \right] \\
x^{3}=\textnormal{arctg}\left(\displaystyle \frac{t_{k}}{\phi_{k}} \right) \\
x^{4}=\pm\; i\; \textnormal{sinh}\; \theta \\
x^{5}=0\equiv \sqrt{\left(x^{1} \right)^{2}+\left(x^{2} \right)^{2}+\left(x^{4} \right)^{2}}
\end{array} \right.
\label{eqc_g}
\end{equation}
\noindent
Ansatz:

\begin{equation}
ds^{2}=Z\cdot \frac{\phi_{k}^{2}}{\left(t_{k}^{2}+\phi_{k}^{2} \right)^{2}}\; dt_{k}^{2}+Z\cdot \frac{t_{k}^{2}}{\left(t_{k}^{2}+\phi_{k}^{2} \right)^{2}}\; d\phi_{k}^{2}-Z\cdot \frac{t_{k}\; \phi_{k}}{\left(t_{k}^{2}+\phi_{k}^{2} \right)^{2}}\; dt_{k}\; d\phi_{k}
\label{eqc_h}
\end{equation}
\noindent
where:
\begin{equation}
Z=\textnormal{cosh}^{2}\theta + \frac{1}{4}\; F^{2}\equiv \textnormal{cosh}^{2}\theta
\label{eqc_i}
\end{equation}

As we saw for the same hyper-surface in {\bf D} case, if we make $\theta=0$ in (\ref{eqc_i}) ansatz, $ds^{2}_{\theta=0}=-\; ds^{2}_{BHIG}$ meaning again that the angle $\theta$ "freezes" to 0 inside the ergosphere and that the object that crosses the ergosphere heading to the BHIG spacetime doesn't violate the causality for the external observer. The object can enter into the BHIG spacetime from the helicoid.

The hidden dimension here is not anymore $x^{4}$ as for {\bf D} case, but $x^{5}$. In this way, the {\bf D} case can be seen as a projection of the $x^{4}$ coordinate of the five-dimensional considered hyper-surface, while the {\bf A} case as a $x^{5}$ projection of the same five-dimensional hyper-surface. The remarque that {\bf A} and {\bf D} cases are $x^{4}$ or $x^{5}$ projections of one and the same hyper-surface (the ansatz are the same) will be valid for all the other considered hyper-surfaces. 

\subsubsection{Ergosphere's Catenoid ($C=1$ ; $\theta=\textnormal{arctg}\left(\displaystyle \frac{t}{\phi} \right)$)}\ 

\noindent
Parameters:

\begin{equation}
\begin{array}{ll}
E=\pm\; 1 \\
F=G=0
\end{array}
\label{eqc_j}
\end{equation}
\noindent
Coordinates:

\begin{equation}
\left\{ \begin{array}{lllll}
x^{1}=\sqrt{t^{2}+\phi^{2}}\; \textnormal{cos}\; \theta \\
x^{2}=\sqrt{t^{2}+\phi^{2}}\; \textnormal{sin}\; \theta \\ \\
x^{3}=A\; \textnormal{cosh}^{-\; 1}\left(\displaystyle \frac{\sqrt{t^{2}+\phi^{2}}}{A} \right) \\
x^{4}=\pm\; i\; \sqrt{t^{2}+\phi^{2}} \\
x^{5}=0\equiv \sqrt{\left(x^{1} \right)^{2}+\left(x^{2} \right)^{2}+\left(x^{4} \right)^{2}}
\end{array} \right.
\label{eqc_k}
\end{equation}
\noindent
Ansatz:

\begin{equation}
ds^{2}=Z\cdot \frac{t^{2}}{t^{2}+\phi^{2}}\; dt^{2}+Z\cdot \frac{\phi^{2}}{t^{2}+\phi^{2}}\; d\phi^{2}+Z\cdot \frac{t\; \phi}{t^{2}+\phi^{2}}\; dt\; d\phi
\label{eqc_l}
\end{equation}
\noindent
where:
\begin{equation}
Z =\frac{\textnormal{sinh}^{2}\left(\displaystyle \frac{\sqrt{t^{2}+\phi^{2}}}{A}\right)}{\textnormal{cosh}^{4}\left(\displaystyle \frac{\sqrt{t^{2}+\phi^{2}}}{A} \right)}+1-E^{2}\equiv \frac{\textnormal{sinh}^{2}\left(\displaystyle \frac{\sqrt{t^{2}+\phi^{2}}}{A}\right)}{\textnormal{cosh}^{4}\left(\displaystyle \frac{\sqrt{t^{2}+\phi^{2}}}{A} \right)}
\label{eqc_m}
\end{equation}

As for our {\bf D} case, by comparing $Z_{eff}$ factor from (\ref{eqb_d}) with {\it Z} from (\ref{eqc_i}) or {\it Z} from (\ref{eqc_m}), for $\chi=\pm\; \pi /2$ and $\varphi$ for helicoid or catenoid, we also get equality ($\theta=0$ in the ergosphere).
Here the projection is on $(x^{1}x^{2}x^{3}x^{4})$ plane.

\subsubsection{Catenoid with $C=1$ and $\theta\neq \textnormal{arctg}\left(\displaystyle \frac{t}{\phi} \right)$}\ 

\noindent
Parameters:

\begin{equation}
\begin{array}{lll}
D=E=0 \\
F=\pm\; i\; \sqrt{t^{2}+\phi^{2}} \\
G=\pm\; \left(t^{2}+\phi^{2} \right)\; \displaystyle \frac{\textnormal{sinh} \left(\displaystyle \frac{\sqrt{t^{2}+\phi^{2}}}{A} \right)}{\textnormal{cosh}^{2} \left(\displaystyle \frac{\sqrt{t^{2}+\phi^{2}}}{A} \right)}\; \displaystyle \frac{1}{\displaystyle \frac{\phi}{t}-\frac{t}{\phi}}
\end{array}
\label{eqc_n}
\end{equation}
\noindent
Coordinates:

\begin{equation}
\left\{ \begin{array}{lllll}
x^{1}=\sqrt{t^{2}+\phi^{2}}\; \textnormal{cos}\; \theta \\
x^{2}=\sqrt{t^{2}+\phi^{2}}\; \textnormal{sin}\; \theta \\
x^{3}=A\; \textnormal{cosh}^{-\; 1}\left(\displaystyle \frac{\sqrt{t^{2}+\phi^{2}}}{A} \right) \\
x^{4}=0 \\
x^{5}=\pm\; i\; \theta \sqrt{t^{2}+\phi^{2}}
\end{array} \right.
\label{eqc_o}
\end{equation}
\noindent
Ansatz:

\begin{equation}
ds^{2}=Z\cdot \frac{t^{2}}{t^{2}+\phi^{2}}\; dt^{2}+Z\cdot \frac{\phi^{2}}{t^{2}+\phi^{2}}\; d\phi^{2}+Z\cdot \frac{t\; \phi}{t^{2}+\phi^{2}}\; dt\; d\phi
\label{eqc_p}
\end{equation}
\noindent
where:
\begin{equation}
\begin{array}{ll}
Z =\displaystyle 1 + \frac{\textnormal{sinh}^{2} \left(\displaystyle \frac{\sqrt{t^{2}+\phi^{2}}}{A} \right)}{\textnormal{cosh}^{4} \left(\displaystyle \frac{\sqrt{t^{2}+\phi^{2}}}{A} \right)} - E^{2} &\equiv \displaystyle 1 + \frac{\textnormal{sinh}^{2} \left(\displaystyle \frac{\sqrt{t^{2}+\phi^{2}}}{A} \right)}{\textnormal{cosh}^{4} \left(\displaystyle \frac{\sqrt{t^{2}+\phi^{2}}}{A} \right)} + \displaystyle \frac{1}{4}\; D^{2}\; \textnormal{cosh}^{2}\theta \\ &\equiv \displaystyle 1 + \frac{\textnormal{sinh}^{2} \left(\displaystyle \frac{\sqrt{t^{2}+\phi^{2}}}{A} \right)}{\textnormal{cosh}^{4} \left(\displaystyle \frac{\sqrt{t^{2}+\phi^{2}}}{A} \right)}
\end{array}
\label{eqc_r}
\end{equation}

Here the projection is on $(x^{1}x^{2}x^{3}x^{5})$ plane.

\subsubsection{Helicoid with $C=1$ and $\theta\neq \textnormal{arctg}\left(\displaystyle \frac{t}{\phi} \right)$}\ 

\noindent
Parameters:

\begin{equation}
\begin{array}{llll}
F=\pm\; 2i\; G=\pm\; 2\sqrt{2}\;i \\
G=\pm\; \sqrt{2} \\
D=\pm\; i \\
E=\pm\; \textnormal{cosh}^{3}\theta\; \displaystyle \frac{1}{\displaystyle \sqrt{t_{k}^{2}+\phi_{k}^{2}}}\; \displaystyle \frac{1}{\displaystyle \frac{t_{k}}{\phi_{k}}-\frac{\phi_{k}}{t_{k}}}
\end{array}
\label{eqc_s}
\end{equation}
\noindent
Coordinates:

\begin{equation}
\left\{ \begin{array}{lllll}
x^{1}=\textnormal{sinh}\; \theta\; \textnormal{sin} \left[\textnormal{arctg}\left(\displaystyle \frac{t_{k}}{\phi_{k}} \right) \right] \\
x^{2}=-\; \textnormal{sinh}\; \theta\; \textnormal{cos} \left[\textnormal{arctg}\left(\displaystyle \frac{t_{k}}{\phi_{k}} \right) \right] \\
x^{3}=\textnormal{arctg}\left(\displaystyle \frac{t_{k}}{\phi_{k}} \right) \\
x^{4}=\pm\; i\; \textnormal{sinh}\; \theta\; \; \textnormal{and}\; \; x^{4}=\mp\; i\; \textnormal{sinh}\; \theta \\
x^{5}=\pm\; i\; \sqrt{2}\; \textnormal{arctg}\left(\displaystyle \frac{t_{k}}{\phi_{k}} \right)\; \; \textnormal{and}\; \; x^{5}=\mp\; i\; \sqrt{2}\; \textnormal{arctg}\left(\displaystyle \frac{t_{k}}{\phi_{k}} \right)
\end{array} \right.
\label{eqc_t}
\end{equation}
\noindent
Ansatz:

\begin{equation}
ds^{2}=Z\cdot \frac{\phi_{k}^{2}}{\left(t_{k}^{2}+\phi_{k}^{2} \right)^{2}}\; dt_{k}^{2}+Z\cdot \frac{t_{k}^{2}}{\left(t_{k}^{2}+\phi_{k}^{2} \right)^{2}}\; d\phi_{k}^{2}-Z\cdot \frac{t_{k}\; \phi_{k}}{\left(t_{k}^{2}+\phi_{k}^{2} \right)^{2}}\; dt_{k}\; d\phi_{k}
\label{eqc_u}
\end{equation}
\noindent
where:
\begin{equation}
Z =\textnormal{cosh}^{2}\theta - G^{2}\equiv \textnormal{cosh}^{2}\theta + \frac{1}{4}\; F^{2}\equiv \textnormal{cosh}^{2}\theta - 2
\label{eqc_v}
\end{equation}

We see that the two different forms for $x^{4}$ and $x^{5}$ permit the existence of four hyper-surfaces, all with the same metric, these (as for the {\bf D} case) being the five-dimensional surfaces which interdict to the external observer to see the BHIG hyper-surface ("no hair" theorem equivalent).

$\\$
\subsubsection{{\bf Pure Catenoidal Hyper-Surface} ($\theta\neq \textnormal{arctg}\left(\displaystyle \frac{t}{\phi} \right)$)}\ 

\noindent
Parameters:

\begin{equation}
\begin{array}{lll}
C=\pm\; \sqrt{\frac{\textnormal{sin}\left[\textnormal{arctg}\left(\displaystyle \frac{t}{\phi} \right) \right]\; \textnormal{cos}\; \displaystyle \theta}{\textnormal{cos}\left[\textnormal{arctg}\left(\displaystyle \frac{t}{\phi} \right) \right]\; \textnormal{sin}\; \displaystyle \theta}} \\
D=E=0 \\
F=\pm\; i\; \frac{\displaystyle A\; \textnormal{cosh}^{-\; 1}\left(\displaystyle \frac{\sqrt{t^{2}+\phi^{2}}}{A} \right)}{\displaystyle \theta} 
\end{array}
\label{eqc_w}
\end{equation}
\noindent
Coordinates:

\begin{equation}
\left\{ \begin{array}{lllll}
x^{1}=\sqrt{t^{2}+\phi^{2}}\; \textnormal{cos}\; \theta \\
x^{2}=\pm\; \sqrt{t^{2}+\phi^{2}}\; \textnormal{sin}\; \theta\; \sqrt{\frac{\textnormal{sin}\left[\textnormal{arctg}\left(\displaystyle \frac{t}{\phi} \right) \right]\; \textnormal{cos}\; \displaystyle \theta}{\textnormal{cos}\left[\textnormal{arctg}\left(\displaystyle \frac{t}{\phi} \right) \right]\; \textnormal{sin}\; \displaystyle \theta}} \\
x^{3}=A\; \textnormal{cosh}^{-\; 1}\left(\displaystyle \frac{\sqrt{t^{2}+\phi^{2}}}{A} \right) \\
x^{4}=0\equiv \sqrt{\left(x^{3} \right)^{2} + \left(x^{5} \right)^{2}} \\
x^{5}=\pm\; i\; A\; \textnormal{cosh}^{-\; 1}\left(\displaystyle \frac{\sqrt{t^{2}+\phi^{2}}}{A} \right)
\end{array} \right.
\label{eqc_x}
\end{equation}
\noindent
Ansatz:

\begin{equation}
ds^{2}=Z\cdot \frac{t^{2}}{t^{2}+\phi^{2}}\; dt^{2}+Z\cdot \frac{\phi^{2}}{t^{2}+\phi^{2}}\; d\phi^{2}+Z\cdot \frac{t\; \phi}{t^{2}+\phi^{2}}\; dt\; d\phi
\label{eqc_y}
\end{equation}
\noindent
where:
\begin{equation}
\begin{array}{lll}
Z =\textnormal{cos}^{2}\theta - \frac{\displaystyle \textnormal{sin}^{2}\theta\; \textnormal{cos}^{2}\theta}{\displaystyle \textnormal{sin}^{2}\theta - \frac{A^{2}\; \textnormal{cosh}^{-\; 2}\left(\displaystyle \frac{\sqrt{t^{2}+\phi^{2}}}{A} \right)}{\theta^{2}\left(t^{2}+\phi^{2} \right)}} + \frac{\displaystyle \textnormal{sinh}^{2} \left(\displaystyle \frac{\sqrt{t^{2}+\phi^{2}}}{A} \right)}{\displaystyle \textnormal{cosh}^{4} \left(\displaystyle \frac{\sqrt{t^{2}+\phi^{2}}}{A} \right)} - E^{2} \equiv \\
\equiv \textnormal{cos}^{2}\theta - \frac{\displaystyle \textnormal{sin}^{2}\theta\; \textnormal{cos}^{2}\theta}{\displaystyle \textnormal{sin}^{2}\theta - \frac{A^{2}\; \textnormal{cosh}^{-\; 2}\left(\displaystyle \frac{\sqrt{t^{2}+\phi^{2}}}{A} \right)}{\theta^{2}\left(t^{2}+\phi^{2} \right)}} + \frac{\displaystyle \textnormal{sinh}^{2} \left(\displaystyle \frac{\sqrt{t^{2}+\phi^{2}}}{A} \right)}{\displaystyle \textnormal{cosh}^{4} \left(\displaystyle \frac{\sqrt{t^{2}+\phi^{2}}}{A} \right)}
+ \displaystyle \frac{1}{4}\; D^{2}\; \textnormal{cosh}^{2}\theta \equiv \\
\equiv \textnormal{cos}^{2}\theta - \frac{\displaystyle \textnormal{sin}^{2}\theta\; \textnormal{cos}^{2}\theta}{\displaystyle \textnormal{sin}^{2}\theta - \frac{A^{2}\; \textnormal{cosh}^{-\; 2}\left(\displaystyle \frac{\sqrt{t^{2}+\phi^{2}}}{A} \right)}{\theta^{2}\left(t^{2}+\phi^{2} \right)}} + \frac{\displaystyle \textnormal{sinh}^{2} \left(\displaystyle \frac{\sqrt{t^{2}+\phi^{2}}}{A} \right)}{\displaystyle \textnormal{cosh}^{4} \left(\displaystyle \frac{\sqrt{t^{2}+\phi^{2}}}{A} \right)}
\end{array}
\label{eqc_z}
\end{equation}

The projection is on $(x^{1}x^{2}x^{3}x^{5})$ plane.

\subsubsection{{\bf Pure Helicoidal Hyper-Surface} ($\theta\neq \textnormal{arctg}\left(\displaystyle \frac{t}{\phi} \right)$)}\ 

\noindent
Parameters:

\begin{equation}
\begin{array}{llll}
C=\pm\; i\; \sqrt{\frac{\textnormal{cos}\left[\textnormal{arctg}\left(\displaystyle \frac{t_{k}}{\phi_{k}} \right) \right]\; \textnormal{cos}\; \displaystyle \theta}{\textnormal{sin}\left[\textnormal{arctg}\left(\displaystyle \frac{t_{k}}{\phi_{k}} \right) \right]\; \textnormal{sin}\; \displaystyle \theta}} \\
D=E=0 \\
G=\pm\; 1 \\
F=\pm\; 2\; i
\end{array}
\label{eqd_a}
\end{equation}
\noindent
Coordinates:

\begin{equation}
\left\{ \begin{array}{lllll}
x^{1}=\textnormal{sinh}\; \theta\; \textnormal{sin} \left[\textnormal{arctg}\left(\displaystyle \frac{t_{k}}{\phi_{k}} \right) \right] \\
x^{2}=\mp\; i\; \textnormal{sinh}\; \theta\; \textnormal{cos} \left[\textnormal{arctg}\left(\displaystyle \frac{t_{k}}{\phi_{k}} \right) \right]\; \sqrt{\frac{\textnormal{cos}\left[\textnormal{arctg}\left(\displaystyle \frac{t_{k}}{\phi_{k}} \right) \right]\; \textnormal{cos}\; \displaystyle \theta}{\textnormal{sin}\left[\textnormal{arctg}\left(\displaystyle \frac{t_{k}}{\phi_{k}} \right) \right]\; \textnormal{sin}\; \displaystyle \theta}} \\
x^{3}=\textnormal{arctg}\left(\displaystyle \frac{t_{k}}{\phi_{k}} \right) \\
x^{4}=0\equiv \sqrt{\left(x^{3} \right)^{2} + \left(x^{5} \right)^{2}} \\
x^{5}=\pm\; i\; \textnormal{arctg}\left(\displaystyle \frac{t_{k}}{\phi_{k}}\right)
\end{array} \right.
\label{eqd_b}
\end{equation}
\noindent
Ansatz:

\begin{equation}
ds^{2}=Z\cdot \frac{\phi_{k}^{2}}{\left(t_{k}^{2}+\phi_{k}^{2} \right)^{2}}\; dt_{k}^{2}+Z\cdot \frac{t_{k}^{2}}{\left(t_{k}^{2}+\phi_{k}^{2} \right)^{2}}\; d\phi_{k}^{2}-Z\cdot \frac{t_{k}\; \phi_{k}}{\left(t_{k}^{2}+\phi_{k}^{2} \right)^{2}}\; dt_{k}\; d\phi_{k}
\label{eqd_c}
\end{equation}
\noindent
where:
\begin{equation}
\!\!\!\!\!\!\!\!\!\!\!\!\!\!\!\!\!\!\!\!\!\!\!\!\!\!\!\!\!\!\!\!\!\!\!\!\!\!\!\!\!\!\!\!\!\begin{array}{lll}
Z =\displaystyle \textnormal{sinh}^{2}\theta\; \left\{\textnormal{cos}^{2}\left[\textnormal{arctg}\left(\displaystyle \frac{t_{k}}{\phi_{k}} \right) \right] - \frac{\textnormal{cos}\; \displaystyle \theta}{\textnormal{sin}\; \displaystyle \theta}\; \frac{\textnormal{cos}\left[\textnormal{arctg}\left(\displaystyle \frac{t_{k}}{\phi_{k}} \right) \right]}{\displaystyle \textnormal{sin}^{3}\left[\textnormal{arctg}\left(\displaystyle \frac{t_{k}}{\phi_{k}} \right) \right]}\; \left\{\textnormal{sin}^{2}\left[\textnormal{arctg}\left(\frac{t_{k}}{\phi_{k}} \right) \right] + \frac{1}{2} \right\}^{2} + 1 - G^{2} \right\}\equiv \\
\displaystyle \equiv \textnormal{sinh}^{2}\theta\; \left\{\textnormal{cos}^{2}\left[\textnormal{arctg}\left(\displaystyle \frac{t_{k}}{\phi_{k}} \right) \right] - \frac{\textnormal{cos}\; \displaystyle \theta}{\textnormal{sin}\; \displaystyle \theta}\; \frac{\textnormal{cos}\left[\textnormal{arctg}\left(\displaystyle \frac{t_{k}}{\phi_{k}} \right) \right]}{\displaystyle \textnormal{sin}^{3}\left[\textnormal{arctg}\left(\displaystyle \frac{t_{k}}{\phi_{k}} \right) \right]}\; \left\{\textnormal{sin}^{2}\left[\textnormal{arctg}\left(\displaystyle \frac{t_{k}}{\phi_{k}} \right) \right] + \frac{1}{2} \right\}^{2} + 1 + \frac{1}{4}\; F^{2} \right\}\equiv \\
\displaystyle \equiv \textnormal{sinh}^{2}\theta\; \left\{\textnormal{cos}^{2}\left[\textnormal{arctg}\left(\displaystyle \frac{t_{k}}{\phi_{k}} \right) \right] - \frac{\textnormal{cos}\; \displaystyle \theta}{\textnormal{sin}\; \displaystyle \theta}\; \frac{\textnormal{cos}\left[\textnormal{arctg}\left(\displaystyle \frac{t_{k}}{\phi_{k}} \right) \right]}{\displaystyle \textnormal{sin}^{3}\left[\textnormal{arctg}\left(\displaystyle \frac{t_{k}}{\phi_{k}} \right) \right]}\; \left\{\textnormal{sin}^{2}\left[\textnormal{arctg}\left(\displaystyle \frac{t_{k}}{\phi_{k}} \right) \right] + \frac{1}{2} \right\}^{2} \right\}
\end{array}
\label{eqd_d}
\end{equation}

For this hyper-surface, the projection is on $(x^{1}x^{2}x^{3}x^{5})$ plane.

\section{Conclusions}

\begin{figure*}
\centering
\leavevmode
\includegraphics[width=15.0cm,height=18.0cm]{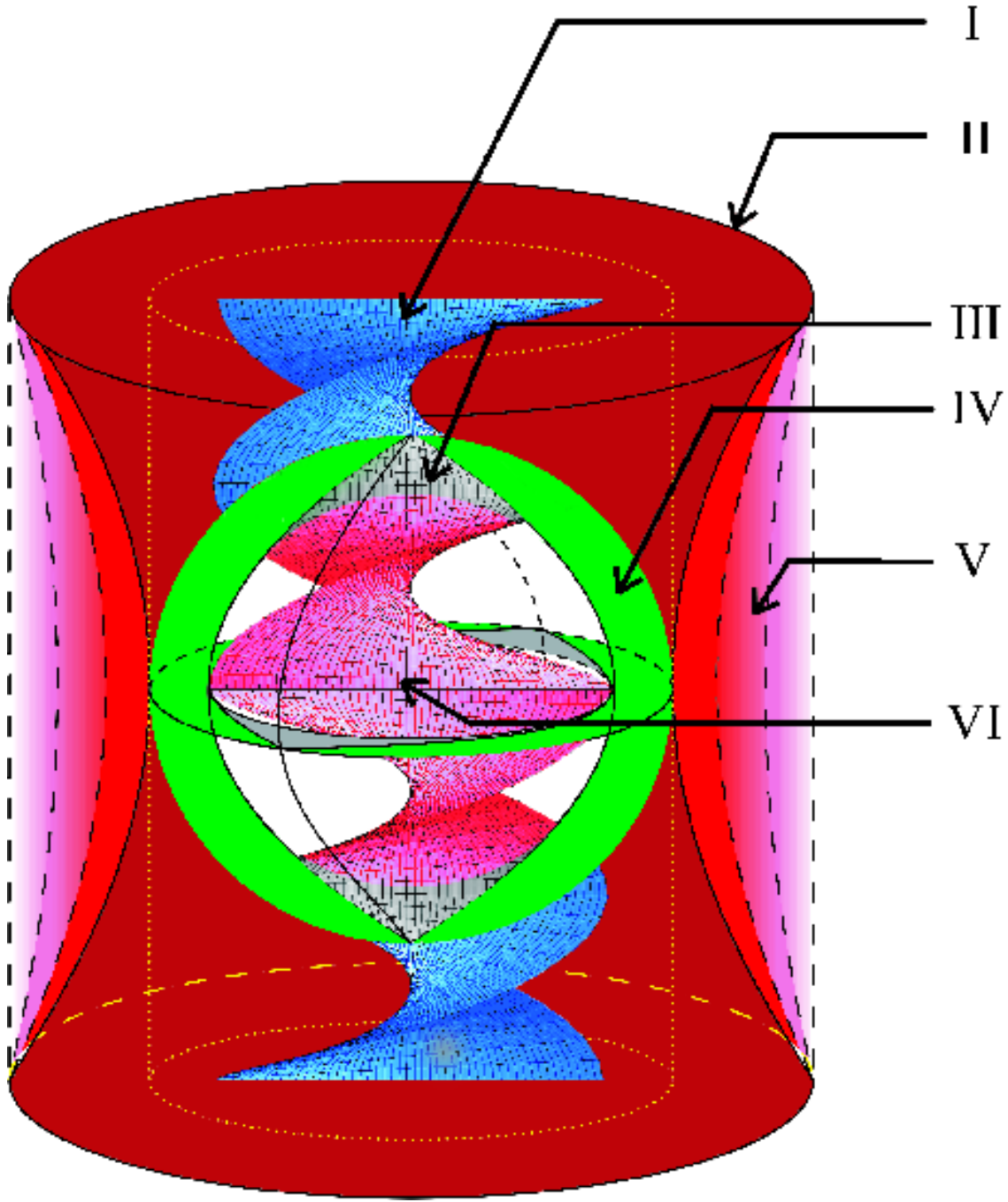}
\caption{Projection of the five-dimensional spacetime topology on ($x^{1}$ $x^{2}$ $x^{3}$) plane. For a better clarity, for having both the catenoids and the helicoids on the same projection, one of them must be in our {\bf D} case while the other in {\bf A} case. On the image are represented:
I - Pure Helicoidal Hyper-Surface with $C\neq 1$ and $\theta\neq \textnormal{arctg}\left(\frac{t}{\phi} \right)$ (blue); II - Pure Catenoidal Hyper-Surface with $C\neq 1$ and $\theta\neq \textnormal{arctg}\left(\frac{t}{\phi} \right)$ (magenta); III - Helicoid with $C=1$ and $\theta\neq \textnormal{arctg}\left(\frac{t}{\phi} \right)$ (gray); IV - Ergosphere (here are living the Ergosphere's Catenoid and the Ergosphere's Helicoid for which $C=1$ and $\theta=\textnormal{arctg}\left(\frac{t}{\phi} \right)$) (green); V - Catenoid with $C=1$ and $\theta\neq \textnormal{arctg}\left(\frac{t}{\phi} \right)$ (from red to pink); VI - BHIG helicoid (pink). The Pure Catenoidal Hyper-Surface [II] will evolve, for the internal observer, together with the objects on it, from a minimal surface representation to the cylindrical geometry, through the empty space where the Catenoids for which $C=1$ and $\theta\neq \textnormal{arctg}\left(\frac{t}{\phi} \right)$ [V] can exist.}
\label{bh}
\end{figure*}

If we pay attention to the way in which the ansatz factors {\it Z} are looking and to the {\it D}, {\it E}, {\it F}, {\it G} parameters modification from one ansatz to the other, we can construct a global image of our five-dimensional spacetimes (Fig. \ref{bh}) and also to see which is the path followed by an object entering/leaving into/from the DEUS object. Because nothing interdict us, we can consider that the five-dimensional timelike case is the {\bf D} case of the BHIG spacetime ($x^{4}=\pm\; i\; \textnormal{arctg}\left(\displaystyle \frac{t_{k}}{\phi_{k}}\right)$\ ) and the five-dimensional spacelike case is the BHIG spacetime's {\bf A} case ($x^{5}=\pm\;  \textnormal{arctg}\left(\displaystyle \frac{t_{k}}{\phi_{k}}\right)$\ ).

If we consider that the object enters in this space as a {\it Pure Catenoidal Hyper-Surface} object in, let say {\bf D} case, it will advance together with this hyper-surface passing through the $C=1$, $\theta\neq \textnormal{arctg}\left(\displaystyle \frac{t}{\phi}\right)$ intermediate region, in which $\theta$ becomes 0 {\bf only} for the external observer which sees the object falling in. This means that the object is seen as frozen in $\theta$ angle, but still falling radially. In its proper coordinate system the object is actually still spinning around the black hole. When the object enters the {\it Ergosphere's Catenoid} its $\theta$ angle becomes equal with $\textnormal{arctg}\left(\displaystyle \frac{t}{\phi}\right)$ and the coordinates $x^{4}$ and $x^{5}$ switch one with the other, meaning that, for the external observer, it becomes a spacelike object, totally frozen on the ergosphere (is "seen" as a {\bf A} case object), while, in its proper frame, the object still remains a {\bf D} case object with two of its five coordinates rotated. The real object remains eternally frozen on the ergosphere. From the five-dimensional external observer point of view, we may interpret intuitively (to the degree in which a five-dimensional manifold may be intuitively described) that the object (see Fig. \ref{bh}) passes instantaneously from the "equatorial" plane of our five-dimensional spacetime to its "pole", from where it becomes an virtual object (spatial fluctuations) on exit path from the black hole, first as an {\it Ergosphere's Helicoid} virtual object in {\bf A} case (here becomes important the strong constraint $\displaystyle \frac{t}{\phi}=-\; \frac{\phi_{k}}{t_{k}}$; $x^{4}$ and $x^{5}$ change again their role) and, after that, to the helicoid with $C=1$, $\theta\neq \textnormal{arctg}\left(\displaystyle \frac{t}{\phi}\right)$ ({\bf A} case) where the radial movement is unfrozen, continuing to the {\it Pure Helicoidal Hyper-Surface} in which also the rotation is unfrozen (also in {\bf A} case). It is very important the fact that, even when, from five-dimensional point of view, the object is spacelike, it is actually a four-dimensional timelike object in his system of reference for all the above hyper-surfaces (with three spatial coordinates and one temporal coordinate), excepting the helicoid with $C=1$ and $\theta\neq \textnormal{arctg}\left(\displaystyle \frac{t}{\phi}\right)$ which makes impossible to the four-dimensional object to see the naked four-dimensional central singularity ("no hair" theorem), and the {\it Pure Helicoidal Hyper-Surface} where we have two spatial and two temporal dimensions (the object is virtual for the four-dimensional internal or external observer).

Independently of what the external observer sees, when the object goes from the {\it Ergosphere's Catenoid} to the {\it Ergosphere's Helicoid}, it will see himself as being inside the BHIG spacetime and leaving it ($ds^{2}_{\theta = 0}=-\; ds^{2}_{BHIG}$ "mirror" effect). If the internal observer is five-dimensional (virtual) it "will be able to see" that its reflection on the inner horizon is real ($x^{4}$ is replaced with $x^{5}$ and reciprocally in {\bf A} case {\bf or} the coordinates remain unchanged but the object passes from the {\it Ergosphere's Helicoid} spacelike {\bf A} case to BHIG's {\bf D} timelike case). 

The external four-dimensional {\bf D} case observer "sees" across the $\theta=0$ singularity an {\bf A} case BHIG helicoid rotated with $90^{\circ}$, this happening because its $x^{3}$ world axis is becoming the BHIG's $x^{5}$ world axis, the internal phenomena seaming uncorrelated with the external ones due to the fake lack of continuity between the BHIG helicoid and the helicoid with $C=1$, $\theta\neq \textnormal{arctg}\left(\displaystyle \frac{t}{\phi}\right)$.

All the above story repeats itself but in reverse order if the object enters the five-dimensional DEUS object represented in Fig. \ref{bh} as {\it Pure Helicoidal Hyper-Surface} virtual object. 

The other two possible situations when a virtual object enters the five-dimensional spacetime (Fig. \ref{bh}) as a {\it Pure Catenoidal Hyper-Surface} or as {\it Pure Helicoidal Hyper-Surface} real object are mainly the same as above, the difference consisting into the fact that the study has to be done by replacing the {\bf A} (respectively {\bf D}) case manifolds the object crosses, with {\bf D} (respectively {\bf A}) case manifolds.

In principle the same remarques are valid also for objects (real or virtual) living inside BHIG and moving toward exterior, as black hole ejecta.

\ack
I want to express all my gratitude to Stefan Sorescu whose computer drawing experience was truly valuable for Fig. \ref{bh} creation.

\noindent
Manuscript registered at the Astronomical Institute, Romanian Academy, record No. 249 from 04/19/07.


\title[D.E.U.S.]{II: Self-Similarity and Implications in Cosmology}

\begin{abstract}
In a five-dimensional background we give an unified model for the Black Hole internal structure and Universe using self-similar minimal hyper-surfaces for the DEUS "particles" distribution inside the supplementary dimensions. Having four basic hypothesis from which the most valuable seams to be the self-similarity of the five-dimensional internal geometry, and using the manifolds derived in paper {\it DEUS I} \cite{deus1}, we show that the DEUS object collapses to a string-like object equivalent with the BHIG spacetime of a bigger DEUS object in which the first one is embedded.

The considered five-dimensional geometry becomes conformal with an external Friedmann-Robertson-Walker (FRW) spacetime (for which the Hubble constant is {\it uniquely} defined as $\sim 72.3$ km/s/Mpc) by evolving the hyper-surfaces composing our five-dimensional DEUS object to a pre-collapse cylinder, the pre-collapse movement of the hyper-surfaces being perceived after the collapse, by an external observer, as dark energy effect.

Because of the considered self-similarity, the SI (or CGS) units, transformed in geometrized units, will describe in a different manner and only locally the spacetime (which evolves to a Minkowski spacetime), while the global spacetime which contains also the local spacetime and its fields (the Friedmann-Robertson-Walker being only the BHIG helicoidal manifold representation of the DEUS object) suffers a transformation through which the Hubble constant will become adimensional and equal to 1 (km/s/Mpc $\equiv$ 1), as a fundamental constant of our model. The units into the local spacetime are the same (for a local observer) as the global ones only because the self-similar structure of the DEUS object, the local observer being able to see at two consecutive scales the Universe and black holes or black holes and their particle-like composing objects.
\end{abstract}


\section{Introduction}

In paper {\it DEUS I} \cite{deus1} we constructed a spacetime geometry departing from the trivial three-dimensional minimal surface representation of the catenoid and its conjugate surface, the helicoid. There we determined the five-dimensional coordinate system that satisfies the Cauchy-Riemann equations, and also the ansatz, the properties and the correlation possible to be made for the DEUS object's composing manifolds.

In this paper we will generalize the geometry of the DEUS Black Hole for Universe formation and evolution, by using self-similar minimal hyper-surfaces for DEUS objects distribution.

Having four basic hypothesis from which the most valuable seams to be the self-similarity of the five-dimensional internal geometry, we show that the DEUS object collapses to a string-like object equivalent with the BHIG spacetime of a bigger DEUS object in which the first one is embedded.

The considered five-dimensional geometry becomes conformal with an external Friedmann-Robertson-Walker (FRW) spacetime (for which the Hubble constant is {\it uniquely} defined as $72.33\pm 0.02$) by evolving the hyper-surfaces composing our five-dimensional DEUS object to a pre-collapse manifold. For the BHIG spacetime written in FRW coordinates we derive the Friedmann equations and some cosmological consequences ($\Omega_{tot}$ and the cosmological distance).

Into the background of a collapsed DEUS object rotating string that creates a BHIG manifold "deformed" in a FRW bubble, the pre-collapse movement of the five-dimensional catenoid hyper-surfaces is perceived by an after-collapse external observer as dark energy effect, while the matter still contained in the collapsed DEUS spacetime as dark matter. For this manifold we give the expression of the cosmological distance at which the external observer sees the collapsing event.

The collapsing helicoidal manifold can be seen by the external timelike observer only considering that the catenoid is having $N$ internal DEUS similar objects, distorted, also collapsing. The helicoids of these $N$-DEUS objects are timelike (five-dimensional) and are giving three-dimensional spacelike and timelike observable effects to the external space. The nature of these events will be discussed in another DEUS paper.

\section{D.E.U.S. Hypotheses}

The hypotheses of the D.E.U.S. model are in number of four:

\begin{enumerate}
\item Inside a Black Hole the information (energy) is not lost and is structured (non-homogeneity and non-isotropy);
\item The geometry of the black hole contains a catenoid-like surface (all the geometry of the DEUS black hole derives from this; see papers {\it DEUS I} \cite{deus1} and \cite{rev});
\item The black hole geometry is self-similar;
\item The spacetime is five-dimensional.
\end{enumerate}

The third hypothesis implies that from Planck scale to the Universe (not just its observed part) we must have the same five-dimensional symmetries, quantified in energy. The self-similarity is having as consequence also that for an external observer of a, let say black hole, when the dimensions collapse to a singularity (as we will see) the symmetry will be local, while for an internal observer the same symmetry will be global (a "new" spacetime unfold around the observer which touches the singularity).

\section{Gaussian Curvatures}

From now on, when we will refer to the hyper-surfaces defined in paper {\it DEUS I} \cite{deus1} we will consider, because the case equivalence, only the {\bf D} case.

\begin{figure}
\centering
\leavevmode
\includegraphics[width=16.0cm,height=10.0cm]{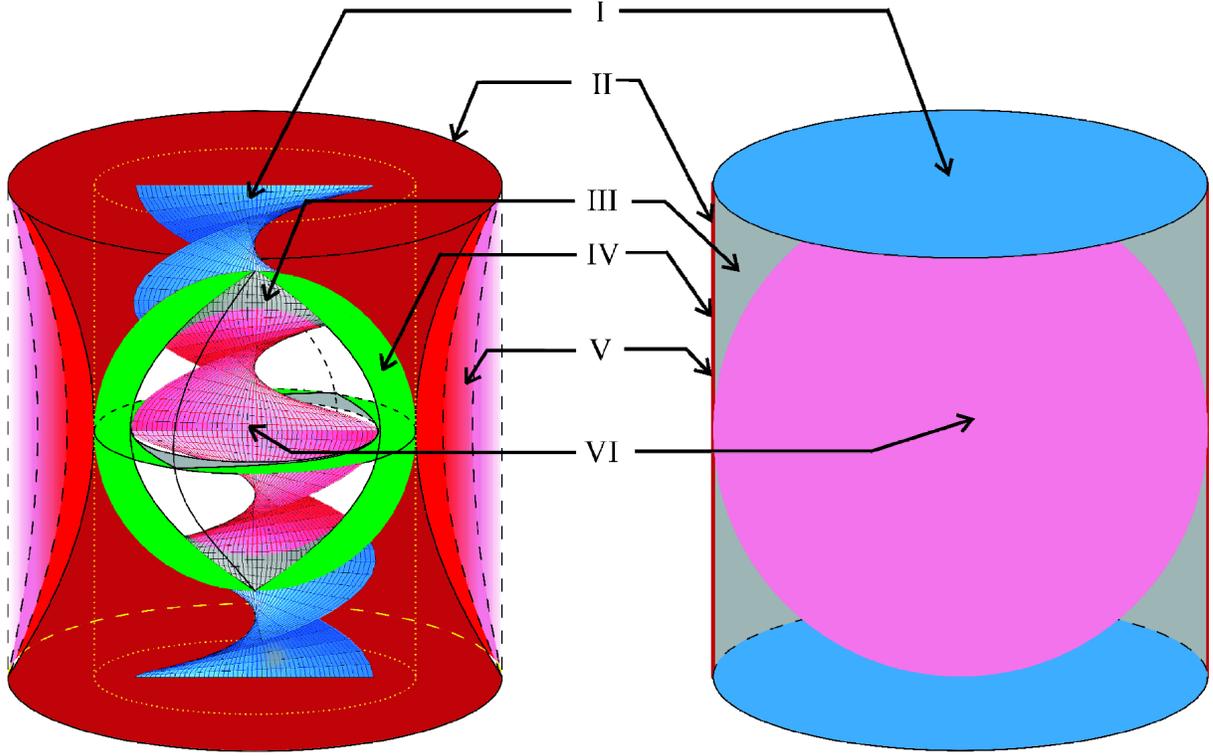}
\caption{Projection of DEUS spacetime topology on ($x^{1}$ $x^{2}$ $x^{3}$) plane. For a better clarity, for having both the catenoids and the helicoids on the same projection, one of them must be in our {\bf D} case while the other in {\bf A} case. On the image are represented:
I - Pure Helicoidal Hyper-Surface with $C\neq 1$ and $\theta\neq \textnormal{arctg}\left(\frac{t}{\phi} \right)$ (blue); II - Pure Catenoidal Hyper-Surface with $C\neq 1$ and $\theta\neq \textnormal{arctg}\left(\frac{t}{\phi} \right)$ (magenta); III - Helicoid with $C=1$ and $\theta\neq \textnormal{arctg}\left(\frac{t}{\phi} \right)$ (gray); IV - Ergosphere (here are living the Ergosphere's Catenoid and the Ergosphere's Helicoid for which $C=1$ and $\theta=\textnormal{arctg}\left(\frac{t}{\phi} \right)$) (green); V - Catenoid with $C=1$ and $\theta\neq \textnormal{arctg}\left(\frac{t}{\phi} \right)$ (from red to pink); VI - BHIG helicoid (pink). The Pure Catenoidal Hyper-Surface [II] will evolve, for the internal observer, together with the objects on it, from a minimal surface representation to the cylindrical geometry, through the empty space where the Catenoids for which $C=1$ and $\theta\neq \textnormal{arctg}\left(\frac{t}{\phi} \right)$ [V] can exist.}
\label{bh2}
\end{figure}

For the external observer of a black hole which accrete material, not only that the mass increases but also the size of the inner horizon, pushing almost all the other DEUS manifolds to an asymptotic flat cylinder spacetime (see the figure \ref{bh2}), for which the Gaussian curvature:

\begin{equation}
K_{\mu}\equiv \frac{1}{2 x^{\mu}}\; \frac{1}{g_{\mu \mu}}\; \frac{\partial g_{\mu \mu}}{\partial x^{\mu}}
\label{eq_1}
\end{equation}
\noindent
after $\phi$ must be equal with the one after {\it t}. For the {\it pure catenoidal hypersurface} ($\theta\neq \textnormal{arctg}\left(\displaystyle \frac{t}{\phi} \right)$ ) evolved to cylinder:

\begin{equation}
(K_{\phi})_{catenoid\rightarrow cylinder} = (K_{t})_{catenoid\rightarrow cylinder} = K_{catenoid\rightarrow cylinder} \; , 
\label{eq_2}
\end{equation}

\noindent
from where results that:

\begin{equation}
t=\mp\; i\; \phi 
\label{eq_3}
\end{equation}
\noindent
or:
\begin{equation}
t=\pm\; \phi 
\label{eq_3a}
\end{equation}

When we substitute (\ref{eq_3}) into the coordinate system of the {\it pure catenoidal hyper-surface} we get:

\begin{equation}
\left\{ \begin{array}{ll}
x^{1}=x^{2}=x^{4}=x^{5}=0 \\
x^{3}=A
\end{array} \right.
\label{eq_4}
\end{equation}

The same substitution made into the coordinate system of the {\it ergosphere's catenoid} ($C=1$; $\theta=\textnormal{arctg}\left(\displaystyle \frac{t}{\phi} \right)$\ ) will look like:

\begin{equation}
\left\{ \begin{array}{ll}
x^{1}=x^{2}=x^{4}=x^{5}=0 \\
x^{3}=A
\end{array} \right.
\label{eq_5}
\end{equation}

So, the observer situated on the cylindrical manifolds sees the {\it pure catenoidal hyper-surface} over-imposed on the {\it ergosphere's catenoid}. But, for the {\it ergosphere's catenoid} we have, as internal rule, the "strong" constraint (see paper {\it DEUS I}):

\begin{equation}
A\; \textnormal{cosh}^{-\; 1} \left(\frac{\sqrt{t^{2}+\phi^{2}}}{A} \right)\pm\; \textnormal{arctg}\left(\frac{t_{k}}{\phi_{k}} \right)=0
\label{eq_6}
\end{equation}
\noindent
from where it results that the {\it pure catenoidal hyper-surface} and the {\it ergosphere's catenoid} will have the same representation, explicitely:

\begin{equation}
\left\{ \begin{array}{ll}
x^{1}=x^{2}=x^{4}=x^{5}=0 \\
x^{3}=\mp\; \textnormal{arctg}\left(\displaystyle \frac{t_{k}}{\phi_{k}} \right)
\end{array} \right.
\label{eq_7}
\end{equation}

After doing the change also for the other manifolds composing the DEUS object we will have:

\begin{quote}
\begin{enumerate}
\item catenoid with $C=1$ and $\theta\neq \textnormal{arctg}\left(\displaystyle \frac{t}{\phi} \right)$

\begin{equation}
\left\{ \begin{array}{ll}
x^{1}=x^{2}=x^{4}=x^{5}=0 \\
x^{3}=\mp\; \textnormal{arctg}\left(\displaystyle \frac{t_{k}}{\phi_{k}} \right)
\end{array} \right.
\label{eq_8}
\end{equation}

\item {\it ergosphere's helicoid} ($\theta_{ext.obs.}=0$; $C=1$; $\theta=\textnormal{arctg}\left(\displaystyle \frac{t}{\phi} \right)$\ )

\begin{equation}
\left\{ \begin{array}{ll}
x^{1}=x^{2}=x^{4}=x^{5}=0 \\
x^{3}=\textnormal{arctg}\left(\displaystyle \frac{t_{k}}{\phi_{k}} \right)
\end{array} \right.
\label{eq_9}
\end{equation}

\item {\it pure helicoidal hyper-surface} ($\theta\neq \textnormal{arctg}\left(\displaystyle \frac{t}{\phi} \right)$\ ):

\begin{equation}
\left\{ \begin{array}{lll}
x^{1}=x^{2}=x^{5}=0 \\
x^{3}=\textnormal{arctg}\left(\displaystyle \frac{t_{k}}{\phi_{k}} \right) \\
x^{4}=\pm\; i\; \textnormal{arctg}\left(\displaystyle \frac{t_{k}}{\phi_{k}} \right)
\end{array} \right.
\label{eq_10}
\end{equation}

\item helicoid with $C=1$ and $\theta\neq \textnormal{arctg}\left(\displaystyle \frac{t}{\phi} \right)$:

\begin{equation}
\left\{ \begin{array}{lll}
x^{1}=x^{2}=x^{5}=0 \\
x^{3}=\textnormal{arctg}\left(\displaystyle \frac{t_{k}}{\phi_{k}} \right) \\
x^{4}=\pm\; i\; \sqrt{2}\; \textnormal{arctg}\left(\displaystyle \frac{t_{k}}{\phi_{k}} \right)\; \; \textnormal{or}\; \; x^{4}=\mp\; i\; \sqrt{2}\; \textnormal{arctg}\left(\displaystyle \frac{t_{k}}{\phi_{k}} \right)
\end{array} \right.
\label{eq_11}
\end{equation}

\end{enumerate}
\end{quote}

From the above equations we see that the helicoid from the ergosphere and all the catenoids are collapsing to a BHIG equivalent rotating rigid string-like object. Exceptions from this behavior are the helicoid with $C=1$ and $\theta\neq \textnormal{arctg}\left(\displaystyle \frac{t}{\phi} \right)$ and the {\it pure helicoidal hyper-surface}, which, as we will see later, can be seen as fields filling the empty space generated by the string collapsed hyper-surfaces. We can view this process as a DEUS space evaporation process into a BHIG helicoidal manifold, part of a another DEUS object in which the evaporated space continues to exist as a center of symmetry.

The four-dimensional observer situated on the cylinder "sees" across the $\theta=0$ singularity a $\pi/2$ rotated BHIG helicoid because his $x^{3}$ world axis is becoming the BHIG's $x^{5}$ world axis, the internal phenomena seaming uncorrelated with the external ones due to the fake lack of continuity between the BHIG helicoid and the helicoid with $C=1$, $\displaystyle \theta\neq \textnormal{arctg}\left(\frac{t}{\phi}\right)$.

Independently of what the external observer sees from the cylinder, when an object goes from the {\it ergosphere's catenoid} to the {\it ergosphere's helicoid} it will see himself as being inside BHIG and leaving it ($ds^{2}_{\theta = 0}=-\; ds^{2}_{BHIG}$ "mirror" effect). In this way, because the inner horizon behaves as a mirror, the internal observer (situated on, let say, the catenoid) sees the hyper-surfaces evolving toward the cylinder, while, the external observer sees the hyper-surfaces evolving away from cylinder, toward a minimal representation (in energy and mean curvature).

\section{BHIG Helicoid as FRW Universe Bubbles}

We will consider the following transformation of coordinates:

\begin{equation}
\begin{array}{ll}
r=\sqrt{t^{2}+\phi^{2}} \\
t_{FRW}=\textnormal{arctg}\left(\displaystyle \frac{t_{k}}{\phi_{k}} \right)
\end{array}
\label{eq_12}
\end{equation}

In the strong constraint $\displaystyle \frac{t_{k}}{\phi_{k}}=-\; \frac{\phi}{t}$ which makes the transformation from catenoid coordinates to helicoid coordinates, when crossing the ergosphere, we have:

\begin{equation}
\begin{array}{ll}
t=a\; \phi_{k} \\
\phi=-\; a\; t_{k}
\end{array}
\label{eq_13}
\end{equation}
\noindent
where {\it a} is a scaling factor inside BHIG manifold and:

\begin{equation}
a_{max}\equiv i\; A
\label{eq_14}
\end{equation}

While the scaling {\it a} is having a spatial significance, the {\it A} scaling factor (the catenoid "neck" width) will be temporal.

With the help of equations (\ref{eq_12}) and (\ref{eq_13}) we can express $t_{k}$ and $\phi_{k}$ as function of {\it r} and $t_{FRW}$:

\begin{equation}
\begin{array}{ll}
\phi_{k}=\displaystyle \frac{r}{a}\; \frac{1}{\sqrt{1+\textnormal{tg}^{2}\; t_{FRW}}} \\
t_{k}=\displaystyle \frac{r}{a}\; \frac{\textnormal{tg}\; t_{FRW}}{\sqrt{1+\textnormal{tg}^{2}\; t_{FRW}}}
\end{array}
\label{eq_15}
\end{equation}
\noindent
and their derivatives:

\begin{equation}
\begin{array}{ll}
d\phi_{k}=\displaystyle \frac{1}{a}\; \frac{1}{\sqrt{1+\textnormal{tg}^{2}\; t_{FRW}}}\; dr - \frac{r}{a}\; \frac{1}{\sqrt{1+\textnormal{tg}^{2}\; t_{FRW}}}\; \left[\frac{\dot{a}}{a} + \textnormal{tg}\; t_{FRW} \right]\; dt_{FRW} \\
dt_{k}=\displaystyle \frac{1}{a}\; \frac{\textnormal{tg}\; t_{FRW}}{\sqrt{1+\textnormal{tg}^{2}\; t_{FRW}}}\; dr - \frac{r}{a}\; \frac{1}{\sqrt{1+\textnormal{tg}^{2}\; t_{FRW}}}\; \left[\frac{\dot{a}}{a}\; \textnormal{tg}\; t_{FRW} - 1 \right]\; dt_{FRW}
\end{array}
\label{eq_16}
\end{equation}

Using (\ref{eq_15}) and (\ref{eq_16}) in the ansatz of BHIG manifold:

\begin{equation}
ds^{2}_{BHIG}=-\; \frac{\phi_{k}^{2}}{\left(t_{k}^{2}+\phi_{k}^{2} \right)^{2}}\; dt_{k}^{2}-\frac{t_{k}^{2}}{\left(t_{k}^{2}+\phi_{k}^{2} \right)^{2}}\; d\phi_{k}^{2}+\frac{t_{k}\; \phi_{k}}{\left(t_{k}^{2}+\phi_{k}^{2} \right)^{2}}\; dt_{k}\; d\phi_{k}
\label{eq_17}
\end{equation}
\noindent
we obtain:

\begin{equation}
\begin{array}{ll}
ds^{2}_{BHIG} = &\displaystyle \frac{3}{4}\; \frac{-1-3H^{2}-2H^{4}+2H^{3}\sqrt{H^{2}+1}+2H\sqrt{H^{2}+1}}{\left[H^{2}-H \sqrt{H^{2}+1}+1 \right]^{2}}\; dt^{2}_{FRW} - \\ &- \displaystyle \frac{1}{r^{2}}\; \frac{2H^{2}-2H \sqrt{H^{2}+1}+1}{4\; \left[H^{2}-H \sqrt{H^{2}+1}+1 \right]^{2}}\; dr^{2}
\end{array}
\label{eq_18}
\end{equation}
with $H\equiv \displaystyle \frac{\dot{a}}{a}$. From (\ref{eq_18}) results:

\begin{equation}
ds^{2}_{BHIG}=\frac{3}{4}\; dt^{2}_{FRW} - \frac{1}{r^{2}}\; \frac{2H^{2}-2H \sqrt{H^{2}+1}+1}{4\; \left[H^{2}-H \sqrt{H^{2}+1}+1 \right]^{2}}\; dr^{2}
\label{eq_19}
\end{equation}

\begin{figure}
\centering
\leavevmode
\includegraphics[width=10.0cm,height=10.0cm]{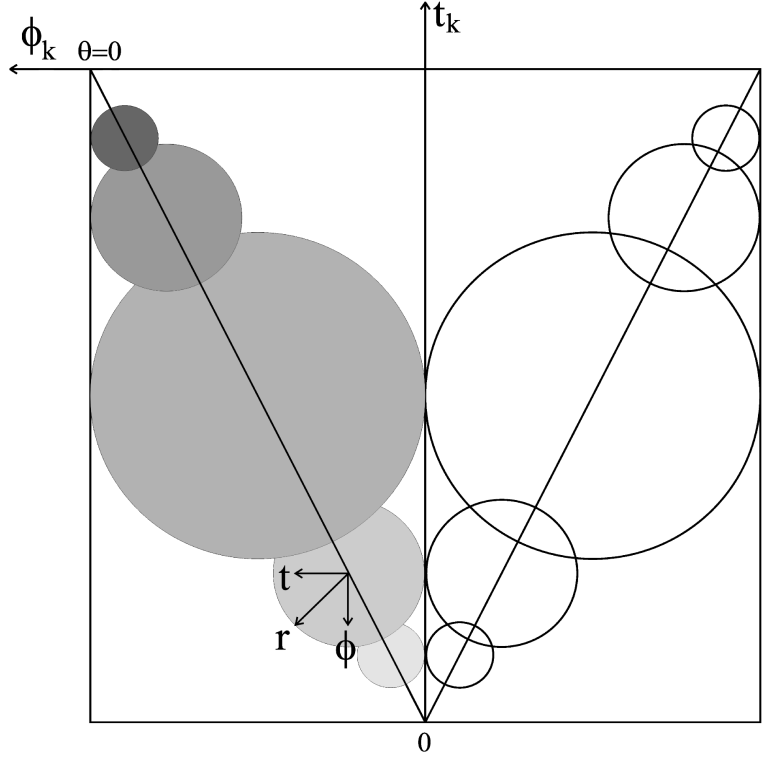}
\caption{The intuitive representation of how BHIG spacetime transforms in FRW spacetimes. For a better understanding, half of the BHIG helicoid is unfolded in a flat torsion-free hyper-surface. The FRW bubble evolves from the center of symmetry (the string-like collapsed object or Big Bang equivalent) to the point where it will exit through the $\theta =0$ point in another DEUS superior spacetime (Big Crunch equivalent). The time $t_{FRW}$ is independent of the sense in which the FRW bubble evolves and this translates in a possibility of having a reversed evolution, from Big Bang to the Big Crunch, from $\theta =0$ to the central string like object.}
\label{frw}
\end{figure}

The intuitive image of how BHIG spacetime transforms in a FRW spacetime can be seen in Fig. \ref{frw}, where the FRW bubble evolves from the center of symmetry (the string-like collapsed object or Big Bang equivalent) to the point where it will exit through the $\theta =0$ point in another DEUS superior spacetime (Big Crunch equivalent). The time $t_{FRW}$ is independent of the sense in which the FRW bubble evolves and this translates in a possibility of having a reversed evolution, from Big Bang to the Big Crunch, so from the point $\theta =0$ to the center of symmetry. We see that for:

\begin{equation}
r^{2}= \frac{2H^{2}-2H\sqrt{H^{2}+1}+1}{3\; \left(H^{2}-H\sqrt{H^{2}+1}+1\right)^{2}}
\label{eq_19a}
\end{equation}
\noindent
the BHIG expressed in FRW coordinates is just a quarter of a external Minkowski space, meaning that each quarter contains a FRW bubble and that all four bubbles generate a global Minkowski external space. While, from the point of view of a black hole, we are in a Minkowski spacetime containing all four FRW spacetimes, on a bigger spacetime scale we are contained in only one quadrant of the BHIG helicoid of a DEUS object (resulting also from the the self-similarity).

The above will have implications on the "type" of FRW Universe contained in each quadrant. If, we place ourselves in a matter dominated FRW bubble, in the second quadrant, the first quadrant will be an Anti-Universe (antimatter dominated) annihilating the Universe at the point $t_{k}=\phi_{k}=0$. Because $t_{FRW}$ begins at the point 0 of the collapsed DEUS object, quadrant three and four are for the past in $t_{k}$, while I and II are for the future. In the frame of the string-like collapsed DEUS object we may see the FRW Universe as coming from the quadrant IV, disappearing ("Big Crunch"; $t_{FRW}=r=0$)  at $t_{k}=0$ and reappearing ("Big Bang") in quadrant II. The same happens with the Anti-Universe, which crosses the $t_{k}=0$ from quadrant III to quadrant I. This, from an external observer perspective situated in a self-similar superior order DEUS object, can be seen as a matter-antimatter annihilation process at the moment $t_{k}$ of observation. This process of creation and annihilation repeats itself indefinitely because the considered chain of self-similar DEUS objects composing a higher order DEUS object catenoidal hyper-surface. 

If we want that the ansatz (\ref{eq_18}) to describe a local Euclidian space, then we must have: 

\begin{equation}
\begin{array}{ll}
2H^{2}-2H \sqrt{H^{2}+1}+1 > 0 \\
-1-3H^{2}-2H^{4}+2H^{3}\sqrt{H^{2}+1}+2H\sqrt{H^{2}+1} > 0
\end{array}
\label{eq_20}
\end{equation}

Let us use the following notations:

\begin{equation}
\begin{array}{llll}
g(H)\equiv -1-3H^{2}-2H^{4}+2H^{3}\sqrt{H^{2}+1}+2H\sqrt{H^{2}+1} \\
h(H)\equiv 2H^{2}-2H \sqrt{H^{2}+1}+1 \\
q(H)\equiv H^{2}-H \sqrt{H^{2}+1}+1 \\
f(H)\equiv 3\; g(H) - [h(H)+1]^{2}
\end{array}
\label{eq_21}
\end{equation}

\begin{figure}
\centering
\leavevmode
\includegraphics[width=13.0cm,height=9.0cm]{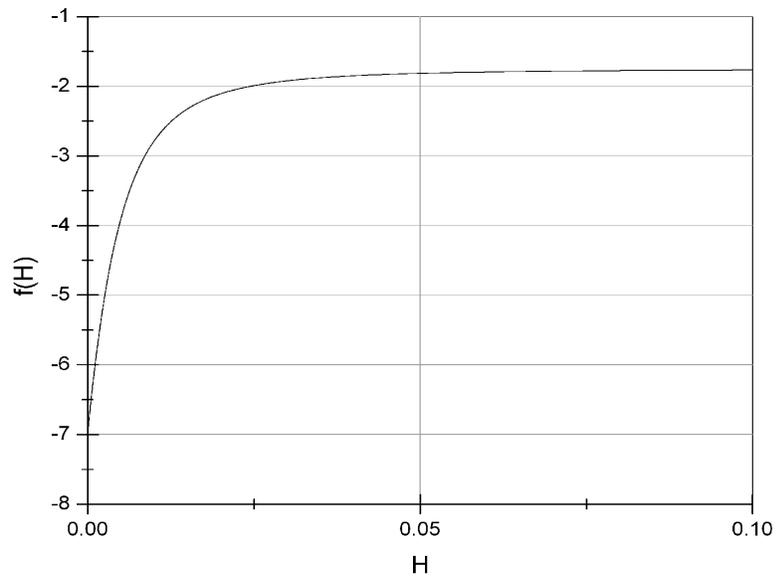}
\includegraphics[width=13.0cm,height=9.0cm]{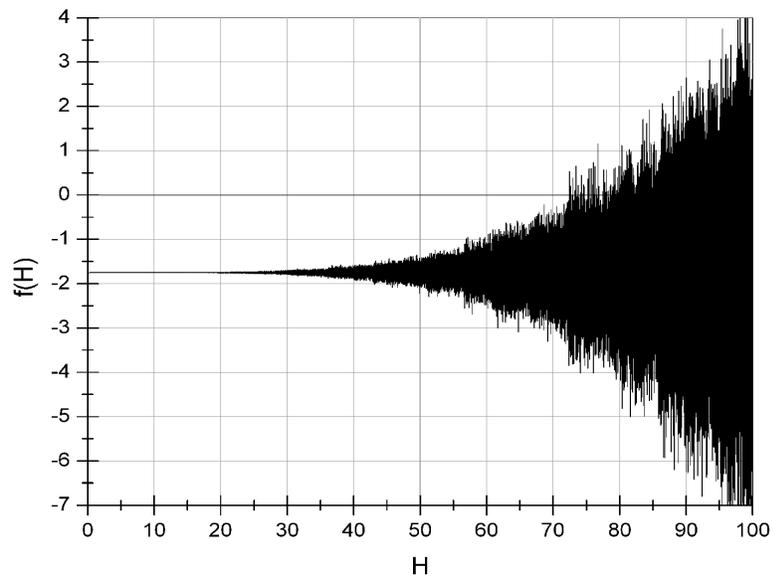}
\caption{Function {\it f(H)} behavior (see (\ref{eq_21})). {\it f(H)} becomes positive for the first time at $H_{0}=72\pm 1$.}
\label{ffunction}
\end{figure}

\begin{figure}
\centering
\leavevmode
\includegraphics[width=13.0cm,height=9.0cm]{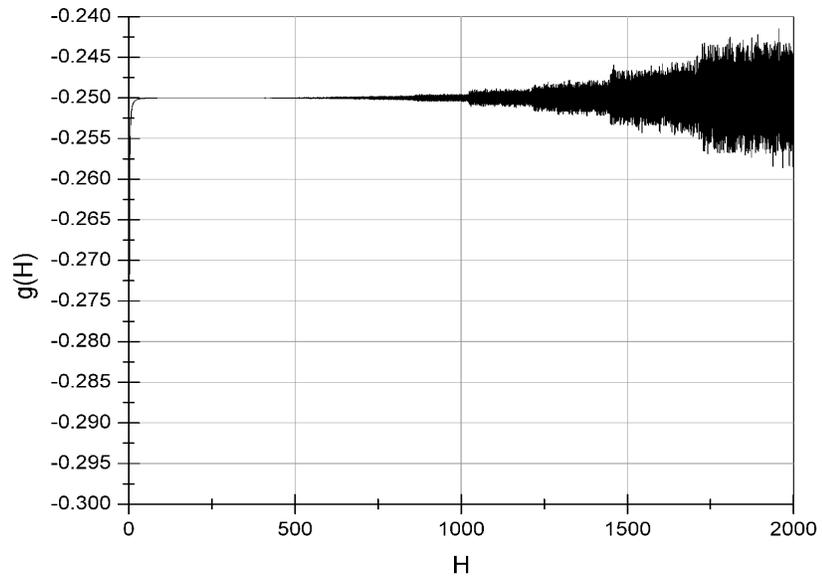}
\includegraphics[width=13.0cm,height=9.0cm]{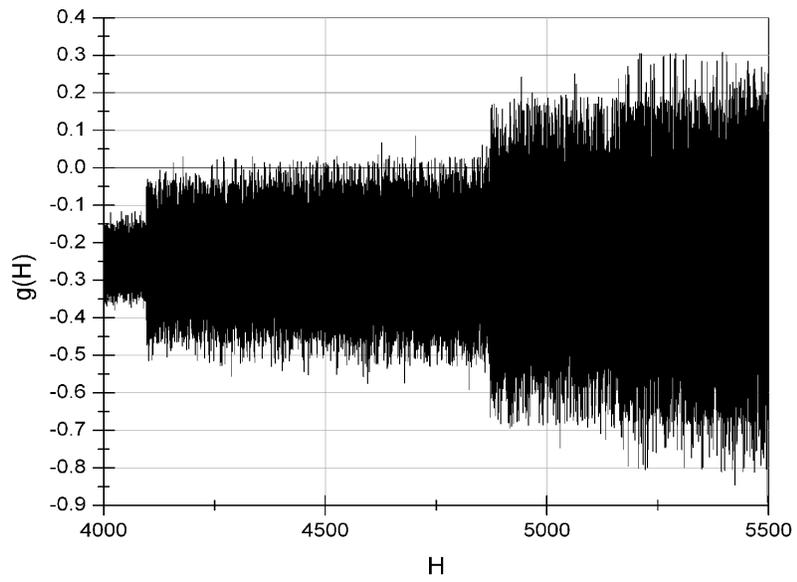}
\caption{Function {\it g(H)} behavior (see (\ref{eq_21})).}
\label{gfunction}
\end{figure}

If we are looking for a $g_{t_{FRW}t_{FRW}}=1$, then we must have $f(H)=0$. For a snapshot of the behavior of {\it f(H)} and {\it g(H)} functions see the figures \ref{ffunction} and \ref{gfunction}. The functions are very sensitive to the values given to {\it H} and their variations are the ones of a dynamical complex system. The first time when {\it f(H)} crosses through zero value, becoming positive, is at $H_{0}=72\pm 1$. A possible value at which {\it g(H)} is positive is $H_{1}=H_{0}^{2}\simeq 5234.52$. At the $H_{1}$ value also $f(H_{1})\geq 0$.


\begin{figure}
\centering
\leavevmode
\includegraphics[width=13.0cm,height=9.0cm]{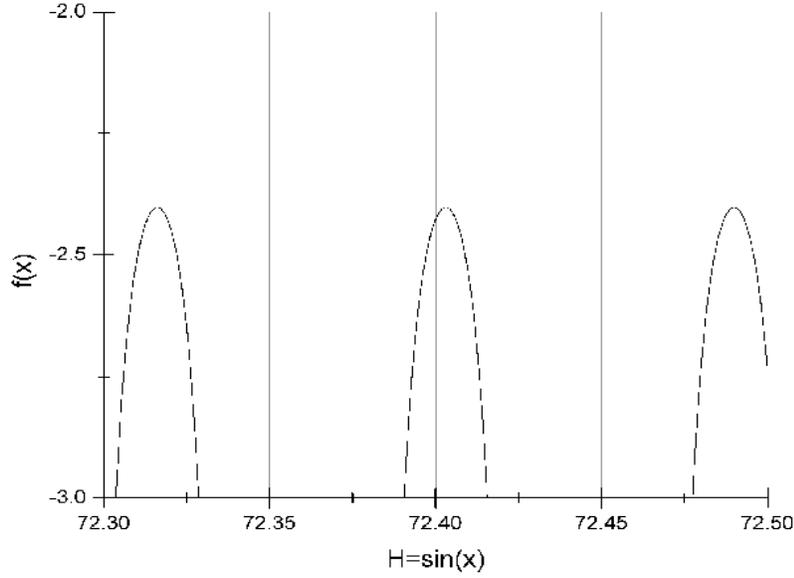}
\caption{Representation of $f(x)$, where $H=\textnormal{sin}\; x$. {\it f(x)} = max. for $H_{0}\simeq 72.31$, where $H=H_{0}\pm n\Delta H$ with $n\in \mathbb{N}$.}
\label{sinus}
\end{figure}

{\it f(H)} and {\it g(H)} are becoming periodic and continuous functions by changing the {\it H} variable with {\it sin x} (see Fig. \ref{sinus}). This periodicity is reflected in {\it H} by $H=H_{0}\pm n\Delta H$ with $n\in \mathbb{N}$. Even that for {\it f(x)}, $\Delta H < 0.1$, for {\it g(x)}, $\Delta H\simeq 6.3$. The value of $H_{0}$ for {\it f(x)} (for which this function is having maximal value) resulted to be $H_{0}\simeq 72.35$. In the same way, for {\it g(x)} it resulted that $H=H_{0}\simeq 72.31$. In consequence, the common value for {\it f} and {\it g} can be written as $H_{0}\simeq 72.33\pm 0.02$. In (\ref{eq_21}), for this $H_{0}$ also {\it h(H)} and {\it q(H)} are positive.

Using the functions from (\ref{eq_21}) and the determined value for $H_{0}$, the (\ref{eq_18}) world-lines of BHIG are:

\begin{equation}
ds^{2}_{BHIG}=dt^{2}_{FRW} - \frac{1}{r^{2}}\; \frac{h(H_{0})}{4\; q^{2}(H_{0})}\; dr^{2}=dt^{2}_{FRW} - \frac{1.9992\times 10^{-4}}{r^{2}}\;  dr^{2}
\label{eq_22}
\end{equation}

If we introduce in (\ref{eq_22}) also the angular term of the FRW spacetime, then:
 
\begin{equation}
ds^{2}_{BHIG}=dt^{2}_{FRW} - \frac{1.9992\times 10^{-4}}{r^{2}}\;  dr^{2} + a^{2}(t_{FRW})\; r^{2}\; \textnormal{sin}^{2}\zeta\; d\xi^{2} + a^{2}(t_{FRW})\; r^{2}\; d\zeta^{2}\; ,
\label{eq_23}
\end{equation}
\noindent
where, at the present time, the scale factor $a(t)\equiv a(t_{0})\equiv a_{0}=1$ and the universe radius $R_{0}=a_{0}\; r$. The general expression for the FRW Universe radius is $R(t_{FRW})\equiv a(t_{FRW})\; r$.

By making the identification of the (\ref{eq_23}) terms in the general expression of the FRW ansatz we get:

\begin{equation}
\frac{a^{2}}{\displaystyle 1-k\frac{r^{2}}{R_{0}^{2}}}=\frac{1.9992\times 10^{-4}}{r^{2}}
\label{eq_24}
\end{equation}
\noindent
where {\it k} is the curvature.

We know from theory that:

\begin{equation}
k=a^{2}R_{0}^{2}H^{2}\left(\Omega_{tot}-1 \right)\; .
\label{eq_25}
\end{equation}

With (\ref{eq_25}) in (\ref{eq_24}) we can write:

\begin{equation}
\frac{1}{a^{2}} - H^{2}\left(\Omega_{tot}-1 \right)\; r^{2} = \frac{r^{2}}{1.9992\times 10^{-4}}\; .
\label{eq_26}
\end{equation}

Having as reference equation (\ref{eq_26}) we can particularize:

\begin{itemize}
\item If $a=a_{0}=1$ and $H=H_{1}$ (a possible value in a primordial, after inflation, Universe) results that $\Omega_{tot}\simeq 0.999817$ of an open Universe. But because our lack of a better precision in $H_{0}$ determination we can say at most that $\Omega_{tot}\sim 1$
\item In the case of $\Omega_{tot}=1$ and $H=H_{1}$ results that $a_{0}^{2}\simeq \displaystyle \frac{2\times 10^{-4}}{r^{2}}$ with which:

\begin{equation}
ds^{2}_{BHIG}\simeq dt^{2}_{FRW} - 2\times 10^{-4}\left(\frac{1}{r^{2}}\; dr^{2} + \textnormal{sin}^{2}\zeta\; d\xi^{2} + d\zeta^{2} \right)
\label{eq_27}
\end{equation}

\item If $\Omega_{tot}\simeq 0.999817$ and $H_{0}\simeq 72.3$ results that $\displaystyle \frac{1}{a_{0}^{2}}\simeq 5001\; r^{2}$ and:

\begin{equation}
ds^{2}_{BHIG}\simeq dt^{2}_{FRW} - \frac{1.9992\times 10^{-4}}{r^{2}}\; dr^{2} + 1.99958\times 10^{-4}\left(\textnormal{sin}^{2}\zeta\; d\xi^{2} + d\zeta^{2}\right)\; .
\label{eq_28}
\end{equation}

\end{itemize}

\section{Cosmological Implications of DEUS Manifolds}

\subsection{BHIG Seen as a FRW Space}

Let us take again the (\ref{eq_19}) ansatz in which we consider also the angular part:

\begin{equation}
ds^{2}_{BHIG}=\frac{3}{4}\; dt_{FRW}^{2} - \frac{2H^{2}-2H \sqrt{H^{2}+1}+1}{4\; \left[H^{2}-H \sqrt{H^{2}+1}+1 \right]^{2}}  \left(\frac{1}{r^{2}}\; dr^{2} + \textnormal{sin}^{2}\zeta\; d\xi^{2} + d\zeta^{2}  \right)
\label{eq_29}
\end{equation}

The affine connections:

\begin{equation}
\Gamma_{ij}^{k}=\frac{1}{2}\; g^{kl}\left(g_{il,\; j} + g_{jl,\; i} - g_{ij,\; l} \right)\; ,
\label{eq_30}
\end{equation}
\noindent
that do not cancel for the ansatz (\ref{eq_29}) are:

\begin{equation}
\begin{array}{lllllll}
\Gamma_{0r}^{r}=\Gamma_{0\zeta}^{\zeta}=\Gamma_{0\xi}^{\xi}=-\; \displaystyle \frac{H\dot{H}}{H^{2}+1} \\
\smallskip \Gamma_{rr}^{0}=\displaystyle \frac{1}{3}\; \frac{H\dot{H}}{(H^{2}+1)^{2}\; r^{2}} \\
\Gamma_{rr}^{r}=-\; \displaystyle \frac{1}{r} \\
\smallskip \Gamma_{\zeta \zeta}^{0}=\displaystyle \frac{1}{3}\; \frac{H\dot{H}}{(H^{2}+1)^{2}} \\
\smallskip \Gamma_{\zeta \xi}^{\xi}=\displaystyle \frac{1}{\textnormal{tg}\; \zeta} \\
\Gamma_{\xi \xi}^{0}=\displaystyle \frac{1}{3}\; \frac{H\dot{H}\; \textnormal{sin}^{2}\zeta}{(H^{2}+1)^{2}} \\
\Gamma_{\xi \xi}^{\zeta}=-\; \textnormal{cos}\; \zeta\; \textnormal{sin}\; \zeta
\end{array}
\label{eq_31}
\end{equation}

With the Christoffel symbol components we can compute the Riemann tensor components:

\begin{equation}
\begin{array}{llllll}
\smallskip R_{0r0r}=\displaystyle \frac{(H^{2}+1)H\ddot{H}+(1-2H^{2})\dot{H}^{2}}{4(H^{6}+3H^{4}+3H^{2}+1)\; r^{2}} \\
\smallskip R_{0\zeta 0\zeta}=\displaystyle \frac{(H^{2}+1)H\ddot{H}+(1-2H^{2})\dot{H}^{2}}{4(H^{6}+3H^{4}+3H^{2}+1)} \\
\smallskip R_{0\xi 0\xi}=\displaystyle \frac{[(H^{2}+1)H\ddot{H}+(1-2H^{2})\dot{H}^{2}]\; \textnormal{sin}^{2}\zeta}{4(H^{6}+3H^{4}+3H^{2}+1)} \\
\smallskip R_{r\zeta r\zeta}=-\; \displaystyle \frac{H^{2}\; \dot{H}^{2}}{12(H^{8}+4H^{6}+6H^{4}+4H^{2}+1)\; r^{2}} \\
\smallskip R_{r\xi r\xi}=-\; \displaystyle \frac{H^{2}\; \dot{H}^{2}\; \textnormal{sin}^{2}\zeta}{12(H^{8}+4H^{6}+6H^{4}+4H^{2}+1)\; r^{2}} \\
R_{\zeta \xi \zeta \xi}=-\; \displaystyle \frac{(H^{2}\dot{H}^{2}-3H^{6}-9H^{4}-9H^{2}-3)\; \textnormal{sin}^{2}\zeta}{12(H^{8}+4H^{6}+6H^{4}+4H^{2}+1)}
\end{array}
\label{eq_32}
\end{equation}

The Ricci curvature tensor components are:

\begin{equation}
\begin{array}{lll}
\smallskip R_{00}=4\; \displaystyle \frac{(H^{2}+1)H\ddot{H}+(1-2H^{2})\dot{H}^{2}}{(H^{2}+1)^{2}} \\
\smallskip R_{rr}=\displaystyle \frac{4}{3}\; \frac{(H^{2}+1)H\ddot{H}+(1-4H^{2})\dot{H}^{2}}{(H^{2}+1)^{2}} \\
R_{\zeta \zeta}=R_{\xi \xi}=\displaystyle \frac{4}{3}\; \frac{(H^{2}+1)H\ddot{H}+(1-4H^{2})\dot{H}^{2}+3H^{6}+9H^{4}+9H^{2}+3}{(H^{2}+1)^{2}}
\end{array}
\label{eq_33}
\end{equation}

Also, the Einstein tensor components are:

\begin{equation}
\begin{array}{lll}
\smallskip G_{0}^{0}=4\; \displaystyle \frac{H^{2}\dot{H}^{2}-H^{6}-3H^{4}-3H^{2}-1}{(H^{2}+1)^{2}} \\
\smallskip G_{r}^{r}=-\; \displaystyle \frac{4}{3}\; \frac{2(H^{2}+1)H\ddot{H}+(2-5H^{2})\dot{H}^{2}+3H^{6}+9H^{4}+9H^{2}+3}{(H^{2}+1)^{2}} \\
G_{\zeta}^{\zeta}=G_{\xi}^{\xi}=-\; \displaystyle \frac{4}{3}\; \frac{2(H^{2}+1)H\ddot{H}+(2-5H^{2})\dot{H}^{2}}{(H^{2}+1)^{2}}
\end{array}
\label{eq_34}
\end{equation}

Because:

\begin{equation}
T_{00}\equiv \rho\; g_{00}=\frac{4}{3}\; \rho \; ,
\label{eq_35}
\end{equation}
\begin{equation}
R=3\; \frac{H\ddot{H}(H^{2}+1)+\dot{H}^{2}(1-2H^{2})}{(H^{2}+1)^{2}} \; ,
\label{eq_36}
\end{equation}
\noindent
and:

\begin{equation}
G_{00}\equiv R_{00} - \frac{1}{2}\; g_{00}\; R=2\; \frac{H\ddot{H}(H^{2}+1)+\dot{H}^{2}(1-2H^{2})}{(H^{2}+1)^{2}}\; ,
\label{eq_37}
\end{equation}
\noindent
from Einstein equations:

\begin{equation}
G_{ij}-\Lambda\; g_{ij}=8\pi G\; T_{ij}\; ,
\label{eq_38}
\end{equation}
\noindent
we will get the first Friedmann equation:

\begin{equation}
\frac{8\pi G\; \rho}{3}=\frac{1}{2}\; \frac{H\ddot{H}}{H^{2}+1} + \frac{1}{2}\; \left(\frac{\dot{H}}{H^{2}+1} \right)^{2} - \left(\frac{H\dot{H}}{H^{2}+1} \right)^{2} - \frac{1}{3}\; \Lambda\; ,
\label{eq_39}
\end{equation}
\noindent
or, by knowing that $H=\dot{a}/a$:

\begin{equation}
\frac{\stackrel{...}{a}\; \dot{a}\; a^{2} - 3\ddot{a}\; \dot{a}^{2}\; a + 2\dot{a}^{4}}{2a^{2}(\dot{a}^{2}+a^{2})} - \frac{\ddot{a}^{2}\; a^{2} + \dot{a}^{4} - 2\ddot{a}\; \dot{a}^{2}\; a}{2(\dot{a}^{2}+a^{2})^{2}} - \frac{1}{3}\; \Lambda=\frac{8\pi G\; \rho}{3}
\label{eq_40}
\end{equation}
\noindent

Also, with:

\begin{equation}
G_{rr}\equiv R_{rr} - \frac{1}{2}\; g_{rr}\; R=\frac{2}{3}\; \frac{H\ddot{H}(H^{2}+1)+\dot{H}^{2}(1-4H^{2})}{(H^{2}+1)^{2}}
\label{eq_41}
\end{equation}
\noindent
and:

\begin{equation}
T_{rr}\equiv -\; p\; g_{rr}=-\; p\; r^{2}\; \frac{4\left(H^{2} - H\sqrt{H^{2}+1} + 1 \right)^{2}}{2H^{2}-2H\sqrt{H^{2}+1}+1}\; ,
\label{eq_42}
\end{equation}
\noindent
in Einstein equations, results the second Friedmann equation:

\begin{equation}
\begin{array}{ll}
\displaystyle -8\pi G\; p\; r^{2}\; \frac{4\left(H^{2} - H\sqrt{H^{2}+1} + 1 \right)^{2}}{2H^{2}-2H\sqrt{H^{2}+1}+1}= &\displaystyle \frac{2}{3}\; \frac{H\ddot{H}(H^{2}+1)+\dot{H}^{2}(1-4H^{2})}{(H^{2}+1)^{2}} - \\ &\displaystyle - \Lambda\; r^{2}\; \frac{4\left(H^{2} - H\sqrt{H^{2}+1} + 1 \right)^{2}}{2H^{2}-2H\sqrt{H^{2}+1}+1}
\end{array}
\label{eq_43}
\end{equation}

If, in the present era, $p\simeq 0$ and $\Lambda \simeq 0$, the equation (\ref{eq_43}) becomes:

\begin{equation}
H\ddot{H}(H^{2}+1)+\dot{H}^{2}-4H^{2}\dot{H}^{2}=0\; ,
\label{eq_43a}
\end{equation}
\noindent
while the equation (\ref{eq_39}) transforms as:

\begin{equation}
\frac{8\pi G\; \rho}{3}=\frac{H^{2}\dot{H}^{2}}{(H^{2}+1)^{2}}\; .
\label{eq_44}
\end{equation}

With the critical density for closing the Universe:

\begin{equation}
\rho_{c}\equiv \frac{3H^{2}}{8\pi G}\; ,
\label{eq_45}
\end{equation}
\noindent
in (\ref{eq_44}) and $\Omega_{tot}\equiv \rho /\rho_{c}$:

\begin{equation}
\Omega_{tot}=\frac{\dot{H}^{2}}{(H^{2}+1)^{2}}=\left(\frac{\ddot{a}\; a -\dot{a}^{2}}{\dot{a}^{2}+a^{2}} \right)^{2} \; .
\label{eq_46}
\end{equation}

If $\Omega_{tot}=1$ in (\ref{eq_46}) it results that:

\begin{equation}
q=2+\frac{1}{H^{2}}\; ,
\label{eq_47}
\end{equation}
\noindent
where the acceleration parameter $q$ is defined as:

\begin{equation}
q\equiv \frac{\ddot{a}a}{\dot{a}^{2}}\; .
\label{eq_48}
\end{equation}

For the present time, with $H_{0}\simeq 72.3$ in (\ref{eq_47}), the acceleration parameter is $q\simeq 2$.

In the general case, from (\ref{eq_46}) we can write that:

\begin{equation}
\Omega_{tot}=\frac{H^{2}}{(H^{2}+1)^{2}}\; \left(\frac{q}{a}-2qH^{2}+H^{2} \right)\; .
\label{eq_49}
\end{equation}

Another important quantity that we can derive from (\ref{eq_29}) is the cosmological distance:

\begin{equation}
d_{BHIG}=\left[\frac{2H^{2}-2H\sqrt{H^{2}+1}+1}{4(H^{2}-H\sqrt{H^{2}+1}+1)^{2}} \right]^{1/2}\; \textnormal{ln}\; r \; .
\label{eq_50}
\end{equation}

\subsection{Catenoid}

The ansatz for this manifold is \cite{deus1}:

\begin{equation}
ds_{catenoid}^{2}=Z_{catenoid}\cdot \frac{t^{2}}{t^{2}+\phi^{2}}\; dt^{2}+Z_{catenoid}\cdot \frac{\phi^{2}}{t^{2}+\phi^{2}}\; d\phi^{2}+Z_{catenoid}\cdot \frac{t\; \phi}{t^{2}+\phi^{2}}\; dt\; d\phi
\label{eq_51}
\end{equation}
\noindent
where:
\begin{equation}
Z_{catenoid} =\textnormal{cos}^{2}\theta - \frac{\textnormal{sin}^{2}\theta\; \textnormal{cos}^{2}\theta}{\textnormal{sin}^{2}\theta - \displaystyle \frac{A^{2}\; \textnormal{cosh}^{-\; 2}\left(\displaystyle \frac{\sqrt{t^{2}+\phi^{2}}}{A} \right)}{\theta^{2}\left(t^{2}+\phi^{2} \right)}} + \frac{\textnormal{sinh}^{2} \left(\displaystyle \frac{\sqrt{t^{2}+\phi^{2}}}{A} \right)}{\textnormal{cosh}^{4} \left(\displaystyle \frac{\sqrt{t^{2}+\phi^{2}}}{A} \right)}
\label{eq_52}
\end{equation}

As for the BHIG manifold we will consider the transformations:

\begin{equation}
\begin{array}{llll}
r=\sqrt{t^{2}+\phi^{2}} \\
t_{FRW}=\textnormal{arctg}\left(\displaystyle \frac{t_{k}}{\phi_{k}} \right) \\
t=a\; \phi_{k} \\
\phi=-\; a\; t_{k} \; ,
\end{array}
\label{eq_53}
\end{equation}
\noindent
from where result the (\ref{eq_15}) and (\ref{eq_16}) transformations with which:

\begin{equation}
ds_{catenoid}^{2}=\frac{Z}{4}\; \left(\frac{\sqrt{H^{2}+1}-H}{H^{2}-H\sqrt{H^{2}+1}+1} \right)^{2}\; \left[r^{2}(H^{2}+1)\; dt_{FRW}^{2}-dr^{2} \right]
\label{eq_54}
\end{equation}
\noindent
and:

\begin{equation}
Z_{catenoid} =\textnormal{cos}^{2}\theta - \frac{\textnormal{sin}^{2}\theta\; \textnormal{cos}^{2}\theta}{\textnormal{sin}^{2}\theta - \displaystyle \frac{A^{2}\; \textnormal{cosh}^{-\; 2}\left(\displaystyle \frac{r}{A} \right)}{\theta^{2}r^{2}}} + \frac{\textnormal{sinh}^{2} \left(\displaystyle \frac{r}{A} \right)}{\textnormal{cosh}^{4} \left(\displaystyle \frac{r}{A} \right)}
\label{eq_55}
\end{equation}

If $H=H_{0}=72.3$ in (\ref{eq_54}):

\begin{equation}
ds_{catenoid}^{2}\simeq \frac{Z}{4}\; \left[r^{2}\; dt_{FRW}^{2} - 1.91\times 10^{-4}\; dr^{2} \right]
\label{eq_56}
\end{equation}

Also, if $H=H_{1}=H_{0}^{2}$ in (\ref{eq_54}):

\begin{equation}
ds_{catenoid}^{2}\simeq \frac{Z}{4}\; \left[r^{2}\; dt_{FRW}^{2} - 3.66\times 10^{-8}\; dr^{2} \right]
\label{eq_57}
\end{equation}

For simplicity, we will consider now the reduced ansatz of (\ref{eq_54}) (with its angular part) as:

\begin{equation}
ds^{2}=(H^{2}+1)r^{2}\; dt_{FRW}^{2}-\left(dr^{2} + r^{2}\; d\theta^{2} + r^{2}\textnormal{sin}^{2}\theta\; d\chi^{2}  \right)
\label{eq_58}
\end{equation}
\noindent

From this ansatz we will obtain the following Christoffel symbols:

\begin{equation}
\begin{array}{lllllll}
\smallskip \Gamma_{00}^{0}=\displaystyle \frac{H\dot{H}}{H^{2}+1} \\
\Gamma_{00}^{r}=(H^{2}+1)r \\
\Gamma_{0r}^{0}=\displaystyle \Gamma_{r\theta}^{\theta}=\Gamma_{r\chi}^{\chi}=\frac{1}{r} \\
\Gamma_{\theta \theta}^{r}=-\; r \\
\Gamma_{\theta \chi}^{\chi}=\displaystyle \frac{1}{\textnormal{tg}\; \theta} \\
\Gamma_{\chi \chi}^{r}=-\; r\; \textnormal{sin}^{2}\theta \\
\Gamma_{\chi \chi}^{\theta}=-\; \textnormal{cos}\; \theta\; \textnormal{sin}\; \theta
\end{array}
\label{eq_59}
\end{equation}

The non-zero Riemann tensor components will be:

\begin{equation}
\begin{array}{ll}
R_{0\theta 0\theta}=-\; (H^{2}+1)r^{2} \\
R_{0\chi 0\chi}=-\; (H^{2}+1)r^{2}\; \textnormal{sin}^{2}\theta\; ,
\end{array}
\label{eq_60}
\end{equation}
\noindent
and the ones of Ricci tensor:

\begin{equation}
\begin{array}{llll}
R_{00}=2(H^{2}+1) \\
R_{rr}=0 \\
R_{\theta \theta}=-\; 1 \\
R_{\chi \chi}=-\; \textnormal{sin}^{2}\theta\; .
\end{array}
\label{eq_61}
\end{equation}

With the Einstein tensor non-zero components determined:

\begin{equation}
\begin{array}{lllll}
G_{r}^{r}=-\; \displaystyle \frac{2}{r^{2}} \\
G_{\theta}^{\theta}=G_{\chi}^{\chi}=-\; \displaystyle \frac{1}{r^{2}} \\
G_{00}=H^{2}+1 \\
G_{rr}=0 \\
G_{\theta \theta}=G_{\chi \chi}=-\; \displaystyle \frac{1}{2}\; ,
\end{array}
\label{eq_62}
\end{equation}
\noindent
and matter-energy tensor components:

\begin{equation}
\begin{array}{ll}
\smallskip T_{00}=\displaystyle \frac{1}{(H^{2}+1)r^{2}}\; \rho \\
T_{\theta \theta}=\displaystyle \frac{p}{r^{2}}\; ,
\end{array}
\label{eq_63}
\end{equation}
\noindent
in Einstein equations we get the first Friedmann equation for this reduced catenoidal manifold ansatz:

\begin{equation}
8\pi G\; \rho = (H^{2}+1)^{2}r^{2} - \Lambda\; ,
\label{eq_64}
\end{equation}
\noindent
and the second Friedmann equation:

\begin{equation}
8\pi G\; p = -\; \frac{r^{2}}{2} + \Lambda\; .
\label{eq_65}
\end{equation}

From (\ref{eq_64}) and (\ref{eq_65}) we are able to determine the cosmological constant:

\begin{equation}
\Lambda = \frac{8\pi G[\rho+2p(H^{2}+1)^{2}]}{2(H^{2}+1)^{2}-1}\; .
\label{eq_66}
\end{equation}

In the case of this manifold ({\it pure catenoidal hyper-surface}), the cosmological distance is:

\begin{equation}
d_{catenoid}=\frac{\sqrt{Z_{catenoid}}}{2}\; \frac{\sqrt{H^{2}+1}-H}{H^{2}-H\sqrt{H^{2}+1}+1}\; r\; .
\label{eq_67}
\end{equation}

If we make use of the first Friedmann equation (\ref{eq_64}) in (\ref{eq_67}):

\begin{equation}
d_{catenoid}=\frac{\sqrt{Z_{catenoid}}}{2}\; \frac{\sqrt{H^{2}+1}-H}{H^{2}-H\sqrt{H^{2}+1}+1}\; \frac{(8\pi G\; \rho + \Lambda)^{1/2}}{H^{2}+1}\; .
\label{eq_68}
\end{equation}

\subsection{Helicoid}

The ansatz for this manifold is \cite{deus1}: 

\begin{equation}
\!\!\!\!\!\!\!\!\!\!\!\!\!\!\!\!\!\!\!\!\!\!\!\!\!ds_{helicoid}^{2}=Z_{helicoid}\cdot \frac{\phi_{k}^{2}}{\left(t_{k}^{2}+\phi_{k}^{2} \right)^{2}}\; dt_{k}^{2}+Z_{helicoid}\cdot \frac{t_{k}^{2}}{\left(t_{k}^{2}+\phi_{k}^{2} \right)^{2}}\; d\phi_{k}^{2}-Z_{helicoid}\cdot \frac{t_{k}\; \phi_{k}}{\left(t_{k}^{2}+\phi_{k}^{2} \right)^{2}}\; dt_{k}\; d\phi_{k}
\label{eq_69}
\end{equation}
\noindent
where:
\begin{equation}
\!\!\!\!\!\!\!\!\!\!\!\!\!\!\!\!\!\!\!\!\!\!\!\!\!Z_{helicoid} =\textnormal{sinh}^{2}\theta\; \left\{\textnormal{cos}^{2}\left[\textnormal{arctg}\left(\frac{t_{k}}{\phi_{k}} \right) \right] - \frac{\textnormal{cos}\; \theta}{\textnormal{sin}\; \theta}\; \frac{\textnormal{cos}\left[\textnormal{arctg}\left(\displaystyle \frac{t_{k}}{\phi_{k}} \right) \right]}{\textnormal{sin}^{3}\left[\textnormal{arctg}\left(\displaystyle \frac{t_{k}}{\phi_{k}} \right) \right]}\; \left\{\textnormal{sin}^{2}\left[\textnormal{arctg}\left(\frac{t_{k}}{\phi_{k}} \right) \right] + \frac{1}{2} \right\}^{2} \right\}
\label{eq_70}
\end{equation}

From (\ref{eq_12}) and (\ref{eq_13}) we get:

\begin{equation}
r=a\sqrt{t_{k}^{2}+\phi_{k}^{2}}
\label{eq_71}
\end{equation} 
\noindent
and:

\begin{equation}
t_{FRW}=\textnormal{arctg}\left(\frac{t_{k}}{\phi_{k}} \right)\; ,
\label{eq_72}
\end{equation} 
\noindent
with which $t_{k}$ and $\phi_{k}$ can be written in (\ref{eq_15}) form and their derivatives as in (\ref{eq_16}). Then, the ansatz will be:

\begin{equation}
\begin{array}{ll}
\!\!\!\!\!\!\!\!\!\!\!\!\!\!\!\!\!\!\!\!\!\!\!\!\!ds_{helicoid}^{2} &=-\; Z_{helicoid}\; ds_{BHIG}^{2} = \\ &=Z_{helicoid} \left[\displaystyle -\frac{3}{4}\; dt_{FRW}^{2} + \frac{2H^{2}-2H \sqrt{H^{2}+1}+1}{4\; \left(H^{2}-H \sqrt{H^{2}+1}+1 \right)^{2}}  \left(\frac{1}{r^{2}}\; dr^{2} + \textnormal{sin}^{2}\theta \; d\chi^{2} + d\theta^{2}  \right)\right]\; ,
\end{array}
\label{eq_73}
\end{equation} 
\noindent
with the (\ref{eq_70}) transformed as:

\begin{equation}
Z_{helicoid} =\textnormal{sinh}^{2}\theta\; \left[\textnormal{cos}^{2}t_{FRW} - \frac{\textnormal{cos}\; \theta}{\textnormal{sin}\; \theta}\; \frac{\textnormal{cos}\; t_{FRW}}{\textnormal{sin}^{3}t_{FRW}}\; \left(\textnormal{sin}^{2}t_{FRW} + \frac{1}{2} \right)^{2} \right] \; .
\label{eq_74}
\end{equation}

When $H=H_{0}\simeq 72.3$ (\ref{eq_73}) behaves as:

\begin{equation}
ds_{helicoid}^{2}\simeq -\; Z_{helicoid} \left[dt_{FRW}^{2} - 2\times 10^{-4}\; \left(\frac{1}{r^{2}}\; dr^{2} + \textnormal{sin}^{2}\theta \; d\chi^{2} + d\theta^{2}  \right)\right]\; .
\label{eq_75}
\end{equation} 

\begin{figure}
\centering
\leavevmode
\includegraphics[width=13.5cm,height=18.0cm]{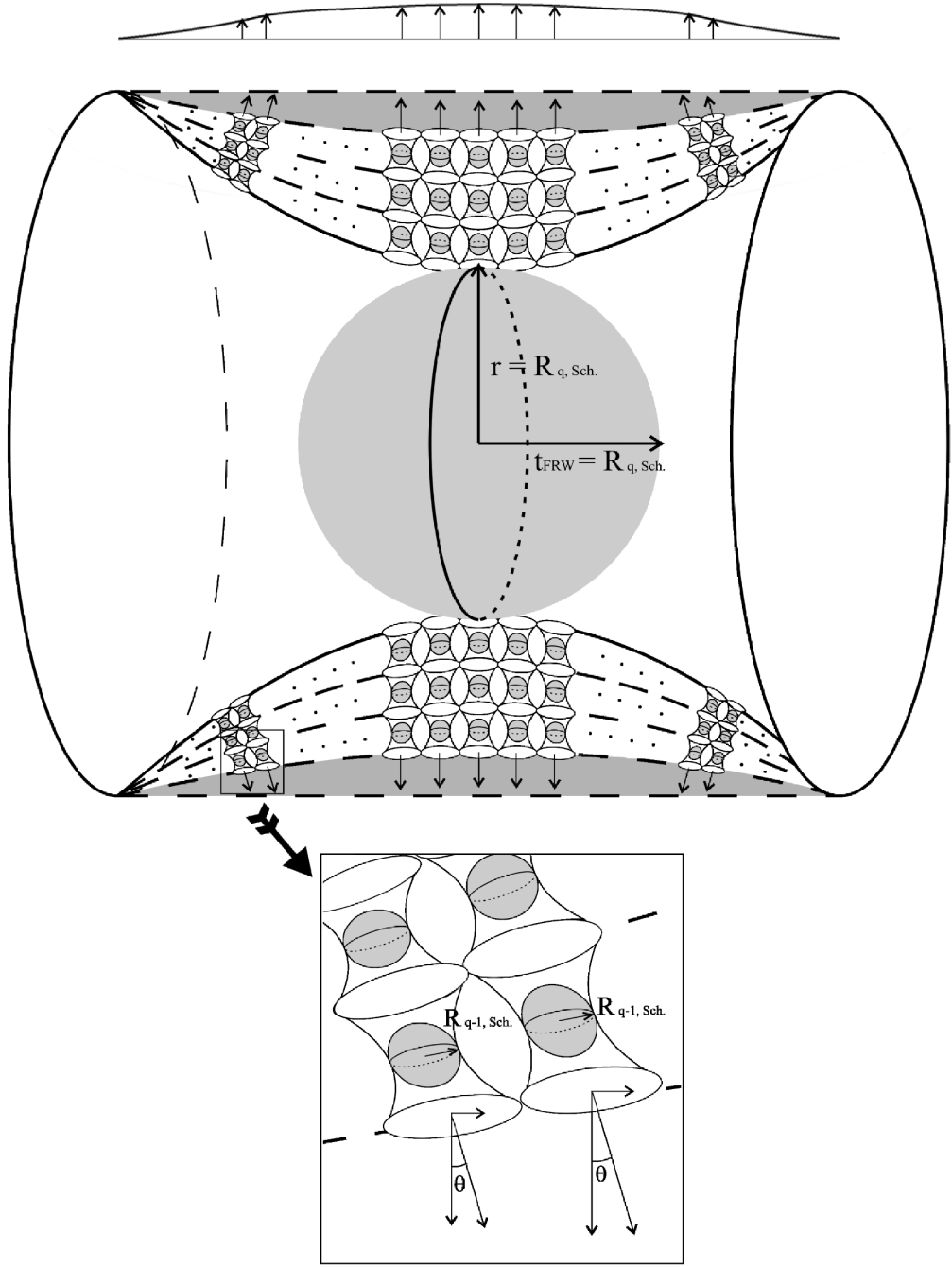}
\caption{Self-similar representation of the before-collapse internal structure (as seen by the external observer) of the DEUS object's catenoidal manifold three-dimensional projection. The symmetry of the $N$-DEUS ($q-1$ order) objects is broken and this perturbation translates in two contributions on the after-collapse $q$ order DEUS string-like rigid object that generates the BHIG helicoidal manifold of a $q+1$ order DEUS object. The zoomed region shows these contributions. The spacelike contribution is perpendicular on the external cylindrical representation of the evolved $q$ order DEUS catenoid, while the timelike contribution is immersed in it. Because, after the collapse of the $q$ order DEUS object, for the BHIG of the $q+1$ DEUS object the time and the space significance are reversed, also these contributions are having reversed significance.}
\label{autosim}
\end{figure}

We will make use here for the second time of the self-similarity (first time was when we observed that the collapsed DEUS object gives an uni-dimensional rotating rigid string-like object which generates the BHIG helicoid of another DEUS space) by considering, apparently for granted, that in the catenoidal wall is existing a limited number $N$ of collapsed DEUS objects (see Fig. \ref{autosim}). We will see that this choice will payoff later. In the figure, the graphical representation is for the way in which the external observer sees the before collapse (into the rotating uni-dimensional string-like BHIG object on which the external observer is situated) $N$-DEUS spacetimes of the DEUS five-dimensional object. By contrast, for the internal observer situated on the pure catenoidal manifold, the space axis of the external observer will be his time axis and the time axis of the external observer his space axis. So, in Fig. \ref{autosim} the $N$-DEUS objects will be represented as rotated with 90$^{\circ}$ from the external observer case. While in the internal observer case the $N$-DEUS objects are linked one to the other at the pure helicoid ends, for the external observer they are linked at the catenoid-cylinder ends. From all above results that, the BHIG external observer in FRW coordinates sees the pure $N$-helicoids, and not the $N$-catenoids (spacelike for him), and these with $t_{FRW}\rightarrow r$ and $r\rightarrow t_{FRW}$. All the previous $ds_{helicoid}^{2}$ where for an internal observer. For the external observer we will have:       

\begin{equation}
ds_{EM}^{2}=Z_{EM} \left[-\frac{3}{4}\; dr^{2} + \frac{2H^{2}-2H \sqrt{H^{2}+1}+1}{4\; \left(H^{2}-H \sqrt{H^{2}+1}+1 \right)^{2}}  \left(\frac{1}{t_{FRW}^{2}}\; dt_{FRW}^{2} + \textnormal{sin}^{2}\theta \; d\chi^{2} + d\theta^{2}  \right)\right]\; ,
\label{eq_76}
\end{equation} 
\noindent
where:

\begin{equation}
Z_{EM} =\textnormal{sinh}^{2}\theta\; \left[\textnormal{cos}^{2}r - \frac{\textnormal{cos}\; \theta}{\textnormal{sin}\; \theta}\; \frac{\textnormal{cos}r}{\textnormal{sin}^{3}r}\; \left(\textnormal{sin}^{2}r + \frac{1}{2} \right)^{2} \right] \; .
\label{eq_77}
\end{equation}

The cosmological distance for an external observer seeing an event that occurs on the $N$-DEUS helicoids will be given by:

\begin{equation}
d_{EM}=\sqrt{\frac{3}{4}}\int_{0}^{r_{\alpha}}\sum_{q=1}^{N} Z_{EM1}\; \textnormal{cos}\; \gamma_{q}\; dr \; ,
\label{eq_78}
\end{equation}
\noindent
where $\gamma_{q}$ is the tilt angle of the $N$-DEUS helicoid from the external timelike case and:

\begin{equation}
Z_{EM1}=\int_{\theta_{1}}^{\theta_{2}} Z_{EM}\; d\theta\; .
\label{eq_79}
\end{equation}

In (\ref{eq_79}), $\Delta \theta\equiv \theta_{2}-\theta_{1}$ is two times the angular diameter of one $N$-DEUS object as seen from the pre-collapse cylinder. The maximal distance $r_{\alpha}$ at which the external observer is still able to see the integral phenomenon given by the $N$-DEUS helicoids after DEUS object's collapse, can be computed as follows.

First of all we will take no angular term in the ansatz. From his location, the BHIG observer will see the $N$-DEUS effects radially. From the redshift formula:

\begin{equation}
z+1=\displaystyle \frac{1}{\displaystyle \sqrt{1-\frac{v_{r}^{2}}{c^{2}}}}\; ,
\label{eq_80}
\end{equation}
and the radial velocity:

\begin{equation}
v_{r}\equiv \frac{dr}{dt}=\left(\frac{g^{00}}{g^{rr}} \right)^{1/2}=r\; \sqrt{3}\; \frac{H^{2}-H\sqrt{H^{2}+1}+1}{(2H^{2}-2H\sqrt{H^{2}+1}+1)^{1/2}}\; ,
\label{eq_81}
\end{equation}
\noindent
it results:

\begin{equation}
r_{\alpha}=\frac{(2H^{2}-2H\sqrt{H^{2}+1}+1)^{1/2}}{\sqrt{3}(H^{2}-H\sqrt{H^{2}+1}+1)}\; \frac{\sqrt{z(z+2)}}{z+1}\; .
\label{eq_82}
\end{equation}

Now, because before collapse, the {\it ergosphere's catenoid} over-imposes on the {\it pure catenoidal hyper-surface}:

\begin{equation}
\theta=\textnormal{arctg}\left(\frac{t_{k}}{\phi{k}} \right)=A\; \textnormal{cosh}^{-1}\left(\frac{\sqrt{t^{2}+\phi^{2}}}{A} \right) \; ,
\label{eq_83}
\end{equation}
\noindent
which, in FRW coordinates gives:

\begin{equation}
t_{FRW}=A\; \textnormal{cosh}^{-1}\left(\frac{r}{A} \right)\; .
\label{eq_84}
\end{equation}

Because the external observer will perceive the inner horizon of the before collapse DEUS object as a Schwarzschild black hole, $r$ will be equivalent with a Schwarzschild radius and the BHIG time from inside the black hole inner horizon also as a $R_{Sch.}$ (spherical symmetry of the inner horizon and $t_{FRW\; int.}\rightarrow r_{ext.}$). Equation (\ref{eq_84}) rewrites as:

\begin{equation}
\frac{R_{Sch.}}{A}-\textnormal{cosh}^{-1}\left(\frac{R_{Sch.}}{A} \right)=0\; ,
\label{eq_85}
\end{equation}
\noindent
having as solution:

\begin{equation}
A=1.307\; R_{Sch.}\; ,
\label{eq_86}
\end{equation}
\noindent
where:

\begin{equation}
R_{Sch.}\equiv \frac{2GM}{c^{2}}\; .
\label{eq_87}
\end{equation}

The relation (\ref{eq_86}) will be valid also for the $N$-DEUS objects (self-similarity of the DEUS objects):

\begin{equation}
A"=1.307\; R"_{Sch.}\; ,
\label{eq_88}
\end{equation}
\noindent
where $A"$ is the "neck" width and $R"_{Sch.}$ the Schwarzschild radius of a $N$-DEUS object.

From geometrical considerations results:

\begin{equation}
R"_{Sch.}=A\; \frac{\Delta \theta}{2}\; ,
\label{eq_89}
\end{equation}
\noindent
from where, with (\ref{eq_86}):

\begin{equation}
\Delta \theta=\frac{2}{1.307}\; \frac{R"_{Sch.}}{R_{Sch.}}\; .
\label{eq_90}
\end{equation}

(\ref{eq_89}) and (\ref{eq_86}) gives:

\begin{equation}
dr"_{Sch.}=1.307\; R_{Sch.}\; \frac{d\theta}{2}\; ,
\label{eq_91}
\end{equation}
\noindent
from where:

\begin{equation}
\theta=\frac{2}{1.307}\; \frac{r"_{Sch.}}{R_{Sch.}}\; .
\label{eq_92}
\end{equation}

Then, with (\ref{eq_90}), the limit angles for the external observer:

\begin{equation}
\begin{array}{ll}
\theta_{1}=\theta_{ext.}-\displaystyle \frac{\Delta \theta}{2}=\theta_{ext.} - \displaystyle \frac{1}{1.307}\; \frac{R"_{Sch.}}{R_{Sch.}} \\
\theta_{2}=\theta_{ext.}+\displaystyle \frac{\Delta \theta}{2}=\theta_{ext.} + \displaystyle \frac{1}{1.307}\; \frac{R"_{Sch.}}{R_{Sch.}}\; ,
\end{array}
\label{eq_93}
\end{equation}
\noindent
where $\theta_{ext.}=0$ from the rotation "freezing" condition for the external observer seeing an event reaching the DEUS ergosphere. 

By doing the substitution of all the above computed quantities in (\ref{eq_77}) and (\ref{eq_79}):

\begin{equation}
\begin{array}{ll}
Z_{EM1}=\textnormal{cos}^{2}r\; \int_{- \frac{1}{1.307}\; \frac{R"_{Sch.}}{R_{Sch.}}}^{+ \frac{1}{1.307}\; \frac{R"_{Sch.}}{R_{Sch.}}} \displaystyle \textnormal{sinh}^{2}\left(\frac{2}{1.307}\; \frac{r"_{Sch.}}{R_{Sch.}} \right)\; \frac{2}{1.307}\; \frac{1}{R_{Sch.}}\; dr"_{Sch.} - \frac{\textnormal{cos}\; r}{\textnormal{sin}^{3}r}\; \left(\textnormal{sin}^{2}r +\frac{1}{2} \right)^{2} \times \\ \times \displaystyle \int_{- \frac{1}{1.307}\; \frac{R"_{Sch.}}{R_{Sch.}}}^{+ \frac{1}{1.307}\; \frac{R"_{Sch.}}{R_{Sch.}}} \textnormal{sinh}^{2}\left(\frac{2}{1.307}\; \frac{r"_{Sch.}}{R_{Sch.}} \right)\; \textnormal{ctg}\left(\frac{2}{1.307}\; \frac{r"_{Sch.}}{R_{Sch.}} \right)\; \frac{2}{1.307}\; \frac{1}{R_{Sch.}}\; dr"_{Sch.}\; .
\end{array}
\label{eq_94}
\end{equation}

When solving (\ref{eq_94}), the second integral is zero and so: 

\begin{equation}
Z_{EM1}=\frac{\textnormal{cos}^{2}r}{2}\; \left[\textnormal{sinh}\left(\frac{2}{1.307}\; \frac{R"_{Sch.}}{R_{Sch.}} \right) - \frac{2}{1.307}\; \frac{R"_{Sch.}}{R_{Sch.}} \right]\; .
\label{eq_95}
\end{equation}

Then, in (\ref{eq_78}), the sum over all the contributions of the $N$-DEUS helicoids will be: 

\begin{equation}
\sum_{q=1}^{N} Z_{EM1}=\frac{\textnormal{cos}^{2}r}{2}\; \sum_{q=1}^{N}\left[\textnormal{sinh}\left(\frac{2}{1.307}\; \frac{R"_{q,\; Sch.}}{R_{q,\; Sch.}} \right) - \frac{2}{1.307}\; \frac{R"_{q,\; Sch.}}{R_{q,\; Sch.}} \right]\; ,
\label{eq_96}
\end{equation}
\noindent
where we considered that the five-dimensional DEUS object (our black hole) is, at its turn, part of a non-collapsed catenoidal bridge of higher order linked $N$-DEUS objects (the total number of black holes existing in Universe from its formation to its death).

The rule for the ratio between the $q-1$ order DEUS embedded objects and the embedding object must be then fractal (self-similar and infinite: see Fig. \ref{autosim}) :

\begin{equation}
\frac{R"_{q,\; Sch.}}{R_{q,\; Sch.}}=p\; \frac{R"_{q-1,\; Sch.}}{R_{q-1,\; Sch.}}\; ,
\label{eq_97}
\end{equation}
\noindent
with {\it p} a scaling factor and the embedded DEUS object Schwarzschild radius $R"_{q,\; Sch.}\equiv R_{q-1,\; Sch.}$.

From (\ref{eq_97}) results that:

\begin{equation}
\frac{R"_{q,\; Sch.}}{R_{q,\; Sch.}}=p^{q-1}\; \frac{R"_{1,\; Sch.}}{R_{1,\; Sch.}}\; .
\label{eq_98}
\end{equation}

In (\ref{eq_96}), when $N\rightarrow \infty$, the sum over $Z_{EM1}$ goes to zero. This means that, either the number of observable DEUS higher or lower orders (levels) is not infinite either, if it is, the cosmological distance for seeing from exterior (from BHIG spacetime or FRW spacetime) an event occurring on the {\it pure catenoidal hyper-surface}'s $N$-{\it pure helicoidal hyper-surfaces} embedded ends of the collapsed DEUS object (our black hole) is zero. We will see later in the model that {\it N} is not infinite and, also, that the number of visible DEUS levels is at least four.

\section{Second Fundamental Form}

The mean curvature for our manifolds can be written as:

\begin{equation}
\mathcal{H} = \frac{1}{2} \frac{Eg - 2Ff + eG}{EG - F^{2}}\; ,
\label{eq_99}
\end{equation}
\noindent
where the coefficients:

\begin{equation}
\begin{array}{lll}
\smallskip e = \displaystyle \frac{\hat{x}_{u}\; \hat{x}_{u}\; \hat{x}_{uu}}{\sqrt{EG-F^{2}}} \\ \smallskip
f = \displaystyle \frac{\hat{x}_{u}\; \hat{x}_{v}\; \hat{x}_{uv}}{\sqrt{EG-F^{2}}} \\
g = \displaystyle \frac{\hat{x}_{v}\; \hat{x}_{v}\; \hat{x}_{vv}}{\sqrt{EG-F^{2}}}\; ,
\end{array}
\label{eq_100}
\end{equation}
\noindent
are in terms of partial derivatives of the local coordinate chart $\hat{x}$ of a Riemannian manifolds and:

\begin{equation}
\begin{array}{llll}
E = g_{11} = \hat{x}_{u}\; \hat{x}_{u} \\
F = g_{12} = \hat{x}_{u}\; \hat{x}_{v} \\
\; \; \; = g_{21} = \hat{x}_{v}\; \hat{x}_{u} \\
G = g_{22} = \hat{x}_{v}\; \hat{x}_{v}\; ,
\end{array}
\label{eq_101}
\end{equation}
\noindent
of the metric:

\begin{equation}
ds^{2} = E\; du^{2} + 2F\; du\; dv + G\; dv^{2}\; .
\label{eq_102}
\end{equation}

When we place ourselves in the five-dimensional {\it catenoidal manifold} (seen as single five-dimensional hyper-surface and without self-similar structure), for which $u = A\; \textnormal{cosh}^{-1}\left(\displaystyle \frac{\sqrt{t^{2}+\phi^{2}}}{A} \right)$, $v=\theta$ \cite{deus1} and:

\begin{equation}
\begin{array}{llll}
g_{11}\equiv g_{tt} \\
g_{22}\equiv g_{\phi \phi} \\
g_{21}\equiv g_{\phi t} \\
g_{12}\equiv g_{t\phi}\; ,
\end{array}
\label{eq_103}
\end{equation}
\noindent
it results that $EG - F^{2} = 0$, meaning that the {\it catenoidal manifolds} in $\hat{x}(t,\phi)$ are behaving as singularities (horizons) with $\mathcal{H}_{catenoid} \rightarrow \infty$. The same hyper-surface, but expressed as $\hat{x}\equiv \hat{x}(r,t_{FRW})$, gives a finite non-zero result for the mean curvature.
 
Instead, if we express the local chart $\hat{x}\equiv \hat{x}(r,t_{FRW})$ of the collapsed {\it catenoidal manifolds} (see (\ref{eq_7})) :

\begin{equation}
\hat{x} = \left(0, 0, \mp\; \textnormal{arctg}\left(\frac{t_{k}}{\phi_{k}} \right), 0, 0 \right) = \left(0, 0, \mp\; t_{FRW}, 0, 0 \right)\; .
\label{eq_104}
\end{equation}
\noindent
we obtain, with the help of (\ref{eq_13}), (\ref{eq_15}) and (\ref{eq_16}), $\hat{x}_{u}\cdot \hat{x}_{u} = 0$ and $\hat{x}_{v} = 0$, from where follows that $\mathcal{H}_{catenoid} = 0$ which is the situation of a {\bf singular minimal surface}. The first observation is that, when the curvature is infinite the {\it catenoidal hyper-surface} is an event horizon for the external observer. This can happen because of the initial self-similar fractal structure (fractality $\equiv$ infinite reproduction of same object at different scales) of the catenoidal bridge, which contains an infinite number of DEUS objects, each one contributing to the total curvature seen by the external observer. Here, the infinity of the DEUS objects does not mean that at one scale we have an infinite number of objects. Even with a finite number of DEUS objects immersed in our {\it catenoidal hyper-surface} we can reach this infinity because the immersed objects have their own {\it catenoidal hyper-surfaces} (scaled at the dimension of this specific DEUS object), and so on.... This means that an object situated on a {\it catenoidal hyper-surface}, is also somewhere on a lower scale DEUS objects' {\it catenoidal hyper-surfaces}... This repeats to infinity. Fortunately we saw that the DEUS objects evaporate when their hyper-surfaces evolve to cylindrical hyper-surfaces. Initially, even that in its reference frame the object moves, from the point of view of the external observer it will move on the first and lower scale (somewhere at infinity), so, it will be seen as {\bf frozen}. Until the lower scale DEUS objects does not evaporate the object cannot be seen as moving. Now, the time in the external observer perspective can be infinite but described by a periodic function (the internal time is correlated with $t_{FRW}$ but not linearly), meaning that, from time to time, we will see (as external observers) the object moving (or becoming a FRW object). Still, this is not the truth because the external observer and his frame is also embedded (as a BHIG spacetime) in a higher scale DEUS object, having a limited "time" to observe the object moving, but not smaller than the lifetime of the lower scale embedded DEUS object which must collapse before or simultaneously the external observer "time" to reach (again) 0. Because after the collapse of a DEUS object the internal time becomes space for the external observer, the internal $t_{FRW}$ will seem uncorrelated with the external one. The external observer will be able to observe at the moment when the mean curvature becomes finite (almost all the lower scale DEUS objects are collapsed) that some objects are moving in his reference frame, but he will not be able to tell what it is about (the space is curved by something invisible), or what moves until the {\it catenoidal hyper-surface} on which these objects live collapses (reaches to conformal flat frame where $\mathcal{H}_{catenoid} = 0$).

In a similar manner, in the five-dimensional {\it helicoidal manifold} cases, where $u=\theta$, $v=\textnormal{arctg}\left(\displaystyle \frac{t_{k}}{\phi_{k}} \right)$ and $\hat{x}\equiv \hat{x}(t_{k},\phi_{k})$, we will have either $EG-F^{2}=0$, either, for $\theta=0$, $Z_{hel}=0$ (see (\ref{eq_70})), which means again singularity in mean curvature: $\mathcal{H}_{helicoid} \rightarrow \infty$.

The situation changes if we compute the mean curvature in the local chart $\hat{x}\equiv \hat{x}(r,t_{FRW})$ of the collapsed {\it helicoidal manifolds} (see (\ref{eq_10})) in FRW coordinates:: 

\begin{equation}
\hat{x} = \left(0, 0, t_{FRW}, \pm\; i\; t_{FRW}, 0 \right)\; ,
\label{eq_105}
\end{equation}
\noindent
for a FRW observer seeing this manifold, where (\ref{eq_13}), (\ref{eq_15}) and (\ref{eq_16}) play an important role in the computation, which gives: 

\begin{equation}
\begin{array}{ll}
\hat{x}_{t_{FRW}} = \left(0, 0, 1, \pm\; i, 0 \right) \\
\hat{x}_{r} = 0\; .
\end{array}
\label{eq_106}
\end{equation}

For the five-dimensional collapsed {\it helicoidal manifold}, expressed in $r$ and $t_{FRW}$, for the external after-collapse observer and using:

\begin{equation}
H\equiv \frac{\dot{a}}{a}=-\; \frac{1}{t_{k}} \; ,
\label{eq_106a}
\end{equation}
\noindent
for the helicoid, when we have the transformation $\phi=-a\; t_{k}$, and:

\begin{equation}
H\equiv \frac{\dot{a}}{a}=\frac{1}{t} \; ,
\label{eq_106b}
\end{equation}
\noindent
for the catenoids of the DEUS lower scale embedded objects, when we have the transformations (\ref{eq_15}) and (\ref{eq_71}) (or $\displaystyle \frac{t}{\phi}=-\; \frac{\phi_{k}}{t_{k}}$), from which we get:

\begin{equation}
\textnormal{tg}\; t_{FRW}= \sqrt{H^{2}+1}-H \; ,
\label{eq_106c}
\end{equation}
\noindent
and also using (\ref{eq_19a}), it results that $\hat{x}_{v}\cdot \hat{x}_{v}=0$ and $\hat{x}_{u}=0$. In consequence, in (\ref{eq_99}), $\mathcal{H}_{helicoid} = 0$, which means that, as for the catenoid, the helicoid is {\bf a minimal surface}. As for the catenoid, an external observer will not see the helicoid (infinite mean curvature) until the evolving {\it catenoidal hyper-surface} in which are embedded the DEUS objects containing these {\it helicoidal hyper-surfaces} (timelike case), does not reach the cylindrical pre-collapse surface. In their turn, these helicoids contain an infinite number of lower scale DEUS objects, first scale being placed as "spikes" on their FRW bubble representation.

As we previously saw, another important manifold of the DEUS geometry is the BHIG five-dimensional hyper-surface in which, for a FRW observer, $\hat{x} = \left(0, 0, t_{FRW}, \pm\; r, 0 \right)$, where we considered the $r_{ext.}=i\; t_{FRW,\; int.}$ transformation from BHIG internal geometry of the $q$ order DEUS spacetime collapsing in a $q + 1$ order DEUS spacetime's BHIG representation. It follows, in a trivial manner that $\hat{x}_{t_{FRW}} = \left(0, 0, 1, 0, 0 \right)$ and $\hat{x}_{r} = \left(0, 0, 0, \pm\; 1, 0 \right)$, with $t_{FRW}\equiv t_{FRW,\; ext.}$ and $r\equiv r_{ext.}$ of the $q+1$ order DEUS spacetime FRW representation of BHIG. In ($t_{k}, \phi_{k}$) coordinates, when (\ref{eq_19a}) and (\ref{eq_106c}) are satisfied, $\displaystyle \frac{\partial\; x^{3}}{\partial\; v} = 4$, $\hat{x}_{v}\cdot \hat{x}_{v}=16$ and $\hat{x}_{u}=0$, from where $\hat{x}_{vv}=0$. The last result in (\ref{eq_99}), will be equivalent with having $\mathcal{H}_{FRW}=0$ of a {\bf minimal surface}, in which the above catenoidal and helicoidal cases evolve.

\section{Number of Real Catenoidal Manifolds and Stability Criterion}

A three-dimensional surface given by the function $u\in \mathbb{R}^{n}$ ({\it n} = 3), in order to be a minimal surface, must satisfy the Dirichlet problem \cite{dor}:

\begin{equation}
\nabla \cdot \left(\frac{\nabla u}{\sqrt{1+|\nabla u|^{2}}} \right) = 0\; ,
\label{eq_107}
\end{equation}
\noindent
for which $u=A\; \textnormal{cosh}^{-1}\left(\displaystyle \frac{u_{1}}{A} \right)$ \cite{deus1} is a solution.

Departing from the above equation, the singular minimal surface equation \cite{dier} is: 

\begin{equation}
\textnormal{div} \left(\frac{Du}{\sqrt{1+|Du|^{2}}} \right) = \mathcal{H}\; ,
\label{eq_108}
\end{equation}
\noindent
with {\it D} the derivatives in the Riemannian manifold and the mean curvature given by:

\begin{equation}
\mathcal{H}=\frac{\alpha}{u\sqrt{1+|Du|^{2}}}\; .
\label{eq_109}
\end{equation}

In (\ref{eq_109}), $|\alpha|\in \mathbb{N}$ is giving the number of possible hyper-surfaces.

There exist entire solutions (defined on $\mathbb{R}^{n}$): 

\begin{equation}
s\equiv s_{n}^{\alpha}(\hat{x})=\sqrt{\frac{\alpha}{n-1}}[(x^{1})^{2}+\cdots+(x^{n})^{2}]^{1/2}\; ,
\label{eq_110}
\end{equation}
\noindent
for any $\alpha > 0$ and $n\geq 2$, named the {\it weak Lipschitz-solutions}. If $n=5$ (five-dimensional surface) and we choose $\alpha=3$ then:
 
\begin{equation}
s=\sqrt{\frac{3}{4}}[(x^{1})^{2}+\cdots+(x^{5})^{2}]^{1/2}\; .
\label{eq_111}
\end{equation}

In our case, for the catenoidal hyper-surfaces, $r_{int.}^{2}\equiv t^{2}+\phi^{2}\equiv (x^{1})^{2}+\cdots+(x^{5})^{2}$, but because after the collapse $r_{int.}\rightarrow t_{FRW,\; ext.}$ and $t_{FRW\; int.}\rightarrow r_{ext.}$ the previous equation rewrites as:

\begin{equation}
s=\sqrt{\frac{3}{4}}\; t_{FRW,\; ext.}\; .
\label{eq_112}
\end{equation}

We observe that the squared derivative of (\ref{eq_112}):

\begin{equation}
ds^{2}=\frac{3}{4}\; dt^{2}_{FRW\; ext.}\; ,
\label{eq_113}
\end{equation}
\noindent
is {\bf identic} with the equation (\ref{eq_29}) temporal part of the FRW representation of the catenoidal hyper-surface. In conclusion, the number of possible real (discrete) {\it pure catenoidal hyper-surfaces} is four (this including also the $\alpha=0$ situation for which $\mathcal{H}=0$ of the minimal catenoid-like hyper-surface adjoint to the ergosphere). There exist the possibility of negative $\alpha$ values ($\alpha = -1,\; -2,\; -3$) interpretable as virtual catenoiodal surfaces inside the ergosphere ({\it ergosphere's catenoids}; \cite{deus1}). Further on, the lack of stability of the catenoidal hyper-surfaces for $\alpha \neq 0$, will determine geometric flows from $\alpha=3$ (FRW conformal space) to $\alpha=0$ ({\it pure catenoidal hyper-surface}), situation in which the tendency is the global minimization of the energy and entropy of the considered hyper-surface (in the catenoid external observer representation; $\mathcal{H}\rightarrow 0$). The excess of energy will be released in the external spacetime as spacelike and timelike effects, pre and after-collapse of the DEUS object. Due to the "mirror effect" of the inner horizon, the pre-collapse internal observer of the BHIG space will see that the catenoidal hyper-surface evolves from $\alpha=0$ to $\alpha=3$ (global increase and local decrease of entropy).

The stability of the catenoidal hyper-surfaces is given through the {\it Vogel's stability criterion} \cite{zhou}. Let $f(u_{1})$ denote the profile of the hyper-surface, $-h\leq u_{1}\leq h$ and $2h$ the hight of the catenoidal bridge. Then {\it f} is a solution of the following Euler-Lagrange equation:

\begin{equation}
-\; \frac{1}{2}\left\{\frac{f"}{\left[1+(f')^{2} \right]^{3/2}} - \frac{1}{f\left[1+(f')^{2} \right]^{1/2}} \right\}=\mathcal{H}\; ,
\label{eq_114}
\end{equation}
\noindent
where $\mathcal{H}$ is the mean curvature of the hyper-surface. When the mean curvature $\mathcal{H}=0$ the solution of (\ref{eq_114}) is a catenoid:

\begin{equation}
f(u_{1})=A\; \textnormal{cosh}\left(\frac{u_{1}}{A} \right)\; ,
\label{eq_115}
\end{equation}
\noindent
where, in our parameterization, $u_{1}=\sqrt{t^{2}+\phi^{2}}=r$.
From the Vogel's stability criterion \cite{zhou} results that we have:

\begin{enumerate}
\item a stable catenoidal hyper-membrane if:

\begin{equation}
\int_{-h}^{h} \Psi(r)\; f(r)\; dr < 0\; ,
\label{eq_116}
\end{equation}
\noindent
with:

\begin{equation}
\Psi(r) = -\; \frac{1}{n}\; \textnormal{arcsin}\left(\frac{E}{r^{n}} \right)\; ,
\label{eq_117}
\end{equation}
\noindent
where $\Psi \in \left(\displaystyle -\frac{\pi}{n},\; 0 \right)$, {\it n} the spacetime dimension and {\it E} the energy of our catenoidal hyper-membrane. For $n=5$ and $A > 0$ results that the catenoid is stable when:

\begin{equation}
\int_{-h}^{h} \textnormal{arcsin}\left(\frac{E}{r^{5}} \right)\; \textnormal{cosh}\left(\frac{r}{A} \right)\; dr > 0 \; ,
\label{eq_118}
\end{equation}
\noindent
with $A=1.307\; R_{Sch.}$ and $R_{Sch.}$ the Schwarzschild radius of the DEUS black hole;

\item an unstable catenoidal hyper-membrane if:

\begin{equation}
\int_{-h}^{h} \textnormal{arcsin}\left(\frac{E}{r^{5}} \right)\; \textnormal{cosh}\left(\frac{r}{A} \right)\; dr < 0 \; .
\label{eq_119}
\end{equation}
\end{enumerate}

The energy {\it E} of the {\it q}-order catenoidal hyper-surface will be equal with the sum over all the internal $q-1$ order, collapsed, self-similar $N$-DEUS objects (see Fig. \ref{autosim}):

\begin{equation}
E = \sum_{t=1}^{P} \left(E_{q-1,\; catenoid} + E_{q-1,\; helicoid} \right) \;.
\label{eq_120}
\end{equation}

\section{Conclusions}

In this paper we generalized the geometry constructed in {\it DEUS I} \cite{deus1} paper for Universe formation and evolution, by using self-similar minimal hyper-surfaces for DEUS objects distribution and we shown that the DEUS spacetime collapses to a string-like object equivalent with the BHIG spacetime of a bigger DEUS object in which the first one is embedded.

The considered five-dimensional geometry becomes conformal with an external Friedmann-Robertson-Walker (FRW) spacetime (for which the Hubble constant is {\it uniquely} defined as $72.33\pm 0.02$) by evolving the hyper-surfaces composing our five-dimensional DEUS object to a conformal flat manifold. For the BHIG spacetime written in FRW coordinates we derived the Friedmann equations and some cosmological consequences ($\Omega_{tot}$ and the cosmological distance). The pre-collapse movement of the five-dimensional catenoid hyper-surfaces is perceived by an after-collapse external observer as {\bf dark energy effect}. 

Because, after the DEUS object collapses, the internal matter and matter distribution will be "hidden" to the FRW external observer, this matter is possible to be considered as a {\bf dark matter component}.

The collapsing helicoidal manifold can be seen by the external timelike observer only considering that the catenoid is having $N$ internal DEUS similar objects, distorted, also collapsing. The helicoids of these $N$-DEUS objects are timelike (five-dimensional) and are giving three-dimensional spacelike and timelike observable effects to the external space.

\ack
I want to express all my gratitude to Stefan Sorescu whose computer drawing experience was truly valuable in this paper.

\noindent
Manuscript registered at the Astronomical Institute, Romanian Academy, record No. 250 from 04/19/07.


\title[D.E.U.S.]{III: Dynamics and Kinematics of DEUS Manifolds}

\begin{abstract}
Departing from the geometry of the DEUS object described in the papers {\it DEUS I} and {\it DEUS II}, we will study the way in which the {\it catenoidal hyper-surface} and the {\it helicoidal hyper-surface} are evolving, which are the energies and the velocities of these evolving hyper-surfaces and how the external observer will perceive them. All this development will be made on the background of geometrical flows (Yamabe flow, Ricci flow, mean curvature flow and inverse mean curvature flow).
\end{abstract}


\section{Introduction}

In paper {\it DEUS I} \cite{deus1} we constructed a five-dimensional spacetime geometry having as starting point the trivial three-dimensional minimal surface representation of the catenoid and its conjugate surface, the helicoid. There we determined the five-dimensional coordinate system that satisfies the Cauchy-Riemann equations, and also the ansatz, the properties and the correlations possible to be made for the DEUS object's composing manifolds.

In {\it DEUS II}  paper \cite{deus2} we generalized the geometry of the DEUS Black Hole for Universe formation and evolution, by using self-similar minimal surfaces for DEUS objects distribution. From the assumed self-similarity of the five-dimensional internal geometry, we shown that the DEUS object collapses to a string-like object equivalent with the BHIG spacetime of a bigger DEUS object in which the first one is embedded.

The considered five-dimensional geometry becomes conformal with an external Friedmann-Robertson-Walker (FRW) spacetime (for which the Hubble constant is {\it uniquely} defined as $72.33\pm 0.02$) by evolving the hyper-surfaces composing our five-dimensional DEUS object to a pre-collapse manifold. For the BHIG spacetime written in FRW coordinates we derived the Friedmann equations and some cosmological consequences ($\Omega_{tot}$ and the cosmological distance).

Into the background of a collapsed DEUS object rotating string that creates a BHIG manifold "deformed" in a FRW bubble, the pre-collapse movement of the five-dimensional catenoid hyper-surfaces is perceived by an after-collapse external observer as dark energy effect, while the matter still contained in the collapsed DEUS spacetime as dark matter.

The collapsing helicoidal manifold can be seen by the external timelike observer only considering that the catenoid is having $N$ internal DEUS similar objects, distorted, also collapsing. The helicoids of these $N$-DEUS objects are timelike (five-dimensional) and are giving three-dimensional spacelike and timelike observable effects to the external space.

In the first section of this paper we find that the energy contained into the {\it catenoidal hyper-surface} and into the {\it helicoidal hyper-surface} of the above geometry can be characterized using the Yamabe energy.

In the second section we show that the energy stored into the curvature of the {\it catenoidal hyper-surface} is released (at the DEUS object evaporation) into the external space as kinetic energy in form of spatial or temporal perturbations. Also, by applying a combination of Ricci flow and mean curvature flow and taking account of the conformity of the {\it reduced catenoidal hyper-surface} \cite{deus2} with the flat external space, we obtain the scalar curvature of the {\it pure catenoidal hyper-surface}, for which the metric was too complicated to be determined in the "classical" way.

Using the mean curvature flow, in section three we determine the velocity with which the {\it catenoidal hyper-surface} evolves to the FRW external space and the velocity of the BHIG's FRW bubble from its formation ("Big Bang") to its disappearance ("Big Crunch"). As we said in {\it DEUS II}, the external observer does not observe directly this movement, perceiving only its inertial effect, as dark energy. In the Lorentz frame of the external observer these velocities are comparable with the infall velocity in the core collapse process of a star becoming a black hole.

\section{Yamabe Flow}

Let ($M,g_{0}$) be a smooth, compact Riemannian manifold without boundary of dimension $n\geq 3$ and with scalar curvature $R_{0}=R_{g_{0}}$. A conformal change of the metric $g_{0}$ produces a metric:

\begin{equation}
g=u^{\frac{4}{n-2}}g_{0}\; ,
\label{eq_0}
\end{equation}
\noindent
having the scalar curvature \cite{sch}:

\begin{equation}
R=R_{g}=u^{-\; \frac{n+2}{n-2}}(-c_{n}\Delta_{0}u + R_{0}u)\; ,
\label{eq_1}
\end{equation}
\noindent
where $c_{n}=4\; \displaystyle\frac{n-1}{n+2}$ and where $\Delta_{0}\equiv \Delta_{g_{0}}$ is the Laplace-Beltrami operator in the metric $g_{0}$. The Laplace-Beltrami operator on a closed {\it n}-dimensional Riemannian manifold is defined in any local coordinates as \cite{safa}:

\begin{equation}
\Delta u(\hat{x})=g^{-1/2}(\hat{x})\sum_{i,j}\partial_{x^{i}}\left[g(\hat{x})\; g^{ij}(\hat{x})\; \partial_{x^{j}}\left(g^{-1/2}(\hat{x})\; u(\hat{x}) \right) \right]\; , 
\label{eq_2}
\end{equation}
\noindent
where $g^{ij}$ is the metric tensor and $g\equiv \left|\textnormal{det}\left\{g^{ij}\right\}\right|^{-1/2}$ is the canonical Riemannian density.

\subsection{Catenoidal Hyper-Surface}

While for the catenoid observer the catenoid evolves from ${\cal H}\neq 0$ toward ${\cal H}= 0$, for the external observer of the DEUS object it will be seen as evolving according to causality fulfillment of the external BHIG-FRW spacetime, so from ${\cal H}= 0$ toward ${\cal H}\neq 0$ (toward the DEUS collapse cylinder).

For a FRW external observer, our catenoidal manifold evolves from the {\it pure catenoidal hyper-surface} minimal representation with $\mathcal{H}= 0$ ($\alpha=0$ discrete hyper-surface) and $(g^{ij})_{0}$, for which \cite{deus2}:

\begin{equation}
ds_{catenoid}^{2}=\frac{Z}{4}\; \left(\frac{\sqrt{H^{2}+1}-H}{H^{2}-H\sqrt{H^{2}+1}+1} \right)^{2}\; \left[r^{2}(H^{2}+1)\; dt_{FRW}^{2}-dr^{2} \right] \; ,
\label{eq_3}
\end{equation}
\noindent
where:

\begin{equation}
Z_{catenoid} =\textnormal{cos}^{2}\theta - \frac{\textnormal{sin}^{2}\theta\; \textnormal{cos}^{2}\theta}{\textnormal{sin}^{2}\theta - \displaystyle \frac{A^{2}\; \textnormal{cosh}^{-\; 2}\left(\displaystyle \frac{r}{A} \right)}{\theta^{2}r^{2}}} + \frac{\textnormal{sinh}^{2} \left(\displaystyle \frac{r}{A} \right)}{\textnormal{cosh}^{4} \left(\displaystyle \frac{r}{A} \right)} \; ,
\label{eq_4}
\end{equation}
\noindent
to the {\it reduced} catenoidal ansatz, conformal with the external BHIG space in FRW coordinates (as we saw in paper \cite{deus1}), for which $\mathcal{H}\neq 0$ ($\alpha=3$ discrete hyper-surface) and the metric tensors $g^{ij}$:

\begin{equation}
ds^{2}=(H^{2}+1)r^{2}\; dt_{FRW}^{2}-\left(dr^{2} + r^{2}\; d\theta^{2} + r^{2}\textnormal{sin}^{2}\theta\; d\chi^{2}  \right)\; .
\label{eq_5}
\end{equation}

The scalar curvature flow converges from $R_{0}$ of the {\it pure catenoidal hyper-surface} to $R=\displaystyle\frac{4}{r^{2}}$ of the {\it reduced catenoidal hyper-surface}. In this case, the Laplace-Beltrami operator writes:

\begin{equation}
\Delta_{0} u(\hat{x})=g_{0}^{-1/2}(\hat{x})\sum_{i,j}\partial_{x^{i}}\left[g_{0}(\hat{x})\; \left[g^{ij}(\hat{x})\right]_{0}\; \partial_{x^{j}}\left(g_{0}^{-1/2}(\hat{x})\; u(\hat{x}) \right) \right]\; ,
\label{eq_6}
\end{equation}
\noindent
where $g_{0}\equiv \left|\textnormal{det}\left\{(g^{ij})_{0}\right\}\right|^{-1/2}$ (for the {\it pure catenoidal hyper-surface}). Also, for the {\it reduced catenoidal hyper-surface}, $g\equiv \left|\textnormal{det}\left\{g^{ij}\right\}\right|^{-1/2}$. From $R=\displaystyle\frac{4}{r^{2}}$ and (\ref{eq_1}), for the five-dimensional case ({\it n} = 5), results that:

\begin{equation}
R_{0}=\frac{4}{r^{2}}\; u^{4/3}+\frac{16}{3}\; \frac{\Delta_{0}u}{u} \; .
\label{eq_7}
\end{equation}

While for $\alpha=3$ hyper-surface $\Delta_{g}u=0$, for $\alpha=0$ hyper-surface (with $\mathcal{H}$=0) $\Delta_{g_{0}}u\equiv \Delta_{0}u\neq 0$. From the expression of the metric tensors we get the canonical Riemannian densities:

\begin{equation}
\begin{array}{ll}
g_{0}=\left[\displaystyle\frac{4}{Z_{catenoid}}\left(\frac{H^{2}-H\sqrt{H^{2}+1}+1}{\sqrt{H^{2}+1}-H} \right)^{2} \right]^{2}\displaystyle\frac{1}{r^{3}\; \sqrt{H^{2}+1}\; \textnormal{sin}\; \theta} \\
g=\displaystyle\frac{1}{r^{3}\; \sqrt{H^{2}+1}\; \textnormal{sin}\; \theta} \; .
\end{array}
\label{eq_8}
\end{equation}

From equations (\ref{eq_0}) and (\ref{eq_8}) follows that ($n=5$):

\begin{equation}
u=\frac{Z_{catenoid}^{3/2}}{8}\; \left(\frac{\sqrt{H^{2}+1}-H}{H^{2}-H\sqrt{H^{2}+1}+1} \right)^{3} \; ,
\label{eq_10}
\end{equation}
\noindent
with $Z_{catenoid}$ as seen from FRW space \cite{deus1,deus2}:

\begin{equation}
Z_{catenoid}=1+\frac{\textnormal{sinh}^{2}\left(\displaystyle\frac{r}{A}\right)}{\textnormal{cosh}^{4}\left(\displaystyle\frac{r}{A} \right)} \; .
\label{eq_9}
\end{equation}

The heat flow \cite{chow,hamilton,ye}, first introduced by Hamilton, for the Yamabe energy $E_{Yamabe}\equiv s$ is:

\begin{equation}
u_{t}=(s-R)u \; .
\label{eq_11}
\end{equation}

It results that:

\begin{equation}
E_{Yamabe}=\frac{1}{u}\; \frac{\partial u}{\partial t}+R \; .
\label{eq_12}
\end{equation}

With the above equation and {\it u} found in (\ref{eq_10}) we get:

\begin{equation}
\begin{centering}
\begin{array}{lll}
E_{Yamabe,\; catenoid}=\displaystyle \frac{4}{r^{2}} + 3\left[1+\frac{\textnormal{sinh}^{2}\left(\displaystyle \frac{r}{A}\right)}{\textnormal{cosh}^{4}\left(\displaystyle \frac{r}{A} \right)} \right]^{-1}\frac{H_{0}^{2}-H_{0}\sqrt{H_{0}^{2}+1}+1}{\sqrt{H_{0}^{2}+1}-H_{0}}\; \times \\ \times\; \left\{\left[1+\frac{\displaystyle \textnormal{sinh}^{2}\left(\displaystyle \frac{r}{A}\right)}{\displaystyle \textnormal{cosh}^{4}\left(\displaystyle \frac{r}{A} \right)} \right]^{2}\; H_{0}^{3}\; \displaystyle\frac{2H_{0}^{2}-2H_{0}\sqrt{H_{0}^{2}+1}+1}{\sqrt{H_{0}^{2}+1}\left(H_{0}^{2}-H_{0}\sqrt{H_{0}^{2}+1}+1\right)^{2}} + \right. \\ \left. +\; \displaystyle\frac{\sqrt{2}}{A}\; \frac{\sqrt{H_{0}^{2}+1}-H_{0}}{\left(H_{0}^{2}-H_{0}\sqrt{H_{0}^{2}+1}+1\right)^{1/2}}\; \frac{\textnormal{sinh}\left(\displaystyle \frac{r}{A} \right)}{\textnormal{cosh}^{5}\left(\displaystyle\frac{r}{A} \right)}\; \left[1-\textnormal{sinh}^{2}\left(\frac{r}{A} \right) \right]\right\} \; ,
\end{array}
\end{centering}
\label{eq_13}
\end{equation}
\noindent
where we considered a present time $H=H_{0}$ value of the Hubble constant.

Because we saw \cite{deus2} that it must exist a correspondence between the temporal scale factor {\it A} of the catenoid and the external spatial scale factor {\it a} of the form $a=i\;A$, we can express the above Yamabe energy for a FRW observer as :

\begin{equation}
E_{Yamabe,\; catenoid}(a,r)=\Re(E_{Yamabe,\; catenoid}(a,r))+i\; \Im(E_{Yamabe,\; catenoid}(a,r)) \; ,
\label{eq_14}
\end{equation}
\noindent
where:

\begin{equation}
\begin{array}{llll}
\Re(E_{Yamabe,\; catenoid}(a,r)) = \displaystyle\frac{4}{r^{2}} + 3\; \left[1+\frac{\textnormal{sin}^{2}\left(\displaystyle\frac{r}{a}\right)}{\textnormal{cos}^{4}\left(\displaystyle\frac{r}{a} \right)} \right]\; \displaystyle\frac{H_{0}^{2}-H_{0}\sqrt{H_{0}^{2}+1}+1}{\sqrt{H_{0}^{2}+1}-H_{0}}\; \times \\ \times\; H_{0}^{3}\; \displaystyle\frac{2H_{0}^{2}-2H_{0}\sqrt{H_{0}^{2}+1}+1}{\sqrt{H_{0}^{2}+1}\; \left(H_{0}^{2}-H_{0}\sqrt{H_{0}^{2}+1}+1\right)^{2}} \\
\Im(E_{Yamabe,\; catenoid}(a,r)) =  \displaystyle\frac{3\sqrt{2}}{a}\left(H_{0}^{2}-H_{0}\sqrt{H_{0}^{2}+1}+1\right)^{1/2} \; \times \\ \displaystyle \times\;\left[1+\frac{\textnormal{sin}^{2}\left(\displaystyle\frac{r}{a}\right)}{\textnormal{cos}^{4}\left(\displaystyle\frac{r}{a} \right)} \right]^{-1} \frac{\textnormal{sin}\left(\displaystyle\frac{r}{a} \right)}{\textnormal{cos}^{5}\left(\displaystyle\frac{r}{a} \right)}\; \left[1-\textnormal{sin}^{2}\left(\displaystyle\frac{r}{a} \right) \right] \; .
\end{array}
\label{eq_15}
\end{equation}

\begin{figure}
\centering
\leavevmode
\includegraphics[width=12.0cm,height=10.0cm]{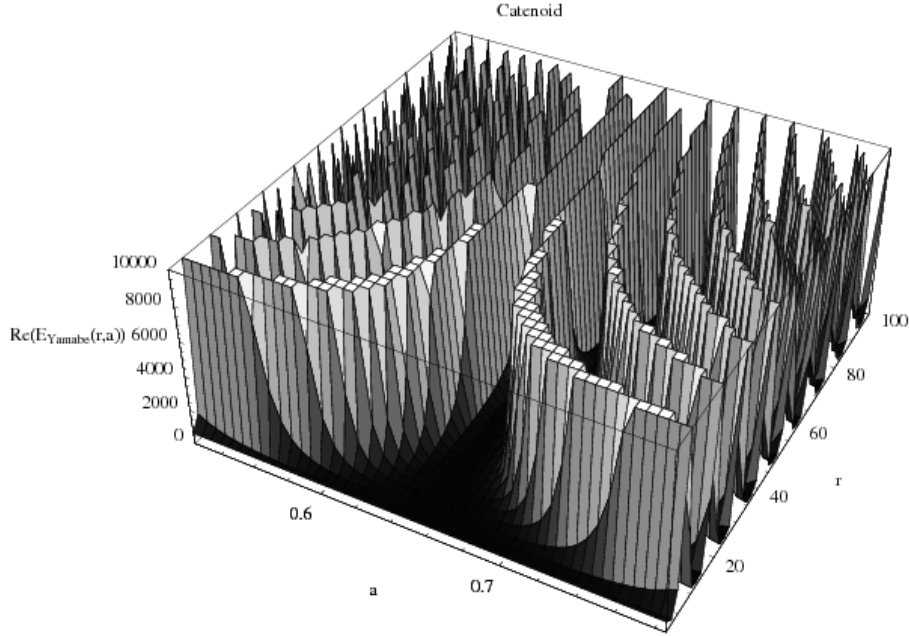}
\caption{Real part of the Yamabe energy (\ref{eq_15}) (timelike emission of energy from the {\it catenoidal hyper-surface} evaporation) seen by the external observer.}
\label{yam1}
\end{figure}

\begin{figure}
\centering
\leavevmode
\includegraphics[width=12.0cm,height=10.0cm]{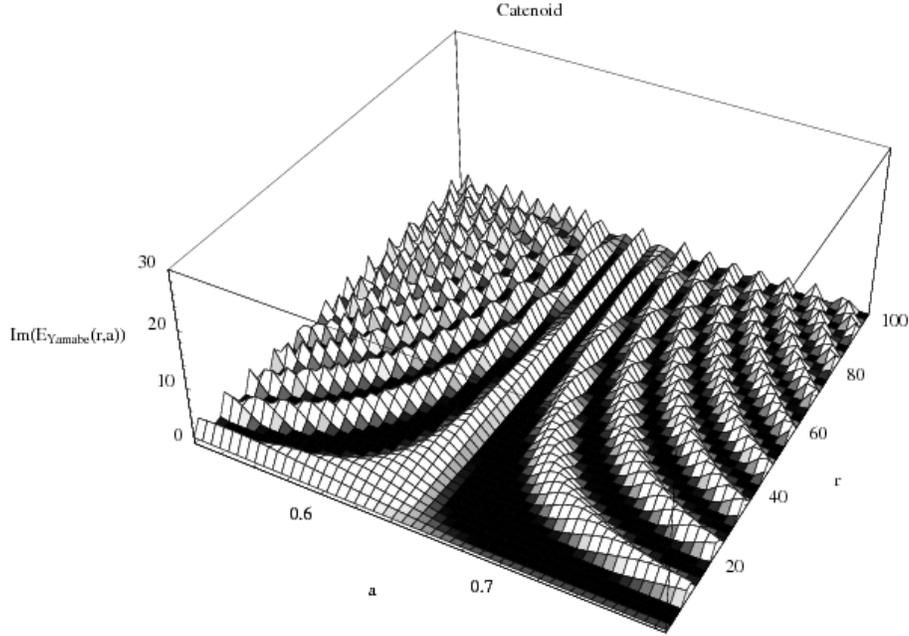}
\caption{Imaginary part of the Yamabe energy (\ref{eq_15}) (spacelike emission of energy from the {\it catenoidal hyper-surface} evaporation) seen by the external observer.}
\label{yam2}
\end{figure}

From the point of view of the external FRW observer, considering the Yamabe energy as being the total energy emitted by the {\it catenoidal hyper-surface}, at its collapse, in FRW external space, the real part $\Re(E_{Yamabe,\; catenoid})$ will be a timelike effect while the imaginary part $\Im(E_{Yamabe,\; catenoid})$ will be a spacelike effect (see Fig. \ref{yam1} and \ref{yam2}).

For the $\Re(E_{Yamabe,\; catenoid})$ representation from Fig. \ref{yam1}, the radial coordinate is arbitrary scaled because we do not dispose {\bf yet} of the "meter-length" to allow us to scale the dimension of the DEUS object in units capable to identify it with a black hole or with the Universe. Still, we can say that the point $r=0$ is the point through which the external FRW observer sees a phenomenon which exits the evaporating DEUS object (the contact point between the {\it catenoidal hyper-surface} and the warped flat spacetime on which this {\it catenoidal hyper-surface} evolves before the collapse to BHIG-FRW spacetime). The {\it catenoidal hyper-surface} is cut at the other end because the other half of the catenoid is across the initial singularity $t_{FRW}=r_{FRW}=0$ at which the FRW is formed, so, outside of the visible Universe (in other worlds, the observer cannot see what happened before the Big Bang or what lies on the "other side" of the black hole). Here, the scale factor {\it A} of the {\it catenoidal hyper-surface} is temporal, being converted to the external observer spatial scale factor by the $a=i\; A$ transformation. The emission from the DEUS object will take place in a very narrow angle (the contact angle between the {\it catenoidal hyper-surface} and its cylindrical boundary, determined from stability constraints on catenoidal bridges) for which $\displaystyle a\in \left[\frac{\pi}{6},\; \frac{\pi}{4} \right]$ times the unit length (the "meter"), but this remains to be proved later, when we will study what is the physics behind these emissions. The same arguments are valid also for $\Im(E_{Yamabe,\; catenoid})$ (Fig. \ref{yam2}). Here, the two major differences from the precedent case are, first that the energy is emitted into the external BHIG-FRW spacetime as spatial perturbation, and second that $10^{8} \leq \displaystyle\frac{\Re(E_{Yamabe,\; catenoid})}{\Im(E_{Yamabe,\; catenoid})} \leq 10^{13}$ for the same pair $(r,a)$.

We saw \cite{deus2} that, at the collapse, the {\it catenoidal  hyper-surface} is over-imposed on the ergosphere (on the same cylindrical hyper-surface). For the ergosphere it is valid the constraint:

\begin{equation}
A^{2}\; \textnormal{cosh}^{-2}\left(\frac{r}{A} \right) + \textnormal{arctg}^{2}\left(\frac{t_{k}}{\phi_{k}} \right)=0\; ,
\label{eqa_1}
\end{equation}
\noindent
with $r=\sqrt{t^{2}+\phi^{2}}$ and the transformation of coordinates $\displaystyle\frac{t}{\phi}=-\; \frac{\phi_{k}}{t_{k}}$. At the cylindrical hyper-surface we have $t_{k}=\pm\; \phi_{k}$, which in (\ref{eqa_1}), gives:

\begin{equation}
A_{max.}=\pm\; \frac{\pi}{4}\; i\; \textnormal{cosh}\left(\frac{r}{A} \right)\; ,
\label{eqa_2}
\end{equation}
\noindent
or, in external external space coordinates $(r,\; t_{FRW})$, {\it r} becoming 0:

\begin{equation}
A=\pm\; \frac{\pi}{4}\; i\; .
\label{eqa_3}
\end{equation} 

For a timelike event in the external spacetime $A_{min.}$ will be $\displaystyle \pm\; i\; \left[\frac{\pi}{4}+\frac{\pi}{12}\right]$, while for a spacelike event it will be $\displaystyle \pm\; i\; \left[\frac{\pi}{4}-\frac{\pi}{12}\right]$, where, again, $a=i\; A$.

\subsection{Helicoidal Hyper-Surface}

Following the same procedure as for the Yamabe energy of the {\it catenoidal hyper-surface}, here the scalar curvature flow converges from the {\it pure helicoidal hyper-surface} with \cite{deus2}:

\begin{equation}
\begin{array}{ll}
ds_{helicoid}^{2}=-\; Z_{helicoid}\; ds_{BHIG}^{2}= \\ = Z_{helicoid} \left[-\; \displaystyle\frac{3}{4}\; dt_{FRW}^{2} + \displaystyle\frac{2H^{2}-2H \sqrt{H^{2}+1}+1}{4\; \left(H^{2}-H \sqrt{H^{2}+1}+1 \right)^{2}}  \left(\frac{1}{r^{2}}\; dr^{2} + \textnormal{sin}^{2}\theta \; d\chi^{2} + d\theta^{2}  \right)\right]\; ,
\end{array}
\label{eq_16}
\end{equation} 
\noindent
where:

\begin{equation}
Z_{helicoid} =\textnormal{sinh}^{2}\theta\; \left[\textnormal{cos}^{2}t_{FRW} - \frac{\textnormal{cos}\; \theta}{\textnormal{sin}\; \theta}\; \frac{\textnormal{cos}\; t_{FRW}}{\textnormal{sin}^{3}t_{FRW}}\; \left(\textnormal{sin}^{2}t_{FRW} + \frac{1}{2} \right)^{2} \right] \; ,
\label{eq_17}
\end{equation}
\noindent
to the BHIG hyper-surface in FRW coordinates ($Z_{helicoid}=1$):

\begin{equation}
\begin{array}{ll}
\!\!\!\!\!\!\!\!\!\!\!\!\!\!\!\!\!\!\!\!\!\!\!\!\!\!\!\!\!\!(ds_{FRW})'\equiv -\; ds_{FRW}^{2}= -\; \displaystyle\frac{3}{4}\; dt_{FRW}^{2} + \displaystyle\frac{2H^{2}-2H \sqrt{H^{2}+1}+1}{4\; \left(H^{2}-H \sqrt{H^{2}+1}+1 \right)^{2}} \left(\frac{1}{r^{2}}\; dr^{2} + \textnormal{sin}^{2}\theta \; d\chi^{2} + d\theta^{2}  \right)\; .
\label{eq_18}
\end{array}
\end{equation}

The scalar curvature is $(R_{FRW})'\equiv -\; R_{FRW}$ and the canonical Riemannian densities are $g_{0}\equiv g_{helicoid}\equiv \left|\textnormal{det}\left\{(g^{ij})_{helicoid}\right\}\right|^{-1/2}$ , $g\equiv (g_{FRW})'=g_{FRW}\equiv \left|\textnormal{det}\left\{(g^{ij})_{FRW}\right\}\right|^{-1/2}$. Results that:

\begin{equation}
\begin{array}{ll}
g_{helicoid}=\displaystyle\frac{\sqrt{3}}{2}\; \left[\displaystyle\frac{\sqrt{H^{2}+1}-H}{2\left(H^{2}-H\sqrt{H^{2}+1}+1 \right)} \right]^{3} Z_{helicoid}^{2} \frac{\textnormal{sin}\; \theta}{r} \\
g_{FRW}= \displaystyle\frac{\sqrt{3}}{2}\; \left[\displaystyle\frac{\sqrt{H^{2}+1}-H}{2\left(H^{2}-H\sqrt{H^{2}+1}+1 \right)} \right]^{3} \frac{\textnormal{sin}\; \theta}{r}\; ,
\end{array}
\label{eq_19}
\end{equation}
\noindent
from where, with the help of (\ref{eq_0}) which, in the present situation (for $n=5$), takes the particular form $(g_{FRW})'=u^{4/3}g_{helicoid}$, we obtain:

\begin{equation}
u=Z_{helicoid}^{-3/2} \; ,
\label{eq_20}
\end{equation}
\noindent
with $Z_{helicoid}$ given by (\ref{eq_17}).

The heat flow for the Yamabe energy $E_{Yamabe,\; helicoid}\equiv s$ is:

\begin{equation}
\frac{\partial u}{\partial t_{k}}=[s-(R_{FRW})']u\; ,
\label{eq_21}
\end{equation}
\noindent
from where:

\begin{equation}
E_{Yamabe,\; helicoid}(u)=\frac{1}{u}\; \frac{\partial u}{\partial t_{k}}+(R_{FRW})' \; .
\label{eq_22}
\end{equation}

By using in the above equation the previously obtained {\it u}, (\ref{eq_20}), the Yamabe energy released by the {\it helicoidal hyper-surface} in the FRW spacetime at the DEUS object collapse will be:

\begin{equation}
\begin{array}{ll}
E_{Yamabe,\; helicoid}(a,r)=\displaystyle\frac{3\sqrt{2}\; a}{r}\; \left(H^{2}-H\sqrt{H^{2}+1}+1 \right)^{-1/2}\; \times \\ \times\; \displaystyle\frac{-4H^{6}+4H^{5}\sqrt{H^{2}+1}-11H^{3}\sqrt{H^{2}+1}+9H^{4}+12H^{2}-6H\sqrt{H^{2}+1}+1}{-20H^{4}\sqrt{H^{2}+1}-21H^{2}\sqrt{H^{2}+1}+20H^{5}+31H^{3}+11H-3\sqrt{H^{2}+1}}\; \times \\ \times \left\{ \begin{array}{ll}
-\; 8\; \displaystyle\frac{\ddot{H}H(H^{2}+1)+\dot{H}^{2}(1-3H^{2})-H^{6}-3H^{4}-3H^{2}-1}{(H^{2}+1)^{2}}\; \; \textnormal{for}\; \;  H=H(t_{FRW}) \\
+ 8(H^{2}+1)\; \; \textnormal{for}\; \; H=const.
\end{array} \right.
\end{array}
\label{eq_23}
\end{equation}

\begin{figure}
\centering
\leavevmode
\includegraphics[width=12.0cm,height=10.0cm]{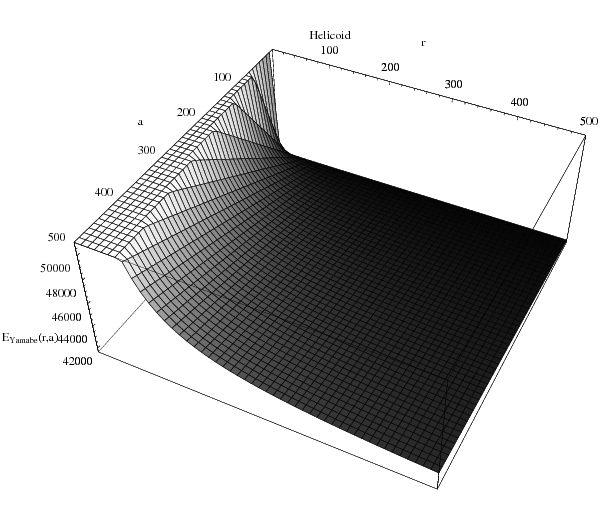}
\caption{The Yamabe energy (\ref{eq_23}) of the {\it helicoidal hyper-surface} seen by the external observer (when $r>0$).}
\label{yam3}
\end{figure}

\begin{figure}
\centering
\leavevmode
\includegraphics[width=12.0cm,height=10.0cm]{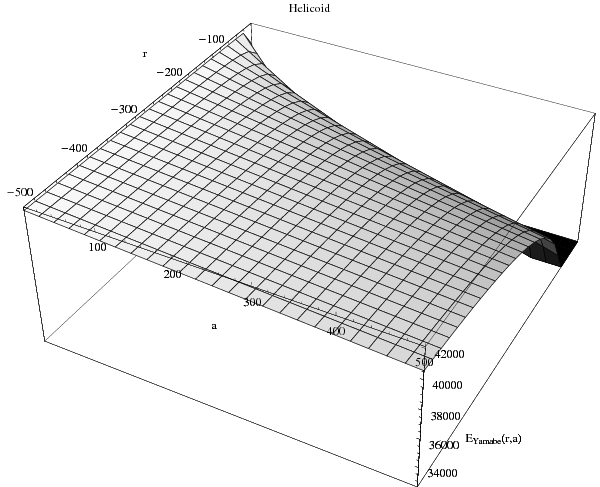}
\caption{The Yamabe energy (\ref{eq_23}) of the {\it helicoidal hyper-surface} seen by the external observer (when $r<0$).}
\label{yam4}
\end{figure}

This energy is represented in Fig. \ref{yam3} and Fig. \ref{yam4}. The Yamabe energy of Fig. \ref{yam3} is just a half of the BHIG helicoid or of the {\it pure helicoidal hyper-surface}, while the other half appears for negative {\it r} (Fig. \ref{yam4}). If initially {\it r} increases from 0 to 500, in the other quarter of period of the {\it helicoid} it will decrease again to 0.  

Here, in the same way as we have done for the {\it catenoidal hyper-surface}, not knowing the total dimension of our spacetime or at least the scale and the units with which to describe our hyper-surface, the best we can do is to give {\it r} and {\it a} (which is here, from the beginning, the spatial scale factor) in arbitrary units.

\section{Ricci Flow and Mean Curvature Flow}

The Ricci flow equation, introduced by Hamilton \cite{hamilton1}, is the evolution equation:

\begin{equation}
\frac{d}{dt}g_{ij}(t)=-2R_{ij} \; ,
\label{eq_24}
\end{equation}
\noindent
for a Riemannian metric $g_{ij}(t)$. Hamilton \cite{hamilton1,hamilton2} proved that the Ricci flow preserves the positivity of the Ricci tensor in dimension three and of the curvature operator in all dimensions.

The Ricci flow has also been discussed in quantum field theory, as an approximation to the renormalization group flow for the two-dimensional nonlinear $\sigma$-model \cite{gaw}. There, the space is described not by some (Riemannian or other) metric, but by a hierarchy of Riemannian metrics, connected by the Ricci flow equation. We can recognize the same apparent paradox as in DEUS model: the regions that appear to be far from each other at large distance scale may become close at smaller distance scale. Moreover, if we allow Ricci flow through singularities, the regions that are in different connected components at larger scale may become neighboring when viewed through microscope \cite{perelman}. In DEUS model this happens, first, because of the collapse of the DEUS object (the global chart becomes a local chart) and second, because of the self-similar construction (the local chart is contained in a similar but global chart) \cite{deus2}.

The connection between the Ricci flow and the renormalization group flow suggests that Ricci flow must be gradient-like \cite{perelman}.

\subsection{Catenoidal Hyper-Surface}

The $g_{ij}$ and $(g_{ij})_{0}$ metric tensors are here the ones defined in the {\it Yamabe Flow} section. In particular, from (\ref{eq_24}) results that:

\begin{equation}
\frac{d}{dt}g_{rr}(t)=-2R_{rr}\equiv -2g_{rr}R
\label{eq_25}
\end{equation}

\begin{equation}
\frac{d}{dt}(g_{rr})_{0}(t)=-2(R_{rr})_{0}\equiv -2(g_{rr})_{0}R_{0} \; ,
\label{eq_26}
\end{equation}
\noindent
but because $R_{rr}=0$, we will have $\displaystyle\frac{d}{dt}g_{rr}$=0, from where we conclude that $g_{rr}$ does not depend on time, which is true and easy to be seen ($g_{rr}=-1$). From equation (\ref{eq_26}) results:

\begin{equation}
R_{0}=-\; \frac{1}{2(g_{rr})_{0}}\; \frac{d}{dt}(g_{rr})_{0} \; .
\label{eq_27}
\end{equation}

We observe that the derivative $\displaystyle\frac{d}{dt}(g_{tt})_{0}=0$, meaning that:

\begin{equation}
R_{0}=-\; \frac{1}{2(g_{tt})_{0}}\; \frac{d}{dt}(g_{tt})_{0}=0 \; .
\label{eq_28}
\end{equation}

But, from (\ref{eq_1}):

\begin{equation}
R=u^{-7/3}\left[\frac{16}{3}\left(-\; \Delta_{g_{0}}u  \right)+R_{0}u \right] \; ,
\label{eq_29}
\end{equation}
\noindent
which for $R_{0}=0$ transforms as:

\begin{equation}
R=-\; \frac{16}{3}\; u^{-7/3} \Delta_{g_{0}}u \; .
\label{eq_30}
\end{equation}

We know from differential geometry that the mean curvature vector $\mathcal{H}\vec{x}$ equals the Laplace-Beltrami operator applied on the immersion {\it x} of a hyper-surface $\mathcal{M}$ \cite{dier}:

\begin{equation}
\mathcal{H}(x)\vec{x}=-\; \Delta_{\mathcal{M}}x \; .
\label{eq_31}
\end{equation}

In our case we will have:

\begin{equation}
\Delta \mathcal{H}_{0}(u)\vec{n}_{0}(u)=-\; \Delta_{g_{0}}u \; ,
\label{eq_32}
\end{equation}
\noindent
where $\Delta \mathcal{H}_{0}$ represents the difference in mean curvature between the minimal case of the {\it pure catenoidal hyper-surface} and the FRW conformal {\it reduced catenoidal hyper-surface}.

With the above equation and equation (\ref{eq_30}) results:

\begin{equation}
\Delta \mathcal{H}_{0}(u)\vec{e}_{t_{FRW}}=\frac{3}{16}\; \frac{4}{r^{2}}\; u^{7/3} \; ,
\label{eq_33}
\end{equation}
\noindent
from where, with (\ref{eq_10}):

\begin{equation}
\Delta \mathcal{H}_{0}(r,a)\vec{e}_{t_{FRW}}=\frac{3}{2^{9}}\; \frac{Z_{catenoid}^{7/2}}{r^{2}}\left(\frac{\sqrt{H^{2}+1}-H}{H^{2}-H\sqrt{H^{2}+1}+1} \right)^{7} \; ,
\label{eq_34}
\end{equation}
which is our final formula for the difference in mean curvature between the minimal case of the {\it pure catenoidal hyper-surface} and the FRW conformal {\it reduced catenoidal hyper-surface} in $t_{FRW}$ direction.

The derivative of $(g_{rr})_{0}$ will be:

\begin{equation}
\frac{d}{dt}(g_{rr})_{0}=-\; \frac{Z_{catenoid}}{2}\; H^{3}\; \frac{\left(\sqrt{H^{2}+1}-H \right)\left(2H^{2}-2H\sqrt{H^{2}+1}+1 \right)}{\sqrt{H^{2}+1}\left(H^{2}-H\sqrt{H^{2}+1}+1 \right)^{3}} \; ,
\label{eq_35}
\end{equation}
\noindent
which, in (\ref{eq_27}), gives:

\begin{equation}
R_{0}=-\; H^{3}\; \frac{2H^{2}-2H\sqrt{H^{2}+1}+1}{\left(H^{2}-H\sqrt{H^{2}+1}+1 \right)^{2}} \; ,
\label{eq_36}
\end{equation}
\noindent
for (- + + +) in the metric tensor $(g_{ij})_{0}$ of the catenoidal hyper-surface ($Z_{catenoid} < 0$ and considering also an angular part in the metric), or:

\begin{equation}
R_{0}= H^{3}\; \frac{2H^{2}-2H\sqrt{H^{2}+1}+1}{\left(H^{2}-H\sqrt{H^{2}+1}+1 \right)^{2}} \; ,
\label{eq_37}
\end{equation}
\noindent
for (+ - - -) sign in the metric tensor $(g_{ij})_{0}$ components ($Z_{catenoid} > 0$ and an angular part in the metric).

From (\ref{eq_32}) and (\ref{eq_1}) (where $\Delta \mathcal{H}_{0}(u)\vec{e}_{r}$ is the mean curvature variation between the minimal case of the {\it pure catenoidal hyper-surface} and the FRW conformal {\it reduced catenoidal hyper-surface} in $r$ direction):

\begin{equation}
\Delta \mathcal{H}_{0}(u)\vec{e}_{r}=\frac{3}{16}\; \left(Ru^{7/3} - R_{0}u \right) \; .
\label{eq_38}
\end{equation}

For $R_{0}$ as in (\ref{eq_37}) and $R=\displaystyle\frac{4}{r^{2}}$:

\begin{equation}
\Delta \mathcal{H}_{0}(r,a)\vec{e}_{r}=\frac{3}{16}\; \left[\frac{4}{r^{2}}\; u^{7/3} + H^{3}\; \frac{2H^{2}-2H\sqrt{H^{2}+1}+1}{\left(H^{2}-H\sqrt{H^{2}+1}+1 \right)^{2}}\; u \right] \; ,
\label{eq_39}
\end{equation}
\noindent
or, with {\it u} from (\ref{eq_10}):

\begin{equation}
\begin{array}{ll}
\Delta \mathcal{H}_{0}(r,a)\vec{e}_{r}=\displaystyle\frac{3}{2^{7}}\; Z_{catenoid}^{3/2}\; \left(\displaystyle\frac{\sqrt{H^{2}+1}-H}{H^{2}-H\sqrt{H^{2}+1}+1}\right)^{5}\; \times \\ \times\;\left[\displaystyle\frac{Z_{catenoid}^{2}}{4r^{2}}\; \left(\displaystyle\frac{\sqrt{H^{2}+1}-H}{H^{2}-H\sqrt{H^{2}+1}+1}\right)^{2} + H^{3} \right] \; .
\end{array}
\label{eq_40}
\end{equation}

\begin{figure}
\centering
\leavevmode
\includegraphics[width=12.0cm,height=10.0cm]{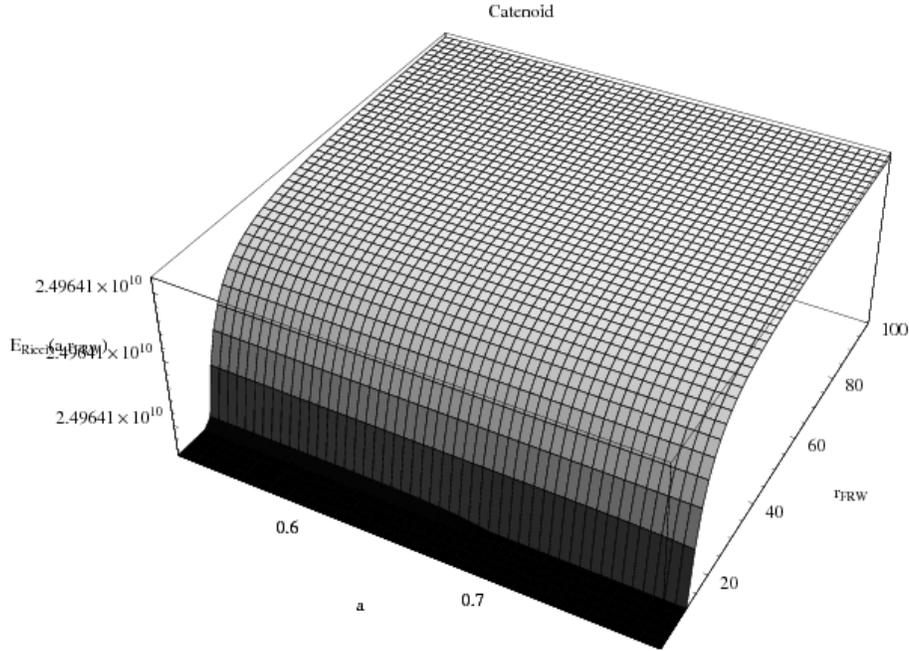}
\caption{Ricci energy (\ref{eq_40a}) spatial perturbation to the external FRW spacetime.}
\label{ricci1}
\end{figure}

\begin{figure}
\centering
\leavevmode
\includegraphics[width=12.0cm,height=10.0cm]{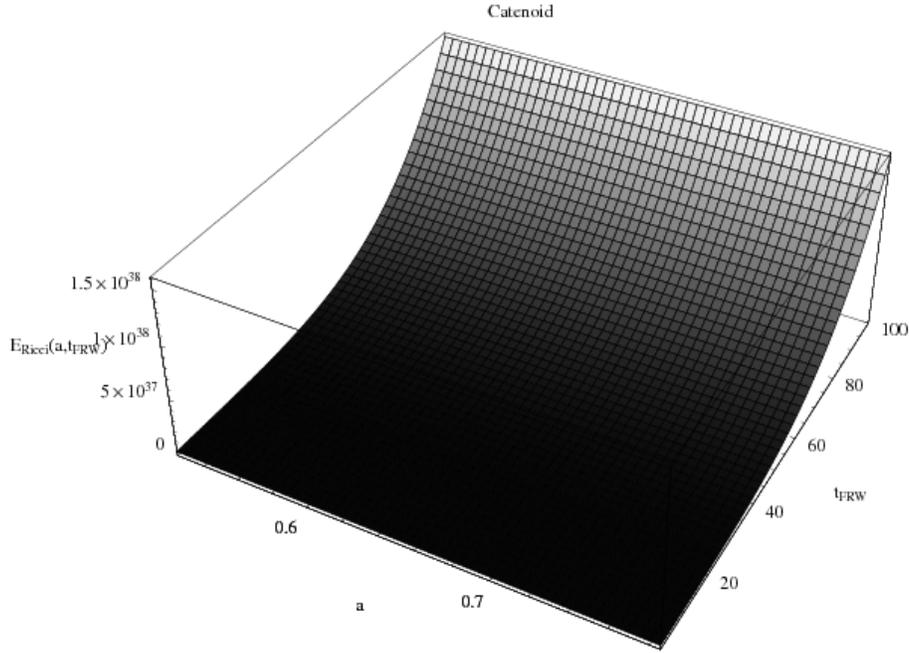}
\caption{Ricci energy (\ref{eq_40a}) temporal perturbation to the external FRW spacetime.}
\label{ricci2}
\end{figure}

The energy stored into the curvature of the {\it catenoidal hyper-surface} is released at the DEUS object evaporation into the external space in form of kinetic energy as spatial perturbations (equation (\ref{eq_40}); Fig. \ref{ricci1}) or as temporal perturbations for which $r_{int}\rightarrow t_{FRW,\; ext.}$ (equation (\ref{eq_34}); Fig. \ref{ricci2}):

\begin{equation}
\begin{array}{ll}
E_{Ricci}(r_{ext.},a)=\Delta \mathcal{H}_{0}(r_{ext.},a) \\
E_{Ricci}(r_{int.},a)\equiv E_{Ricci}(t_{FRW,\; ext.},a)=\Delta \mathcal{H}_{0}(t_{FRW.},a) \; .
\end{array}
\label{eq_40a}
\end{equation}

The timelike energy release (see Fig. \ref{ricci2}) continously decreases with the $\Delta \mathcal{H}$ decrease (from the internal hyper-surface evolution point of view) {\bf or}, it can be seen as increasing in $t_{FRW}$ time (from the external FRW perspective).

The interesting behavior is for the spacelike energy release (see Fig. \ref{ricci1}) where we observe constant levels (plateaus). This is like, even that the scale factor or the radius are varying, the spatial energy release of the DEUS objects evaporating in the FRW space remains constant and unmodified by the radius variation of the FRW bubble. As for the Yamabe energy, $a=i\;A$. In figure \ref{ricci1} we can also observe that the spatial perturbation is released in a slowly varying (almost constant) $(r_{ext.},\; a)$ pair for the external background when the pair $(r_{int.},\; A)$ varies rapidly (black holes in a flat external space), and conclude that this emission occurs in pulses, the initial one (near the $r_{int.}\rightarrow 0$ contact point between the {\it catenoidal hyper-surface} and the external flat space) being the weakest, while the final one, the strongest (the curvature is maximal at large $r_{int.}$).

What is here spatial will be temporal and what temporal will be spatial for the case of effects as coming not, as above, from the catenoidal hyper-surface without embeddings, but from the DEUS objects (for which the time and space are reversed) embedded in this hyper-surface.

\subsection{Helicoidal Hyper-Surface}

This case is more complicate because not the helicoid itself evolves but the matter filling the BHIG represented as a FRW bubble (with $Z_{helicoid}=1$), all this information being contained in the {\it pure helicoidal hyper-surface}. We saw \cite{deus2} in a intuitive way how the {\it pure helicoidal hyper-surface} becomes a FRW bubble with the Hubble constant $H_{0}\sim 72.3$ but, until now, we gave no description of how this happens. Also, because of the complexity of the {\it pure helicoidal hyper-surface} ansatz we were not able to compute directly the Ricci tensor components in $Z_{helicoid}\neq 1$ case. In this section we will analyze by Ricci flow seen as a gradient flow the dynamic process through which the {\it pure helicoidal hyper-surface} (in FRW coordinates) becomes a FRW spacetime, this making possible also the computation of the initial Ricci tensor components of the {\it pure helicoidal hyper-surface} (without angular part).

First consider the functional $\mathcal{F} = \int_{\mathcal{M}} [(R)_{FRW}'+|\nabla f|^{2}]e^{-f}\; dV$ for the Riemannian metric $(g_{ij})_{FRW}'$ and a function {\it f} on the closed non-empty FRW manifold ($\mathcal{M}$). The symmetric tensor $-\; [(R_{ij})_{FRW}'+\nabla_{i}\nabla_{j}f]$ is the $L^{2}$ gradient of the functional $\mathcal{F}^{m}=\int_{\mathcal{M}}[(R)_{FRW}'+|\nabla f|^{2}]\; dm$, where now $\it f$ denotes log($dV/dm$) \cite{perelman}. Thus given a measure {\it m}, we may consider the gradient flow:

\begin{equation}
\frac{d}{dt_{k}}(g_{ij})'_{FRW}=-2(R_{ij}^{m})_{FRW}'=-2[(R_{ij})_{FRW}'+\nabla_{i}\nabla_{j}f] \; .
\label{eq_41}
\end{equation}

The second evolution equation for the gradient flow of $\mathcal{F}^{m}$ is:

\begin{equation}
f_{t_{k}}=-\; (R)_{FRW}'-\Delta f = -\; (R)_{FRW}'-\nabla_{i}\nabla^{i}f \; .
\label{eq_42}
\end{equation}

In this case, $\mathcal{F}^{m}$ satisfies:

\begin{equation}
\mathcal{F}^{m}_{t_{k}}=2\int \left|(R_{ij})_{FRW}'+\nabla_{i}\nabla_{j}f \right|^{2}\; dm \; .
\label{eq_42a}
\end{equation}

For general {\it m} this flow may not exist even for short time. However, when it exists, it is just a Ricci flow modified by a diffeomorphism. Different choices of {\it m} lead to the same flow, up to a diffeomorphism, meaning that the choice of {\it m} is analogous to the choice of {\bf gauge} \cite{perelman}.

From paper {\it DEUS II} \cite{deus2} we know (the "mirror effect" of the inner horizon changes the signs of the BHIG spacetime's metric tensor components):

\begin{equation}
\begin{array}{ll}
(R_{t_{FRW}t_{FRW}})_{FRW}'=-\; \displaystyle4\; \frac{(H^{2}+1)H\ddot{H}+(1-2H^{2})\dot{H}^{2}}{(H^{2}+1)^{2}} \\
(R_{rr})_{FRW}'=-\; \displaystyle\frac{4}{3}\; \frac{(H^{2}+1)H\ddot{H}+(1-4H^{2})\dot{H}^{2}}{(H^{2}+1)^{2}} \; .
\end{array}
\label{eq_43}
\end{equation}

With the above Ricci tensor components and the derivatives of $(g_{rr})'_{FRW}$:

\begin{equation}
\frac{d}{dt_{k}}(g_{rr})_{FRW}'= \left\{ \begin{array}{lllll}
-\; \displaystyle\frac{a}{\sqrt{2}\; r^{3}}\; \left(2H^{2}-2H\sqrt{H^{2}+1}+1 \right)\; \times \\ \; \times \displaystyle\frac{H\dot{H}\left(2H^{2}-2H\sqrt{H^{2}+1}+1 \right) + \sqrt{H^{2}+1}\left(H^{2}-H\sqrt{H^{2}+1}+1 \right)^{2}}{\left(H^{2}-H\sqrt{H^{2}+1}+1 \right)^{9/2}} \\ \; \; \textnormal{when}\; \; H=H(t_{FRW}) \\
-\; \displaystyle\frac{a}{\sqrt{2}\; r^{3}}\; \frac{2H_{0}^{2}-2H_{0}\sqrt{H_{0}^{2}+1}+1}{\left(\sqrt{H_{0}^{2}+1}-H_{0} \right)\left(H_{0}^{2}-H_{0}\sqrt{H_{0}^{2}+1}+1 \right)^{3/2}}\\ \; \; \textnormal{when}\; \; H=H_{0}=const.
\end{array} \right.
\label{eq_44}
\end{equation}
\noindent
and of $(g_{t_{FRW}t_{FRW}})'_{FRW}$:

\begin{equation}
\frac{d}{dt_{k}}(g_{t_{t_{FRW}t_{FRW}}})_{FRW}'=0
\label{eq_45}
\end{equation}
\noindent
in (\ref{eq_41}) we can form a system of equations which solves for:

\begin{equation}
\begin{array}{lll}
f(r,a)=\displaystyle\frac{a}{4\sqrt{2}\; r}\; \left(2H^{2}-2H\sqrt{H^{2}+1}+1 \right)\; \times \\ \times \; \displaystyle\frac{H\dot{H}\left(2H^{2}-2H\sqrt{H^{2}+1}+1 \right) + \sqrt{H^{2}+1}\left(H^{2}-H\sqrt{H^{2}+1}+1 \right)^{2}}{\left(H^{2}-H\sqrt{H^{2}+1}+1 \right)^{9/2}} + \\ + \displaystyle\frac{2}{3}\; r^{2}\; \frac{\ddot{H}H(H^{2}+1)+\dot{H}^{2}(1-4H^{2})}{(H^{2}+1)^{2}} \; .
\end{array}
\label{eq_46}
\end{equation}

When $H=H_{0}=const.$:

\begin{equation}
f(r,a)=\frac{a}{4\sqrt{2}\; r}\; \frac{\sqrt{H_{0}^{2}+1}\left(2H_{0}^{2}-2H_{0}\sqrt{H_{0}^{2}+1}+1 \right)}{\left(H_{0}^{2}-H_{0}\sqrt{H_{0}^{2}+1}+1 \right)^{5/2}} \; ,
\label{eq_47}
\end{equation}
\noindent
or, for $r^{2}= \displaystyle\frac{2H^{2}-2H\sqrt{H^{2}+1}+1}{3\; \left(H^{2}-H\sqrt{H^{2}+1}+1\right)^{2}}$ (which results when writing the BHIG-FRW metric as a Minkowski metric):

\begin{equation}
f(r,a)=\frac{3a}{4\sqrt{2}}\; r\; \frac{\sqrt{H_{0}^{2}+1}}{\left(H_{0}^{2}-H_{0}\sqrt{H_{0}^{2}+1}+1 \right)^{1/2}} \; .
\label{eq_48}
\end{equation}

The function $f(r,a)$ can be regarded as a complex scalar field of mass {\it m} (the mass of the BHIG hyper-surface "particle" which exits from the inner horizon of the DEUS object and is seen by an external FRW observer) obeying the equation \cite{pavlov}:  

\begin{equation}
\left[\nabla_{i}\nabla^{i}+\xi\; (R)_{FRW}' + m^{2} \right]f(r,a)=0 \; ,
\label{eq_49}
\end{equation}
\noindent
with the coupling constant $\xi$. The value $\xi=\xi_{c}=\displaystyle\frac{n-2}{4(n-1)}$ correspond to conformal coupling in spacetime of dimension {\it n}. In our five-dimensional spacetime situation $\xi=\xi_{c}=3/16$.

(\ref{eq_42}) and (\ref{eq_49}) allows us to determine the quantity {\it m}:

\begin{equation}
m^{2}=\frac{1}{f}\; [(R)_{FRW}'+\nabla_{t}f]-\frac{3}{16}(R)_{FRW}' \; ,
\label{eq_50}
\end{equation}
\noindent
for (\ref{eq_47}) $f(r,a)$. $H=H_{0}=const.$ gives $\nabla_{t}f=0$ and $(R)_{FRW}'=0$, in this case the mass {\it m} of the "particle" being 0.

\section{Inverse Mean Curvature Flow}

The inverse mean curvature flow is defined as \cite{anciaux,huisken}:

\begin{equation}
\left|\vec{V}\right|\equiv \frac{1}{\left|\Delta \vec{\mathcal{H}}\right|} \; ,
\label{eq_51}
\end{equation}
\noindent
where {\it V} is the way in which the local chart describing the manifold evolves with time (velocity of the manifold) and $\Delta \mathcal{H}$ is the variation of the mean curvature between two moments of time in this evolution.

In the particular case of the {\it catenoidal hyper-surface}:

\begin{equation}
\left|\vec{V}_{catenoid}\right|\equiv \frac{1}{\left|\Delta \vec{\mathcal{H}}_{catenoid}\right|} \; ,
\label{eq_52}
\end{equation}
\noindent
or:

\begin{equation}
\left|{V}_{catenoid}\right|\; \vec{e}_{i}\equiv \frac{1}{\left|\Delta \mathcal{H}_{catenoid}\right|\; \vec{e}_{i}} \; ,
\label{eq_53}
\end{equation}
\noindent
in the orthonormal set of basis vectors ${e_{i}}$ of the external flat space, where the direction after which the mean curvature is modified is $r$ or $t_{FRW}$. 

Analogous, for the BHIG {\it helicoidal hyper-surface} evolving FRW bubble \cite{deus2} we will have:

\begin{equation}
\left|\vec{V}_{FRW}\right|\equiv \frac{1}{\left|\Delta \vec{\mathcal{H}}_{FRW}\right|} \; ,
\label{eq_54}
\end{equation}
\noindent
or:

\begin{equation}
\left|{V}_{FRW}\right|\; \vec{e}_{j}\equiv \frac{1}{\left|\Delta \mathcal{H}_{FRW}\right|\; \vec{e}_{j}} \; ,
\label{eq_55}
\end{equation}
\noindent
in the orthonormal set of basis vectors ${e_{j}}$ of the external flat space.

But, because we have the initial relation between $(t_{k},\phi_{k})$ of the {\it helicoidal hyper-surface} and $(t,\phi)$ of the of the {\it catenoidal hyper-surface} \cite{deus1}:

\begin{equation}
\frac{\phi_{k}}{t_{k}}=-\; \frac{t}{\phi} \; ,
\label{eq_56}
\end{equation} 
\noindent
and, by definition:

\begin{equation}
\begin{array}{ll}
|V_{catenoid}|\equiv \displaystyle\frac{d\phi_{k}}{dt_{k}} \\
|V_{FRW}|\equiv \displaystyle\frac{d\phi}{dt} \; ,
\end{array}
\label{eq_57}
\end{equation}
\noindent
it results that:

\begin{equation}
|V_{FRW}|=-\; \frac{1}{|V_{catenoid}|} \; ,
\label{eq_58}
\end{equation}
\noindent
or, with (\ref{eq_53}) and (\ref{eq_55}), for $i\neq j$:

\begin{equation}
|\Delta \mathcal{H}_{catenoid}|\; \vec{e}_{t_{FRW}}=-\; \frac{1}{|\Delta \mathcal{H}_{helicoid}|\; \vec{e}_{r}}
\label{eq_59}
\end{equation}
\noindent
and:

\begin{equation}
|\Delta \mathcal{H}_{catenoid}|\; \vec{e}_{r}=-\; \frac{1}{|\Delta \mathcal{H}_{helicoid}|\; \vec{e}_{t_{FRW}}} \; .
\label{eq_60}
\end{equation}

In FRW coordinates $r$ and $t_{FRW}$:

\begin{equation}
|V_{FRW}|\equiv \frac{d\phi}{dt}=-\; \frac{r}{V} \; ,
\label{eq_61}
\end{equation}
\noindent
where {\it V} is the velocity with which the catenoidal manifold evolves as seen by an external, FRW observer. If this velocity is equal with the velocity of light ($V=c$; $c=1$ in geometrized units) and the radius before the collapse cannot surpass $R_{Sch.}$ of the inner horizon of the DEUS collapsing object into the external FRW spacetime of radius {\it r} ($r=2R_{Sch.}$), the {\it catenoidal hyper-surface} velocity for the internal observer will be given as $|V_{catenoid}|=-2R_{Sch.}$ and, in consequence, the velocity of the FRW bubble between the central string-like singularity and $\theta =0$ point (Fig. \ref{frw}), will be $|V_{FRW}|=\displaystyle\frac{1}{2R_{Sch.}}$.

\section{Velocity}

\subsection{Catenoidal Hyper-Surface}

In the specific Lorentz reference frame of the external observer the components of the velocity will be \cite{misner}:

\begin{equation}
U_{catenoid}^{0}= \frac{dt_{FRW}}{dt}=\frac{1}{\sqrt{1-\vec{V}^{2}}} \; ,
\label{eq_62}
\end{equation}
\noindent
with {\it V} the velocity of the catenoidal manifold as seen by an external observer. In {\it DEUS II} paper \cite{deus2} we saw that:

\begin{equation}
\begin{array}{ll}
\phi_{k}=\displaystyle\frac{r}{a}\; \frac{1}{\sqrt{1+\textnormal{tg}^{2}\; t_{FRW}}} \\
t_{k}=\displaystyle\frac{r}{a}\; \frac{\textnormal{tg}\; t_{FRW}}{\sqrt{1+\textnormal{tg}^{2}\; t_{FRW}}}
\end{array}
\label{eq_63}
\end{equation}
\noindent
and:

\begin{equation}
\begin{array}{ll}
t=a\; \phi_{k} \\
\phi=-\; a\; t_{k} \; ,
\end{array}
\label{eq_64}
\end{equation}
\noindent
from where we will have:

\begin{equation}
\frac{dt}{dt_{FRW}}=-\; r\; \frac{\textnormal{tg}\; t_{FRW}}{\sqrt{\textnormal{tg}^{2}t_{FRW}+1}} \; ,
\label{eq_65}
\end{equation}
\noindent
which, in (\ref{eq_62}), gives:

\begin{equation}
\sqrt{1-\vec{V}^{2}}=-\; r\; \frac{\textnormal{tg}\; t_{FRW}}{\sqrt{\textnormal{tg}^{2}t_{FRW}+1}} \; .
\label{eq_66}
\end{equation}

\begin{figure}
\centering
\leavevmode
\includegraphics[width=10.0cm,height=10.0cm]{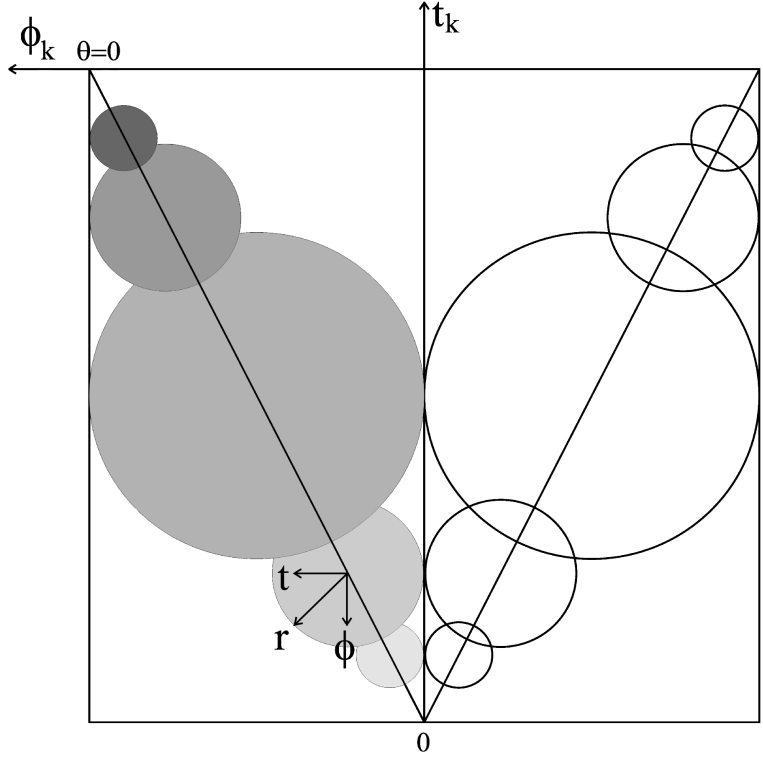}
\caption{The intuitive representation of how the BHIG spacetime transforms in a FRW spacetime. For a better understanding, the represented half of the BHIG helicoid is unfolded in a flat torsion-free hyper-surface. The FRW bubble evolves from the center of symmetry (the string-like collapsed object or Big Bang equivalent) to the point where it will exit through the $\theta =0$ point in another DEUS, superior, object (Big Crunch equivalent). The time $t_{FRW}$ is independent of the sense in which the FRW bubble evolves and this translates into the possibility of having a reversed evolution: from Big Bang to the Big Crunch than from $\theta =0$ to the central string-like object.}
\label{frw}
\end{figure}

Because when we computed the mean curvatures in paper {\it DEUS II} \cite{deus2} we obtained:

\begin{equation}
\textnormal{tg}\; t_{FRW}= \sqrt{H^{2}+1}-H \; ,
\label{eq_69}
\end{equation}
\noindent
which gives in (\ref{eq_66}):

\begin{equation}
\sqrt{1-\vec{V}^{2}}=-\; \frac{r}{\sqrt{2}}\; \frac{\sqrt{H^{2}+1}-H}{\left(H^{2}-H\sqrt{H^{2}+1}+1 \right)^{1/2}} \; .
\label{eq_70}
\end{equation}

Now, we can write also the other component of the velocity in the Lorentz frame as:

\begin{equation}
U_{catenoid}^{1}=\frac{dt_{FRW}}{d\phi}=\frac{V_{catenoid}^{1}}{\sqrt{1-\vec{V}^{2}}} \; .
\label{eq_71}
\end{equation}

By computing $d\phi/dt_{FRW}$ from (\ref{eq_63}) and (\ref{eq_64}):

\begin{equation}
\frac{d\phi}{dt_{FRW}}=-\; r\; \frac{1}{\sqrt{\textnormal{tg}^{2}t_{FRW}+1}} \; ,
\label{eq_72}
\end{equation}
\noindent
and inserting it together with (\ref{eq_66}) in (\ref{eq_71}), results:

\begin{equation}
V_{catenoid}^{1}=\textnormal{tg}\; t_{FRW} \; .
\label{eq_73}
\end{equation} 

For the particular case, when $V^{2}=(V_{catenoid}^{1})^{2}=\textnormal{tg}^{2} t_{FRW}$, and considering (\ref{eq_69}) we obtain, in geometrized units ($c=1$):

\begin{equation}
V=\left(2H^{2}-2H\sqrt{H^{2}+1}+1 \right)^{1/2} \; .
\label{eq_74}
\end{equation}

As well, we can write:

\begin{equation}
V=c\; \left(2H^{2}-2H\sqrt{H^{2}+1}+1 \right)^{1/2} \; .
\label{eq_75}
\end{equation}

If the velocity is expressed in km/s, the velocity of light $c=3\times 10^{8}$ m/s and $H=H_{0}\simeq 72.35$, then:

\begin{equation}
V=2074\; \;  \textnormal{km/s} \; ,
\label{eq_76}
\end{equation}
\noindent
velocity comparable with the core collapse infall velocity of a star becoming a black hole. In this way the infall velocity is constrained to this limit by the value of the Hubble constant $H_{0}$ and the frame invariance of the velocity of light. However, a higher infall velocity value is possible but only in a curved space with $H\neq H_{0}$.

\subsection{Helicoidal Hyper-Surface}

As for the {\it catenoidal hyper-surface} we will have also here, in a specific Lorentz reference frame:

\begin{equation}
U_{helicoid}^{0}=\frac{dt_{FRW}}{dt_{k}}
\label{eq_77}
\end{equation}
\noindent
and:

\begin{equation}
U_{helicoid}^{0}=\frac{1}{\sqrt{1-V^{2}}} \; .
\label{eq_78}
\end{equation}

Here {\it V} is the helicoid velocity as seen by the FRW external observer. From (\ref{eq_77}) and (\ref{eq_78}) results:

\begin{equation}
\sqrt{1-V^{2}}=\frac{dt_{k}}{dt_{FRW}} \; .
\label{eq_79}
\end{equation}

From (\ref{eq_63}) we get almost directly:

\begin{equation}
\frac{dt_{k}}{dt_{FRW}}=-\; \frac{r}{a}\; \frac{1}{\sqrt{\textnormal{tg}^{2}t_{FRW}+1}}\; \left(\frac{\dot{a}}{a}\; \textnormal{tg}\; t_{FRW}-1 \right) \; ,
\label{eq_80}
\end{equation}
\noindent
which, in (\ref{eq_79}), gives:

\begin{equation}
\sqrt{1-V^{2}}=-\; \frac{r}{a}\; \frac{1}{\sqrt{\textnormal{tg}^{2}t_{FRW}+1}}\; \left(\frac{\dot{a}}{a}\; \textnormal{tg}\; t_{FRW}-1 \right) \; .
\label{eq_81}
\end{equation}

The second component of the velocity in this Lorentz frame will be:

\begin{equation}
U_{helicoid}^{1}=\frac{dt_{FRW}}{d\phi_{k}}
\label{eq_82}
\end{equation}
\noindent
and:

\begin{equation}
U_{helicoid}^{1}=\frac{V_{helicoid}^{1}}{\sqrt{1-V^{2}}} \; ,
\label{eq_83}
\end{equation}
\noindent
resulting that:

\begin{equation}
V_{helicoid}^{1}=\sqrt{1-V^{2}}\; \frac{dt_{FRW}}{d\phi_{k}} \; ,
\label{eq_84}
\end{equation}
\noindent
or, with the substitution of (\ref{eq_81}) and of the computed derivative $dt_{FRW}/d\phi_{k}$ from (\ref{eq_63}):

\begin{equation}
V_{helicoid}^{1}=\frac{\displaystyle\frac{\dot{a}}{a}\; \textnormal{tg}\; t_{FRW}-1}{\displaystyle\frac{\dot{a}}{a}+\textnormal{tg}\; t_{FRW}} \; .
\label{eq_85}
\end{equation}

The equation (\ref{eq_85}), written as function of the Hubble parameter (see (\ref{eq_69})), takes the following form:

\begin{equation}
V_{helicoid}^{1}=-\; \frac{H^{2}-H\sqrt{H^{2}+1}+1}{\sqrt{H^{2}+1}} \; .
\label{eq_86}
\end{equation}

If we consider, in the same way we have done for the {\it catenoidal hyper-surface}, that $V^{2}=(V^{1}_{helicoid})^{2}$ we get, in geometrized units:

\begin{equation}
V^{2}=\left[\frac{H^{2}-H\sqrt{H^{2}+1}+1}{\sqrt{H^{2}+1}}\right]^{2} \; ,
\label{eq_87}
\end{equation}
\noindent
or, as seen by the external observer:

\begin{equation}
V^{2}=c\; \left[\frac{H^{2}-H\sqrt{H^{2}+1}+1}{\sqrt{H^{2}+1}}\right]^{2} \; .
\label{eq_88}
\end{equation}

When $H=H_{0}\simeq 72.35$ and $c=10^{8}$ m/s, $V=2074$ km/s. First of all, this means that, obtaining the same result as for the {\it catenoidal hyper-surface}, the {\it helicoidal hyper-surfaces} can be seen as the timelike helicoids of the self-similar DEUS objects embedded in the {\it catenoidal hyper-surface} \cite{deus2}. Second, for the external observer, the {\it helicoidal hyper-surfaces} are "particle"-like objects engaged in the movement of the core collapse process. The "local" frame of these "particles" is immersed in the "global" frame of the {\it catenoidal hyper-surface} movement.

\section{Equations of Motion: Preliminaries}

A dynamic and kinematic study cannot be complete without the equations of motion. Consider the Lorentz equation of motion:

\begin{equation}
m\; \left[\frac{d^{2}x^{\alpha}}{dt^{2}} + \Gamma_{\mu \nu}^{\alpha}\; \frac{dx^{\mu}}{dt}\; \frac{dx^{\nu}}{dt} \right] = q\; F_{\beta}^{\alpha}\; \; \frac{dx^{\beta}}{dt} \; ,
\label{eq_89}
\end{equation}
\noindent
with $F_{\beta}^{\alpha}$, the electromagnetic tensor components and $q$ the charge. 

In $(t,\phi)$ and $(t_{k},\phi_{k})$ coordinates, for all the hyper-surfaces composing the DEUS object \cite{deus1}, the Christoffel symbols of the second kind, $\Gamma_{\mu \nu}^{\alpha} = 0$, meaning that, the equation of motion reduces to:

\begin{equation}
m_{catenoid}\; \frac{d^{2}x_{catenoid}^{\alpha}}{dt^{2}} = q\; (F_{\beta}^{\alpha})_{catenoid}\; \; \frac{dx_{catenoid}^{\beta}}{dt} \; ,
\label{eq_90}
\end{equation}
\noindent
for the {\it catenoidal hyper-surfaces}, or:

\begin{equation}
m_{helicoid}\; \frac{d^{2}x_{helicoid}^{\alpha}}{dt_{k}^{2}} = q\; (F_{\beta}^{\alpha})_{helicoid}\; \; \frac{dx_{helicoid}^{\beta}}{dt_{k}} \; ,
\label{eq_91}
\end{equation}
\noindent
for the {\it helicoidal hyper-surfaces}. The $q$ charge is the total charge and $m$ the total mass of the particles that live in a specific manifold.

\section{Conclusions}

Because in the previous {\it DEUS} paper \cite{deus2} we saw that the geometry of the hyper-surfaces contained in a DEUS object evolves, in order to elucidate their dynamics and kinematics, in this paper we successfully engaged in a geometrical flows study. The results of this paper can be summarized as follows:

\begin{enumerate}
\item The Yamabe energy applied on the {\it catenoidal hyper-surface} proved to be equivalent with half of the total energy (mass) contained into this hyper-surface and released into the external FRW spacetime. We found that the real part of the Yamabe energy can be interpreted as timelike effect given by the DEUS object evaporation into the external spacetime, while its imaginary part as spacelike effect to the same external FRW spacetime. The ratio of these energies was found to be $10^{8} \leq \displaystyle\frac{\Re(E_{Yamabe,\; catenoid})}{\Im(E_{Yamabe,\; catenoid})} \leq 10^{13}$.

For the {\it helicoidal hyper-surface} we saw that the Yamabe energy represents only a quarter of the energy (mass) contained in this surface and that in FRW coordinates the half-period BHIG helicoid FRW bubble increases from zero to a maximal radius, after that decreasing again to zero. 

\item From the conformity of the {\it reduced catenoidal hyper-surface} with a flat external space \cite{deus2}, by applying a combination of Ricci flow and mean curvature flow, we obtained the scalar curvature of the {\it pure catenoidal hyper-surface}, too complicated to be determined in a "classical" way from the metric.

We also observed that the energy stored into the curvature of the {\it catenoidal hyper-surface} is released (at the DEUS object evaporation) into the external space as kinetic energy in form of spatial or temporal perturbations. The timelike energy release continously decreases with the curvature decrease (from the internal hyper-surface evolution point of view) {\bf or}, equivalently, increases in $t_{FRW}$ time (from the external FRW perspective). For the spacelike energy release, even when the scale factor or the radius are varying, the spatial energy release of the DEUS objects evaporating in the FRW spacetime remains constant and unmodified by the radius variation of the FRW bubble. If the spatial perturbation is released in a slowly varying (almost constant) external background, when the DEUS object dimension is very small or very short living, this emission is in pulses, the initial one being the weakest, while the final one, the strongest.

Because the {\it pure helicoidal hyper-surface} is released in the FRW as matter "particles" (from the self-similar DEUS object embedded in the catenoidal hyper-surface; black hole-DEUS object equivalence) or as non-empty FRW bubbles (non-conformal; Universe-DEUS object equivalence), we analyzed the dynamic process through which the {\it pure helicoidal hyper-surface} becomes a FRW spacetime by Ricci flow seen as a gradient flow. This made possible also the computation of the initial Ricci tensor components of the {\it pure helicoidal hyper-surface}.

\item Using the mean curvature flow we were able to determine the velocity with which the {\it catenoidal hyper-surface} evolves to the FRW external space and the one of the BHIG's FRW bubble from its formation ("Big Bang") to its disappearance ("Big Crunch"). As we said in {\it DEUS II} paper, this movement is not possible to be directly observed by the external observer which perceives only its inertial effect as dark energy effect. In the Lorentz frame of the external observer these velocities are comparable with the infall velocity in the core collapse process of a star becoming a black hole.

\item In the Lorentz equations of motion, for all the hyper-surfaces composing the DEUS object expressed in $(t,\phi)$ or $(t_{k},\phi_{k})$ coordinates, the Christoffel symbols of the second kind are zero.
\end{enumerate}

\ack
Manuscript registered at the Astronomical Institute, Romanian Academy, record No. 251 from 04/19/07.


\title[D.E.U.S.]{IV: Fields and their Cosmological Meaning}

\begin{abstract}
We will prove here that, when a DEUS geometry collapses in a conformal flat external spacetime (Minkowski or Friedmann-Robertson-Walker - FRW) \cite{deus1,deus2}, only when the hyper-surfaces composing it are seen as having self-similar internal structure (immersed DEUS objects), the timelike observer ({\bf D} case of paper {\it DEUS I} ) of the external space will perceive, apart from the spacetime generated by them, effects originating in the {\it catenoidal hyper-surface} of the self-similar DEUS immersed objects and released into the observer's flat space.
While, for the five-dimensional timelike projection observer, the self-similar DEUS objects are the internal building blocks of the collapsing {\it catenoidal
hyper-surface}, they are the building blocks of the collapsing {\it helicoidal hyper-surface} of the five-dimensional spacelike projection observer ({\bf A} case). For the effects of the self-similar DEUS objects contained into the {\it helicoidal hyper-surfaces} it can be given the meaning of particle-like FRW bubbles released into the external observer's Minkowski spacetime. Also, the {\it catenoidal hyper-surfaces} can be seen (and this will remain to be proven into another paper dealing with the quantum implications of the DEUS objects) as waves associated to the {\it helicoidal hyper-surfaces} FRW "particle" bubble's representation, but not valid in the same space or at the same moment of time with the first mentioned.
  
In this paper we will analyze only the five-dimensional timelike projection observer case, which, in consequence sees only the {\it catenoidal hyper-surface} of the collapsing DEUS object and this {\it catenoidal hyper-surface}'s DEUS self-similar immersed objects, and, to be more exact, from these immersed objects, only the interpretation which can be given to their {\it helicoidal hyper-surfaces}. We will see that they can be interpreted by the FRW or Minkowski spacetime observers as dark energy, dark matter, baryonic matter and electromagnetic radiation.

Because the {\it helicoidal hyper-surfaces} of the studied DEUS objects are given a "particle" meaning but also because these DEUS objects are contained into the {\it catenoidal hyper-surfaces} (for which we gave a "wave" interpretation) of the higher scale DEUS object seen by our external observer, results that we have a duality "wave"-"particles" for the external observer perspective on the effects.
\end{abstract}


\section{Introduction}

Having as starting point the trivial three-dimensional minimal surface representation of the catenoid and its conjugate surface, the helicoid, in paper {\it DEUS I} \cite{deus1} we constructed a five-dimensional spacetime geometry. There we determined the five-dimensional coordinate system that satisfies the Cauchy-Riemann equations, and also the ansatz, the properties and the correlation possible to be made for the DEUS object's composing manifolds.

In {\it DEUS II}  paper \cite{deus2} we generalized the geometry of the DEUS Black Hole for Universe formation and evolution, by using self-similar minimal surfaces for DEUS objects distribution. From the assumed self-similarity of the five-dimensional internal geometry, we shown that the DEUS object collapses to a string-like object equivalent with the BHIG spacetime of a bigger DEUS object in which the first one is embedded.

The considered five-dimensional geometry becomes conformal with an external Friedmann-Robertson-Walker (FRW) spacetime (for which the Hubble constant is {\it uniquely} defined as $72.33\pm 0.02$) by evolving the hyper-surfaces composing our five-dimensional DEUS object to a pre-collapse manifold. For the BHIG spacetime written in FRW coordinates we derived the Friedmann equations and some cosmological consequences ($\Omega_{tot}$ and the cosmological distance).

Into the background of a collapsed DEUS object rotating string that creates a BHIG manifold "deformed" in a FRW bubble, the pre-collapse movement of the five-dimensional catenoid hyper-surfaces is perceived by an after-collapse external observer as dark energy effect, while the matter still contained in the collapsed DEUS spacetime as dark matter.

The collapsing helicoidal manifold can be seen by the external timelike observer only considering that the catenoid is having $N$ internal DEUS similar objects, distorted, also collapsing. The helicoids of these $N$-DEUS objects are timelike (five-dimensional) and are giving three-dimensional spacelike and timelike observable effects to the external space.

In the above geometry framework, in {\it DEUS III} \cite{deus3} we found that the energy contained into the {\it catenoidal hyper-surface} and into the {\it helicoidal hyper-surface} can be characterized using the Yamabe energy. We shown that the energy stored into the curvature of the {\it catenoidal hyper-surface} is released (at the DEUS object evaporation) into the external space as kinetic energy in form of spatial or temporal perturbations. Also, by applying a combination of Ricci flow and mean curvature flow and taking account of the conformity of the {\it reduced catenoidal hyper-surface} \cite{deus2} with the flat external space, we obtained the scalar curvature of the {\it pure catenoidal hyper-surface}, for which the metric was too complicated to be determined in the "classical" way.

Using the mean curvature flow, we determined \cite{deus3} the velocity with which the {\it catenoidal hyper-surface} evolves to the FRW external space and the one of the BHIG's FRW bubble from its formation ("Big Bang") to its disappearance ("Big Crunch"). As we said in {\it DEUS II}, this movement is not possible to be directly observed by the external observer, which perceives only its inertial effect as gravitational effect (dark energy in helicoidal frame). In the Lorentz frame of the external observer these velocities are comparable with the infall velocity in the core collapse process of a star becoming a black hole.

Here, by using the Lorentz equations of motion we will show that the {\it catenoidal hyper-surface}'s embedded self-similar DEUS objects' {\it helicoidal hyper-surfaces} have non-zero electromagnetic tensor components. The electromagnetic tensor is having two possible forms. The external observer will perceive the matter or energy fields residing from each of these forms as: inertial effects coming from a previous ("before time") Big Bang - Big Crunch cycle of evolution, hidden by the initial string-like singularity (dark energy in catenoidal frame); particles or radiation fields originating from the present ("time") Big Bang - Big Crunch cycle of the evolution (observer in a global FRW observable Universe bubble) or from Black Holes (observer in a local Minkowski spacetime) (dark matter, baryonic matter and electromagnetic radiation). For both field classes we find, in the perfect fluid approximation, the associated densities and pressures, and also the density dependence of the different fields with the scale factor and the observed Universe radius.

In the local Minkowski spacetime we obtain an extremely interesting behavior, the matter and the radiation fields being emitted from the evaporating black hole DEUS object at $\displaystyle \zeta=\pm\; \frac{\pi}{2}$ polar angle ({\bf black hole jets}).

\section{Catenoidal Hyper-Surface Wave Interpretation}

An exact analytic solution of the Korteweg-de Vries (KdV) equation:

\begin{equation}
\frac{\partial u(x,\; t)}{\partial t} + u(x,\; t)\; \frac{\partial u(x,\; t)}{\partial x}\; + \beta\; \frac{\partial^{3} u(x,\; t)}{\partial x^{3}}=0
\label{eqa_1}
\end{equation}
\noindent
for a particle of mass $\beta$, is, in its general form:

\begin{equation}
u(x,\; t)=u_{max.}\; \textnormal{cosh}^{-2}\left(\frac{x-Vt}{\Delta} \right)\; ,
\label{eqa_2}
\end{equation}
\noindent
where {\it V} is the velocity of the the solitonic wave, and:

\begin{equation}
\begin{array}{ll}
4\; \beta=V\; \Delta^{2} \\
12\; \beta=\Delta^{2}\; u_{max.} 
\end{array} \; .
\label{eqa_3}
\end{equation}

In equation (\ref{eqa_1}) by making the change of variable $u(x,\; t)=u(\xi=x-Vt)$, we can write:

\begin{equation}
[V-u(\xi)]\; \frac{d u(\xi)}{d \xi} + \beta\; \frac{d^{3} u(\xi)}{d \xi^{3}}=0\; ,
\label{eqa_4}
\end{equation}
\noindent
having as solution:

\begin{equation}
u(\xi)=u_{max.}\; \textnormal{cosh}^{-2}\left(\frac{\xi}{\Delta} \right) \; .
\label{eqa_5}
\end{equation}

It is remarkable the similarity between the function that generates the catenoid:

\begin{equation}
u(r,\; A)=A\; \textnormal{cosh}^{-1}\left(\frac{r}{A} \right)\; ,
\label{eqa_6}
\end{equation}
\noindent
where $r=\sqrt{t^{2}+\phi^{2}}$ \cite{deus1,deus2}, and the $u(\xi)$ solution from (\ref{eqa_5}). Here we see that we must have the 
identities $u(\xi)=u^{2}(r,\; A)$, $u_{max.}=A^{2}$ and $\displaystyle \xi=\Delta\; \frac{r}{A}$. When we substitute these identities into (\ref{eqa_4}) (taking into account also (\ref{eqa_3}) relations) and solving the KdV equation we obtain:

\begin{equation}
\textnormal{cosh}^{-3}\left(\frac{r}{A} \right)\; \textnormal{sinh}\left(\frac{r}{A} \right)=0\; ,
\label{eqa_7}
\end{equation}
or:

\begin{equation}
\frac{A}{3}-8\; \frac{\beta}{A^{3}} - \left[A-12\; \frac{\beta}{A^{3}}\; \textnormal{sinh}^{2}\left(\frac{r}{A} \right) \right]\; \textnormal{cosh}^{-2}\left(\frac{r}{A} \right)=0\; .
\label{eqa_8}
\end{equation}

An interesting behavior is achieved when (\ref{eqa_8}) is satisfied. Before its collapse into a conformal external flat Minkowski spacetime \cite{deus2}, the {\it catenoidal hyper-surface} neck width varies from {\it pure catenoidal hyper-surface} to the {\it reduced catenoidal hyper-surface}, "dragging" all the other hyper-surfaces of the DEUS object with it. In this process the radius $r\equiv r_{int.}$ of the BHIG spacetime will increase until it reaches the radius where the hyper-surfaces are over-imposed on the pre-collapse cylinder. From this point it begins the perception of the observer of the Minkowski spacetime ($r_{ext.}=0$ or $t_{k}=\pm\; \phi_{k}$). But, because of the "mirror effect" of the inner horizon of the DEUS object black hole (which is tangent to the pre-collapse cylinder), $ds^{2}=-ds^{2}_{BHIG,\; int.}$ \cite{deus1}, the collapse of the DEUS object will be "seen" by the external observer as happening in reverse, towards an infinite curvature \cite{deus2} and energy (the Yamabe energy of the {\it catenoidal hyper-surface} goes to infinity as sum of all the energies of the infinite scale self-similar DEUS objects immersed into it), and disappearing from the spacetime of the external observer at the Schwarzschild radius, where the ergosphere (also on the cylinder) of the DEUS object "freeze" the perception of motion. From (\ref{eqa_8}), in the frame of the DEUS collapsing object, the $\beta$ mass contained into the {\it catenoidal hyper-surface} (the Yamabe energy of its immersed DEUS objects's helicoids) will decrease with the increase of $A$ to the point where $t_{k}=\pm\; \phi_{k}$ or $r_{ext.}=0$, after which it transfers to the empty external BHIG spacetime the mass-energy of the FRW bubble:

\begin{equation}
\beta=\frac{a^{3}(a-3)}{24}\; ,
\label{eqa_9}
\end{equation}
\noindent
where $a$ is the scale factor of the FRW bubble and $a=i\; A$ \cite{deus3}. For $A\leq r_{cylinder}$ and $r_{ext}\neq 0$ the energy emitted by the DEUS object's {\it catenoidal hyper-surface} will be perceived by the external as fields or waves of different nature.

So, even that the external perception of the collapse is frozen on the inner horizon of the DEUS object, we will still have some perception of what is happening with the DEUS object through the effects given by the mirrored collapse.

\section{Electromagnetic Tensor}

As we saw in paper {\it DEUS III} \cite{deus3} the Christoffel symbols of the second kind $\Gamma_{\mu \nu}^{\alpha}$ are 0 for all the DEUS hyper-surfaces expressed in $(t,\; \phi)$ or in $(t_{k},\; \phi_{k})$. Then, the Lorentz equation of motion can be written:

\begin{equation}
m_{catenoid}\; \frac{d^{2}x_{catenoid}^{\alpha}}{dt^{2}} = q\; (F_{\beta}^{\alpha})_{catenoid}\; \frac{dx_{catenoid}^{\beta}}{dt} \; ,
\label{eq_1}
\end{equation}
\noindent
for the {\it catenoidal hyper-surfaces}, or:

\begin{equation}
m_{helicoid}\; \frac{d^{2}x_{helicoid}^{\alpha}}{dt_{k}^{2}} = q\; (F_{\beta}^{\alpha})_{helicoid}\; \frac{dx_{helicoid}^{\beta}}{dt_{k}} \; ,
\label{eq_2}
\end{equation}
\noindent
for the {\it helicoidal hyper-surfaces}. In the above relations $F_{\beta}^{\alpha}$ are the electromagnetic tensor components, $q$ the total charge and $m$ the total mass of the particles which live in a specific manifold. We will concentrate only on the five-dimensional timelike {\bf D} case from paper {\it DEUS I} \cite{deus1}. Let us study now the possibility of having fields constrained into DEUS objects hyper-surfaces and which might give observable effects to an external FRW or Minkowski spacetime.

\subsection{Ergosphere's Catenoid $\left(C=1;\; \theta=0\right)$}

We saw that for the ergosphere of DEUS object we must have, from the point of view of the external observer, $\theta=0$ \cite{deus1}. Then, for this hyper-surface:

\begin{equation}
\left\{ \begin{array}{lllll}
x^{1}=\sqrt{t^{2}+\phi^{2}} \\
x^{2}=0 \\
\displaystyle x^{3}=A\; \textnormal{cosh}^{-\; 1}\left(\frac{\sqrt{t^{2}+\phi^{2}}}{A} \right) \\
x^{4}=0 \\
x^{5} \equiv i\; x_{0}=\pm\; i\; \sqrt{t^{2}+\phi^{2}}
\end{array} \right.
\label{eq_3}
\end{equation}

From the equation of motion (\ref{eq_1}) with the derivatives from the above chart, and taking $(F_{\beta}^{\alpha})_{catenoid}=0$ for $\alpha=\beta$, we obtain 16 equations. But, because $x^{2}=x^{4}=0$, the number of independent equations reduces to 9. After solving this systems of equations we obtain $(F_{\beta}^{\alpha})_{catenoid}=0$, no matter the choice of $\alpha$ and $\beta$. Because we know that ($\left(x^{1},\; x^{2},\; x^{3},\; x^{4},\; x^{5}\right)_{catenoid.}\rightarrow \left(x^{1},\; x^{2},\; x^{3},\; x^{0}\right)_{FRW}$ in the external observer frame):

\begin{equation}
(F_{\beta}^{\alpha})_{catenoid} = \left( \begin{array}{cccc}
0 & E_{x^{1}} & E_{x^{2}} & E_{x^{3}} \\
E_{x^{1}} & 0 & B_{x^{3}} & -\; B_{x^{2}} \\
E_{x^{2}} & -\; B_{x^{3}} & 0 & B_{x^{1}} \\
E_{x^{3}} & B_{x^{2}} & -\; B_{x^{1}} & 0 
\end{array} \right) \; ,
\label{eq_4}
\end{equation}
\noindent
there is no electric or magnetic field component emitted by this hyper-surface ($E_{x^{i}}=B_{x^{i}}=0$) into the external observer's spacetime.

\subsection{Ergosphere's Helicoid $\left(C=1;\; \theta=0\right)$}

Here, when $\theta=0$ (as we saw that we must have when we consider the ergosphere as containing both the {\it catenoidal hyper-surface} and the {\it helicoidal hyper-surface}):

\begin{equation}
\left\{ \begin{array}{lllll}
x^{1}=0 \\
x^{2}=0 \\
x^{3}=\displaystyle \textnormal{arctg}\left(\frac{t_{k}}{\phi_{k}} \right) \\
x^{4}=0 \\
x^{5}=0
\end{array} \right.
\label{eq_5}
\end{equation}

From (\ref{eq_2}) with the derivatives of the above chart (and when $(F_{\beta}^{\alpha})_{helicoid}=0$ for $\alpha=\beta$), all we can say is that:

\begin{equation}
(F_{3}^{3})_{helicoid}=-\; \frac{m_{helicoid}}{q}\; \frac{t_{k}}{t_{k}^{2}+\phi_{k}^{2}}=0 \; ,
\label{eq_6}
\end{equation}
\noindent
from where it results that either $t_{k}=0$, either $t_{k}=\pm \; i\; \phi_{k}$, not sufficient to determine the electromagnetic tensor components. 

\subsection{BHIG Spacetime}

Because we concentrate only on timelike cases (which, for the catenoidal or helicoidal hyper-surfaces, is the {\bf D} case from paper {\it DEUS I}) for this manifold \cite{deus1} we have:

\begin{equation}
\left\{ \begin{array}{ll}
x^{1}=x^{2}=x^{3}=x^{5}=0 \\
\displaystyle x^{4}=\pm\; i\; \textnormal{arctg}\left(\frac{t_{k}}{\phi_{k}} \right) \\
\end{array} \right.
\label{eq_7}
\end{equation}
\noindent
meaning that the only thing we can get from this is:

\begin{equation}
m_{BHIG}\; \frac{d^{2}x_{BHIG}^{4}}{dt_{k}^{2}} = q\; (F_{4}^{4})_{BHIG}\; \; \frac{dx_{BHIG}^{4}}{dt_{k}} \; ,
\label{eq_8}
\end{equation}
\noindent
from where:

\begin{equation}
(F_{4}^{4})_{BHIG}=-\; 2\; \frac{m_{BHIG}}{q}\; \frac{t_{k}}{t_{k}^{2}+\phi_{k}^{2}}=0 \; ,
\label{eq_9}
\end{equation}
\noindent
satisfied for (as we saw also for the {\it ergosphere's helicoid}) $t_{k}=0$, or $t_{k}=\pm \; i\; \phi_{k}$.

\subsection{Pure Catenoidal Hyper-Surface $\displaystyle \left(C\neq 1;\; \theta\neq \textnormal{arctg}\left(\frac{t}{\phi} \right)\right)$}

\begin{equation}
\left\{ \begin{array}{lllll}
x^{1}=\sqrt{t^{2}+\phi^{2}}\; \textnormal{cos}\; \theta \\
x^{2}=\displaystyle \pm\; \sqrt{t^{2}+\phi^{2}}\; \textnormal{sin}\; \theta\; \sqrt{\displaystyle \frac{\textnormal{sin}\left[\textnormal{arctg}\left(\displaystyle \frac{t}{\phi} \right) \right]\; \textnormal{cos}\; \theta}{\textnormal{cos}\left[\textnormal{arctg}\left(\displaystyle \frac{t}{\phi} \right) \right]\; \textnormal{sin}\; \theta}} \\
x^{3}=\displaystyle A\; \textnormal{cosh}^{-\; 1}\left(\frac{\sqrt{t^{2}+\phi^{2}}}{A} \right) \\
x^{4}=\displaystyle \pm\; i\; A\; \textnormal{cosh}^{-\; 1}\left(\frac{\sqrt{t^{2}+\phi^{2}}}{A} \right) \\
x^{5}=0
\end{array} \right.
\label{eq_10}
\end{equation}

When we take the derivatives of (\ref{eq_10}) in (\ref{eq_1}) we obtain 16 equations. From $(F_{\beta}^{\alpha})_{catenoid}=0$ for $\alpha=\beta$ we get:

\begin{equation}
\frac{\phi}{t\; \left(t^{2}+\phi^{2}\right)}=0 \; ,
\label{eq_11}
\end{equation}

\begin{equation}
\frac{3t^{4}+6t^{2}\phi^{2}-\phi^{4}}{2t\; (t^{2}+\phi^{2})(3t^{2}+\phi^{2})}=0 \; ,
\label{eq_12}
\end{equation}

\begin{equation}
\displaystyle \left[\frac{t}{A}\; \frac{1-\textnormal{sinh}^{2}\left(\displaystyle \frac{\sqrt{t^{2}+\phi^{2}}}{A} \right)}{\sinh \left(\displaystyle \frac{\sqrt{t^{2}+\phi^{2}}}{A} \right)\; \cosh \left(\displaystyle \frac{\sqrt{t^{2}+\phi^{2}}}{A} \right)} + \frac{\phi^{2}}{t\; (t^{2}+\phi^{2})^{1/2}} \right]\; \frac{1}{(t^{2}+\phi^{2})^{1/2}}=0 \; .
\label{eq_13}
\end{equation}

From (\ref{eq_11}) and the equations of motion, for $t\neq \pm\; i\; \phi$ and $t\neq 0$, results that $(F_{3}^{1})_{catenoid}=(F_{4}^{1})_{catenoid}=0$. Also, when $t\neq \pm\; i\; \phi$, $t\neq \mp\; i\; \displaystyle \frac{1}{\sqrt{3}}\; \phi$ and $t\neq 0$, from (\ref{eq_12}) results that $(F_{3}^{2})_{catenoid}=(F_{4}^{2})_{catenoid}=(F_{1}^{2})_{catenoid}=0$, while, from (\ref{eq_13}) results $(F_{4}^{3})_{catenoid}=(F_{3}^{4})_{catenoid}=0$.

Looking at equation (\ref{eq_11}) we see that it is satisfied for $\phi=0$ and, because of this, also  $(F_{2}^{1})_{catenoid}=(F_{2}^{4})_{catenoid}=0$. $\phi=0$ in (\ref{eq_13}) is having as consequence (when $\sinh \left(\displaystyle \frac{t}{A} \right)\neq 0$ and $\cosh \left(\displaystyle \frac{t}{A} \right)\neq 0$) $\sinh^{2}\left(\displaystyle \frac{t}{A} \right)=1$. This makes also $(F_{1}^{3})_{catenoid}=(F_{1}^{4})_{catenoid}=0$.

The conclusion of this analysis is that, for this hyper-surface, $(F_{\beta}^{\alpha})_{catenoid}=0$.

When $\phi=0$ in (\ref{eq_12}), the time $t\rightarrow \infty$, which means that for the external observer an object situated on this {\it catenoidal hyper-surface} will be seen as forever frozen on it (it behaves as an event horizon).

\subsection{Pure Helicoidal Hyper-Surface $\left(C\neq 1;\; \theta\neq \textnormal{arctg}\left(\displaystyle \frac{t}{\phi} \right)\right)$}

The chart for this surface will be:  

\begin{equation}
\left\{ \begin{array}{lllll}
x^{1}=\sinh\; \theta\; \sin \left[\textnormal{arctg}\left(\displaystyle \frac{t_{k}}{\phi_{k}} \right) \right] \\
x^{2}=\mp\; i\; \sinh\; \theta\; \cos \left[\textnormal{arctg}\left(\displaystyle \frac{t_{k}}{\phi_{k}} \right) \right]\; \sqrt{\displaystyle \frac{\cos\left[\textnormal{arctg}\left(\displaystyle \frac{t_{k}}{\phi_{k}} \right) \right]\; \cos\; \theta}{\sin\left[\textnormal{arctg}\left(\displaystyle \frac{t_{k}}{\phi_{k}} \right) \right]\; \sin\; \theta}} \\
x^{3}=\textnormal{arctg}\left(\displaystyle \frac{t_{k}}{\phi_{k}} \right) \\
x^{4}=\pm\; i\; \textnormal{arctg}\left(\displaystyle \frac{t_{k}}{\phi_{k}}\right) \\
x^{5}=0
\end{array} \right.
\label{eq_14}
\end{equation}

With the above and (\ref{eq_2}) we can construct again a system of 16 equations. 

\begin{enumerate}
\item From $(F_{1}^{1})_{helicoid}=0$ we get:

\begin{equation}
\frac{t_{k}}{t_{k}^{2}+\phi_{k}^{2}}=0 \; ,
\label{eq_15}
\end{equation}
\noindent
for which, when $t_{k}\neq \pm\; i\; \phi_{k}$, we obtain $t_{k}=0$. With $t_{k}=0$ results $(F_{2}^{1})_{helicoid}=(F_{3}^{1})_{helicoid}=(F_{4}^{1})_{helicoid}=0$.
\item From $(F_{2}^{2})_{helicoid}=0$ we get:

\begin{equation}
\frac{\phi_{k}^{4}+5t_{k}^{4}+2t_{k}^{2}\phi_{k}^{2}}{t_{k}\; (t_{k}^{2}+\phi_{k}^{2})\; (3t_{k}^{2}+\phi_{k}^{2})}=0 \; ,
\label{eq_16}
\end{equation}
\noindent
from which we obtain (when $t_{k}\neq 0$, $t_{k}\neq \pm\; i\; \phi_{k}$ and $\phi_{k}\neq \pm\; i\; \sqrt{3}\; t_{k}$) $\phi_{k}^{4}+5t_{k}^{4}+2t_{k}^{2}\phi_{k}^{2}=0$, resulting that $(F_{1}^{2})_{helicoid}=(F_{3}^{2})_{helicoid}=(F_{4}^{2})_{helicoid}=0$.

\item From $(F_{3}^{3})_{helicoid}=0$ we get again (\ref{eq_15}), so, $t_{k}=0$ valid if $t_{k}\neq \pm\; i\; \phi_{k}$. In consequence $(F_{1}^{3})_{helicoid}=(F_{2}^{3})_{helicoid}=(F_{4}^{3})_{helicoid}=0$.

\item The same (\ref{eq_15}) comes also from $(F_{4}^{4})_{helicoid}=0$. Here, $t_{k}=0$ for $t_{k}\neq \pm\; i\; \phi_{k}$. In consequence it makes $(F_{1}^{4})_{helicoid}=(F_{2}^{4})_{helicoid}=(F_{4}^{4})_{helicoid}=0$.
\end{enumerate}

If all the above cases are to be taken separately, then the electromagnetic tensor components $(F_{\beta}^{\alpha})_{helicoid}=0$. But, because $t_{k}=0$ and $t_{k}\neq 0$ situations are contradicting each other, we cannot consider an electromagnetic tensor having these components. However, as we will see later, we will be able to talk about an electromagnetic tensor that satisfies only one of the above conditions (and that being $t_{k}\neq 0$) plus {\it another tensor}, having a different physical meaning.

Actually, if we think better how the external observer sees the ergosphere, we will come to the conclusion that this will happen {\it only} on the pre-collapse cylindrical hyper-surface, conformal with the external space FRW. The ergosphere is not possible to be seen also because between it and the external observer interpose the {\it pure catenoidal hyper-surface} (which behaves as an event horizon), and the spacetime of the Catenoid with $C=1$ and $\theta\neq \textnormal{arctg}\left(\displaystyle \frac{t}{\phi} \right)$ \cite{deus1} (void until the {\it pure catenoidal hyper-surface} passes through it). At the cylindrical hyper-surfaces all the differences between the above named hyper-surfaces erase and the $\theta_{int.}$ angle of the ergosphere becomes the $\theta_{ext.}$ angle of the {\it reduced catenoidal} surface, conformal with the external flat space (the rotation is unfrozen) and, in consequence, the angle after which the external observer sees a phenomenon occurring (the contact angle between the cylinder and the {\it reduced catenoidal hyper-surface}). In paper {\it DEUS III} \cite{deus3} we saw that the {\it pure catenoidal hyper-surface} and the {\it pure helicoidal hyper-surface} evolve to the {\it reduced catenoidal hyper-surface}, respectively, to FRW bubbles. For the {\it ergosphere's catenoid} and {\it ergosphere's helicoid} we will take also the evolved cases in which, for now, $\theta_{int.}=0\rightarrow \theta_{ext.}=\textnormal{arctg} \left(\displaystyle \frac{t}{\phi} \right) $. The behavior of the {\it pure catenoidal hyper-surface} and of the {\it pure helicoidal hyper-surface} will be the one we saw above. For simplicity we will use the notation $\theta\equiv\theta_{ext.}$.

While until now we took in the above hyper-surfaces no self-similar geometry \cite{deus2,deus3}, in the next analysis we will consider it, together with the effects coming from lower scales helicoids or catenoids of the DEUS objects embedded in the previous studied simple hyper-surfaces.  

\subsection{Evolved Ergosphere's Catenoid $\displaystyle \left(C=1;\; \theta=\textnormal{arctg}\left(\displaystyle \frac{t}{\phi} \right)\right)$}

The chart for this hyper-surface will be given by:

\begin{equation}
\left\{ \begin{array}{lllll}
x^{1}=\sqrt{t^{2}+\phi^{2}}\; \cos \left[\textnormal{arctg}\left(\displaystyle \frac{t}{\phi} \right)\right] \\
x^{2}=\sqrt{t^{2}+\phi^{2}}\; \sin \left[\textnormal{arctg}\left(\displaystyle \frac{t}{\phi} \right)\right] \\
x^{3}=A\; \textnormal{cosh}^{-\; 1}\left(\displaystyle \frac{\sqrt{t^{2}+\phi^{2}}}{A} \right) \\
x^{4}=0 \\
x^{5}=\pm\; i\; \sqrt{t^{2}+\phi^{2}}
\end{array} \right.
\label{eq_17}
\end{equation}

Still under the influence of the contradiction found ($t_{k}=0$ or $t_{k}\neq 0$), we will solve the system of equations formed from  
(\ref{eq_1}) with the derivatives of the above coordinates, in two particular cases: $t=0$ and $t\neq 0$.

\begin{itemize}
\item $t\neq 0$ \\
As in the previous studied situations we obtain $(F_{\beta}^{\alpha})_{catenoid}=0$ with no infinities or contradictions. We also remarque that:

\begin{equation}
(F_{1}^{1})_{catenoid}\equiv (F_{4}^{4})_{BHIG}=-\; 2\; \frac{m_{BHIG}}{q}\; \frac{t_{k}}{t_{k}^{2}+\phi_{k}^{2}}=0 \; ,
\label{eq_18}
\end{equation}
\noindent
having as significance a transformation $x^{4}_{BHIG}\rightarrow x^{1}_{catenoid}$ (in the five-dimensional timelike case or $x^{5}_{BHIG}\rightarrow x^{1}_{catenoid}$ in the five-dimensional spacelike case), as an object "exits" the BHIG spacetime and goes to the external pre-collapse cylinder together with the ergosphere in which it lives, result which will become relevant later. 
 
\item $t=0$ \\
This situation is not possible because, when $m_{catenoid}\neq 0$ and $q_{catenoid}\neq \infty$, we obtain $(F_{3}^{3})_{catenoid}\rightarrow \infty$.
\end{itemize}

\subsection{Evolved Ergosphere's Helicoid $\displaystyle \left(C=1;\; \theta=\textnormal{arctg}\left(\displaystyle \frac{t}{\phi} \right)\right)$}

For this hyper-surface, where $\theta \neq 0$:

\begin{equation}
\left\{ \begin{array}{lllll}
x^{1}=\textnormal{sinh}\; \theta\; \sin \left[\textnormal{arctg}\left(\displaystyle \frac{t_{k}}{\phi_{k}} \right) \right] \\
x^{2}=-\; \textnormal{sinh}\; \theta\; \cos \left[\textnormal{arctg}\left(\displaystyle \frac{t_{k}}{\phi_{k}} \right) \right] \\
x^{3}=\textnormal{arctg}\left(\displaystyle \frac{t_{k}}{\phi_{k}} \right) \\
x^{4}=0 \\
x^{5}=\pm\; i\; \textnormal{sinh}\; \theta
\end{array} \right.
\label{eq_19}
\end{equation}
\noindent
or, knowing that it exists the transformation $\displaystyle \frac{t}{\phi}=-\; \frac{\phi_{k}}{t_{k}}$ \cite{deus1} in $\theta=\textnormal{arctg}\left(\displaystyle \frac{t}{\phi} \right)$:

\begin{equation}
\left\{ \begin{array}{lllll}
x^{1}=-\; \sinh \left[\textnormal{arctg}\left(\displaystyle \frac{\phi_{k}}{t_{k}} \right) \right]\; \sin \left[\textnormal{arctg}\left(\displaystyle \frac{t_{k}}{\phi_{k}} \right) \right] \\
x^{2}=\sinh \left[\textnormal{arctg}\left(\displaystyle \frac{\phi_{k}}{t_{k}} \right) \right]\; \cos \left[\textnormal{arctg}\left(\displaystyle \frac{t_{k}}{\phi_{k}} \right) \right] \\
x^{3}=\textnormal{arctg}\left(\displaystyle \frac{t_{k}}{\phi_{k}} \right) \\
x^{4}=0 \\
x^{5}=\mp\; i\; \sinh \left[\textnormal{arctg}\left(\displaystyle \frac{\phi_{k}}{t_{k}} \right) \right]
\end{array} \right.
\label{eq_20}
\end{equation}

We obtain the following results:

\begin{equation}
\!\!\!\!\!\!\!\!\!\!\!\!\!\!\!\!\!\!\!\!\!\!\!\!\!\begin{array}{llllllllllll}
\displaystyle (F_{2}^{1})_{helicoid}=-\; 2\; \frac{m_{helicoid}}{q}\; \frac{1}{t_{k}^{2}+\phi_{k}^{2}}\; \left\{\phi_{k} - \frac{t_{k}^{2}}{\phi_{k}+t_{k}\; \textnormal{tanh}\left[\textnormal{arctg}\left(\displaystyle \frac{\phi_{k}}{t_{k}} \right) \right] } \right\} \\
\displaystyle (F_{3}^{1})_{helicoid}=2\; \frac{m_{helicoid}}{q}\; \frac{1}{\left(t_{k}^{2}+\phi_{k}^{2}\right)^{3/2}}\; \left\{\left(\phi_{k}^{2}-t_{k}^{2} \right)\; \textnormal{cosh}\left[\textnormal{arctg}\left(\displaystyle \frac{\phi_{k}}{t_{k}} \right) \right] + t_{k}\phi_{k}\; \textnormal{sinh}\left[\textnormal{arctg}\left(\displaystyle \frac{\phi_{k}}{t_{k}} \right) \right] \right\} \\
\displaystyle (F_{5}^{1})_{helicoid}=\pm\; 2i\; \frac{m_{helicoid}}{q}\; \frac{1}{\left(t_{k}^{2}+\phi_{k}^{2}\right)^{3/2}}\; \left\{\phi_{k}^{2}-t_{k}^{2} + t_{k}\phi_{k}\; \textnormal{tanh}\left[\textnormal{arctg}\left(\displaystyle \frac{\phi_{k}}{t_{k}} \right) \right] \right\} \\
\displaystyle (F_{1}^{2})_{helicoid}=2\; \frac{m_{helicoid}}{q}\; \frac{t_{k}}{t_{k}^{2}+\phi_{k}^{2}}\; \frac{2\phi_{k}\; \textnormal{cosh}\left[\textnormal{arctg}\left(\displaystyle \frac{\phi_{k}}{t_{k}} \right) \right] + t_{k}\; \textnormal{sinh}\left[\textnormal{arctg}\left(\displaystyle \frac{\phi_{k}}{t_{k}} \right) \right]}{t_{k}\; \textnormal{cosh}\left[\textnormal{arctg}\left(\displaystyle \frac{\phi_{k}}{t_{k}} \right) \right] - \phi_{k}\; \textnormal{sinh}\left[\textnormal{arctg}\left(\displaystyle \frac{\phi_{k}}{t_{k}} \right) \right]} \\
\displaystyle (F_{3}^{2})_{helicoid}=2\; \frac{m_{helicoid}}{q}\; \frac{t_{k}}{\left(t_{k}^{2}+\phi_{k}^{2}\right)^{3/2}}\; \left\{2\phi_{k}\;  \textnormal{cosh}\left[\textnormal{arctg}\left(\displaystyle \frac{\phi_{k}}{t_{k}} \right) \right] + t_{k}\; \textnormal{sinh}\left[\textnormal{arctg}\left(\displaystyle \frac{\phi_{k}}{t_{k}} \right) \right] \right\} \\
\displaystyle (F_{5}^{2})_{helicoid}=\pm\; 2i\; \frac{m_{helicoid}}{q}\; \frac{t_{k}}{\left(t_{k}^{2}+\phi_{k}^{2}\right)^{3/2}}\; \left\{2\phi_{k}\;  + t_{k}\; \textnormal{tanh}\left[\textnormal{arctg}\left(\displaystyle \frac{\phi_{k}}{t_{k}} \right) \right] \right\} \\
\displaystyle (F_{1}^{3})_{helicoid}=-\; 2\; \frac{m_{helicoid}}{q}\; \frac{t_{k}}{\left(t_{k}^{2}+\phi_{k}^{2}\right)^{1/2}}\; \frac{1}{t_{k}\; \textnormal{cosh}\left[\textnormal{arctg}\left(\displaystyle \frac{\phi_{k}}{t_{k}} \right) \right] - \phi_{k}\; \textnormal{sinh}\left[\textnormal{arctg}\left(\displaystyle \frac{\phi_{k}}{t_{k}} \right) \right]} \\
\displaystyle (F_{2}^{3})_{helicoid}=2\; \frac{m_{helicoid}}{q}\; \frac{t_{k}}{\left(t_{k}^{2}+\phi_{k}^{2}\right)^{1/2}}\; \frac{1}{\phi_{k}\; \textnormal{cosh}\left[\textnormal{arctg}\left(\displaystyle \frac{\phi_{k}}{t_{k}} \right) \right] + t_{k}\; \textnormal{sinh}\left[\textnormal{arctg}\left(\displaystyle \frac{\phi_{k}}{t_{k}} \right) \right]} \\
\displaystyle (F_{5}^{3})_{helicoid}=\mp\; 2i\; \frac{m_{helicoid}}{q}\; \frac{t_{k}}{t_{k}^{2}+\phi_{k}^{2}}\; \frac{1}{\textnormal{cosh}\left[\textnormal{arctg}\left(\displaystyle \frac{\phi_{k}}{t_{k}} \right) \right]}\\
\displaystyle (F_{1}^{5})_{helicoid}=\mp i\; \frac{m_{helicoid}}{q}\; \frac{1}{\left(t_{k}^{2}+\phi_{k}^{2}\right)^{1/2}}\; \frac{2t_{k}\; \textnormal{cosh}\left[\textnormal{arctg}\left(\displaystyle \frac{\phi_{k}}{t_{k}} \right) \right] + \phi_{k}\; \textnormal{sinh}\left[\textnormal{arctg}\left(\displaystyle \frac{\phi_{k}}{t_{k}} \right) \right]}{t_{k}\; \textnormal{cosh}\left[\textnormal{arctg}\left(\displaystyle \frac{\phi_{k}}{t_{k}} \right) \right] - \phi_{k}\; \textnormal{sinh}\left[\textnormal{arctg}\left(\displaystyle \frac{\phi_{k}}{t_{k}} \right) \right]}\\
\displaystyle (F_{2}^{5})_{helicoid}=\pm i\; \frac{m_{helicoid}}{q}\; \frac{1}{\left(t_{k}^{2}+\phi_{k}^{2}\right)^{1/2}}\; \frac{2t_{k}\; \textnormal{cosh}\left[\textnormal{arctg}\left(\displaystyle \frac{\phi_{k}}{t_{k}} \right) \right] + \phi_{k}\; \textnormal{sinh}\left[\textnormal{arctg}\left(\displaystyle \frac{\phi_{k}}{t_{k}} \right) \right]}{\phi_{k}\; \textnormal{cosh}\left[\textnormal{arctg}\left(\displaystyle \frac{\phi_{k}}{t_{k}} \right) \right] + t_{k}\; \textnormal{sinh}\left[\textnormal{arctg}\left(\displaystyle \frac{\phi_{k}}{t_{k}} \right) \right]}\\
\displaystyle (F_{3}^{5})_{helicoid}=\mp\; i\; \frac{m_{helicoid}}{q}\; \frac{1}{t_{k}^{2}+\phi_{k}^{2}}\; \left\{2t_{k}\;  \textnormal{cosh}\left[\textnormal{arctg}\left(\displaystyle \frac{\phi_{k}}{t_{k}} \right) \right] + \phi_{k}\; \textnormal{sinh}\left[\textnormal{arctg}\left(\displaystyle \frac{\phi_{k}}{t_{k}} \right) \right] \right\} \; .
\end{array}
\label{eq_21}
\end{equation}

The electromagnetic tensor can be either a symmetric, either an antisymmetric tensor. Applying this to the above components we get:

\begin{enumerate}
\item From $(F_{1}^{2})_{helicoid}=(\pm)\; (F_{2}^{1})_{helicoid}$ results nothing useful for the final, simplified form of $(F_{\beta}^{\alpha})_{helicoid}$
\item From $(F_{3}^{5})_{helicoid}=(\pm)\; (F_{5}^{3})_{helicoid}$:

\begin{equation}
2t_{k}\; \textnormal{cosh}\left[\textnormal{arctg}\left(\frac{\phi_{k}}{t_{k}} \right) \right] + \phi_{k}\; \textnormal{sinh}\left[\textnormal{arctg}\left(\frac{\phi_{k}}{t_{k}} \right) \right]=(\pm)\; \frac{t_{k}}{\textnormal{cosh}\left[\textnormal{arctg}\left(\displaystyle \frac{\phi_{k}}{t_{k}} \right) \right]} \; .
\label{eq_22}
\end{equation}

\item From $(F_{2}^{5})_{helicoid}=(\pm)\; (F_{5}^{2})_{helicoid}$

\begin{equation}
\!\!\!\!\!\!\!\!\!\!\!\!\!\!\!\!\!\!\!\!\!\!\!\!\!\!\!\!\!\!\!\!\!\frac{2t_{k}\; \textnormal{cosh}\left[\textnormal{arctg}\left(\displaystyle \frac{\phi_{k}}{t_{k}} \right) \right] + \phi_{k}\; \textnormal{sinh}\left[\textnormal{arctg}\left(\displaystyle \frac{\phi_{k}}{t_{k}} \right) \right]}{\phi_{k}\; \textnormal{cosh}\left[\textnormal{arctg}\left(\displaystyle \frac{\phi_{k}}{t_{k}} \right) \right] + t_{k}\; \textnormal{sinh}\left[\textnormal{arctg}\left(\displaystyle \frac{\phi_{k}}{t_{k}} \right) \right]}=(\pm)\; 2\; \frac{t_{k}}{t_{k}^{2}+\phi_{k}^{2}}\; \left\{2\phi_{k}\;  + t_{k}\; \textnormal{tanh}\left[\textnormal{arctg}\left(\displaystyle \frac{\phi_{k}}{t_{k}} \right) \right] \right\} \; .
\label{eq_23}
\end{equation}

\item From $(F_{3}^{2})_{helicoid}=(\pm)\; (F_{2}^{3})_{helicoid}$:

\begin{equation}
\begin{array}{ll}
\displaystyle \left\{2\phi_{k}\;  \textnormal{cosh}\left[\textnormal{arctg}\left(\frac{\phi_{k}}{t_{k}} \right) \right] + t_{k}\; \textnormal{sinh}\left[\textnormal{arctg}\left(\frac{\phi_{k}}{t_{k}} \right) \right]\right\}\; \times \\
\displaystyle \times \left\{\phi_{k}\; \textnormal{cosh}\left[\textnormal{arctg}\left(\frac{\phi_{k}}{t_{k}} \right) \right] + t_{k}\; \textnormal{sinh}\left[\textnormal{arctg}\left(\frac{\phi_{k}}{t_{k}} \right) \right]\right\}=(\pm)\; (t_{k}^{2}+\phi_{k}^{2}) \; .
\end{array}
\label{eq_24}
\end{equation}

\end{enumerate}

From the equations of motion we also get the electromagnetic tensor components for which $\alpha=\beta$:

\begin{equation}
\!\!\!\!\!\!\!\!\!\!\!\!\!(F_{1}^{1})_{helicoid}=2\; \frac{m_{helicoid}}{q}\; \frac{1}{t_{k}^{2}+\phi_{k}^{2}}\; \frac{\left(\phi_{k}^{2}-t_{k}^{2}\right)\; \textnormal{cosh}\left[\textnormal{arctg}\left(\displaystyle \frac{\phi_{k}}{t_{k}} \right) \right] + t_{k}\phi_{k}\; \textnormal{sinh}\left[\textnormal{arctg}\left(\displaystyle \frac{\phi_{k}}{t_{k}} \right) \right]}{t_{k}\; \textnormal{cosh}\left[\textnormal{arctg}\left(\displaystyle \frac{\phi_{k}}{t_{k}} \right) \right] - \phi_{k}\; \textnormal{sinh}\left[\textnormal{arctg}\left(\displaystyle \frac{\phi_{k}}{t_{k}} \right) \right]} \; ,
\label{eq_25}
\end{equation}
\noindent
which can be written in a more convenient form for our purpose as:

\begin{equation}
(F_{1}^{1})_{helicoid}=-\; 2\; \frac{m_{helicoid}}{q}\; \frac{t_{k}}{t_{k}^{2}+\phi_{k}^{2}} + 2\; \frac{m_{helicoid}}{q}\; \frac{\phi_{k}^{2}}{t_{k}^{2}+\phi_{k}^{2}}\; \frac{1}{t_{k} - \phi_{k}\; \textnormal{tanh}\left[\textnormal{arctg}\left(\displaystyle \frac{\phi_{k}}{t_{k}} \right) \right]}
\label{eq_26}
\end{equation}
\noindent
and:

\begin{equation}
(F_{2}^{2})_{helicoid}=-\; 2\; \frac{m_{helicoid}}{q}\; \frac{t_{k}}{t_{k}^{2}+\phi_{k}^{2}}\; \frac{2\phi_{k}\; \textnormal{cosh}\left[\textnormal{arctg}\left(\displaystyle \frac{\phi_{k}}{t_{k}} \right) \right] + t_{k}\; \textnormal{sinh}\left[\textnormal{arctg}\left(\displaystyle \frac{\phi_{k}}{t_{k}} \right) \right]}{\phi_{k}\; \textnormal{cosh}\left[\textnormal{arctg}\left(\displaystyle \frac{\phi_{k}}{t_{k}} \right) \right] + t_{k}\; \textnormal{sinh}\left[\textnormal{arctg}\left(\displaystyle \frac{\phi_{k}}{t_{k}} \right) \right]} \; ,
\label{eq_27}
\end{equation}
\noindent
written also as:

\begin{equation}
(F_{2}^{2})_{helicoid}=-\; 2\; \frac{m_{helicoid}}{q}\; \frac{t_{k}}{t_{k}^{2}+\phi_{k}^{2}} - 2\; \frac{m_{helicoid}}{q}\; \frac{t_{k}\phi_{k}}{t_{k}^{2}+\phi_{k}^{2}}\; \frac{1}{\phi_{k} + t_{k}\; \textnormal{tanh}\left[\textnormal{arctg}\left(\displaystyle \frac{\phi_{k}}{t_{k}} \right) \right]} 
\label{eq_28}
\end{equation}
\noindent
and:

\begin{equation}
(F_{3}^{3})_{helicoid}=-\; 2\; \frac{m_{helicoid}}{q}\; \frac{t_{k}}{t_{k}^{2}+\phi_{k}^{2}}
\label{eq_29}
\end{equation}
\noindent
and:

\begin{equation}
(F_{5}^{5})_{helicoid}=2\; \frac{m_{helicoid}}{q}\; \frac{t_{k}}{t_{k}^{2}+\phi_{k}^{2}} + \frac{m_{helicoid}}{q}\; \frac{\phi_{k}}{t_{k}^{2}+\phi_{k}^{2}}\; \textnormal{tanh}\left[\textnormal{arctg}\left(\displaystyle \frac{\phi_{k}}{t_{k}} \right) \right] \; .
\label{eq_30}
\end{equation}

\section{Method}

\subsection{Step 1}

From the matrix formulation of the tensor $(F_{\beta}^{\alpha})_{helicoid}$ we can extract a tensor $h_{\beta}^{\alpha}$ for which:

\begin{equation}
\!\!\!\!\!\!\!\!\!\!\!\!\!\!\!\!\!\!\!\!\!\!\!\!\!\!\left( \begin{array}{cccc}
\displaystyle -\; 2\; \frac{m_{helicoid}}{q}\; \frac{t_{k}}{t_{k}^{2}+\phi_{k}^{2}} & 0 & 0 & 0 \\
0 & \displaystyle -\; 2\; \frac{m_{helicoid}}{q}\; \frac{t_{k}}{t_{k}^{2}+\phi_{k}^{2}} & 0 & 0 \\
0 & 0 & 0 & 0 \\
0 & 0 & 0 & \displaystyle 2\; \frac{m_{helicoid}}{q}\; \frac{t_{k}}{t_{k}^{2}+\phi_{k}^{2}} 
\end{array} \right) = 2\; \frac{m_{helicoid}}{q}\; h_{\beta}^{\alpha} \; ,
\label{eq_31}
\end{equation}
\noindent
where the tensor $h_{\beta}^{\alpha}$ is defined by the diagonal matrix:

\begin{equation}
h_{\beta}^{\alpha}=\left( \begin{array}{cccc}
\displaystyle -\; \frac{t_{k}}{t_{k}^{2}+\phi_{k}^{2}} & 0 & 0 & 0 \\
0 & \displaystyle -\; \frac{t_{k}}{t_{k}^{2}+\phi_{k}^{2}} & 0 & 0 \\
0 & 0 & 0 & 0 \\
0 & 0 & 0 & \displaystyle \frac{t_{k}}{t_{k}^{2}+\phi_{k}^{2}} 
\end{array} \right) \; .
\label{eq_32}
\end{equation}

In the matrix formulation of $(F_{\beta}^{\alpha})_{helicoid}$, we will add to the first two lines $\displaystyle 2\; \frac{m_{helicoid}}{q}\; \frac{t_{k}}{t_{k}^{2}+\phi_{k}^{2}}$ and we will substract the same amount from the fifth. In this way we will eliminate the tensor $h_{\beta}^{\alpha}$ from $(F_{\beta}^{\alpha})_{helicoid}$, putting aside, for the moment, the {\it pure catenoidal hyper-surface} (and its associated effects) of the self-similar first lower scale DEUS objects which are immersed in the {\it pure catenoidal hyper-surface} of the studied DEUS object.

\subsection{Step 2}

When the {\it catenoidal hyper-surfaces} and {\it ergosphere's helicoid} over-impose on the DEUS pre-collapse cylinder we saw \cite{deus2} that for the last named surface we will have $x^{1}=x^{2}=x^{4}=x^{5}=0$ and $x^{3}=\textnormal{arctg} \left(\displaystyle \frac{t_{k}}{\phi_{k}} \right)$. This can be achieved in (\ref{eq_20}) for:

\begin{equation}
\begin{array}{ll}
\textnormal{sinh} \left[\textnormal{arctg} \left(\displaystyle \frac{\phi_{k}}{t_{k}} \right) \right]=0 \\
\textnormal{cosh} \left[\textnormal{arctg}\left(\displaystyle \frac{\phi_{k}}{t_{k}} \right) \right]=1 \; .
\end{array}
\label{eq_33}
\end{equation}

With the (\ref{eq_33}) considerations and the changes {\it Step 1} in (\ref{eq_21}), (\ref{eq_26}), (\ref{eq_28}), (\ref{eq_29}) and (\ref{eq_30}) (for convenience and simplicity, even that the tensor changes from one computational step to another we will keep our notation of $(F_{\beta}^{\alpha})_{helicoid}$):

\begin{equation}
\begin{array}{llllllllllllllll}
\displaystyle (F_{1}^{1})_{helicoid}=2\; \frac{m_{helicoid}}{q}\; \frac{\phi_{k}^{2}}{t_{k}\; \left(t_{k}^{2}+\phi_{k}^{2}\right)} \\
\displaystyle (F_{2}^{2})_{helicoid}=-\; 2\; \frac{m_{helicoid}}{q}\; \frac{t_{k}}{t_{k}^{2}+\phi_{k}^{2}} \\
\displaystyle (F_{3}^{3})_{helicoid}=-\; 2\; \frac{m_{helicoid}}{q}\; \frac{t_{k}}{t_{k}^{2}+\phi_{k}^{2}} \\
(F_{5}^{5})_{helicoid}=0 \\
\displaystyle (F_{2}^{1})_{helicoid}=-\; 2\; \frac{m_{helicoid}}{q}\; \frac{1}{t_{k}^{2}+\phi_{k}^{2}}\; \left\{\phi_{k}-\frac{t_{k}^{2}}{\phi_{k}}-t_{k}\right\} \\
\displaystyle (F_{3}^{1})_{helicoid}=2\; \frac{m_{helicoid}}{q}\; \left(\phi_{k}^{2}-t_{k}^{2} \right)\; \frac{1}{\left(t_{k}^{2}+\phi_{k}^{2}\right)^{3/2}} + 2\; \frac{m_{helicoid}}{q}\; \frac{t_{k}}{t_{k}^{2}+\phi_{k}^{2}} \\
\displaystyle (F_{5}^{1})_{helicoid}=\pm\; 2i\; \frac{m_{helicoid}}{q}\; \left(\phi_{k}^{2}-t_{k}^{2} \right)\; \frac{1}{\left(t_{k}^{2}+\phi_{k}^{2}\right)^{3/2}} + 2\; \frac{m_{helicoid}}{q}\; \frac{t_{k}}{t_{k}^{2}+\phi_{k}^{2}} \\
\displaystyle (F_{1}^{2})_{helicoid}=2\; \frac{m_{helicoid}}{q}\; \left(2\phi_{k}+t_{k} \right)\; \frac{1}{t_{k}^{2}+\phi_{k}^{2}} \\
\displaystyle (F_{3}^{2})_{helicoid}=2\; \frac{m_{helicoid}}{q}\; \frac{2t_{k}\phi_{k}}{\left(t_{k}^{2}+\phi_{k}^{2}\right)^{3/2}} + 2\; \frac{m_{helicoid}}{q}\; \frac{t_{k}}{t_{k}^{2}+\phi_{k}^{2}} \\
\displaystyle (F_{5}^{2})_{helicoid}=\pm\; 2i\; \frac{m_{helicoid}}{q}\; \frac{2t_{k}\phi_{k}}{\left(t_{k}^{2}+\phi_{k}^{2}\right)^{3/2}} + 2\; \frac{m_{helicoid}}{q}\; \frac{t_{k}}{t_{k}^{2}+\phi_{k}^{2}} \\
\displaystyle (F_{1}^{3})_{helicoid}=-\; 2\; \frac{m_{helicoid}}{q}\; \frac{1}{\left(t_{k}^{2}+\phi_{k}^{2}\right)^{1/2}} \\
\displaystyle (F_{2}^{3})_{helicoid}=2\; \frac{m_{helicoid}}{q}\; \frac{t_{k}}{\phi_{k}\; \left(t_{k}^{2}+\phi_{k}^{2}\right)^{1/2}} \\
\displaystyle (F_{5}^{3})_{helicoid}=\mp\; 2i\; \frac{m_{helicoid}}{q}\; \frac{t_{k}}{t_{k}^{2}+\phi_{k}^{2}} \\
\displaystyle (F_{1}^{5})_{helicoid}=\mp\; 2i\; \frac{m_{helicoid}}{q}\; \frac{1}{\left(t_{k}^{2}+\phi_{k}^{2}\right)^{1/2}} - 2\; \frac{m_{helicoid}}{q}\; \frac{t_{k}}{t_{k}^{2}+\phi_{k}^{2}} \\
\displaystyle (F_{2}^{5})_{helicoid}=\pm\; 2i\; \frac{m_{helicoid}}{q}\; \frac{t_{k}}{\phi_{k}\; \left(t_{k}^{2}+\phi_{k}^{2}\right)^{1/2}} - 2\; \frac{m_{helicoid}}{q}\; \frac{t_{k}}{t_{k}^{2}+\phi_{k}^{2}} \\
\displaystyle (F_{3}^{5})_{helicoid}=\mp\; 2i\; \frac{m_{helicoid}}{q}\; \frac{t_{k}}{t_{k}^{2}+\phi_{k}^{2}} - 2\; \frac{m_{helicoid}}{q}\; \frac{t_{k}}{t_{k}^{2}+\phi_{k}^{2}} \; .
\end{array}
\label{eq_34}
\end{equation}

Is the moment to remember that for the {\it pure catenoidal hyper-surface}, when it evolves to BHIG spacetime, we had the transformation $x^{1}_{catenoid}\rightarrow x^{4}_{BHIG}$ ($(F_{1}^{1})_{catenoid}\equiv (F_{4}^{4})_{BHIG}$), as first step towards the collapse, when the {\it catenoidal hyper-surface} "losses" one dimension. But, because the {\it pure catenoidal hyper-surface} contains DEUS objects (self-similarity) which have their own helicoids we can study what happens with their axis. We observe from the above tensor components, by comparing them with the components of the electromagnetic tensor for the {\it pure catenoidal hyper-surface}, that the axis $x^{3}_{helicoid}\rightarrow x^{4}_{catenoid}$ ($(F_{3}^{3})_{helicoid}\equiv (F_{4}^{4})_{catenoid}$) (in a three-dimensional projection, the helicoid "rotates" from the $z$ symmetry axis to the plane $(x,\; y)$).

\subsection{Step 3}

Even if we want it or not, we must consider that the {\it catenoidal hyper-surfaces} (pure or from ergosphere) are self-similar. This means that the effects for which we are looking for are given by the lower scale DEUS objects' helicoids (see Fig. 6 from {\it DEUS II} paper) immersed in that {\it catenoidal hyper-surface}. Also, because the external observer sees the effects only after the DEUS object collapse he will not be able to locate the place (hyper-surface) from where they come. He will see an effect from a string-like singularity which, in order to exist has to have the DEUS history. This observer will see a "global" effect, as if all the surfaces contribute to it. So, we must take out {\it by hand} the hyper-surfaces whom topology give the main contributions to the geometry of the external flat space, keeping only the effects in which we are interested. In other words, we must substract the pre-collapse {\it catenoidal hyper-surface} ({\it reduced catenoidal hyper-surface} \cite{deus3} seen as having no self-similar structure) which gives the external flat space, and keep its self-similar internal DEUS objects. For this, because we saw at {\it Step 2} that $x^{3}_{helicoid}\rightarrow x^{4}_{catenoid}$ ($(F_{3}^{3})_{helicoid}\equiv (F_{4}^{4})_{catenoid}$), it results that the new $(F_{3}^{3})_{helicoid}$ component must be 0; We add $\displaystyle 2\; \frac{m_{helicoid}}{q}\; \frac{t_{k}}{t_{k}^{2}+\phi_{k}^{2}}$ to the third line of $(F_{\beta}^{\alpha})_{helicoid}$. Relative to (\ref{eq_34}), the four modified components will be:

\begin{equation}
\begin{array}{llll}
(F_{3}^{3})_{helicoid}=0 \\
\displaystyle (F_{1}^{3})_{helicoid}=-\; 2\; \frac{m_{helicoid}}{q}\; \frac{1}{\left(t_{k}^{2}+\phi_{k}^{2}\right)^{1/2}} + \; 2\; \frac{m_{helicoid}}{q}\; \frac{t_{k}}{t_{k}^{2}+\phi_{k}^{2}} \\
\displaystyle (F_{2}^{3})_{helicoid}=2\; \frac{m_{helicoid}}{q}\; \frac{t_{k}}{\phi_{k}\; \left(t_{k}^{2}+\phi_{k}^{2}\right)^{1/2}} + \; 2\; \frac{m_{helicoid}}{q}\; \frac{t_{k}}{t_{k}^{2}+\phi_{k}^{2}} \\
\displaystyle (F_{5}^{3})_{helicoid}=\mp\; 2i\; \frac{m_{helicoid}}{q}\; \frac{t_{k}}{t_{k}^{2}+\phi_{k}^{2}} + \; 2\; \frac{m_{helicoid}}{q}\; \frac{t_{k}}{t_{k}^{2}+\phi_{k}^{2}} \; .
\end{array}
\label{eq_35}
\end{equation}

\subsection{Step 4}

Trying to identify how the other axes transform when the {\it helicoidal hyper-surfaces} are expelled from DEUS object into the external BHIG (initial) spacetime and how this modifies the topology of our radiation field, we observe that:

\begin{equation}
\begin{array}{ll}
\displaystyle (F_{1}^{1})_{helicoid}\; \frac{t_{k}}{t_{k}^{2}+\phi_{k}^{2}}\equiv 2\; \frac{m_{helicoid}}{q}\; (g_{t_{k}t_{k}})_{BHIG} \\
\displaystyle (F_{2}^{2})_{helicoid}\; \frac{t_{k}}{t_{k}^{2}+\phi_{k}^{2}}\equiv -\; 2\; \frac{m_{helicoid}}{q}\; (g_{\phi_{k} \phi_{k}})_{BHIG} \; ,
\end{array}
\label{eq_36}
\end{equation}
\noindent
meaning that $x^{1}_{helicoid}\rightarrow t_{k}$ and $x^{2}_{helicoid}\rightarrow \phi_{k}$.

When we multiply with $\displaystyle \frac{t_{k}}{t_{k}^{2}+\phi_{k}^{2}}$ all the $(F_{\beta}^{\alpha})_{helicoid}$ components coming from {\it Steps} 2 and 3 we get:

\begin{equation}
\begin{array}{lllllll}
\displaystyle (F_{1}^{1})_{helicoid}=2\; \frac{m_{helicoid}}{q}\; \frac{\phi_{k}^{2}}{\left(t_{k}^{2}+\phi_{k}^{2}\right)^{2}} \\
\displaystyle (F_{2}^{2})_{helicoid}=-\; 2\; \frac{m_{helicoid}}{q}\; \frac{t_{k}^{2}}{\left(t_{k}^{2}+\phi_{k}^{2}\right)^{2}} \\
\displaystyle (F_{3}^{3})_{helicoid}=(F_{5}^{5})_{helicoid}=0 \\
\displaystyle (F_{2}^{1})_{helicoid}=-\; 2\; \frac{m_{helicoid}}{q}\; \frac{t_{k}\phi_{k}}{\left(t_{k}^{2}+\phi_{k}^{2}\right)^{2}}\; + \; 2\; \frac{m_{helicoid}}{q}\; \frac{t_{k}}{\left(t_{k}^{2}+\phi_{k}^{2}\right)^{2}}\; \left\{\frac{t_{k}^{2}}{\phi_{k}}+t_{k}\right\} \\
\displaystyle (F_{3}^{1})_{helicoid}=2\; \frac{m_{helicoid}}{q}\; \left(\phi_{k}^{2}-t_{k}^{2} \right)\; \frac{t_{k}}{\left(t_{k}^{2}+\phi_{k}^{2}\right)^{5/2}} + 2\; \frac{m_{helicoid}}{q}\; \frac{t_{k}^{2}}{\left(t_{k}^{2}+\phi_{k}^{2}\right)^{2}} \\
\displaystyle (F_{5}^{1})_{helicoid}=\pm\; 2i\; \frac{m_{helicoid}}{q}\; \left(\phi_{k}^{2}-t_{k}^{2} \right)\; \frac{t_{k}}{\left(t_{k}^{2}+\phi_{k}^{2}\right)^{5/2}} + 2\; \frac{m_{helicoid}}{q}\; \frac{t_{k}^{2}}{\left(t_{k}^{2}+\phi_{k}^{2}\right)^{2}} \\
\displaystyle (F_{1}^{2})_{helicoid}=2\; \frac{m_{helicoid}}{q}\; \frac{t_{k}\phi_{k}}{\left(t_{k}^{2}+\phi_{k}^{2}\right)^{2}} + 2\; \frac{m_{helicoid}}{q}\; \frac{t_{k}}{\left(t_{k}^{2}+\phi_{k}^{2}\right)^{2}}\; \left(\phi_{k} + t_{k} \right) \\
\displaystyle (F_{3}^{2})_{helicoid}=4\; \frac{m_{helicoid}}{q}\; \frac{t_{k}^{2}\phi_{k}}{\left(t_{k}^{2}+\phi_{k}^{2}\right)^{5/2}} + 2\; \frac{m_{helicoid}}{q}\; \frac{t_{k}^{2}}{\left(t_{k}^{2}+\phi_{k}^{2}\right)^{2}}
\end{array}
\label{eq_37}
\end{equation}

$$\displaystyle (F_{5}^{2})_{helicoid}=\pm\; 4i\; \frac{m_{helicoid}}{q}\; \frac{t_{k}^{2}\phi_{k}}{\left(t_{k}^{2}+\phi_{k}^{2}\right)^{5/2}} + 2\; \frac{m_{helicoid}}{q}\; \frac{t_{k}^{2}}{\left(t_{k}^{2}+\phi_{k}^{2}\right)^{2}}$$
$$\displaystyle (F_{1}^{3})_{helicoid}=-\; 2\; \frac{m_{helicoid}}{q}\; \frac{t_{k}}{\left(t_{k}^{2}+\phi_{k}^{2}\right)^{3/2}} + 2\; \frac{m_{helicoid}}{q}\; \frac{t_{k}^{2}}{\left(t_{k}^{2}+\phi_{k}^{2}\right)^{2}}$$
$$\displaystyle (F_{2}^{3})_{helicoid}=2\; \frac{m_{helicoid}}{q}\; \frac{t_{k}^{2}}{\phi_{k}\; \left(t_{k}^{2}+\phi_{k}^{2}\right)^{3/2}} + 2\; \frac{m_{helicoid}}{q}\; \frac{t_{k}^{2}}{\left(t_{k}^{2}+\phi_{k}^{2}\right)^{2}}$$
$$\displaystyle (F_{5}^{3})_{helicoid}=\mp\; 2i\; \frac{m_{helicoid}}{q}\; \frac{t_{k}^{2}}{\left(t_{k}^{2}+\phi_{k}^{2}\right)^{2}} + 2\; \frac{m_{helicoid}}{q}\; \frac{t_{k}^{2}}{\left(t_{k}^{2}+\phi_{k}^{2}\right)^{2}}$$
$$\displaystyle (F_{1}^{5})_{helicoid}=\mp\; 2i\; \frac{m_{helicoid}}{q}\; \frac{t_{k}}{\left(t_{k}^{2}+\phi_{k}^{2}\right)^{3/2}} - 2\; \frac{m_{helicoid}}{q}\; \frac{t_{k}^{2}}{\left(t_{k}^{2}+\phi_{k}^{2}\right)^{2}}$$
$$\displaystyle (F_{2}^{5})_{helicoid}=\pm\; 2i\; \frac{m_{helicoid}}{q}\; \frac{t_{k}^{2}}{\phi_{k}\; \left(t_{k}^{2}+\phi_{k}^{2}\right)^{3/2}} - 2\; \frac{m_{helicoid}}{q}\; \frac{t_{k}^{2}}{\left(t_{k}^{2}+\phi_{k}^{2}\right)^{2}}$$
$$\displaystyle (F_{3}^{5})_{helicoid}=\mp\; 2i\; \frac{m_{helicoid}}{q}\; \frac{t_{k}^{2}}{\left(t_{k}^{2}+\phi_{k}^{2}\right)^{2}} - 2\; \frac{m_{helicoid}}{q}\; \frac{t_{k}^{2}}{\left(t_{k}^{2}+\phi_{k}^{2}\right)^{2}} \; .$$

\subsection{Step 5}

We eliminate the "mirror effect" \cite{deus1} of the ergosphere by changing the sign in the $(F_{\beta}^{\alpha})_{helicoid}$ second column ($x^{2}_{helicoid}\rightarrow x^{2}_{BHIG}$):

\begin{equation}
\begin{array}{llll}
\displaystyle (F_{2}^{1})_{helicoid}=2\; \frac{m_{helicoid}}{q}\; \frac{t_{k}\phi_{k}}{\left(t_{k}^{2}+\phi_{k}^{2}\right)^{2}}\; - \; 2\; \frac{m_{helicoid}}{q}\; \frac{t_{k}}{\left(t_{k}^{2}+\phi_{k}^{2}\right)^{2}}\; \left\{\frac{t_{k}^{2}}{\phi_{k}}+t_{k}\right\} \\
\displaystyle (F_{2}^{2})_{helicoid}=2\; \frac{m_{helicoid}}{q}\; \frac{t_{k}^{2}}{\left(t_{k}^{2}+\phi_{k}^{2}\right)^{2}} \\
\displaystyle (F_{2}^{3})_{helicoid}=-\; 2\; \frac{m_{helicoid}}{q}\; \frac{t_{k}^{2}}{\phi_{k}\; \left(t_{k}^{2}+\phi_{k}^{2}\right)^{3/2}} - 2\; \frac{m_{helicoid}}{q}\; \frac{t_{k}^{2}}{\left(t_{k}^{2}+\phi_{k}^{2}\right)^{2}} \\
\displaystyle (F_{2}^{5})_{helicoid}=\mp\; 2i\; \frac{m_{helicoid}}{q}\; \frac{t_{k}^{2}}{\phi_{k}\; \left(t_{k}^{2}+\phi_{k}^{2}\right)^{3/2}} + 2\; \frac{m_{helicoid}}{q}\; \frac{t_{k}^{2}}{\left(t_{k}^{2}+\phi_{k}^{2}\right)^{2}} \; .
\end{array}
\label{eq_38}
\end{equation}

\subsection{Step 6}

We must eliminate the BHIG hyper-surface which lives inside the inner horizon of the collapsing DEUS object for which we expect to find electromagnetic effects. We will substract from the present form of $(F_{\beta}^{\alpha})_{helicoid}$ the quantity:

\begin{equation}
\!\!\!\!\!\!\!\!\!\!\!\!\!\!\!\!\!\!\!\!2\; \frac{m_{helicoid}}{q}\; \left(g_{\beta \alpha}\right)_{BHIG,\; int.}=\left( \begin{array}{cccc}
\displaystyle 2\; \frac{m_{helicoid}}{q}\; \frac{\phi_{k}^{2}}{\left(t_{k}^{2}+\phi_{k}^{2}\right)^{2}} & \displaystyle \frac{m_{helicoid}}{q}\; \frac{t_{k}\phi_{k}}{\left(t_{k}^{2}+\phi_{k}^{2}\right)^{2}} & 0 & 0 \\
\displaystyle \frac{m_{helicoid}}{q}\; \frac{t_{k}\phi_{k}}{\left(t_{k}^{2}+\phi_{k}^{2}\right)^{2}} & \displaystyle 2\; \frac{m_{helicoid}}{q}\; \frac{t_{k}^{2}}{\left(t_{k}^{2}+\phi_{k}^{2}\right)^{2}} & 0 & 0 \\
0 & 0 & 0 & 0 \\
0 & 0 & 0 & 0 
\end{array} \right) \; ,
\label{eq_39}
\end{equation}
\noindent
the new components of $(F_{\beta}^{\alpha})_{helicoid}$ becoming:

\begin{equation}
\begin{array}{lllllllllllll}
\displaystyle (F_{1}^{1})_{helicoid}=(F_{2}^{2})_{helicoid}=(F_{3}^{3})_{helicoid}=(F_{5}^{5})_{helicoid}=0 \\
\displaystyle (F_{2}^{1})_{helicoid}=\frac{m_{helicoid}}{q}\; \frac{t_{k}\phi_{k}}{\left(t_{k}^{2}+\phi_{k}^{2}\right)^{2}}\; - \; 2\; \frac{m_{helicoid}}{q}\; \frac{t_{k}}{\left(t_{k}^{2}+\phi_{k}^{2}\right)^{2}}\; \left\{\frac{t_{k}^{2}}{\phi_{k}}+t_{k}\right\} \\
\displaystyle (F_{3}^{1})_{helicoid}=2\; \frac{m_{helicoid}}{q}\; \left(\phi_{k}^{2}-t_{k}^{2} \right)\; \frac{t_{k}}{\left(t_{k}^{2}+\phi_{k}^{2}\right)^{5/2}} + 2\; \frac{m_{helicoid}}{q}\; \frac{t_{k}^{2}}{\left(t_{k}^{2}+\phi_{k}^{2}\right)^{2}} \\
\displaystyle (F_{5}^{1})_{helicoid}=\pm\; 2i\; \frac{m_{helicoid}}{q}\; \left(\phi_{k}^{2}-t_{k}^{2} \right)\; \frac{t_{k}}{\left(t_{k}^{2}+\phi_{k}^{2}\right)^{5/2}} + 2\; \frac{m_{helicoid}}{q}\; \frac{t_{k}^{2}}{\left(t_{k}^{2}+\phi_{k}^{2}\right)^{2}} \\
\displaystyle (F_{1}^{2})_{helicoid}=\frac{m_{helicoid}}{q}\; \frac{t_{k}\phi_{k}}{\left(t_{k}^{2}+\phi_{k}^{2}\right)^{2}} + 2\; \frac{m_{helicoid}}{q}\; \frac{t_{k}}{\left(t_{k}^{2}+\phi_{k}^{2}\right)^{2}}\; \left(\phi_{k} + t_{k} \right) \\
\displaystyle (F_{3}^{2})_{helicoid}=4\; \frac{m_{helicoid}}{q}\; \frac{t_{k}^{2}\phi_{k}}{\left(t_{k}^{2}+\phi_{k}^{2}\right)^{5/2}} + 2\; \frac{m_{helicoid}}{q}\; \frac{t_{k}^{2}}{\left(t_{k}^{2}+\phi_{k}^{2}\right)^{2}} \\
\displaystyle (F_{5}^{2})_{helicoid}=\pm\; 4i\; \frac{m_{helicoid}}{q}\; \frac{t_{k}^{2}\phi_{k}}{\left(t_{k}^{2}+\phi_{k}^{2}\right)^{5/2}} + 2\; \frac{m_{helicoid}}{q}\; \frac{t_{k}^{2}}{\left(t_{k}^{2}+\phi_{k}^{2}\right)^{2}} \\
\displaystyle (F_{1}^{3})_{helicoid}=-\; 2\; \frac{m_{helicoid}}{q}\; \frac{t_{k}}{\left(t_{k}^{2}+\phi_{k}^{2}\right)^{3/2}} + 2\; \frac{m_{helicoid}}{q}\; \frac{t_{k}^{2}}{\left(t_{k}^{2}+\phi_{k}^{2}\right)^{2}} \\
\displaystyle (F_{2}^{3})_{helicoid}=-\; 2\; \frac{m_{helicoid}}{q}\; \frac{t_{k}^{2}}{\phi_{k}\; \left(t_{k}^{2}+\phi_{k}^{2}\right)^{3/2}} - 2\; \frac{m_{helicoid}}{q}\; \frac{t_{k}^{2}}{\left(t_{k}^{2}+\phi_{k}^{2}\right)^{2}} \\
\displaystyle (F_{5}^{3})_{helicoid}=\mp\; 2i\; \frac{m_{helicoid}}{q}\; \frac{t_{k}^{2}}{\left(t_{k}^{2}+\phi_{k}^{2}\right)^{2}} + 2\; \frac{m_{helicoid}}{q}\; \frac{t_{k}^{2}}{\left(t_{k}^{2}+\phi_{k}^{2}\right)^{2}} \\
\displaystyle (F_{1}^{5})_{helicoid}=\mp\; 2i\; \frac{m_{helicoid}}{q}\; \frac{t_{k}}{\left(t_{k}^{2}+\phi_{k}^{2}\right)^{3/2}} - 2\; \frac{m_{helicoid}}{q}\; \frac{t_{k}^{2}}{\left(t_{k}^{2}+\phi_{k}^{2}\right)^{2}} \\
\displaystyle (F_{2}^{5})_{helicoid}=\mp\; 2i\; \frac{m_{helicoid}}{q}\; \frac{t_{k}^{2}}{\phi_{k}\; \left(t_{k}^{2}+\phi_{k}^{2}\right)^{3/2}} + 2\; \frac{m_{helicoid}}{q}\; \frac{t_{k}^{2}}{\left(t_{k}^{2}+\phi_{k}^{2}\right)^{2}} \\
\displaystyle (F_{3}^{5})_{helicoid}=\mp\; 2i\; \frac{m_{helicoid}}{q}\; \frac{t_{k}^{2}}{\left(t_{k}^{2}+\phi_{k}^{2}\right)^{2}} - 2\; \frac{m_{helicoid}}{q}\; \frac{t_{k}^{2}}{\left(t_{k}^{2}+\phi_{k}^{2}\right)^{2}} \; .
\end{array}
\label{eq_40}
\end{equation}

\subsection{Step 7}

We saw in {\it DEUS III} \cite{deus3} that the conformity of a {\it helicoidal hyper-surface} with the external observer's flat hyper-surface is assured only for: 

\begin{equation}
(ds^{2}_{BHIG})'\equiv -\; ds^{2}_{BHIG}=\frac{\phi_{k}^{2}}{\left(t_{k}^{2}+\phi_{k}^{2} \right)^{2}}\; dt_{k}^{2}+\frac{t_{k}^{2}}{\left(t_{k}^{2}+\phi_{k}^{2} \right)^{2}}\; d\phi_{k}^{2}-\frac{t_{k}\phi_{k}}{\left(t_{k}^{2}+\phi_{k}^{2} \right)^{2}}\; dt_{k}\; d\phi_{k} \; .
\label{eq_41}
\end{equation}

Until now, after 6 steps, we remained only with the {\it helicoidal hyper-surfaces} (pure + ergosphere's) of the self-similar DEUS objects immersed into {\it catenoidal hyper-surface} of the initial DEUS object from which we departed at the beginning of this section. We operated on the self-similar {\it pure helicoidal hyper-surface} but we neglected the second "mirror effect", of its corresponding self-similar {\it ergosphere's helicoids}, seen by the external observer. This can be corrected now because:

\begin{equation}
\!\!\!\!\!\!\!\!\!\!\!\!\!\!\!\!\!\!\!\!2\; \frac{m_{helicoid}}{q}\; \left(g_{\beta \alpha}\right)'_{BHIG}=\left( \begin{array}{cccc}
\displaystyle 2\; \frac{m_{helicoid}}{q}\; \frac{\phi_{k}^{2}}{\left(t_{k}^{2}+\phi_{k}^{2}\right)^{2}} & \displaystyle -\; \frac{m_{helicoid}}{q}\; \frac{t_{k}\phi_{k}}{\left(t_{k}^{2}+\phi_{k}^{2}\right)^{2}} & 0 & 0 \\
\displaystyle -\; \frac{m_{helicoid}}{q}\; \frac{t_{k}\phi_{k}}{\left(t_{k}^{2}+\phi_{k}^{2}\right)^{2}} & \displaystyle 2\; \frac{m_{helicoid}}{q}\; \frac{t_{k}^{2}}{\left(t_{k}^{2}+\phi_{k}^{2}\right)^{2}} & 0 & 0 \\
0 & 0 & 0 & 0 \\
0 & 0 & 0 & 0 
\end{array} \right) \; ,
\label{eq_42}
\end{equation}
\noindent
which means that it is enough to change the signs of the $(F_{2}^{1})_{helicoid}$ and $(F_{1}^{2})_{helicoid}$ components from {\it Step 6} to make it right:

\begin{equation}
\begin{array}{ll}
\displaystyle (F_{2}^{1})_{helicoid}=-\; \frac{m_{helicoid}}{q}\; \frac{t_{k}\phi_{k}}{\left(t_{k}^{2}+\phi_{k}^{2}\right)^{2}}\; + \; 2\; \frac{m_{helicoid}}{q}\; \frac{t_{k}}{\left(t_{k}^{2}+\phi_{k}^{2}\right)^{2}}\; \left\{\frac{t_{k}^{2}}{\phi_{k}}+t_{k}\right\} \\
\displaystyle (F_{1}^{2})_{helicoid}=-\; \frac{m_{helicoid}}{q}\; \frac{t_{k}\phi_{k}}{\left(t_{k}^{2}+\phi_{k}^{2}\right)^{2}} - 2\; \frac{m_{helicoid}}{q}\; \frac{t_{k}}{\left(t_{k}^{2}+\phi_{k}^{2}\right)^{2}}\; \left(\phi_{k} + t_{k} \right) \; .
\end{array}
\label{eq_43}
\end{equation}

\section{Fields}

When $t_{k}\neq \pm\; i\phi_{k}$, from the $(F_{2}^{1})_{helicoid}=(F_{1}^{2})_{helicoid}$ results that we can have either $t_{k}=0$, either $t_{k}=\pm\; \phi_{k}$.

Analyzing the $(F_{\beta}^{\alpha})_{helicoid}$ components for $t_{k}=0$ we get:

\begin{equation}
\begin{array}{llllll}
E_{1}\equiv (F_{5}^{1})_{helicoid}=(F_{1}^{5})_{helicoid}=0 \\
E_{2}\equiv (F_{3}^{1})_{helicoid}=(F_{1}^{3})_{helicoid}=0 \\
E_{3}\equiv (F_{1}^{2})_{helicoid}=(F_{2}^{1})_{helicoid}=0 \\
B_{1}\equiv (F_{5}^{3})_{helicoid}=-\; (F_{3}^{5})_{helicoid}=0 \\
B_{2}\equiv (F_{5}^{2})_{helicoid}=-\; (F_{2}^{5})_{helicoid}=0 \\
B_{3}\equiv (F_{3}^{2})_{helicoid}=-\; (F_{2}^{3})_{helicoid}=0 \; ,
\end{array}
\label{eq_44}
\end{equation}
\noindent
where $E_{x^{i}}\equiv E_{i}$ are the components of an electric field and $B_{x^{i}}\equiv B_{i}$ the components of a magnetic field seen by the three-dimensional external observer. From (\ref{eq_44}) $(F_{\beta}^{\alpha})_{helicoid}=0$ result for $t_{k}=0$, we have no fields coming from the collapsed DEUS object.

However, when $t_{k}=\pm\; \phi_{k}$ we have:

\begin{equation}
\begin{array}{ccc}
(F_{1}^{1})_{helicoid}=(F_{2}^{2})_{helicoid}=(F_{3}^{3})_{helicoid}=(F_{5}^{5})_{helicoid}=0 \\
(F_{3}^{1})_{helicoid}=(F_{1}^{3})_{helicoid}=(F_{5}^{1})_{helicoid}=(F_{1}^{5})_{helicoid}=(F_{2}^{5})_{helicoid}= \\
= (F_{5}^{2})_{helicoid}=(F_{5}^{3})_{helicoid}=(F_{3}^{5})_{helicoid}=0 \; ,
\end{array}
\label{eq_45}
\end{equation}
\noindent
but also:

\begin{equation}
\begin{array}{ll}
\displaystyle E_{3} =\frac{m_{helicoid}}{q}\; \frac{1}{t_{k}^{2}}\; \left(\pm\; \frac{1}{4} - \frac{1}{2} \right)\equiv \frac{m_{helicoid}}{q}\; \frac{1}{\phi_{k}^{2}}\; \left(\pm\; \frac{1}{4} - \frac{1}{2} \right) \\
\displaystyle B_{3}=\frac{m_{helicoid}}{q}\; \frac{1}{t_{k}^{2}}\; \left(\frac{1}{\sqrt{2}} + \frac{1}{2} \right)\equiv \frac{m_{helicoid}}{q}\; \frac{1}{\phi_{k}^{2}}\; \left(\frac{1}{\sqrt{2}} + \frac{1}{2} \right) \; ,
\end{array}
\label{eq_46}
\end{equation}
\noindent
where $E_{3}$ and $B_{3}$ components are given by:

\begin{equation}
\begin{array}{ll}
E_{3}\equiv (F_{1}^{2})_{helicoid}=(F_{2}^{1})_{helicoid} \\
B_{3}\equiv (F_{3}^{2})_{helicoid}=-\; (F_{2}^{3})_{helicoid} \; .
\end{array}
\label{eq_47}
\end{equation}

The electromagnetic tensor will be:

\begin{equation}
(F_{\beta}^{\alpha})_{helicoid}=\left( \begin{array}{cccc}
0 & 0 & 0 & E_{3} \\
0 & 0 & B_{3} & 0 \\
0 & -\; B_{3} & 0 & 0 \\
E_{3} & 0 & 0 & 0 
\end{array} \right) \; ,
\label{eq_48}
\end{equation}
\noindent
or explicitly (as function of $\phi_{k}$):

\begin{equation}
\!\!\!\!\!\!\!\!\!\!\!\!\!\!\!\!\!\!\!\!\!\!\!\!\!\!\!\!\!\!\!\!\!\!\!\!\!\!\!\!\!\!\!\!\!\!\!\!\!\!\!\!\!\!\!\!\!\!\!\!\!(F_{\beta}^{\alpha})_{helicoid}=\left( \begin{array}{cccc}
0 & 0 & 0 & \displaystyle \frac{m_{helicoid}}{q}\; \frac{1}{\phi_{k}^{2}}\; \left(\pm\; \frac{1}{4} - \frac{1}{2} \right) \\
0 & 0 & \displaystyle \frac{m_{helicoid}}{q}\; \frac{1}{\phi_{k}^{2}}\; \left(\frac{1}{\sqrt{2}} + \frac{1}{2} \right) & 0 \\
0 & \displaystyle -\; \frac{m_{helicoid}}{q}\; \frac{1}{\phi_{k}^{2}}\; \left(\frac{1}{\sqrt{2}} + \frac{1}{2} \right) & 0 & 0 \\
\displaystyle \frac{m_{helicoid}}{q}\; \frac{1}{\phi_{k}^{2}}\; \left(\pm\; \frac{1}{4} - \frac{1}{2} \right) & 0 & 0 & 0 
\end{array} \right) \; ,
\label{eq_49}
\end{equation}
\noindent
where $t_{k}=\pm\; \phi_{k}$. We know \cite{deus2} that we can make the transformation from $(t_{k},\; \phi_{k})$ coordinates of the helicoid to FRW $(t_{FRW},\; r)$ coordinates:

\begin{equation}
\begin{array}{ll}
\displaystyle \phi_{k}=\frac{r}{a}\; \frac{1}{\sqrt{1+\textnormal{tg}^{2}\; t_{FRW}}} \\
\displaystyle t_{k}=\frac{r}{a}\; \frac{\textnormal{tg}\; t_{FRW}}{\sqrt{1+\textnormal{tg}^{2}\; t_{FRW}}} \; ,
\end{array}
\label{eq_50}
\end{equation}
\noindent
which, when $t_{k}=\pm\; \phi_{k}$, gives $\textnormal{tg}\; t_{FRW}=\pm\; 1$, meaning that the (\ref{eq_50}) relations transform into:

\begin{equation}
\begin{array}{ll}
\displaystyle \phi_{k}=\frac{r}{a}\; \frac{1}{\sqrt{2}} \\
\displaystyle t_{k}=\pm\; \frac{r}{a}\; \frac{1}{\sqrt{2}} \; ,
\end{array}
\label{eq_51}
\end{equation}
\noindent
where $a$ is the spatial scale factor of the FRW Universe.

With (\ref{eq_51}) in (\ref{eq_49}) we find the contribution of an individual self-similar DEUS object ("particle") which was initially part of the {\it catenoidal hyper-surface} and now, after the collapse of the DEUS object ("black hole") containing this {\it catenoidal hyper-surface}, is released into the external FRW spacetime as timelike effect:

\begin{equation}
\!\!\!\!\!\!\!\!\!\!\!\!\!\!\!\!\!\!\!\!\!\!\!\!\!\!\!\!\!\!\!\!\!\!\!\!\!\!\!\!\!\!\!\!\!\!\!\!\!\!\!\!\!\!\!\!\!\!\!\!\!\!\!\!\!\!\!\!(F_{\beta}^{\alpha})_{helicoid}=\left( \begin{array}{cccc}
0 & 0 & 0 & \displaystyle 2\; \frac{m_{helicoid}}{q}\; \frac{a^{2}}{r^{2}}\; \left(\pm\; \frac{1}{4} - \frac{1}{2} \right) \\
0 & 0 & \displaystyle 2\; \frac{m_{helicoid}}{q}\; \frac{a^{2}}{r^{2}}\; \left(\frac{1}{\sqrt{2}} + \frac{1}{2} \right) & 0 \\
0 & \displaystyle -\; 2\; \frac{m_{helicoid}}{q}\; \frac{a^{2}}{r^{2}}\; \left(\frac{1}{\sqrt{2}} + \frac{1}{2} \right) & 0 & 0 \\
\displaystyle 2\; \frac{m_{helicoid}}{q}\; \frac{a^{2}}{r^{2}}\; \left(\pm\; \frac{1}{4} - \frac{1}{2} \right) & 0 & 0 & 0 
\end{array} \right) \; .
\label{eq_52}
\end{equation}

From (\ref{eq_52}) we will have two distinct situations:

\begin{equation}
\!\!\!\!\!\!\!\!\!\!\!\!\!\!\!\!\!\!\!\!\!\!\!\!\!\!\!\!\!\!\!\!\!\!\!\!\!(F_{\beta}^{\alpha})^{I}=\left( \begin{array}{cccc}
0 & 0 & 0 & \displaystyle -\; \frac{3}{2}\; \frac{m_{helicoid}}{q}\; \frac{a^{2}}{r^{2}} \\
0 & 0 & \displaystyle \frac{2+\sqrt{2}}{\sqrt{2}}\; \frac{m_{helicoid}}{q}\; \frac{a^{2}}{r^{2}} & 0 \\
0 & \displaystyle -\; \frac{2+\sqrt{2}}{\sqrt{2}}\; \frac{m_{helicoid}}{q}\; \frac{a^{2}}{r^{2}} & 0 & 0 \\
\displaystyle -\; \frac{3}{2}\; \frac{m_{helicoid}}{q}\; \frac{a^{2}}{r^{2}} & 0 & 0 & 0 
\end{array} \right)
\label{eq_53}
\end{equation}
for "-" sign in $E_{3}$ expression, or:

\begin{equation}
\!\!\!\!\!\!\!\!\!\!\!\!\!\!\!\!\!\!\!\!\!\!\!\!\!\!\!\!\!\!\!\!\!\!\!\!\!\!\!\!(F_{\beta}^{\alpha})^{II}=\left( \begin{array}{cccc}
0 & 0 & 0 & \displaystyle -\; \frac{1}{2}\; \frac{m_{helicoid}}{q}\; \frac{a^{2}}{r^{2}} \\
0 & 0 & \displaystyle \frac{2+\sqrt{2}}{\sqrt{2}}\; \frac{m_{helicoid}}{q}\; \frac{a^{2}}{r^{2}} & 0 \\
0 & \displaystyle -\; \frac{2+\sqrt{2}}{\sqrt{2}}\; \frac{m_{helicoid}}{q}\; \frac{a^{2}}{r^{2}} & 0 & 0 \\
\displaystyle -\; \frac{1}{2}\; \frac{m_{helicoid}}{q}\; \frac{a^{2}}{r^{2}} & 0 & 0 & 0 
\end{array} \right)
\label{eq_54}
\end{equation}
\noindent
for "+" sign in $E_{3}$ expression.

As we saw \cite{deus1,deus2}, in a three-dimensional projection of a DEUS object, the spacetime contained inside its inner horizon  is composed of two helicoidal surfaces generated and separated by a rigid string-like singularity. One of these helicoids can contain two FRW bubbles that cross through the initial string-like singularity, from a "before time" to a "time" region of evolution \cite{deus3}, as they are annihilated and are created again. The fields resulting from $(F_{\beta}^{\alpha})^{I}$ propagate in the "before time" of the BHIG helicoid's FRW bubble and are perceived by the "time" FRW observer as ghosts of the movement of an ante-annihilation FRW bubble (inertial frame) relative to the fixed string-like singularity (fixed by the moment $t_{FRW}=0$; non-inertial frame) - in a helicoidal interpretation, or as a pre-collapse DEUS' {\it catenoidal hyper-surface} movement (inertial frame) relative to the collapsed string-like singularity (fixed by the moment $t_{FRW}=0$; non-inertial frame) - in a catenoidal interpretation. The catenoidal and the helicoidal interpretations are part of a single dual image of the same process. The fields resulted from $(F_{\beta}^{\alpha})^{II}$ propagate into the "time" region of the FRW observer. In "time" FRW observer approach, the "before time" fields inertial movement is equivalent with the dark energy effect.

\subsection{Situation I}

For the fields resulting from $(F_{\beta}^{\alpha})^{I}$ we have:

\begin{equation}
\begin{array}{cc}
\!\!\!\!\!\!\!\!\!\!\!\!\!\!\!\!\!\!\!\!\!\!\!\!\!\!\!\!\!\!\!\!\!\!\!\!\!\!\!\!\!\!\!\!\!\!\!\!\!\!\!\!\!\!\!\!\!\!\!\!\!\!\!\!\!\!\!\!\!\!(F_{\beta \alpha})^{I}=(g_{\beta \alpha})^{I}\; (F_{\beta}^{\alpha})^{I}= \\
\!\!\!\!\!\!\!\!\!\!\!\!\!\!\!\!\!\!\!\!\!\!\!\!\!\!\!\!\!\!\!\!\!\!\!\!\!\!\!\!\!\!\!\!\!\!\!\!\!\!\!\!\!\!\!\!\!\!\!\!\!\!\!\!\!\!\!\!\!\!=\left( \begin{array}{cccc}
0 & 0 & 0 & -\; \frac{3}{8}\; \frac{m_{helicoid}}{q}\; \frac{a^{2}}{r^{4}}\; \frac{2H^{2}-2H\sqrt{H^{2}+1}+1}{\left(H^{2}-H\sqrt{H^{2}+1}+1 \right)^{2}} \\
0 & 0 & \frac{2+\sqrt{2}}{4\sqrt{2}}\; \frac{m_{helicoid}}{q}\; \frac{a^{2}}{r^{2}}\; \frac{2H^{2}-2H\sqrt{H^{2}+1}+1}{\left(H^{2}-H\sqrt{H^{2}+1}+1 \right)^{2}} & 0 \\
0 & -\; \frac{2+\sqrt{2}}{4\sqrt{2}}\; \frac{m_{helicoid}}{q}\; \frac{a^{2}}{r^{2}}\; \frac{2H^{2}-2H\sqrt{H^{2}+1}+1}{\left(H^{2}-H\sqrt{H^{2}+1}+1 \right)^{2}}\; \textnormal{sin}^{2}\zeta  & 0 & 0 \\
\frac{9}{8}\; \frac{m_{helicoid}}{q}\; \frac{a^{2}}{r^{2}} & 0 & 0 & 0 
\end{array} \right) \; ,
\end{array}
\label{eq_55}
\end{equation}
\noindent
where $(g_{\beta \alpha})^{I}$ is the metric tensor components of the "before time" BHIG helicoid's FRW representation:

\begin{equation}
\!\!\!\!\!\!\!\!\!\!\!\!\!\!\!\!\!(g_{\beta \alpha})^{I}=\left( \begin{array}{cccc}
\frac{2H^{2}-2H\sqrt{H^{2}+1}+1}{4\; \left(H^{2}-H\sqrt{H^{2}+1}+1 \right)^{2}}\; \frac{1}{r^{2}} & 0 & 0 & 0 \\
0 & \frac{2H^{2}-2H\sqrt{H^{2}+1}+1}{4\; \left(H^{2}-H\sqrt{H^{2}+1}+1 \right)^{2}} & 0 & 0 \\
0 & 0 & \frac{2H^{2}-2H\sqrt{H^{2}+1}+1}{4\; \left(H^{2}-H\sqrt{H^{2}+1}+1 \right)^{2}}\; \textnormal{sin}^{2}\zeta & 0 \\
0 & 0 & 0 & -\; \frac{3}{4} 
\end{array} \right) \; .
\label{eq_56}
\end{equation}

But:

\begin{equation}
(F_{\beta \alpha})^{I}=\left( \begin{array}{cccc}
0 & 0 & 0 & -\; E_{3} \\
0 & 0 & B_{3}  & 0 \\
0 & -\; B_{3} & 0 & 0 \\
E_{3} & 0 & 0 & 0 
\end{array} \right) \; .
\label{eq_57}
\end{equation}

If we identify (\ref{eq_56}) in (\ref{eq_57}) we obtain, first:

\begin{equation}
E_{3} =\frac{3}{8}\; \frac{m_{helicoid}}{q}\; \frac{a^{2}}{r^{4}}\; \frac{2H^{2}-2H\sqrt{H^{2}+1}+1}{\left(H^{2}-H\sqrt{H^{2}+1}+1 \right)^{2}} \; ,
\label{eq_58}
\end{equation}
\noindent
but also:

\begin{equation}
E_{3} =\frac{9}{8}\; \frac{m_{helicoid}}{q}\; \frac{a^{2}}{r^{2}} \; ,
\label{eq_59}
\end{equation}
\noindent
which, must be simultaneously satisfied, from where results that:

\begin{equation}
r^{2} =\frac{2H^{2}-2H\sqrt{H^{2}+1}+1}{3\; \left(H^{2}-H\sqrt{H^{2}+1}+1 \right)^{2}}
\label{eq_60}
\end{equation}
\noindent
of a Minkowski spacetime (four FRW bubbles; see {\it DEUS II} paper); and second:

\begin{equation}
B_{3} =\frac{2+\sqrt{2}}{4\sqrt{2}}\; \frac{m_{helicoid}}{q}\; \frac{a^{2}}{r^{2}}\; \frac{2H^{2}-2H\sqrt{H^{2}+1}+1}{\left(H^{2}-H\sqrt{H^{2}+1}+1 \right)^{2}} \; ,
\label{eq_61}
\end{equation}
\noindent
but also:

\begin{equation}
B_{3} =\frac{2+\sqrt{2}}{4\sqrt{2}}\; \frac{m_{helicoid}}{q}\; \frac{a^{2}}{r^{2}}\; \frac{2H^{2}-2H\sqrt{H^{2}+1}+1}{\left(H^{2}-H\sqrt{H^{2}+1}+1 \right)^{2}}\; \textnormal{sin}^{2}\zeta \; ,
\label{eq_62}
\end{equation}
\noindent
from whose simultaneous validity we get:

\begin{equation}
\zeta=\pm\; \frac{\pi}{2} \; .
\label{eq_63}
\end{equation}

The importance of (\ref{eq_63}) is {\bf huge} because in spherical coordinates, the external observer that sees the DEUS collapsed object as a black hole, and for which, even that the DEUS object is collapsed, the image of this collapse is cut at the radius of the inner horizon (the inner horizon and the pre-collapse cylinder that contains the ergosphere's hyper-surfaces are tangent), at least $\vec{B}$ field is released along the rotational symmetry axis (from the polar region) of the observed "black hole".

Also, if in (\ref{eq_60}) $r=1$, it results (see the function $f(H)$ from paper {\it DEUS II}) that $H=H_{0}\simeq 72.3$ as an unique possible value for a local Euclidian space.

From general relativity we know that the stress-energy tensor can be written as:

\begin{equation}
T^{\mu \nu}=\frac{1}{4\pi}\left(F^{\mu \alpha}\; F_{\alpha}^{\nu}-\frac{1}{4}\; g^{\mu \nu}\; F_{\beta \alpha}\; F^{\beta \alpha} \right) \; .
\label{eq_64}
\end{equation}

In our case:

\begin{equation}
(T^{\beta \alpha})^{I}=\frac{1}{4\pi}\left[(F^{\beta \alpha})^{I}\; (F_{\beta}^{\alpha})^{I}-\frac{1}{4}\; (g^{\beta \alpha})^{I}\; (F_{\beta \alpha})^{I}\; (F^{\beta \alpha})^{I} \right] \; ,
\label{eq_65}
\end{equation}
\noindent
with $(F_{\beta \alpha})^{I}$ from (\ref{eq_55}), $(g_{\beta \alpha})^{I}$ from (\ref{eq_56}) and $(F^{\beta \alpha})^{I}=(g^{\beta \gamma})^{I}\; (F_{\gamma}^{\beta})^{I}$, or explicitely:

\begin{equation}
\begin{array}{cc}
\!\!\!\!\!\!\!\!\!\!\!\!\!\!\!\!\!\!\!\!\!\!\!\!\!\!\!\!\!\!\!\!\!\!\!\!\!\!\!\!\!\!\!\! \displaystyle (T_{\beta \alpha})^{I}=-\; \frac{27}{128\; \pi}\; \frac{m^{2}_{helicoid}}{q^{2}}\; \frac{\left(H^{2}-H\sqrt{H^{2}+1}+1\right)^{4}}{\left(2H^{2}-2H\sqrt{H^{2}+1}+1 \right)^{2}}\; a^{4}\; \times \\
\!\!\!\!\!\!\!\!\!\!\!\!\!\!\!\!\!\!\!\!\!\!\!\!\!\!\!\!\!\!\!\!\!\!\!\!\!\!\!\!\!\!\!\! \times \left( \begin{array}{cccc}
\displaystyle \frac{27}{2} & 0 & 0 & 0 \\
0 & \displaystyle \left(2+\sqrt{2} \right)^{2}\; \frac{2H^{2}-2H\sqrt{H^{2}+1}+1}{\left(H^{2}-H\sqrt{H^{2}+1}+1 \right)^{2}} & 0 & 0 \\
0 & 0 & \displaystyle \left(2+\sqrt{2} \right)^{2}\; \frac{2H^{2}-2H\sqrt{H^{2}+1}+1}{\left(H^{2}-H\sqrt{H^{2}+1}+1 \right)^{2}} & 0 \\
0 & 0 & 0 & \displaystyle \frac{27}{2}
\end{array} \right)\; ,
\end{array}
\label{eq_66}
\end{equation}
\noindent
where we had used the (\ref{eq_60}) and (\ref{eq_63}) constraints.

Now, in the rest frame of an external observer of a "time" Minkowski spacetime (DEUS object with all the four FRW bubbles seen together as Minkowski spacetime) described by the metric tensor:

\begin{equation}
(\eta_{\beta \alpha})^{II}=\left( \begin{array}{cccc}
-\; 1 & 0 & 0 & 0 \\
0 & 1 & 0 & 0 \\
0 & 0 & 1 & 0 \\
0 & 0 & 0 & 1 
\end{array} \right) \; ,
\label{eq_67}
\end{equation}
\noindent
which sees the "black hole" effect released into the "before time" part of the Minkowski spacetime of metric tensor:

\begin{equation}
(\eta_{\beta \alpha})^{I}=\left( \begin{array}{cccc}
1 & 0 & 0 & 0 \\
0 & 1 & 0 & 0 \\
0 & 0 & 1 & 0 \\
0 & 0 & 0 & -\; 1 
\end{array} \right) \; ,
\label{eq_68}
\end{equation}
\noindent
the stress-energy tensor is:

\begin{equation}
(\mathcal{T}_{\beta \alpha})^{I}\equiv (\eta_{\beta \alpha})^{I}\; (T_{\beta \alpha})^{I}=\left( \begin{array}{cccc}
p_{\Lambda} & 0 & 0 & 0 \\
0 & p_{\Lambda} & 0 & 0 \\
0 & 0 & p_{\Lambda} & 0 \\
0 & 0 & 0 & \rho_{\Lambda} 
\end{array} \right) \; ,
\label{eq_69}
\end{equation}
\noindent
where $p_{\Lambda}$ is the kinetic pressure and $\rho_{\Lambda}$ the density for the considered dark energy (seen as coming from a perfect fluid) effect.

Because from (\ref{eq_66}) and (\ref{eq_69}) results that we must have $(\mathcal{T}_{11})^{I}=-\; (\mathcal{T}_{44})^{I}$, the equation of state for dark energy (in a perfect fluid approximation) will be:

\begin{equation}
\rho_{\Lambda}=-\; p_{\Lambda} \; ,
\label{eq_70}
\end{equation}
\noindent
where:

\begin{equation}
\rho_{\Lambda}=\frac{729}{256\; \pi}\; \frac{m^{2}_{helicoid}}{q^{2}}\; \frac{\left(H^{2}-H\sqrt{H^{2}+1}+1\right)^{4}}{\left(2H^{2}-2H\sqrt{H^{2}+1}+1 \right)^{2}}\; a^{4}
\label{eq_71}
\end{equation}
\noindent
or, as function of $r$ (from (\ref{eq_60})):

\begin{equation}
\rho_{\Lambda}=\frac{81}{256\; \pi}\; \frac{m^{2}_{helicoid}}{q^{2}}\; \frac{a^{4}}{r^{4}}
\label{eq_72}
\end{equation}
\noindent
and so appears the law:

\begin{equation}
r\propto \rho_{\Lambda}^{-\; 1/4}
\label{eq_73}
\end{equation}
\noindent
that gives the way in which the dark energy density varies with the radius of the FRW bubble in which the observer and its local Minkowski spacetime is globally immersed (self-similarity between the DEUS object properties seen as a FRW Universe or as a black hole).

Also, from the $(\mathcal{T}_{11})^{I}=(\mathcal{T}_{22})^{I}$ equality resides (here we do not consider a Hubble constant but a Hubble parameter, case in which it is not necessary to have $r=1$):

\begin{equation}
r=\frac{3}{\sqrt{2}\; (2+\sqrt{2})}\; ,
\label{eq_74}
\end{equation}
\noindent
with which, in (\ref{eq_72}):

\begin{equation}
\rho_{\Lambda}=\frac{\left(2+\sqrt{2}\right)^{4}}{64\; \pi}\; \frac{m^{2}_{helicoid}}{q^{2}}\; a^{4}
\label{eq_75}
\end{equation}
\noindent
and, when demanding to be satisfied also the equation of state (\ref{eq_70}):

\begin{equation}
p_{\Lambda}=-\; \frac{\left(2+\sqrt{2}\right)^{4}}{64\; \pi}\; \frac{m^{2}_{helicoid}}{q^{2}}\; a^{4} \; ,
\label{eq_76}
\end{equation}
\noindent
which states that, when the Hubble parameter is not constant in time ($t_{FRW}$), but the observation distance of the FRW observer is kept constant for a long enough time, the pressure and density of the dark energy component will be seen as increasing with the fourth power of the scale factor {\it a}.

With (\ref{eq_75}) and (\ref{eq_76}) in \cite{deus2}:

\begin{equation}
\Lambda=\frac{8\pi\; G\; \left[\rho + 2\; p\; \left(H^{2}+1 \right)^{2} \right]}{ 2\; \left(H^{2}+1 \right)^{2} - 1} \; ,
\label{eqa_76}
\end{equation}
\noindent
and considering a dark energy dominated era ($\rho\equiv \rho_{\Lambda}$, $p\equiv p_{\Lambda}$) we obtain a negative cosmological constant:

\begin{equation}
\Lambda=-\; 8\pi\; G\; \rho_{\Lambda} \; .
\label{eqa_77}
\end{equation}

\subsection{Situation II}

As said before, from the FRW observer perspective, the effects given at the DEUS collapse (black hole formation or Big Bang) are here directly observable as "time" effects. The metric tensor of this spacetime is the same one that forms the FRW bubble (global perspective of the Universe) or the Minkowski spacetime (local perspective of the Universe):

\begin{equation}
\!\!\!\!\!\!\!\!\!\!\!\!\!\!\!\!\!\!\!\!\!\!\!\!(g_{\beta \alpha})^{II}=\left( \begin{array}{cccc}
-\; \frac{2H^{2}-2H\sqrt{H^{2}+1}+1}{4\; \left(H^{2}-H\sqrt{H^{2}+1}+1 \right)^{2}}\; \frac{1}{r^{2}} & 0 & 0 & 0 \\
0 & -\; \frac{2H^{2}-2H\sqrt{H^{2}+1}+1}{4\; \left(H^{2}-H\sqrt{H^{2}+1}+1 \right)^{2}} & 0 & 0 \\
0 & 0 & -\; \frac{2H^{2}-2H\sqrt{H^{2}+1}+1}{4\; \left(H^{2}-H\sqrt{H^{2}+1}+1 \right)^{2}}\; \sin^{2}\zeta & 0 \\
0 & 0 & 0 & \frac{3}{4}
\end{array} \right) \; .
\label{eq_77}
\end{equation}

From the above metric tensor and (\ref{eq_54}) we obtain:

\begin{equation}
\begin{array}{cc}
\!\!\!\!\!\!\!\!\!\!\!\!\!\!\!\!\!\!\!\!\!\!\!\!\!\!\!\!\!\!\!\!\!\!\!\!\!\!\!\!\!\!\!\!\!\!\!\!\!\!\!\!\!\!\!\!\!\!\!\!\!\!\!\!\!\!\!\!\!\! (F_{\beta \alpha})^{II}=(g_{\beta \alpha})^{II}\; (F_{\beta}^{\alpha})^{II}= \\
\!\!\!\!\!\!\!\!\!\!\!\!\!\!\!\!\!\!\!\!\!\!\!\!\!\!\!\!\!\!\!\!\!\!\!\!\!\!\!\!\!\!\!\!\!\!\!\!\!\!\!\!\!\!\!\!\!\!\!\!\!\!\!\!\!\!\!\!\!\! =\left( \begin{array}{cccc}
0 & 0 & 0 & \frac{1}{8}\; \frac{m_{helicoid}}{q}\; \frac{a^{2}}{r^{4}}\; \frac{2H^{2}-2H\sqrt{H^{2}+1}+1}{\left(H^{2}-H\sqrt{H^{2}+1}+1 \right)^{2}} \\
0 & 0 & -\; \frac{2+\sqrt{2}}{4\sqrt{2}}\; \frac{m_{helicoid}}{q}\; \frac{a^{2}}{r^{2}}\; \frac{2H^{2}-2H\sqrt{H^{2}+1}+1}{\left(H^{2}-H\sqrt{H^{2}+1}+1 \right)^{2}} & 0 \\
0 & \frac{2+\sqrt{2}}{4\sqrt{2}}\; \frac{m_{helicoid}}{q}\; \frac{a^{2}}{r^{2}}\; \frac{2H^{2}-2H\sqrt{H^{2}+1}+1}{\left(H^{2}-H\sqrt{H^{2}+1}+1 \right)^{2}}\; \textnormal{sin}^{2}\zeta  & 0 & 0 \\
-\; \frac{3}{8}\; \frac{m_{helicoid}}{q}\; \frac{a^{2}}{r^{2}} & 0 & 0 & 0 
\end{array} \right) \; .
\end{array}
\label{eq_78}
\end{equation}

But, because of the way of defining the fields:

\begin{equation}
(F_{\beta \alpha})^{II}=\left( \begin{array}{cccc}
0 & 0 & 0 & -\; E_{3} \\
0 & 0 & B_{3}  & 0 \\
0 & -\; B_{3} & 0 & 0 \\
E_{3} & 0 & 0 & 0 
\end{array} \right) \; ,
\label{eq_78a}
\end{equation}
\noindent
we should have:

\begin{equation}
E_{3} =-\; \frac{3}{8}\; \frac{m_{helicoid}}{q}\; \frac{a^{2}}{r^{2}} \; ,
\label{eq_79}
\end{equation}
\noindent
identical with:

\begin{equation}
E_{3} =-\; \frac{1}{8}\; \frac{m_{helicoid}}{q}\; \frac{a^{2}}{r^{4}}\; \frac{2H^{2}-2H\sqrt{H^{2}+1}+1}{\left(H^{2}-H\sqrt{H^{2}+1}+1 \right)^{2}} \; ,
\label{eq_80}
\end{equation}
\noindent
which, makes us find again the {\it Situation I} result (same as {\it DEUS II} paper's and necessary for the transformation of four FRW bubbles in a Minkowski spacetime):

\begin{equation}
r^{2} =\frac{2H^{2}-2H\sqrt{H^{2}+1}+1}{3\; \left(H^{2}-H\sqrt{H^{2}+1}+1 \right)^{2}} \; ,
\label{eq_81}
\end{equation}
\noindent
which for $r=1$ gives $H=H_{0}=72.3$.

Identifying also the other two components from (\ref{eq_78a}) in (\ref{eq_78}) we get:

\begin{equation}
B_{3} =-\; \frac{2+\sqrt{2}}{4\sqrt{2}}\; \frac{m_{helicoid}}{q}\; \frac{a^{2}}{r^{2}}\; \frac{2H^{2}-2H\sqrt{H^{2}+1}+1}{\left(H^{2}-H\sqrt{H^{2}+1}+1 \right)^{2}} \; ,
\label{eq_82}
\end{equation}
\noindent
but also:

\begin{equation}
B_{3} =-\; \frac{2+\sqrt{2}}{4\sqrt{2}}\; \frac{m_{helicoid}}{q}\; \frac{a^{2}}{r^{2}}\; \frac{2H^{2}-2H\sqrt{H^{2}+1}+1}{\left(H^{2}-H\sqrt{H^{2}+1}+1 \right)^{2}}\; \textnormal{sin}^{2}\zeta \; ,
\label{eq_83}
\end{equation}
\noindent
with the same result as in {\it Situation I} section:

\begin{equation}
\zeta=\pm\; \frac{\pi}{2} \; .
\label{eq_84}
\end{equation}

Again, as in the dark energy effect case, the emission takes place from the rotation symmetry axis of the DEUS object seen as black hole, the only difference being that, here, the external observer really sees the effect occurring in his proper time.  

Applying (\ref{eq_65}), but for the {\it II} tensors, it follows that:

\begin{equation}
\begin{array}{cc}
\!\!\!\!\!\!\!\!\!\!\!\!\!\!\!\!\!\!\!\!\!\!\!\!\!\!\!\!\!\!\!\!\!\!\!\!\!\!\!\!\!\!\!\!\!\!\!\!\!\!\!\!\!\!\!\!\!\!\!\!\!\!\!\!\!\!\!\!\!\! \displaystyle (T_{\beta \alpha})^{II}=\frac{1}{4\pi}\; \frac{m^{2}_{helicoid}}{q^{2}}\; \times \\
\!\!\!\!\!\!\!\!\!\!\!\!\!\!\!\!\!\!\!\!\!\!\!\!\!\!\!\!\!\!\!\!\!\!\!\!\!\!\!\!\!\!\!\!\!\!\!\!\!\!\!\!\!\!\!\!\!\!\!\!\!\!\!\!\!\!\!\!\!\!\times \left( \begin{array}{cccc}
-\; \frac{15}{4}\; \frac{9}{16}\; \frac{\left(H^{2}-H\sqrt{H^{2}+1}+1\right)^{4}}{\left(2H^{2}-2H\sqrt{H^{2}+1}+1 \right)^{2}}\; a^{4} & 0 & 0 & 0 \\
0 & \frac{27}{16}\; \left(\frac{2+\sqrt{2}}{\sqrt{2}}\right)^{2}\;  \frac{\left(H^{2}-H\sqrt{H^{2}+1}+1\right)^{2}}{2H^{2}-2H\sqrt{H^{2}+1}+1}\; a^{4} & 0 & 0 \\
0 & 0 & \frac{27}{16}\; \left(\frac{2+\sqrt{2}}{\sqrt{2}}\right)^{2}\;  \frac{\left(H^{2}-H\sqrt{H^{2}+1}+1\right)^{2}}{2H^{2}-2H\sqrt{H^{2}+1}+1}\; a^{4} & 0 \\
0 & 0 & 0 & \frac{15}{4}\; \frac{9}{16}\; \frac{\left(H^{2}-H\sqrt{H^{2}+1}+1\right)^{4}}{\left(2H^{2}-2H\sqrt{H^{2}+1}+1 \right)^{2}}\; a^{4}
\end{array} \right) \; ,
\end{array}
\label{eq_85}
\end{equation}
\noindent
where we used also (\ref{eq_81}) and (\ref{eq_84}).

In the rest frame of an external observer of a "time" Minkowski spacetime with the (\ref{eq_67}) metric tensor, the stress-energy tensor of the effects (into the same "time" part of the Minkowski spacetime as the one of the observer) given by the evolution of the "black hole" toward collapse, is:

\begin{equation}
(\mathcal{T}_{\beta \alpha})^{II}\equiv (\eta_{\beta \alpha})^{II}\; (T_{\beta \alpha})^{II}=\left( \begin{array}{llll}
\rho & 0 & 0 & 0 \\
0 & p & 0 & 0 \\
0 & 0 & p & 0 \\
0 & 0 & 0 & p 
\end{array} \right) \; ,
\label{eq_86}
\end{equation}
\noindent
where $p$ stands for the kinetic pressure and $\rho$ for the density (in the assumption of a perfect fluid) of {\bf all} the "time" effects. These effects are: dark matter, normal (baryonic) matter and radiation (photons).

When we establish the correspondence between (\ref{eq_85}) and (\ref{eq_86}) we find that $p=(\mathcal{T}_{33})^{II}=(\mathcal{T}_{44})^{II}$, from where we obtain:

\begin{equation}
\rho=\frac{27\; \left(2+\sqrt{2}\right)^{4}}{320\; \pi}\; \frac{m^{2}_{helicoid}}{q^{2}}\; a^{4} \; .
\label{eq_87}
\end{equation}

As an interesting result of (\ref{eq_87}) and (\ref{eq_75}) is:

\begin{equation}
\frac{\rho_{\Lambda}}{\rho}=\frac{5}{27}\simeq 0.185 \; .
\label{eq_88}
\end{equation}

Also, when the equation of state is described as sum over the "time" effects (dark matter, baryonic matter and radiation):

\begin{equation}
p=\rho=w_{1}\; \rho_{DM} + w_{2}\; \rho_{M} + w_{3}\; \rho_{rad} \; ,
\label{eq_89}
\end{equation}
\noindent
with $w_{2}=0$ for matter, $w_{3}=\displaystyle \frac{1}{3}$ for radiation, we get (in a matter dominated era):

\begin{equation}
p_{DM}=\frac{2}{3}\; \rho_{DM} \; .
\label{eq_90}
\end{equation}

(\ref{eq_81}) substitution in $p=(\mathcal{T}_{44})^{II}$ brings us the:

\begin{equation}
\rho=\frac{15}{256\; \pi}\; \frac{m^{2}_{helicoid}}{q^{2}}\; \frac{a^{4}}{r^{4}}
\label{eq_91}
\end{equation}
\noindent
dependency $r\propto \rho^{-\; 1/4}$ with same interpretation as for the one obtained in {\it Situation I} section, but this time applicable for dark matter, matter and energy (photons).

\section{Conclusions}

In this paper, using the Lorentz equations of motion we shown that the {\it catenoidal hyper-surface}'s embedded self-similar DEUS objects' 
{\it helicoidal hyper-surfaces} have non-zero electromagnetic tensor components. The electromagnetic tensor is having two possible 
forms. The external observer will perceive the matter or energy fields residing from each of these forms as: inertial effects coming 
from a previous ("before time") Big Bang - Big Crunch cycle of evolution, hidden by the initial string-like singularity (dark energy);
particles or fields from the present ("time") Big Bang - Big Crunch cycle of evolution (observer in a global FRW observable
Universe bubble) or from Black Holes (observer in a local Minkowski spacetime) (dark matter, baryonic matter and electromagnetic 
radiation). For both field classes we found the associated densities and pressures, and also the density dependence of the 
different fields with the scale factor and the observed Universe radius.

In the Minkowski local spacetime we obtained an extremely interesting behavior: the fields are emitted from the evaporating black hole 
DEUS object at $\displaystyle \zeta=\pm\; \frac{\pi}{2}$ polar angle ({\bf black hole jets}).

\ack
Manuscript registered at the Astronomical Institute, Romanian Academy, record No. 252 from 04/19/07.


\title[D.E.U.S.]{V: Fields and Waves}

\begin{abstract}
We continuate the study from paper {\it DEUS IV} considering instead the {\it catenoidal hyper-surface}'s embedded self-similar DEUS objects' {\it helicoidal hyper-surfaces} their {\it catenoidal hyper-surfaces} counterparts. By releasing these objects into the external spacetime, in the process of DEUS collapse, the {\it catenoidal hyper-surfaces} of the embedded DEUS objects will behave as matter, antimatter and dark energy, filling the empty spacetime created by the {\it reduced catenoidal hyper-surface} of the collapsing DEUS object.
\end{abstract}


\section{Introduction}

Having as starting point the trivial three-dimensional minimal surface representation of the catenoid and its conjugate surface, the helicoid, in paper {\it DEUS I} \cite{deus1} we constructed a five-dimensional spacetime geometry. There we determined the five-dimensional coordinate system that satisfies the Cauchy-Riemann equations, and also the ansatz, the properties and the correlation possible to be made for the DEUS object's composing manifolds.

In {\it DEUS II}  paper \cite{deus2} we generalized the geometry of the DEUS Black Hole for Universe formation and evolution, by using self-similar minimal surfaces for DEUS objects distribution. From the assumed self-similarity of the five-dimensional internal geometry, we shown that the DEUS object collapses to a string-like object equivalent with the BHIG spacetime of a bigger DEUS object in which the first one is embedded.

The considered five-dimensional geometry becomes conformal with an external Friedmann-Robertson-Walker (FRW) spacetime (for which the Hubble constant is {\it uniquely} defined as $72.33\pm 0.02$) by evolving the hyper-surfaces composing our five-dimensional DEUS object to a pre-collapse manifold. For the BHIG spacetime written in FRW coordinates we derived the Friedmann equations and some cosmological consequences ($\Omega_{tot}$ and the cosmological distance).

Into the background of a collapsed DEUS object rotating string that creates a BHIG manifold "deformed" in a FRW bubble, the pre-collapse movement of the five-dimensional catenoid hyper-surfaces is perceived by an after-collapse external observer as dark energy effect, while the matter still contained in the collapsed DEUS spacetime as dark matter.

The collapsing helicoidal manifold can be seen by the external timelike observer only considering that the catenoid is having $N$ internal DEUS similar objects, distorted, also collapsing. The helicoids of these $N$-DEUS objects are timelike (five-dimensional) and are giving three-dimensional spacelike and timelike observable effects to the external space.

In the above geometry framework, in {\it DEUS III} \cite{deus3} we found that the energy contained into the {\it catenoidal hyper-surface} and into the {\it helicoidal hyper-surface} can be characterized using the Yamabe energy. We shown that the energy stored into the curvature of the {\it catenoidal hyper-surface} is released (at the DEUS object evaporation) into the external space as kinetic energy in form of spatial or temporal perturbations. Also, by applying a combination of Ricci flow and mean curvature flow and taking account of the conformity of the {\it reduced catenoidal hyper-surface} \cite{deus2} with the flat external space, we obtained the scalar curvature of the {\it pure catenoidal hyper-surface}, for which the metric was too complicated to be determined in the "classical" way.

Using the mean curvature flow, we determined \cite{deus3} the velocity with which the {\it catenoidal hyper-surface} evolves to the FRW external space and the one of the BHIG's FRW bubble from its formation ("Big Bang") to its disappearance ("Big Crunch"). As we said in {\it DEUS II}, this movement is not possible to be directly observed by the external observer, which perceives only its inertial effect as gravitational effect (dark energy in helicoidal frame). In the Lorentz frame of the external observer these velocities are comparable with the infall velocity in the core collapse process of a star becoming a black hole.

Because of the fact that the self-similar DEUS objects are embedded into the {\it 
catenoidal hyper-surface} of a higher scale DEUS object, they will be highly distorted from the perfect symmetric case, and with them 
all the contained surfaces. This perturbation of symmetry makes them evolve to a minimal case in curvature and energy, dragging 
also the surface in which they are contained (which are components of a DEUS object also distorted by an embedding in a superior scale 
{\it catenoidal hyper-surface}). The distortion increases from the neck of the catenoid to its edge, where it gives a more intense 
effect to the external flat space. The given effects are different not only because of the distortion, but also because of the 
different {\it catenoidal hyper-surface}'s immersed DEUS object dimension, which decreases from the neck to the edge of the catenoid (the Yamabe energy describing the energy of the DEUS hyper-surfaces decreases in the same direction). In a non-perturbed symmetry we were supposed to have no effect to the external space. But the distortion makes the effects of same nature to not compensate each other, their resultant being perceived by the external observer while the DEUS object evolves toward collapse.

We had shown into paper {\it DEUS IV} \cite{deus4} that the {\it catenoidal hyper-surface}'s embedded self-similar DEUS objects' 
{\it helicoidal hyper-surfaces} have non-zero electromagnetic tensor components. The electromagnetic tensor is having two possible 
forms. The external observer will perceive the matter or energy fields residing from each of these forms as: inertial effects coming 
from a previous ("before time") Big Bang - Big Crunch evolution cycle, hidden by the initial string-like singularity (dark energy);
particles or radiation fields from the present ("time") Big Bang - Big Crunch evolution cycle (observer in a global FRW observable Universe bubble) or from black holes (observer in a local Minkowski spacetime) (dark matter, baryonic matter and electromagnetic radiation). For finding the densities and pressures associated to these fields we used the perfect fluid approximation.

In the Minkowski local spacetime we observed an extremely interesting behavior, the fields being emitted from the evaporating 
black hole DEUS object at $\displaystyle \zeta=\pm\; \frac{\pi}{2}$ polar angle ({\bf black hole jets}).

While in paper {\it DEUS IV} we considered only the effects given by the embedded self-similar DEUS objects' 
{\it helicoidal hyper-surfaces} (timelike perturbations of the external spacetime), here we will analyze also their {\it catenoidal 
hyper-surfaces} counterparts which, as it will be seen, give again the matter and dark energy fields but also 
their dual representation as waves, respectively gravitational waves into the external local Minkowski spacetime.

\section{Stability of Catenoidal Bridges}

\begin{figure}
\centering
\leavevmode
\includegraphics[width=7.0cm,height=7.0cm]{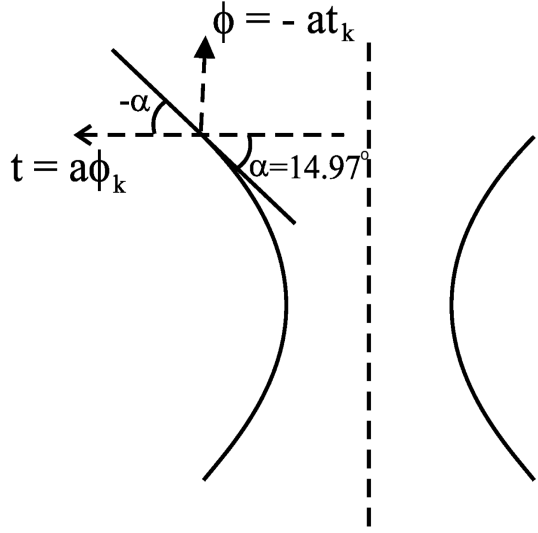}
\caption{The $\alpha$ angle through which we can describe the stability of the {\it catenoidal hyper-surface}.}
\label{deus4d}
\end{figure}

Zhou \cite{zhou} proved that for a catenoidal bridge trapped between two plates (situated in the plane perpendicular on the rotation 
symmetry axis of the catenoid) and making equal contact angles $\alpha$ with them, it can be established a stability criterion 
which states that there exist a $\alpha_{0}$ such that a stable catenoid forms if $\alpha > \alpha_{0}$. The bridge will not be 
formed or it will be unstable if $\alpha < \alpha_{0}$. Zhou showed that this angle is $\alpha_{0}=14.97^{\circ}$. For our 
three-dimensional projection DEUS object's {\it catenoidal hyper-surface} we see this angle in Fig. \ref{deus4d}. Into 
$(t,\; \phi)$ catenoidal coordinates we will have:

\begin{equation}
\textnormal{tg}\; \alpha = \frac{\phi}{t}\; .
\label{eq_1}
\end{equation}

The external observer is able to perceive \cite{deus4} only the effects given by the catenoids and helicoids of the self-similar 
DEUS objects immersed into the collapsing {\it catenoidal hyper-surface}. While in paper {\it DEUS IV} we analyzed the timelike effects (helicoids) given to the external BHIG spacetime seen as a local Minkowski 
or global FRW spacetime, here we will study the form taken by the spacelike effects (catenoids). These spacelike effects 
will perturb the Minkowski or the FRW metric in a reversible (black hole DEUS object), respectively, irreversible manner (giving a permanent curvature to the external conformal flat BHIG helicoidal spacetime; Universe DEUS object). The irreversible effects will curve the BHIG spacetime as a FRW bubble and through its matter and radiation fields seen in the "wave" representation of the duality {\it catenoid-helicoid into collapsing DEUS object}$\rightarrow$ {\it wave-particle into FRW or Minkowski} \cite{deus4}. While the visible FRW matter bubble Universe forms into a $\pi/2$ sector of the string-like collapsed DEUS object, the Minkowski local spacetime will contain all the $2\pi$. But, in the same time, the local Minkowski is part of a global FRW spacetime representation of a BHIG spacetime contained in a higher scale DEUS object. This makes that, into the infinite levels of self-similar DEUS objects topology, only the irreversible effects to be possible.

\section{Effects}

In $(t_{k},\; \phi_{k})$ coordinates of the external observer situated into the BHIG helicoid generated by the rotating string-like 
collapsed {\it catenoidal hyper-surface} \cite{deus1,deus2}:

\begin{equation}
\alpha_{k} = \frac{\pi}{2} - 2\; \alpha=\frac{\pi}{2} - 2\; \frac{\pi}{12}=\frac{\pi}{3} \; ,
\label{eq_2}
\end{equation}
\noindent
where the emission takes place into the $\pi/2$ sector of one FRW bubble in which one of the self-similar immersed DEUS object's 
catenoidal bridges covers an overall angle of $2\times \pi/12$ (the angle made by the catenoidal bridge 
with the "plates" between which it is contained).

In this way, translating the above stability criterion in $\alpha_{k}$, we will have a stable catenoidal bridge for $\alpha_{k}" > 
\alpha_{k}$ and a unstable catenoidal bridge for $\alpha_{k}" < \alpha_{k}$.

The favorable case will be the unstable {\it catenoidal hyper-surface} which evolves until reaches stability and zero mean curvature 
\cite{deus2}. This will happen for $\displaystyle \alpha_{k}" < \frac{\pi}{3}$.

\begin{figure}
\centering
\leavevmode
\includegraphics[width=10.0cm,height=10.0cm]{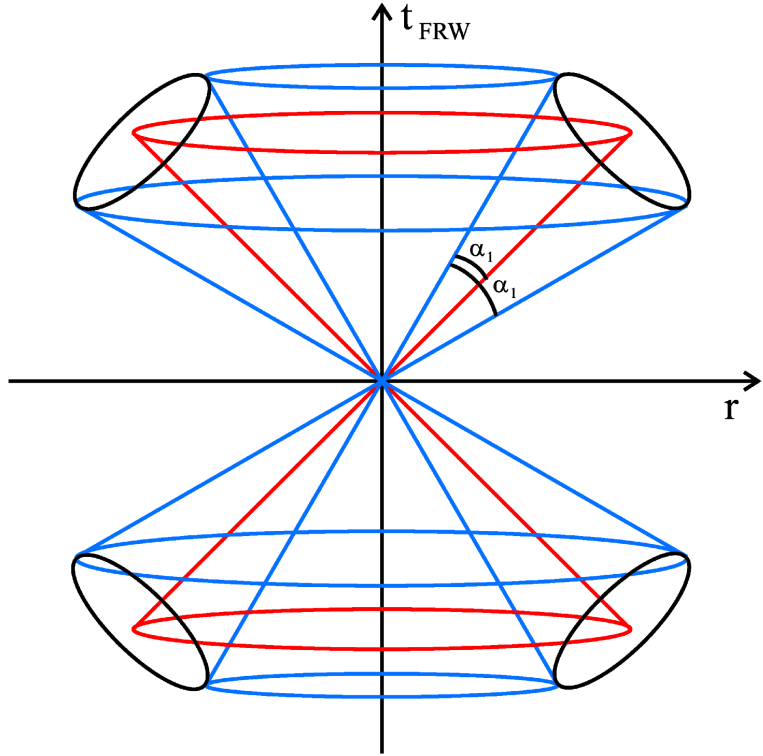}
\caption{The light cone of the external observer is represented in red. The spacelike perturbations resulting from the DEUS collapse 
are emitted in the angle contained between the light cone ($\alpha_{k}"=\frac{\pi}{4}$) and its external blue cone 
($\frac{\pi}{6}\leq \alpha_{k}" < \frac{\pi}{4}$), while the timelike perturbations, between the light cone and its internal 
blue cone ($\frac{\pi}{4} < \alpha_{k}"\leq  \frac{\pi}{3}$). The external observer situated into a FRW spacetime will be able to 
see only one of the four cones as coming from the DEUS object that generated its Universe. If the observer sees a DEUS black hole 
it will be in a local Minkowski spacetime part of a global FRW Universe. In this case, it will see all the four emissions of the 
DEUS black hole as events into his spacetime.}
\label{deus4a}
\end{figure}

\begin{figure}
\centering
\leavevmode
\includegraphics[width=8.0cm,height=8.0cm]{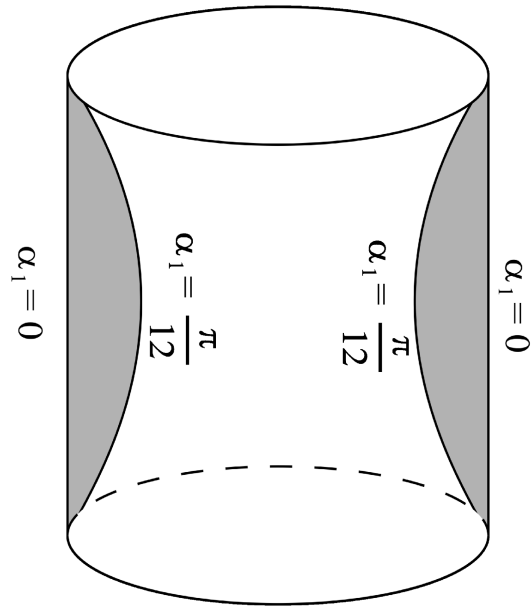}
\caption{The {\it catenoidal hyper-surface} begins to lose its self-similar internal structure at $\alpha_{1}=\frac{\pi}{12}$ and evolves until $\alpha_{1}=0$ where is a {\it reduced catenoidal hyper-surface}, without internal structure that forms the empty external BHIG spacetime.}
\label{deus4c}
\end{figure}

Into the light cone of the external FRW spacetime frame the spatial perturbation effects will be exclusively spacelike for 
$\displaystyle \frac{\pi}{6}\leq \alpha_{k}" < \frac{\pi}{4}$ and exclusively timelike for $\displaystyle \frac{\pi}{4} < \alpha_{k}"\leq \frac{\pi}{3}$ 
(see Fig. \ref{deus4a}); $\displaystyle \alpha_{k}"= \frac{\pi}{4}\pm \alpha_{1}$ and $\alpha_{1}\in \left[0,\; \alpha_{0} \right]$, where $\alpha_{0}$ is the angle found by Zhou \cite{zhou} for the stability of the catenoidal bridge. Fig. \ref{deus4c} gives a pictorial representation of how the catenoidal bridge will look like for the DEUS pre-collapse object $\alpha_{1}$ angle limiting values.  

The timelike effects will be the "dual" representation of the fields studied into {\it DEUS IV} paper \cite{deus4}, of the DEUS objects {\it helicoidal hyper-surfaces} contained into the {\it catenoidal hyper-surface} of the higher scale collapsing DEUS object, while the spacelike perturbations will give the curvature modifications of the BHIG external spacetime to the FRW bubble and the "waves" that locally (Minkowski) perturb this FRW bubble.

From (\ref{eq_1}) and knowing that at the DEUS object collapse $\displaystyle \frac{\phi}{t}=-\; \frac{t_{k}}{\phi_{k}}$ \cite{deus1,deus2} we 
observe that:

\begin{equation}
\textnormal{tg}\; \alpha =-\; \frac{t_{k}}{\phi_{k}} \; .
\label{eq_3}
\end{equation}

But because we have \cite{deus2} the:

\begin{equation}
\textnormal{tg}\; t_{FRW} = \frac{t_{k}}{\phi_{k}}
\label{eq_3a}
\end{equation}
\noindent
transformation from helicoidal to FRW coordinates, results:

\begin{equation}
\textnormal{tg}\; t_{FRW} = -\; \textnormal{tg}\; \alpha \; .
\label{eq_3b}
\end{equation}

It follows that, with (\ref{eq_3b}) and:

\begin{equation}
\begin{array}{ll}
\displaystyle \phi_{k}=\frac{r}{a}\; \frac{1}{\sqrt{1+\textnormal{tg}^{2}\; t_{FRW}}} \\
\displaystyle t_{k}=\frac{r}{a}\; \frac{\textnormal{tg}\; t_{FRW}}{\sqrt{1+\textnormal{tg}^{2}\; t_{FRW}}}
\end{array}
\label{eq_4}
\end{equation}
\noindent
relations \cite{deus2}, we can write:

\begin{equation}
\frac{t_{k}}{t_{k}^{2}+\phi_{k}^{2}}=-\; \frac{a}{r}\; \frac{\textnormal{tg}\; \alpha}{\sqrt{\textnormal{tg}^{2}\alpha +1}}\; ,
\label{eq_5}
\end{equation}
\noindent
or, in the $\alpha_{k}"$ angle of the BHIG external observer ($\alpha\rightarrow \alpha_{k}"$):

\begin{equation}
\frac{t_{k}}{t_{k}^{2}+\phi_{k}^{2}}=-\; \frac{a}{r}\; \frac{\textnormal{tg}\; \alpha_{k}"}{\sqrt{\textnormal{tg}^{2}\alpha_{k}" +1}}\; .
\label{eq_6}
\end{equation}

The {\it catenoidal hyper-surface} of the immersed DEUS objects will give into the external spacetime ansatz the perturbation:

\begin{equation}
h_{\beta \alpha}=\left( \begin{array}{cccc}
\displaystyle -\; \frac{t_{k}}{t_{k}^{2}+\phi_{k}^{2}} & 0 & 0 & 0 \\
0 & \displaystyle -\; \frac{t_{k}}{t_{k}^{2}+\phi_{k}^{2}} & 0 & 0 \\
0 & 0 & \displaystyle -\; \frac{t_{k}}{t_{k}^{2}+\phi_{k}^{2}} & 0 \\
0 & 0 & 0 & \displaystyle \frac{t_{k}}{t_{k}^{2}+\phi_{k}^{2}} 
\end{array} \right) \; ,
\label{eq_7}
\end{equation}
\noindent
that we had to exclude when we studied \cite{deus4} the effects of the self-similar immersed DEUS objects' {\it helicoidal 
hyper-surfaces}.

We saw in paper {\it DEUS IV} \cite{deus4} that the $x^{3}$ in the pre-collapse local chart of catenoid's immersed DEUS objects goes to $x^{4}$ in the global chart of the BHIG remaining after the DEUS collapse. So, $h_{33}$ component from (\ref{eq_7}) is the one which actually generates the external BHIG helicoidal symmetry. Writing (\ref{eq_7}) with the help of (\ref{eq_6}) we obtain:

\begin{equation}
\!\!\!\!\!\!\!\!\!\!\!\!\!\!\!\!\!\!h_{\beta \alpha}=\left( \begin{array}{cccc}
\displaystyle \frac{a"}{r"}\; \frac{\textnormal{tg}\; \alpha_{k}"}{\sqrt{\textnormal{tg}^{2}\alpha_{k} +1}} & 0 & 0 & 0 \\
0 & \displaystyle \frac{a"}{r"}\; \frac{\textnormal{tg}\; \alpha_{k}"}{\sqrt{\textnormal{tg}^{2}\alpha_{k} +1}} & 0 & 0 \\
0 & 0 & \displaystyle \frac{a}{r}\; \frac{\textnormal{tg}\; \alpha_{k}}{\sqrt{\textnormal{tg}^{2}\alpha_{k} +1}} & 0 \\
0 & 0 & 0 & \displaystyle -\; \frac{a"}{r"}\; \frac{\textnormal{tg}\; \alpha_{k}"}{\sqrt{\textnormal{tg}^{2}\alpha_{k} +1}}
\end{array} \right) \; ,
\label{eq_8}
\end{equation}
\noindent
where {\it a} and {\it r} are the scale factor, respectively the radial coordinate of the external FRW spacetime created at 
the collapse of the DEUS object and, $a"$ and $r"$ are the scale factor, respectively the radial coordinate of the 
perturbation propagating inside the space having {\it a} and {\it r}. {\it a"} and {\it r"} can be pictured as properties of the FRW "particles" exiting from $\theta=0$ of the BHIG spacetime of the immersed into the {\it catenoidal hyper-surface} 
self-similar DEUS objects. We saw in {\it DEUS IV} paper that, at the moment of collapse, the studied DEUS object (Universe or 
black hole) must satisfy the relation $t_{k}=\pm\; \phi_{k}$ \cite{deus4}. For 
$\displaystyle \textnormal{tg}\; \alpha_{k}=\textnormal{tg}\; t_{FRW}=\frac{t_{k}}{\phi_{k}}$, the last relation makes $\textnormal{tg}\; 
\alpha_{k}=\mp\; 1$ and, in consequence $\displaystyle \alpha_{k}=\frac{\pi}{4}$. Also, $\displaystyle \alpha_{k}"= \alpha_{k}\pm \alpha_{1}=\frac{\pi}{4}\pm 
\alpha_{1}$. The tensor (\ref{eq_8}) becomes:

\begin{equation}
\!\!\!\!\!\!\!\!\!\!\!\!\!\!\!\!\!\!\!\!\!\!\!\!\!\!\!\!\!\!\!\!\!\!\!\!\!\!\!\!\!\!\!h_{\beta \alpha}=\left( \begin{array}{cccc}
\displaystyle \frac{a"}{r"}\; \frac{\textnormal{tg}\; \left(\displaystyle \frac{\pi}{4}\pm \alpha_{1}\right)}{\sqrt{\textnormal{tg}^{2}\left(\displaystyle \frac{\pi}{4}\pm \alpha_{1}\right)+1}} & 0 & 0 & 0 \\
0 & \displaystyle \frac{a"}{r"}\; \frac{\textnormal{tg}\; \left(\displaystyle \frac{\pi}{4}\pm \alpha_{1}\right)}{\sqrt{\textnormal{tg}^{2}\left(\displaystyle \frac{\pi}{4}\pm \alpha_{1}\right)+1}} & 0 & 0 \\
0 & 0 & \displaystyle \mp \frac{a}{r}\; \frac{\textnormal{tg}\; \displaystyle \frac{\pi}{4}}{\sqrt{\displaystyle \textnormal{tg}^{2}\frac{\pi}{4}+1}} & 0 \\
0 & 0 & 0 & \displaystyle -\; \frac{a"}{r"}\; \frac{\textnormal{tg}\; \left(\displaystyle \frac{\pi}{4}\pm \alpha_{1}\right)}{\sqrt{\textnormal{tg}^{2}\left(\displaystyle \frac{\pi}{4}\pm \alpha_{1}\right)+1}}
\end{array} \right) \; ,
\label{eq_9}
\end{equation}
\noindent
with $\displaystyle \alpha_{1}\in \left[0,\; \frac{\pi}{12} \right]$.

From (\ref{eq_9}) we distinguish two main possible forms of $h_{\beta \alpha}$:

\begin{enumerate}
\item
\begin{equation}
\!\!\!\!\!\!\!\!\!\!\!\!\!\!\!\!\!\!\!\!\!\!\!\!\!\!\!\!\!\!\!\!\!\!\!\!\!\!\!\!\!\!\!(h_{\beta \alpha})^{I}=\left( \begin{array}{cccc}
\displaystyle \frac{a"}{r"}\; \frac{\textnormal{tg}\; \left(\displaystyle \frac{\pi}{4}\pm \alpha_{1}\right)}{\sqrt{\textnormal{tg}^{2}\left(\displaystyle \frac{\pi}{4}\pm \alpha_{1}\right)+1}} & 0 & 0 & 0 \\
0 & \displaystyle \frac{a"}{r"}\; \frac{\textnormal{tg}\; \left(\displaystyle \frac{\pi}{4}\pm \alpha_{1}\right)}{\sqrt{\textnormal{tg}^{2}\left(\displaystyle \frac{\pi}{4}\pm \alpha_{1}\right)+1}} & 0 & 0 \\
0 & 0 & \displaystyle -\; \frac{a}{r}\; \frac{\textnormal{tg}\; \displaystyle \frac{\pi}{4}}{\sqrt{\displaystyle \textnormal{tg}^{2}\frac{\pi}{4}+1}} & 0 \\
0 & 0 & 0 & \displaystyle -\; \frac{a"}{r"}\; \frac{\textnormal{tg}\; \left(\displaystyle \frac{\pi}{4}\pm \alpha_{1}\right)}{\sqrt{\textnormal{tg}^{2}\left(\displaystyle \frac{\pi}{4}\pm \alpha_{1}\right)+1}}
\end{array} \right) \; ,
\label{eq_10}
\end{equation}
\noindent

\item
\begin{equation}
\!\!\!\!\!\!\!\!\!\!\!\!\!\!\!\!\!\!\!\!\!\!\!\!\!\!\!\!\!\!\!\!\!\!\!\!\!\!\!\!\!\!\!(h_{\beta \alpha})^{II}=\left( \begin{array}{cccc}
\displaystyle \frac{a"}{r"}\; \frac{\textnormal{tg}\; \left(\displaystyle \frac{\pi}{4}\pm \alpha_{1}\right)}{\sqrt{\textnormal{tg}^{2}\left(\displaystyle \frac{\pi}{4}\pm \alpha_{1}\right)+1}} & 0 & 0 & 0 \\
0 & \displaystyle \frac{a"}{r"}\; \frac{\textnormal{tg}\; \left(\displaystyle \frac{\pi}{4}\pm \alpha_{1}\right)}{\sqrt{\textnormal{tg}^{2}\left(\displaystyle \frac{\pi}{4}\pm \alpha_{1}\right)+1}} & 0 & 0 \\
0 & 0 & \displaystyle \frac{a}{r}\; \frac{\textnormal{tg}\; \displaystyle \frac{\pi}{4}}{\sqrt{\displaystyle \textnormal{tg}^{2}\frac{\pi}{4}+1}} & 0 \\
0 & 0 & 0 & \displaystyle -\; \frac{a"}{r"}\; \frac{\textnormal{tg}\; \left(\displaystyle \frac{\pi}{4}\pm \alpha_{1}\right)}{\sqrt{\textnormal{tg}^{2}\left(\displaystyle \frac{\pi}{4}\pm \alpha_{1}\right)+1}}
\end{array} \right) \; .
\label{eq_11}
\end{equation}
\noindent

\end{enumerate}

\begin{figure}
\centering
\leavevmode
\includegraphics[width=14.0cm,height=8.0cm]{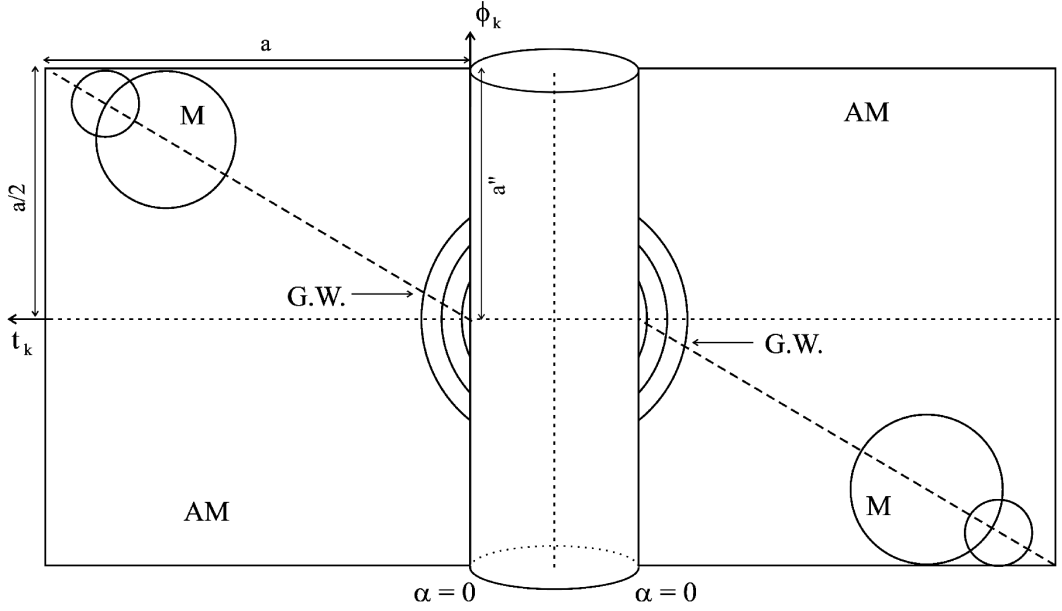}
\caption{DEUS object collapse at $\alpha_{1}=0$. The four matter-antimatter FRW bubbles are emitted after their associated gravitational wave emission ($\alpha_{1}>0$).}
\label{deus4b}
\end{figure}

First, we will focus our attention on $(h_{\beta \alpha})^{II}$, which we can also write as:

\begin{equation}
(h_{\beta \alpha})^{II}=\left( \begin{array}{cccc}
\displaystyle \frac{a"}{r"}\; \textnormal{sin}\; \left(\displaystyle \frac{\pi}{4}\pm \alpha_{1}\right) & 0 & 0 & 0 \\
0 & \displaystyle \frac{a"}{r"}\; \textnormal{sin}\; \left(\frac{\pi}{4}\pm \alpha_{1}\right) & 0 & 0 \\
0 & 0 & \displaystyle \frac{a}{\sqrt{2}\; r} & 0 \\
0 & 0 & 0 & \displaystyle -\; \frac{a"}{r"}\; \textnormal{sin}\; \left(\frac{\pi}{4}\pm \alpha_{1}\right)
\end{array} \right) \; .
\label{eq_12}
\end{equation}

If $\alpha_{1}=0$ in (\ref{eq_12}):

\begin{equation}
(h_{\beta \alpha})^{II}=\left( \begin{array}{cccc}
\displaystyle \frac{a"}{\sqrt{2}\; r"} & 0 & 0 & 0 \\
0 & \displaystyle \frac{a"}{\sqrt{2}\; r"} & 0 & 0 \\
0 & 0 & \displaystyle \frac{a}{\sqrt{2}\; r} & 0 \\
0 & 0 & 0 & \displaystyle -\; \frac{a"}{\sqrt{2}\; r"}
\end{array} \right) \; ,
\label{eq_13}
\end{equation}
\noindent
and when the perturbation is of the same magnitude as the external, after-collapse, created spacetime ({\it catenoidal hyper-surface} with only one immersed 
DEUS object having the same dimension as the collapsing DEUS object to which this {\it catenoidal hyper-surface} appertains), 
$r"=r$ and $a"=a$:

\begin{equation}
(h_{\beta \alpha})^{II}=\frac{a}{\sqrt{2}\; r}\; \left( \begin{array}{cccc}
1 & 0 & 0 & 0 \\
0 & 1 & 0 & 0 \\
0 & 0 & 1 & 0 \\
0 & 0 & 0 & -\; 1
\end{array} \right)\equiv \frac{a}{\sqrt{2}\; r}\; \eta_{\beta \alpha} \; ,
\label{eq_14}
\end{equation}
\noindent
which is the perturbation in the Minkowski local external spacetime of the observer that sees the DEUS collapsing object as a 
black hole.

If $\displaystyle \alpha_{1}=\frac{\pi}{12}$ in (\ref{eq_12}) then:

\begin{equation}
\!\!\!\!\!\!\!\!\!\!\!\!\!\!\!\!\!\!\!\!\!\!\!\!\!\!\!\!\!\!\!\!\!\!\!\!\!\!\!\!\!\!\! (h_{\beta \alpha})^{II}=\left( \begin{array}{cccc}
\displaystyle \frac{a"}{\sqrt{2}\; r"}\; \left(\displaystyle \textnormal{cos}\; \frac{\pi}{12}\pm \textnormal{sin}\; \frac{\pi}{12} \right) & 0 & 0 & 0 \\
0 & \displaystyle \frac{a"}{\sqrt{2}\; r"}\; \left(\displaystyle \textnormal{cos}\; \frac{\pi}{12}\pm \textnormal{sin}\; \frac{\pi}{12} \right) & 0 & 0 \\
0 & 0 & \displaystyle \frac{a}{\sqrt{2}\; r} & 0 \\
0 & 0 & 0 & \displaystyle -\; \frac{a"}{\sqrt{2}\; r"}\; \left(\displaystyle \textnormal{cos}\; \frac{\pi}{12}\pm \textnormal{sin}\; \frac{\pi}{12} \right)
\end{array} \right).
\label{eq_15}
\end{equation}

We can decompose the $(h_{\beta \alpha})^{II}$ tensor as:

\begin{equation}
\begin{array}{cc}
2\; (h_{\beta \alpha})^{II}=2\; \left[(h_{\beta \alpha})^{II,\; +} + (h_{\beta \alpha})^{II,\; -} \right]= \\
=\left\{ \begin{array}{ll}
2\; \left[(h_{\beta \alpha})^{II,\; +}_{\alpha_{1}=0} + (h_{\beta \alpha})^{II,\; -}_{\alpha_{1}=0} \right]\; \; \textnormal{for}\; \; \alpha_{1}=0 \\
2\; \left[(h_{\beta \alpha})^{II,\; +}_{\alpha_{1}=\frac{\pi}{12}} + (h_{\beta \alpha})^{II,\; -}_{\alpha_{1}=\frac{\pi}{12}} \right]=(h_{\beta \alpha})^{II,\; +}_{\alpha_{1}=\frac{\pi}{12}} + (h_{\beta \alpha})^{I,\; -}_{\alpha_{1}=\frac{\pi}{12}} + 2\; (h_{\beta \alpha})^{II,\; -}_{\alpha_{1}=\frac{\pi}{12}}\; \; \textnormal{for}\; \; \alpha_{1}=\frac{\pi}{12}
\end{array} \right.
\end{array}
\label{eq_16}
\end{equation}
\noindent
where:

\begin{equation}
(h_{\beta \alpha})^{II,\; +}=\left( \begin{array}{cccc}
\displaystyle \frac{a"}{r"}\; \textnormal{sin}\; \left(\displaystyle \frac{\pi}{4} + \alpha_{1}\right) & 0 & 0 & 0 \\
0 & \displaystyle \frac{a"}{r"}\; \textnormal{sin}\; \left(\displaystyle \frac{\pi}{4} + \alpha_{1}\right) & 0 & 0 \\
0 & 0 & \displaystyle \frac{a}{r}\; \textnormal{sin}\; \frac{\pi}{4} & 0 \\
0 & 0 & 0 & 0
\end{array} \right) \; ,
\label{eq_17}
\end{equation}

\begin{equation}
(h_{\beta \alpha})^{II,\; -}=\left( \begin{array}{cccc}
0 & 0 & 0 & 0 \\
0 & 0 & 0 & 0 \\
0 & 0 & \displaystyle \frac{a}{r}\; \textnormal{sin}\; \frac{\pi}{4} & 0 \\
0 & 0 & 0 & \displaystyle -\; \frac{a"}{r"}\; \textnormal{sin}\; \left(\displaystyle \frac{\pi}{4} - \alpha_{1}\right)
\end{array} \right) \; ,
\label{eq_18}
\end{equation}

\begin{equation}
(h_{\beta \alpha})^{I,\; +}=\left( \begin{array}{cccc}
\displaystyle \frac{a"}{r"}\; \textnormal{sin}\; \left(\displaystyle \frac{\pi}{4} + \alpha_{1}\right) & 0 & 0 & 0 \\
0 & \displaystyle \frac{a"}{r"}\; \textnormal{sin}\; \left(\displaystyle \frac{\pi}{4} + \alpha_{1}\right) & 0 & 0 \\
0 & 0 & \displaystyle -\; \frac{a}{r}\; \textnormal{sin}\; \frac{\pi}{4} & 0 \\
0 & 0 & 0 & 0
\end{array} \right) \; ,
\label{eq_19}
\end{equation}

\begin{equation}
(h_{\beta \alpha})^{I,\; -}=\left( \begin{array}{cccc}
0 & 0 & 0 & 0 \\
0 & 0 & 0 & 0 \\
0 & 0 & \displaystyle -\; \frac{a}{r}\; \textnormal{sin}\; \frac{\pi}{4} & 0 \\
0 & 0 & 0 & \displaystyle -\; \frac{a"}{r"}\; \textnormal{sin}\; \left(\displaystyle \frac{\pi}{4} - \alpha_{1}\right)
\end{array} \right) \; .
\label{eq_20}
\end{equation}

\subsection{$(h_{\beta \alpha})^{II,\; +}$ Effects}

When $\alpha_{1}=0$ in (\ref{eq_17}):

\begin{equation}
(h_{\beta \alpha})^{II,\; +}_{\alpha_{1}=0}=\left( \begin{array}{cccc}
\displaystyle \frac{a"}{r"\; \sqrt{2}} & 0 & 0 & 0 \\
0 & \displaystyle \frac{a"}{r"\; \sqrt{2}} & 0 & 0 \\
0 & 0 & \displaystyle \frac{a}{r\; \sqrt{2}} & 0 \\
0 & 0 & 0 & 0
\end{array} \right) \; .
\label{eq_21}
\end{equation}

Because the perturbation propagates into the FRW spacetime of metric: 

\begin{equation}
\!\!\!\!\!\!\!\!\!\!\!\!\!\!\!\!\!\!\!\!\!\!\!\!\!\!\!\!\!\!\!\!\!\!\!\!\!\!\!\!\!\!\!\!\!\!\!\!\!\!\!\!\!\!\!\!\!\!\!(g_{\beta \alpha})_{FRW}=\left( \begin{array}{cccc}
\displaystyle -\; \frac{2H^{2}-2H\sqrt{H^{2}+1}+1}{4\; \left(H^{2}-H\sqrt{H^{2}+1}+1 \right)^{2}}\; \frac{1}{r^{2}} & 0 & 0 & 0 \\
0 & \displaystyle -\; \frac{2H^{2}-2H\sqrt{H^{2}+1}+1}{4\; \left(H^{2}-H\sqrt{H^{2}+1}+1 \right)^{2}} & 0 & 0 \\
0 & 0 & \displaystyle -\; \frac{2H^{2}-2H\sqrt{H^{2}+1}+1}{4\; \left(H^{2}-H\sqrt{H^{2}+1}+1 \right)^{2}}\; \textnormal{sin}^{2}\zeta & 0 \\
0 & 0 & 0 & \displaystyle \frac{3}{4}
\end{array} \right) \; ,
\label{eq_22}
\end{equation}
\noindent
we can write:

\begin{equation}
\begin{array}{ll}
(h_{\beta}^{\alpha})^{II,\; +}_{\alpha_{1}=0}=(g^{\beta \alpha})_{FRW}\; (h_{\beta \alpha})^{II,\; +}_{\alpha_{1}=0} \\
(h^{\beta \alpha})^{II,\; +}_{\alpha_{1}=0}=(g^{\beta \alpha})_{FRW}\; (h_{\beta}^{\alpha})^{II,\; +}_{\alpha_{1}=0}\; .
\end{array} 
\label{eq_23}
\end{equation}

The stress-energy tensor will be:

\begin{equation}
(T^{\beta \alpha})^{II,\; +}_{\alpha_{1}=0}=\frac{1}{4\; \pi}\; \left[(h^{\beta \alpha})^{II,\; +}_{\alpha_{1}=0}\; 
(h_{\beta}^{\alpha})^{II,\; +}_{\alpha_{1}=0} - \frac{1}{4}\; (g^{\beta \alpha})_{FRW}\; (h_{\beta \alpha})^{II,\; +}_{\alpha_{1}=0}\; 
(h^{\beta \alpha})^{II,\; +}_{\alpha_{1}=0} \right]\; .
\label{eq_24}
\end{equation}

After few computations we get:

\begin{equation}
\!\!\!\!\!\!\!\!\!\!\!\!\!\!\!\!\!\!\!\!\!\!\!\!\!\!\!\!\!\!\!\!\!\!\!\!\!\!\!\!\!\!\!\!\!\!\!\!\!\!\!\!\!\!\!\!\!\!\!\!\!\!\!\!\!\!\!\!\!\!\!\!\!\!\!\!\!\!\!\!\!\!\!\!\!\!\!(T_{\beta \alpha})^{II,\; +}_{\alpha_{1}=0}=\frac{1}{4\; \pi}\left( \begin{array}{cccc}
\displaystyle \frac{(a")^{2}}{2}\; \frac{3\; \left(H^{2}-H\sqrt{H^{2}+1}+1\right)^{2}}{2H^{2}-2H\sqrt{H^{2}+1}+1} & 0 & 0 & 0 \\
0 & \displaystyle \frac{(a")^{2}}{2}\; \frac{3\; \left(H^{2}-H\sqrt{H^{2}+1}+1\right)^{2}}{2H^{2}-2H\sqrt{H^{2}+1}+1}\; \frac{1}{(r")^{2}} & 0 & 0 \\
0 & 0 & \displaystyle \frac{a^{2}}{2}\; \frac{3\; \left(H^{2}-H\sqrt{H^{2}+1}+1\right)^{2}}{2H^{2}-2H\sqrt{H^{2}+1}+1}\; \frac{1}{r^{2}} & 0 \\
0 & 0 & 0 & 0
\end{array} \right) \; ,
\label{eq_25}
\end{equation}
\noindent
where we used the {\it DEUS IV} \cite{deus4} $\displaystyle \zeta=\pm\; \frac{\pi}{2}$ result for the external observer that perceives the 
effect as coming from the spin axis of the black hole DEUS object, and:

\begin{equation}
(T_{\beta \alpha})^{II,\; +}_{\alpha_{1}=0}=(g_{\beta \alpha})_{FRW}\; 
(g_{\beta \alpha})_{FRW}\; (T^{\beta \alpha})^{II,\; +}_{\alpha_{1}=0}\; .
\label{eq_26}
\end{equation}

In the frame of the local "time" Minkowski spacetime \cite{deus4} we have:

\begin{equation}
(\eta_{\beta \alpha})^{II}=\left( \begin{array}{cccc}
-\; 1 & 0 & 0 & 0 \\
0 & 1 & 0 & 0 \\
0 & 0 & 1 & 0 \\
0 & 0 & 0 & 1 
\end{array} \right) \; .
\label{eq_26a}
\end{equation}

Considering the perfect fluid case:

\begin{equation}
(\mathcal{T}_{\beta \alpha})^{II,\; +}_{\alpha_{1}=0}\equiv (\eta_{\beta \alpha})^{II}\; (T_{\beta \alpha})^{II,\; 
+}_{\alpha_{1}=0}=\left( \begin{array}{llll}
\rho & 0 & 0 & 0 \\
0 & p & 0 & 0 \\
0 & 0 & p & 0 \\
0 & 0 & 0 & p 
\end{array} \right) \; ,
\label{eq_27}
\end{equation}
\noindent
we observe that we must have:

\begin{equation}
\rho=(\mathcal{T}_{11})^{II,\; +}_{\alpha_{1}=0}=(\mathcal{T}_{22})^{II,\; +}_{\alpha_{1}=0}\; ,
\label{eq_28}
\end{equation}
\noindent
and asking to be fulfilled also the condition that we found in paper {\it DEUS II} 
\cite{deus2}:

\begin{equation}
r^{2}=(r")^{2}= \frac{2H^{2}-2H\sqrt{H^{2}+1}+1}{3\; \left(H^{2}-H\sqrt{H^{2}+1}+1\right)^{2}}\; ,
\label{eq_29}
\end{equation}
\noindent
results that $p=-\; \rho$. Also, from (\ref{eq_25}) and (\ref{eq_27}):

\begin{equation}
(\mathcal{T}_{22})^{II,\; +}_{\alpha_{1}=0}=(\mathcal{T}_{33})^{II,\; +}_{\alpha_{1}=0}\; ,
\label{eq_30}
\end{equation}
\noindent
which is assured by the self-similarity of DEUS objects at any scale ($a=a"$ 
and $H=H"$) and by (\ref{eq_29}).

But, because $(T_{44})^{II,\; +}_{\alpha_{1}=0}=0$, we have 
$p=0$ and $w=-1$ of a dark energy effect coming from the "before time" matter component 
movement relative to the zero reference frame string-like object  and hidden because of it (see also the comments from {\it 
DEUS IV} paper).

Now, when $\displaystyle \alpha_{1}=\frac{\pi}{12}$ in (\ref{eq_17}):

\begin{equation}
(h_{\beta \alpha})^{II,\; +}_{\alpha_{1}=\frac{\pi}{12}}=\left( \begin{array}{cccc}
\displaystyle \frac{a"}{r"}\; \frac{\sqrt{3}}{2} & 0 & 0 & 0 \\
0 & \displaystyle \frac{a"}{r"}\; \frac{\sqrt{3}}{2}  & 0 & 0 \\
0 & 0 & \displaystyle \frac{a}{r\; \sqrt{2}} & 0 \\
0 & 0 & 0 & 0
\end{array} \right) \; .
\label{eq_31}
\end{equation}

With $(g_{\beta \alpha})_{FRW}$ from (\ref{eq_22}):

\begin{equation}
\begin{array}{ll}
(h_{\beta}^{\alpha})^{II,\; +}_{\alpha_{1}=\frac{\pi}{12}}=(g^{\beta \alpha})_{FRW}\; (h_{\beta \alpha})^{II,\; +}_{\alpha_{1}=\frac{\pi}{12}} \\
(h^{\beta \alpha})^{II,\; +}_{\alpha_{1}=\frac{\pi}{12}}=(g^{\beta \alpha})_{FRW}\; (h_{\beta}^{\alpha})^{II,\; +}_{\alpha_{1}=\frac{\pi}{12}}\; ,
\end{array} 
\label{eq_32}
\end{equation}
\noindent
the stress-energy tensor will be:

\begin{equation}
(T^{\beta \alpha})^{II,\; +}_{\alpha_{1}=\frac{\pi}{12}}=\frac{1}{4\; \pi}\; \left[(h^{\beta \alpha})^{II,\; +}_{\alpha_{1}=\frac{\pi}{12}}\; 
(h_{\beta}^{\alpha})^{II,\; +}_{\alpha_{1}=\frac{\pi}{12}} - \frac{1}{4}\; (g^{\beta \alpha})_{FRW}\; (h_{\beta \alpha})^{II,\; +}_{\alpha_{1}=\frac{\pi}{12}}\; 
(h^{\beta \alpha})^{II,\; +}_{\alpha_{1}=\frac{\pi}{12}} \right]\; ,
\label{eq_33}
\end{equation}
\noindent
and:

\begin{equation}
(T_{\beta \alpha})^{II,\; +}_{\alpha_{1}=\frac{\pi}{12}}=(g_{\beta \alpha})_{FRW}\; 
(g_{\beta \alpha})_{FRW}\; (T^{\beta \alpha})^{II,\; +}_{\alpha_{1}=\frac{\pi}{12}}\; ,
\label{eq_34}
\end{equation}
\noindent
and we can conclude that the stress-energy tensor can be written as:

\begin{equation}
\!\!\!\!\!\!\!\!\!\!\!\!\!\!\!\!\!\!\!\!\!\!\!\!\!\!\!\!\!\!\!\!\!\!\!\!\!\!\!\!\!\!\!\!\!\!\!\!\!\!\!\!\!\!\!\!\!\!\!\!\!\!\!\!\!\!\!\!\!\!\!\!\!\!(T_{\beta \alpha})^{II,\; +}_{\alpha_{1}=\frac{\pi}{12}}=\frac{1}{4\; \pi}\left( \begin{array}{cccc}
\displaystyle \frac{9}{4}\; \frac{\left(H^{2}-H\sqrt{H^{2}+1}+1\right)^{2}}{2H^{2}-2H\sqrt{H^{2}+1}+1}\; \left(\frac{a"}{r"}\right)^{2} & 0 & 0 & 0 \\
0 & \displaystyle \frac{9}{4}\; \frac{\left(H^{2}-H\sqrt{H^{2}+1}+1\right)^{2}}{2H^{2}-2H\sqrt{H^{2}+1}+1}\; \left(\frac{a"}{r"}\right)^{2} & 0 & 0 \\
0 & 0 & \displaystyle \frac{3}{2}\; \frac{\left(H^{2}-H\sqrt{H^{2}+1}+1\right)^{2}}{2H^{2}-2H\sqrt{H^{2}+1}+1}\; \left(\frac{a}{r}\right)^{2} & 0 \\
0 & 0 & 0 & 0
\end{array} \right) \; .
\label{eq_35}
\end{equation}

In the approximation of the perfect fluid in a local Minkowski spacetime and, in the same time, in a global FRW spacetime:

\begin{equation}
(\mathcal{T}_{\beta \alpha})^{II,\; +}_{\alpha_{1}=\frac{\pi}{12}}\equiv (\eta_{\beta \alpha})^{II}\; (T_{\beta \alpha})^{II,\; 
+}_{\alpha_{1}=\frac{\pi}{12}}=\left( \begin{array}{llll}
\rho & 0 & 0 & 0 \\
0 & p & 0 & 0 \\
0 & 0 & p & 0 \\
0 & 0 & 0 & p 
\end{array} \right) \; ,
\label{eq_36}
\end{equation}
\noindent
we are able to see that we must have the identity:

\begin{equation}
(\mathcal{T}_{11})^{II,\; +}_{\alpha_{1}=\frac{\pi}{12}}=(\mathcal{T}_{22})^{II,\; +}_{\alpha_{1}=\frac{\pi}{12}}\; ,
\label{eq_37}
\end{equation}
\noindent
or, equivalentely: 

\begin{equation}
p=w\; \rho\; ,
\label{eq_39}
\end{equation}
\noindent
where the parameter from the equation of state proves to be $w=-1$ of the previously obtained dark energy effect.
But, because $(T_{44})^{II,\; +}_{\alpha_{1}=\frac{\pi}{12}}=0$, results that $p=0$ (matter in the "before time" evolving as a 
FRW bubble Universe toward Big Crunch or, equivalently, toward a "time" Big Bang at the central string-like singularity).

We must have also the equality:

\begin{equation}
(T_{33})^{II,\; +}_{\alpha_{1}=\frac{\pi}{12}}=(T_{22})^{II,\; +}_{\alpha_{1}=\frac{\pi}{12}}\; ,
\label{eq_38}
\end{equation}
\noindent
from which, when $H=H"$ and $a=a"$ as for self-similar DEUS objects at any scale, we get $r^{2}=\displaystyle \frac{2}{3}\; (r")^{2}$. If at the present 
time the FRW bubble is having a radius $R$, it means that the dark energy effect will be at $R"=\sqrt{\displaystyle \frac{3}{2}}\; R$, becoming completely visible (not only through its influence) only at the $\alpha_{1}=0$ DEUS collapse moment (Big Bang or Big Crunch moment).

\subsection{$(h_{\beta\alpha})^{II,\; -}$ Effects}

In (\ref{eq_18}), following the same mathematical reasoning we had done for $(h_{\beta \alpha})^{II,\; +}$, when $\alpha_{1}=0$:

\begin{equation}
(h_{\beta \alpha})^{II,\; -}_{\alpha_{1}=0}=\left( \begin{array}{cccc}
0 & 0 & 0 & 0 \\
0 & 0 & 0 & 0 \\
0 & 0 & \displaystyle \frac{a}{r\; \sqrt{2}} & 0 \\
0 & 0 & 0 & \displaystyle -\; \frac{a"}{r"\; \sqrt{2}}
\end{array} \right) \; ,
\label{eq_41}
\end{equation}
\noindent
it results ($(g_{\beta \alpha})_{FRW}$ is the same one as in (\ref{eq_22})):

\begin{equation}
(T_{\beta \alpha})^{II,\; -}_{\alpha_{1}=0}=\frac{1}{4\; \pi}\; \frac{3}{4}\;\left( \begin{array}{cccc}
0 & 0 & 0 & 0 \\
0 & 0 & 0 & 0 \\
0 & 0 & \displaystyle \frac{a^{2}}{2}\; \frac{4\; \left(H^{2}-H\sqrt{H^{2}+1}+1\right)^{2}}{2H^{2}-2H\sqrt{H^{2}+1}+1}\; \frac{1}{r^{2}} & 0 \\
0 & 0 & 0 & \displaystyle -\; \frac{4}{6}\; \left(\frac{a"}{r"} \right)^{2}
\end{array} \right) \; .
\label{eq_42}
\end{equation}

For a perfect fluid:

\begin{equation}
(\mathcal{T}_{\beta \alpha})^{II,\; -}_{\alpha_{1}=0}\equiv (\eta_{\beta \alpha})^{I}\; 
(T_{\beta \alpha})^{II,\; -}_{\alpha_{1}=0}=\left( \begin{array}{llll}
p & 0 & 0 & 0 \\
0 & p & 0 & 0 \\
0 & 0 & p & 0 \\
0 & 0 & 0 & \rho 
\end{array} \right) \; ,
\label{eq_43}
\end{equation}
\noindent
where the emission takes place in a Minkowski spacetime described by \cite{deus4}:

\begin{equation}
(\eta_{\beta \alpha})^{I}=\left( \begin{array}{cccc}
1 & 0 & 0 & 0 \\
0 & 1 & 0 & 0 \\
0 & 0 & 1 & 0 \\
0 & 0 & 0 & -\; 1 
\end{array} \right) \; ,
\label{eq_43a}
\end{equation}
\noindent
the stress-energy tensor will be:

\begin{equation}
(\mathcal{T}_{\beta \alpha})^{II,\; -}_{\alpha_{1}=0}=\frac{1}{4\; \pi}\; \frac{3}{4}\;\left( \begin{array}{cccc}
0 & 0 & 0 & 0 \\
0 & 0 & 0 & 0 \\
0 & 0 & \displaystyle \frac{a^{2}}{2}\; \frac{4\; \left(H^{2}-H\sqrt{H^{2}+1}+1\right)^{2}}{2H^{2}-2H\sqrt{H^{2}+1}+1}\; \frac{1}{r^{2}} & 0 \\
0 & 0 & 0 & \displaystyle \frac{4}{6}\; \left(\frac{a"}{r"} \right)^{2}
\end{array} \right) \; .
\label{eq_42b}
\end{equation}

When we describe the same DEUS object effect as for $(h_{\beta \alpha})^{II,\; +}_{\alpha_{1}=0}$, but emitted in an opposite 
Minkowski spacetime, for $\alpha_{1}=0$ we have $a=a"$, $H=H"$, $r=r"$. In (\ref{eq_42b}) we must have:

\begin{equation}
(\mathcal{T}_{33})^{II,\; -}_{\alpha_{1}=0}=(\mathcal{T}_{44})^{II,\; -}_{\alpha_{1}=0}\; .
\label{eq_43c}
\end{equation}

In this situation we obtain the equation of state $p=\rho$. Also:

\begin{equation}
(\mathcal{T}_{11})^{II,\; -}_{\alpha_{1}=0}=(\mathcal{T}_{22})^{II,\; -}_{\alpha_{1}=0}=(\mathcal{T}_{33})^{II,\; -}_{\alpha_{1}=0}\; ,
\label{eq_44}
\end{equation}
\noindent
implies that $p=0$. $(h_{\beta \alpha})^{II,\; -}_{\alpha_{1}=0}$ will describe a matter component emission into the "time" part 
of the Universe (the after Big Bang Universe). In the $(h_{\beta \alpha})^{II,\; -}_{\alpha_{1}=\frac{\pi}{12}}$ situation $a=a"$, $H=H"$, $r=2\; r"$: emission of matter.

Following the same type of mathematical development, we get for $(h_{\beta \alpha})^{I,\; +}$:

\begin{itemize}
\item When $\alpha_{1}=0$

$a=\pm\; i\; a"$ and $p=0$ of an antimatter effect (also $p=\rho$).

\item When $\displaystyle \alpha_{1}=\frac{\pi}{12}$

$a=\pm\; i\; a"$, $\displaystyle r^{2}=\frac{2}{3}\; (r")^{2}$, and $p=0$ of an antimatter effect (also $p=\rho$). The antimatter and matter FRW bubbles are separated by $\phi_{k}=0$ of the BHIG helicoid. Relative to the matter FRW bubble the equivalent antimatter FRW bubble is outside its horizon, becoming visible at Big Bang or Big Crunch, when they annihilate each other (at the string-like central object). The only antimatter contributions to the FRW bubble comes from the black hole DEUS object, that emits (and annihilates) both type of FRW "particles" into the local Minkowski spacetime.

\end{itemize}
\noindent
and in the case of $(h_{\beta \alpha})^{I,\; -}$ for $\alpha_{1}=0$: $a=\pm\; i\; a"$, $p=-\; \rho$, but also $p=0$ of a dark energy 
effect coming from the relative motion of the matter (or antimatter) FRW bubble of a previous Big Bang - Big Crunch cycle. 

The final conclusion that follows is if, for example, $(h_{\beta \alpha})^{II}$ tensor will give the matter FRW bubble and the 
dark energy component, $(h_{\beta \alpha})^{I}$ will create the antimatter FRW bubble and its dark energy component. 

{\bf We predict} that the gravitational wave tensor is given by the 
$\left|(h_{\beta \alpha})^{II,\; -}_{\alpha_{1}=0}-(h_{\beta \alpha})^{II,\; -}_{0<\alpha_{1}\leq \frac{\pi}{12}}\right|$ in the case of baryonic matter, or by $\left|(h_{\beta \alpha})^{I,\; +}_{0<\alpha_{1}\leq \frac{\pi}{12}}-(h_{\beta \alpha})^{I,\; +}_{\alpha_{1}=0}\right|$ difference in the case of baryonic antimatter, as an "anticipation" of the matter emission FRW "particle" bubbles at $\alpha_{1}=0$ moment of the DEUS object (black hole) 
collapse into the local Minkowski spacetime, effect that curves the empty BHIG spacetime.

Also, from the matter-antimatter annihilation into a local Minkowski perception, at $\alpha_{1}=0$, we expect an electromagnetic perturbation given by $\left|(h_{\beta \alpha})^{II,\; -}_{\alpha_{1}=0} + (h_{\beta \alpha})^{I,\; +}_{\alpha_{1}=0} \right|$ and spatial perturbations, at $0<\alpha_{1}\leq \displaystyle \frac{\pi}{12}$, $\left|(h_{\beta \alpha})^{II,\; -}_{0<\alpha_{1}\leq \frac{\pi}{12}} + (h_{\beta \alpha})^{I,\; +}_{0<\alpha_{1}\leq \frac{\pi}{12}}\right|$, with their corresponding stress-energy tensors.

\section{Conclusions}

While in paper {\it DEUS IV} we considered only the effects given by the embedded self-similar DEUS objects' 
{\it helicoidal hyper-surfaces} (timelike perturbations of the external spacetime), here we analyzed also their {\it catenoidal 
hyper-surfaces} counterparts, that give again the matter and dark energy fields, but also 
their dual representation as waves (at $\alpha_{1}=0$), respectively gravitational waves (for $\displaystyle 0<\alpha_{1}\leq \frac{\pi}{12}$)
into the external FRW or Minkowski spacetime.

In this paper we predict the form of the gravitational wave tensor, which remains to be verified experimentally, associated with the DEUS object collapse.

\ack
I want to express all my gratitude to Stefan Sorescu whose computer drawing experience was truly valuable for this paper.

\noindent
Manuscript registered at the Astronomical Institute, Romanian Academy, record No. 253 from 04/19/07.


\title[D.E.U.S.]{VI: Electromagnetic and Gravitational Radiation from Black Holes}

\begin{abstract}
In the framework of the Dimension Embedded in Unified Symmetry (DEUS) model described by five-dimensional minimal energy hypersurfaces we observe that, when the considered black hole internal geometry, which is Riemannian, evolves to the asymptotic flat external spacetime (which can be seen as a local Euclidean spacetime), the energy contained in curvature is released as electromagnetic and gravitational radiation. The electromagnetic release in an external observer perception of a DEUS black hole is a jet-like emission from the black hole polar regions, with a spectrum consistent with the observed SED for blazars. Also, considering the self-similarity of the DEUS model, with the Friedmann-Robertson-Walker (FRW) metric of the Universe as another description of the a higher level DEUS object's helicoidal hypersurface, the extragalactic background light (EBL) spectrum must be consistent with our spectrum. Through these fittings we are able to determine the age of the Universe at present time.
\end{abstract}


\section{Introduction}

Having as starting point the trivial three-dimensional minimal surface representation of the catenoid and its conjugate surface, the helicoid, in paper {\it DEUS I} \cite{deus1} we constructed a five-dimensional spacetime geometry. There we determined the five-dimensional coordinate system that satisfies the Cauchy-Riemann equations, and also the ansatz, the properties and the correlation possible to be made for the DEUS object's composing manifolds.

In {\it DEUS II}  paper \cite{deus2} we generalized the geometry of the DEUS Black Hole for Universe formation and evolution, by using self-similar minimal surfaces for DEUS objects distribution. From the assumed self-similarity of the five-dimensional internal geometry, we shown that the DEUS object collapses to a string-like object equivalent with the BHIG spacetime of a bigger DEUS object in which the first one is embedded.

The considered five-dimensional geometry becomes conformal with an external Friedmann-Robertson-Walker (FRW) spacetime (for which the Hubble constant is {\it uniquely} defined as $72.33\pm 0.02$) by evolving the hyper-surfaces composing our five-dimensional DEUS object to a pre-collapse manifold. For the BHIG spacetime written in FRW coordinates we derived the Friedmann equations and some cosmological consequences ($\Omega_{tot}$ and the cosmological distance).

Into the background of a collapsed DEUS object rotating string that creates a BHIG manifold "deformed" in a FRW bubble, the pre-collapse movement of the five-dimensional catenoid hyper-surfaces is perceived by an after-collapse external observer as dark energy effect, while the matter still contained in the collapsed DEUS spacetime as dark matter.

The collapsing helicoidal manifold can be seen by the external timelike observer only considering that the catenoid is having $N$ internal DEUS similar objects, distorted, also collapsing. The helicoids of these $N$-DEUS objects are timelike (five-dimensional) and are giving three-dimensional spacelike and timelike observable effects to the external space.

In the above geometry framework, in {\it DEUS III} \cite{deus3} we found that the energy contained into the {\it catenoidal hyper-surface} and into the {\it helicoidal hyper-surface} can be characterized using the Yamabe energy. We shown that the energy stored into the curvature of the {\it catenoidal hyper-surface} is released (at the DEUS object evaporation) into the external space as kinetic energy in form of spatial or temporal perturbations. Also, by applying a combination of Ricci flow and mean curvature flow and taking account of the conformity of the {\it reduced catenoidal hyper-surface} \cite{deus2} with the flat external space, we obtained the scalar curvature of the {\it pure catenoidal hyper-surface}, for which the metric was too complicated to be determined in the "classical" way.

Using the mean curvature flow, we determined \cite{deus3} the velocity with which the {\it catenoidal hyper-surface} evolves to the FRW external space and the one of the BHIG's FRW bubble from its formation ("Big Bang") to its disappearance ("Big Crunch"). As we said in {\it DEUS II}, this movement is not possible to be directly observed by the external observer, which perceives only its inertial effect as gravitational effect (dark energy in helicoidal frame). In the Lorentz frame of the external observer these velocities are comparable with the infall velocity in the core collapse process of a star becoming a black hole.

Because of the fact that the self-similar DEUS objects are embedded into the {\it 
catenoidal hyper-surface} of a higher scale DEUS object, they will be highly distorted from the perfect symmetric case, and with them 
all the contained surfaces. This perturbation of symmetry makes them evolve to a minimal case in curvature and energy, dragging 
also the surface in which they are contained (which are components of a DEUS object also distorted by an embedding in a superior scale 
{\it catenoidal hyper-surface}). The distortion increases from the neck of the catenoid to its edge, where it gives a more intense 
effect to the external flat space. The given effects are different not only because of the distortion, but also because of the 
different {\it catenoidal hyper-surface}'s immersed DEUS object dimension, which decreases from the neck to the edge of the catenoid (the Yamabe energy describing the energy of the DEUS hyper-surfaces decreases in the same direction). In a non-perturbed symmetry we were supposed to have no effect to the external space. But the distortion makes the effects of same nature to not compensate each other, their resultant being perceived by the external observer while the DEUS object evolves toward collapse.

In the local Minkowski spacetime we obtain an extremely interesting behavior, the matter and the radiation fields being emitted from the evaporating black hole DEUS object at $\displaystyle \zeta=\pm\; \frac{\pi}{2}$ polar angle ({\bf black hole jets}).

\section{Spectral Energy Distribution}

Because the external observer sees the catenoids of the DEUS objects immersed into {\it catenoidal hypersurface} as timelike effects to the local Minkowski spacetime (part of a global FRW spacetime) while the {\it catenoidal hypersurface} itself as spacelike (as it creates the external flat space), the Yamabe energy \cite{deus3} (half of the total energy contained into the hypersurface) of the DEUS immersed {\it catenoidal hypersurfaces} will be:

\begin{equation}
E"_{Yamabe,\; catenoid}(a",\; r)=i\; E_{Yamabe,\; catenoid}(a,\; r)\; ,
\label{eqa_3}
\end{equation}
\noindent
the energy transmitted as a spatial (gravitational) perturbation being:

\begin{equation}
\!\!\!\!\!\!\!\!\!\!\!\!\!\!\!\!\!\!\!\!\!\!\!\!\!\!\!\!\!\!\!\!\!\begin{array}{cc}
i\; \Im(E_{Yamabe,\; catenoid}(a,\; r))=i^{2}\; \Im(E"_{Yamabe,\; catenoid}(a",\; r))\equiv -\; E_{GW}(a",\; r)) = \\
=-\displaystyle\frac{4}{r^{2}} - 3\;
\left[1+\frac{\textnormal{sin}^{2}\left(\displaystyle r/a"
    \right)}{\textnormal{cos}^{4}\left(\displaystyle r/a" \right)} \right]\; \displaystyle\frac{H_{0}^{2}-H_{0}\sqrt{H_{0}^{2}+1}+1}{\sqrt{H_{0}^{2}+1}-H_{0}}\; H_{0}^{3}\; \frac{2H_{0}^{2}-2H_{0}\sqrt{H_{0}^{2}+1}+1}{\sqrt{H_{0}^{2}+1}\; \left(H_{0}^{2}-H_{0}\sqrt{H_{0}^{2}+1}+1\right)^{2}} \; ,
\end{array}
\label{eqa_4}
\end{equation}
\noindent
while the electromagnetic perturbation:

\begin{equation}
\!\!\!\!\!\!\!\!\!\!\!\!\!\!\!\!\!\!\begin{array}{cc}
\Re(E_{Yamabe,\; catenoid}(a,\; r))=i\; \Re(E"_{Yamabe,\; catenoid}(a",\; r))\equiv  E_{EM}(a",\; r))= \\
=
\displaystyle\frac{3\sqrt{2}}{a"}\left(H_{0}^{2}-H_{0}\sqrt{H_{0}^{2}+1}+1\right)^{1/2}\left[1+\frac{\textnormal{sin}^{2}\left(\displaystyle
      r/a" \right)}{\textnormal{cos}^{4}\left(\displaystyle r/a" \right)}
\right]^{-1} \frac{\textnormal{sin}\left(\displaystyle r/a"
  \right)}{\textnormal{cos}^{5}\left(\displaystyle r/a" \right)}\;
\left[1-\textnormal{sin}^{2}\left(\displaystyle r/a" \right) \right] \; .
\end{array}
\label{eqa_5}
\end{equation}

At the DEUS object collapse we will have an overlapping between the ergosphere and the {\it catenoidal hypersurface} and, because the same behavior must be present at any scale as a characteristic of before collapse DEUS objects, we will have:

\begin{equation}
(A")^{2}\; \textnormal{cosh}^{-2}\left( \left. \sqrt{t^{2}+\phi^{2}}\right/ A"
\right)+\textnormal{arctg}^{2}\left( t_{k}/\phi_{k} \right)=0 \; ,
\label{eqa_6}
\end{equation}
\noindent
for the immersed DEUS objects which are collapsing before (into the DEUS
internal frame description) the higher scale DEUS object collapse in which they are immersed,
its {\it catenoidal hypersurface} will remain (the immersed DEUS objects are
already collapsed) without internal structure and will give the after-collapse external BHIG hypersurface. With $r"=\sqrt{t^{2}+\phi^{2}}$ and
$t"_{FRW}=\textnormal{arctg}^{2}\left(\displaystyle t_{k}/\phi_{k} \right)$ into the external observer's coordinates, (\ref{eqa_6}) will become:

\begin{equation}
(A")^{2}\; \textnormal{cosh}^{-2}\left( r_{int}/A" \right)+(t"_{FRW})^{2}=0 \; .
\label{eqa_7}
\end{equation}

At the collapse $t=\pm\; i\; \phi$  ($r_{int}=0$ and, previously achieved, $r"=0$ for the DEUS's BHIG internal spacetimes are different from the newly created $r\equiv r_{ext}$ of the external observer spacetime) and so, in (\ref{eqa_7}):

\begin{equation}
A"=\pm\; i\; t"_{FRW}\; ,
\label{eqa_8}
\end{equation}
\noindent
but, because we saw that $a=i\; A$, results that $a"=\pm\; t"_{FRW}$. Here $t"_{FRW}$ plays the role of half of the rotation period $T$ of the string-like object that generates the helicoids of the immersed  DEUS objects ($T\in [-1,\; 1]$) and varies between the creation and collapse moments from 0 to 1. We will be interested only in the effect given in the direction of "time" (after Big Bang) spacetime BHIG region, so only on $a"= t"_{FRW}$.

Because the ergosphere of the DEUS collapsing black hole/Universe is also satisfying:

\begin{equation}
A^{2}\; \textnormal{cosh}^{-2}\left( r_{ext}/A \right)+(t_{FRW})^{2}=0 \; ,
\label{eqa_10}
\end{equation}
\noindent
from where results:

\begin{equation}
t_{FRW}=a\; \textnormal{cos}^{-1}\left( r_{ext}/a \right) \; ,
\label{eqa_11}
\end{equation}
\noindent
where we used $a=i\; A$, we obtain:

\begin{equation}
r_{ext}=a\; \textnormal{arccos}\left( a/t_{FRW}\right) \; .
\label{eqa_12}
\end{equation}

But, knowing that the DEUS objects are the same at any scale (by definition),
meaning that $a=\pm\; t_{FRW}$, we get in (\ref{eqa_12}):

\begin{equation}
r\equiv r_{ext}=\pm\; a\; (\pi/2) \; ,
\label{eqa_14}
\end{equation}
\noindent
from where we will consider only the positive {\it r} of the "time" BHIG region.

Now, we know that the luminosity associated with the GW energy or EM energy from (\ref{eqa_4}) and (\ref{eqa_5}) is in the global FRW spacetime:

\begin{equation}
L\equiv dE/dt_{FRW} \; ,
\label{eqa_16}
\end{equation}
\noindent
and the flux $F\equiv L/(4\; \pi\; d^{2})$, where the distance at which the event is taking place is the one computed for the observer situated into the external BHIG spacetime (seen as a global FRW spacetime) of the collapsed DEUS object:

\begin{equation}
d=d_{BHIG}=\left[\frac{2H^{2}-2H\sqrt{H^{2}+1}}{4\; \left(H^{2}-H\sqrt{H^{2}+1}+1 \right)^{2}} \right]^{1/2}\; \textnormal{ln}\; r \; ,
\label{eqa_18}
\end{equation}
\noindent
which becomes, when we use the condition for an emission occurring into a local Minkowski spacetime:

\begin{equation}
r^{2}=\frac{2H^{2}-2H\sqrt{H^{2}+1}}{3\; \left(H^{2}-H\sqrt{H^{2}+1}+1 \right)^{2}} \; ,
\label{eqa_19}
\end{equation}

\begin{equation}
d=\sqrt{3/4}\; r\; \textnormal{ln}\; r \; ,
\label{eqa_20}
\end{equation}
\noindent
where {\it r} is the one from (\ref{eqa_14}).

With {\it r} from (\ref{eqa_14}) and $a"=t"_{FRW}$ in (\ref{eqa_4}) and (\ref{eqa_5}) we can simulate  the dependence of the luminosity (\ref{eqa_16}) and flux of the 
electromagnetic (Fig. \ref{sed1}) and gravitational emission as function of
energy.

Scaling one of the peak values of our spectrum with the GeV-TeV X-ray spectral
energy distribution (SED) data for the quasars (Giommi et al. catalog - see,
for example, PKS 048-097 for which we have enough 
data points to construct a complete SED), \cite{ghi}, it proves that all the
other data points in our spectra are fitting well the observational SED data points. Taking into account that we must have only one and not 
multiple spectra in the SED, the match was found to be at age of the Universe $t_{FRW}\simeq 0.524$, where $t_{FRW}\in (0,1)$. The obtained age of the Universe is in 
agreement with observations which say that the Universe "just" passed by its
half lifetime.

\begin{figure}
\begin{center}
\includegraphics[scale=0.11,angle=90]{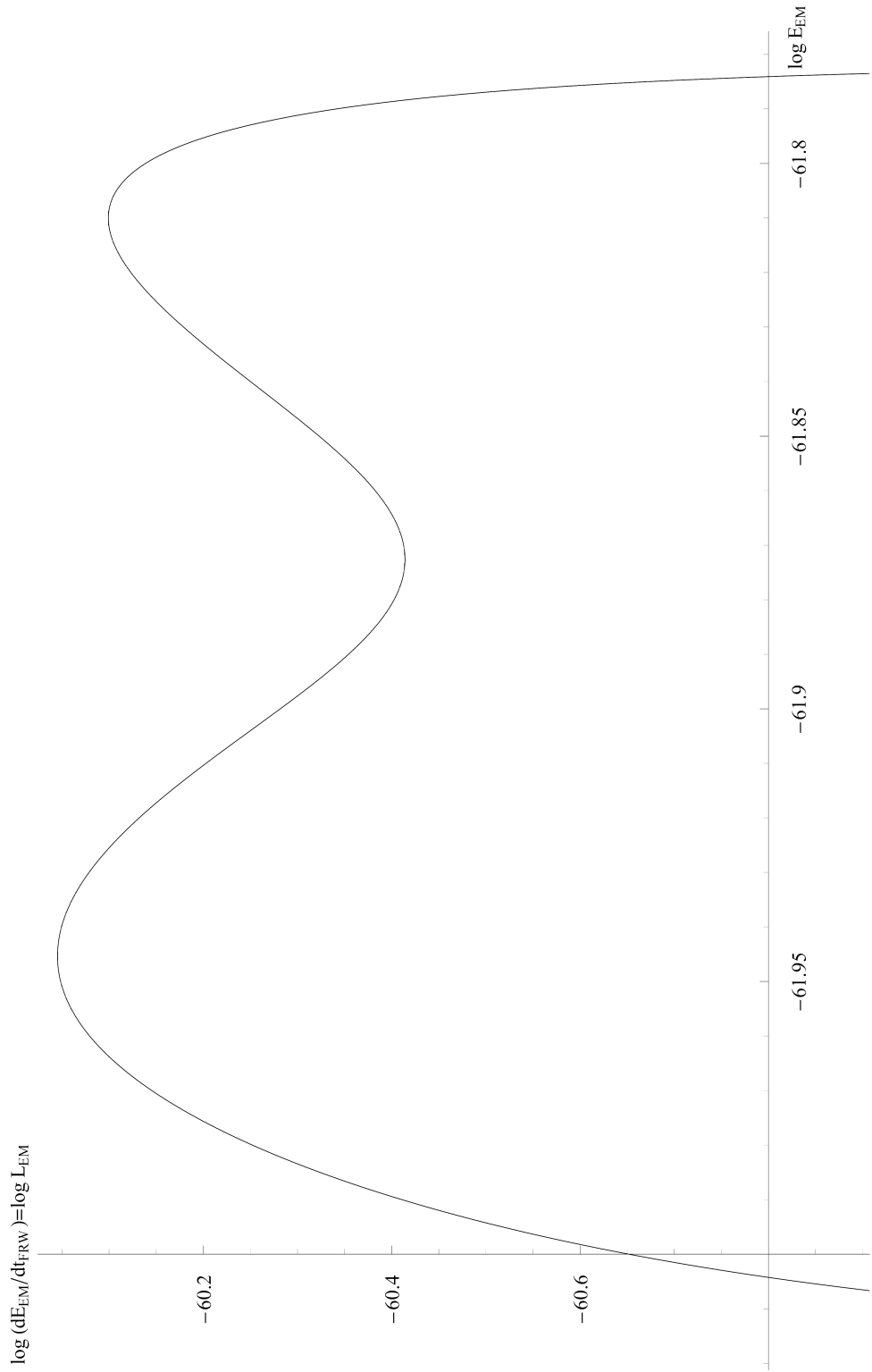}
\caption{Electromagnetic radiation luminosity spectrum emitted by a DEUS black hole and observed from the external Minkowski spacetime.}
\end{center}
\label{sed1}
\end{figure}

For the Universe seen as a FRW bubble representation of DEUS object's helicoid
the spectrum must fit the observed extragalactic background light (EBL) \cite{dejag,prim}. This happens for $t_{FRW}\simeq
0.48$ or for $t_{FRW}\simeq 0.524$, where the second value is, surprisingly,
the same as the one previously obtained for the SED of the quasars (emission from a supermassive black hole).

In Fig. \ref{sed2} we represented the power spectrum for the gravitational waves, where:
\begin{equation}
P_{GW}\propto d^{2}E_{GW}/dt_{FRW}^{2}\; .
\label{eqa_24}
\end{equation}

At $\nu_{EM}\simeq 5\times 10^{10}$ [Hz] and $\nu_{EM}F_{EM}\simeq 9\times 10^{-13}$ [ergs/s $\cdot$ cm$^{2}$] in electromagnetic emission will correspond 
$\nu_{GW}\simeq 1.3\times 10^{-29}$ [Hz] and $\nu_{GW}F_{GW}\simeq 3.7\times 10^{-56}$ [ergs/s $\cdot$ cm$^{2}$] in gravitational emission, while for 
$\nu_{EM}\simeq 5\times 10^{14}$ [Hz] and $\nu_{EM}F_{EM}\simeq 9\times 10^{-10}$ [ergs/s $\cdot$ cm$^{2}$] we predict to have an GW emission with 
$\nu_{GW}\simeq 3\times 10^{-32}$ [Hz] and $\nu_{GW}F_{GW}\simeq 1.4\times 10^{-58}$ [ergs/s $\cdot$ cm$^{2}$].

In the future of our Universe ($t_{FRW}\geq 0.524$) we should expect the emergence of a second, sharper and more luminous spectrum in the SED for blazars and also in the EBL.

\begin{figure}
\begin{center}
\includegraphics[scale=0.11,angle=90]{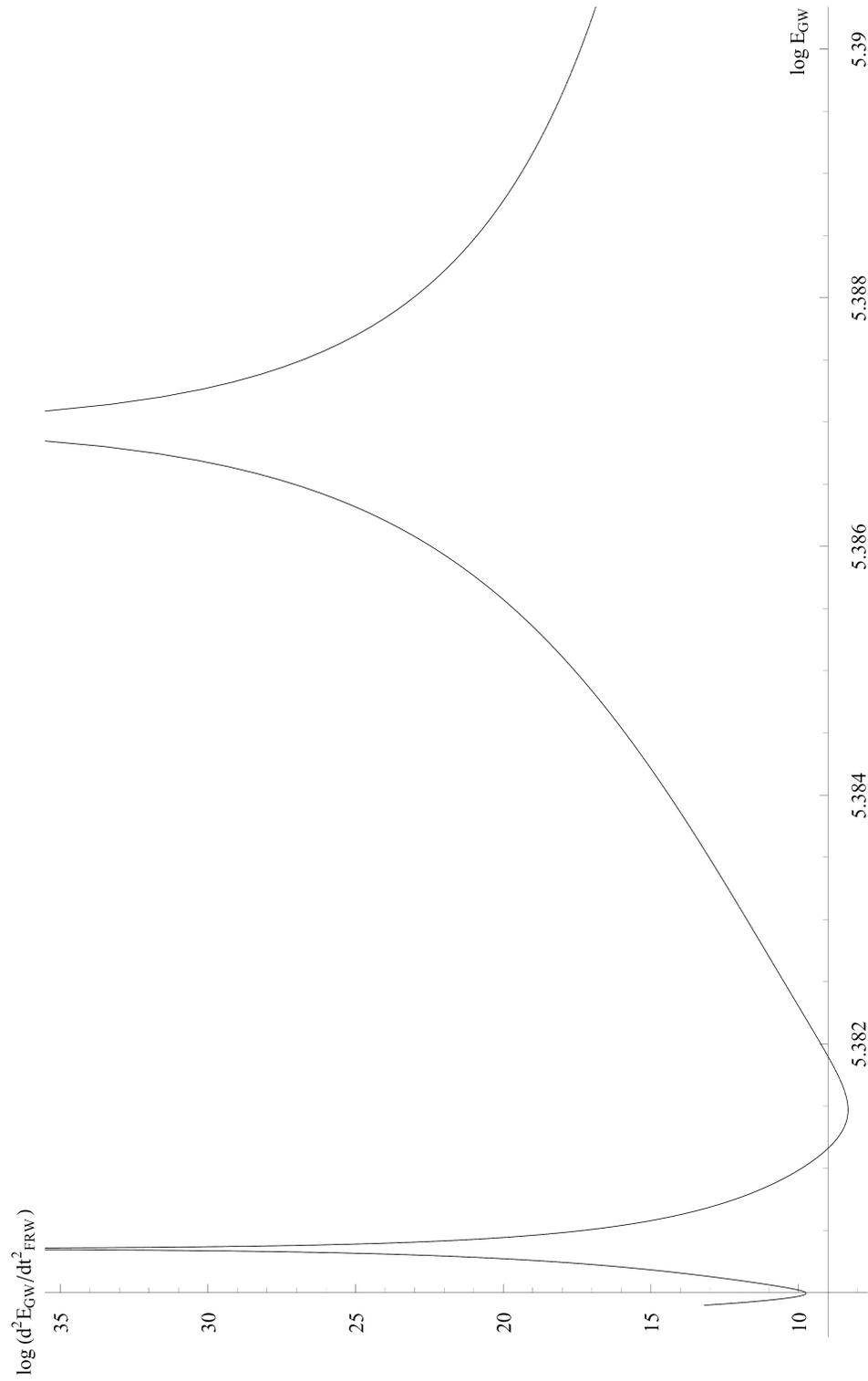}
\caption{Gravitational radiation power spectrum emitted by a DEUS black hole and observed from the external Minkowski spacetime.}
\end{center}
\label{sed2}
\end{figure}

\ack
Manuscript registered at the Astronomical Institute, Romanian Academy, record No. 254 from 04/19/07.


\title[D.E.U.S.]{VII: Global Energy Spectra}

\begin{abstract}
In paper {\it DEUS VI} \cite{deus6} we explained {\it why} the emissions from a DEUS object take place and {\it how} they are perceived by the external observer. In this paper we will give an answer to the {\it when} question: the moment of time from the external observer history at which this emission occurs.
\end{abstract}


\section{Introduction}
In paper {\it DEUS VI} \cite{deus6} we analyzed the emission spectrum from individual DEUS black holes, both electromagnetic and gravitational, without taking into account the moment in the external's observer proper time when the emission occurs, this being the main topic of the present paper.

In the framework of the Dimension Embedded in Unified Symmetry (DEUS) model described by five-dimensional minimal energy hyper-surfaces \cite{deus1} we observed that, when the considered DEUS internal geometry, which is Riemannian, evolves to the asymptotic flat external spacetime (which can be seen as a local Euclidean spacetime) \cite{deus2}, the energy contained in curvature is released as electromagnetic and gravitational radiation \cite{deus4, deus5}. The electromagnetic release in an external observer perception of a DEUS black hole is a jet-like emission from the black hole's polar regions \cite{deus4}, with a spectrum consistent with the observed SED for blazars \cite{deus6}. Also, considering the self-similarity of the DEUS model \cite{deus3}, with the Friedmann-Robertson-Walker (FRW) metric of the Universe as another description of the a higher level DEUS object's helicoidal hyper-surface \cite{deus2}, we see that the extragalactic background light (EBL) spectrum is consistent with our spectrum \cite{deus6}.

Through these fittings we were able to determine the "age" of the Universe at present time \cite{deus6}.

\section{Energy Spectrum}

For answering to {\it when} question we must consider that the emitting self-similar DEUS objects are still on the catenoid of the collapsing superior DEUS object (which is seen by the external observer as a black hole, Big Bang or atom). This means that the catenoid's DEUS objects (having as coordinates $r"$ and $t"_{FRW}$) are not rotated yet (in the catenoid frame) with $\displaystyle \frac{\pi}{2}$. For observable effects to the external spacetime, the collapsing DEUS object in whose catenoid the emitting DEUS objects are immersed must have $r'\rightarrow 0$ (as seen by the external observer). The temporal coordinate for it is $t'_{FRW}$. For the external observer situated, in his frame, at $r$ and $t_{FRW}$, the emission from the catenoid's DEUS objects will be, at $r'\rightarrow 0$ (or, equivalently, $t'_{FRW}\rightarrow \infty$, or on the horizon of the DEUS object): 

\begin{equation}
\begin{array}{ll}
E_{GW}(A, r) = \displaystyle\frac{4}{r^{2}} + 3\;
\left[1+\frac{\textnormal{sin}^{2}\left(\displaystyle r/A
    \right)}{\textnormal{cos}^{4}\left(\displaystyle r/A \right)} \right]\; H^{3}\; \frac{2H^{2}-2H\sqrt{H^{2}+1}+1}{\left(H^{2}-H\sqrt{H^{2}+1}+1\right)^{2}} \\
E_{EM}(A, r)= \displaystyle\frac{3\sqrt{2}}{A}\left[1+\frac{\textnormal{sin}^{2}\left(\displaystyle
      r/A \right)}{\textnormal{cos}^{4}\left(\displaystyle r/A \right)}
\right]^{-1} \frac{\textnormal{sin}\left(\displaystyle r/A
  \right)}{\textnormal{cos}^{3}\left(\displaystyle r/A \right)}\left(H^{2}-H\sqrt{H^{2}+1}+1\right)^{1/2}\; ,
\end{array}
\label{eq_0}
\end{equation}
\noindent
where {\it H} is the Hubble parameter, the temporal scale factor:

\begin{equation}
A=\textnormal{exp}\left[ H\;t_{FRW} \right]\; ,
\label{eq_1}
\end{equation}
\noindent
and the radial coordinate:

\begin{equation}
r=A\; \textnormal{arccosh}\left( A'/t'_{FRW} \right)\; .
\label{eq_2}
\end{equation}

But, because $A'=const.$, while $t'_{FRW}\rightarrow \infty$, results that $r=A\; \displaystyle \frac{\pi}{2}$.

Knowing that the luminosity $L=dE/dt_{FRW}$, the flux $F=L\left/ \left( 4\pi d_{FRW}^{2} \right) \right.$ and the intensity:

\begin{equation}
I=(h\nu)\; F=E\;F=\frac{1}{4\pi\; d_{FRW}^{2}}E\frac{dE}{dt_{FRW}}\; ,
\label{eq_3}
\end{equation}
\noindent
we can see {\it when} a DEUS object emits and at what energy and intensity it does. The distance at which the event is taking place is the one computed for the observer situated into the external Black Hole Internal Geometry (BHIG) spacetime \cite{deus1} (seen as a global FRW spacetime \cite{deus2}) of the collapsed DEUS object:

\begin{equation}
d_{FRW}=\sqrt{\frac{3}{4}}\; r\; \textnormal{ln}\;r\; .
\label{eq_4}
\end{equation}

\begin{figure}
\centering
\leavevmode
\includegraphics[width=14.5cm,height=9cm]{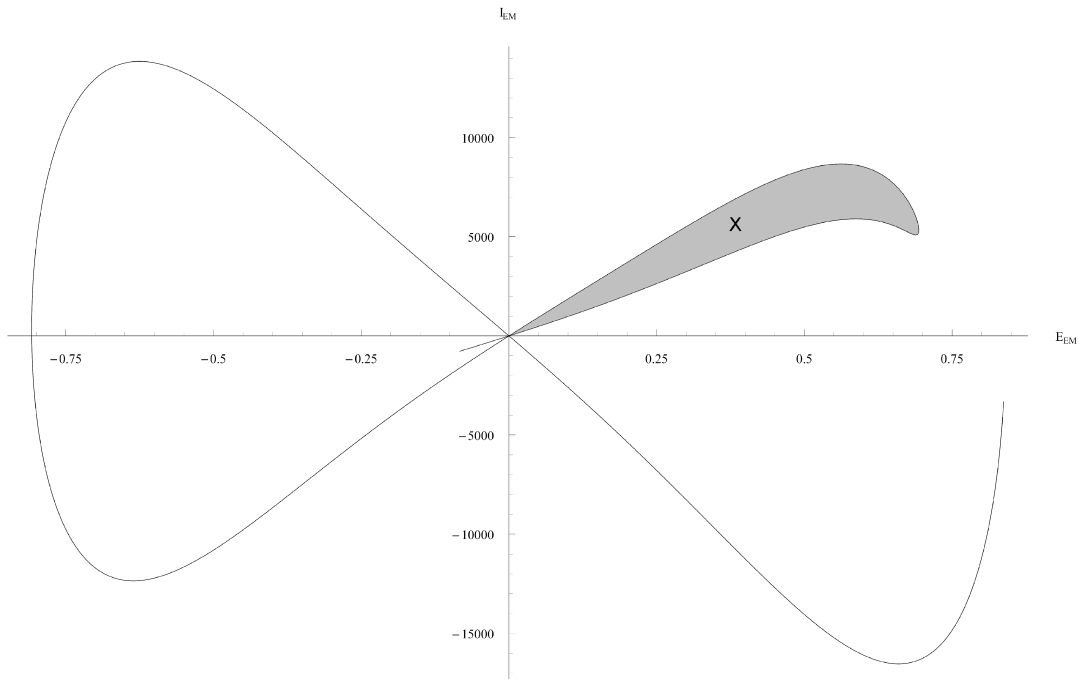}
\caption{Electromagnetic radiation power spectrum of a DEUS black hole as "observed" from the external FRW spacetime at the time $t_{FRW}\in [0.012, 0.0148]$.}
\label{sed3}
\end{figure}

\begin{figure}
\centering
\leavevmode
\includegraphics[width=14.5cm,height=9cm]{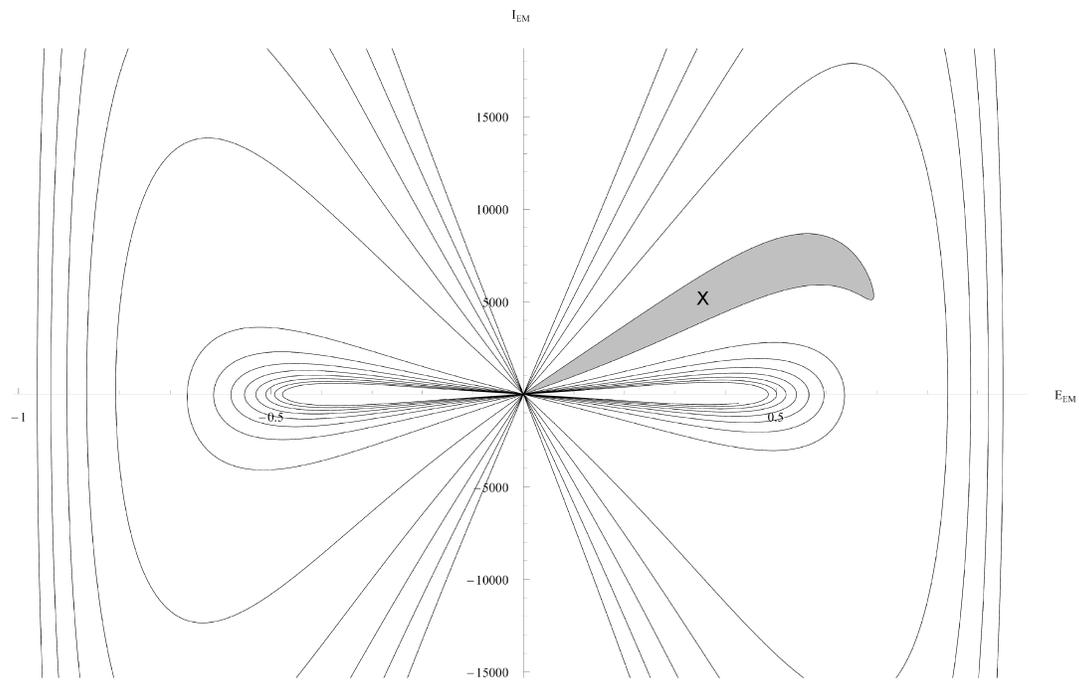}
\caption{Electromagnetic radiation power spectrum of a DEUS black hole as "observed" from the external FRW spacetime at the time $t_{FRW}\in [0.01, 0.02]$.}
\label{sed4}
\end{figure}

In the above background, in the figures Fig. \ref{sed3} ($t_{FRW}\in [0.012, 0.0148]$) and Fig. \ref{sed4} ($t_{FRW}\in [0.01, 0.02]$) we represented the emitted electromagnetic spectrum. We observe that the emission, departing from zero, reaches to a maximal value in {\it E} and {\it I}, after which the intensity decreases rapidly to zero ($E_{EM}\neq 0$) and becomes negative. Here ($E_{EM} > 0$, $I_{EM} < 0$), the DEUS object absorbs radiation from the external FRW spacetime, the catenoid evolving away from the cylinder. After that, the energy and the intensity are going again to zero, from where the emission starts again (but this time toward the BHIG region of the DEUS black hole object), $E_{EM} < 0$ and $I_{EM} > 0$, the catenoid of the emitting DEUS object evolving toward a cylindrical geometry. This increase in emission is followed by a rapid decrease to $I_{EM} = 0$ and a maximal (negative) energy. From here on we have an absorption ($E_{EM} < 0$, $I_{EM} < 0$) inside the catenoid which evolves again away from the collapse cylinder. This absorption is from the BHIG direction. All this cycle repeats itself with a period $T\simeq 0.0148 - 0.012 = 0.0028$ from the total $t_{FRW}$ Universe lifetime, between a Big Bang and a Big Crunch, where $t_{FRW}=1\equiv 100\%$. Results that one complete cycle as the one from figure Fig. \ref{sed3} takes place in 0.28$\%$ of the physical lifetime of our Universe.

At each new cycle we have a bigger energy and radiation intensity, the catenoid oscillating inside the DEUS object's ergosphere with bigger and bigger amplitude each time. This increase is due to the fact that to the catenoid can be given a wave interpretation and, inside the ergoregion, this wave is a soliton (satisfying the Korteweg-de Vries equation \cite{deus4}), which explains why the amplitude and the energy increases.

Still, because at the external boundary of the ergosphere we have the outer horizon of the DEUS object, all the particles and the radiation "freeze" on it, not having the chance to cross this boundary between the internal structure of the DEUS object and the observer situated in the external spacetime until the catenoid reaches finally to the cylinder and when the DEUS object evaporates. Then, in its proper $t_{FRW}$ time, the external observer perceives the periodic (0.28 $\%$ of $t_{FRW}$) emissions and absorptions. From his point of view is like piling radiation on the outer horizon of the black hole until a critical value is reached, the radiation bursting in the external spacetime (the {\bf X} region in Fig. \ref{sed3} and Fig. \ref{sed4}). This burst is specific for only {\bf one} level of self-similar DEUS objects embedded in the catenoid, each level of DEUS object having its proper timing for bursting in the external spacetime. After an evaporation from one such level, the higher level catenoid, in which its embedded DEUS object evaporated, "relax" to the ergosphere limit ($E = 0$ and $I = 0$). This limit is different from the previous one (the limit before the evaporation of the embedded DEUS object, or of one DEUS catenoid "particle") the catenoid being "shorter" with one "particle" and closer to cylinder. After this cycle finishes all the process repeats for the next catenoid particle, meaning a new set of periods which culminate with another {\bf X} region emission (see Fig. \ref{sed4}).

\begin{figure}
\centering
\leavevmode
\includegraphics[width=14.5cm,height=9cm]{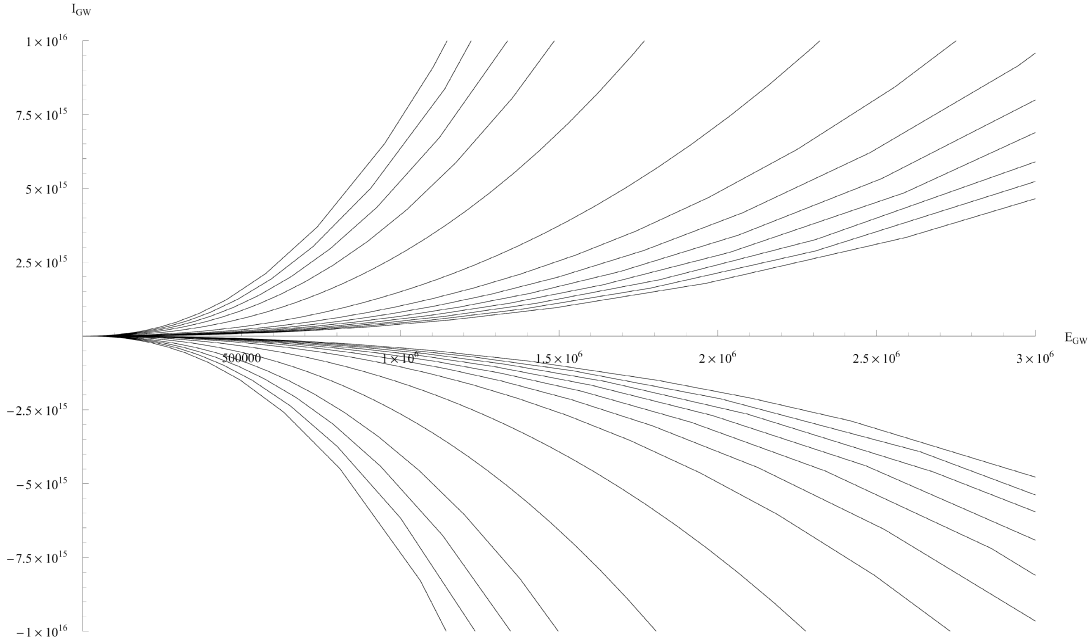}
\caption{Gravitational radiation power spectrum of a DEUS black hole as "observed" from the external FRW spacetime at the time $t_{FRW}\in [0.01, 0.02]$.}
\label{sed5}
\end{figure}

Simultaneously with the electromagnetic (EM) emission from the evaporating DEUS "particle" we have a gravitational wave (GW) emission (see Fig. \ref{sed5}). For the representation of the GW spectrum we used the same temporal range $t_{FRW}\in [0.01, 0.02]$ as for Fig. \ref{sed4} EM spectrum. We observe that here we have only emissions, one towards the external spacetime (the upper branches in Fig. \ref{sed5}) and, in the same time, one towards DEUS' BHIG. The intensity of both these GW emissions increases in $t_{FRW}$ time, covering a smaller energy range than their predecessors.

Because the external horizon plays no role in stopping the GW emission, in Fig. \ref{sed5} we have no {\bf X} region as the one we observed in the EM spectrum case. The gravitational waves do not "pile" on the horizon, being emitted in the external spacetime and towards the BHIG spacetime in a continuous manner and not as burst.

While, until now we placed our reference system on the catenoid ($A = \textnormal{exp}[H\; t_{FRW}]$) observing the emission of only one DEUS black hole object, we can set now our system of reference on the external FRW spacetime ($A = t_{FRW}$), analyzing the global emission spectrum from all the existing DEUS black holes, at a particular moment of the external observer's $t_{FRW}$ time.

We will keep the {\it A} and $t_{FRW}$ notations for the external observer's space and time and change the notations for the DEUS black hole's evaporating "particle" space and time (lifetime) to $A"$ and $t"_{FRW}$ ($t_{FRW}\in (0, 1)$ and $t"_{FRW}\in (0, 1)$). For the above energy and intensity formulas we obtain, as function of $t_{FRW}/t"_{FRW}$ ratio, the Fig. \ref{sed1} EM spectrum and the Fig. \ref{sed2} GW spectrum.

We see that, in the range $t_{FRW}/t"_{FRW}\in (5, 7.5)$, the EM spectrum (Fig. \ref{sed1}) presents, at specific energy values, intensity "spikes" which are related to the EM bursts from Fig. \ref{sed3} and Fig. \ref{sed4}, this time being possible to identify them with photon emissions from DEUS "atom" objects. These bursts begin when the lifetime of the DEUS "particle" is roughly 1/5 of the age of FRW Universe. We can identify the moment $t_{FRW}/t"_{FRW}\simeq 5$ with the re-ionization era of our Universe.

In Fig. \ref{sed2} we give the global GW spectrum prediction which can be checked in the future by the gravitational wave detectors (GEO 600, LIGO, LISA, etc.).

\begin{figure}
\centering
\leavevmode
\includegraphics[width=10cm,height=21.5cm]{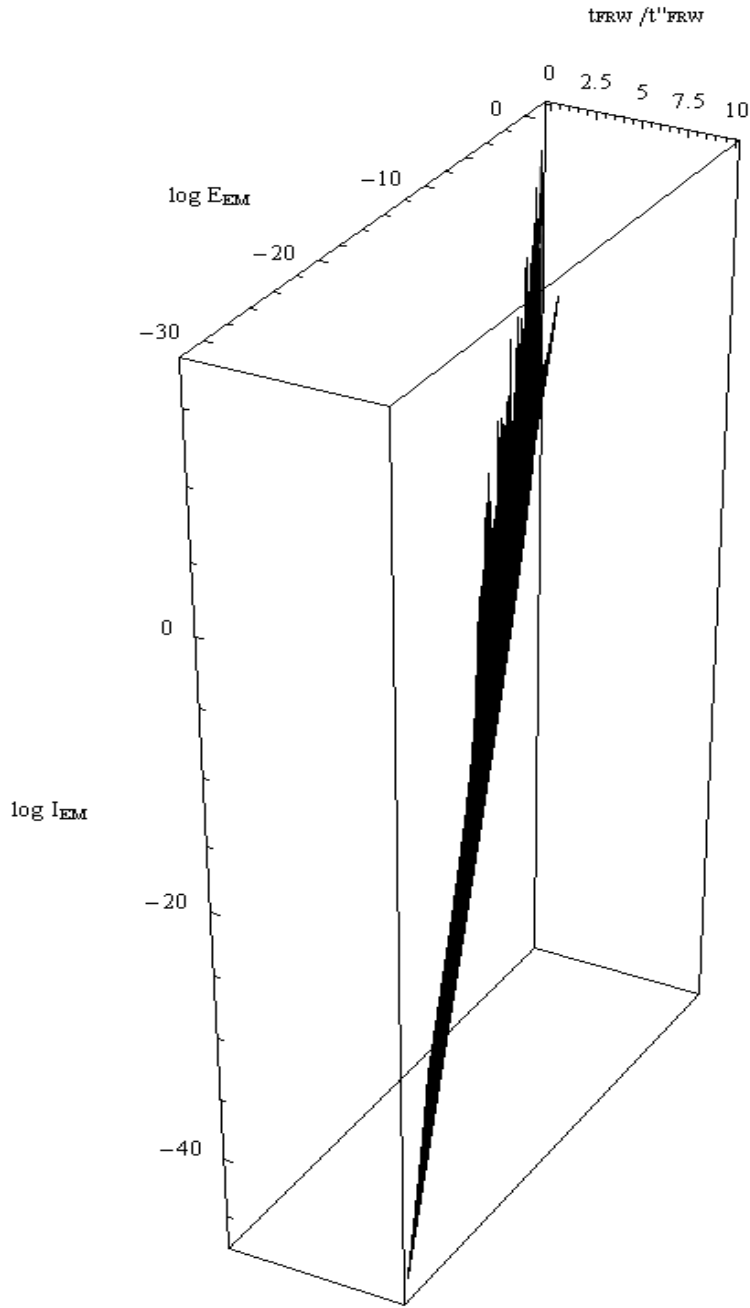}
\caption{Electromagnetic radiation global spectrum emitted by the DEUS black holes and observed from the external FRW spacetime.}
\label{sed1}
\end{figure}

\begin{figure}
\centering
\leavevmode
\includegraphics[width=7cm,height=21.5cm]{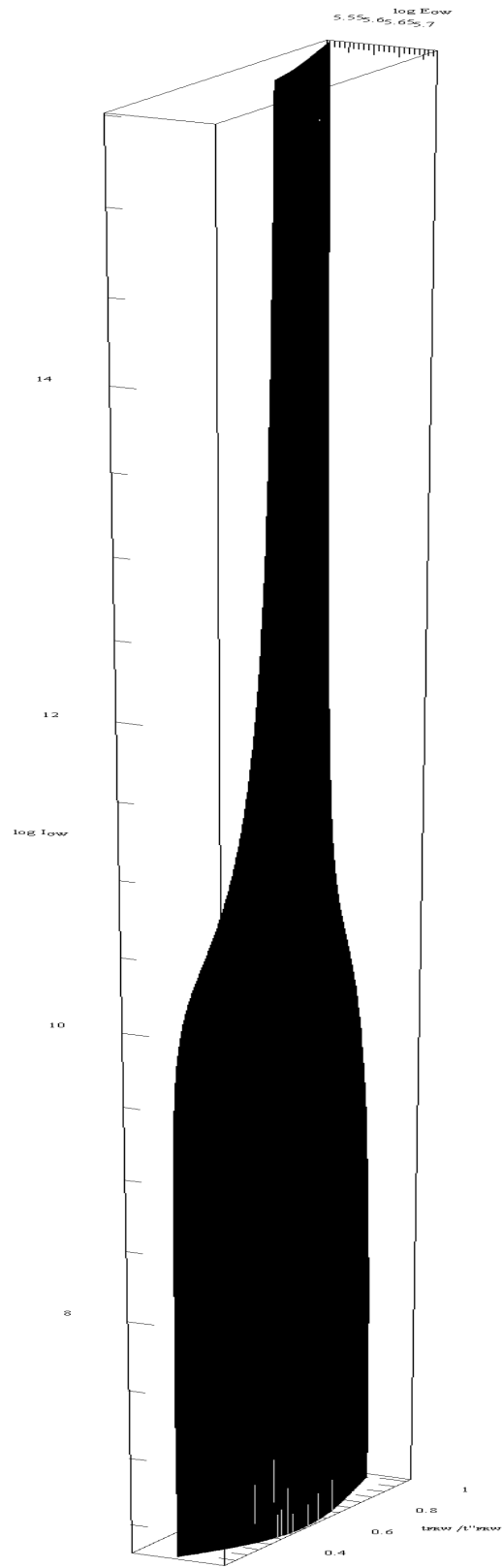}
\caption{Gravitational radiation global spectrum emitted by the DEUS black holes and observed from the external FRW spacetime.}
\label{sed2}
\end{figure}

\ack
Manuscript registered at the Astronomical Institute, Romanian Academy, record No. 255 from 04/19/07.


\title[D.E.U.S.]{VIII: The Mass of a DEUS Black Hole}

\begin{abstract}
With the help of classical Gauss laws for electricity and magnetism we will compute here the DEUS black hole's emitted electromagnetic radiation density.

Based on DEUS objects self-similarity hypothesis and on the "wave" interpretation of the catenoid hyper-surface \cite{deus4} and, also, knowing the radiation density and, as well, the magnetic vector potential and the electric scalar potential, we will be able derive the mass of a DEUS black hole and its growth rate as function of the distance from the observer.
\end{abstract}


\section{Introduction}

Using the Lorentz equation of motion we had shown in paper \cite{deus4} that the {\it catenoidal hyper-surface}'s embedded self-similar DEUS objects' 
{\it helicoidal hyper-surfaces} have non-zero electromagnetic tensor components. One of the two possible forms of this electromagnetic tensor will be perceived by the external observer as particles or fields originating from the present ("time") Big Bang - Big Crunch cycle of Universe evolution (observer in a global Friedmann-Robertson-Walker observable Universe bubble) or from black holes (observer in a local Minkowski spacetime). The electromagnetic radiation energy was computed in paper \cite{deus3} and the electromagnetic flux was compared with observations in papers \cite{deus6,deus7}.

\section{Magnetic Vector Potential and Electric Scalar Potential}

In paper \cite{deus4} we observed that for the {\it evolved ergosphere's helicoid} we have non-zero 
magnetic and electric field components which propagate in the external observer spacetime. The local coordinates 
of this hyper-surface are:

\begin{equation}
\left\{
\begin{array}{llll}
x^{1}=\textnormal{sinh}\; \theta\; \textnormal{sin}\left[ \textnormal{arctg}\left(\displaystyle \frac{t_{k}}{\phi_{k}}\right)\right]=-\textnormal{sinh}\left[ \textnormal{arctg} \left(\displaystyle \frac{\phi_{k}}{t_{k}}\right)\right]\; \textnormal{sin}\left[ \textnormal{arctg} \left(\displaystyle \frac{t_{k}}{\phi_{k}}\right)\right] \\
x^{2}=-\textnormal{sinh}\; \theta\; \textnormal{cos}\left[ \textnormal{arctg}\left(\displaystyle \frac{t_{k}}{\phi_{k}}\right)\right]=\textnormal{sinh}\left[ \textnormal{arctg}\; \left(\displaystyle \frac{\phi_{k}}{t_{k}}\right)\right]\; \textnormal{cos}\left[ \textnormal{arctg}\left(\displaystyle \frac{t_{k}}{\phi_{k}}\right)\right] \\
x^{3}=\textnormal{arctg}\left(\displaystyle \frac{t_{k}}{\phi_{k}}\right) \\
x^{5}\equiv i\; x^{0}=\pm\; i\; \textnormal{sinh}\; \theta=\mp\; i\; \textnormal{sinh}\left[ \textnormal{arctg} \left(\displaystyle \frac{\phi_{k}}{t_{k}}\right) \right]\; .
\end{array}
\right.
\label{eq_1}
\end{equation}

We know that the magnetic vector potential {\bf A} of electromagnetic theory generates the electromagnetic field tensor via the geometric equations:

\begin{equation}
F_{\beta\alpha}=\frac{\partial A_{\alpha}}{\partial x^{\beta}}-\frac{\partial A_{\beta}}{\partial x^{\alpha}} ,
\label{eq_2}
\end{equation}
\noindent
where $x^{\alpha}$ and $x^{\beta}$ are the equation (\ref{eq_1}) coordinates and the component $A_{5}\equiv A_{0}=f$ is the electric scalar potential.

Before the collapse of the DEUS object, at the moment at which the hyper-surfaces composing the object are evolved on the pre-collapse cylinder, from the equality of the Gaussian curvatures we saw that $t_{k}=\pm \phi_{k}$ \cite{deus2}. Then, using the results obtained in \cite{deus4} for the electromagnetic emission of the {\it evolved ergosphere's helicoid} (of the DEUS objects contained in the ergosphere's catenoid of the collapsing DEUS object) we can write (\ref{eq_2}) as:

\begin{equation}
\left\{
\begin{array}{llllll}
\displaystyle \frac{\partial A_{1}}{\partial x^{5}}-\frac{\partial f}{\partial x^{1}}=\frac{m_{helicoid}}{q}\; \frac{1}{\phi_{k}^{2}}\left( \pm\; \frac{1}{4}-\frac{1}{2} \right) \\
\displaystyle \frac{\partial A_{2}}{\partial x^{5}}-\frac{\partial f}{\partial x^{2}}=0 \Leftrightarrow \frac{\partial A_{2}}{\partial x^{5}}=\frac{\partial f}{\partial x^{2}} \\
\displaystyle \frac{\partial f}{\partial x^{3}}-\frac{\partial A_{3}}{\partial x^{5}}=0 \Leftrightarrow \frac{\partial f}{\partial x^{3}}=\frac{\partial A_{3}}{\partial x^{5}} \\
\displaystyle \frac{\partial A_{3}}{\partial x^{2}}-\frac{\partial A_{2}}{\partial x^{3}}=\frac{m_{helicoid}}{q}\; \frac{1}{\phi_{k}^{2}}\left( \frac{1}{\sqrt{2}}+\frac{1}{2} \right) \\
\displaystyle -\frac{\partial A_{2}}{\partial x^{1}}+\frac{\partial A_{1}}{\partial x^{2}}=0 \Leftrightarrow \frac{\partial A_{1}}{\partial x^{2}}=\frac{\partial A_{2}}{\partial x^{1}} \\
\displaystyle -\frac{\partial A_{3}}{\partial x^{1}}+\frac{\partial A_{1}}{\partial x^{3}}=0 \Leftrightarrow \frac{\partial A_{1}}{\partial x^{3}}=\frac{\partial A_{3}}{\partial x^{1}}\; ,
\end{array}
\right.
\label{eq_3}
\end{equation}
\noindent
$m_{helicoid}$ being the mass and {\it q} the charge contained into the helicoid.

After solving this system of equations we obtain:

\begin{equation}
\frac{\partial A_{1}}{\partial t_{k}}=\frac{\partial A_{2}}{\partial t_{k}}=\frac{\partial A_{3}}{\partial t_{k}}=0\; ,
\label{eq_4}
\end{equation}
\noindent
or, taking the pre-collapse condition $t_{k}=\pm \phi_{k}$:

\begin{equation}
\frac{\partial A_{1}}{\partial \phi_{k}}=\frac{\partial A_{2}}{\partial \phi_{k}}=\frac{\partial A_{3}}{\partial \phi_{k}}=0\; ,
\label{eq_5}
\end{equation}
\noindent
which, together with:

\begin{equation}
\frac{\partial f}{\partial t_{k}}=\frac{\partial f}{\partial \phi_{k}}=0
\label{eq_6}
\end{equation}
\noindent
gives the {\bf gauge transformation} for the DEUS object:

\begin{equation}
\partial_{t_{k}} A_{\mu}=\partial_{\phi_{k}} A_{\mu}=0\; .
\label{eq_7}
\end{equation}

From the system (\ref{eq_3}) we obtain also the vector potential components and the scalar potential:

\begin{equation}
\begin{array}{llll}
f=\displaystyle const.\; \frac{\sqrt{2}}{16}\left( \frac{1}{2}\mp \frac{1}{4}\right)\frac{m_{helicoid}}{q}\left( \textnormal{cosh}\frac{\pi}{4}-\textnormal{sinh}\frac{\pi}{4}\right) \\
A_{1}=-i\; \displaystyle \left[const.\; \frac{1}{8}\left( \frac{1}{2}\mp \frac{1}{4}\right)\frac{m_{helicoid}}{q}\; \textnormal{cosh}\frac{\pi}{4} \right] \\
A_{2}=i\; \displaystyle \left[const.\; \frac{1}{8}\left( \frac{1}{4}\mp \frac{1}{2}\right)\frac{m_{helicoid}}{q}\; \textnormal{cosh}\frac{\pi}{4}\; \frac{\displaystyle  \textnormal{cosh}\frac{\pi}{4}-\textnormal{sinh}\frac{\pi}{4}}{\displaystyle \textnormal{cosh}\frac{\pi}{4}+\textnormal{sinh}\frac{\pi}{4}} \right] \\
A_{3}=\displaystyle i\; \left[ const.\; \frac{\sqrt{2}}{16}\left( -\frac{1}{4}\pm \frac{1}{2}\right)\frac{m_{helicoid}}{q}\; \textnormal{cosh}\frac{\pi}{4} \left( \textnormal{cosh}\frac{\pi}{4}-\textnormal{sinh}\frac{\pi}{4}\right)\right]\; ,
\end{array}
\label{eq_8}
\end{equation}
\noindent
for $t_{k}=\pm \phi_{k}$.

In order that the new helicoidal hyper-surface (generated by the rotation of the remnant string-like object of the collapsed DEUS object, source of electromagnetic emission) and all the other hyper-surfaces composing the external DEUS object to mantain their defined local chart of coordinates meaning, the system of coordinates for the observer situated outside the collapsed DEUS object is rotated. Here "external" and "internal" are used just to point out the causality of the processes. In reality, for the external spacetime the internal spacetime does not exist, being hidden behind the singularity that generates the external time, space and the effects from the "internal" topological evolution from a curved to a flat spacetime \cite{deus1, deus2, deus3}. In this way, when we rotate the chart to the external observer's spacetime proper and unique system of DEUS coordinates, the collapsed DEUS object emitted electric and magnetic field components will "flip" to the new coordinates, the ones of the external observer (which we will denote with prime). Then we will have $E_{1}\rightarrow E_{3}'$, $E_{2}\rightarrow E_{2}'$ and $E_{3}\rightarrow E_{1}'$, where the only non-zero electric component is $E_{3}$:

\begin{equation}
E_{3}=\frac{m_{helicoid}}{q}\; \frac{1}{\phi_{k}^{2}}\left(\pm\; \frac{1}{4}-\frac{1}{2}\right)\; .
\label{eq_8a}
\end{equation}

In the internal coordinates $E_{3} < 0$, while in the external's space coordinates it is oriented toward the external observer (on the "new" $x^{1}$).

In the external observer's perception of the collapsing DEUS object's catenoid containing the self-similar DEUS objects whose helicoids give the timelike effects, $B_{3}$ "flips", being trapped (from the external observer point of view) in the "collapse" of the DEUS objects ($B_{1}=B_{2}=B_{3}=0$). It exists only in the collapsing catenoid frame. The magnetic fields in the external observer frame are induced by the "escaping" electric current.

\section{Maxwell Equations in Classical Electrodynamics}

We will work in curvilinear coordinates using the classical notation $h_{i}\equiv \sqrt{(g_{ii})_{helicoid}}$, where $(g_{ii})_{helicoid}$ are the metric tensor components of the helicoid \cite{deus2}. This means that:

\begin{equation}
\begin{array}{lll}
h_{1}=\displaystyle \frac{\displaystyle \left(2H^{2}-2H\sqrt{H^{2}+1}+1 \right)^{1/2}}{\displaystyle 2\left(H^{2}-H\sqrt{H^{2}+1}+1 \right)}\; \frac{1}{r} \\
h_{2}=\frac{\displaystyle \left(2H^{2}-2H\sqrt{H^{2}+1}+1 \right)^{1/2}}{\displaystyle 2\left(H^{2}-H\sqrt{H^{2}+1}+1 \right)} \\
h_{3}=\frac{\displaystyle \left(2H^{2}-2H\sqrt{H^{2}+1}+1 \right)^{1/2}}{\displaystyle 2\left(H^{2}-H\sqrt{H^{2}+1}+1 \right)}\; \textnormal{sin}\; \zeta \; ,
\end{array}
\label{eq_9}
\end{equation}
\noindent
where {\it r}, $\zeta$ and $\varphi$ are the Friedmann-Robertson-Walker (FRW) bubble coordinates and {\it H} is the Hubble parameter.
We will write the Gauss law for electricity:

\begin{equation}
\nabla_{\cdot} \textnormal{{\bf E}}=4\pi \rho_{catenoid\; ergosphere}
\label{eq_10}
\end{equation}
\noindent
in curvilinear coordinates:

\begin{equation}
\nabla_{\cdot} \textnormal{{\bf E'}}=\frac{1}{h_{1}h_{2}h_{3}}\left[\frac{\partial}{\partial r} (h_{2}h_{3}E_{3}')+\frac{\partial}{\partial \varphi} (h_{1}h_{2}E_{1}')+\frac{\partial}{\partial \zeta} (h_{1}h_{3}E_{2}') \right]\; ,
\label{eq_11}
\end{equation}
\noindent
where, as we saw in \cite{deus3}, in the external observer's DEUS object $E_{1}'=E_{2}'=0$ and:

\begin{equation}
E_{3}'=\frac{m_{helicoid}}{q}\frac{a^{2}}{r^{2}}\left(\pm \frac{1}{2}-1 \right) \; .
\label{eq_12}
\end{equation}

From (\ref{eq_11}) and (\ref{eq_12}) results that:

\begin{equation}
\nabla_{\cdot} \textnormal{{\bf E'}}=\frac{\displaystyle 2\left(H^{2}-H\sqrt{H^{2}+1}+1 \right)}{\displaystyle \left(2H^{2}-2H\sqrt{H^{2}+1}+1 \right)^{1/2}}\; (2\mp 1)\; \frac{m_{helicoid}}{q}\; \frac{a^{2}}{r^{2}}\; .
\label{eq_13}
\end{equation}

With (\ref{eq_13}) in (\ref{eq_10}) we obtain the radiation density as perceived by the external observer situated in the FRW bubble ($\rho_{catenoid\; ergosphere}\rightarrow \rho_{FRW}$):

\begin{equation}
\rho_{FRW}=\frac{1}{2\pi}\; (2\mp 1)\; \frac{\displaystyle H^{2}-H\sqrt{H^{2}+1}+1}{\displaystyle \left(2H^{2}-2H\sqrt{H^{2}+1}+1 \right)^{1/2}}\; \frac{m_{helicoid}}{q}\; \frac{a^{2}}{r^{2}}\; ,
\label{eq_14}
\end{equation}
\noindent
which, because the spacetime is locally a Minkowski spacetime \cite{deus2} with:

\begin{equation}
\frac{1}{r^{2}}=\frac{\displaystyle 3\left(H^{2}-H\sqrt{H^{2}+1}+1 \right)^{2}}{2H^{2}-2H\sqrt{H^{2}+1}+1}\; ,
\label{eq_15}
\end{equation}
\noindent
makes it equal to ($\rho_{FRW}\rightarrow \rho_{Minkowski})$:

\begin{equation}
\rho_{Minkowski}=\frac{1}{2\sqrt{3}\; \pi}\; (2\mp 1)\; \frac{m_{helicoid}}{q}\; \frac{a^{2}}{r^{3}}\; ,
\label{eq_16}
\end{equation}

For the Gauss law of magnetism we will have:

\begin{equation}
\nabla_{\cdot} \textnormal{{\bf B'}}=\frac{1}{h_{1}h_{2}h_{3}}\left[\frac{\partial}{\partial \varphi} (h_{1}h_{2}B_{3}')+\frac{\partial}{\partial \zeta} (h_{1}h_{3}B_{2}')+\frac{\partial}{\partial r} (h_{2}h_{3}B_{1}') \right]=\frac{1}{h_{3}}\; \frac{\partial B_{3}'}{\partial \varphi}\; ,
\label{eq_17}
\end{equation}
\noindent
where $B_{1}'=B_{2}'=0$ and, for $t_{k}=\pm \phi_{k}$:

\begin{equation}
B_{3}'=\left(\sqrt{2}+1\right)\; \frac{m_{helicoid}}{q}\; \frac{a^{2}}{r^{2}}\; ,
\label{eq_18}
\end{equation}
\noindent
resulting that $\displaystyle \frac{\partial B'_{3}}{\partial \varphi}=0$. In consequence, $\nabla_{\cdot} \textnormal{{\bf B'}}=0$ is satisfied.

Now, for the Faraday's law of induction we have, in curvilinear coordinates:

\begin{equation}
\!\!\!\!\!\!\!\!\!\!\!\!\!\!\!\!\!\!\!\!\!\begin{array}{ll}
\displaystyle \nabla \times \textnormal{{\bf E'}}=\frac{1}{h_{2}h_{3}}\left[ \frac{\partial}{\partial \zeta}(h_{3}E_{3}')-\frac{\partial}{\partial \varphi}(h_{2}E_{2}') \right]\; x^{1}\textnormal{{\bf e}}_{1} + \frac{1}{h_{1}h_{3}}\left[ \frac{\partial}{\partial \varphi}(h_{1}E_{1}')-\frac{\partial}{\partial r}(h_{3}E_{3}') \right]\; x^{2}\textnormal{{\bf e}}_{2} + \\
\displaystyle + \frac{1}{h_{1}h_{2}}\left[ \frac{\partial}{\partial r}(h_{2}E_{2}')-\frac{\partial}{\partial \zeta}(h_{1}E_{1}') \right]\; x^{3}\textnormal{{\bf e}}_{3}=-\frac{1}{h_{1}}\frac{\partial E_{3}'}{\partial r}\; x^{2} \textnormal{{\bf e}}_{2}\; ,
\end{array}
\label{eq_19}
\end{equation}
\noindent
where we used the "flip" of the electrical field components from the collapsed DEUS object to the external observer's BHIG-FRW spacetime and we made a basis transformation from $\{ \textnormal{{\bf e}}_{1}, \textnormal{{\bf e}}_{2}, \textnormal{{\bf e}}_{3}\}$ to the orthonormal basis $\{ \textnormal{{\bf e}}_{r}, \textnormal{{\bf e}}_{\zeta}, \textnormal{{\bf e}}_{\varphi}\}$.

When we compute:

\begin{equation}
\frac{\partial \textnormal{{\bf B'}}}{\partial x^{0}}=\frac{\partial B_{3}'}{\partial x^{0}}x^{3}\textnormal{{\bf e}}_{3}=-\frac{16}{\displaystyle \textnormal{cosh}\frac{\pi}{4}}\; \left( \frac{1}{\sqrt{2}}+\frac{1}{2}\right)\; \frac{m_{helicoid}}{q}\; \frac{a^{2}}{r^{2}}\; ,
\label{eq_20}
\end{equation}
\noindent
at $t_{k}=\pm \phi_{k}$ we observe that the Faraday's law is satisfied only if all the field components are 0. But, as we saw in \cite{deus4}, $B_{3}'\neq 0$ and $E_{3}'\neq 0$, meaning that the Faraday's law is not satisfied. We also observed that Ampere's law is not satisfied because of the same reasons. The reason of this inconsistency in the DEUS model is the self-similarity at any level of the emitting DEUS objects, from the "classical" ones as the DEUS object whose collapse (Big Bang) generates the FRW Universe bubble or the DEUS black holes, to quantum DEUS objects (primordial black holes, particles and atoms), having emissions that can not be explained through classical Maxwell equations. We will need a quantum treatment of our objects and of their emissions and, also, a local description in the Minkowski spacetime. The complete set of classical Maxwell equations will be satisfied for solar and stellar atmospheres, as it will be seen in a later paper.

\section{Maxwell's Equations Considering Quantum Field Theory}

In quantum field theory the extended Maxwell's equations (or the so-called Proca's equations) are: 

\begin{equation}
\begin{array}{llll}
rot\; \textnormal{{\bf H'}}=\textnormal{{\bf J}}+\displaystyle \frac{d \textnormal{{\bf D}}}{dx^{0}}-\frac{\kappa^{2}}{\mu_{0}} \textnormal{{\bf A}} \\
rot\; \textnormal{{\bf E'}}=-\displaystyle \frac{d \textnormal{{\bf B}}}{dx^{0}} \\
div\; \textnormal{{\bf D'}}=\rho_{Minkowski}-\epsilon_{0} \kappa^{2}f \\
div\; \textnormal{{\bf B'}}=0
\end{array}
\label{eq_21}
\end{equation}

The difference of the above quantum field theory Maxwell's equations from the classsical ones consists in additional terms containing the magnetic vector potential {\bf A} and the electric scalar potential {\it f}. In these terms appears also the vacuum permeability $\mu_{0}$ and the vacuum permittivity $\epsilon_{0}$.

$\kappa$ is the Einstein's energy and is equal with the Compton frequency divided by the velocity of light {\it c}, or:

\begin{equation}
\kappa=\frac{m_{0}c}{\hbar}\; ,
\label{eq_22}
\end{equation}
\noindent
with $m_{0}$ the rest mass and $\hbar$ the Planck's constant.

Extending the homogeneous Klein-Gordon wave equation ({\it s} = 0) and applying a probability function $\Psi$ we get the non homogeneous Proca wave equations ($s\neq 0$):

\begin{equation}
\Delta \Psi - \frac{1}{c^{2}}\frac{\partial^{2} \Psi}{\left( \partial x^{0}\right)^{2}}-\left( \frac{m_{0}c}{\hbar}\right)^{2}\Psi=s\; .
\label{eq_23}
\end{equation}

Introducing the magnetic vector potential {\bf A} ($s=s(\textnormal{{\bf A}})$) and the electric scalar potential {\it f} ($s=s(f)$) we can derive the following wave equations:

\begin{equation}
\Delta \textnormal{{\bf A}} - \frac{1}{c^{2}}\frac{\partial^{2} \textnormal{{\bf A}}}{\left( \partial x^{0}\right)^{2}}-\kappa^{2}\textnormal{{\bf A}}=-\mu_{0}\textnormal{{\bf J}}\; ,
\label{eq_24}
\end{equation}
\noindent
respectively:

\begin{equation}
\Delta f - \frac{1}{c^{2}}\frac{\partial^{2} f}{\left( \partial x^{0}\right)^{2}}-\kappa^{2}f=-\frac{1}{\epsilon_{0}}\rho_{Minkowski}\; .
\label{eq_25}
\end{equation}

In natural units $\hbar=c=1$ making $\kappa^{2}=m_{0}^{2}$ (see (\ref{eq_22})).

In the Riemannian spacetime we have to substitute the Laplace operator from (\ref{eq_23}) and (\ref{eq_24}) with the Laplace-Beltrami operator ($\Delta\rightarrow \Delta_{LB}$), which, applied on the scalar potential {\it f} (or, in the same way, on the vector potential {\bf A}):

\begin{equation}
\Delta_{LB}f = g^{-1/2} \sum_{i,j} \partial_{x^{i}}\left[g\; g^{ij}\; \partial_{x^{j}}\left( g^{-1/2}f\right) \right]\; ,
\label{eq_26}
\end{equation}
\noindent
where $\displaystyle g\equiv \left| det \left\{ g^{ij}\right\}\right|^{-1/2}$ and:

\begin{equation}
\begin{array}{llll}
g_{11}=(g_{11})_{helicoid}=\displaystyle \frac{\displaystyle 2H^{2}-2H\sqrt{H^{2}+1}+1}{\displaystyle 4\left(H^{2}-H\sqrt{H^{2}+1}+1 \right)^{2}}\; \frac{1}{r^{2}} \\
g_{22}=(g_{22})_{helicoid}=\displaystyle \frac{\displaystyle 2H^{2}-2H\sqrt{H^{2}+1}+1}{\displaystyle 4\left(H^{2}-H\sqrt{H^{2}+1}+1 \right)^{2}} \\
g_{33}=(g_{33})_{helicoid}=\displaystyle \frac{\displaystyle 2H^{2}-2H\sqrt{H^{2}+1}+1}{\displaystyle 4\left(H^{2}-H\sqrt{H^{2}+1}+1 \right)^{2}}\; \textnormal{sin}^{2}\zeta \\
g_{44}=(g_{44})_{helicoid}=-\displaystyle \frac{3}{4}
\end{array}
\label{eq_27}
\end{equation}

With (\ref{eq_15}) in (\ref{eq_27}), after the index raising operation, we get:

\begin{equation}
\begin{array}{llll}
g^{11}=(g^{11})_{Minkowski}=\displaystyle \frac{4}{3} \\
g^{22}=(g^{22})_{Minkowski}=\displaystyle \frac{4}{3}\; \frac{1}{r^{2}} \\
g^{33}=(g^{33})_{Minkowski}=\displaystyle \frac{4}{3}\; \frac{1}{r^{2}\; \textnormal{sin}^{2}\zeta} \\
g^{44}=(g^{44})_{Minkowski}=-\displaystyle \frac{4}{3}
\end{array}
\label{eq_28}
\end{equation}
\noindent
making possible the computation of {\it g}:

\begin{equation}
g=\frac{9}{16}\; r^{2}\; \textnormal{sin}\; \zeta
\label{eq_29}
\end{equation}
\noindent
and, finally, of $\Delta_{LB}$ applied on {\it f} and {\bf A}.

In the Proca's wave equation we will have $\Delta_{LB}f=0$, $\displaystyle \frac{\partial f}{\partial x^{0}}=0$ and, if we bear in mind that in natural units the units of charge are defined by choosing $\epsilon_{0}=1$:

\begin{equation}
m_{0}=\left[\frac{\rho_{Minkowski}}{f}\right]^{1/2}=\left[\displaystyle \frac{1}{const.}\; 16\sqrt{\frac{2}{3}}\; \frac{1}{\pi}\; \frac{1}{\displaystyle \textnormal{cosh}\frac{\pi}{4}-\textnormal{sinh}\frac{\pi}{4}} \right]^{1/2}\; \frac{a}{r^{3/2}}\; .
\label{eq_30}
\end{equation}

The result (\ref{eq_30}) means that in a Minkowski Universe the mass of the observed DEUS black holes decreases with the spatial distance from the observer. In other words, the mass of the black hole increases toward the present observer's epoch ($a=1$). The derivative of $m_{0}$ to FRW bubble (and Minkowski) proper time:

\begin{equation}
\frac{dm_{0}}{dt_{FRW}}=\left[\displaystyle \frac{1}{const.}\; 16\sqrt{\frac{2}{3}}\; \frac{1}{\pi}\; \frac{1}{\displaystyle \textnormal{cosh}\frac{\pi}{4}-\textnormal{sinh}\frac{\pi}{4}} \right]^{1/2}\; a\; \left[ H_{0}r^{-3/2}-\frac{3}{2}\; r^{-5/2}\; v_{r} \right]\; ,
\label{eq_31}
\end{equation}
\noindent
where the Hubble parameter $H=\dot{a}/a$ for the present epoch is constant (the Hubble constant), $a=1$ and $v_{r}=dr/dt_{FRW}$ is the relative velocity of the black hole.

\section{Conclusions}

In this paper we derived the emitted electromagnetic radiation density, the magnetic vector potential and the electric scalar potential which will prove very useful further on, in the next paper, where we will explicitly compute the Lagrangian for our DEUS black hole fields in the quantum electrodynamic formalism, as seen from the local Minkowski spacetime.

The other important quantity computed here is the mass of a DEUS black hole and its growth rate as function of the radial coordinate of the FRW Universe which, in the local Minkowki Universe, is the distance separating the source from the observer. For nearby objects moving toward the observer ($v_{r}<0$) this rate is increasing while, for moving away objects ($v_{r}>0$), situated at a higher distance from the observer, it decreases as the distance increase. This happens because the farther objects are moving away faster than the nearby ones (the objects are redshifted).

\ack
Manuscript registered at the Astronomical Institute, Romanian Academy, record No. 256 from 04/19/07.


\title[D.E.U.S.]{IX: Level Three of Self-Similarity}

\begin{abstract}
This paper is dedicated to the description of processes taking place at the third DEUS level of self-similarity, as seen by a Friedmann-Robertson-Walker (FRW) global and Minkowski local observer: fermion interaction with formation of bosons in a background of vacuum energy.
\end{abstract}


\section{Introduction}

While in the second self-similar DEUS level we have effects as the ones described in papers \cite{deus4,deus5}, at level three of self-similarity we have DEUS objects embedded in the catenoids and the helicoid of level two of self-similarity. The helicoids of this third level must generate at collapse the catenoid of level two and observable effects for the external spacetime. The external observer will see from the evaporated level one of self-similarity (generated FRW Universe bubble from collapsed pre-Big Bang DEUS object) the FRW conformal image (FRW particle bubbles) of these DEUS objects as atoms (non-evaporated level two and non-evaporated level three), bosons (evaporated level two, but non-evaporated level three), or fermions (evaporated level two and evaporated level three).

In this paper we will compute the energy of a level three object, evaporated and non-evaporated, as seen by the external Minkowski observer. For exemplification we will derive these energies for an interaction between an electron and a positron with creation of photons.

Between the energies (potentials) of the evaporated and non-evaporated DEUS level three is a difference representing the vacuum energy.

Quantities, as the electric {\it E} and the magnetic {\it B} fields \cite{deus4}, the four-potential $A_{\mu}$, the radiation density $\rho_{Minkowski}$, or the relation between the $m_{0}$ fermion mass and {\it const} \cite{deus8}, will play an important role in the following computations. 

\section{The Lagrangian in Quantum Electrodynamic Formalism}

Quantizing the Maxwell field equations, the quantum electrodynamics (QED) Lagrangian density corresponding to a spin 1/2 particle coupled to the electromagnetic field of the collapsed one and two DEUS levels (Big Bang and black holes) writes:

\begin{equation}
{\cal L}=-\textnormal{{\bf E}}\; \textnormal{\bf {\.{A}}}-A_{0}\; \nabla_{\cdot} \textnormal{{\bf E}}-\frac{1}{2}\; \textnormal{{\bf E}}^{2}-\frac{1}{4}\; F_{ij}^{2}+\overline{\Psi}(i\not\!\! D-m )\Psi ,
\label{eq_1}
\end{equation}
\noindent
where $F_{ij}$ is the electromagnetic tensor (derived in \cite{deus4}), $A_{0}\equiv f$ is the electric scalar potential, {\bf A} is the magnetic vector potential, $\Psi$ is the wave function of the DEUS object and the operator:

\begin{equation}
\not\!\! D\equiv D_{\mu}\gamma^{\mu} \; ,
\label{eq_2}
\end{equation}
\noindent
with $D_{\mu}\equiv \partial_{\mu}+i\; eA_{\mu}$ and $\gamma^{\mu}$ the Dirac matrices for the four-dimensional Minkowski space:

\begin{equation}
\gamma^{0}=\left( \begin{array}{cc}
0 & -I \\
-I & 0 \\
\end{array} \right) \;\;\; ; \;\;\;
\gamma^{i}=\left( \begin{array}{cc}
0 & \sigma^{i} \\
-\sigma^{i} & 0 \\
\end{array} \right) \;\;\; ; \;\;\;
\gamma_{5}=\left( \begin{array}{cc}
I & 0 \\
0 & -I \\
\end{array} \right)
\label{eq_3}
\end{equation}
\noindent
as function of the identity matrix {\it I} and of the Pauli sigma matrices:

\begin{equation}
\sigma^{x}=\left( \begin{array}{cc}
0 & 1 \\
1 & 0 \\
\end{array} \right) \;\;\; ; \;\;\;
\sigma^{y}=\left( \begin{array}{cc}
0 & -i \\
i & 0 \\
\end{array} \right) \;\;\; ; \;\;\;
\sigma^{z}=\left( \begin{array}{cc}
1 & 0 \\
0 & -1 \\
\end{array} \right)
\label{eq_4}
\end{equation}

The (\ref{eq_1}) Lagrangian can be expressed as:

\begin{equation}
{\cal L}=-\textnormal{{\bf E}}\; \textnormal{\bf {\.{A}}}-\frac{1}{2}\left(\textnormal{{\bf E}}^{2}+\textnormal{{\bf B}}^{2} \right)+{\cal L}(\Psi,\textnormal{{\bf A}})- A_{0}\left(\nabla_{\cdot} \textnormal{{\bf E}}-e\; \overline{\Psi}\gamma^{0}\Psi\right)\; ,
\label{eq_5}
\end{equation}
\noindent
where:

\begin{equation}
{\cal L}(\Psi,\textnormal{{\bf A}})=i\; \overline{\Psi}\gamma^{\mu}\partial_{\mu} \Psi - q\; \overline{\Psi}A_{\mu}\gamma^{\mu}\Psi-\overline{\Psi}m\Psi\; .
\label{eq_6}
\end{equation}

In (\ref{eq_6}) we made the substitution of the elementary charge {\it e} of the electron with the DEUS object charge {\it q} and we took into account the Gauss law for electricity as an additional constraint to ${\cal L}$ :

\begin{equation}
\nabla_{\cdot} \textnormal{{\bf E}}=\rho_{Minkowski}= q\; \overline{\Psi}\gamma_{0} \Psi\; .
\label{eq_7}
\end{equation}

Because of the self-similarity of the DEUS objects the elementary charge will be, for each DEUS object, independently of its observed Minkowski value, the charge of its first lower level's embedded DEUS object.

We know also that:

\begin{equation}
\overline{\Psi}\equiv \Psi^{+}\gamma^{0}\; ,
\label{eq_8}
\end{equation}
\noindent
which, in (\ref{eq_7}), gives:

\begin{equation}
\rho_{Minkowski}= q\; \Psi^{+} \Psi\; .
\label{eq_9}
\end{equation}

In ${\cal L}(\Psi,\textnormal{{\bf A}})$, the electromagnetic four-current of the spin 1/2 particle is:

\begin{equation}
q\; \overline{\Psi}\gamma^{\mu}\Psi= J^{\mu}\equiv \left(\rho_{Minkowski},\textnormal{{\bf J}} \right)\; ,
\label{eq_10}
\end{equation}
\noindent
where $\textnormal{{\bf J}}$ is the electric current in the Minkowski space. The product $\overline{\Psi}\Psi$ must be a Lorentz scalar \cite{bjo}:

\begin{equation}
\overline{\Psi}\Psi= \left| \Psi_{1} \right|^{2}+\left| \Psi_{2} \right|^{2}-\left| \Psi_{3} \right|^{2}-\left| \Psi_{4} \right|^{2}\; .
\label{eq_11}
\end{equation}

The wave function $\Psi$ is described by the Dirac bispinor form:

\begin{equation}
\Psi=\left(
\begin{array}{llll}
\Psi_{1} \\
\Psi_{2} \\
\Psi_{3} \\
\Psi_{4}
\end{array}
\right)
\label{eq_12}
\end{equation}
\noindent
and its Dirac adjoint (as function of the Hermitian-conjugate wave function $\Psi^{+}$) \cite{norb}:

\begin{equation}
\overline{\Psi}\equiv \Psi^{+}\gamma^{0}=\left(\Psi^{*}_{1}\;\;\; \Psi^{*}_{2}\;\;\; -\Psi^{*}_{3}\;\;\; -\Psi^{*}_{4}\right)\; .
\label{eq_13}
\end{equation}

When $E_{3}\neq 0$ and $B_{3}\neq 0$ having the \cite{deus3} expressions, and with the four-potential $A_{\mu}$ components from \cite{deus8}, we can determine that:

\begin{equation}
\begin{array}{lll}
\displaystyle -\textnormal{{\bf E}}\; \textnormal{\bf {\.{A}}}=const\; \frac{1}{2\sqrt{2}}\; \left(\frac{1}{2}\mp \frac{1}{4} \right)^{2}\left( \cosh \frac{\pi}{4}-\sinh \frac{\pi}{4} \right)\; \left(\frac{m_{helicoid}}{q} \right)^{2}\; \frac{a^{2}}{r^{2}} \\
\displaystyle -\frac{1}{2}\; \textnormal{{\bf E}}^{2}=-\frac{1}{2}\; E_{3}^{2}=-2\; \left(\frac{1}{2}\mp \frac{1}{4} \right)^{2}\; \left(\frac{m_{helicoid}}{q} \right)^{2}\; \frac{a^{4}}{r^{4}} \\
\displaystyle -\frac{1}{2}\; \textnormal{{\bf B}}^{2}=-\frac{1}{2}\; B_{3}^{2}=-\frac{1}{2}\; \left(\sqrt{2}+1 \right)^{2}\; \left(\frac{m_{helicoid}}{q} \right)^{2}\; \frac{a^{4}}{r^{4}}
\end{array}
\label{eq_14}
\end{equation}

\section{Explicit Computation of the ${\cal L}(\Psi,\textnormal{{\bf A}})$ Terms}

We will take a tridimensional wave function:

\begin{equation}
\Psi=\Psi_{0}\exp \left[-i\; \left(\omega x^{5}\pm k_{i}x^{i}\right)\right]\; ,
\label{eq_15}
\end{equation}
\noindent
from where:

\begin{equation}
\Psi^{+}=\Psi_{0}\exp \left[i\; \left(\omega x^{5}\pm k_{i}x^{i}\right)\right]\; .
\label{eq_16}
\end{equation}

So:

\begin{equation}
\overline{\Psi}=\Psi^{+}\gamma^{0}=-\Psi_{0}\; \left(e^{\pm i\; k_{1}x^{1}}\;\;\; e^{\pm i\; k_{2}x^{2}}\;\;\; e^{\pm i\; k_{3}x^{3}}\;\;\; e^{i\; \omega x^{5}}\right)\; .
\label{eq_17}
\end{equation}

We can identify in (\ref{eq_13}), resulting:

\begin{equation}
\begin{array}{llll}
\displaystyle \Psi^{*}_{1}=-\Psi_{0}\; e^{\pm i\; k_{1}x^{1}} \\
\displaystyle \Psi^{*}_{2}=-\Psi_{0}\; e^{\pm i\; k_{2}x^{2}} \\
\displaystyle \Psi^{*}_{3}=\Psi_{0}\; e^{\pm i\; k_{3}x^{3}} \\
\displaystyle \Psi^{*}_{4}=\Psi_{0}\; e^{i\; \omega x^{5}}
\end{array}
\label{eq_18}
\end{equation}

In (\ref{eq_17}) we see that:

\begin{equation}
\Psi^{+}=\Psi_{0}\; \left(e^{\pm i\; k_{3}x^{3}}\;\;\; e^{i\; \omega x^{5}}\;\;\; e^{\pm i\; k_{1}x^{1}}\;\;\; e^{\pm i\; k_{2}x^{2}}\right)\; ,
\label{eq_19}
\end{equation}
\noindent
which leads us to conclude that:

\begin{equation}
\begin{array}{llll}
\displaystyle \Psi_{1}=\Psi_{0}\; e^{\mp i\; k_{3}x^{3}} \\
\displaystyle \Psi_{2}=\Psi_{0}\; e^{- i\; \omega x^{5}} \\
\displaystyle \Psi_{3}=\Psi_{0}\; e^{\mp i\; k_{1}x^{1}} \\
\displaystyle \Psi_{4}=\Psi_{0}\; e^{\mp i\; k_{2}x^{2}}
\end{array}
\label{eq_20}
\end{equation}

\subsection{$q\; \overline{\Psi}A_{\mu}\gamma^{\mu}\Psi$}

For $\mu=0$ we have:

\begin{equation}
q\; \overline{\Psi}f\gamma^{0}\Psi=q\;f\left(\Psi_{3}^{*}\Psi_{1}+\Psi_{4}^{*}\Psi_{2}-\Psi_{1}^{*}\Psi_{3}-\Psi_{2}^{*}\Psi_{4} \right)\; .
\label{eq_21}
\end{equation}

But, from (\ref{eq_7}):

\begin{equation}
q\; \overline{\Psi}f\gamma^{0}\Psi=f\; \rho_{Minkowki}\; ,
\label{eq_22}
\end{equation}
\noindent
which means that:

\begin{equation}
\Psi_{3}^{*}\Psi_{1}+\Psi_{4}^{*}\Psi_{2}-\Psi_{1}^{*}\Psi_{3}-\Psi_{2}^{*}\Psi_{4}=\frac{\rho_{Minkowki}}{q}=\frac{1}{2\sqrt{3}\;\pi}\; (2\mp1)\; \frac{m_{helicoid}}{q^{2}}\; \frac{a^{2}}{r^{3}}\; ,
\label{eq_23}
\end{equation}
\noindent
where we used the radiation density obtained in \cite{deus8}.

In the same way, for $\mu=1$:

\begin{equation}
q\; \overline{\Psi}A_{1}\gamma^{1}\Psi=q\;A_{1}\left(\Psi_{4}^{*}\Psi_{1}+\Psi_{3}^{*}\Psi_{2}+\Psi_{2}^{*}\Psi_{3}+\Psi_{1}^{*}\Psi_{4} \right)\; .
\label{eq_24}
\end{equation}

But from (\ref{eq_10}), and because $J^{1}=J^{2}=0$, we have:

\begin{equation}
q\; \overline{\Psi}\gamma^{1}\Psi=J^{1}=0\; ,
\label{eq_25}
\end{equation}
\noindent
meaning that:

\begin{equation}
\Psi_{4}^{*}\Psi_{1}+\Psi_{3}^{*}\Psi_{2}+\Psi_{2}^{*}\Psi_{3}+\Psi_{1}^{*}\Psi_{4}=0\; .
\label{eq_26}
\end{equation}

For $\mu=2$:

\begin{equation}
q\; \overline{\Psi}A_{2}\gamma^{2}\Psi=i\; q\;A_{2}\left(\Psi_{4}^{*}\Psi_{1}-\Psi_{3}^{*}\Psi_{2}+\Psi_{2}^{*}\Psi_{3}-\Psi_{1}^{*}\Psi_{4} \right)
\label{eq_27}
\end{equation}
\noindent
and, from (\ref{eq_10}):

\begin{equation}
q\; \overline{\Psi}\gamma^{2}\Psi=J^{2}=0\; ,
\label{eq_28}
\end{equation}
\noindent
resulting:

\begin{equation}
\Psi_{4}^{*}\Psi_{1}-\Psi_{3}^{*}\Psi_{2}+\Psi_{2}^{*}\Psi_{3}-\Psi_{1}^{*}\Psi_{4}=0\; .
\label{eq_29}
\end{equation}

Finally, for $\mu=3$:

\begin{equation}
q\; \overline{\Psi}A_{3}\gamma^{3}\Psi=q\;A_{3}\left(\Psi_{3}^{*}\Psi_{1}-\Psi_{4}^{*}\Psi_{2}+\Psi_{1}^{*}\Psi_{3}-\Psi_{2}^{*}\Psi_{4} \right)
\label{eq_30}
\end{equation}
\noindent
and, because also:

\begin{equation}
q\; \overline{\Psi}\gamma^{3}\Psi=J^{3}\; ,
\label{eq_31}
\end{equation}
\noindent
we obtain:

\begin{equation}
\Psi_{3}^{*}\Psi_{1}-\Psi_{4}^{*}\Psi_{2}+\Psi_{1}^{*}\Psi_{3}-\Psi_{2}^{*}\Psi_{4}=\frac{J^{3}}{q}\; .
\label{eq_32}
\end{equation}

By making the substitution of $\Psi_{\mu}^{*}$ and $\Psi_{\mu}$ ((\ref{eq_18}) and (\ref{eq_20}))

\begin{itemize}
\item in (\ref{eq_23}):

\begin{equation}
\Psi_{0}^{2}=\frac{1}{8\sqrt{3}\; \pi}\; (2\mp1)\; \frac{m_{helicoid}}{q^{2}}\; \frac{a^{2}}{r^{3}} \; ;
\label{eq_33}
\end{equation}

\item in (\ref{eq_26}) and (\ref{eq_29}):

\begin{equation}
k_{1}x^{1}-k_{2}x^{2}=k_{3}x^{3}\mp \omega x^{5} \; ;
\label{eq_34}
\end{equation}

\item in (\ref{eq_32}):

\begin{equation}
J^{3}=0 \; .
\label{eq_35}
\end{equation}
In conclusion, for the Minkowski observer, the electric current from the DEUS object is $\textnormal{{\bf J}}=0$.
\end{itemize}

So, the term:

\begin{equation}
q\; \overline{\Psi}A_{\mu}\gamma^{\mu}\Psi=f\; \rho_{Minkowski} \; .
\label{eq_36}
\end{equation}

\subsection{$i\; \overline{\Psi}\gamma^{\mu}\partial_{\mu}\Psi$}

The derivative of the DEUS wave function will be:

\begin{equation}
\partial_{\mu}\Psi=\frac{\partial \Psi}{\partial x^{\mu}}=\frac{\partial \Psi}{\partial t_{FRW}}\; \frac{\partial t_{FRW}}{\partial x^{\mu}} \; .
\label{eq_37}
\end{equation}

Knowing the expressions of $\Psi_{\mu}^{*}$ and $\Psi_{\mu}$ ((\ref{eq_18}) and (\ref{eq_20})) and using (\ref{eq_34}), we will have:

\begin{itemize}
\item for $\mu=0$:

\begin{equation}
\!\!\!\!\!\!\!\!\!\!\!\!\!\!\!\!\!\!\!\!i\; \overline{\Psi}\gamma^{0}\partial_{x^{5}}\Psi=\Psi_{0}^{2}\left[\omega\; \left(2+\frac{\displaystyle \sinh \frac{\pi}{4}}{\displaystyle \cosh \frac{\pi}{4}} \right)-i\; k_{3}\; \left(\frac{1}{\displaystyle \cosh \frac{\pi}{4}} +\frac{\pi}{4}\; \frac{1}{\displaystyle \sinh \frac{\pi}{4}}+\frac{\pi}{4}\; \frac{1}{\displaystyle \cosh \frac{\pi}{4}}\right) \right] \; ;
\label{eq_38}
\end{equation}

\item for $\mu=1$:

\begin{equation}
\!\!\!\!\!\!\!\!\!\!\!\!\!\!\!\!\!\!\!\!\begin{array}{cc}
i\; \overline{\Psi}\gamma^{1}\partial_{x^{1}}\Psi=\displaystyle \frac{\sqrt{2}\; \Psi_{0}^{2}}{\displaystyle \cosh \frac{\pi}{4}-\sinh \frac{\pi}{4}}\; \left\{e^{i\left(\omega x^{5}\mp k_{3}x^{3} \right)}\left[k_{3}\mp \frac{1}{\sqrt{2}}\; k_{1}\left( \cosh \frac{\pi}{4}-\sinh \frac{\pi}{4}\right) \right]+ \right. \\
\displaystyle \left. + e^{-i\left(\omega x^{5}\mp k_{3}x^{3} \right)}\left[i\; \omega\; \cosh \frac{\pi}{4} + \frac{1}{\sqrt{2}}\; k_{2}\left( \cosh \frac{\pi}{4}+\sinh \frac{\pi}{4}\right) \right] \right\} \; ;
\end{array}
\label{eq_39}
\end{equation}

\item for $\mu=2$:

\begin{equation}
\!\!\!\!\!\!\!\!\!\!\!\!\!\!\!\!\!\!\!\!\begin{array}{cc}
i\; \overline{\Psi}\gamma^{2}\partial_{x^{2}}\Psi=\pm \displaystyle i\; \frac{\sqrt{2}\; \Psi_{0}^{2}}{\displaystyle \cosh \frac{\pi}{4}+\sinh \frac{\pi}{4}}\; \left\{e^{i\left(\omega x^{5}\mp k_{3}x^{3} \right)}\left[-k_{3}\pm \frac{1}{\sqrt{2}}\; k_{1}\left( \cosh \frac{\pi}{4}-\sinh \frac{\pi}{4}\right) \right]+ \right. \\
\displaystyle \left. + e^{-i\left(\omega x^{5}\mp k_{3}x^{3} \right)}\left[i\; \omega\; \cosh \frac{\pi}{4} + \frac{1}{\sqrt{2}}\; k_{2}\left( \cosh \frac{\pi}{4}+\sinh \frac{\pi}{4}\right) \right] \right\} \; ;
\end{array}
\label{eq_40}
\end{equation}

\item for $\mu=3$:

\begin{equation}
\!\!\!\!\!\!\!\!\!\!\!\!\!\!\!\!\!\!\!\!\begin{array}{cc}
i\; \overline{\Psi}\gamma^{3}\partial_{x^{3}}\Psi=\displaystyle \mp \; \Psi_{0}^{2}\left[-k_{3}+i\; \omega\; \cosh \frac{\pi}{4}\pm \frac{1}{\sqrt{2}}\; k_{1}\left( \cosh \frac{\pi}{4}-\sinh \frac{\pi}{4}\right) + \right. \\
\left. \displaystyle + \frac{1}{\sqrt{2}}\; k_{2}\left( \cosh \frac{\pi}{4}+\sinh \frac{\pi}{4}\right) \right] \; .
\end{array}
\label{eq_41}
\end{equation}

\end{itemize}

If we take into account the symmetry of the DEUS catenoidal hyper-surface, the probability to find the "particle" on $x^{1}$ is equal with the probability of finding it on $x^{2}$. In other words $\left| \Psi_{1} \right|^{2}=\left| \Psi_{2} \right|^{2}$. From here we get that $\pm k_{3}x^{3}=\omega x^{5}$, which, into (\ref{eq_34}), makes:

\begin{equation}
k_{1}x^{1}=k_{2}x^{2}\; .
\label{eq_42}
\end{equation}

Before the DEUS collapse the local chart of the helicoid will have to satisfy the $t_{k}=\pm \phi_{k}$ condition for which its coordinates are seen by the external observer as:

\begin{equation}
\begin{array}{llll}
x^{1}=-\displaystyle \frac{1}{\sqrt{2}}\; \sinh \frac{\pi}{4} \\
x^{2}=\pm\; \displaystyle \frac{1}{\sqrt{2}}\; \sinh \frac{\pi}{4} \\
x^{3}=\pm\; \displaystyle \frac{\pi}{4} \\
x^{5}=-\;i\; \displaystyle \sinh \frac{\pi}{4} \; ,
\end{array}
\label{eq_43}
\end{equation}
\noindent
which, in (\ref{eq_42}), gives:

\begin{equation}
k_{1}=\mp k_{2}\; .
\label{eq_44}
\end{equation}

Also, from $\pm k_{3}x^{3}=\omega x^{5}$ with (\ref{eq_43}) we have:

\begin{equation}
k_{3}=-i\; \omega\; \frac{\displaystyle \sinh \frac{\pi}{4}}{\displaystyle \frac{\pi}{4}}
\label{eq_45}
\end{equation}
\noindent
and:

\begin{equation}
e^{i\left(\omega x^{5}\mp k_{3}x^{3} \right)}=e^{-i\left(\omega x^{5}\mp k_{3}x^{3} \right)}=1 \; .
\label{eq_46}
\end{equation}

With the above symmetry considerations and with (\ref{eq_33}) we will be able now to write $i\; \overline{\Psi}\gamma^{\mu}\partial_{\mu}\Psi$ as function of $k_{2}$ and $\omega$. Taking into account that, when the DEUS object collapses, $k_{1}=k_{2}=0$:

\begin{equation}
\!\!\!\!\!\!\!\!\!\!\!\!\!\!\!\!\!\!\!\!\!\!\!\!\!\!\displaystyle i\; \overline{\Psi}\gamma^{\mu}\partial_{\mu}\Psi=\frac{1}{8\sqrt{3}\; \pi}\; (2\mp1)\; \frac{m_{helicoid}}{q^{2}}\; \frac{a^{2}}{r^{3}}\; \omega \left(
\begin{array}{cccc}
\displaystyle 1-\frac{1}{\displaystyle \pi/4}\; \frac{\displaystyle \sinh \frac{\pi}{4}}{\displaystyle \cosh \frac{\pi}{4}} \\
i\; \frac{\displaystyle \sqrt{2}}{\displaystyle \cosh \frac{\pi}{4}-\sinh \frac{\pi}{4}}\left[ -\frac{\displaystyle 1}{\displaystyle \pi/4}\; \sinh \frac{\displaystyle \pi}{\displaystyle 4} +\cosh \frac{\displaystyle \pi}{\displaystyle 4} \right] \\
\mp\; \frac{\displaystyle \sqrt{2}}{\displaystyle \cosh \frac{\pi}{4}+\sinh \frac{\pi}{4}}\left[\frac{\displaystyle 1}{\displaystyle \pi/4}\; \sinh \frac{\displaystyle \pi}{\displaystyle 4} +\cosh \frac{\displaystyle \pi}{\displaystyle 4} \right] \\
\mp\; i\; \left[\frac{1}{\displaystyle \pi/4}\; \sinh \frac{\displaystyle \pi}{\displaystyle 4} +\cosh \frac{\displaystyle \pi}{\displaystyle 4} \right]
\end{array}
\right)
\label{eq_47}
\end{equation}

Now, the wave function will be:

\begin{equation}
\Psi=\Psi_{0}\left(
\begin{array}{cccc}
e^{-i\omega x^{5}} \\
e^{-i\omega x^{5}} \\
1 \\
1
\end{array}
\right)
\label{eq_47a}
\end{equation}

\subsection{$\overline{\Psi}m\Psi$}

With the above results for $k_{i}$, $x^{\mu}$ and $\Psi$ we have:

\begin{equation}
\begin{array}{lll}
\overline{\Psi}m\Psi=m\left(\left| \Psi_{1} \right|^{2}+\left| \Psi_{2} \right|^{2}-\left| \Psi_{3} \right|^{2}-\left| \Psi_{4} \right|^{2} \right)= \\
=m\; \Psi_{0}^{2}\; \left(e^{\mp 2i\; k_{3}x^{3}}+e^{-2i\; \omega x^{5}}-e^{\mp 2i\; k_{1}x^{1}}-e^{\mp 2i\; k_{2}x^{2}} \right)= \\
=\frac{\displaystyle 1}{\displaystyle 4\sqrt{3}\; \pi}\; (2\mp1)\; \frac{\displaystyle m_{helicoid}}{\displaystyle q^{2}}\; m\; \frac{\displaystyle a^{2}}{\displaystyle r^{3}}\; \left(e^{-2i\; \omega x^{5}}-1 \right)\; .
\end{array}
\label{eq_48}
\end{equation}

\section{The Lagrangian of the Self-Similar DEUS Objects Embedded in the Catenoid (DEUS Fermions)}

From all the computed terms we can conclude that the (\ref{eq_5}) Lagrangian will be:

\begin{equation}
\!\!\!\!\!\!\!\!\!\!\!\begin{array}{ccccc}
{\cal L}\stackrel{not}{=}{\cal L}_{F}=\displaystyle const\; \frac{1}{2\sqrt{2}}\; \left(\frac{1}{2}\mp \frac{1}{4} \right)^{2}\left( \cosh \frac{\pi}{4}-\sinh \frac{\pi}{4} \right)\; \left(\frac{m_{helicoid}}{q} \right)^{2}\; \frac{a^{2}}{r^{2}}- \\
\displaystyle -2\; \left(\frac{1}{2}\mp \frac{1}{4} \right)^{2}\; \left(\frac{m_{helicoid}}{q} \right)^{2}\; \frac{a^{4}}{r^{4}}-\frac{1}{2}\; \left(\sqrt{2}+1 \right)^{2}\; \left(\frac{m_{helicoid}}{q} \right)^{2}\; \frac{a^{4}}{r^{4}}- \\
\displaystyle -const\; \frac{1}{16\sqrt{6}\; \pi}\; \left(\frac{1}{2}\mp \frac{1}{4} \right)\; \left(2\mp 1 \right)\; \left( \cosh \frac{\pi}{4}-\sinh \frac{\pi}{4} \right)\; \left(\frac{m_{helicoid}}{q} \right)^{2}\; \frac{a^{2}}{r^{3}}+ \\
\displaystyle + \frac{\displaystyle 1}{\displaystyle 4\sqrt{3}\; \pi}\; (2\mp1)\; \frac{\displaystyle m_{helicoid}}{\displaystyle q^{2}}\; m\; \frac{\displaystyle a^{2}}{\displaystyle r^{3}}\; \left(1-e^{-2i\; \omega x^{5}} \right)+ \\
\displaystyle + \frac{1}{8\sqrt{3}\; \pi}\; (2\mp1)\; \frac{m_{helicoid}}{q^{2}}\; \frac{a^{2}}{r^{3}}\; \omega \left(
\begin{array}{cccc}
\displaystyle 1-\frac{1}{\displaystyle \pi/4}\; \frac{\displaystyle \sinh \frac{\pi}{4}}{\displaystyle \cosh \frac{\pi}{4}} \\
i\; \frac{\displaystyle \sqrt{2}}{\displaystyle \cosh \frac{\pi}{4}-\sinh \frac{\pi}{4}}\left[ -\frac{\displaystyle 1}{\displaystyle \pi/4}\; \sinh \frac{\displaystyle \pi}{\displaystyle 4} +\cosh \frac{\displaystyle \pi}{\displaystyle 4} \right] \\
\mp\; \frac{\displaystyle \sqrt{2}}{\displaystyle \cosh \frac{\pi}{4}+\sinh \frac{\pi}{4}}\left[\frac{\displaystyle 1}{\displaystyle \pi/4}\; \sinh \frac{\displaystyle \pi}{\displaystyle 4} +\cosh \frac{\displaystyle \pi}{\displaystyle 4} \right] \\
\mp\; i\; \left[\frac{1}{\displaystyle \pi/4}\; \sinh \frac{\displaystyle \pi}{\displaystyle 4} +\cosh \frac{\displaystyle \pi}{\displaystyle 4} \right]
\end{array}
\right)
\end{array}
\label{eq_49}
\end{equation}
\noindent
where $m=m_{0}$ is mass of the particle \cite{deus8} and, at collapse limit, $x^{5}=-i\; \sinh \displaystyle \frac{\pi}{4}$.

\section{The Proca Lagrangian of a Created/Annihilated DEUS Object (DEUS Bosons)}

On the same basis of self-similarity of DEUS objects, the evaporating DEUS object has to "expel" in the external observer's Minkowski local spacetime its matter content (in its frame, DEUS objects), objects which in the external spacetime behave as particles and atoms. In this way, from the evaporating object into the external spacetime we have to have not only spin {\it n}/2 particles but also spin {\it n} particles, described by the known physics.

We will focus our attention on spin 0 or 1 vector fields (in interaction with the electromagnetic fields of the evaporated DEUS level one), which we will describe through the Proca Lagrangian for vector fields:

\begin{equation}
{\cal L}_{B}=-\frac{1}{16\; \pi}\left(\partial^{\mu}A^{\nu}-\partial^{\nu}A^{\mu} \right)\left(\partial_{\mu}A_{\nu}-\partial_{\nu}A_{\mu} \right)+\frac{1}{8\; \pi}\left(\frac{M_{0}c}{\hbar} \right)^{2}A^{\nu}A_{\nu} \; ,
\label{eq_50}
\end{equation}
\noindent
with the $A_{\mu}$ determined in \cite{deus8}. For two interacting particles (fermions) we will have $M_{0}\simeq \; m_{01}+m_{02}$, or, for the particular case of a particle interacting with its antiparticle, $M_{0}\simeq 2\; m_{0}$.

When working in natural units ($\hbar=c=1$) and when:

\begin{equation}
F^{\mu\nu}\equiv \partial^{\mu}A^{\nu}-\partial^{\nu}A^{\mu} \; ,
\label{eq_51}
\end{equation}
\noindent
we will be able to write (\ref{eq_50}) as:

\begin{equation}
{\cal L}_{B}=-\frac{1}{16\; \pi}\; F^{\mu\nu}F_{\mu\nu}+\frac{1}{8\; \pi}\; M_{0}^{2}\; A^{\nu}A_{\nu} \; .
\label{eq_52}
\end{equation}

In \cite{deus8} we determined that the gauge for the DEUS object is $\partial_{\mu}A_{\nu}=\partial_{\nu}A_{\mu}=0$, which makes:

\begin{equation}
{\cal L}_{B}=\frac{1}{8\; \pi}\; M_{0}^{2}\; A^{\nu}A_{\nu} \; .
\label{eq_53}
\end{equation}

As function of the Lagrangian and the wave function (see (\ref{eq_12}) and (\ref{eq_20})), the electromagnetic tensor is given by:

\begin{equation}
T_{\mu\nu}=-g_{\mu\nu}{\cal L}_{B}+\frac{\partial {\cal L}_{B}}{\partial \left(\partial \Psi/\partial x_{\mu} \right)}\; \frac{\partial \Psi}{\partial x^{\nu}} \; .
\label{eq_54}
\end{equation}

Because $\displaystyle \frac{\partial {\cal L}_{B}}{\partial \left(\partial \Psi/\partial x_{\mu} \right)}=0$, results:

\begin{equation}
\!\!\!\!\!\!\!\!\!\!\!\!\!\!\!\!\!\begin{array}{cc}
T_{\nu\nu}=\displaystyle -g_{\nu\nu}{\cal L}_{B}=-\frac{1}{8\; \pi}\; M_{0}^{2}\; g_{\nu\nu}g^{\nu\nu}\left(A_{\nu}\right)^{2}= \\
\displaystyle =-const^{2}\; \frac{1}{512\; \pi}\; M_{0}^{2}\; \left(\frac{m_{helicoid}}{q}\right)^{2} \left(
\begin{array}{cccc}
\displaystyle \frac{1}{2}\; \left(\frac{1}{2}\mp \frac{1}{4} \right)^{2}\left( \cosh \frac{\pi}{4}-\sinh \frac{\pi}{4} \right)^{2} \\
-\displaystyle \left(\frac{1}{2}\mp \frac{1}{4} \right)^{2}\cosh^{2} \frac{\pi}{4} \\
-\displaystyle \left(\frac{1}{4}\mp \frac{1}{2} \right)^{2}\cosh^{2} \frac{\pi}{4}\left(\frac{\displaystyle \cosh \frac{\pi}{4}-\sinh \frac{\pi}{4}}{\displaystyle \cosh \frac{\pi}{4}+\sinh \frac{\pi}{4}}\right)^{2} \\
\displaystyle - \frac{1}{2}\;\left(-\frac{1}{4}\pm \frac{1}{2} \right)^{2}\cosh^{2} \frac{\pi}{4}\left(\cosh \frac{\pi}{4}-\sinh \frac{\pi}{4}\right)^{2}
\end{array}
\right)
\end{array}
\label{eq_55}
\end{equation}

We can express {\it const} as function of $m_{0}$ using the relation determined in \cite{deus8} for the mass of the particle. Then:

\begin{equation}
T_{\nu\nu}=-\displaystyle \frac{4}{3\; \pi^{3}}\; \frac{1}{m_{0}^{2}}\; \left(\frac{m_{helicoid}}{q}\right)^{2}\; \frac{a^{4}}{r^{6}} \left(
\begin{array}{cccc}
\displaystyle \frac{1}{2}\; \left(\frac{1}{2}\mp \frac{1}{4} \right)^{2} \\
-\displaystyle \left(\frac{1}{2}\mp \frac{1}{4} \right)^{2}\left(\frac{\displaystyle \cosh \frac{\pi}{4}}{\displaystyle \cosh \frac{\pi}{4}-\sinh \frac{\pi}{4}}\right)^{2} \\
-\displaystyle \left(\frac{1}{4}\mp \frac{1}{2} \right)^{2}\left(\frac{\displaystyle \cosh \frac{\pi}{4}}{\displaystyle \cosh \frac{\pi}{4}+\sinh \frac{\pi}{4}}\right)^{2} \\
\displaystyle - \frac{1}{2}\;\left(-\frac{1}{4}\pm \frac{1}{2} \right)^{2}\cosh^{2} \frac{\pi}{4}
\end{array}
\right)
\label{eq_56}
\end{equation}

In the above formula:

\begin{equation}
T_{\nu\nu}=\left(
\begin{array}{cccc}
\displaystyle T_{00} \\
\displaystyle T_{11} \\
\displaystyle T_{22} \\
\displaystyle T_{33}
\end{array}
\right)
\label{eq_57}
\end{equation}

With, $(g^{00})_{Minkowski}\rightarrow (g^{55})_{helicoid}$ \cite{deus1,deus2} and $\displaystyle (g^{55})_{helicoid}=-(g^{00})_{Minkowski}=-\frac{4}{3}$ (see the (29) system of coordinates from \cite{deus8}) we can determine the energy of the resulting Minkowski boson (the (\ref{eq_56}) $T_{\nu\nu}$ is already converted to the Minkowski frame):

\begin{equation}
E_{B}=\int_{V} (g^{55})_{helicoid}T_{00}\; d^{3}r=\frac{1}{45\; \pi^{3}}\; \frac{1}{m_{0}^{2}}\; \left(\frac{m_{helicoid}}{q}\right)^{2}\; \frac{a^{4}}{r^{3}}\left[ \displaystyle \frac{1}{2}\; \left(\frac{1}{2}\mp \frac{1}{4} \right)^{2}\right] \; .
\label{eq_58}
\end{equation}

The integration has to be done after the volume {\it V} of the DEUS boson. As part of catenoidal hyper-surface of the collapsing DEUS object, the observed scale factor will be  $a=\pi/4$  \cite{deus3}. The $1/\epsilon^{2}$ factor was not written because in natural units it is $\simeq 1$.

\section{Fermion Interaction in Vacuum}

In an interaction between two DEUS fermions (in the presence of vacuum) with the production of a DEUS boson, the binding energy will be:

\begin{equation}
E_{\Lambda 0}=2\; E_{F,\Lambda}-E_{B} \; ,
\label{eq_59}
\end{equation}
\noindent
or, in general, for fermions having a different $m_{0}$, $E_{\Lambda 0}=(E_{F1,\Lambda}+E_{F2,\Lambda})-E_{B}$.

We will compute the $E_{F,\Lambda}$ energy of the fermions in the same way as we have done it in the previous section for the energy of the boson. For that we need to know the $T_{00}$ component:

\begin{equation}
T_{00}=-(g_{55})_{helicoid}{\cal L}_{F}+\frac{\partial {\cal L}_{F}}{\partial \left(\partial \Psi/\partial x_{5} \right)}\; \frac{\partial \Psi}{\partial x^{5}} \; ,
\label{eq_60}
\end{equation}
\noindent
where $\displaystyle (g_{00})_{Minkowski}\rightarrow \left(g_{55} \right)_{helicoid}=-\frac{3}{4}$ and ${\cal L}_{F}$ from (\ref{eq_49}). From (\ref{eq_47a}) we have:

\begin{equation}
\frac{\partial \Psi}{\partial x_{5}}=\frac{\partial \Psi}{\partial t_{FRW}}\; \frac{\partial t_{FRW}}{\partial x^{5}}\; \frac{1}{(g_{55})_{helicoid}}=\frac{4}{3}\;(i\omega)\; \Psi_{0}\left(
\begin{array}{cccc}
\displaystyle e^{-i\omega x^{5}} \\
\displaystyle e^{-i\omega x^{5}} \\
\displaystyle 0 \\
\displaystyle 0
\end{array}
\right) \; ,
\label{eq_61}
\end{equation}
\noindent
where:

\begin{equation}
\frac{\partial x^{5}}{\partial t_{FRW}}=\pm i\; \cosh\; \frac{\pi}{4}\; ,
\label{eq_62}
\end{equation}
\noindent
after the substitution of $t_{k}=\pm \phi_{k}$. In (\ref{eq_62}), $\omega=2\pi\nu$, with the frequency $\nu=E_{helicoid}/\hbar=E_{helicoid}$ and $E_{helicoid}=m_{helicoid}$ ($\hbar=c=1$).

Then $\displaystyle \frac{\partial {\cal L}_{F}}{\partial \left(\partial \Psi/\partial x_{5} \right)}=0$ and:

\begin{equation}
T_{00}=-(g_{55})_{helicoid}{\cal L}_{F}=\frac{3}{4}\; {\cal L}_{F} \; .
\label{eq_63}
\end{equation}

In the Hamilton formalism, $\displaystyle H=\int_{V} {\cal H}\; d^{3}r$, where ${\cal H}=T_{00}$ and $H=E_{F,\Lambda}$. So:

\begin{equation}
E_{F,\Lambda}=\int_{V} T_{00}\; d^{3}r=\frac{3}{4}\; \int_{V} {\cal L}_{F}\; d^{3}r \; .
\label{eq_64}
\end{equation}

\section{Fermion Interaction without Vacuum}

Until now we observed only the behavior of the DEUS particles in a Minkowski Universe as manifestations of the {\it pure helicoidal hyper-surface} of the DEUS objects contained into the catenoid of the collapsing DEUS object. But all these particles can be seen equivalently as parts of the BHIG spacetime which is the central helicoid of the considered DEUS object \cite{deus1}. The BHIG spacetime is, in the acceptance of the external observer, "frozen in time" (spacelike) while, for the {\it pure helicoidal hyper-surface}, it is a "reversed mirror".

In this case, the BHIG spacetimes will not contain the vacuum component because there are no FRW bubbles (which exist only for external FRW observers) that through their inertial motion between their birth and death (see Fig. 2 from \cite{deus2}) to create the vacuum energy effect. It will exist only as a helicoid, invisible to the external spacetime. Between it and the external spacetime there are the ergosphere's hyper-surfaces that do not allow to access it directly because the space and time role are reversed at their crossing.

The third coordinate of the BHIG helicoids of the DEUS objects embedded in the DEUS' catenoid rotate their temporal coordinate to the third coordinate of the {\it pure helicoidal hyper-surface} \cite{deus1}. Only after the collapse and transformation of the DEUS object in a global, conformal, FRW Universe (and local Minkowski) the time $t_{FRW}$ begins to exist and, in consequence, the rotation gains a meaning. In a mathematical description this means that the $x^{5}=\pm i\; \textnormal{arctg}\; (t_{k}/\phi_{k})$ rotates to $x^{3}=\textnormal{arctg}\; (t_{k}/\phi_{k})$. In a physical description this happens because a rigid string-like collapsed DEUS object generates in its rotation (so, in time) a helicoidal sheet with symmetry after the $x^{3}$ axis.

In conclusion, $(g^{55})_{BHIG}\rightarrow (g^{33})_{helicoid}$ where:

\begin{equation}
(g_{33})_{helicoid}=-\frac{\displaystyle 2H^{2}-2H\sqrt{H^{2}+1}+1}{\displaystyle 4\; \left(H^{2}-H\sqrt{H^{2}+1}+1 \right)^{2}}\; \sin^{2}\; \zeta \equiv -\frac{3}{4}\; r^{2} \; .
\label{eq_65}
\end{equation}

Here we have:

\begin{equation}
T_{00}=-(g_{33})_{helicoid}{\cal L}_{F}+\frac{\partial {\cal L}_{F}}{\partial \left(\partial \Psi/\partial x_{3} \right)}\; \frac{\partial \Psi}{\partial x^{3}} \; ,
\label{eq_66}
\end{equation}
\noindent
where:

\begin{equation}
\frac{\partial \Psi}{\partial x_{3}}=\frac{\partial \Psi}{\partial t_{FRW}}\; \frac{\partial t_{FRW}}{\partial x^{3}}\; \frac{1}{(g_{33})_{helicoid}}=\mp \frac{4}{3}\;\omega\; \Psi_{0}\; \cosh\; \frac{\pi}{4}\; e^{-i\omega x^{5}}\; \left(
\begin{array}{cccc}
\displaystyle 1 \\
\displaystyle 1 \\
\displaystyle 0 \\
\displaystyle 0
\end{array}
\right)\; \frac{1}{r^{2}}
\label{eq_67}
\end{equation}
\noindent
and:

\begin{equation}
\!\!\!\!\!\!\!\!\!\!\!\!\!\!\!\!\!\begin{array}{ccccc}
\displaystyle \frac{\partial {\cal L}_{F}}{\partial \left(\partial \Psi/\partial x_{3} \right)}\; \frac{\partial \Psi}{\partial x^{3}}=-const\; \frac{3}{4\sqrt{2}}\; \left(\frac{1}{2}\mp \frac{1}{4} \right)^{2}\left(\cosh \frac{\pi}{4}-\sinh \frac{\pi}{4} \right)\left(\frac{m_{helicoid}}{q} \right)^{2} a^{2}+ \\
\displaystyle + 6 \left(\frac{1}{2}\mp \frac{1}{4} \right)^{2}\left(\frac{m_{helicoid}}{q} \right)^{2} \frac{a^{4}}{r^{2}}+\frac{3}{2}\left(\sqrt{2}+1 \right)^{2}\left(\frac{m_{helicoid}}{q} \right)^{2} \frac{a^{4}}{r^{2}}+ \\
\displaystyle + const\; \frac{9\sqrt{2}}{128\sqrt{3}\; \pi}\left(\frac{1}{2}\mp \frac{1}{4} \right)\; (2\mp 1)\left(\cosh \frac{\pi}{4}-\sinh \frac{\pi}{4} \right) \left(\frac{m_{helicoid}}{q} \right)^{2}\frac{a^{2}}{r}- \\
\displaystyle -\frac{9}{16\sqrt{3}\; \pi}(2\mp 1) \left(\frac{m_{helicoid}}{q} \right)^{2}\frac{a^{2}}{r}\left(1-e^{-2i\omega x^{5}} \right)- \\
\displaystyle -\frac{9}{32\sqrt{3}\; \pi}(2\mp 1)\frac{m_{helicoid}}{q^{2}}\omega\; \frac{a^{2}}{r}
 \left(
\begin{array}{cccc}
\displaystyle \frac{1}{2}\; \left(\frac{1}{2}\mp \frac{1}{4} \right)^{2} \\
-\displaystyle \left(\frac{1}{2}\mp \frac{1}{4} \right)^{2}\left(\frac{\displaystyle \cosh \frac{\pi}{4}}{\displaystyle \cosh \frac{\pi}{4}-\sinh \frac{\pi}{4}}\right)^{2} \\
-\displaystyle \left(\frac{1}{4}\mp \frac{1}{2} \right)^{2}\left(\frac{\displaystyle \cosh \frac{\pi}{4}}{\displaystyle \cosh \frac{\pi}{4}+\sinh \frac{\pi}{4}}\right)^{2} \\
\displaystyle - \frac{1}{2}\;\left(-\frac{1}{4}\pm \frac{1}{2} \right)^{2}\cosh^{2} \frac{\pi}{4}
\end{array}
\right)
\end{array}
\label{eq_68}
\end{equation}

As in the precedent case, the energy of the fermion will be:

\begin{equation}
E_{F}=\int_{V} T_{00}\; d^{3}r \; .
\label{eq_69}
\end{equation}

Then the binding energy is given by:

\begin{equation}
E_{0}=2\; E_{F}-E_{B} \; .
\label{eq_70}
\end{equation}
\noindent
or, for the general case, by $E_{0}=(E_{F1}+E_{F2})-E_{B}$.

\section{Graphical Representations}

For the exemplification of the interaction potential we used the interaction between an electron and a positron with creation of photons. For this situation, in natural units, $m_{0}=m_{e}=(9.10938\times 10^{-31})\times (5.61\times 10^{26})$ [GeV], quantity that enters (excepting $E_{B}$) into the expression of {\it const} \cite{deus8}. Because the mass of the {\it pure helicoidal hyper-surface} and of the BHIG spacetime of the evaporating DEUS object is dependent on the number of contained DEUS "particles" and taking the charge $q$ as being the description of one such embedded "particle", $m_{helicoid}=q$. This is possible to consider because, in the four-dimensional frame of the DEUS object, we can see only one level of self-similarity (in Minkowski frame we can see down to level four), each embedded object being an elementary particle for the upper level.
As we said above $a=\pi/4$, $\epsilon=1$ and $\omega=2\pi\; m_{helicoid}$.

\begin{figure}
\centering
\leavevmode
\includegraphics[width=14.5cm,height=9.5cm]{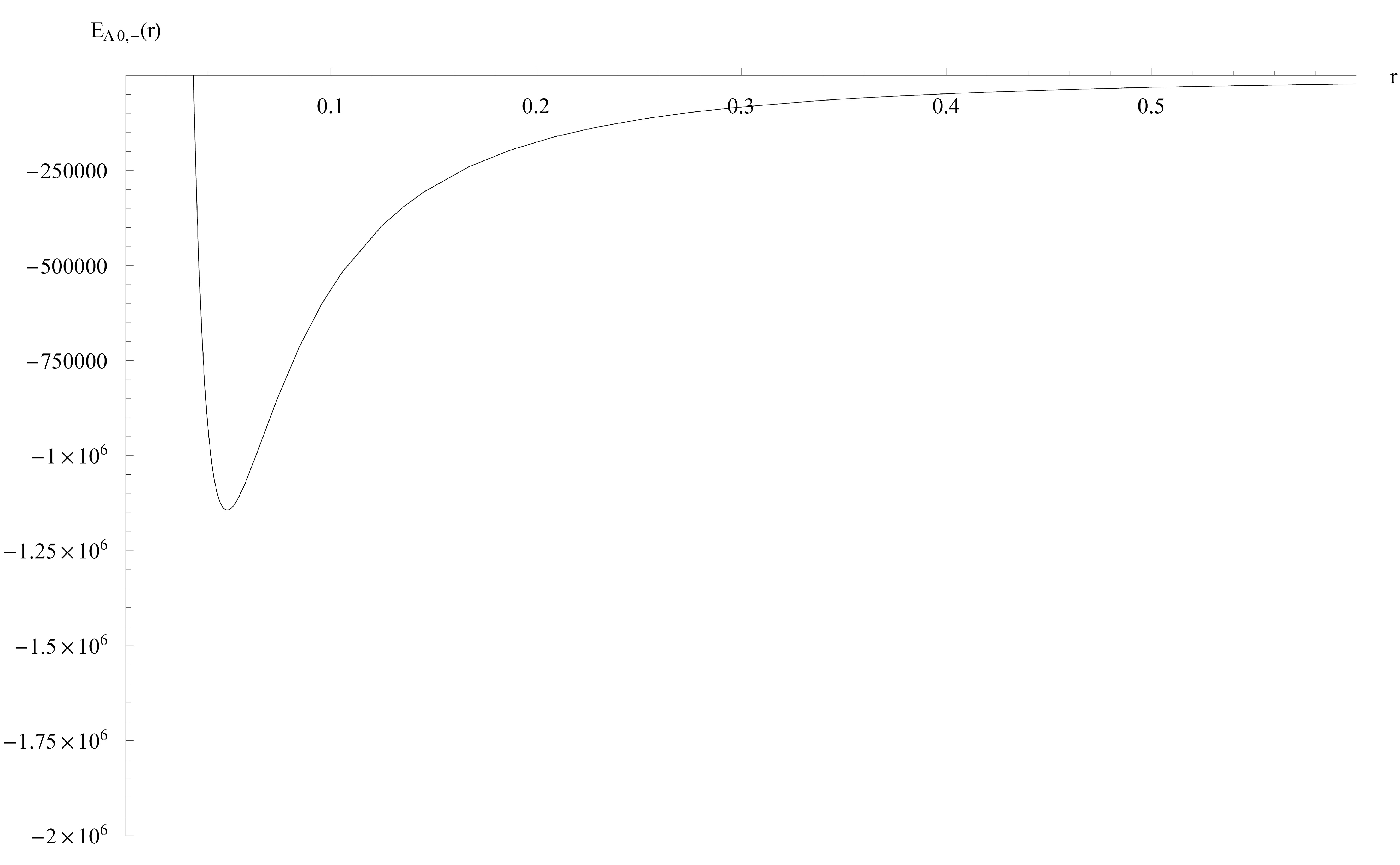}
\caption{Graphical representation of the DEUS binding energy for an electron and a positron in the presence of vacuum ($E_{\Lambda 0,\; -}$).}
\label{fermi4}
\end{figure}

\begin{figure}
\centering
\leavevmode
\includegraphics[width=14.5cm,height=9.5cm]{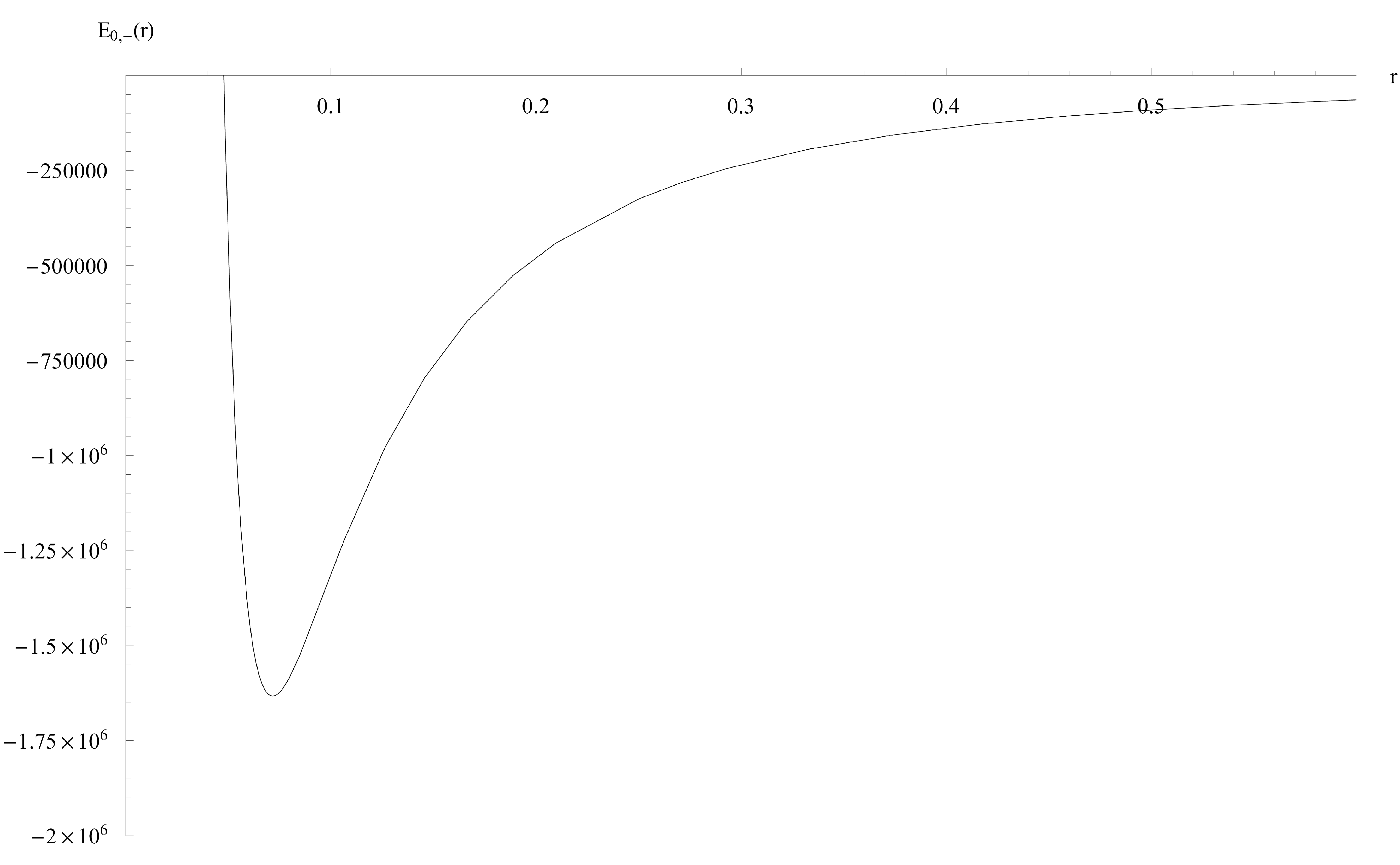}
\caption{Graphical representation of the DEUS binding energy for an electron and a positron in the absence of vacuum ($E_{0,\; -}$).}
\label{fermi3}
\end{figure}

\begin{figure}
\centering
\leavevmode
\includegraphics[width=14.5cm,height=9.5cm]{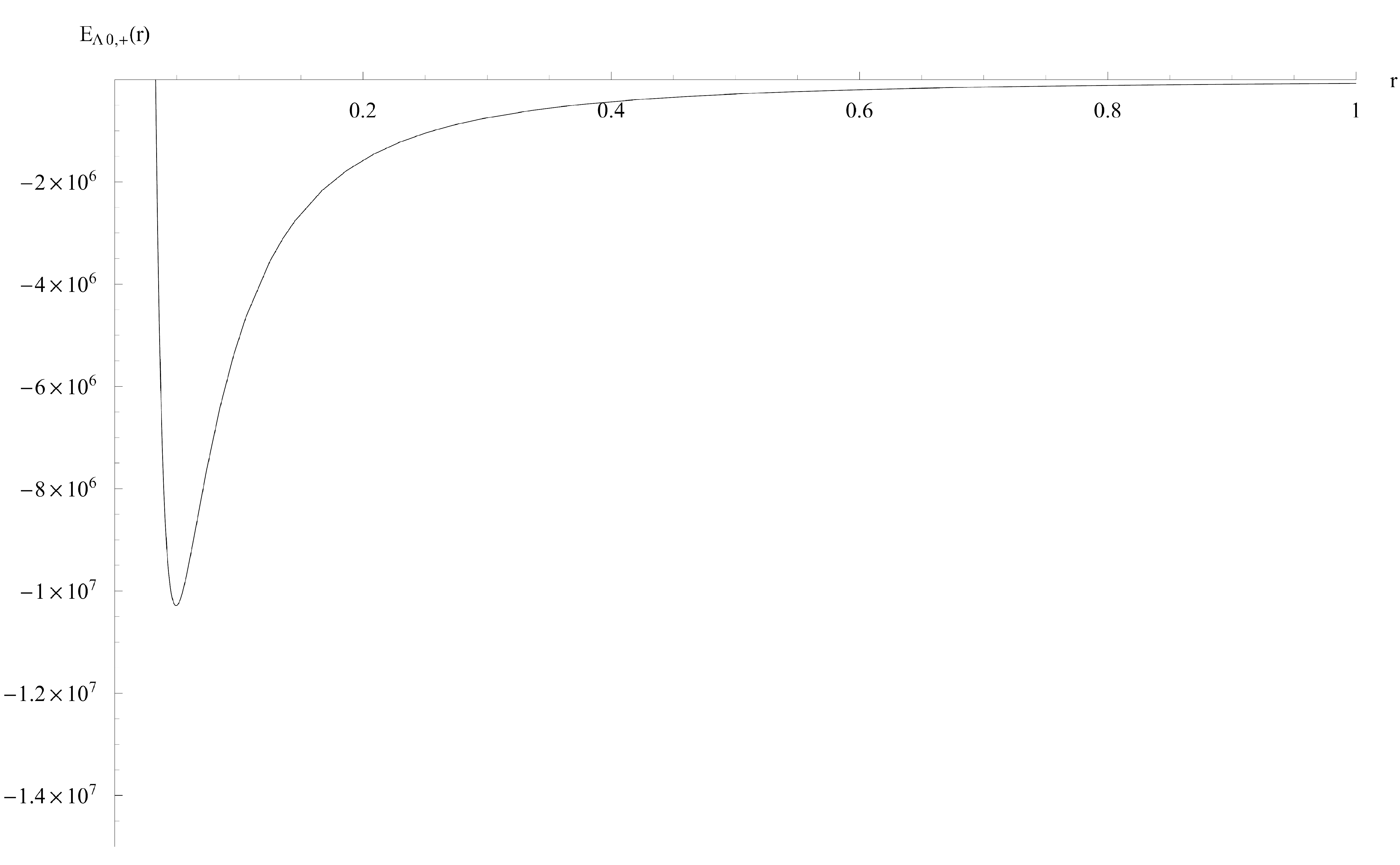}
\caption{Graphical representation of the DEUS binding energy for an electron and a positron in the presence of vacuum ($E_{\Lambda 0,\; +}$).}
\label{fermi2}
\end{figure}

\begin{figure}
\centering
\leavevmode
\includegraphics[width=14.5cm,height=9.5cm]{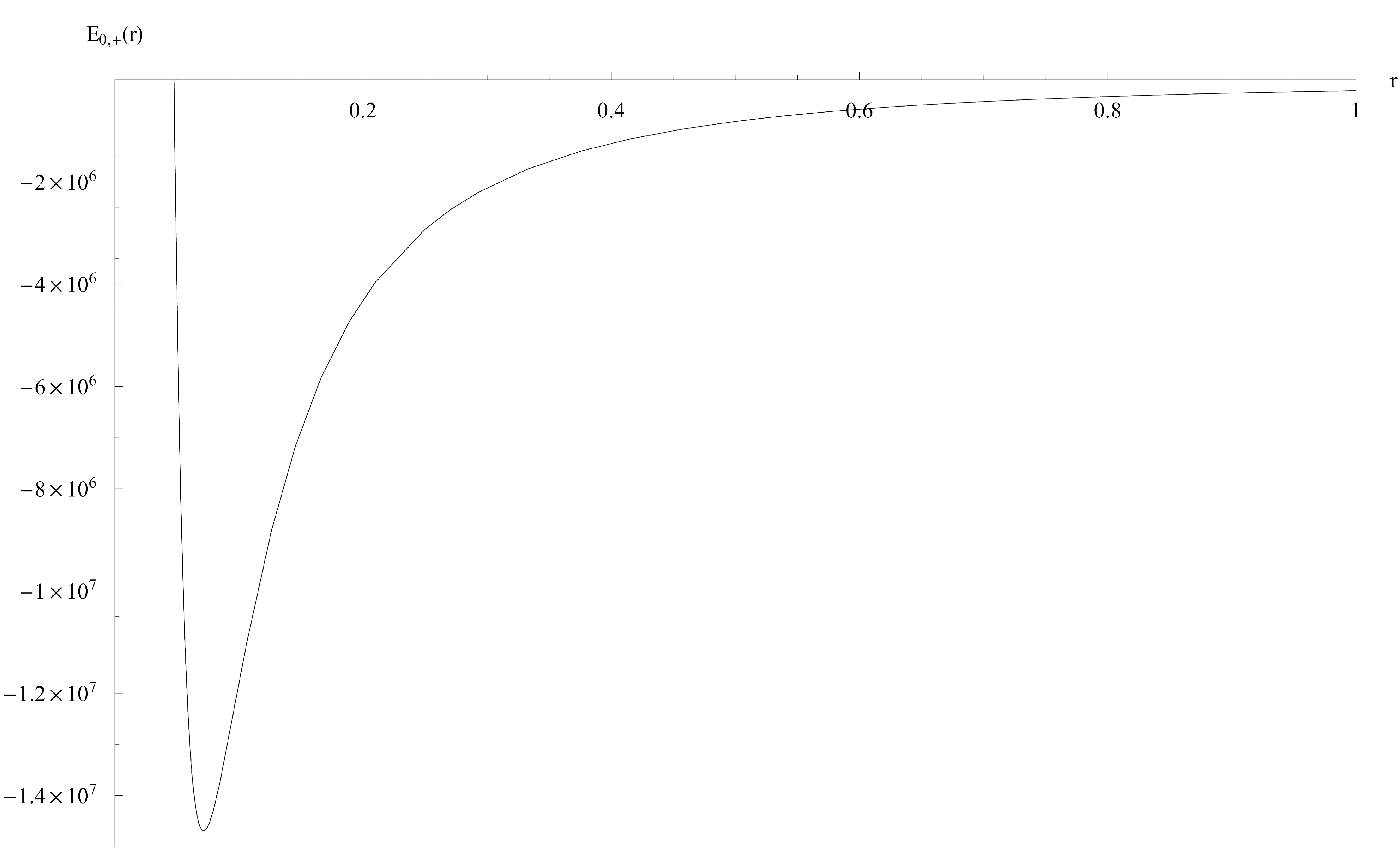}
\caption{Graphical representation of the DEUS binding energy for an electron and a positron in the absence of vacuum ($E_{0,\; +}$).}
\label{fermi1}
\end{figure}

The cases under discussion are $E_{\Lambda 0}(r)$ and $E_{0}(r)$, with {\it r} varying between 0 and 1, the limits for the FRW particle bubbles creation and annihilation.
These two energies are different if we use the up sign in our formulas (as, for example, in $\displaystyle \left(\frac{1}{2} - \frac{1}{4} \right)$) or if we use the down sign (as, for example, in $\displaystyle \left(\frac{1}{2} + \frac{1}{4} \right)$). We will note the energies containing the "up sign" with $E_{\Lambda 0,\; -}(r)$ (Fig. \ref{fermi4}) and $E_{0,\; -}(r)$ (Fig. \ref{fermi3}) and the ones with the "down sign" with $E_{\Lambda 0,\; +}(r)$ (Fig. \ref{fermi2}) and $E_{0,\; +}(r)$ (Fig. \ref{fermi1}). In these figures we represented only the electron and positron bound in the DEUS object's potential.

\begin{figure}
\centering
\leavevmode
\includegraphics[width=14.5cm,height=9.5cm]{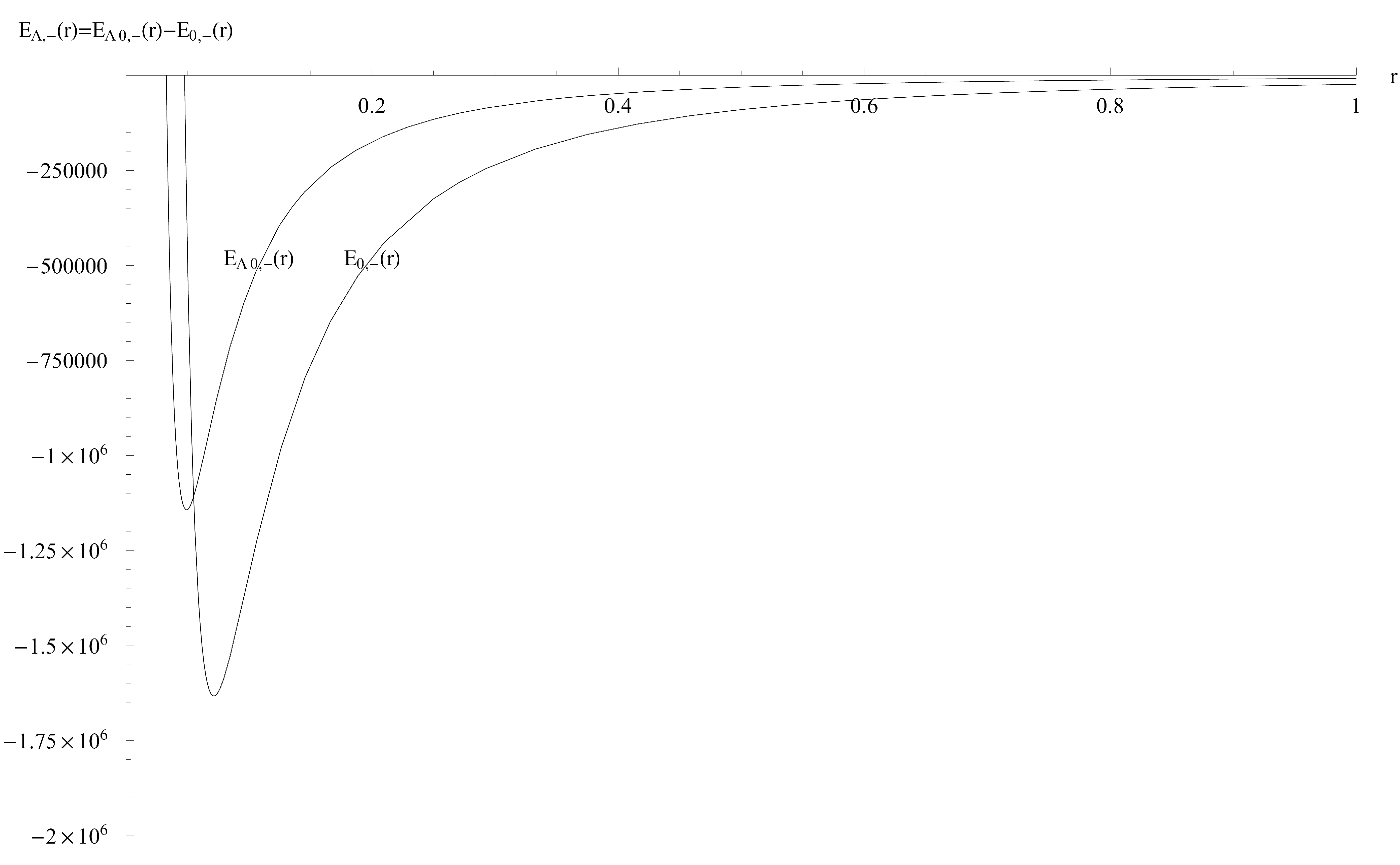}
\caption{Comparative representation of $E_{\Lambda 0,\; -}$ and $E_{0,\; -}$, their difference representing the vacuum energy.}
\label{grav1}
\end{figure}

\begin{figure}
\centering
\leavevmode
\includegraphics[width=14.5cm,height=9.5cm]{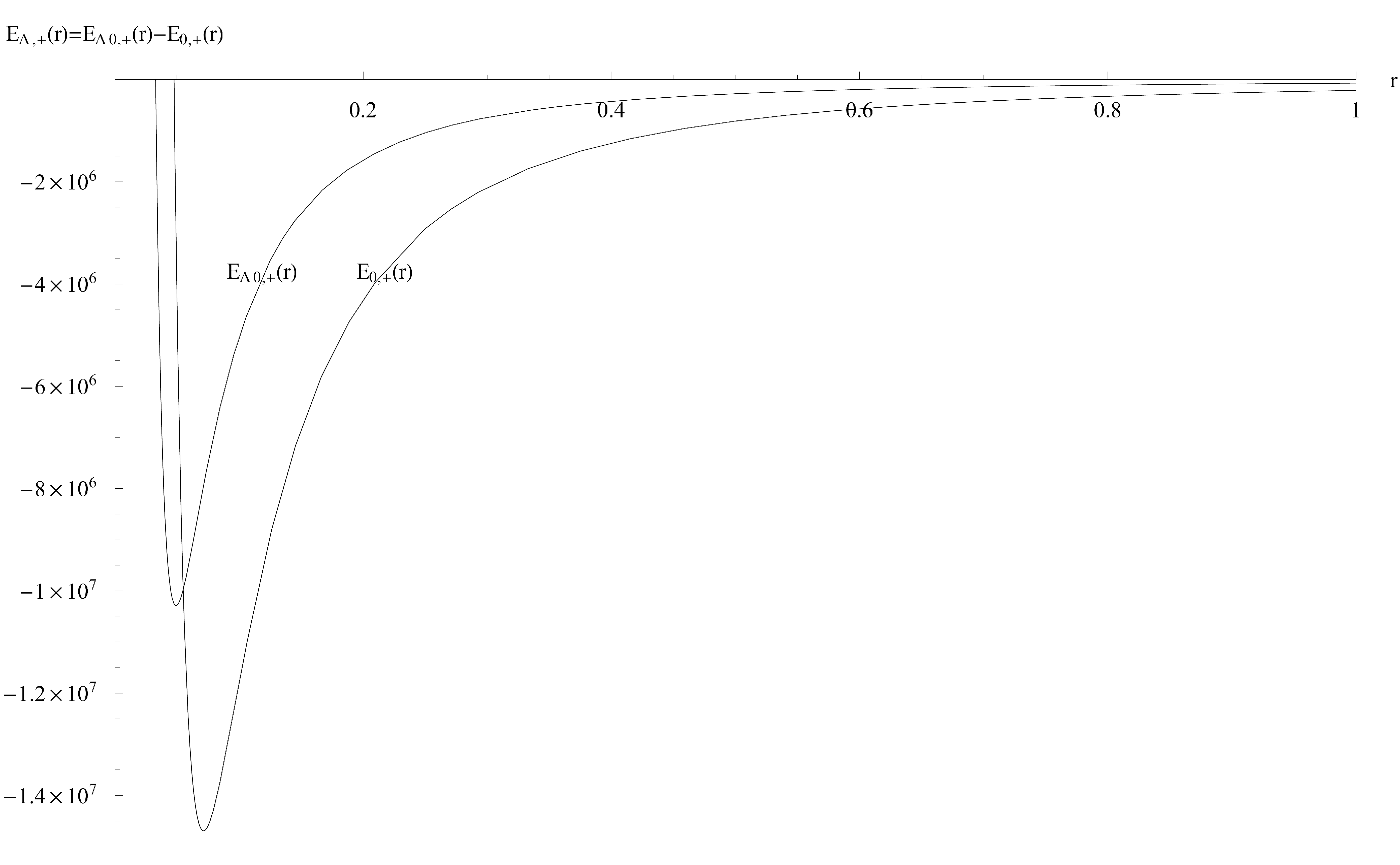}
\caption{Comparative representation of $E_{\Lambda 0,\; +}$ and $E_{0,\; +}$, their difference representing the vacuum energy.}
\label{grav}
\end{figure}

The difference between the interaction in the {\it pure helicoidal hyper-surface} and the BHIG spacetime will give the vacuum energy for "up" (Fig. \ref{grav1}):

\begin{equation}
E_{\Lambda,\; -}=E_{\Lambda 0,\; -}-E_{0,\; -}
\label{eq_71}
\end{equation}
\noindent
and "down" (Fig. \ref{grav}) signs:

\begin{equation}
E_{\Lambda,\; +}=E_{\Lambda 0,\; +}-E_{0,\; +}
\label{eq_72}
\end{equation}

Between the equilibrium energy for $E_{0,\; -}$ and the one for $E_{\Lambda 0,\; -}$, we have $E_{\Lambda,\; -}\simeq (1.632\times 10^{6}-1.143\times 10^{6})$ [GeV] $=489$ [TeV] (Fig. \ref{grav1}) while, between the equilibrium energy for $E_{0,\; +}$ and the one for $E_{\Lambda 0,\; +}$, we have $E_{\Lambda,\; +}\simeq (1.469\times 10^{7}-1.028\times 10^{7})$ [GeV] $=4410$ [TeV] (Fig. \ref{grav}). For observing these energies we should be able to achieve an experimental precision $\displaystyle R_{\Lambda}=\frac{\hbar c}{E_{\Lambda}}=\frac{1}{E_{\Lambda}}$ with $R_{\Lambda , -}\simeq 4.028\times 10^{-22}$ [m] for $E_{\Lambda,\; -}$ and $R_{\Lambda , +}\simeq 4.467\times 10^{-23}$ [m] for $E_{\Lambda,\; +}$. In other words, we have to reach these energies or lengths in order to enter in the "DEUS dimension".

The $\Delta r$ horizontal displacement between $E_{\Lambda 0,\; -}$ and $E_{0,\; -}$ in Fig. \ref{grav1} and between $E_{\Lambda 0,\; +}$ and $E_{0,\; +}$ in Fig. \ref{grav} is a spacelike particle generated in the process: {\bf a gravitational instanton}. In both Fig. \ref{grav1} and \ref{grav}, the dimension of the gravitational instanton is $\Delta r=0.0225\times r$, where {\it r} is the interaction action range. This gravitational instanton is the imaginary mass (Yamabe energy \cite{deus3}) of the catenoidal hyper-surface conjugate of the studied helicodal hyper-surface of the DEUS boson.

\section{Conclusions}

At the third DEUS level of self-similarity we determined the explicit wave function of an evaporated DEUS level, the Lagrangian of the observed Minkowski fermions (FRW particle bubbles) and their energies in the DEUS frame of the external observer perception. In the same way we determined the energy of a Minkowski boson.

The binding energy (two Minkowski fermions forming a Minkowski boson or a Minkowski boson forming two Minkowski fermions) difference between the fermions as part of the BHIG-FRW spacetime or of its DEUS static "image", the {\it pure helicoidal hyper-surface}, gave us the vacuum energy, particularized for an electron-positron interaction.

It is not by chance that we had chosen to study the case of equal mass particle-antiparticle pair interaction. The interaction of a matter with an antimatter FRW bubble, at any self-similarity level, should follow the same algorithm. For example, having the mass of the Universe and of the anti-Universe (Fig. 2 in \cite{deus2}) as $m_{0}$ and the right $A_{\mu}$ fields at this level, we should be able to "reconstruct" the pre-Big Bang DEUS object. The corresponding value for the vacuum energy increases from the $m_{0}$ mass of electrons to the mass of the FRW Universe. Also, the vacuum energy from levels two-four of self-similarity piles up on the background of vacuum energy from the creation of the FRW Universe, the result being a vacuum sea, differentiated (by energy) as function of DEUS level, but still the same through its nature.

\ack
Manuscript registered at the Astronomical Institute, Romanian Academy, record No. 257 from 04/19/07.


\title[D.E.U.S.]{X: Neutrinos}

\begin{abstract}
This paper is a try through which we intend to verify if we can obtain neutrino masses, in the Dirac or Majorana formalisms, with our previous results of the DEUS model. 
\end{abstract}


\section{Neutrino Mass?}

With the wave function of the DEUS object \cite{deus9}:

\begin{equation}
\Psi=\Psi_{0}\left(
\begin{array}{cccc}
e^{-i\omega x^{0}} \\
e^{-i\omega x^{0}} \\
1 \\
1
\end{array}
\right)
\label{eq_1}
\end{equation}
\noindent
where, at the collapse of the DEUS object, $x^{0}=\displaystyle \pm \textnormal{arctg} \left(\frac{t_{k}}{\phi_{k}}\right)$, and the Dirac matrix:

\begin{equation}
\gamma_{5}=\left( \begin{array}{cc}
I & 0 \\
0 & -I
\end{array} \right)
\label{eq_2}
\end{equation}
\noindent
with {\it I} the identity matrix, we can compute the chiral components of $\Psi$, for which:

\begin{equation}
\Psi=\Psi_{L}+\Psi_{R}
\label{eq_3}
\end{equation}
\noindent
and:

\begin{equation}
\Psi_{L}=\frac{1-\gamma_{5}}{2}\; \Psi \;\;\;\;\;\; , \;\;\;\;\;\; \Psi_{R}=\frac{1+\gamma_{5}}{2}\; \Psi
\label{eq_4}
\end{equation}

For our $\Psi$:

\begin{equation}
\Psi_{L}=\Psi_{0}\left(
\begin{array}{cccc}
0 \\
0 \\
1 \\
1
\end{array}
\right) \;\;\;\; , \;\;\;\;
\Psi_{R}=\Psi_{0}\left(
\begin{array}{cccc}
e^{-i\omega x^{0}} \\
e^{-i\omega x^{0}} \\
0 \\
0
\end{array}
\right)
\label{eq_5}
\end{equation}

The (\ref{eq_4}) components are also eigenstates of chirality, satisfying the $\gamma_{5}\Psi_{L}=-\Psi_{L}$ and $\gamma_{5}\Psi_{R}=\Psi_{R}$ relations.

In the QED Lagrangian, the Dirac mass term writes as:

\begin{equation}
\!\!\!\!\!\!\!\!\!\!\!\!\!\!\!\!\!\!\!\!{\cal L}_{D}\equiv -m_{D}\overline{\Psi}\Psi= -m_{D}(\overline{\Psi_{L}+\Psi_{R}})(\Psi_{L}+\Psi_{R})=-m_{D}\overline{\Psi}_{R}\Psi_{L}-m_{D}\overline{\Psi}_{L}\Psi_{R}=-m_{D}\overline{\Psi}_{R}\Psi_{L}+h.c.
\label{eq_6}
\end{equation}

Because the Dirac adjoind of $\Psi$ is \cite{deus9}:

\begin{equation}
\overline{\Psi}\equiv \left(\Psi^{*}_{1}\;\;\; \Psi^{*}_{2}\;\;\; -\Psi^{*}_{3}\;\;\; -\Psi^{*}_{4}\right)\; ,
\label{eq_7}
\end{equation}
\noindent
results:

\begin{equation}
\overline{\Psi}_{R}= \Psi_{0}^{*}\left(e^{i\omega x^{0}}\;\;\; e^{i\omega x^{0}}\;\;\; 0\;\;\; 0\right)\; .
\label{eq_8}
\end{equation}

With \cite{deus9}:

\begin{equation}
\Psi_{0}^{2}=\frac{1}{8\sqrt{3}\; \pi}\; (2\mp1)\; \frac{m_{helicoid}}{q^{2}}\; \frac{a^{2}}{r^{3}} \; ,
\label{eq_9}
\end{equation}
\noindent
we can compute that $\overline{\Psi}_{R}\Psi_{L}=0$, resulting:

\begin{equation}
{\cal L}_{D}=0
\label{eq_10}
\end{equation}
\noindent
and:

\begin{equation}
\overline{\Psi}_{R}\Psi_{L}=\overline{\Psi}_{L}\Psi_{R}=0 \; .
\label{eq_11}
\end{equation}

In the Majorana formalism, the Majorana mass terms (invariant under Lorentz transformations) are:

\begin{equation}
\begin{array}{cc}
{\cal L}_{L}^{M}=-m_{L}\Psi_{L}^{T}C\Psi_{L}+h.c. \\
{\cal L}_{R}^{M}=-m_{R}\Psi_{R}^{T}C\Psi_{R}+h.c.
\end{array}
\label{eq_12}
\end{equation}

The {\it C} matrix is obeying:

\begin{equation}
C\gamma_{\mu}C^{-1}=-\gamma_{\mu}^{T} \; .
\label{eq_13}
\end{equation}

In Dirac representation of $\gamma$'s:

\begin{equation}
\begin{array}{ccc}
C=i\gamma^{0}\gamma^{2} \\
C^{T}=C^{+}=-C \\
C^{+}C=CC^{+}=1
\end{array}
\label{eq_14}
\end{equation}

One often defines a charge-conjugated spinor:

\begin{equation}
\Psi^{C}\equiv C\overline{\Psi}^{T}=C\left(\Psi^{+}\gamma^{0} \right)^{T} \; .
\label{eq_15}
\end{equation}

In terms of it:

\begin{equation}
\begin{array}{cc}
{\cal L}_{L}^{M}=-m_{L}\overline{\left(\Psi^{C} \right)_{R}} \Psi_{L}+h.c. \\
{\cal L}_{R}^{M}=-m_{R}\overline{\left(\Psi^{C} \right)_{L}}\Psi_{R}+h.c.
\end{array}
\label{eq_16}
\end{equation}

From (\ref{eq_5}), we have:

\begin{equation}
\begin{array}{cc}
\Psi^{T}_{L}=\Psi_{0}\left(0\;\;\; 0\;\;\; 1\;\;\; 1\right) \\
\Psi^{T}_{R}=\Psi_{0}\left(e^{-i\omega x^{0}}\;\;\; e^{-i\omega x^{0}}\;\;\; 0\;\;\; 0\right)
\end{array}
\label{eq_17}
\end{equation}
\noindent
with which, results:

\begin{equation}
\begin{array}{cc}
\Psi^{T}_{L}C\Psi_{L}=0 \\
\Psi^{T}_{R}C\Psi_{R}=0
\end{array}
\label{eq_18}
\end{equation}
\noindent
meaning that, in the first approximation (without higher correction terms):

\begin{equation}
{\cal L}_{L}^{M}={\cal L}_{R}^{M}=0 \; .
\label{eq_19}
\end{equation}

The most general mass term for a single neutrino flavor is:

\begin{equation}
-{\cal L}_{m}=\frac{1}{2}\; m_{L}\; \nu_{L}^{T}\; C\; \nu_{L}+\frac{1}{2}\; m_{R}^{*}\; \nu_{R}^{T}\; C\; \nu_{R}+m_{D}\; \overline{\nu}_{R}\; \nu_{L}+h.c.
\label{eq_20}
\end{equation}
\noindent
where the first two terms from the right side are the Majorana masses, while the third is the Dirac mass term.

(\ref{eq_20}) can be written in a convenient matrix by using the identities:

\begin{equation}
\begin{array}{cc}
\left(\nu_{R}^{T}\; C\; \nu_{R} \right)^{T}=\left(\nu^{C} \right)_{L}^{T}\; C\; \left(\nu^{C} \right)_{L} \\
\overline{\nu}_{R}\; \nu_{L}=\left(\nu^{C} \right)_{L}^{T}\; C\; \nu_{L}=\nu_{L}^{T}\; C\; \left(\nu^{C} \right)_{L}
\end{array}
\label{eq_21}
\end{equation}

Then:

\begin{equation}
\begin{array}{cc}
\displaystyle -{\cal L}_{m}=\frac{1}{2}\left(\nu_{L}^{T}\;\;\; \left(\nu^{C} \right)_{L}^{T} \right)C\left(
\begin{array}{cc}
m_{L} & m_{D} \\
m_{D} & m_{R}
\end{array} \right) \left(
\begin{array}{cc}
\nu_{L} \\
\left(\nu^{C} \right)_{L}
\end{array} \right) +h.c.= \\
\displaystyle =\frac{1}{2}\left(\overline{\left(\nu^{C}\right)_{R}}\;\;\; \overline{\nu_{R}} \right)\left(
\begin{array}{cc}
m_{L} & m_{D} \\
m_{D} & m_{R}
\end{array} \right) \left(
\begin{array}{cc}
\nu_{L} \\
\left(\nu^{C} \right)_{L}
\end{array} \right)+h.c.
\end{array}
\label{eq_22}
\end{equation}
\noindent
where:

\begin{equation}
M_{\nu}=\left(
\begin{array}{cc}
m_{L} & m_{D} \\
m_{D} & m_{R}
\end{array} \right)
\label{eq_23}
\end{equation}
\noindent
is the neutrino mass (symmetric) matrix.

The effects of the fields on $\nu$ and $\overline{\nu}$ can be summarized as in table \ref{unu}.

\begin{table}[ht]
\begin{center}
\begin{tabular}{ccc} \hline\hline
\noalign{\smallskip}
Field & Effect on $\nu$ & Effect on $\overline{\nu}$ \\
\noalign{\smallskip} \hline
\noalign{\smallskip}
$\Psi$ & annihilates & creates \\
$\Psi^{C}$ & creates & annihilates \\
$\overline{\Psi}$ & creates & annihilates \\
$\overline{\Psi^{C}}$ & annihilates & creates \\
\noalign{\smallskip} \hline
\noalign{\smallskip}
\end{tabular}
\label{unu}
\end{center}
\end{table}

In conlusion, because, in the first approximation, also ${\cal L}_{m}=0$, it is impossible to say something about $m_{D}$ or about $m_{L}$ and $m_{R}$. For the neutrino existence, when $\overline{\nu}_{R}\nu_{L}\neq 0$, $\nu_{L}^{T}C\nu_{L}\neq 0$ and $\nu_{R}^{T}C\nu_{R}\neq 0$, we must have $m_{D}=0$ (for Dirac neutrinos) and $m_{L}=m_{R}=0$ (for Majorana neutrinos).

\section{Conclusions}

We tried to see if its possible, at this development stage of the DEUS model, to determine the mass of neutrinos. Our conclusion is that, in both Dirac and Majorana formalisms, the masses of neutrinos are zero. Later, in another DEUS paper, we will show that the neutrino, if bound in the catenoids of the third DEUS self-similarity level (three flavors for each real state of the catenoid between equilibrium and collapse and, for $\alpha=3$, a mixed state - see section 7 in \cite{deus2}), will have mass and, also, flavors clearly separated. If the third self-similarity DEUS objects are seen as evaporated in the BHIG helicoid of second or first self-similarity DEUS level, the "free" neutrinos are having mixed flavors. They still have flavors because having the third level evaporated does not mean that also the second or the first self-similarity DEUS level is evaporated.

\ack
Manuscript registered at the Astronomical Institute, Romanian Academy, record No. 258 from 04/19/07.


\title[D.E.U.S.]{XII: Scalar Fields}

\begin{abstract}
With the generators determined for the SU(2) and SU(3) symmetries, we will analyze the situations that minimize the action for a scalar DEUS field, as function of the chosen covariant derivative. We will also obtain detailed and differentiated expressions for the $A_{\mu}$ components, depending at which self-similarity DEUS level we are looking and if this level is evaporated or not.
\end{abstract}


\section{Electromagnetic Field Trapped in the Curvature}

The DEUS objects trapped inside the non-evaporated self-similar DEUS level three or four will not have an electromagnetic component (as seen from level one of self-similarity). The electromagnetic component is translated in curvature of the DEUS hyper-surfaces, being released together with the other fields in the external observer's FRW spacetime at the DEUS level evaporation. Then, inside the non-evaporated catenoid of self-similarity level three (or four), the Lagrangian of the DEUS objects, that will be seen later by the external observer as DEUS fermion FRW bubbles (${\cal L}_{F}$ in \cite{deus9}), will be ${\cal L}_{F}={\cal L}(\Psi,\textnormal{{\bf A}})$, the full Lagrangian for charged Dirac fermions in interaction with the electromagnetic field from evaporated (Big Bang) level one of self-similarity. The terms representing the interaction with the electromagnetic field of the evaporated level three (or four) of self-similarity will be zero:

\begin{equation}
-\textnormal{{\bf E}}\; \textnormal{\bf {\.{A}}}-\frac{1}{2}\left(\textnormal{{\bf E}}^{2}+\textnormal{{\bf B}}^{2} \right)=0\; .
\label{eq_1}
\end{equation}

The above relation together with the relation between {\it const} and the rest mass $m_{0}$ gives:

\begin{equation}
const=\sqrt{2}\; \frac{1}{\displaystyle \left(\frac{1}{2}\mp \frac{1}{4} \right)^{2}}\left[4\left(\frac{1}{2}\mp \frac{1}{4} \right)^{2}+\left(\sqrt{2}+1 \right)^{2} \right]\frac{1}{\cosh \alpha_{k}-\sinh \alpha_{k}}\frac{a^{2}}{r^{2}}\; ,
\label{eq_2}
\end{equation}
\noindent
where, instead of the angle $\pi/4$ under which the external observer sees the effect as coming from the evaporated DEUS object, we used the angle $\alpha_{k}$ \cite{deus5} for the helicoid of the objects embedded in the catenoid of the non-collapsed DEUS level. Also, the non-constant behavior of {\it const} is related to the $m_{0}$ dependence of {\it a} and {\it r}. (\ref{eq_2}) makes the \cite{deus8} magnetic vector potential and the electric scalar potential to be:

\begin{equation}
\begin{array}{cccc}
f\stackrel{not}{=}\Phi=\displaystyle \frac{1}{8}\left[4\left(\frac{1}{2}\mp \frac{1}{4} \right)+\frac{\left(\sqrt{2}+1 \right)^{2}}{\displaystyle \frac{1}{2}\mp \frac{1}{4}}\right]\frac{m_{helicoid}}{q}\; \frac{a^{2}}{r^{2}} \\
A_{1}=\displaystyle -i\; \frac{\sqrt{2}}{8}\; \left[4\left(\frac{1}{2}\mp \frac{1}{4} \right)+\frac{\left(\sqrt{2}+1 \right)^{2}}{\displaystyle \frac{1}{2}\mp \frac{1}{4}}\right]\frac{m_{helicoid}}{q}\; \frac{1}{1-\tanh \alpha_{k}}\; \frac{a^{2}}{r^{2}} \\
A_{2}=\displaystyle \mp i\; \frac{\sqrt{2}}{8}\; \left[4\left(\frac{1}{2}\mp \frac{1}{4} \right)+\frac{\left(\sqrt{2}+1 \right)^{2}}{\displaystyle \frac{1}{2}\mp \frac{1}{4}}\right]\frac{m_{helicoid}}{q}\; \frac{1}{1+\tanh \alpha_{k}}\; \frac{a^{2}}{r^{2}} \\
A_{3}=\displaystyle \pm i\; \frac{1}{8}\; \left[4\left(\frac{1}{2}\mp \frac{1}{4} \right)+\frac{\left(\sqrt{2}+1 \right)^{2}}{\displaystyle \frac{1}{2}\mp \frac{1}{4}}\right]\frac{m_{helicoid}}{q}\; \cosh \alpha_{k}\; \frac{a^{2}}{r^{2}}\; .
\end{array}
\label{eq_3}
\end{equation}

\section{DEUS Scalar Fields}

For the external observer, the dynamics of the scalar component of the non-collapsed catenoid's DEUS objects (and non-interacting), observed from the external spacetime, will be described by the action (in {\it r} and $t_{FRW}$ coordinates):

\begin{equation}
S=\int{d^{4}x\; \sqrt{-g}}\left\{\frac{1}{2}\; g^{\mu\nu}\; {\cal D}_{\mu}\Phi\; {\cal D}_{\nu}\Phi-V(\Phi) \right\}\; ,
\label{eq_4}
\end{equation}
\noindent
where for the BHIG-FRW helicoid:

\begin{equation}
g^{\mu\nu}=\frac{4}{3}\left(
\begin{array}{cc}
-1 & 0 \\
0 & 1 \\
\end{array}
\right)
\label{eq_5}
\end{equation}

The gauge transformation of $A_{\mu}$ field under the SU(2) or the SU(3) symmetry \cite{deus11} requires, instead of the $\partial_{\mu}$ derivative, the covariant derivative ${\cal D}_{\mu}$:

\begin{equation}
\begin{array}{ccc}
\displaystyle {\cal D}_{\mu}\equiv \partial_{\mu}+i\; \frac{q}{\hbar c}\; c_{i}\; A_{\mu}=\partial_{\mu}+i\; q\; c_{i}\; A_{\mu}  \\ \\
\displaystyle {\cal D}_{\mu}\equiv \partial_{\mu}+i\; \frac{q}{\hbar c}\; \lambda_{i}\; A_{\mu}=\partial_{\mu}+i\; q\; \lambda_{i}\; A_{\mu}\; ,
\end{array}
\label{eq_6}
\end{equation}
\noindent
written in natural units, $\hbar=c=1$, with \cite{deus11} $c_{i}$ (for SU(2)) and $\lambda_{i}$ (for SU(3)) generators, and the (\ref{eq_3}) $A_{\mu}$'s.

For a flat (empty) spacetime ${\cal D}_{\mu}\equiv \partial_{\mu}$. In this situation, when we minimise the (\ref{eq_4}) action $\delta S=0$, we obtain:

\begin{equation}
-\frac{1}{2}\; \ddot{\Phi}+\frac{1}{2}\; \frac{\partial}{\partial r}\; \frac{\partial \Phi}{\partial r}-V'(\Phi)=0\; .
\label{eq_7}
\end{equation}

With (\ref{eq_3}), we can rewrite (\ref{eq_7}) as:

\begin{equation}
\frac{1}{8}\left[4\left(\frac{1}{2}\mp \frac{1}{4} \right)+\frac{\left(\sqrt{2}+1 \right)^{2}}{\displaystyle \frac{1}{2}\mp \frac{1}{4}}\right]\frac{m_{helicoid}}{q}\left[-\frac{\dot{a}^{2}+a\; \ddot{a}}{r^{2}}+3\; \frac{a^{2}}{r^{4}} \right]-V'(\Phi)=0\; .
\label{eq_8}
\end{equation}

When $V(\Phi)$ is constant, in (\ref{eq_8}):

\begin{equation}
\dot{a}^{2}+a\; \ddot{a}-3\; \frac{a^{2}}{r^{2}}=0\; .
\label{eq_9}
\end{equation}

From the Friedmann equations for BHIG-FRW spacetime we saw that \cite{deus2}:

\begin{equation}
a\; \ddot{a}-a^{2}-2\; \dot{a}^{2}=0\; .
\label{eq_10}
\end{equation}

Then, with (\ref{eq_9}) and (\ref{eq_10}), in the case of empty space, results:

\begin{equation}
\frac{a^{2}}{r^{2}}=a^{2}\left(H_{0}^{2}+\frac{1}{3} \right)\; ,
\label{eq_11}
\end{equation}
\noindent
where $H_{0}$ is the Hubble constant local particularization of the Hubble parameter $\displaystyle H\equiv \frac{\dot{a}}{a}$.

Now, for a non-empty spacetime, the objects that populate it have the metric tensor of their BHIG-FRW hyper-surface described by:

\begin{equation}
g^{\mu\nu}=\frac{4}{3}\left(
\begin{array}{cc}
1 & 0 \\
0 & -1 \\
\end{array}
\right)
\label{eq_12}
\end{equation}
\noindent
their space coordinate being rotated to time (and reciprocal) from the (empty) spacetime in which they are embedded.

For SU(2), we can write:

\begin{equation}
\displaystyle {\cal D}_{\mu}=\partial_{\mu}+i\; q\; c_{i}\; \left(
\begin{array}{ll}
A_{j} \\
A_{k}
\end{array}
\right)\; .
\label{eq_13}
\end{equation}

In this situation, $\delta S=0$ gives a system of two equations that has solution only if:

\begin{equation}
\displaystyle {\cal D}_{\mu}=\partial_{\mu}+i\; q\; c_{i}\; \left(
\begin{array}{cc}
\Phi \\
A_{3}
\end{array}
\right)\; ,
\label{eq_14}
\end{equation}
\noindent
or:

\begin{equation}
\displaystyle {\cal D}_{\mu}=\partial_{\mu}+i\; q\; c_{i}\; \left(
\begin{array}{cc}
A_{3} \\
\Phi
\end{array}
\right)\; ,
\label{eq_15}
\end{equation}
\noindent
when:

\begin{equation}
\begin{array}{ccc}
\displaystyle H_{0}+\frac{1}{r}=0 \\
\alpha_{k}=0 \\
V'(\Phi)=0
\end{array}
\label{eq_16}
\end{equation}

For SU(3), we have:

\begin{equation}
\displaystyle {\cal D}_{\mu}=\partial_{\mu}+i\; q\; \lambda_{i}\; \left(
\begin{array}{ccc}
A_{j} \\
A_{k} \\
A_{l}
\end{array}
\right)\; .
\label{eq_17}
\end{equation}

In the same way as we analyzed the SU(2) situation, from $\delta S=0$ results a system of three equations that has solution when we have:

\begin{enumerate}
\item
For $\lambda_{7}$ and {\it sin} and {\it cos} (catenoid) instead of {\it sinh} and {\it cosh} (helicoid) in $A_{\mu}$:

\begin{equation}
\displaystyle {\cal D}_{\mu}=\partial_{\mu}+i\; q\; \lambda_{i}\; \left(
\begin{array}{ccc}
A_{1} \\
A_{3} \\
A_{2}
\end{array}
\right)
\label{eq_18}
\end{equation}
\noindent
and:

\begin{equation}
\begin{array}{cc}
\displaystyle \alpha=\frac{\pi}{4} \\
\displaystyle H_{0}+\frac{1}{r}=0
\end{array} \; ;
\label{eq_19}
\end{equation}

\item
For $\lambda_{7}$, $\alpha_{k}=0$, and {\it sinh} and {\it cosh} (helicoid) in $A_{\mu}$:
\begin{equation}
\displaystyle {\cal D}_{\mu}=\partial_{\mu}+i\; q\; \lambda_{i}\; \left(
\begin{array}{ccc}
A_{1} \\
A_{3} \\
\Phi
\end{array}
\right)\; ,
\label{eq_20}
\end{equation}
\noindent
or:

\begin{equation}
\displaystyle {\cal D}_{\mu}=\partial_{\mu}+i\; q\; \lambda_{i}\; \left(
\begin{array}{ccc}
A_{1} \\
\Phi \\
A_{3}
\end{array}
\right)\; ,
\label{eq_21}
\end{equation}
\noindent
or:

\begin{equation}
\displaystyle {\cal D}_{\mu}=\partial_{\mu}+i\; q\; \lambda_{i}\; \left(
\begin{array}{ccc}
A_{2} \\
A_{3} \\
\Phi
\end{array}
\right)\; ,
\label{eq_22}
\end{equation}
\noindent
or:

\begin{equation}
\displaystyle {\cal D}_{\mu}=\partial_{\mu}+i\; q\; \lambda_{i}\; \left(
\begin{array}{ccc}
A_{2} \\
\Phi \\
A_{3}
\end{array}
\right)\; ;
\label{eq_23}
\end{equation}

\item
For $\lambda_{2}$ or $\lambda_{3}$, and {\it sin} and {\it cos} (catenoid) in $A_{\mu}$:
\begin{equation}
\displaystyle {\cal D}_{\mu}=\partial_{\mu}+i\; q\; \lambda_{i}\; \left(
\begin{array}{ccc}
A_{1} \\
\Phi \\
A_{3}
\end{array}
\right)\; ,
\label{eq_24}
\end{equation}
\noindent
where $\displaystyle \alpha=-\frac{\pi}{4}$ and:

\begin{equation}
\begin{array}{cc}
A_{2}=\displaystyle - i\; \frac{\sqrt{2}}{8}\; \left[4\left(\frac{1}{2}- \frac{1}{4} \right)+\frac{\left(\sqrt{2}+1 \right)^{2}}{\displaystyle \frac{1}{2}- \frac{1}{4}}\right]\frac{m_{helicoid}}{q}\; \frac{1}{1+\textnormal{tg}\; \alpha}\; \frac{a^{2}}{r^{2}} \\
A_{3}=\displaystyle - i\; \frac{1}{8}\; \left[4\left(\frac{1}{2}+ \frac{1}{4} \right)+\frac{\left(\sqrt{2}+1 \right)^{2}}{\displaystyle \frac{1}{2}+ \frac{1}{4}}\right]\frac{m_{helicoid}}{q}\; \cos \alpha\; \frac{a^{2}}{r^{2}}\; ,
\end{array}
\label{eq_25}
\end{equation}
\noindent
or:

\begin{equation}
\displaystyle {\cal D}_{\mu}=\partial_{\mu}+i\; q\; \lambda_{i}\; \left(
\begin{array}{ccc}
A_{2} \\
\Phi \\
A_{3}
\end{array}
\right)\; ,
\label{eq_26}
\end{equation}
\noindent
where $\displaystyle \alpha=\frac{\pi}{4}$ and:

\begin{equation}
\begin{array}{cc}
A_{2}=\displaystyle i\; \frac{\sqrt{2}}{8}\; \left[4\left(\frac{1}{2}+ \frac{1}{4} \right)+\frac{\left(\sqrt{2}+1 \right)^{2}}{\displaystyle \frac{1}{2}+ \frac{1}{4}}\right]\frac{m_{helicoid}}{q}\; \frac{1}{1+\textnormal{tg}\; \alpha}\; \frac{a^{2}}{r^{2}} \\
A_{3}=\displaystyle i\; \frac{1}{8}\; \left[4\left(\frac{1}{2}- \frac{1}{4} \right)+\frac{\left(\sqrt{2}+1 \right)^{2}}{\displaystyle \frac{1}{2}- \frac{1}{4}}\right]\frac{m_{helicoid}}{q}\; \cos \alpha\; \frac{a^{2}}{r^{2}}\; ;
\end{array}
\label{eq_27}
\end{equation}

\item
For $\lambda_{2}$ or $\lambda_{3}$, and {\it sinh} and {\it cosh} (helicoid) in $A_{\mu}$:
\begin{equation}
\displaystyle {\cal D}_{\mu}=\partial_{\mu}+i\; q\; \lambda_{i}\; \left(
\begin{array}{ccc}
A_{1} \\
A_{3} \\
A_{2}
\end{array}
\right)\; ,
\label{eq_28}
\end{equation}
\noindent
where $\displaystyle \alpha_{k}=0$ and:

\begin{equation}
\begin{array}{cc}
A_{2}=\displaystyle i\; \frac{\sqrt{2}}{8}\; \left[4\left(\frac{1}{2}+ \frac{1}{4} \right)+\frac{\left(\sqrt{2}+1 \right)^{2}}{\displaystyle \frac{1}{2}+ \frac{1}{4}}\right]\frac{m_{helicoid}}{q}\; \frac{1}{1+\tanh \alpha_{k}}\; \frac{a^{2}}{r^{2}} \\
A_{3}=\displaystyle \pm i\; \frac{1}{8}\; \left[4\left(\frac{1}{2}\mp \frac{1}{4} \right)+\frac{\left(\sqrt{2}+1 \right)^{2}}{\displaystyle \frac{1}{2}\mp \frac{1}{4}}\right]\frac{m_{helicoid}}{q}\; \cosh \alpha_{k}\; \frac{a^{2}}{r^{2}}\; ;
\end{array}
\label{eq_29}
\end{equation}

\item
For $\lambda_{4}$ or $\lambda_{5}$, and {\it sin} and {\it cos} (catenoid) in $A_{\mu}$:
\begin{equation}
\displaystyle {\cal D}_{\mu}=\partial_{\mu}+i\; q\; \lambda_{i}\; \left(
\begin{array}{ccc}
A_{1} \\
A_{3} \\
\Phi
\end{array}
\right)\; ,
\label{eq_30}
\end{equation}
\noindent
or:

\begin{equation}
\displaystyle {\cal D}_{\mu}=\partial_{\mu}+i\; q\; \lambda_{i}\; \left(
\begin{array}{ccc}
A_{2} \\
A_{3} \\
\Phi
\end{array}
\right)\; ,
\label{eq_31}
\end{equation}
\noindent
where $\displaystyle \alpha=0$ or, in {\it sinh} and {\it cosh} (helicoid) in $A_{\mu}$:

\begin{equation}
\displaystyle {\cal D}_{\mu}=\partial_{\mu}+i\; q\; \lambda_{i}\; \left(
\begin{array}{ccc}
A_{1} \\
\Phi \\
A_{3}
\end{array}
\right)\; ,
\label{eq_32}
\end{equation}
\noindent
or:

\begin{equation}
\displaystyle {\cal D}_{\mu}=\partial_{\mu}+i\; q\; \lambda_{i}\; \left(
\begin{array}{ccc}
A_{2} \\
\Phi \\
A_{3}
\end{array}
\right)\; ,
\label{eq_33}
\end{equation}
\noindent
where $\displaystyle \alpha_{k}=0$.
\end{enumerate}

\section{Conclusions}

{\bf a)} The global representation of the FRW Universe bubble will be characterized by the $\lambda_{7}$, $\lambda_{2}$, $\lambda_{3}$ and $\lambda_{1}$, $\lambda_{6}$, $\lambda_{8}$ generators, where with $\displaystyle \alpha=\frac{\pi}{4}$ we can describe the {\bf matter Universe} (with the $A_{\mu}$ fields in {\it sin} and {\it cos}), or, with $\alpha_{k}=0$ we can describe the {\bf pre-Big Bang Universe} DEUS object (with the $A_{\mu}$ fields in {\it sinh} and {\it cosh}).

\noindent
{\bf b)} The local representation of the Universe, containing its global representation, will be characterized by the $\lambda_{2}$, $\lambda_{3}$, $\lambda_{4}$, $\lambda_{5}$, $\lambda_{7}$ and $\lambda_{0}\equiv \lambda_{1}$, $\lambda_{6}$, or $\lambda_{8}$ generators, where with $\alpha_{k}=0$ we can describe the {\bf interaction} of the {\bf FRW Universe} bubble with the {\bf FRW Anti-Universe} bubble, or, with $\displaystyle \alpha=\frac{\pi}{4}$ their {\bf particle} contents and with $\displaystyle \alpha=-\frac{\pi}{4}$ their {\bf antiparticle} contents in non-evaporated self-similarity level three or four.

\noindent
{\bf c)} Local representation, characterized by $\lambda_{7}$, $\lambda_{4}$, $\lambda_{5}$ and $\lambda_{0}\equiv \lambda_{1}$, $\lambda_{6}$, or $\lambda_{8}$ generators, where with $\alpha=0$ we can describe the {\bf free particles and antiparticles in Universe and Anti-Universe}.

The covariant derivative for the {\bf a)} situation is:

\begin{equation}
\displaystyle {\cal D}_{\mu}=\partial_{\mu}+i\; q\; \lambda_{i}\; \left(
\begin{array}{ccc}
A_{1} \\
A_{3} \\
A_{2}
\end{array}
\right)\; .
\label{eq_34}
\end{equation}

For the {\bf b)} situation with $\displaystyle \alpha=-\frac{\pi}{4}$ ({\bf antiparticles}) and $\lambda_{2}$, $\lambda_{3}$, $\lambda_{7}$, $\lambda_{0}$, the covariant derivative is:

\begin{equation}
\displaystyle {\cal D}_{\mu}=\partial_{\mu}+i\; q\; \lambda_{i}\; \left(
\begin{array}{ccc}
A_{1} \\
\Phi \\
A_{3}
\end{array}
\right)\; .
\label{eq_35}
\end{equation}

For the {\bf b)} situation with $\displaystyle \alpha=\frac{\pi}{4}$ ({\bf particles}) and $\lambda_{2}$, $\lambda_{3}$, $\lambda_{7}$, $\lambda_{0}$, the covariant derivative is:

\begin{equation}
\displaystyle {\cal D}_{\mu}=\partial_{\mu}+i\; q\; \lambda_{i}\; \left(
\begin{array}{ccc}
A_{2} \\
\Phi \\
A_{3}
\end{array}
\right)\; .
\label{eq_36}
\end{equation}

For the {\bf b)} situation with $\alpha_{k}=0$ ({\bf annihilation} of Universe with Anti-Universe or, of particles with antiparticles) and $\lambda_{4}$, $\lambda_{5}$, $\lambda_{0}$, the covariant derivative is:

\begin{equation}
\displaystyle {\cal D}_{\mu}=\partial_{\mu}+i\; q\; \lambda_{i}\; \left(
\begin{array}{ccc}
A_{1} \\
\Phi \\
A_{3}
\end{array}
\right)\; ,
\label{eq_37}
\end{equation}
\noindent
or:

\begin{equation}
\displaystyle {\cal D}_{\mu}=\partial_{\mu}+i\; q\; \lambda_{i}\; \left(
\begin{array}{ccc}
A_{2} \\
\Phi \\
A_{3}
\end{array}
\right)\; .
\label{eq_38}
\end{equation}

For the {\bf c)} situation with $\alpha=0$ ({\bf annihilation} of particles with antiparticles), the covariant derivative is:

\begin{equation}
\displaystyle {\cal D}_{\mu}=\partial_{\mu}+i\; q\; \lambda_{i}\; \left(
\begin{array}{ccc}
A_{1} \\
A_{3} \\
\Phi
\end{array}
\right)\; ,
\label{eq_39}
\end{equation}
\noindent
or:

\begin{equation}
\displaystyle {\cal D}_{\mu}=\partial_{\mu}+i\; q\; \lambda_{i}\; \left(
\begin{array}{ccc}
A_{2} \\
A_{3} \\
\Phi
\end{array}
\right)\; .
\label{eq_40}
\end{equation}

$\left(
\begin{array}{cc}
A_{3} \\
\Phi
\end{array}
\right)$ and $\left(
\begin{array}{cc}
\Phi \\
A_{3}
\end{array}
\right)$ for SU(2) are contained into the {\bf b)} and {\bf c)} $\left(
\begin{array}{ccc}
A_{1} \\
A_{3} \\
\Phi
\end{array}
\right)$, $\left(
\begin{array}{ccc}
A_{2} \\
A_{3} \\
\Phi
\end{array}
\right)$, $\left(
\begin{array}{ccc}
A_{1} \\
\Phi \\
A_{3}
\end{array}
\right)$ and $\left(
\begin{array}{ccc}
A_{2} \\
\Phi \\
A_{3}
\end{array}
\right)$ representations for SU(3), as the $A_{1}$ and $A_{2}$ fields of level four of self-similarity become trapped in this level when it collapses in the level three of self-similarity.

\ack
Manuscript registered at the Astronomical Institute, Romanian Academy, record No. 260 from 04/19/07.


\title[D.E.U.S.]{XIII: Interactions}

\begin{abstract}
With a beyond Standard Model approach, we will analyze the interactions resulting from the three possible situations to which we referred in the conclusions of paper {\it DEUS XII}.
\end{abstract}


\section{Introduction}

In paper \cite{deus12} we obtained three possible situations for the weak and strong interactions in Universe (or Anti-Universe), each one with its specific fields, generators and covariant derivatives. We will develop this subject considering the DEUS objects, at any level, as the source of all the FRW particle representations and the beginning of space and time. In our usual temporal acception ($t_{FRW}$) and relative to the DEUS object as source of space and time, any event occurs either "before time" (before our FRW perception of DEUS object formation and existence), either in "time" (after our FRW perception of DEUS object formation and existence). Then, before our FRW bubble existence or before the event existence, the currents and the charge in the DEUS object (with self-similar structure) must be zero. Any non-zero current or charge will not be allowed to exist at the DEUS "formation" or "evaporation".

Because of the large amount of interactions to be analyzed we will use for reviewing out most important results, instead of an unical final conclusion, intermediary conclusion sections.

\section{Interactions in the {\bf b)} Case from \cite{deus12}}

The Noether currents associated to the SO(2) group are:

\begin{equation}
J_{\mu}=i\; A_{\mu}^{+}\; \partial_{\nu}A_{\mu}-i\; A_{\mu}\; \partial_{\nu}A_{\mu}^{+} \; .
\label{eq_1}
\end{equation}

The corresponding charge is:

\begin{equation}
Q=\int{d^{3}x\; i\left(A_{\mu}^{+}\dot{A}_{\mu}-\dot{A}_{\mu}^{+}A_{\mu} \right)} \; ,
\label{eq_2}
\end{equation}
\noindent
where $\mu=\overline{0, 3}$ and $\nu\in\{r,\; t_{FRW}\}$.

The above physical quantities are defined in this way for empty space (${\cal D}_{\mu}\equiv \partial_{\mu}$). In general, for a non-empty spacetime:

\begin{equation}
J_{\mu}=i\; A_{\mu}^{+}\; {\cal D}_{\nu}A_{\mu}-i\; A_{\mu}\; {\cal D}_{\nu}A_{\mu}^{+} \; .
\label{eq_3}
\end{equation}
\noindent
where the covariant derivatives will be the ones determined for SU(2) and SU(3) (here for {\bf b)} case) in \cite{deus12}.

We saw that, for $\alpha_{k}=0$, we have:

\begin{equation}
A_{2}=\displaystyle i\; \frac{\sqrt{2}}{8}\; \left[4\left(\frac{1}{2}+ \frac{1}{4} \right)+\frac{\left(\sqrt{2}+1 \right)^{2}}{\displaystyle \frac{1}{2}+ \frac{1}{4}}\right]\frac{m_{helicoid}}{q}\; \frac{1}{1+\tanh \alpha_{k}}\; \frac{a^{2}}{r^{2}}\; ,
\label{eq_4}
\end{equation}
\noindent
where $A_{2}=A_{2}^{+}$. Then, the (\ref{eq_3}) electric current $J_{2}=0$ and, also, the charge $Q=0$. The $A_{3}$ component:

\begin{equation}
A_{3}=\displaystyle \pm i\; \frac{1}{8}\; \left[4\left(\frac{1}{2}\mp \frac{1}{4} \right)+\frac{\left(\sqrt{2}+1 \right)^{2}}{\displaystyle \frac{1}{2}\mp \frac{1}{4}}\right]\frac{m_{helicoid}}{q}\; \cosh \alpha_{k}\; \frac{a^{2}}{r^{2}}
\label{eq_5}
\end{equation}
\noindent
splits in two possible forms. The first one is:

\begin{equation}
A_{3}=\displaystyle i\; \frac{1}{8}\; \left[4\left(\frac{1}{2}+ \frac{1}{4} \right)+\frac{\left(\sqrt{2}+1 \right)^{2}}{\displaystyle \frac{1}{2}+ \frac{1}{4}}\right]\frac{m_{helicoid}}{q}\; \cosh \alpha_{k}\; \frac{a^{2}}{r^{2}}\; ,
\label{eq_6}
\end{equation}
\noindent
and it applies to {\bf matter in a matter Universe}. For (\ref{eq_6}) we have:

\begin{equation}
A_{3}^{+}=\displaystyle -i\; \frac{1}{8}\; \left[4\left(\frac{1}{2}- \frac{1}{4} \right)+\frac{\left(\sqrt{2}+1 \right)^{2}}{\displaystyle \frac{1}{2}- \frac{1}{4}}\right]\frac{m_{helicoid}}{q}\; \cosh \alpha_{k}\; \frac{a^{2}}{r^{2}}\; ,
\label{eq_7}
\end{equation}
\noindent
applying to {\bf antimatter in a matter Universe}.

The second possibility for $A_{3}$ is:

\begin{equation}
A_{3}=\displaystyle -i\; \frac{1}{8}\; \left[4\left(\frac{1}{2}- \frac{1}{4} \right)+\frac{\left(\sqrt{2}+1 \right)^{2}}{\displaystyle \frac{1}{2}- \frac{1}{4}}\right]\frac{m_{helicoid}}{q}\; \cosh \alpha_{k}\; \frac{a^{2}}{r^{2}}\; ,
\label{eq_8}
\end{equation}
\noindent
for {\bf matter in an Anti-Universe}, having:

\begin{equation}
A_{3}^{+}=\displaystyle i\; \frac{1}{8}\; \left[4\left(\frac{1}{2}+ \frac{1}{4} \right)+\frac{\left(\sqrt{2}+1 \right)^{2}}{\displaystyle \frac{1}{2}+ \frac{1}{4}}\right]\frac{m_{helicoid}}{q}\; \cosh \alpha_{k}\; \frac{a^{2}}{r^{2}}\; ,
\label{eq_9}
\end{equation}
\noindent
for {\bf antimatter in an Anti-Universe}.

The last two components of $A_{\mu}$ are:

\begin{equation}
\begin{array}{cc}
A_{1}=A_{1}^{+}=\displaystyle -i\; \frac{\sqrt{2}}{8}\; \left[4\left(\frac{1}{2}\mp \frac{1}{4} \right)+\frac{\left(\sqrt{2}+1 \right)^{2}}{\displaystyle \frac{1}{2}\mp \frac{1}{4}}\right]\frac{m_{helicoid}}{q}\; \frac{1}{1-\tanh \alpha_{k}}\; \frac{a^{2}}{r^{2}} \\
\Phi=\Phi^{+}=\displaystyle \frac{1}{8}\left[4\left(\frac{1}{2}\mp \frac{1}{4} \right)+\frac{\left(\sqrt{2}+1 \right)^{2}}{\displaystyle \frac{1}{2}\mp \frac{1}{4}}\right]\frac{m_{helicoid}}{q}\; \frac{a^{2}}{r^{2}}\; ,
\end{array}
\label{eq_10}
\end{equation}
\noindent
resulting that $J_{1}=J_{4}=0$ and $Q=0$.

When ${\cal D}_{\mu}=\partial_{\mu}$ we observe that also $J_{3}=i\; A_{3}^{+}\; \partial_{r}A_{3}-i\; A_{3}\; \partial_{r}A_{3}^{+}=i\; A_{3}^{+}\; \partial_{t_{FRW}}A_{3}-i\; A_{3}\; \partial_{t_{FRW}}A_{3}^{+}=0$ and $Q=0$ (the global charge over the DEUS object is conserved independently of the $A_{\mu}$ component choice - the number of "positive" charges is equal with the number of "negative" charges).

On the above field background over-impose the particle fields ($\displaystyle \alpha=\frac{\pi}{4}$):

\begin{equation}
\begin{array}{cc}
{\cal A}_{2}^{P}=\displaystyle i\; \frac{\sqrt{2}}{8}\; \left[4\left(\frac{1}{2}+ \frac{1}{4} \right)+\frac{\left(\sqrt{2}+1 \right)^{2}}{\displaystyle \frac{1}{2}+ \frac{1}{4}}\right]\frac{m_{helicoid}}{q}\; \frac{1}{1+\textnormal{tg}\; \alpha}\; \frac{a^{2}}{r^{2}} \\
{\cal A}_{3}^{P}=\displaystyle i\; \frac{1}{8}\; \left[4\left(\frac{1}{2}- \frac{1}{4} \right)+\frac{\left(\sqrt{2}+1 \right)^{2}}{\displaystyle \frac{1}{2}- \frac{1}{4}}\right]\frac{m_{helicoid}}{q}\; \cos \alpha\; \frac{a^{2}}{r^{2}}\; ,
\end{array}
\label{eq_11}
\end{equation}
\noindent
having:

\begin{equation}
\begin{array}{ccc}
J_{1}=J_{2}=J_{4}=0 \\
J_{3}=i\left( A_{3}^{+}\; {\cal D}_{\mu}A_{3}-A_{3}\; {\cal D}_{\mu}A_{3}^{+}\right) \\
Q=0
\end{array}\; ,
\label{eq_12}
\end{equation}
\noindent
where, in SU(3) \cite{deus12}:

\begin{equation}
\displaystyle {\cal D}_{\mu}=\partial_{\mu}+i\; q\; \lambda\; \left(
\begin{array}{ccc}
{\cal A}_{2}^{P} \\
{\cal A}_{4}^{P} \\
{\cal A}_{3}^{P}
\end{array}
\right)\; ,
\label{eq_13}
\end{equation}
\noindent
and antiparticle fields ($\displaystyle \alpha=-\frac{\pi}{4}$):

\begin{equation}
\begin{array}{cc}
{\cal A}_{2}^{AP}=\displaystyle - i\; \frac{\sqrt{2}}{8}\; \left[4\left(\frac{1}{2}- \frac{1}{4} \right)+\frac{\left(\sqrt{2}+1 \right)^{2}}{\displaystyle \frac{1}{2}- \frac{1}{4}}\right]\frac{m_{helicoid}}{q}\; \frac{1}{1+\textnormal{tg}\; \alpha}\; \frac{a^{2}}{r^{2}} \\
{\cal A}_{3}^{AP}=\displaystyle - i\; \frac{1}{8}\; \left[4\left(\frac{1}{2}+ \frac{1}{4} \right)+\frac{\left(\sqrt{2}+1 \right)^{2}}{\displaystyle \frac{1}{2}+ \frac{1}{4}}\right]\frac{m_{helicoid}}{q}\; \cos \alpha\; \frac{a^{2}}{r^{2}}\; ,
\end{array}
\label{eq_14}
\end{equation}
\noindent
having:

\begin{equation}
\begin{array}{ccc}
J_{1}=J_{2}=J_{4}=0 \\
J_{3}=i\left( A_{3}^{+}\; {\cal D}_{\mu}A_{3}-A_{3}\; {\cal D}_{\mu}A_{3}^{+}\right) \\
Q=0
\end{array}\; ,
\label{eq_15}
\end{equation}
\noindent
where, in SU(3) \cite{deus12}:

\begin{equation}
\displaystyle {\cal D}_{\mu}=\partial_{\mu}+i\; q\; \lambda\; \left(
\begin{array}{ccc}
{\cal A}_{1}^{AP} \\
{\cal A}_{4}^{AP} \\
{\cal A}_{3}^{AP}
\end{array}
\right)\; .
\label{eq_16}
\end{equation}

For $\displaystyle \alpha=\frac{\pi}{4}$ we have also:

\begin{equation}
\begin{array}{cc}
{\cal A}_{2}^{+P}=\displaystyle - i\; \frac{\sqrt{2}}{8}\; \left[4\left(\frac{1}{2}- \frac{1}{4} \right)+\frac{\left(\sqrt{2}+1 \right)^{2}}{\displaystyle \frac{1}{2}- \frac{1}{4}}\right]\frac{m_{helicoid}}{q}\; \frac{1}{1+\textnormal{tg}\; \alpha}\; \frac{a^{2}}{r^{2}} \\
{\cal A}_{3}^{+P}=\displaystyle - i\; \frac{1}{8}\; \left[4\left(\frac{1}{2}+ \frac{1}{4} \right)+\frac{\left(\sqrt{2}+1 \right)^{2}}{\displaystyle \frac{1}{2}+ \frac{1}{4}}\right]\frac{m_{helicoid}}{q}\; \cos \alpha\; \frac{a^{2}}{r^{2}}\; ,
\end{array}
\label{eq_17}
\end{equation}
\noindent
where:

\begin{equation}
\displaystyle {\cal D}_{\mu}=\partial_{\mu}+i\; q\; \lambda\; \left(
\begin{array}{ccc}
{\cal A}_{2}^{+P} \\
{\cal A}_{4}^{+P} \\
{\cal A}_{3}^{+P}
\end{array}
\right)\; ,
\label{eq_18}
\end{equation}
\noindent
and, for $\displaystyle \alpha=-\frac{\pi}{4}$:

\begin{equation}
\begin{array}{cc}
{\cal A}_{2}^{+AP}=\displaystyle i\; \frac{\sqrt{2}}{8}\; \left[4\left(\frac{1}{2}+ \frac{1}{4} \right)+\frac{\left(\sqrt{2}+1 \right)^{2}}{\displaystyle \frac{1}{2}+ \frac{1}{4}}\right]\frac{m_{helicoid}}{q}\; \frac{1}{1+\textnormal{tg}\; \alpha}\; \frac{a^{2}}{r^{2}} \\
{\cal A}_{3}^{+AP}=\displaystyle i\; \frac{1}{8}\; \left[4\left(\frac{1}{2}- \frac{1}{4} \right)+\frac{\left(\sqrt{2}+1 \right)^{2}}{\displaystyle \frac{1}{2}- \frac{1}{4}}\right]\frac{m_{helicoid}}{q}\; \cos \alpha\; \frac{a^{2}}{r^{2}}\; ,
\end{array}
\label{eq_19}
\end{equation}
\noindent
where:

\begin{equation}
\displaystyle {\cal D}_{\mu}=\partial_{\mu}+i\; q\; \lambda\; \left(
\begin{array}{ccc}
{\cal A}_{1}^{+AP} \\
{\cal A}_{4}^{+AP} \\
{\cal A}_{3}^{+AP}
\end{array}
\right)\; .
\label{eq_20}
\end{equation}

For the moment:

\begin{equation}
\!\!\!\!\!\!\!\!\!\!\!\!\!\!\!\!\!\!{\cal A}_{1}^{P}={\cal A}_{1}^{+P}={\cal A}_{1}^{AP}={\cal A}_{1}^{+AP}=\displaystyle -i\; \frac{\sqrt{2}}{8}\; \left[4\left(\frac{1}{2}\mp \frac{1}{4} \right)+\frac{\left(\sqrt{2}+1 \right)^{2}}{\displaystyle \frac{1}{2}\mp \frac{1}{4}}\right]\frac{m_{helicoid}}{q}\; \frac{1}{1-\textnormal{tg}\; \alpha}\; \frac{a^{2}}{r^{2}}\; .
\label{eq_20a}
\end{equation}

\subsection{Gluon and Quark Strong Interactions in Universe}

We will summarize the quark generations as $\displaystyle \left(
\begin{array}{cc}
q_{A} \\
q'_{A}
\end{array}\right)$, where {\it A} indexes the generation. The fourth self-similarity DEUS level will have on its possible real levels the {\it up}, {\it charm} and {\it top} quarks (the helicoids of the DEUS objects embedded in the catenoid of the fourth self-similar DEUS level) and the {\it down}, {\it strange} and {\it bottom} quarks (the catenoids of the DEUS objects embedded in the catenoid of the fourth self-similar DEUS level). While the catenoid is the conjugate surface of the helicoid, also, for their FRW conformal particle fields, we have, for example, the {\it down} ("correct" state) quarks as particle complement field of the {\it up} ("physical" state) particle field. In general, this can be written: $q_{A}\leftrightarrow q'_{A}$.

\subsubsection{Interaction Between Two Particle Fields in SU(3)} $\\$

For this interaction we have:

\begin{equation}
\displaystyle {\cal D}_{\mu}A_{3}=\partial_{\mu}A_{3}+i\; q_{1}\; \lambda_{i}\; \left(
\begin{array}{ccc}
{\cal A}_{2}^{P} \\
{\cal A}_{4}^{P} \\
{\cal A}_{3}^{P}
\end{array}
\right)A_{3}
\label{eq_21}
\end{equation}
\noindent
and:

\begin{equation}
\displaystyle {\cal D}_{\mu}A_{3}^{+}=\partial_{\mu}A_{3}^{+}+i\; q_{2}\; \lambda_{j}\; \left(
\begin{array}{ccc}
{\cal A}_{2}^{P} \\
{\cal A}_{4}^{P} \\
{\cal A}_{3}^{P}
\end{array}
\right)A_{3}^{+}\; .
\label{eq_22}
\end{equation}

Then:

\begin{equation}
J_{3}=A_{3}A_{3}^{+}\displaystyle \left[q_{2}\; \lambda_{j}\; \left(
\begin{array}{ccc}
{\cal A}_{2}^{P} \\
{\cal A}_{4}^{P} \\
{\cal A}_{3}^{P}
\end{array}
\right) -q_{1}\; \lambda_{i}\; \left(
\begin{array}{ccc}
{\cal A}_{2}^{P} \\
{\cal A}_{4}^{P} \\
{\cal A}_{3}^{P}
\end{array}
\right) \right]\; .
\label{eq_23}
\end{equation}

But, because we require the conservation of the {\it Q} charge, and, from $J_{1}=J_{2}=J_{4}=0$, we have $Q=0$, also $J_{3}=0$. We will express the result of (\ref{eq_23}) as:

\begin{equation}
J_{3}\stackrel{not}{=}A_{3}A_{3}^{+}\displaystyle \left(
\begin{array}{ccc}
{\cal Z}_{1} \\
{\cal Z}_{2} \\
{\cal Z}_{3}
\end{array}
\right)\; .
\label{eq_24}
\end{equation}

For an unique $J_{3}$ we require that ${\cal Z}_{1}={\cal Z}_{2}={\cal Z}_{3}$.

The first generators that satisfy the above requirements are $\lambda_{i}=\lambda_{j}=\lambda_{0}$ \cite{deus11}. This will be seen by the external observer as {\bf interactions between gluons}: \newpage

\begin{figure}[!h]
\centering
\leavevmode
\includegraphics[width=3.5cm,height=3cm]{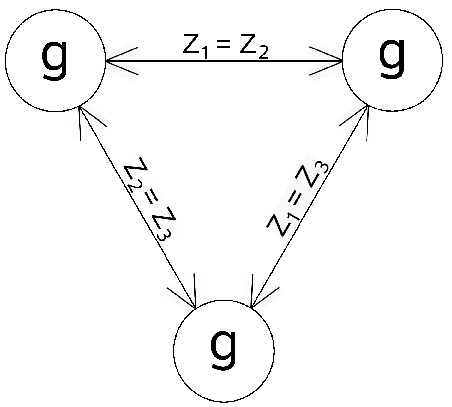}
\label{int1}
\end{figure}

We will use the following conventions: prime for particles in the "correct" states while, without prime, we will denote the particles in the physical states; overline for their antiparticles.

Now, if $\lambda_{i}=\lambda_{j}=\lambda_{2}$, $\lambda_{i}=\lambda_{j}=\lambda_{3}$, $\lambda_{i}=\lambda_{j}=\lambda_{4}$, $\lambda_{i}=\lambda_{j}=\lambda_{5}$, or $\lambda_{i}=\lambda_{j}=\lambda_{7}$ \cite{deus11}, we also have ${\cal Z}_{1}={\cal Z}_{2}={\cal Z}_{3}$, $J_{3}=0$ and $Q=0$. It will be seen by the external observer as {\bf interactions between quarks in physical state} ({\it u}, {\it c}, or {\it t}):

\begin{figure}[h]
\centering
\leavevmode
\includegraphics[width=3.5cm,height=3cm]{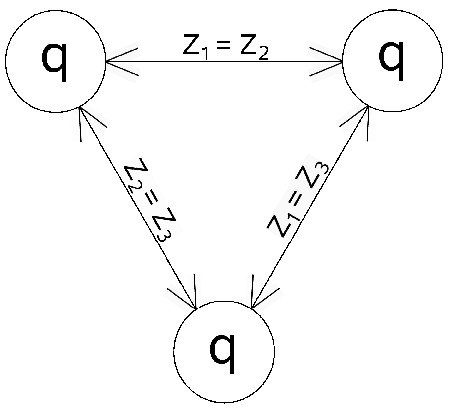}
\label{int2}
\end{figure}

For a different $\lambda$ choice we may have $J_{3}\neq 0$ and $Q\neq 0$.

Because ${\cal A}_{2}^{+AP}={\cal A}_{2}^{P}$, ${\cal A}_{3}^{+AP}={\cal A}_{3}^{P}$ and ${\cal A}_{4}^{+AP}={\cal A}_{4}^{P}$, the same result is obtained for $\overline{g'}$ ($\equiv g$) and for the "correct" anti-quark states $\overline{q'}$ ($\overline{d'}$, $\overline{s'}$, or $\overline{b'}$), where:

\begin{equation}
\displaystyle {\cal D}_{\mu}A_{3}=\partial_{\mu}A_{3}+i\; q_{1}\; \lambda_{i}\; \left(
\begin{array}{ccc}
{\cal A}_{2}^{+AP} \\
{\cal A}_{4}^{+AP} \\
{\cal A}_{3}^{+AP}
\end{array}
\right)A_{3}
\label{eq_25}
\end{equation}
\noindent
and:

\begin{equation}
\displaystyle {\cal D}_{\mu}A_{3}^{+}=\partial_{\mu}A_{3}^{+}+i\; q_{2}\; \lambda_{j}\; \left(
\begin{array}{ccc}
{\cal A}_{2}^{+AP} \\
{\cal A}_{4}^{+AP} \\
{\cal A}_{3}^{+AP}
\end{array}
\right)A_{3}^{+}\; .
\label{eq_26}
\end{equation}

\subsubsection{Interaction Between Two Antiparticle Fields in SU(3)} $\\$

For this case we have:

\begin{equation}
\displaystyle {\cal D}_{\mu}A_{3}=\partial_{\mu}A_{3}+i\; q_{1}\; \lambda_{i}\; \left(
\begin{array}{ccc}
{\cal A}_{1}^{AP} \\
{\cal A}_{4}^{AP} \\
{\cal A}_{3}^{AP}
\end{array}
\right)A_{3} \; ,
\label{eq_27}
\end{equation}

\begin{equation}
\displaystyle {\cal D}_{\mu}A_{3}^{+}=\partial_{\mu}A_{3}^{+}+i\; q_{2}\; \lambda_{j}\; \left(
\begin{array}{ccc}
{\cal A}_{1}^{AP} \\
{\cal A}_{4}^{AP} \\
{\cal A}_{3}^{AP}
\end{array}
\right)A_{3}^{+}
\label{eq_28}
\end{equation}
\noindent
and:

\begin{equation}
J_{3}=A_{3}A_{3}^{+}\displaystyle \left[q_{2}\; \lambda_{j}\; \left(
\begin{array}{ccc}
{\cal A}_{1}^{AP} \\
{\cal A}_{4}^{AP} \\
{\cal A}_{3}^{AP}
\end{array}
\right) -q_{1}\; \lambda_{i}\; \left(
\begin{array}{ccc}
{\cal A}_{1}^{AP} \\
{\cal A}_{4}^{AP} \\
{\cal A}_{3}^{AP}
\end{array}
\right) \right]\stackrel{not}{=}A_{3}A_{3}^{+}\displaystyle \left(
\begin{array}{ccc}
{\cal Z}_{1} \\
{\cal Z}_{2} \\
{\cal Z}_{3}
\end{array}
\right) \; .
\label{eq_29}
\end{equation}

If $\lambda_{i}=\lambda_{j}=\lambda_{0}$, then ${\cal Z}_{1}={\cal Z}_{2}={\cal Z}_{3}$ and, with its help, we can describe {\bf interactions between anti-gluons} ($\equiv$ gluons): \newpage

\begin{figure}[h]
\centering
\leavevmode
\includegraphics[width=3.5cm,height=3cm]{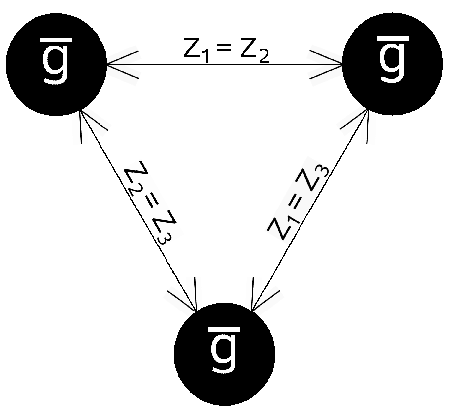}
\label{int3}
\end{figure}
\noindent
where $J_{3}=0$ and $Q=0$.

When $\lambda_{i}=\lambda_{j}=\lambda_{2}$, $\lambda_{i}=\lambda_{j}=\lambda_{3}$, $\lambda_{i}=\lambda_{j}=\lambda_{4}$, $\lambda_{i}=\lambda_{j}=\lambda_{5}$, or $\lambda_{i}=\lambda_{j}=\lambda_{7}$, we get ${\cal Z}_{1}={\cal Z}_{2}={\cal Z}_{3}$ describing {\bf interactions between anti-quarks in physical state} ($\overline{u}$, $\overline{c}$, $\overline{t}$):

\begin{figure}[h]
\centering
\leavevmode
\includegraphics[width=3.5cm,height=3cm]{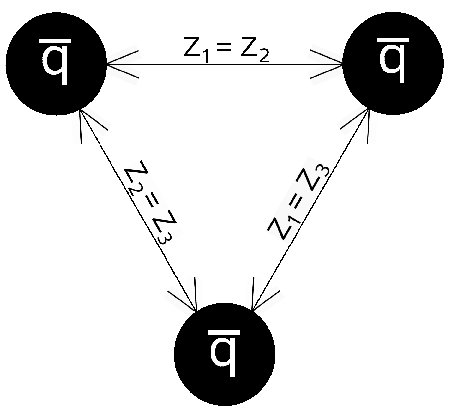}
\label{int4}
\end{figure}
\noindent
where, for our $\lambda$'s, $J_{3}=0$ and $Q=0$, or, for a different choice of $\lambda$'s, we may have $J_{3}\neq 0$ and $Q\neq 0$.

Because ${\cal A}_{3}^{+P}={\cal A}_{3}^{AP}$, ${\cal A}_{1}^{+P}={\cal A}_{1}^{AP}$ and ${\cal A}_{4}^{+P}={\cal A}_{4}^{AP}$, we obtain the same result for interactions between $g'$ ($\equiv \overline{g}\equiv \overline{g'}\equiv g$) and for the "correct" $q'$ quark states ($d'$, $s'$ or $b'$), where:

\begin{equation}
\displaystyle {\cal D}_{\mu}A_{3}=\partial_{\mu}A_{3}+i\; q_{1}\; \lambda_{i}\; \left(
\begin{array}{ccc}
{\cal A}_{1}^{+P} \\
{\cal A}_{4}^{+P} \\
{\cal A}_{3}^{+P}
\end{array}
\right)A_{3}
\label{eq_30}
\end{equation}
\noindent
and:

\begin{equation}
\displaystyle {\cal D}_{\mu}A_{3}^{+}=\partial_{\mu}A_{3}^{+}+i\; q_{2}\; \lambda_{j}\; \left(
\begin{array}{ccc}
{\cal A}_{1}^{+P} \\
{\cal A}_{4}^{+P} \\
{\cal A}_{3}^{+P}
\end{array}
\right)A_{3}^{+}\; .
\label{eq_31}
\end{equation}

\subsubsection{Interaction Between One Particle Field and One Antiparticle Conjugate Field in SU(3)} $\\$

For this case we have:

\begin{equation}
\displaystyle {\cal D}_{\mu}A_{3}=\partial_{\mu}A_{3}+i\; q_{1}\; \lambda_{i}\; \left(
\begin{array}{ccc}
{\cal A}_{2}^{P} \\
{\cal A}_{4}^{P} \\
{\cal A}_{3}^{P}
\end{array}
\right)A_{3} \; ,
\label{eq_32}
\end{equation}

\begin{equation}
\displaystyle {\cal D}_{\mu}A_{3}^{+}=\partial_{\mu}A_{3}^{+}+i\; q_{2}\; \lambda_{j}\; \left(
\begin{array}{ccc}
{\cal A}_{1}^{+AP} \\
{\cal A}_{4}^{+AP} \\
{\cal A}_{3}^{+AP}
\end{array}
\right)A_{3}^{+}
\label{eq_33}
\end{equation}
\noindent
and:

\begin{equation}
J_{3}=A_{3}A_{3}^{+}\displaystyle \left[q_{2}\; \lambda_{j}\; \left(
\begin{array}{ccc}
{\cal A}_{1}^{+AP} \\
{\cal A}_{4}^{+AP} \\
{\cal A}_{3}^{+AP}
\end{array}
\right) -q_{1}\; \lambda_{i}\; \left(
\begin{array}{ccc}
{\cal A}_{2}^{P} \\
{\cal A}_{4}^{P} \\
{\cal A}_{3}^{P}
\end{array}
\right) \right]\stackrel{not}{=}A_{3}A_{3}^{+}\displaystyle \left(
\begin{array}{ccc}
{\cal Z}_{1} \\
{\cal Z}_{2} \\
{\cal Z}_{3}
\end{array}
\right) \; .
\label{eq_34}
\end{equation}

\newpage
If $\lambda_{i}=\lambda_{j}=\lambda_{0}$, then ${\cal Z}_{1}={\cal Z}_{2}={\cal Z}_{3}$ and, with its help, we can describe {\bf interactions between gluons}:

\begin{figure}[h]
\centering
\leavevmode
\includegraphics[width=3.5cm,height=3cm]{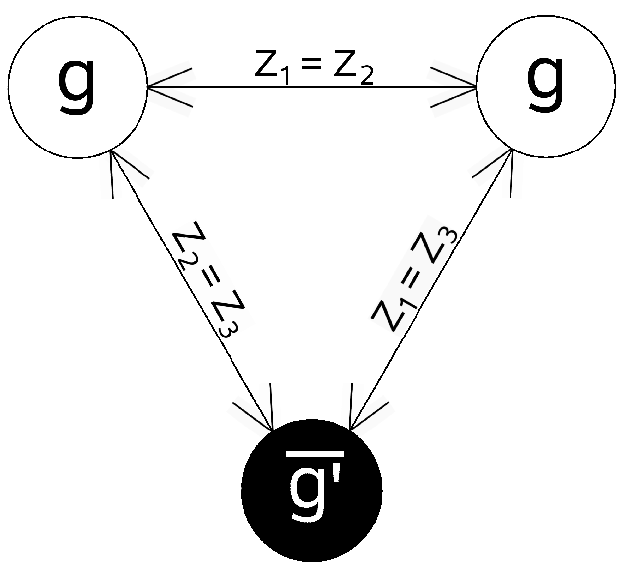}\;\;\;\;\; or \;\;\;\;\;
\includegraphics[width=3.5cm,height=3cm]{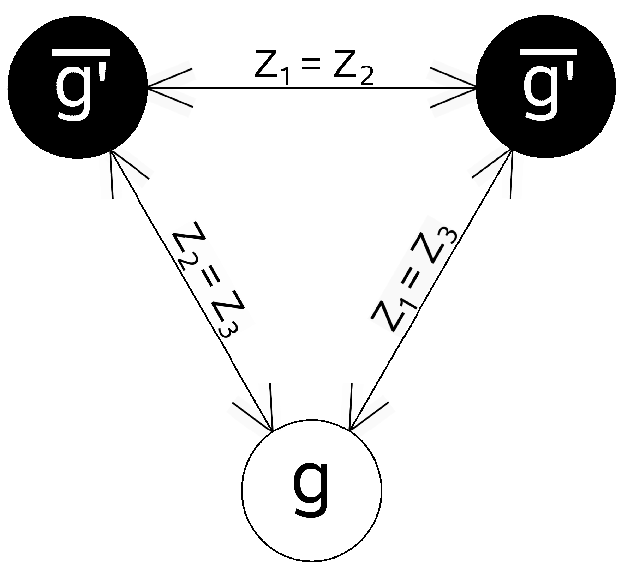}
\label{int5-6}
\end{figure}
\noindent
where $J_{3}=0$ and $Q=0$.

If $\lambda_{i}=\lambda_{j}=\lambda_{7}$, we obtain ${\cal Z}_{1}={\cal Z}_{2}={\cal Z}_{3}$ for the {\bf interactions between quarks in physical state and anti-quarks in "correct" state} (for example, $(u\;u\;\overline{d'})$ or $(u\; \overline{d'}\; \overline{d'})$):

\begin{figure}[h]
\centering
\leavevmode
\includegraphics[width=3.5cm,height=3cm]{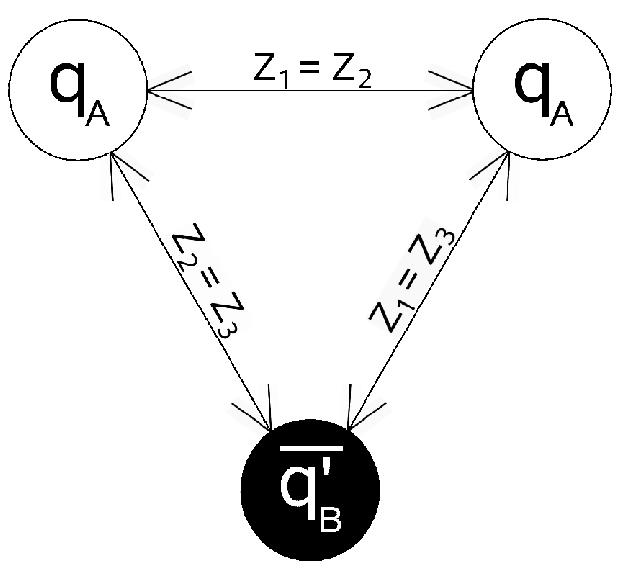}\;\;\;\;\; or \;\;\;\;\;
\includegraphics[width=3.5cm,height=3cm]{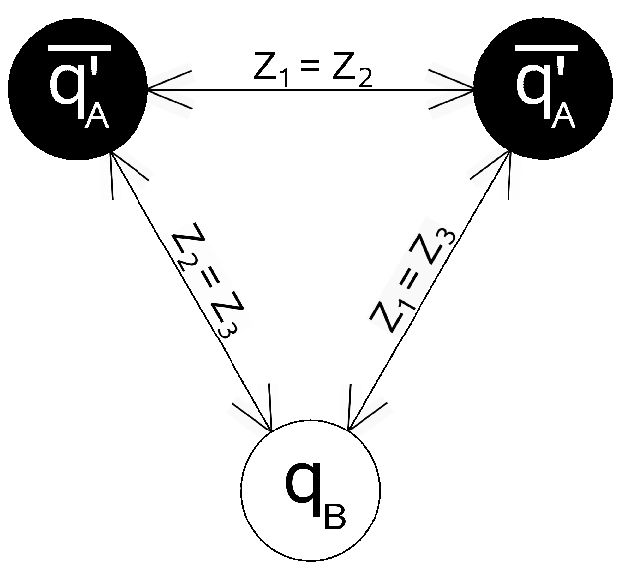}
\label{int7-8}
\end{figure}

The above interaction schemes work also for $\lambda_{i}=\lambda_{j}=\lambda_{4}$ and $\lambda_{i}=\lambda_{j}=\lambda_{5}$, but here $J_{3}\neq 0$ and $Q\neq 0$.

The situation repeats itself for:

\begin{equation}
\displaystyle {\cal D}_{\mu}A_{3}=\partial_{\mu}A_{3}+i\; q_{1}\; \lambda_{i}\; \left(
\begin{array}{ccc}
{\cal A}_{1}^{+AP} \\
{\cal A}_{4}^{+AP} \\
{\cal A}_{3}^{+AP}
\end{array}
\right)A_{3}
\label{eq_35}
\end{equation}

\begin{equation}
\displaystyle {\cal D}_{\mu}A_{3}^{+}=\partial_{\mu}A_{3}^{+}+i\; q_{2}\; \lambda_{j}\; \left(
\begin{array}{ccc}
{\cal A}_{2}^{P} \\
{\cal A}_{4}^{P} \\
{\cal A}_{3}^{P}
\end{array}
\right)A_{3}^{+}\; .
\label{eq_36}
\end{equation}

\subsubsection{Interaction Between One Antiparticle Field and One Particle Conjugate Field in SU(3)} $\\$

Here we have:

\begin{equation}
\displaystyle {\cal D}_{\mu}A_{3}=\partial_{\mu}A_{3}+i\; q_{1}\; \lambda_{i}\; \left(
\begin{array}{ccc}
{\cal A}_{2}^{+P} \\
{\cal A}_{4}^{+P} \\
{\cal A}_{3}^{+P}
\end{array}
\right)A_{3} \; ,
\label{eq_37}
\end{equation}

\begin{equation}
\displaystyle {\cal D}_{\mu}A_{3}^{+}=\partial_{\mu}A_{3}^{+}+i\; q_{2}\; \lambda_{j}\; \left(
\begin{array}{ccc}
{\cal A}_{1}^{AP} \\
{\cal A}_{4}^{AP} \\
{\cal A}_{3}^{AP}
\end{array}
\right)A_{3}^{+}
\label{eq_38}
\end{equation}
\noindent
and:

\begin{equation}
J_{3}=A_{3}A_{3}^{+}\displaystyle \left[q_{2}\; \lambda_{j}\; \left(
\begin{array}{ccc}
{\cal A}_{1}^{AP} \\
{\cal A}_{4}^{AP} \\
{\cal A}_{3}^{AP}
\end{array}
\right) -q_{1}\; \lambda_{i}\; \left(
\begin{array}{ccc}
{\cal A}_{2}^{+P} \\
{\cal A}_{4}^{+P} \\
{\cal A}_{3}^{+P}
\end{array}
\right) \right]\stackrel{not}{=}A_{3}A_{3}^{+}\displaystyle \left(
\begin{array}{ccc}
{\cal Z}_{1} \\
{\cal Z}_{2} \\
{\cal Z}_{3}
\end{array}
\right) \; .
\label{eq_39}
\end{equation}

\newpage
If $\lambda_{i}=\lambda_{j}=\lambda_{0}$, then ${\cal Z}_{1}={\cal Z}_{2}={\cal Z}_{3}$ and we can describe {\bf interactions between gluons}:

\begin{figure}[h]
\centering
\leavevmode
\includegraphics[width=3.5cm,height=3cm]{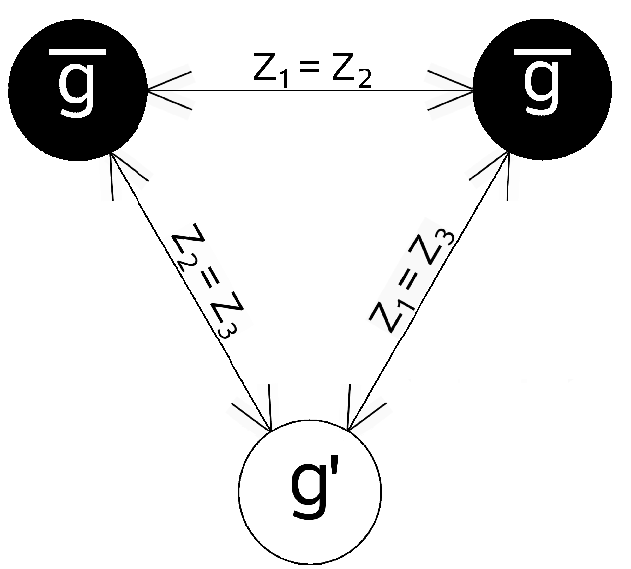}\;\;\;\;\; or \;\;\;\;\;
\includegraphics[width=3.5cm,height=3cm]{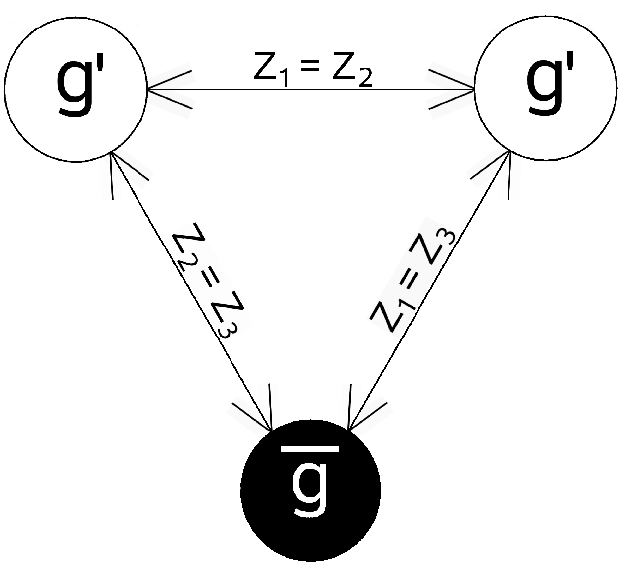}
\label{int10-11}
\end{figure}
\noindent
where $J_{3}=0$ and $Q=0$.

When $\lambda_{i}=\lambda_{j}=\lambda_{2}$, $\lambda_{i}=\lambda_{j}=\lambda_{3}$, $\lambda_{i}=\lambda_{j}=\lambda_{4}$, $\lambda_{i}=\lambda_{j}=\lambda_{5}$, or $\lambda_{i}=\lambda_{j}=\lambda_{7}$, we get ${\cal Z}_{1}={\cal Z}_{2}={\cal Z}_{3}$ describing {\bf interactions between anti-quarks in physical state and quarks in "correct" state} (for example, $(d'\; \overline{u}\; \overline{u})$ or $(\overline{u}\; d'\; d')$):

\begin{figure}[h]
\centering
\leavevmode
\includegraphics[width=3.5cm,height=3cm]{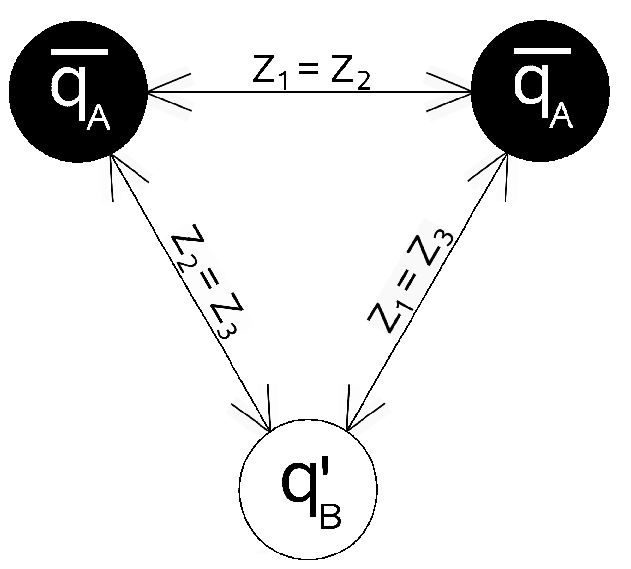}\;\;\;\;\; or \;\;\;\;\;
\includegraphics[width=3.5cm,height=3cm]{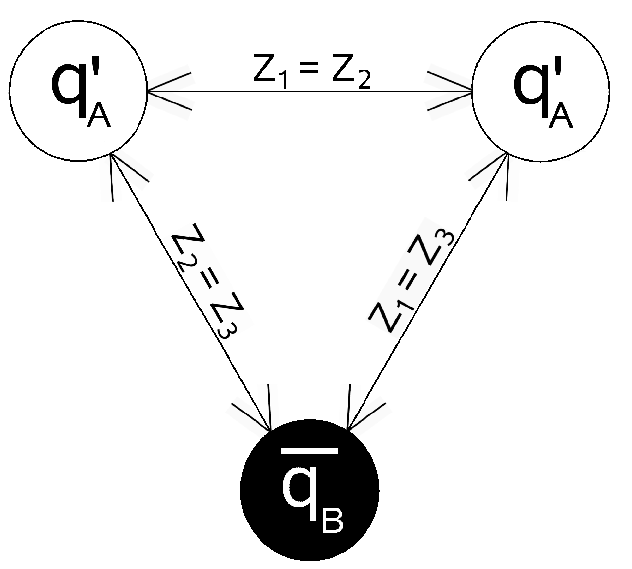}
\label{int12-13}
\end{figure}
\noindent
only if:

\begin{equation}
{\cal A}_{1}^{AP}=\displaystyle -i\; \frac{\sqrt{2}}{8}\; \left[4\left(\frac{1}{2}- \frac{1}{4} \right)+\frac{\left(\sqrt{2}+1 \right)^{2}}{\displaystyle \frac{1}{2}- \frac{1}{4}}\right]\frac{m_{helicoid}}{q}\; \frac{1}{1-\textnormal{tg}\; \alpha}\; \frac{a^{2}}{r^{2}}\; ,
\label{eq_40}
\end{equation}
\noindent
for which $J_{3}=0$ and $Q=0$. For other $\lambda$ choices $J_{3}\neq 0$ and $Q\neq 0$.

The situation is the same for:

\begin{equation}
\displaystyle {\cal D}_{\mu}A_{3}=\partial_{\mu}A_{3}+i\; q_{1}\; \lambda_{i}\; \left(
\begin{array}{ccc}
{\cal A}_{1}^{AP} \\
{\cal A}_{4}^{AP} \\
{\cal A}_{3}^{AP}
\end{array}
\right)A_{3}
\label{eq_41}
\end{equation}

\begin{equation}
\displaystyle {\cal D}_{\mu}A_{3}^{+}=\partial_{\mu}A_{3}^{+}+i\; q_{2}\; \lambda_{j}\; \left(
\begin{array}{ccc}
{\cal A}_{2}^{+P} \\
{\cal A}_{4}^{+P} \\
{\cal A}_{3}^{+P}
\end{array}
\right)A_{3}^{+}\; .
\label{eq_42}
\end{equation}

\subsection{Quark Weak Interactions in Universe}

\subsubsection{Interaction Between Two Particle Fields in SU(2)} $\\$

In the same way we had done for interactions in the SU(3) (the level one of self-similarity observer that sees the fourth level's quark-DEUS objects as embedded in level four), we will proceed with the quarks obeying the SU(2) symmetry (the level one of self-similarity observer that sees the fourth level's quark-DEUS objects as embedded in level three). The gluons existence is limited only at the fourth self-similar DEUS level.

With the $c$ generators determined in \cite{deus11} and the $A_{\mu}$ expected field combinations \cite{deus12}, we can write:

\begin{equation}
\displaystyle {\cal D}_{\mu}A_{3}=\partial_{\mu}A_{3}+i\; q_{1}\; c_{i}\; \left(
\begin{array}{cc}
{\cal A}_{4}^{P} \\
{\cal A}_{3}^{P}
\end{array}
\right)A_{3} \; ,
\label{eq_43}
\end{equation}

\begin{equation}
\displaystyle {\cal D}_{\mu}A_{3}^{+}=\partial_{\mu}A_{3}^{+}+i\; q_{2}\; c_{j}\; \left(
\begin{array}{cc}
{\cal A}_{4}^{P} \\
{\cal A}_{3}^{P}
\end{array}
\right)A_{3}^{+}
\label{eq_44}
\end{equation}
\noindent
and:

\begin{equation}
J_{3}=A_{3}A_{3}^{+}\displaystyle \left[q_{2}\; c_{j}\; \left(
\begin{array}{cc}
{\cal A}_{4}^{P} \\
{\cal A}_{3}^{P}
\end{array}
\right) -q_{1}\; c_{i}\; \left(
\begin{array}{cc}
{\cal A}_{4}^{P} \\
{\cal A}_{3}^{P}
\end{array}
\right) \right]\stackrel{not}{=}A_{3}A_{3}^{+}\displaystyle \left(
\begin{array}{cc}
{\cal Z}_{1} \\
{\cal Z}_{2}
\end{array}
\right) \; .
\label{eq_45}
\end{equation}

If $c_{i}=c_{j}=c_{1}$, $c_{i}=c_{j}=c_{2}$, or $c_{i}=c_{j}=c_{3}$, we get ${\cal Z}_{1}={\cal Z}_{2}$, $J_{3}=0$ and $Q=0$ for {\bf interactions between two quarks in physical state} (for example $(uc)$, $(uu)$, $(ut)$, etc.):

\begin{figure}[h]
\centering
\leavevmode
\includegraphics[width=3cm,height=0.8cm]{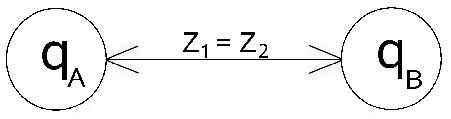}
\label{int14}
\end{figure}

For $c_{i}\neq c_{j}$ generators, this interaction has $J_{3}\neq 0$ and $Q\neq 0$. 

Because ${\cal A}_{3}^{P}={\cal A}_{3}^{+AP}$ and ${\cal A}_{4}^{P}={\cal A}_{4}^{+AP}$, we get (for the above $c$'s) the same result with:

\begin{equation}
\displaystyle {\cal D}_{\mu}A_{3}=\partial_{\mu}A_{3}+i\; q_{1}\; c_{i}\; \left(
\begin{array}{cc}
{\cal A}_{4}^{+AP} \\
{\cal A}_{3}^{+AP}
\end{array}
\right)A_{3} \; ,
\label{eq_46}
\end{equation}

\begin{equation}
\displaystyle {\cal D}_{\mu}A_{3}^{+}=\partial_{\mu}A_{3}^{+}+i\; q_{2}\; c_{j}\; \left(
\begin{array}{cc}
{\cal A}_{4}^{+AP} \\
{\cal A}_{3}^{+AP}
\end{array}
\right)A_{3}^{+}
\label{eq_47}
\end{equation}
\noindent
and:

\begin{equation}
J_{3}=A_{3}A_{3}^{+}\displaystyle \left[q_{2}\; c_{j}\; \left(
\begin{array}{cc}
{\cal A}_{4}^{+AP} \\
{\cal A}_{3}^{+AP}
\end{array}
\right) -q_{1}\; c_{i}\; \left(
\begin{array}{cc}
{\cal A}_{4}^{+AP} \\
{\cal A}_{3}^{+AP}
\end{array}
\right) \right]\stackrel{not}{=}A_{3}A_{3}^{+}\displaystyle \left(
\begin{array}{cc}
{\cal Z}_{1} \\
{\cal Z}_{2}
\end{array}
\right) \; .
\label{eq_48}
\end{equation}

As we observed previously for the SU(3) (strong) interactions, {\bf the quark in physical state is the same particle as the anti-quark in "correct" state}, meaning that, for (\ref{eq_46})-(\ref{eq_48}) we have a SU(2) (weak) {\bf interaction between two anti-quarks in "correct" state} (for example, $(\overline{d'}\; \overline{d'})$, $(\overline{d'}\; \overline{s'})$, $(\overline{u'}\; \overline{c'})$, etc.):

\begin{figure}[h]
\centering
\leavevmode
\includegraphics[width=3cm,height=0.8cm]{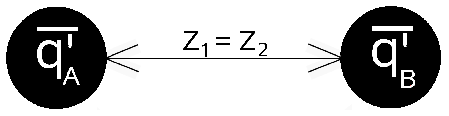}
\label{int15}
\end{figure}

\subsubsection{Interaction Between One Particle Field and One Antiparticle Conjugate Field in SU(2)} $\\$

With the $c$ generators determined in \cite{deus11} and the $A_{\mu}$ expected field combinations \cite{deus12}, we can write:

\begin{equation}
\displaystyle {\cal D}_{\mu}A_{3}=\partial_{\mu}A_{3}+i\; q_{1}\; c_{i}\; \left(
\begin{array}{cc}
{\cal A}_{4}^{P} \\
{\cal A}_{3}^{P}
\end{array}
\right)A_{3} \; ,
\label{eq_49}
\end{equation}

\begin{equation}
\displaystyle {\cal D}_{\mu}A_{3}^{+}=\partial_{\mu}A_{3}^{+}+i\; q_{2}\; c_{j}\; \left(
\begin{array}{cc}
{\cal A}_{4}^{+AP} \\
{\cal A}_{3}^{+AP}
\end{array}
\right)A_{3}^{+}
\label{eq_50}
\end{equation}
\noindent
and:

\begin{equation}
J_{3}=A_{3}A_{3}^{+}\displaystyle \left[q_{2}\; c_{j}\; \left(
\begin{array}{cc}
{\cal A}_{4}^{+AP} \\
{\cal A}_{3}^{+AP}
\end{array}
\right) -q_{1}\; c_{i}\; \left(
\begin{array}{cc}
{\cal A}_{4}^{P} \\
{\cal A}_{3}^{P}
\end{array}
\right) \right]\stackrel{not}{=}A_{3}A_{3}^{+}\displaystyle \left(
\begin{array}{cc}
{\cal Z}_{1} \\
{\cal Z}_{2}
\end{array}
\right) \; .
\label{eq_51}
\end{equation}

If $c_{i}=c_{j}=c_{1}$, $c_{i}=c_{j}=c_{2}$, or $c_{i}=c_{j}=c_{3}$, we get ${\cal Z}_{1}={\cal Z}_{2}$, $J_{3}=0$ and $Q=0$. We saw previously that the implications of:

\begin{equation}
\left(
\begin{array}{cc}
{\cal A}_{4}^{P} \\
{\cal A}_{3}^{P}
\end{array} \right) \equiv \left(
\begin{array}{cc}
{\cal A}_{4}^{+AP} \\
{\cal A}_{3}^{+AP}
\end{array}\right)
\label{eq_52}
\end{equation}
\noindent
are the reciprocity of the interaction, being possible to interchange ${\cal D}A_{\mu}$ with ${\cal D}A_{\mu}^{+}$, and that the {\bf interactions between one quark in physical state with an anti-quark in "correct" state} (for example, $(u\overline{d'})$, or $(u\overline{s'})$, etc.) is equivalent with the {\bf interaction between two quarks in physical state} (the same example but with $(ud)$, or $(us)$, etc.):
\newpage
\begin{figure}[h]
\centering
\leavevmode
\includegraphics[width=3cm,height=0.8cm]{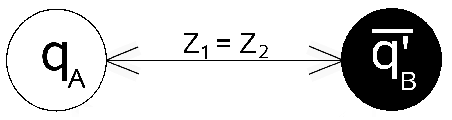}
\label{int16}
\end{figure}

Again, for our SU(2) generators, $J_{3}=0$ and the {\bf charge is conserved} ($Q=0$). For other $c$ choices we may have $J_{3}\neq 0$ and $Q\neq 0$.

\subsubsection{Interaction Between Two Antiparticle Fields in SU(2)} $\\$

For this situation, at the third DEUS level of self-similarity, we can write:

\begin{equation}
\displaystyle {\cal D}_{\mu}A_{3}=\partial_{\mu}A_{3}+i\; q_{1}\; c_{i}\; \left(
\begin{array}{cc}
{\cal A}_{4}^{AP} \\
{\cal A}_{3}^{AP}
\end{array}
\right)A_{3} \; ,
\label{eq_53}
\end{equation}

\begin{equation}
\displaystyle {\cal D}_{\mu}A_{3}^{+}=\partial_{\mu}A_{3}^{+}+i\; q_{2}\; c_{j}\; \left(
\begin{array}{cc}
{\cal A}_{4}^{AP} \\
{\cal A}_{3}^{AP}
\end{array}
\right)A_{3}^{+}
\label{eq_54}
\end{equation}
\noindent
and:

\begin{equation}
J_{3}=A_{3}A_{3}^{+}\displaystyle \left[q_{2}\; c_{j}\; \left(
\begin{array}{cc}
{\cal A}_{4}^{AP} \\
{\cal A}_{3}^{AP}
\end{array}
\right) -q_{1}\; c_{i}\; \left(
\begin{array}{cc}
{\cal A}_{4}^{AP} \\
{\cal A}_{3}^{AP}
\end{array}
\right) \right]\stackrel{not}{=}A_{3}A_{3}^{+}\displaystyle \left(
\begin{array}{cc}
{\cal Z}_{1} \\
{\cal Z}_{2}
\end{array}
\right) \; .
\label{eq_55}
\end{equation}

The {\bf interaction between two anti-quarks in physical state} (for example, $(\overline{u}\; \overline{u})$, $(\overline{u}\; \overline{c})$, $(\overline{u}\; \overline{t})$, etc.) can be achieved if $c_{i}=c_{j}=c_{1}$, $c_{i}=c_{j}=c_{2}$, or $c_{i}=c_{j}=c_{3}$:

\begin{figure}[h]
\centering
\leavevmode
\includegraphics[width=3cm,height=0.8cm]{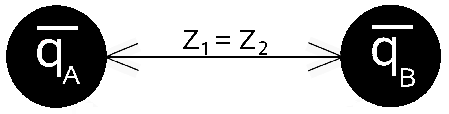}
\label{int17}
\end{figure}
\noindent
with $J_{3}=0$ and $Q=0$ or, for a different $c$'s, $J_{3}\neq 0$ and $Q\neq 0$. The remark regarding the quark equivalences is valid also here, for $c_{i}=c_{j}=c_{1}$, $c_{i}=c_{j}=c_{2}$, or $c_{i}=c_{j}=c_{3}$, and because ${\cal A}_{3}^{AP}={\cal A}_{3}^{+P}$ and ${\cal A}_{4}^{AP}={\cal A}_{4}^{+P}$, we obtain the scheme of an {\bf interaction between two quarks in "correct" state} (for example, $(d'd')$, $(d's')$, $(u'u')$, etc.):

\begin{figure}[h]
\centering
\leavevmode
\includegraphics[width=3cm,height=0.8cm]{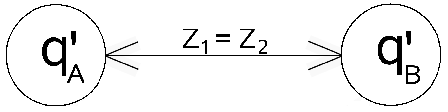}
\label{int18}
\end{figure}

For this interaction, in the given $c$ conditions, $J_{3}$ and $Q$ are zero. For different $c$'s, $J_{3}\neq 0$ and $Q\neq 0$.

For:

\begin{equation}
\displaystyle {\cal D}_{\mu}A_{3}=\partial_{\mu}A_{3}+i\; q_{1}\; c_{i}\; \left(
\begin{array}{cc}
{\cal A}_{4}^{+P} \\
{\cal A}_{3}^{+P}
\end{array}
\right)A_{3} \; ,
\label{eq_56}
\end{equation}

\begin{equation}
\displaystyle {\cal D}_{\mu}A_{3}^{+}=\partial_{\mu}A_{3}^{+}+i\; q_{2}\; c_{j}\; \left(
\begin{array}{cc}
{\cal A}_{4}^{AP} \\
{\cal A}_{3}^{AP}
\end{array}
\right)A_{3}^{+}
\label{eq_57}
\end{equation}
\noindent
and:

\begin{equation}
J_{3}=A_{3}A_{3}^{+}\displaystyle \left[q_{2}\; c_{j}\; \left(
\begin{array}{cc}
{\cal A}_{4}^{AP} \\
{\cal A}_{3}^{AP}
\end{array}
\right) -q_{1}\; c_{i}\; \left(
\begin{array}{cc}
{\cal A}_{4}^{+P} \\
{\cal A}_{3}^{+P}
\end{array}
\right) \right]\stackrel{not}{=}A_{3}A_{3}^{+}\displaystyle \left(
\begin{array}{cc}
{\cal Z}_{1} \\
{\cal Z}_{2}
\end{array}
\right) \; .
\label{eq_58}
\end{equation}
\noindent
we have the {\bf interaction between one quark in "correct" state and one anti-quark in physical state} (for example, $(d'\overline{u})$, $(s'\overline{u})$, $(u'\overline{u})$, etc.):

\begin{figure}[h]
\centering
\leavevmode
\includegraphics[width=3cm,height=0.8cm]{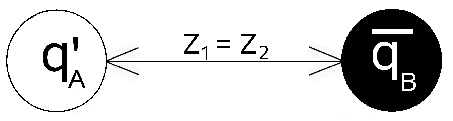}
\label{int19}
\end{figure}
\noindent
where $J_{3}=0$ and $Q=0$. Different $c$'s from $c_{i}=c_{j}=c_{1}$, $c_{i}=c_{j}=c_{2}$, or $c_{i}=c_{j}=c_{3}$ give non-zero $J_{3}$ and $Q$.

\subsubsection{Interaction Between One Particle Field and One Antiparticle Field, or Between One Particle Conjugate Field and One Antiparticle Conjugate Field, in SU(2)} $\\$

Here we have:

\begin{equation}
\displaystyle {\cal D}_{\mu}A_{3}=\partial_{\mu}A_{3}+i\; q_{1}\; c_{i}\; \left(
\begin{array}{cc}
{\cal A}_{4}^{P} \\
{\cal A}_{3}^{P}
\end{array}\right)A_{3}\equiv \partial_{\mu}A_{3}+i\; q_{1}\; c_{i}\; \left(
\begin{array}{cc}
{\cal A}_{4}^{+AP} \\
{\cal A}_{3}^{+AP}
\end{array}
\right)A_{3} \; ,
\label{eq_59}
\end{equation}

\begin{equation}
\displaystyle {\cal D}_{\mu}A_{3}^{+}=\partial_{\mu}A_{3}^{+}+i\; q_{2}\; c_{j}\; \left(
\begin{array}{cc}
{\cal A}_{4}^{AP} \\
{\cal A}_{3}^{AP}
\end{array}
\right)A_{3}^{+}\equiv \partial_{\mu}A_{3}^{+}+i\; q_{2}\; c_{j}\; \left(
\begin{array}{cc}
{\cal A}_{4}^{+P} \\
{\cal A}_{3}^{+P}
\end{array}
\right)A_{3}^{+}
\label{eq_60}
\end{equation}
\noindent
and:

\begin{equation}
\begin{array}{cc}
J_{3}=A_{3}A_{3}^{+}\displaystyle \left[q_{2}\; c_{j}\; \left(
\begin{array}{cc}
{\cal A}_{4}^{AP} \\
{\cal A}_{3}^{AP}
\end{array}
\right) -q_{1}\; c_{i}\; \left(
\begin{array}{cc}
{\cal A}_{4}^{P} \\
{\cal A}_{3}^{P}
\end{array}
\right) \right]\equiv \\
\equiv A_{3}A_{3}^{+}\displaystyle \left[q_{2}\; c_{j}\; \left(
\begin{array}{cc}
{\cal A}_{4}^{+P} \\
{\cal A}_{3}^{+P}
\end{array}
\right) -q_{1}\; c_{i}\; \left(
\begin{array}{cc}
{\cal A}_{4}^{+AP} \\
{\cal A}_{3}^{+AP}
\end{array}
\right) \right]\stackrel{not}{=}A_{3}A_{3}^{+}\displaystyle \left(
\begin{array}{cc}
{\cal Z}_{1} \\
{\cal Z}_{2}
\end{array}
\right)
\end{array} \; .
\label{eq_61}
\end{equation}

{\bf There is no interaction of this type for the b) situation} \cite{deus12} (${\cal Z}_{1}\neq {\cal Z}_{2}$, or ${\cal A}_{3}^{P}={\cal A}_{3}^{AP}={\cal A}_{3}^{+P}={\cal A}_{3}^{+AP}=0$, or the same for ${\cal A}_{4}$'s).

\subsubsection{Intermediary Conclusions} $\\$

The quarks in the "correct" state are the antiparticles of the quarks in physical state, and the quarks in the physical state are the antiparticles of the quarks in "correct" state. This can account for the apparent matter over antimatter excess in the Universe, the antimatter particles being seen as matter particles in a different state. The second reason for the matter excess, let say in an Universe, is that the missing antimatter is to be found in the Anti-Universe, somewhere across the initial singularity (Big Bang).

\subsection{Lepton Weak Interaction (SU(2) with $\alpha_{k}=0$ for the global fields) in Universe}

We will summarize the lepton generations as:

\begin{equation}
\left(
\begin{array}{cc}
l_{A} \\
l'_{A}
\end{array}\right) \equiv
\left(
\begin{array}{cc}
l_{A} \\
\nu_{A}
\end{array}\right) \; ,
\label{eq_62}
\end{equation}
\noindent
where {\it A} indexes the generation. The third self-similarity DEUS level will have on its possible real levels electrons, muons and tauons (the helicoids of the DEUS objects embedded in the catenoid of the third self-similar DEUS level) and neutrinos (the catenoids of the DEUS objects embedded in the catenoid of the third self-similar DEUS level). While the catenoid is the conjugate surface of the helicoid, also, for their FRW conformal particle fields, we have the neutrinos as particle conjugate fields of their associated lepton fields: $l_{A}\leftrightarrow \nu_{A}$.

\subsubsection{Interaction Between Two Particle Fields} $\\$

In a similar way as for the previous cases dealing with quarks, we have:

\begin{equation}
\displaystyle {\cal D}_{\mu}A_{3}=\partial_{\mu}A_{3}+i\; q_{1}\; c_{i}\; \left(
\begin{array}{cc}
{\cal A}_{3}^{P} \\
{\cal A}_{4}^{P}
\end{array}\right)A_{3} \; ,
\label{eq_63}
\end{equation}

\begin{equation}
\displaystyle {\cal D}_{\mu}A_{3}^{+}=\partial_{\mu}A_{3}^{+}+i\; q_{2}\; c_{j}\; \left(
\begin{array}{cc}
{\cal A}_{3}^{P} \\
{\cal A}_{4}^{P}
\end{array}
\right)A_{3}^{+}
\label{eq_64}
\end{equation}
\noindent
and:

\begin{equation}
J_{3}=A_{3}A_{3}^{+}\displaystyle \left[q_{2}\; c_{j}\; \left(
\begin{array}{cc}
{\cal A}_{3}^{P} \\
{\cal A}_{4}^{P}
\end{array}
\right) -q_{1}\; c_{i}\; \left(
\begin{array}{cc}
{\cal A}_{3}^{P} \\
{\cal A}_{4}^{P}
\end{array}
\right) \right]\stackrel{not}{=}A_{3}A_{3}^{+}\displaystyle \left(
\begin{array}{cc}
{\cal Z}_{1} \\
{\cal Z}_{2}
\end{array}
\right)\; .
\label{eq_65}
\end{equation}

If $c_{i}=c_{j}=c_{1}$, $c_{i}=c_{j}=c_{2}$, or $c_{i}=c_{j}=c_{3}$, then ${\cal Z}_{1}= {\cal Z}_{2}$ and $J_{3}=0$, $Q=0$, for the {\bf interaction between two leptons}:

\begin{figure}[h]
\centering
\leavevmode
\includegraphics[width=3cm,height=0.8cm]{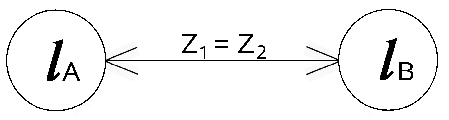}
\label{int20}
\end{figure}

In case of $c_{i}\neq c_{j}$ we can  obtain $J_{3}\neq 0$ and $Q\neq 0$.

Because ${\cal A}_{3}^{P}={\cal A}_{3}^{+AP}$ and ${\cal A}_{4}^{P}={\cal A}_{4}^{+AP}$, when $c_{i}=c_{j}=c_{1}$, $c_{i}=c_{j}=c_{2}$, or $c_{i}=c_{j}=c_{3}$, we obtain the same result when:

\begin{equation}
\displaystyle {\cal D}_{\mu}A_{3}=\partial_{\mu}A_{3}+i\; q_{1}\; c_{i}\; \left(
\begin{array}{cc}
{\cal A}_{3}^{+AP} \\
{\cal A}_{4}^{+AP}
\end{array}\right)A_{3} \; ,
\label{eq_66}
\end{equation}

\begin{equation}
\displaystyle {\cal D}_{\mu}A_{3}^{+}=\partial_{\mu}A_{3}^{+}+i\; q_{2}\; c_{j}\; \left(
\begin{array}{cc}
{\cal A}_{3}^{+AP} \\
{\cal A}_{4}^{+AP}
\end{array}
\right)A_{3}^{+}
\label{eq_67}
\end{equation}
\noindent
and:

\begin{equation}
J_{3}=A_{3}A_{3}^{+}\displaystyle \left[q_{2}\; c_{j}\; \left(
\begin{array}{cc}
{\cal A}_{3}^{+AP} \\
{\cal A}_{4}^{+AP}
\end{array}
\right) -q_{1}\; c_{i}\; \left(
\begin{array}{cc}
{\cal A}_{3}^{+AP} \\
{\cal A}_{4}^{+AP}
\end{array}
\right) \right]\stackrel{not}{=}A_{3}A_{3}^{+}\displaystyle \left(
\begin{array}{cc}
{\cal Z}_{1} \\
{\cal Z}_{2}
\end{array}
\right)\; .
\label{eq_68}
\end{equation}
\noindent
describing an {\bf interaction between two anti-neutrinos} (${\cal Z}_{1}= {\cal Z}_{2}$, $J_{3}=0$, $Q=0$):

\begin{figure}[h]
\centering
\leavevmode
\includegraphics[width=3cm,height=0.8cm]{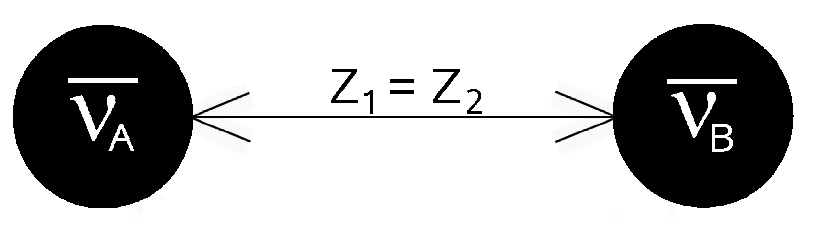}
\label{int21}
\end{figure}

For different generators we can obtain $J_{3}\neq 0$ and $Q\neq 0$.

Based on the same considerents (${\cal A}_{3}^{P}={\cal A}_{3}^{+AP}$ and ${\cal A}_{4}^{P}={\cal A}_{4}^{+AP}$), we have the {\bf interaction between one lepton and one anti-neutrino} (${\cal Z}_{1}= {\cal Z}_{2}$, $J_{3}=0$, $Q=0$ for $c_{i}=c_{j}=c_{1}$, $c_{i}=c_{j}=c_{2}$, or $c_{i}=c_{j}=c_{3}$):

\begin{figure}[h]
\centering
\leavevmode
\includegraphics[width=3cm,height=0.8cm]{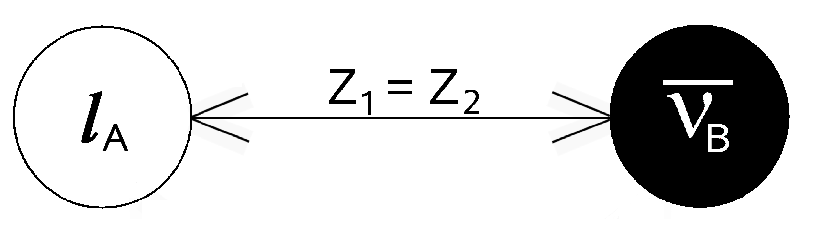}
\label{int22}
\end{figure}
\noindent
where:

\begin{equation}
\displaystyle {\cal D}_{\mu}A_{3}=\partial_{\mu}A_{3}+i\; q_{1}\; c_{i}\; \left(
\begin{array}{cc}
{\cal A}_{3}^{P} \\
{\cal A}_{4}^{P}
\end{array}\right)A_{3} \; ,
\label{eq_69}
\end{equation}

\begin{equation}
\displaystyle {\cal D}_{\mu}A_{3}^{+}=\partial_{\mu}A_{3}^{+}+i\; q_{2}\; c_{j}\; \left(
\begin{array}{cc}
{\cal A}_{3}^{+AP} \\
{\cal A}_{4}^{+AP}
\end{array}
\right)A_{3}^{+}
\label{eq_70}
\end{equation}
\noindent
and:

\begin{equation}
\begin{array}{cc}
J_{3}=A_{3}A_{3}^{+}\displaystyle \left[q_{2}\; c_{j}\; \left(
\begin{array}{cc}
{\cal A}_{3}^{+AP} \\
{\cal A}_{4}^{+AP}
\end{array}
\right) -q_{1}\; c_{i}\; \left(
\begin{array}{cc}
{\cal A}_{3}^{P} \\
{\cal A}_{4}^{P}
\end{array}
\right) \right]= \\
=A_{3}A_{3}^{+}\displaystyle \left[q_{2}\; c_{j}\; \left(
\begin{array}{cc}
{\cal A}_{3}^{P} \\
{\cal A}_{4}^{P}
\end{array}
\right) -q_{1}\; c_{i}\; \left(
\begin{array}{cc}
{\cal A}_{3}^{+AP} \\
{\cal A}_{4}^{+AP}
\end{array}
\right) \right] \stackrel{not}{=}A_{3}A_{3}^{+}\displaystyle \left(
\begin{array}{cc}
{\cal Z}_{1} \\
{\cal Z}_{2}
\end{array}
\right)\; .
\end{array}
\label{eq_71}
\end{equation}

\subsubsection{Interaction Between Two Antiparticle Fields} $\\$

With:

\begin{equation}
\displaystyle {\cal D}_{\mu}A_{3}=\partial_{\mu}A_{3}+i\; q_{1}\; c_{i}\; \left(
\begin{array}{cc}
{\cal A}_{3}^{AP} \\
{\cal A}_{4}^{AP}
\end{array}\right)A_{3} \; ,
\label{eq_72}
\end{equation}

\begin{equation}
\displaystyle {\cal D}_{\mu}A_{3}^{+}=\partial_{\mu}A_{3}^{+}+i\; q_{2}\; c_{j}\; \left(
\begin{array}{cc}
{\cal A}_{3}^{AP} \\
{\cal A}_{4}^{AP}
\end{array}
\right)A_{3}^{+}
\label{eq_73}
\end{equation}
\noindent
and:

\begin{equation}
J_{3}=A_{3}A_{3}^{+}\displaystyle \left[q_{2}\; c_{j}\; \left(
\begin{array}{cc}
{\cal A}_{3}^{AP} \\
{\cal A}_{4}^{AP}
\end{array}
\right) -q_{1}\; c_{i}\; \left(
\begin{array}{cc}
{\cal A}_{3}^{AP} \\
{\cal A}_{4}^{AP}
\end{array}
\right) \right]\stackrel{not}{=}A_{3}A_{3}^{+}\displaystyle \left(
\begin{array}{cc}
{\cal Z}_{1} \\
{\cal Z}_{2}
\end{array}
\right)\; ,
\label{eq_74}
\end{equation}
\newpage
\noindent
where with $c_{i}=c_{j}=c_{1}$, $c_{i}=c_{j}=c_{2}$, or $c_{i}=c_{j}=c_{3}$, we obtain $J_{3}=0$ and $Q=0$ for an {\bf interaction between two anti-leptons}:

\begin{figure}[h]
\centering
\leavevmode
\includegraphics[width=3cm,height=0.8cm]{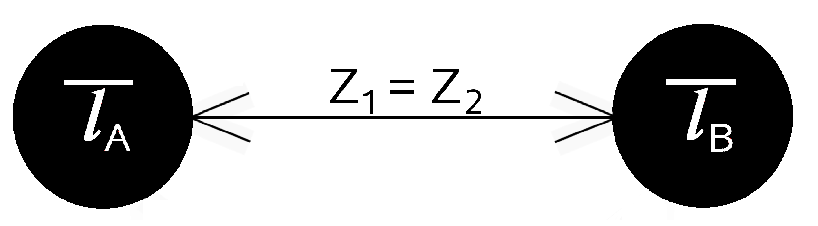}
\label{int23}
\end{figure}

With the same $c$'s (and ${\cal A}_{3}^{+P}={\cal A}_{3}^{AP}$, ${\cal A}_{4}^{+P}={\cal A}_{4}^{AP}$), we obtain the same result (${\cal Z}_{1}= {\cal Z}_{2}$, $J_{3}=0$, $Q=0$) for:

\begin{equation}
\displaystyle {\cal D}_{\mu}A_{3}=\partial_{\mu}A_{3}+i\; q_{1}\; c_{i}\; \left(
\begin{array}{cc}
{\cal A}_{3}^{+P} \\
{\cal A}_{4}^{+P}
\end{array}\right)A_{3} \; ,
\label{eq_75}
\end{equation}

\begin{equation}
\displaystyle {\cal D}_{\mu}A_{3}^{+}=\partial_{\mu}A_{3}^{+}+i\; q_{2}\; c_{j}\; \left(
\begin{array}{cc}
{\cal A}_{3}^{+P} \\
{\cal A}_{4}^{+P}
\end{array}
\right)A_{3}^{+}
\label{eq_76}
\end{equation}
\noindent
and:

\begin{equation}
J_{3}=A_{3}A_{3}^{+}\displaystyle \left[q_{2}\; c_{j}\; \left(
\begin{array}{cc}
{\cal A}_{3}^{+P} \\
{\cal A}_{4}^{+P}
\end{array}
\right) -q_{1}\; c_{i}\; \left(
\begin{array}{cc}
{\cal A}_{3}^{+P} \\
{\cal A}_{4}^{+P}
\end{array}
\right) \right]\stackrel{not}{=}A_{3}A_{3}^{+}\displaystyle \left(
\begin{array}{cc}
{\cal Z}_{1} \\
{\cal Z}_{2}
\end{array}
\right)\; ,
\label{eq_77}
\end{equation}
\noindent
representing the {\bf interaction between two neutrinos}:

\begin{figure}[h]
\centering
\leavevmode
\includegraphics[width=3cm,height=0.8cm]{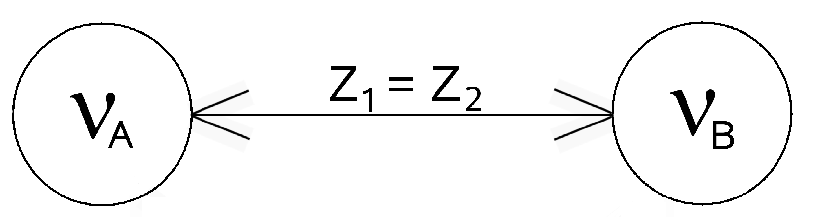}
\label{int24}
\end{figure}
\noindent
or for:

\begin{equation}
\displaystyle {\cal D}_{\mu}A_{3}=\partial_{\mu}A_{3}+i\; q_{1}\; c_{i}\; \left(
\begin{array}{cc}
{\cal A}_{3}^{AP} \\
{\cal A}_{4}^{AP}
\end{array}\right)A_{3}\equiv \partial_{\mu}A_{3}+i\; q_{1}\; c_{i}\; \left(
\begin{array}{cc}
{\cal A}_{3}^{+P} \\
{\cal A}_{4}^{+P}
\end{array}
\right)A_{3} \; ,
\label{eq_78}
\end{equation}

\begin{equation}
\displaystyle {\cal D}_{\mu}A_{3}^{+}=\partial_{\mu}A_{3}^{+}+i\; q_{2}\; c_{j}\; \left(
\begin{array}{cc}
{\cal A}_{3}^{+P} \\
{\cal A}_{4}^{+P}
\end{array}
\right)A_{3}^{+}\equiv \partial_{\mu}A_{3}^{+}+i\; q_{2}\; c_{j}\; \left(
\begin{array}{cc}
{\cal A}_{3}^{AP} \\
{\cal A}_{4}^{AP}
\end{array}
\right)A_{3}^{+}
\label{eq_79}
\end{equation}
\noindent
and:

\begin{equation}
\begin{array}{cc}
J_{3}=A_{3}A_{3}^{+}\displaystyle \left[q_{2}\; c_{j}\; \left(
\begin{array}{cc}
{\cal A}_{3}^{+P} \\
{\cal A}_{4}^{+P}
\end{array}
\right) -q_{1}\; c_{i}\; \left(
\begin{array}{cc}
{\cal A}_{3}^{AP} \\
{\cal A}_{4}^{AP}
\end{array}
\right) \right]\equiv \\
\equiv A_{3}A_{3}^{+}\displaystyle \left[q_{2}\; c_{j}\; \left(
\begin{array}{cc}
{\cal A}_{3}^{AP} \\
{\cal A}_{4}^{AP}
\end{array}
\right) -q_{1}\; c_{i}\; \left(
\begin{array}{cc}
{\cal A}_{3}^{+P} \\
{\cal A}_{4}^{+P}
\end{array}
\right) \right]\stackrel{not}{=}A_{3}A_{3}^{+}\displaystyle \left(
\begin{array}{cc}
{\cal Z}_{1} \\
{\cal Z}_{2}
\end{array}
\right)
\end{array} \; ,
\label{eq_80}
\end{equation}
\noindent
representing the {\bf interaction between one neutrino and one anti-lepton}:

\begin{figure}[h]
\centering
\leavevmode
\includegraphics[width=3cm,height=0.8cm]{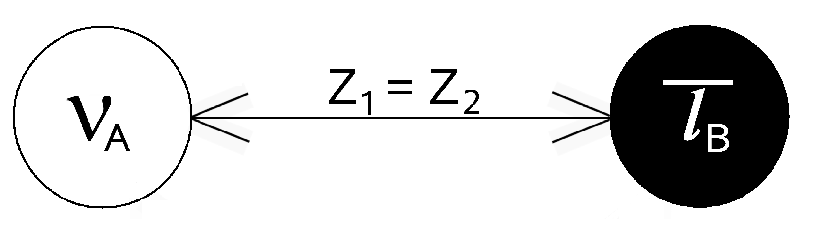}
\label{int25}
\end{figure}
\noindent
where, as for the quark interactions, we have the possibility to interchange ${\cal D}A_{\mu}$ with ${\cal D}A_{\mu}^{+}$, the {\bf interactions between two neutrinos} being equivalent with the {\bf interaction between one neutrino and one anti-lepton}, as an argument for the perception of the external observer that sees an increased amount of neutrinos in the detriment of anti-leptons.

For all the above situations we have, for our $c$ generators, $J_{3}=0$ and $Q=0$.

\subsubsection{Interaction Between One Particle Field and One Antiparticle Field, or Between One Particle Conjugate Field and One Antiparticle Conjugate Field} $\\$

In this case we have:

\begin{equation}
\displaystyle {\cal D}_{\mu}A_{3}=\partial_{\mu}A_{3}+i\; q_{1}\; c_{i}\; \left(
\begin{array}{cc}
{\cal A}_{3}^{P} \\
{\cal A}_{4}^{P}
\end{array}\right)A_{3}\equiv \partial_{\mu}A_{3}+i\; q_{1}\; c_{i}\; \left(
\begin{array}{cc}
{\cal A}_{3}^{+AP} \\
{\cal A}_{4}^{+AP}
\end{array}
\right)A_{3} \; ,
\label{eq_81}
\end{equation}

\begin{equation}
\displaystyle {\cal D}_{\mu}A_{3}^{+}=\partial_{\mu}A_{3}^{+}+i\; q_{2}\; c_{j}\; \left(
\begin{array}{cc}
{\cal A}_{3}^{AP} \\
{\cal A}_{4}^{AP}
\end{array}
\right)A_{3}^{+}\equiv \partial_{\mu}A_{3}^{+}+i\; q_{2}\; c_{j}\; \left(
\begin{array}{cc}
{\cal A}_{3}^{+P} \\
{\cal A}_{4}^{+P}
\end{array}
\right)A_{3}^{+}
\label{eq_82}
\end{equation}
\noindent
and:

\begin{equation}
\begin{array}{cc}
J_{3}=A_{3}A_{3}^{+}\displaystyle \left[q_{2}\; c_{j}\; \left(
\begin{array}{cc}
{\cal A}_{3}^{AP} \\
{\cal A}_{4}^{AP}
\end{array}
\right) -q_{1}\; c_{i}\; \left(
\begin{array}{cc}
{\cal A}_{3}^{P} \\
{\cal A}_{4}^{P}
\end{array}
\right) \right]\equiv \\
\equiv A_{3}A_{3}^{+}\displaystyle \left[q_{2}\; c_{j}\; \left(
\begin{array}{cc}
{\cal A}_{3}^{+P} \\
{\cal A}_{4}^{+P}
\end{array}
\right) -q_{1}\; c_{i}\; \left(
\begin{array}{cc}
{\cal A}_{3}^{+AP} \\
{\cal A}_{4}^{+AP}
\end{array}
\right) \right]\stackrel{not}{=}A_{3}A_{3}^{+}\displaystyle \left(
\begin{array}{cc}
{\cal Z}_{1} \\
{\cal Z}_{2}
\end{array}
\right)
\end{array} \; .
\label{eq_83}
\end{equation}

Because we have ${\cal Z}_{1}\neq {\cal Z}_{2}$ for any $c$ combination, an interaction of this type is not possible here (in the {\bf b)} case). For making this interaction possible we need either ${\cal A}_{3}^{P}={\cal A}_{3}^{AP}=0$, either ${\cal A}_{4}^{P}={\cal A}_{4}^{AP}=0$, which, in both cases is impossible. This interaction will be possible, as we will see, in the {\bf c)} and {\bf a)} cases.

\subsubsection{Intermediary Conclusions} $\\$

The neutrinos (in physical state) are the same particles anti-leptons (in "correct" state). This can account for the apparent matter over antimatter excess in the Universe, the antimatter particles being seen as matter particles in a different state. The second reason for the matter excess, let say in an Universe, is that the missing antimatter is to be found in the Anti-Universe, somewhere across the initial singularity (Big Bang).

\section{Interactions in the {\bf c)} Case from \cite{deus12}}

In the {\bf c)} situation, the global fields ($\alpha=0$) on which the particle and the antiparticle fields over-impose are:

\begin{itemize}
\item
\begin{equation}
\begin{array}{cc}
A_{1}=\displaystyle -i\; \frac{\sqrt{2}}{8}\; \left[4\left(\frac{1}{2}\mp \frac{1}{4} \right)+\frac{\left(\sqrt{2}+1 \right)^{2}}{\displaystyle \frac{1}{2}\mp \frac{1}{4}}\right]\frac{m_{helicoid}}{q}\; \frac{1}{1-\textnormal{tg}\; \alpha}\; \frac{a^{2}}{r^{2}} \; ,
\end{array}
\label{eq_84}
\end{equation}
\noindent
having as particularizations:

\begin{equation}
A_{1}=\displaystyle -i\; \frac{\sqrt{2}}{8}\; \left[4\left(\frac{1}{2}- \frac{1}{4} \right)+\frac{\left(\sqrt{2}+1 \right)^{2}}{\displaystyle \frac{1}{2}- \frac{1}{4}}\right]\frac{m_{helicoid}}{q}\; \frac{1}{1-\textnormal{tg}\; \alpha}\; \frac{a^{2}}{r^{2}} \; ,
\label{eq_85}
\end{equation}
\noindent
for {\bf matter in a matter Universe}, and:

\begin{equation}
A_{1}^{+}=\displaystyle -i\; \frac{\sqrt{2}}{8}\; \left[4\left(\frac{1}{2}- \frac{1}{4} \right)+\frac{\left(\sqrt{2}+1 \right)^{2}}{\displaystyle \frac{1}{2}- \frac{1}{4}}\right]\frac{m_{helicoid}}{q}\; \frac{1}{1-\textnormal{tg}\; \alpha}\; \frac{a^{2}}{r^{2}} \; ,
\label{eq_86}
\end{equation}
\noindent
for {\bf antimatter in an Anti-Universe}.

From $A_{1}=A_{1}^{+}$ for the empty space (${\cal D}_{\mu}=\partial_{\mu}$) results that the current (\ref{eq_1}) is $J_{1}=0$ and the charge (\ref{eq_2}) is $Q=0$.

\item

\begin{equation}
A_{2}=\displaystyle i\; \frac{\sqrt{2}}{8}\; \left[4\left(\frac{1}{2}+ \frac{1}{4} \right)+\frac{\left(\sqrt{2}+1 \right)^{2}}{\displaystyle \frac{1}{2}+ \frac{1}{4}}\right]\frac{m_{helicoid}}{q}\; \frac{1}{1+\textnormal{tg}\; \alpha}\; \frac{a^{2}}{r^{2}}
\label{eq_87}
\end{equation}
\noindent
and, as for $A_{1}$, we have $A_{2}=A_{2}^{+}$. For the empty Universe (and Anti-Universe) the current is $J_{2}=0$ and the charge $Q=0$.

\item

Into a {\bf matter Universe} we have:

\begin{equation}
A_{3}=\displaystyle \pm i\; \frac{1}{8}\; \left[4\left(\frac{1}{2}\mp \frac{1}{4} \right)+\frac{\left(\sqrt{2}+1 \right)^{2}}{\displaystyle \frac{1}{2}\mp \frac{1}{4}}\right]\frac{m_{helicoid}}{q}\; \cos \alpha\; \frac{a^{2}}{r^{2}}\; ,
\label{eq_88}
\end{equation}
\noindent
with the particularizations:

\begin{equation}
A_{3}=\displaystyle i\; \frac{1}{8}\; \left[4\left(\frac{1}{2}+ \frac{1}{4} \right)+\frac{\left(\sqrt{2}+1 \right)^{2}}{\displaystyle \frac{1}{2}+ \frac{1}{4}}\right]\frac{m_{helicoid}}{q}\; \cos \alpha\; \frac{a^{2}}{r^{2}}\; ,
\label{eq_89}
\end{equation}
\noindent
for {\bf matter}, and:

\begin{equation}
A_{3}^{+}=\displaystyle -i\; \frac{1}{8}\; \left[4\left(\frac{1}{2}- \frac{1}{4} \right)+\frac{\left(\sqrt{2}+1 \right)^{2}}{\displaystyle \frac{1}{2}- \frac{1}{4}}\right]\frac{m_{helicoid}}{q}\; \cos \alpha\; \frac{a^{2}}{r^{2}}\; ,
\label{eq_90}
\end{equation}
\noindent
for {\bf antimatter}. This will be the background of the pre-Big Bang (empty) DEUS non-collapsed surviving fields $A_{3}\neq A^{+}_{3}$ on which the interactions will over-impose, and will need our further attention. We will see that the initial current $J_{3}$ and the initial charge $Q$ will be conserved (zero for the FRW) at the formation/annihilation of the FRW Universe and Anti-Universe bubbles due to the matter and antimatter content of the Universe (and the same for the Anti-Universe). In this idea, at the formation of the Universe and of the Anti-Universe, the first fields that are created in both bubbles are $A_{1}$ (and $A_{1}^{+}$), $A_{2}$ (and $A_{2}^{+}$) and $\Phi$ (and $\Phi^{+}$), of the empty FRW bubble spacetimes, followed by the release of the field $A_{3}$ and $A_{3}^{+}$ and the matter (or antimatter) annihilating effect that differentiate the Universe from the Anti-Universe. Being at reach, we will resume ourselves to the Universe fields and matter. Still, for the unrealistic empty space (${\cal D}_{\mu}=\partial_{\mu}$) we have the current $J_{3}=0$ and the charge $Q=0$.

\item
\begin{equation}
\Phi=\Phi^{+}=\displaystyle \frac{1}{8}\left[4\left(\frac{1}{2}\mp \frac{1}{4} \right)+\frac{\left(\sqrt{2}+1 \right)^{2}}{\displaystyle \frac{1}{2}\mp \frac{1}{4}}\right]\frac{m_{helicoid}}{q}\; \frac{a^{2}}{r^{2}}\; ,
\label{eq_91}
\end{equation}
\noindent
with the same results for an empty space: $J_{4}=0$ and $Q=0$.
\end{itemize}

While in the {\bf b)} case we analyzed, from the the external FRW observer perception point of view, the particles of the non-collapsed ($\alpha\neq 0$) DEUS level four and three, in the {\bf c)} case we will study the particles and their interactions for a FRW observation of the same self-similar DEUS objects, but collapsed ($\alpha=0$). As we saw in \cite{deus12}, the generators for the {\bf c)} are $\lambda_{4}$, $\lambda_{5}$, $\lambda_{7}$ and $\lambda_{0}=\{\lambda_{1},\; \lambda_{6},\; \lambda_{8} \}$ for SU(3) and $c_{1}$, $c_{2}$, $c_{3}$ for SU(2).

As said before, on the above $A_{\mu}$ fields over-impose the particle and antiparticle ${\cal A}_{\nu}$ fields (with $\alpha=0$):

\begin{equation}
\left\{ \begin{array}{cc}
{\cal A}_{2}^{N}=\displaystyle i\; \frac{\sqrt{2}}{8}\; \left[4\left(\frac{1}{2}+ \frac{1}{4} \right)+\frac{\left(\sqrt{2}+1 \right)^{2}}{\displaystyle \frac{1}{2}+ \frac{1}{4}}\right]\frac{m_{helicoid}}{q}\; \frac{1}{1+\textnormal{tg}\; \alpha}\; \frac{a^{2}}{r^{2}} \\
{\cal A}_{3}^{N}=\displaystyle \mp i\; \frac{1}{8}\; \left[4\left(\frac{1}{2}\pm \frac{1}{4} \right)+\frac{\left(\sqrt{2}+1 \right)^{2}}{\displaystyle \frac{1}{2}\pm \frac{1}{4}}\right]\frac{m_{helicoid}}{q}\; \cos \alpha\; \frac{a^{2}}{r^{2}}\; ,
\end{array}\right.
\label{eq_92}
\end{equation}

\begin{equation}
\left\{ \begin{array}{cc}
{\cal A}_{2}^{+N}=\displaystyle -i\; \frac{\sqrt{2}}{8}\; \left[4\left(\frac{1}{2}- \frac{1}{4} \right)+\frac{\left(\sqrt{2}+1 \right)^{2}}{\displaystyle \frac{1}{2}- \frac{1}{4}}\right]\frac{m_{helicoid}}{q}\; \frac{1}{1+\textnormal{tg}\; \alpha}\; \frac{a^{2}}{r^{2}}\equiv \left. {\cal A}_{1}^{AN}\right|_{\alpha=0} \\
{\cal A}_{3}^{+N}=\displaystyle \pm i\; \frac{1}{8}\; \left[4\left(\frac{1}{2}\mp \frac{1}{4} \right)+\frac{\left(\sqrt{2}+1 \right)^{2}}{\displaystyle \frac{1}{2}\mp \frac{1}{4}}\right]\frac{m_{helicoid}}{q}\; \cos \alpha\; \frac{a^{2}}{r^{2}}\equiv {\cal A}_{3}^{AN}\; ,
\end{array}\right.
\label{eq_93}
\end{equation}

\begin{equation}
\left\{ \begin{array}{cc}
{\cal A}_{1}^{AN}=\displaystyle -i\; \frac{\sqrt{2}}{8}\; \left[4\left(\frac{1}{2}\mp \frac{1}{4} \right)+\frac{\left(\sqrt{2}+1 \right)^{2}}{\displaystyle \frac{1}{2}\mp \frac{1}{4}}\right]\frac{m_{helicoid}}{q}\; \frac{1}{1-\textnormal{tg}\; \alpha}\; \frac{a^{2}}{r^{2}} \\
{\cal A}_{3}^{AN}=\displaystyle \pm i\; \frac{1}{8}\; \left[4\left(\frac{1}{2}\mp \frac{1}{4} \right)+\frac{\left(\sqrt{2}+1 \right)^{2}}{\displaystyle \frac{1}{2}\mp \frac{1}{4}}\right]\frac{m_{helicoid}}{q}\; \cos \alpha\; \frac{a^{2}}{r^{2}}\; ,
\end{array}\right.
\label{eq_94}
\end{equation}
\noindent
and, more accurately, by comparison with $A_{1}^{+}$ of the global field and with the result obtained for ${\cal A}_{1}^{AP}$ in {\bf b)} situation, with ${\cal A}_{1}^{AN}=\displaystyle -i\; \frac{\sqrt{2}}{8}\; \left[4\left(\frac{1}{2}- \frac{1}{4} \right)+\frac{\left(\sqrt{2}+1 \right)^{2}}{\displaystyle \frac{1}{2}- \frac{1}{4}}\right]\frac{m_{helicoid}}{q}\; \frac{1}{1-\textnormal{tg}\; \alpha}\; \frac{a^{2}}{r^{2}}$\; ,

\begin{equation}
\left\{ \begin{array}{cc}
{\cal A}_{1}^{+AN}=\displaystyle i\; \frac{\sqrt{2}}{8}\; \left[4\left(\frac{1}{2}+ \frac{1}{4} \right)+\frac{\left(\sqrt{2}+1 \right)^{2}}{\displaystyle \frac{1}{2}+ \frac{1}{4}}\right]\frac{m_{helicoid}}{q}\; \frac{1}{1-\textnormal{tg}\; \alpha}\; \frac{a^{2}}{r^{2}}\equiv \left. {\cal A}_{2}^{N}\right|_{\alpha=0} \\
{\cal A}_{3}^{+AN}=\displaystyle \mp i\; \frac{1}{8}\; \left[4\left(\frac{1}{2}\pm \frac{1}{4} \right)+\frac{\left(\sqrt{2}+1 \right)^{2}}{\displaystyle \frac{1}{2}\pm \frac{1}{4}}\right]\frac{m_{helicoid}}{q}\; \cos \alpha\; \frac{a^{2}}{r^{2}}\equiv {\cal A}_{3}^{N}\; .
\end{array}\right.
\label{eq_95}
\end{equation}

\subsection{Baryon Interactions in SU(3)}

\subsubsection{Interaction Between Two Particle Fields in Universe}

\begin{equation}
\displaystyle {\cal D}_{\mu}A_{3}=\partial_{\mu}A_{3}+i\; q_{1}\; \lambda_{i}\; \left(
\begin{array}{ccc}
{\cal A}_{2}^{N} \\
{\cal A}_{3}^{N} \\
{\cal A}_{4}^{N}
\end{array}\right)A_{3} \; ,
\label{eq_96}
\end{equation}

\begin{equation}
\displaystyle {\cal D}_{\mu}A_{3}^{+}=\partial_{\mu}A_{3}^{+}+i\; q_{2}\; \lambda_{j}\; \left(
\begin{array}{ccc}
{\cal A}_{2}^{N} \\
{\cal A}_{3}^{N} \\
{\cal A}_{4}^{N}
\end{array}
\right)A_{3}^{+}
\label{eq_97}
\end{equation}
\noindent
and:

\begin{equation}
\begin{array}{cc}
J_{3}\equiv i\left(A_{3}^{+}{\cal D}_{\mu}A_{3}-A_{3}{\cal D}_{\mu}A_{3}^{+} \right) =A_{3}A_{3}^{+}\displaystyle \left[q_{2}\; \lambda_{j}\; \left(
\begin{array}{ccc}
{\cal A}_{2}^{N} \\
{\cal A}_{3}^{N} \\
{\cal A}_{4}^{N}
\end{array}
\right) -q_{1}\; \lambda_{i}\; \left(
\begin{array}{ccc}
{\cal A}_{2}^{N} \\
{\cal A}_{3}^{N} \\
{\cal A}_{4}^{N}
\end{array}
\right) \right]\equiv \\
\equiv A_{3}A_{3}^{+}\displaystyle \left[q_{2}\; \lambda_{j}\; \left(
\begin{array}{ccc}
{\cal A}_{2}^{N} \\
{\cal A}_{3}^{N} \\
{\cal A}_{4}^{N}
\end{array}
\right) -q_{1}\; \lambda_{i}\; \left(
\begin{array}{ccc}
{\cal A}_{1}^{+AN} \\
{\cal A}_{3}^{+AN} \\
{\cal A}_{4}^{+AN}
\end{array}
\right) \right]\stackrel{not}{=}A_{3}A_{3}^{+}\displaystyle \left(
\begin{array}{cc}
{\cal Z}_{1} \\
{\cal Z}_{2} \\
{\cal Z}_{3}
\end{array}
\right)
\end{array} \; .
\label{eq_98}
\end{equation}

If $\lambda_{i}=\lambda_{j}=\lambda_{0}$ then, naturally, we have $J_{3}=0$ and $Q=0$ for the {\bf interaction between one baryon and a neutrino or an anti-neutrino}. The baryons are described as in the following diagrams:

\begin{figure}[h]
\centering
\leavevmode
\includegraphics[width=4cm,height=4cm]{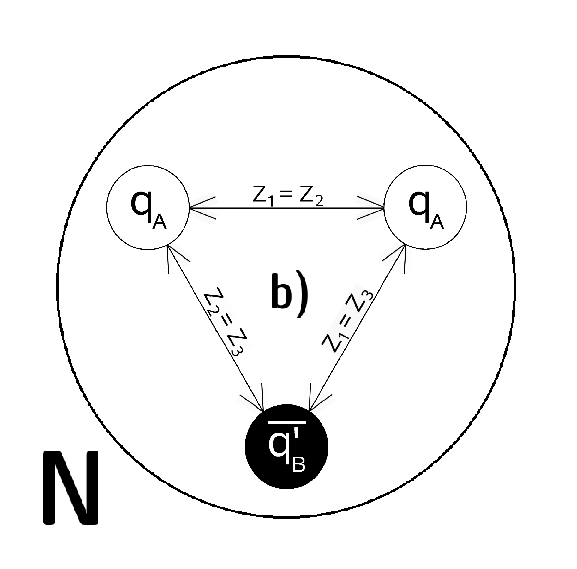}
\label{int34}
\end{figure}
\newpage
\noindent
and:

\begin{figure}[h]
\centering
\leavevmode
\includegraphics[width=4.6cm,height=4cm]{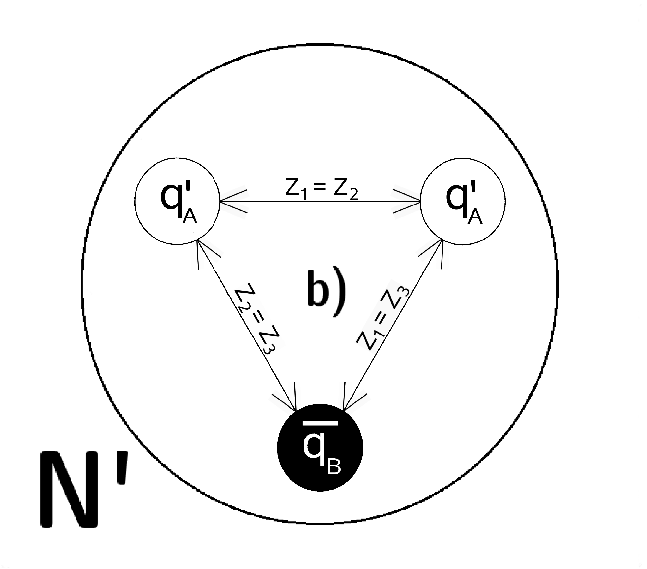}
\label{int35}
\end{figure}
\noindent
where the internal interactions between quarks satisfy the {\bf b)} situation. At the third self-similar (collapsed) DEUS level, the neutrinos are seen in the same way as the fourth self-similar (collapsed) DEUS level's gluons: massles particles.

If $\lambda_{i}=\lambda_{j}=\lambda_{4}$, $\lambda_{i}=\lambda_{j}=\lambda_{5}$, or $\lambda_{i}=\lambda_{j}=\lambda_{7}$, then the condition ${\cal Z}_{1}= {\cal Z}_{2}={\cal Z}_{3}$ is satisfied and we have {\bf interactions between baryons seen as nucleons in a nucleus}:

\begin{figure}[h]
\centering
\leavevmode
\includegraphics[width=4.5cm,height=4cm]{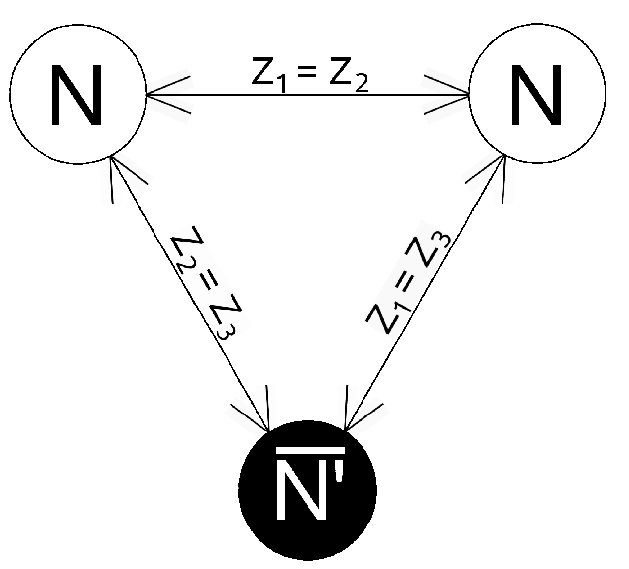}
\label{int27}
\end{figure}
\noindent
or, equivalently:

\begin{figure}[h]
\centering
\leavevmode
\includegraphics[width=8.7cm,height=9cm]{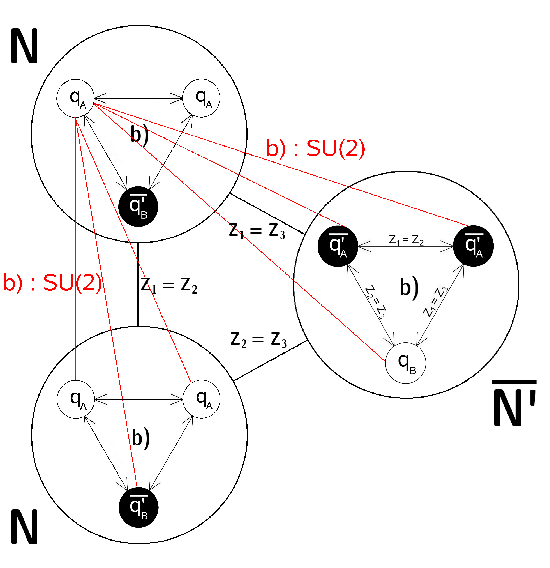}
\label{int26}
\end{figure}
\noindent
and:

\begin{figure}[h]
\centering
\leavevmode
\includegraphics[width=4.5cm,height=4cm]{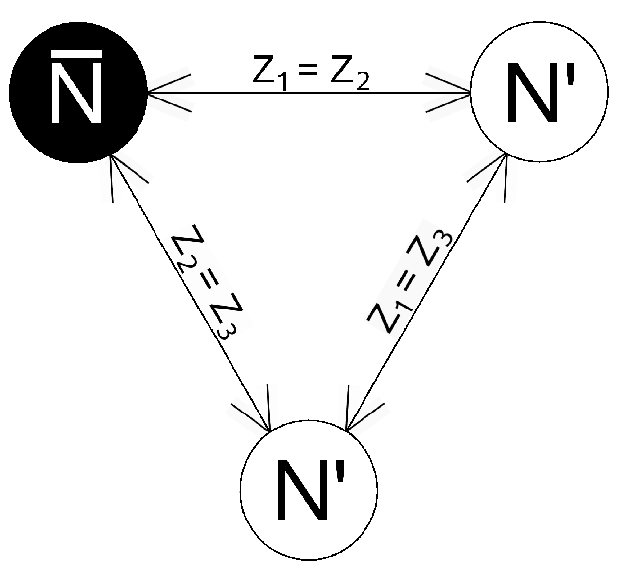}
\label{int29}
\end{figure}
\noindent
or, equivalently:

\begin{figure}[h]
\centering
\leavevmode
\includegraphics[width=8.7cm,height=9cm]{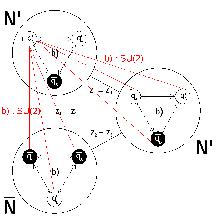}
\label{int28}
\end{figure}
\noindent
where, for the given conditions for $\lambda_{4}$, $\lambda_{5}$ and $\lambda_{7}$, we have $J_{3}=0$ and $Q=0$. In the above representations, the quarks interact with each other as in the {\bf b)} case (the interactions in which one of the quarks enters are marked in red). By adding $N$, $\overline{N}$, $N'$ or $\overline{N'}$ (as, for example, protons and neutrons) in the above schemes we can construct heavy nuclei. The departure from stability of these nuclei is characterized by $J_{3}\neq 0$ and $Q\neq 0$, in consequence the nucleus disintegrating.

\subsubsection{Interaction Between Two Antiparticle Fields in Universe}

\begin{equation}
\displaystyle {\cal D}_{\mu}A_{3}=\partial_{\mu}A_{3}+i\; q_{1}\; \lambda_{i}\; \left(
\begin{array}{ccc}
{\cal A}_{1}^{AN} \\
{\cal A}_{3}^{AN} \\
{\cal A}_{4}^{AN}
\end{array}\right)A_{3} \; ,
\label{eq_99}
\end{equation}

\begin{equation}
\displaystyle {\cal D}_{\mu}A_{3}^{+}=\partial_{\mu}A_{3}^{+}+i\; q_{2}\; \lambda_{j}\; \left(
\begin{array}{ccc}
{\cal A}_{1}^{AN} \\
{\cal A}_{3}^{AN} \\
{\cal A}_{4}^{AN}
\end{array}
\right)A_{3}^{+}
\label{eq_100}
\end{equation}
\noindent
and:

\begin{equation}
\begin{array}{cc}
J_{3}\equiv i\left(A_{3}^{+}{\cal D}_{\mu}A_{3}-A_{3}{\cal D}_{\mu}A_{3}^{+} \right) =A_{3}A_{3}^{+}\displaystyle \left[q_{2}\; \lambda_{j}\; \left(
\begin{array}{ccc}
{\cal A}_{1}^{AN} \\
{\cal A}_{3}^{AN} \\
{\cal A}_{4}^{AN}
\end{array}
\right) -q_{1}\; \lambda_{i}\; \left(
\begin{array}{ccc}
{\cal A}_{1}^{AN} \\
{\cal A}_{3}^{AN} \\
{\cal A}_{4}^{AN}
\end{array}
\right) \right]\equiv \\
\equiv A_{3}A_{3}^{+}\displaystyle \left[q_{2}\; \lambda_{j}\; \left(
\begin{array}{ccc}
{\cal A}_{1}^{AN} \\
{\cal A}_{3}^{AN} \\
{\cal A}_{4}^{AN}
\end{array}
\right) -q_{1}\; \lambda_{i}\; \left(
\begin{array}{ccc}
{\cal A}_{2}^{+N} \\
{\cal A}_{3}^{+N} \\
{\cal A}_{4}^{+N}
\end{array}
\right) \right]\stackrel{not}{=}A_{3}A_{3}^{+}\displaystyle \left(
\begin{array}{ccc}
{\cal Z}_{1} \\
{\cal Z}_{2} \\
{\cal Z}_{3}
\end{array}
\right)
\end{array} \; .
\label{eq_101}
\end{equation}

If $\lambda_{i}=\lambda_{j}=\lambda_{0}$, we obtain $J_{3}=0$ and $Q=0$ for the {\bf interaction between one anti-baryon and a neutrino or an anti-neutrino}. The anti-baryons are described as in the following diagrams:

\begin{figure}[h]
\centering
\leavevmode
\includegraphics[width=4.3cm,height=4cm]{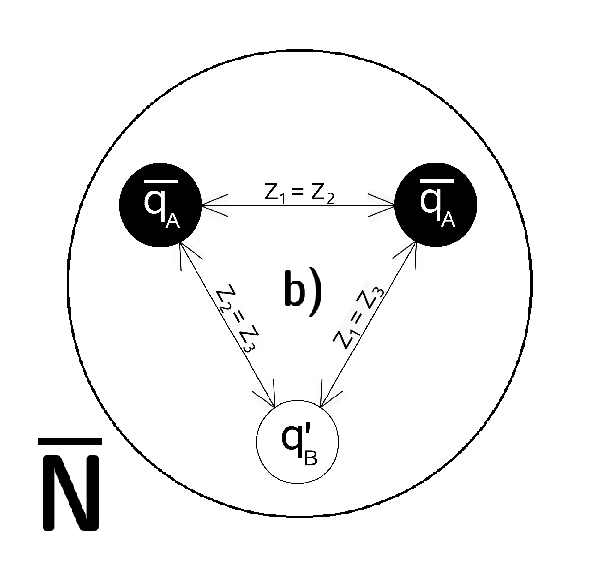}
\label{int36}
\end{figure}
\noindent
and:

\begin{figure}[h]
\centering
\leavevmode
\includegraphics[width=4.2cm,height=4cm]{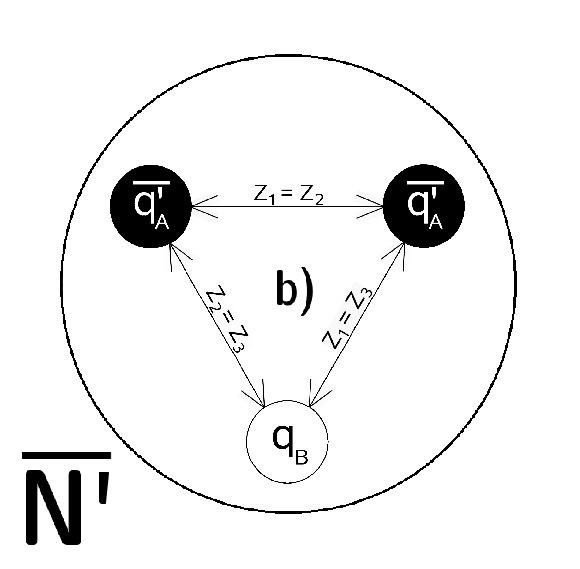}
\label{int37}
\end{figure}
\noindent
where the internal interactions between quarks satisfy the {\bf b)} situation.

If $\lambda_{i}=\lambda_{j}=\lambda_{4}$, $\lambda_{i}=\lambda_{j}=\lambda_{5}$, or $\lambda_{i}=\lambda_{j}=\lambda_{7}$, then the condition ${\cal Z}_{1}= {\cal Z}_{2}={\cal Z}_{3}$ is satisfied and we have {\bf interactions between anti-baryons seen as anti-nucleons in an anti-nucleus}:

\begin{figure}[h]
\centering
\leavevmode
\includegraphics[width=4.5cm,height=4cm]{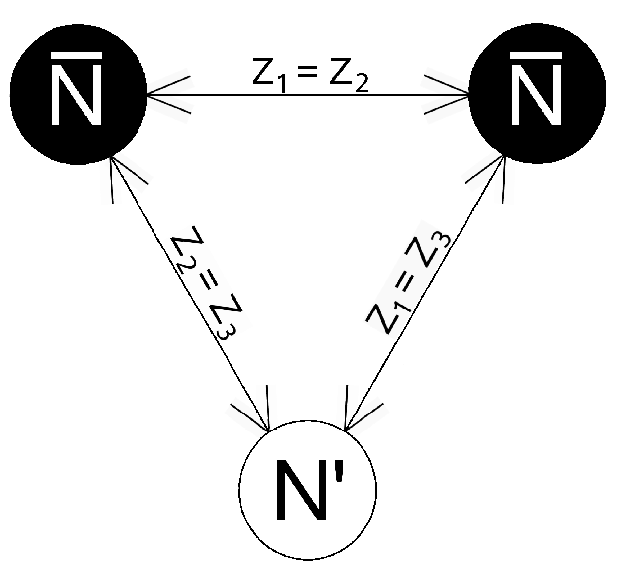}
\label{int31}
\end{figure}
\newpage
\noindent
or, equivalently:

\begin{figure}[h]
\centering
\leavevmode
\includegraphics[width=8.7cm,height=9cm]{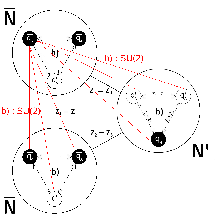}
\label{int30}
\end{figure}
\noindent
and:

\begin{figure}[h]
\centering
\leavevmode
\includegraphics[width=4.5cm,height=4cm]{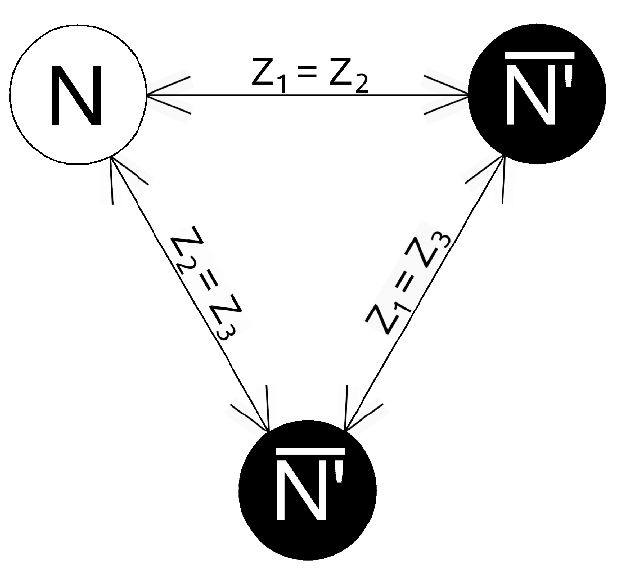}
\label{int33}
\end{figure}
\newpage
\noindent
or, equivalently:

\begin{figure}[h]
\centering
\leavevmode
\includegraphics[width=8.7cm,height=9cm]{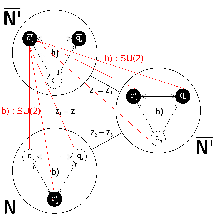}
\label{int32}
\end{figure}
\noindent
where for established $\lambda$ conditions, $J_{3}=0$ and $Q=0$. In the above representations, the anti-quarks interact with each other as in the {\bf b)} case (the interactions in which one of the anti-quarks enters are marked in red). By adding $N$, $\overline{N}$, $N'$ or $\overline{N'}$ (as, for example, anti-protons and anti-neutrons) in the above schemes we can construct heavy anti-nuclei. The departure from stability of this anti-nuclei is characterized by $J_{3}\neq 0$ and $Q\neq 0$, in consequence the anti-nucleus disintegrating.

\subsubsection{Interaction Between One Particle Field and One Antiparticle Field in Universe}

\begin{equation}
\displaystyle {\cal D}_{\mu}A_{3}=\partial_{\mu}A_{3}+i\; q_{1}\; \lambda_{i}\; \left(
\begin{array}{ccc}
{\cal A}_{2}^{N} \\
{\cal A}_{3}^{N} \\
{\cal A}_{4}^{N}
\end{array}\right)A_{3}\equiv \partial_{\mu}A_{3}+i\; q_{1}\; \lambda_{i}\; \left(
\begin{array}{ccc}
{\cal A}_{1}^{+AN} \\
{\cal A}_{3}^{+AN} \\
{\cal A}_{4}^{+AN}
\end{array}
\right)A_{3} \; ,
\label{eq_102}
\end{equation}

\begin{equation}
\displaystyle {\cal D}_{\mu}A_{3}^{+}=\partial_{\mu}A_{3}^{+}+i\; q_{2}\; \lambda_{j}\; \left(
\begin{array}{ccc}
{\cal A}_{1}^{AN} \\
{\cal A}_{3}^{AN} \\
{\cal A}_{4}^{AN}
\end{array}
\right)A_{3}^{+}\equiv \partial_{\mu}A_{3}^{+}+i\; q_{2}\; \lambda_{j}\; \left(
\begin{array}{ccc}
{\cal A}_{2}^{+N} \\
{\cal A}_{3}^{+N} \\
{\cal A}_{4}^{+N}
\end{array}
\right)A_{3}^{+}
\label{eq_103}
\end{equation}
\noindent
and:

\begin{equation}
\begin{array}{ccc}
\!\!\!\!\!\!\!\!\!\!\!\!\!\!\!\!\!\!\!\!\!\!\!\!\!\!\!\!\!\!\!\!\!\!\!\!\!\!\!\!\!J_{3}=A_{3}A_{3}^{+}\displaystyle \left[q_{2}\; \lambda_{j}\; \left(
\begin{array}{ccc}
{\cal A}_{1}^{AN} \\
{\cal A}_{3}^{AN} \\
{\cal A}_{4}^{AN}
\end{array}
\right) -q_{1}\; \lambda_{i}\; \left(
\begin{array}{ccc}
{\cal A}_{2}^{N} \\
{\cal A}_{3}^{N} \\
{\cal A}_{4}^{N}
\end{array}
\right) \right]\equiv
A_{3}A_{3}^{+}\displaystyle \left[q_{2}\; \lambda_{j}\; \left(
\begin{array}{ccc}
{\cal A}_{2}^{+N} \\
{\cal A}_{3}^{+N} \\
{\cal A}_{4}^{+N}
\end{array}
\right) -q_{1}\; \lambda_{i}\; \left(
\begin{array}{ccc}
{\cal A}_{2}^{N} \\
{\cal A}_{3}^{N} \\
{\cal A}_{4}^{N}
\end{array}
\right) \right] \equiv \\
\!\!\!\!\!\!\!\!\!\!\!\!\!\!\!\!\!\!\!\!\!\!\!\!\!\!\!\!\!\!\!\!\!\!\!\!\!\!\!\!\!\equiv A_{3}A_{3}^{+}\displaystyle \left[q_{2}\; \lambda_{j}\; \left(
\begin{array}{ccc}
{\cal A}_{1}^{AN} \\
{\cal A}_{3}^{AN} \\
{\cal A}_{4}^{AN}
\end{array}
\right) -q_{1}\; \lambda_{i}\; \left(
\begin{array}{ccc}
{\cal A}_{1}^{+AN} \\
{\cal A}_{3}^{+AN} \\
{\cal A}_{4}^{+AN}
\end{array}
\right) \right] \equiv
A_{3}A_{3}^{+}\displaystyle \left[q_{2}\; \lambda_{j}\; \left(
\begin{array}{ccc}
{\cal A}_{2}^{+N} \\
{\cal A}_{3}^{+N} \\
{\cal A}_{4}^{+N}
\end{array}
\right) -q_{1}\; \lambda_{i}\; \left(
\begin{array}{ccc}
{\cal A}_{1}^{+AN} \\
{\cal A}_{3}^{+AN} \\
{\cal A}_{4}^{+AN}
\end{array}
\right) \right]\stackrel{not}{=} \\
\!\!\!\!\!\!\!\!\!\!\!\!\!\!\!\!\!\!\!\!\!\!\!\!\!\!\!\!\!\!\!\!\!\!\!\!\!\!\!\!\!\stackrel{not}{=}A_{3}A_{3}^{+}\displaystyle \left(
\begin{array}{ccc}
{\cal Z}_{1} \\
{\cal Z}_{2} \\
{\cal Z}_{3}
\end{array}
\right)
\end{array} \; .
\label{eq_104}
\end{equation}

If $\lambda_{i}=\lambda_{j}=\lambda_{0}$, we obtain $J_{3}=0$ and $Q=0$ for the {\bf interaction between one baryon or anti-baryon with a neutrino or an anti-neutrino}.

If $\lambda_{i}=\lambda_{j}=\lambda_{4}$, $\lambda_{i}=\lambda_{j}=\lambda_{5}$, or $\lambda_{i}=\lambda_{j}=\lambda_{7}$, we have the previous two situations: {\bf interaction between baryons} or {\bf interaction between anti-baryons}.

\subsection{Baryon Interactions in SU(2)}

\subsubsection{Interaction Between Two Particle Fields in Universe}

\begin{equation}
\displaystyle {\cal D}_{\mu}A_{3}=\partial_{\mu}A_{3}+i\; q_{1}\; c_{i}\; \left(
\begin{array}{cc}
{\cal A}_{4}^{N} \\
{\cal A}_{3}^{N}
\end{array}\right)A_{3} \; ,
\label{eq_105}
\end{equation}

\begin{equation}
\displaystyle {\cal D}_{\mu}A_{3}^{+}=\partial_{\mu}A_{3}^{+}+i\; q_{2}\; c_{j}\; \left(
\begin{array}{cc}
{\cal A}_{4}^{N} \\
{\cal A}_{3}^{N}
\end{array}
\right)A_{3}^{+}
\label{eq_106}
\end{equation}
\noindent
and:

\begin{equation}
\begin{array}{cc}
J_{3}\equiv i\left(A_{3}^{+}{\cal D}_{\mu}A_{3}-A_{3}{\cal D}_{\mu}A_{3}^{+} \right) =A_{3}A_{3}^{+}\displaystyle \left[q_{2}\; c_{j}\; \left(
\begin{array}{cc}
{\cal A}_{4}^{N} \\
{\cal A}_{3}^{N}
\end{array}
\right) -q_{1}\; c_{i}\; \left(
\begin{array}{cc}
{\cal A}_{4}^{N} \\
{\cal A}_{3}^{N}
\end{array}
\right) \right]\equiv \\
\equiv A_{3}A_{3}^{+}\displaystyle \left[q_{2}\; c_{j}\; \left(
\begin{array}{cc}
{\cal A}_{4}^{N} \\
{\cal A}_{3}^{N}
\end{array}
\right) -q_{1}\; c_{i}\; \left(
\begin{array}{cc}
{\cal A}_{4}^{+AN} \\
{\cal A}_{3}^{+AN}
\end{array}
\right) \right]\stackrel{not}{=}A_{3}A_{3}^{+}\displaystyle \left(
\begin{array}{cc}
{\cal Z}_{1} \\
{\cal Z}_{2}
\end{array}
\right)
\end{array} \; .
\label{eq_107}
\end{equation}

If $c_{i}=c_{j}=c_{1}$, $c_{i}=c_{j}=c_{2}$, or $c_{i}=c_{j}=c_{3}$, results that ${\cal Z}_{1}={\cal Z}_{2}=0$ and, in consequence, $J_{3}=0$ and $Q=0$. In this case we have the {\bf interaction between two baryons}:

\begin{figure}[h]
\centering
\leavevmode
\includegraphics[width=3cm,height=0.8cm]{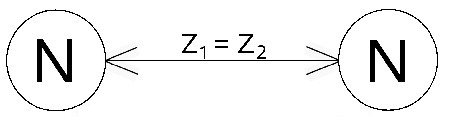}
\label{int39}
\end{figure}
\noindent
or, equivalently:

\begin{figure}[h]
\centering
\leavevmode
\includegraphics[width=8.3cm,height=4cm]{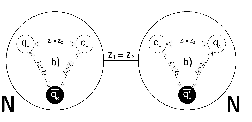}
\label{int38}
\end{figure}
\noindent
or, the {\bf interaction between one baryon (for example, a proton) and one anti-baryon conjugate (for example, the anti-proton conjugate, an anti-neutron)}:

\begin{figure}[h]
\centering
\leavevmode
\includegraphics[width=3cm,height=0.8cm]{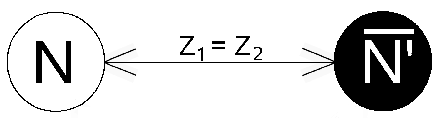}
\label{int41}
\end{figure}
\noindent
or, equivalently:

\begin{figure}[h]
\centering
\leavevmode
\includegraphics[width=8.3cm,height=4cm]{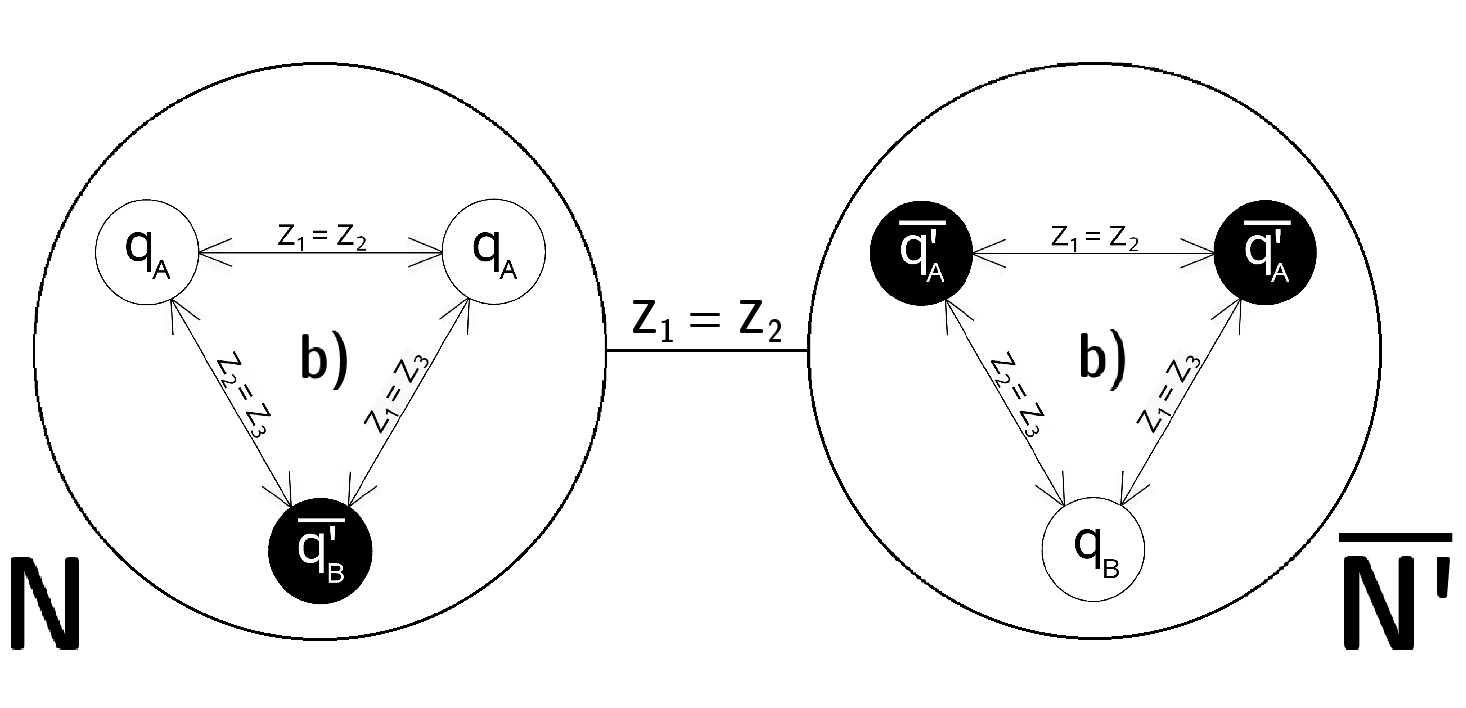}
\label{int40}
\end{figure}

Here we have also the way in which, for example, one supplementary proton or neutron "couples" in the process of heavy nuclei formation. At the previous SU(3) strong interacting nucleons we can couple one proton or neutron that interacts weakly with each of the other nucleons. In this way we can construct the heavy nuclei that do not satisfy the rules $k\times \left(2 \times \textnormal{neutron}+\textnormal{proton} \right)$ or $k\times \left(2 \times \textnormal{proton} + \textnormal{neutron}\right)$, where $k\in\mathbb{N}^{*}$.

\subsubsection{Interaction Between Two Antiparticle Fields in Universe}

\begin{equation}
\displaystyle {\cal D}_{\mu}A_{3}=\partial_{\mu}A_{3}+i\; q_{1}\; c_{i}\; \left(
\begin{array}{cc}
{\cal A}_{4}^{AN} \\
{\cal A}_{3}^{AN}
\end{array}\right)A_{3} \; ,
\label{eq_108}
\end{equation}

\begin{equation}
\displaystyle {\cal D}_{\mu}A_{3}^{+}=\partial_{\mu}A_{3}^{+}+i\; q_{2}\; c_{j}\; \left(
\begin{array}{cc}
{\cal A}_{4}^{AN} \\
{\cal A}_{3}^{AN}
\end{array}
\right)A_{3}^{+}
\label{eq_109}
\end{equation}
\noindent
and:

\begin{equation}
\begin{array}{cc}
J_{3}\equiv i\left(A_{3}^{+}{\cal D}_{\mu}A_{3}-A_{3}{\cal D}_{\mu}A_{3}^{+} \right) =A_{3}A_{3}^{+}\displaystyle \left[q_{2}\; c_{j}\; \left(
\begin{array}{cc}
{\cal A}_{4}^{AN} \\
{\cal A}_{3}^{AN}
\end{array}
\right) -q_{1}\; c_{i}\; \left(
\begin{array}{cc}
{\cal A}_{4}^{AN} \\
{\cal A}_{3}^{AN}
\end{array}
\right) \right]\equiv \\
\equiv A_{3}A_{3}^{+}\displaystyle \left[q_{2}\; c_{j}\; \left(
\begin{array}{cc}
{\cal A}_{4}^{AN} \\
{\cal A}_{3}^{AN}
\end{array}
\right) -q_{1}\; c_{i}\; \left(
\begin{array}{cc}
{\cal A}_{4}^{+N} \\
{\cal A}_{3}^{+N}
\end{array}
\right) \right]\stackrel{not}{=}A_{3}A_{3}^{+}\displaystyle \left(
\begin{array}{cc}
{\cal Z}_{1} \\
{\cal Z}_{2}
\end{array}
\right)
\end{array} \; .
\label{eq_110}
\end{equation}

When $c_{i}=c_{j}=c_{1}$, $c_{i}=c_{j}=c_{2}$, or $c_{i}=c_{j}=c_{3}$, results ${\cal Z}_{1}={\cal Z}_{2}=0$ and, from this, $J_{3}=0$ and $Q=0$. In this case we have the {\bf interaction between two anti-baryons}:

\begin{figure}[h]
\centering
\leavevmode
\includegraphics[width=3cm,height=0.8cm]{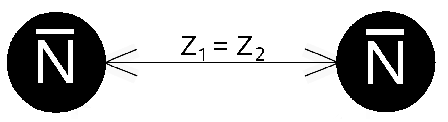}
\label{int43}
\end{figure}
\noindent
or, equivalently:

\begin{figure}[h]
\centering
\leavevmode
\includegraphics[width=8.3cm,height=4cm]{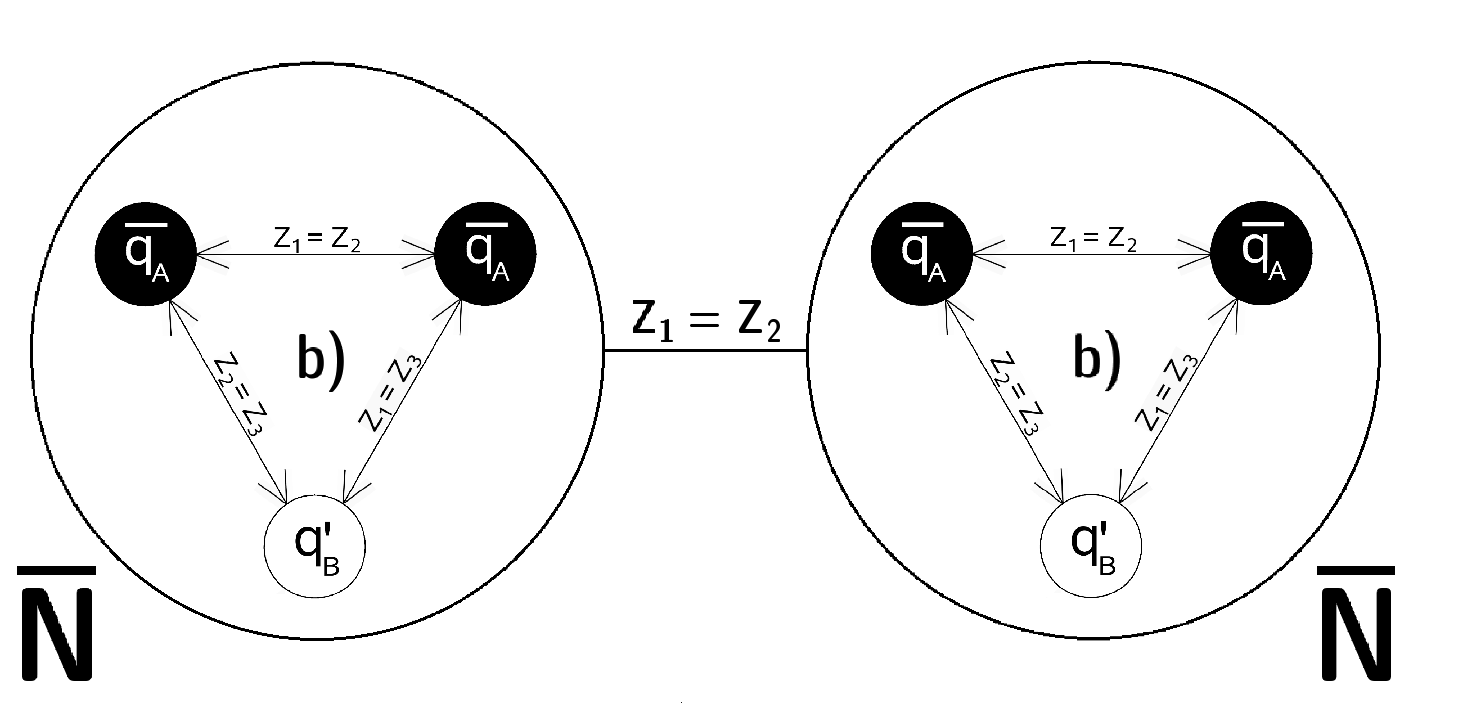}
\label{int42}
\end{figure}
\noindent
or, the {\bf interaction between one anti-baryon (for example, an anti-neutron) and one baryon conjugate (for example, the neutron conjugate, a proton)}:

\begin{figure}[h]
\centering
\leavevmode
\includegraphics[width=3cm,height=0.8cm]{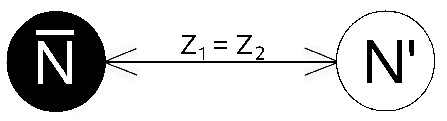}
\label{int45}
\end{figure}
\noindent
or, equivalently:

\begin{figure}[h]
\centering
\leavevmode
\includegraphics[width=8.3cm,height=4cm]{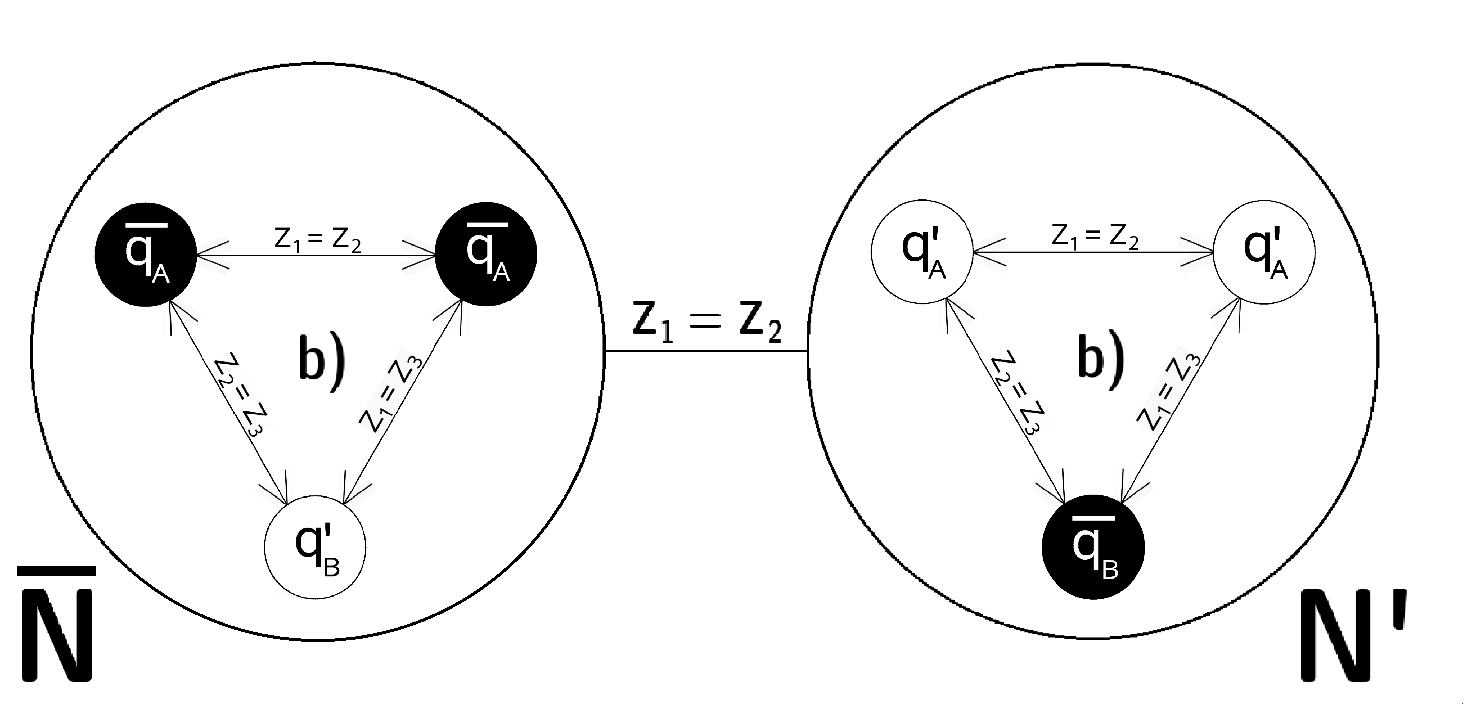}
\label{int44}
\end{figure}
\noindent
or, the {\bf interaction between two baryon conjugates}.

The comments valid here are the same as for the previous case but for anti-baryons and baryon conjugates instead of baryons and anti-baryon conjugates.

\subsubsection{Interaction Between One Particle Field and One Antiparticle Field in Universe}

\begin{equation}
\displaystyle {\cal D}_{\mu}A_{3}=\partial_{\mu}A_{3}+i\; q_{1}\; c_{i}\; \left(
\begin{array}{cc}
{\cal A}_{4}^{N} \\
{\cal A}_{3}^{N}
\end{array}\right)A_{3} \; ,
\label{eq_111}
\end{equation}

\begin{equation}
\displaystyle {\cal D}_{\mu}A_{3}^{+}=\partial_{\mu}A_{3}^{+}+i\; q_{2}\; c_{j}\; \left(
\begin{array}{cc}
{\cal A}_{4}^{AN} \\
{\cal A}_{3}^{AN}
\end{array}
\right)A_{3}^{+}
\label{eq_112}
\end{equation}
\noindent
and:

\begin{equation}
\begin{array}{ccc}
\!\!\!\!\!\!\!\!\!\!\!\!\!\!\!\!\!\!\!\!\!\!\!\!\!\!\!\!\!\!\!\!\!\!\!\!\!\!\!\!\!J_{3}=A_{3}A_{3}^{+}\displaystyle \left[q_{2}\; c_{j}\; \left(
\begin{array}{ccc}
{\cal A}_{4}^{AN} \\
{\cal A}_{3}^{AN}
\end{array}
\right) -q_{1}\; c_{i}\; \left(
\begin{array}{cc}
{\cal A}_{4}^{N} \\
{\cal A}_{3}^{N}
\end{array}
\right) \right]\equiv
A_{3}A_{3}^{+}\displaystyle \left[q_{2}\; c_{j}\; \left(
\begin{array}{cc}
{\cal A}_{4}^{AN} \\
{\cal A}_{3}^{AN}
\end{array}
\right) -q_{1}\; c_{i}\; \left(
\begin{array}{cc}
{\cal A}_{4}^{+AN} \\
{\cal A}_{3}^{+AN}
\end{array}
\right) \right] \equiv \\
\!\!\!\!\!\!\!\!\!\!\!\!\!\!\!\!\!\!\!\!\!\!\!\!\!\!\!\!\!\!\!\!\!\!\!\!\!\!\!\!\!\equiv A_{3}A_{3}^{+}\displaystyle \left[q_{2}\; c_{j}\; \left(
\begin{array}{cc}
{\cal A}_{4}^{+N} \\
{\cal A}_{3}^{+N}
\end{array}
\right) -q_{1}\; c_{i}\; \left(
\begin{array}{cc}
{\cal A}_{4}^{N} \\
{\cal A}_{3}^{N}
\end{array}
\right) \right] \equiv
A_{3}A_{3}^{+}\displaystyle \left[q_{2}\; c_{j}\; \left(
\begin{array}{cc}
{\cal A}_{4}^{+N} \\
{\cal A}_{3}^{+N}
\end{array}
\right) -q_{1}\; c_{i}\; \left(
\begin{array}{ccc}
{\cal A}_{4}^{+AN} \\
{\cal A}_{3}^{+AN}
\end{array}
\right) \right]\stackrel{not}{=} \\
\!\!\!\!\!\!\!\!\!\!\!\!\!\!\!\!\!\!\!\!\!\!\!\!\!\!\!\!\!\!\!\!\!\!\!\!\!\!\!\!\!\stackrel{not}{=}A_{3}A_{3}^{+}\displaystyle \left(
\begin{array}{cc}
{\cal Z}_{1} \\
{\cal Z}_{2}
\end{array}
\right)
\end{array} \; .
\label{eq_113}
\end{equation}

Is not possible to achieve this interaction for neither of our generators (${\cal Z}_{1}\neq {\cal Z}_{2}$). For this interaction to take place we need that ${\cal A}_{3}^{N}={\cal A}_{3}^{+N}={\cal A}_{3}^{AN}={\cal A}_{3}^{+AN}=0$, or ${\cal A}_{4}^{N}={\cal A}_{4}^{+N}={\cal A}_{4}^{AN}={\cal A}_{4}^{+AN}=0$, which, in both cases is impossible. This interaction is possible, as we will see, for the {\bf c)} case fermions and for the {\bf a)} case global interactions. 

\subsection{Intermediary Conclusions (for Baryons)}

Without supplementary rules, from the above cases results that the anti-neutron is the same particle as the proton and the anti-proton is the same particle as the neutron, where the particle is a "time" object in our Universe, while its antiparticle conjugate is a "before time" (pre-Big Bang) object of a previous FRW bubble evolution. So, from FRW point of view, at the moment of initial/final singularity crossing, the antiparticles conjugates are transformed in particles (and the Universe to an Anti-Universe "mirror"). The "before time" objects are dark matter effects for the "time" Universe.  

\subsection{Meson and Fermion Interactions in SU(2)}

\subsubsection{Interaction Between Two Particle Fields in Universe}

\begin{equation}
\displaystyle {\cal D}_{\mu}A_{3}=\partial_{\mu}A_{3}+i\; q_{1}\; c_{i}\; \left(
\begin{array}{cc}
{\cal A}_{3}^{N} \\
{\cal A}_{4}^{N}
\end{array}\right)A_{3} \; ,
\label{eq_114}
\end{equation}

\begin{equation}
\displaystyle {\cal D}_{\mu}A_{3}^{+}=\partial_{\mu}A_{3}^{+}+i\; q_{2}\; c_{j}\; \left(
\begin{array}{cc}
{\cal A}_{3}^{N} \\
{\cal A}_{4}^{N}
\end{array}
\right)A_{3}^{+}
\label{eq_115}
\end{equation}
\noindent
and:

\begin{equation}
\begin{array}{cc}
J_{3}\equiv i\left(A_{3}^{+}{\cal D}_{\mu}A_{3}-A_{3}{\cal D}_{\mu}A_{3}^{+} \right) =A_{3}A_{3}^{+}\displaystyle \left[q_{2}\; c_{j}\; \left(
\begin{array}{cc}
{\cal A}_{3}^{N} \\
{\cal A}_{4}^{N}
\end{array}
\right) -q_{1}\; c_{i}\; \left(
\begin{array}{cc}
{\cal A}_{3}^{N} \\
{\cal A}_{4}^{N}
\end{array}
\right) \right]\equiv \\
\equiv A_{3}A_{3}^{+}\displaystyle \left[q_{2}\; c_{j}\; \left(
\begin{array}{cc}
{\cal A}_{3}^{N} \\
{\cal A}_{4}^{N}
\end{array}
\right) -q_{1}\; c_{i}\; \left(
\begin{array}{cc}
{\cal A}_{3}^{+AN} \\
{\cal A}_{4}^{+AN}
\end{array}
\right) \right]\stackrel{not}{=}A_{3}A_{3}^{+}\displaystyle \left(
\begin{array}{cc}
{\cal Z}_{1} \\
{\cal Z}_{2}
\end{array}
\right)
\end{array} \; .
\label{eq_116}
\end{equation}

Only for $c_{i}=c_{j}=c_{1}$, $c_{i}=c_{j}=c_{2}$, or $c_{i}=c_{j}=c_{3}$, results ${\cal Z}_{1}={\cal Z}_{2}=0$ and, from this, $J_{3}=0$ and $Q=0$. In this case we have:
\begin{itemize}
\item {\bf interaction between two mesons} (if the particles are composed of {\bf b)} case weakly interacting quarks from the fourth self-similar DEUS level):

\begin{figure}[h]
\centering
\leavevmode
\includegraphics[width=3cm,height=0.8cm]{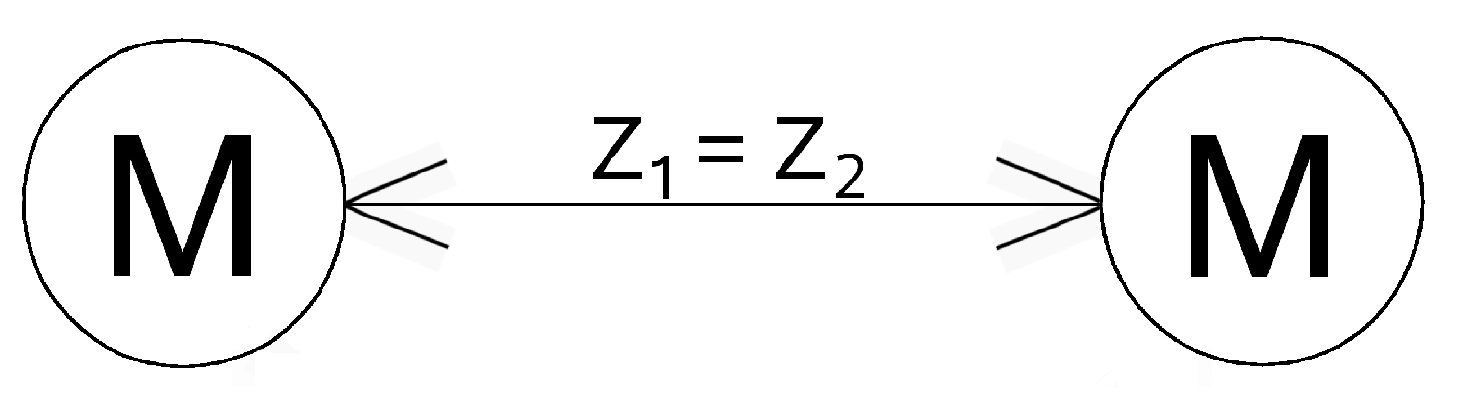}
\label{int46}
\end{figure}
\newpage
\noindent
or, equivalently:

\begin{figure}[h]
\centering
\leavevmode
\includegraphics[width=8.2cm,height=4cm]{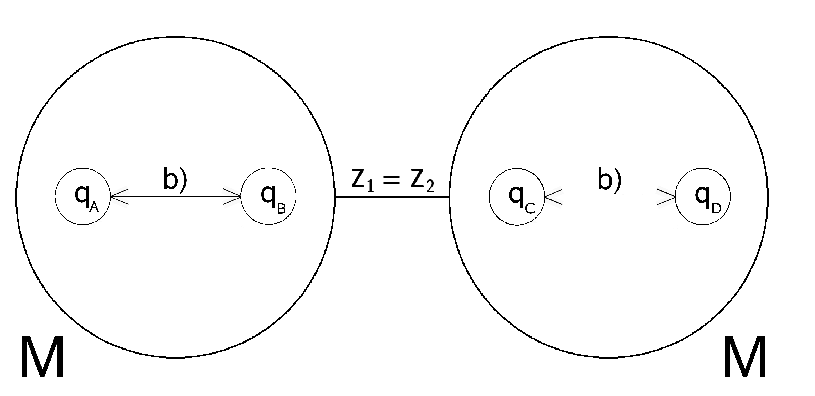}
\label{int47}
\end{figure}
\noindent
or, from the {\bf b)} case field-conjugate field equivalence:

\begin{figure}[h]
\centering
\leavevmode
\includegraphics[width=8.2cm,height=4cm]{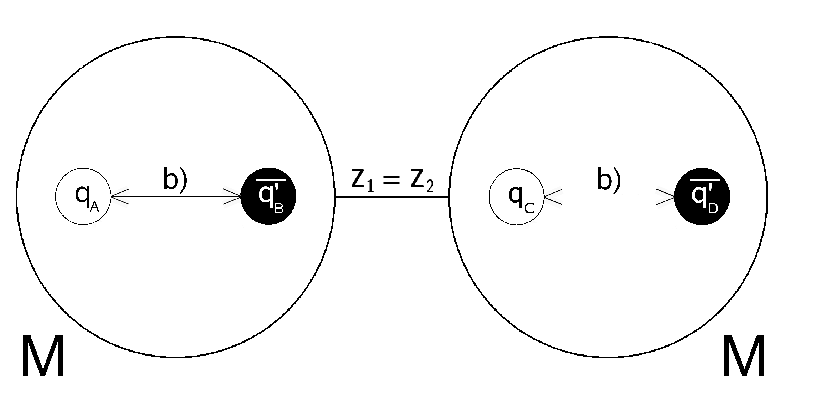}
\label{int48}
\end{figure}

\item {\bf interaction between two "before time" quarks with production of other two "time" quarks} (if the particles are non-bounded quarks from the fourth self-similar DEUS level): $q_{A}+q_{B}\rightarrow q_{C}+q_{D}$, or $q_{A}+\overline{q'_{B}}\rightarrow q_{C}+\overline{q'_{D}}$, etc.;

\item {\bf interaction between two lepton pairs} (if the particles are composed of two by two {\bf b)} case weakly interacting leptons from the third self-similar DEUS level; in the FRW spacetime the link between two leptons on the same level three real state catenoid is symbolized by the Pauli exclusion principle):

\begin{figure}[h]
\centering
\leavevmode
\includegraphics[width=8.2cm,height=4cm]{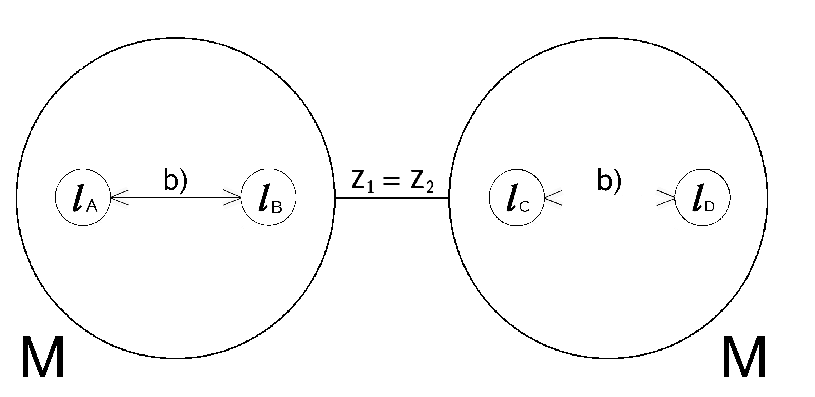}
\label{int49}
\end{figure}
\newpage
\noindent
or, from the {\bf b)} case field-anti-field conjugate equivalence:

\begin{figure}[h]
\centering
\leavevmode
\includegraphics[width=8.2cm,height=4cm]{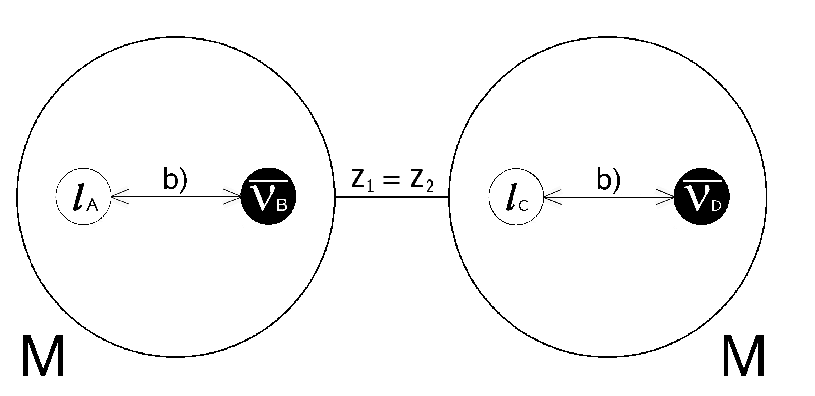}
\label{int50}
\end{figure}

\item {\bf interaction between two "before time" leptons with production of two "time" leptons} (if the particles are non-bounded leptons from the third self-similar DEUS level): $l_{A}+l_{B}\rightarrow l_{C}+l_{D}$, or $l_{A}+\overline{\nu_{B}}\rightarrow l_{C}+\overline{\nu_{D}}$, etc.
\end{itemize}

From the equivalence of the {\bf c)} case field-anti-field conjugate we can have:

\begin{itemize}
\item {\bf interaction between one meson and one anti-meson conjugate} (if the particles are composed of {\bf b)} case weakly interacting quarks and anti-quarks from the fourth self-similar DEUS level):

\begin{figure}[h]
\centering
\leavevmode
\includegraphics[width=3cm,height=0.8cm]{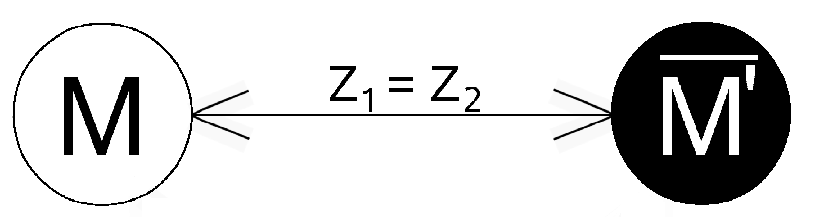}
\label{int51}
\end{figure}
\noindent
or, equivalently, $(q_{A}\; q_{B})\rightarrow (\overline{q'_{C}}\; \overline{q'_{D}})$, or $(q_{A}\; \overline{q'_{B}})\rightarrow (q_{C}\; \overline{q'_{D}})$, etc.;

\item {\bf interaction between two "before time" quarks with production of other two "time" anti-quarks} (if the particles are non-bounded quarks and anti-quarks from the fourth self-similar DEUS level): $q_{A}+q_{B}\rightarrow \overline{q'_{C}}+\overline{q'_{D}}$, or $q_{A}+ \overline{q'_{B}}\rightarrow q_{C}+ \overline{q'_{D}}$, etc.;

\item {\bf interaction between two lepton pairs} (if the particles are composed of two by two {\bf b)} case weakly interacting leptons from the third self-similar DEUS level; in the FRW spacetime the link between two leptons or anti-leptons on the same level three real state catenoid is symbolized by the Pauli exclusion principle): $(l_{A}\;l_{B})\rightarrow (\overline{\nu_{C}}\; \overline{\nu_{D}})$, etc.;

\item {\bf interaction between one "before time" lepton pair with production of one "time" anti-neutrino pair} (if the particles are non-bounded leptons from the third self-similar DEUS level): $l_{A}+l_{B}\rightarrow \overline{\nu_{C}}+\overline{\nu_{D}}$, etc.
\end{itemize}

\subsubsection{Interaction Between Two Antiparticle Fields in Universe}

\begin{equation}
\displaystyle {\cal D}_{\mu}A_{3}=\partial_{\mu}A_{3}+i\; q_{1}\; c_{i}\; \left(
\begin{array}{cc}
{\cal A}_{3}^{AN} \\
{\cal A}_{4}^{AN}
\end{array}\right)A_{3} \; ,
\label{eq_117}
\end{equation}

\begin{equation}
\displaystyle {\cal D}_{\mu}A_{3}^{+}=\partial_{\mu}A_{3}^{+}+i\; q_{2}\; c_{j}\; \left(
\begin{array}{cc}
{\cal A}_{3}^{AN} \\
{\cal A}_{4}^{AN}
\end{array}
\right)A_{3}^{+}
\label{eq_118}
\end{equation}
\noindent
and:

\begin{equation}
\begin{array}{cc}
J_{3}\equiv i\left(A_{3}^{+}{\cal D}_{\mu}A_{3}-A_{3}{\cal D}_{\mu}A_{3}^{+} \right) =A_{3}A_{3}^{+}\displaystyle \left[q_{2}\; c_{j}\; \left(
\begin{array}{cc}
{\cal A}_{3}^{AN} \\
{\cal A}_{4}^{AN}
\end{array}
\right) -q_{1}\; c_{i}\; \left(
\begin{array}{cc}
{\cal A}_{3}^{AN} \\
{\cal A}_{4}^{AN}
\end{array}
\right) \right]\equiv \\
\equiv A_{3}A_{3}^{+}\displaystyle \left[q_{2}\; c_{j}\; \left(
\begin{array}{cc}
{\cal A}_{3}^{AN} \\
{\cal A}_{4}^{AN}
\end{array}
\right) -q_{1}\; c_{i}\; \left(
\begin{array}{cc}
{\cal A}_{3}^{+N} \\
{\cal A}_{4}^{+N}
\end{array}
\right) \right]\stackrel{not}{=}A_{3}A_{3}^{+}\displaystyle \left(
\begin{array}{cc}
{\cal Z}_{1} \\
{\cal Z}_{2}
\end{array}
\right)
\end{array} \; .
\label{eq_119}
\end{equation}

Only for $c_{i}=c_{j}=c_{1}$, $c_{i}=c_{j}=c_{2}$, or $c_{i}=c_{j}=c_{3}$, results ${\cal Z}_{1}={\cal Z}_{2}=0$ and, from this, $J_{3}=0$ and $Q=0$. In this case we have:

\begin{itemize}
\item {\bf interaction between two anti-mesons} (if the particles are composed of {\bf b)} case weakly interacting anti-quarks from the fourth self-similar DEUS level):
\newpage
\begin{figure}[h]
\centering
\leavevmode
\includegraphics[width=3cm,height=0.8cm]{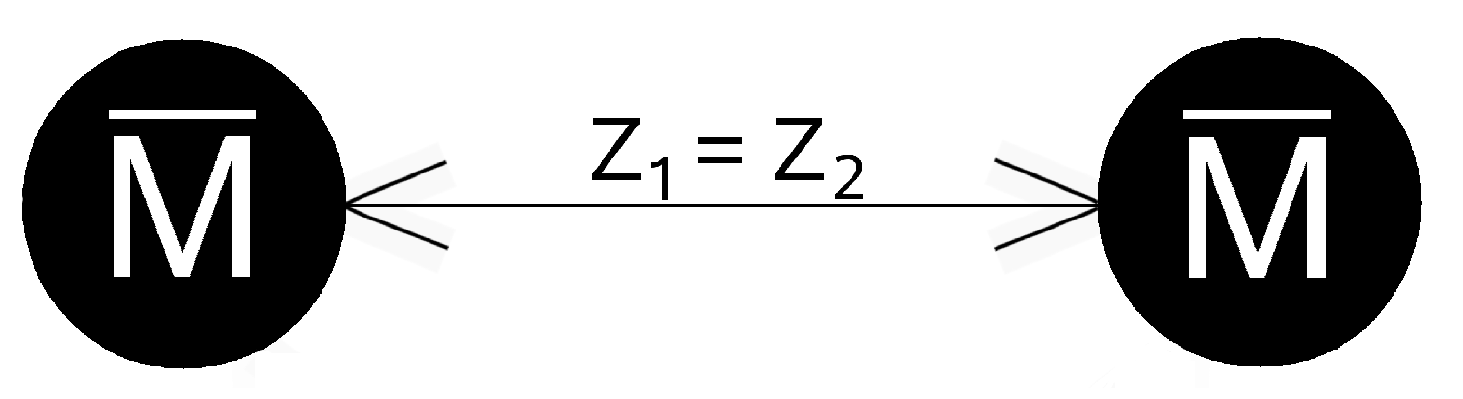}
\label{int53}
\end{figure}
\noindent
or, equivalently: $(\overline{q_{A}}\; \overline{q_{B}})\rightarrow (\overline{q_{C}}\; \overline{q_{D}})$, or, from the {\bf b)} case anti-field-field conjugate equivalence, $(\overline{q_{A}}\; q'_{B})\rightarrow (\overline{q_{C}}\; q'_{D})$;

\item {\bf interaction between two "before time" anti-quarks with production of other two "time" anti-quarks} (if the particles are non-bounded anti-quarks from the fourth self-similar DEUS level): $\overline{q_{A}}+ \overline{q_{B}}\rightarrow \overline{q_{C}}+ \overline{q_{D}}$, or $\overline{q_{A}}+ q'_{B}\rightarrow \overline{q_{C}}+ q'_{D}$, etc.;

\item {\bf interaction between two anti-lepton pairs} (if the particles are composed of two by two {\bf b)} case weakly interacting anti-leptons from the third self-similar DEUS level; in the FRW spacetime the link between two anti-leptons on the same level three real state catenoid is symbolized by the Pauli exclusion principle): $(\overline{l_{A}}\; \overline{l_{B}})\rightarrow (\overline{l_{C}}\; \overline{l_{D}})$, or, from the {\bf b)} case anti-field--field conjugate equivalence, $(\overline{l_{A}}\; \nu_{B})\rightarrow (\overline{l_{C}}\; \nu_{D})$, etc.;

\item {\bf interaction between two "before time" anti-lepton with production of two "time" leptons} (if the particles are non-bounded leptons and anti-leptons from the third self-similar DEUS level): $\overline{l_{A}}+\overline{l_{B}}\rightarrow \overline{l_{C}}+\overline{l_{D}}$, or $\overline{l_{A}}+\nu_{B}\rightarrow \overline{l_{C}}+\nu_{D}$, etc.
\end{itemize}

From the equivalence of the {\bf c)} case anti-field-field conjugate we can have:

\begin{itemize}
\item {\bf interaction between one anti-meson and one meson conjugate} (if the particles are composed of {\bf b)} case weakly interacting anti-quarks from the fourth self-similar DEUS level):

\begin{figure}[h]
\centering
\leavevmode
\includegraphics[width=3cm,height=0.8cm]{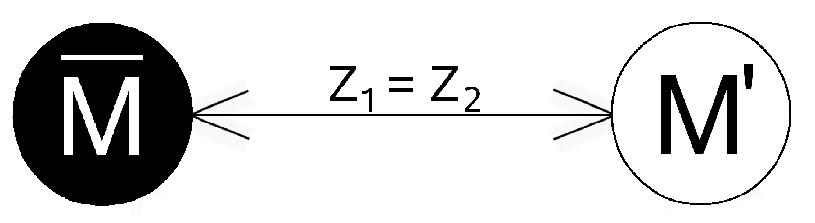}
\label{int54}
\end{figure}
\noindent
or, equivalently: $(\overline{q_{A}}\; \overline{q_{B}})\rightarrow (q'_{C}\; q'_{D})$;

\item {\bf interaction between two "before time" anti-quarks with production of other two "time" quark conjugates} (if the particles are non-bounded anti-quarks and quarks from the fourth self-similar DEUS level): $\overline{q_{A}}+\overline{q_{B}}\rightarrow q'_{C}+q'_{D}$;

\item {\bf interaction between two anti-lepton pairs} (if the particles are composed of two by two {\bf b)} case weakly interacting anti-leptons and leptons from the third self-similar DEUS level; in the FRW spacetime the link between two anti-leptons or leptons on the same level three real state catenoid is symbolized by the Pauli exclusion principle): $(\overline{l_{A}}\; \overline{l_{B}})\rightarrow (\nu_{C}\; \nu_{D})$, etc.;

\item {\bf interaction between one "before time" anti-lepton pair with production of one "time" neutrino pair} (if the particles are non-bounded leptons and anti-leptons from the third self-similar DEUS level): $\overline{l_{A}}+\overline{l_{B}}\rightarrow \nu_{C}+\nu_{D}$.
\end{itemize}

\subsubsection{Interaction Between One Particle Field and One Antiparticle Field in Universe}

\begin{equation}
\displaystyle {\cal D}_{\mu}A_{3}=\partial_{\mu}A_{3}+i\; q_{1}\; c_{i}\; \left(
\begin{array}{cc}
{\cal A}_{3}^{N} \\
{\cal A}_{4}^{N}
\end{array}\right)A_{3} \; ,
\label{eq_120}
\end{equation}

\begin{equation}
\displaystyle {\cal D}_{\mu}A_{3}^{+}=\partial_{\mu}A_{3}^{+}+i\; q_{2}\; c_{j}\; \left(
\begin{array}{cc}
{\cal A}_{3}^{AN} \\
{\cal A}_{4}^{AN}
\end{array}
\right)A_{3}^{+}
\label{eq_121}
\end{equation}
\noindent
and:

\begin{equation}
\begin{array}{ccc}
\!\!\!\!\!\!\!\!\!\!\!\!\!\!\!\!\!\!\!\!\!\!\!\!\!\!\!\!\!\!\!\!\!\!\!\!\!\!\!\!\!J_{3}=A_{3}A_{3}^{+}\displaystyle \left[q_{2}\; c_{j}\; \left(
\begin{array}{ccc}
{\cal A}_{3}^{AN} \\
{\cal A}_{4}^{AN}
\end{array}
\right) -q_{1}\; c_{i}\; \left(
\begin{array}{cc}
{\cal A}_{3}^{N} \\
{\cal A}_{4}^{N}
\end{array}
\right) \right]\equiv
A_{3}A_{3}^{+}\displaystyle \left[q_{2}\; c_{j}\; \left(
\begin{array}{cc}
{\cal A}_{3}^{AN} \\
{\cal A}_{4}^{AN}
\end{array}
\right) -q_{1}\; c_{i}\; \left(
\begin{array}{cc}
{\cal A}_{3}^{+AN} \\
{\cal A}_{4}^{+AN}
\end{array}
\right) \right] \equiv \\
\!\!\!\!\!\!\!\!\!\!\!\!\!\!\!\!\!\!\!\!\!\!\!\!\!\!\!\!\!\!\!\!\!\!\!\!\!\!\!\!\!\equiv A_{3}A_{3}^{+}\displaystyle \left[q_{2}\; c_{j}\; \left(
\begin{array}{cc}
{\cal A}_{3}^{+N} \\
{\cal A}_{4}^{+N}
\end{array}
\right) -q_{1}\; c_{i}\; \left(
\begin{array}{cc}
{\cal A}_{3}^{N} \\
{\cal A}_{4}^{N}
\end{array}
\right) \right] \equiv
A_{3}A_{3}^{+}\displaystyle \left[q_{2}\; c_{j}\; \left(
\begin{array}{cc}
{\cal A}_{3}^{+N} \\
{\cal A}_{4}^{+N}
\end{array}
\right) -q_{1}\; c_{i}\; \left(
\begin{array}{ccc}
{\cal A}_{3}^{+AN} \\
{\cal A}_{4}^{+AN}
\end{array}
\right) \right]\stackrel{not}{=} \\
\!\!\!\!\!\!\!\!\!\!\!\!\!\!\!\!\!\!\!\!\!\!\!\!\!\!\!\!\!\!\!\!\!\!\!\!\!\!\!\!\!\stackrel{not}{=}A_{3}A_{3}^{+}\displaystyle \left(
\begin{array}{cc}
{\cal Z}_{1} \\
{\cal Z}_{2}
\end{array}
\right)
\end{array} \; .
\label{eq_122}
\end{equation}

This interaction is possible only if $c_{j}=c_{2}$ and $c_{i}=c_{3}$ and:

\begin{equation}
\left\{ \begin{array}{cc}
{\cal A}_{3}^{AN}=\displaystyle i\; \frac{1}{8}\; \left[4\left(\frac{1}{2}\mp \frac{1}{4} \right)+\frac{\left(\sqrt{2}+1 \right)^{2}}{\displaystyle \frac{1}{2}\mp \frac{1}{4}}\right]\frac{m_{helicoid}}{q}\; \cos \alpha\; \frac{a^{2}}{r^{2}} \\
{\cal A}_{3}^{N}=\displaystyle - i\; \frac{1}{8}\; \left[4\left(\frac{1}{2}\pm \frac{1}{4} \right)+\frac{\left(\sqrt{2}+1 \right)^{2}}{\displaystyle \frac{1}{2}\pm \frac{1}{4}}\right]\frac{m_{helicoid}}{q}\; \cos \alpha\; \frac{a^{2}}{r^{2}}\; ,
\end{array}\right.
\label{eq_123}
\end{equation}
\noindent
resulting ${\cal Z}_{1}={\cal Z}_{2}\neq 0$ and, from this, $Q\neq 0$ and:

\begin{equation}
J_{3}=A_{3}A_{3}^{+}\displaystyle \left[-\frac{1}{2}\; q_{2}\; {\cal A}_{3}^{AN}+i\; \frac{1}{2}\; q_{1}\; {\cal A}_{3}^{N}\right]
\label{eq_124}
\end{equation}
\noindent
for the {\bf simultaneous creation of two charged particles intermediating the interaction, one timelike and one spacelike}. These current and charge have in the Anti-Universe case opposed signs, canceling each other at the Big Crunch/Big Bang moment of Universe-Anti-Universe "collision".

\section{Interactions in the {\bf a)} Case from \cite{deus12}}

For the empty space, ${\cal D}_{\mu}=\partial_{\mu}$, we obtain $J_{3}=0$ and $Q=0$.

When on the global field $A_{\mu}$ (with $\alpha_{k}=0$) we over-impose a particle field with the unique SU(3) representation (with $\displaystyle \alpha=\frac{\pi}{4}$) given by $\displaystyle {\cal D}_{\mu}=\partial_{\mu}+i\; q\; \lambda\; \left(
\begin{array}{ccc}
{\cal A}_{1}^{U} \\
{\cal A}_{3}^{U} \\
{\cal A}_{2}^{U}
\end{array}
\right)$, we observe that from the covariant derivative are missing the SU(2)'s blocks $\displaystyle \left(
\begin{array}{cc}
{\cal A}_{3}^{U} \\
{\cal A}_{4}^{U}
\end{array}
\right)$ or $\displaystyle \left(
\begin{array}{cc}
{\cal A}_{4}^{U} \\
{\cal A}_{3}^{U}
\end{array}
\right)$. In consequence, for the {\bf a)} case there is no SU(2) representation.

Also, because the only possible $\alpha$ value for SU(3) is $\displaystyle \frac{\pi}{4}$ (for matter), and no $\displaystyle \alpha=-\frac{\pi}{4}$, results that we may have observations only over the matter Universe {\bf global} representation. The Anti-Universe (in the helicoidal, or particle, representation) is beyond the DEUS central singularity at which the two FRW bubbles (Universe and Anti-Universe) annihilate ($\alpha_{k}=0$). The complement $\left(
\begin{array}{ccc}
{\cal A}_{1}^{+U} \\
{\cal A}_{3}^{+U} \\
{\cal A}_{2}^{+U}
\end{array}
\right)$ stands for the pre-Big Bang effects ("before time"). This is seen in the helicoidal representation of the "primordial" DEUS object as {\bf dark matter} interacting into and with the "time" Universe.

The matter fields for this case are:

\begin{equation}
\left\{ \begin{array}{ccc}
{\cal A}_{1}^{U}=\displaystyle -i\; \frac{\sqrt{2}}{8}\; \left[4\left(\frac{1}{2}\mp \frac{1}{4} \right)+\frac{\left(\sqrt{2}+1 \right)^{2}}{\displaystyle \frac{1}{2}\mp \frac{1}{4}}\right]\frac{m_{helicoid}}{q}\; \frac{1}{1-\textnormal{tg}\; \alpha}\; \frac{a^{2}}{r^{2}} \\
{\cal A}_{2}^{U}=\displaystyle i\; \frac{\sqrt{2}}{8}\; \left[4\left(\frac{1}{2}+ \frac{1}{4} \right)+\frac{\left(\sqrt{2}+1 \right)^{2}}{\displaystyle \frac{1}{2}+ \frac{1}{4}}\right]\frac{m_{helicoid}}{q}\; \frac{1}{1+\textnormal{tg}\; \alpha}\; \frac{a^{2}}{r^{2}} \\
{\cal A}_{3}^{U}=\displaystyle \pm i\; \frac{1}{8}\; \left[4\left(\frac{1}{2}\mp \frac{1}{4} \right)+\frac{\left(\sqrt{2}+1 \right)^{2}}{\displaystyle \frac{1}{2}\mp \frac{1}{4}}\right]\frac{m_{helicoid}}{q}\; \cos \alpha\; \frac{a^{2}}{r^{2}}
\end{array}\right.
\label{eq_125}
\end{equation}
\noindent
and:

\begin{equation}
\left\{ \begin{array}{ccc}
{\cal A}_{1}^{+U}=\displaystyle i\; \frac{\sqrt{2}}{8}\; \left[4\left(\frac{1}{2}\pm \frac{1}{4} \right)+\frac{\left(\sqrt{2}+1 \right)^{2}}{\displaystyle \frac{1}{2}\pm \frac{1}{4}}\right]\frac{m_{helicoid}}{q}\; \frac{1}{1-\textnormal{tg}\; \alpha}\; \frac{a^{2}}{r^{2}} \\
{\cal A}_{2}^{+U}=\displaystyle -i\; \frac{\sqrt{2}}{8}\; \left[4\left(\frac{1}{2}- \frac{1}{4} \right)+\frac{\left(\sqrt{2}+1 \right)^{2}}{\displaystyle \frac{1}{2}- \frac{1}{4}}\right]\frac{m_{helicoid}}{q}\; \frac{1}{1+\textnormal{tg}\; \alpha}\; \frac{a^{2}}{r^{2}} \\
{\cal A}_{3}^{+U}=\displaystyle \mp i\; \frac{1}{8}\; \left[4\left(\frac{1}{2}\pm \frac{1}{4} \right)+\frac{\left(\sqrt{2}+1 \right)^{2}}{\displaystyle \frac{1}{2}\pm \frac{1}{4}}\right]\frac{m_{helicoid}}{q}\; \cos \alpha\; \frac{a^{2}}{r^{2}}
\end{array}\right.
\label{eq_126}
\end{equation}
\noindent
over the global fields $A_{1}\equiv \left. {\cal A}_{1}^{U}\right|_{\alpha_{k}=0}$ and $A_{1}=A_{1}^{+}$ , \; $A_{2}\equiv \left. {\cal A}_{2}^{U}\right|_{\alpha_{k}=0}$ and $A_{2}=A_{2}^{+}$ , \; $A_{3}\equiv \left. {\cal A}_{3}^{U}\right|_{\alpha_{k}=0}$ and $A_{3}=A_{3}^{+}$.

As we saw in paper \cite{deus12}, the $\lambda$ generators for the {\bf a)} case are $\lambda_{2}$, $\lambda_{3}$, $\lambda_{7}$ and $\lambda_{0}=\{\lambda_{1},\; \lambda_{6},\; \lambda_{8}\}$.

\subsection{Interaction Between the "Time" Universe with Itself}

\begin{equation}
\displaystyle {\cal D}_{\mu}A_{3}=\partial_{\mu}A_{3}+i\; q_{1}\; \lambda_{i}\; \left(
\begin{array}{ccc}
{\cal A}_{1}^{U} \\
{\cal A}_{3}^{U} \\
{\cal A}_{2}^{U}
\end{array}\right)A_{3} \; ,
\label{eq_127}
\end{equation}

\begin{equation}
\displaystyle {\cal D}_{\mu}A_{3}^{+}=\partial_{\mu}A_{3}^{+}+i\; q_{2}\; \lambda_{j}\; \left(
\begin{array}{ccc}
{\cal A}_{1}^{U} \\
{\cal A}_{3}^{U} \\
{\cal A}_{2}^{U}
\end{array}
\right)A_{3}^{+}
\label{eq_128}
\end{equation}
\noindent
and:

\begin{equation}
J_{3}=A_{3}A_{3}^{+}\displaystyle \left[q_{2}\; \lambda_{j}\; \left(
\begin{array}{ccc}
{\cal A}_{1}^{U} \\
{\cal A}_{3}^{U} \\
{\cal A}_{2}^{U}
\end{array}
\right) -q_{1}\; \lambda_{i}\; \left(
\begin{array}{ccc}
{\cal A}_{1}^{U} \\
{\cal A}_{3}^{U} \\
{\cal A}_{2}^{U}
\end{array}
\right) \right]\stackrel{not}{=}A_{3}A_{3}^{+}\displaystyle \left(
\begin{array}{ccc}
{\cal Z}_{1} \\
{\cal Z}_{2} \\
{\cal Z}_{3}
\end{array}
\right) \; .
\label{eq_129}
\end{equation}

We obtain ${\cal Z}_{1}={\cal Z}_{2}={\cal Z}_{3}=0$, situation in which $J_{3}=0$ and $Q=0$, for:

\begin{itemize}
\item $\lambda_{i}=\lambda_{j}=\lambda_{0}$ associated to the {\bf fifth self-similar DEUS level FRW wave representation as part of the Universe} and in interaction with it;
\item $\lambda_{i}=\lambda_{j}=\lambda_{2}$ associated to the {\bf fourth self-similar DEUS level FRW quarks as part of the Universe} and in interaction with it;
\item $\lambda_{i}=\lambda_{j}=\lambda_{3}$ associated to the {\bf third self-similar DEUS level FRW leptons as part of the Universe} and in interaction with it;
\item $\lambda_{i}=\lambda_{j}=\lambda_{7}$ associated to the {\bf second self-similar DEUS level FRW black holes as part of the Universe} and in interaction with it;
\end{itemize}

Also, when $\lambda_{j}=\lambda_{7}$ and $\lambda_{i}=\lambda_{0}$ results ${\cal Z}_{1}={\cal Z}_{2}={\cal Z}_{3}=0$ and $J_{3}=0$, $Q=0$, only if:

\begin{equation}
{\cal A}_{3}^{U}=\displaystyle - i\; \frac{1}{8}\; \left[4\left(\frac{1}{2}+ \frac{1}{4} \right)+\frac{\left(\sqrt{2}+1 \right)^{2}}{\displaystyle \frac{1}{2}+ \frac{1}{4}}\right]\frac{m_{helicoid}}{q}\; \cos \alpha\; \frac{a^{2}}{r^{2}}\; ,
\label{eq_130}
\end{equation}
\noindent
meaning that {\bf a black hole can be seen as a sub-quark wave} or interacting at this scale with the surrounding Universe.

\subsection{Interaction Between "Before Time" Universe with Itself}

\begin{equation}
\displaystyle {\cal D}_{\mu}A_{3}=\partial_{\mu}A_{3}+i\; q_{1}\; \lambda_{i}\; \left(
\begin{array}{ccc}
{\cal A}_{1}^{+U} \\
{\cal A}_{3}^{+U} \\
{\cal A}_{2}^{+U}
\end{array}\right)A_{3} \; ,
\label{eq_131}
\end{equation}

\begin{equation}
\displaystyle {\cal D}_{\mu}A_{3}^{+}=\partial_{\mu}A_{3}^{+}+i\; q_{2}\; \lambda_{j}\; \left(
\begin{array}{ccc}
{\cal A}_{1}^{+U} \\
{\cal A}_{3}^{+U} \\
{\cal A}_{2}^{+U}
\end{array}
\right)A_{3}^{+}
\label{eq_132}
\end{equation}
\noindent
and:

\begin{equation}
J_{3}=A_{3}A_{3}^{+}\displaystyle \left[q_{2}\; \lambda_{j}\; \left(
\begin{array}{ccc}
{\cal A}_{1}^{+U} \\
{\cal A}_{3}^{+U} \\
{\cal A}_{2}^{+U}
\end{array}
\right) -q_{1}\; \lambda_{i}\; \left(
\begin{array}{ccc}
{\cal A}_{1}^{+U} \\
{\cal A}_{3}^{+U} \\
{\cal A}_{2}^{+U}
\end{array}
\right) \right]\stackrel{not}{=}A_{3}A_{3}^{+}\displaystyle \left(
\begin{array}{ccc}
{\cal Z}_{1} \\
{\cal Z}_{2} \\
{\cal Z}_{3}
\end{array}
\right) \; .
\label{eq_133}
\end{equation}

In the "before time" Universe we obtain ${\cal Z}_{1}={\cal Z}_{2}={\cal Z}_{3}=0$, resulting $J_{3}=0$ and $Q=0$, for:

\begin{itemize}
\item $\lambda_{i}=\lambda_{j}=\lambda_{0}$ associated to the {\bf fifth self-similar DEUS level FRW wave representation as part of the "before time" Universe} and in interaction with it;
\item $\lambda_{i}=\lambda_{j}=\lambda_{2}$ associated to the {\bf fourth self-similar DEUS level FRW quarks as part of the "before time" Universe} and in interaction with it;
\item $\lambda_{i}=\lambda_{j}=\lambda_{3}$ associated to the {\bf third self-similar DEUS level FRW leptons as part of the "before time" Universe} and in interaction with it;
\item $\lambda_{i}=\lambda_{j}=\lambda_{7}$ associated to the {\bf second self-similar DEUS level FRW black holes as part of the "before time" Universe} and in interaction with it;
\end{itemize}

As for the previous analyzed "time" Universe case, ${\cal Z}_{1}={\cal Z}_{2}={\cal Z}_{3}=0$ (resulting $J_{3}=0$ and $Q=0$) also for $\lambda_{j}=\lambda_{7}$ and $\lambda_{i}=\lambda_{0}$, where we must have:

\begin{equation}
{\cal A}_{3}^{+U}=\displaystyle i\; \frac{1}{8}\; \left[4\left(\frac{1}{2}- \frac{1}{4} \right)+\frac{\left(\sqrt{2}+1 \right)^{2}}{\displaystyle \frac{1}{2}- \frac{1}{4}}\right]\frac{m_{helicoid}}{q}\; \cos \alpha\; \frac{a^{2}}{r^{2}}\; .
\label{eq_134}
\end{equation}

This means that, in the "before time" Universe, {\bf a black hole can be seen as a sub-quark wave} or interacting at this scale with the surrounding "before time" Universe.

\subsection{Interaction Between "Before Time" Universe with the "Time" Universe}

\begin{equation}
\displaystyle {\cal D}_{\mu}A_{3}=\partial_{\mu}A_{3}+i\; q_{1}\; \lambda_{i}\; \left(
\begin{array}{ccc}
{\cal A}_{1}^{U} \\
{\cal A}_{3}^{U} \\
{\cal A}_{2}^{U}
\end{array}\right)A_{3} \; ,
\label{eq_135}
\end{equation}

\begin{equation}
\displaystyle {\cal D}_{\mu}A_{3}^{+}=\partial_{\mu}A_{3}^{+}+i\; q_{2}\; \lambda_{j}\; \left(
\begin{array}{ccc}
{\cal A}_{1}^{+U} \\
{\cal A}_{3}^{+U} \\
{\cal A}_{2}^{+U}
\end{array}
\right)A_{3}^{+}
\label{eq_136}
\end{equation}
\noindent
and:

\begin{equation}
J_{3}=A_{3}A_{3}^{+}\displaystyle \left[q_{2}\; \lambda_{j}\; \left(
\begin{array}{ccc}
{\cal A}_{1}^{+U} \\
{\cal A}_{3}^{+U} \\
{\cal A}_{2}^{+U}
\end{array}
\right) -q_{1}\; \lambda_{i}\; \left(
\begin{array}{ccc}
{\cal A}_{1}^{U} \\
{\cal A}_{3}^{U} \\
{\cal A}_{2}^{U}
\end{array}
\right) \right]\stackrel{not}{=}A_{3}A_{3}^{+}\displaystyle \left(
\begin{array}{ccc}
{\cal Z}_{1} \\
{\cal Z}_{2} \\
{\cal Z}_{3}
\end{array}
\right) \; .
\label{eq_137}
\end{equation}

We obtain ${\cal Z}_{1}={\cal Z}_{2}={\cal Z}_{3}=0$ (resulting $J_{3}=0$ and $Q=0$) for ${\cal A}_{3}^{U}$ as in (\ref{eq_130}), ${\cal A}_{3}^{+U}$ as in (\ref{eq_134}) and:

\begin{itemize}
\item $\lambda_{i}=\lambda_{j}=\lambda_{0}$ associated to the interaction between "time" Universe and "before time" Universe at the {\bf fifth self-similar DEUS level FRW wave representation};
\item $\lambda_{i}=\lambda_{j}=\lambda_{7}$ associated to the interaction between "time" Universe and "before time" Universe at the {\bf second self-similar DEUS level FRW black holes};
\end{itemize}
\noindent
at each of the above levels (and only for them) the "before time" Universe being perceived from the "time" Universe as dark matter effects (evaporated from level one of self-similarity).

If $\lambda_{j}=\lambda_{7}$ and $\lambda_{i}=\lambda_{0}$, or $\lambda_{j}=\lambda_{2}$ and $\lambda_{i}=\lambda_{0}$, or $\lambda_{j}=\lambda_{3}$ and $\lambda_{i}=\lambda_{0}$, we get ${\cal Z}_{1}={\cal Z}_{2}={\cal Z}_{3}=0$. Results $J_{3}=0$ and $Q=0$. The levels three and four of DEUS self-similarity contain their dark matter effects as non-evaporated, at the sub-quark level (non-evaporated level five of self-similarity).

\section{Sub-Quark Particles in SU(6)}

At the fifth self-similar DEUS level we have (in the helicoidal representation) sub-quarks. The fifth DEUS level can be represented in the SU(6) group symmetry, breaking into the quark (or anti-quark) SU(3) symmetry of the self-similar DEUS level four and in a massless particle field (electroweak field in which, at another symmetry breaking, ${\cal A}_{1}$ or ${\cal A}_{2}$ separates from the weak field). Considering the same invariants for the stability of this DEUS level ($J_{\mu}=0$ and $Q=0$) and the previously described {\bf b)} (non-collapsed DEUS levels particles and antiparticles) and {\bf c)} (collapsed DEUS levels "free" particles and antiparticles) cases' global $A_{\alpha}$ and local ${\cal A}_{\beta}$ fields, we can write the covariant derivative for the interactions at the sub-quark level:

\begin{itemize}
\item between sub-quark particles (and sub-quark conjugate antiparticles) as:

\begin{equation}
\displaystyle {\cal D}_{\mu}A_{3}=\partial_{\mu}A_{3}+i\; q_{1}\; \lambda_{i,\; SU(6)}\; \left(
\begin{array}{cccccc}
\textcolor{red}{{\cal A}_{1}^{+AP}} \\
\textcolor{red}{{\cal A}_{4}^{+AP}} \\
\textcolor{red}{{\cal A}_{3}^{+AP}} \\
\textcolor{blue}{{\cal A}_{1}^{+AP}} \\
\textcolor{blue}{{\cal A}_{4}^{+AP}} \\
\textcolor{blue}{{\cal A}_{3}^{+AP}}
\end{array}\right)A_{3} \; ,
\label{eq_138}
\end{equation}

\begin{equation}
\displaystyle {\cal D}_{\mu}A_{3}^{+}=\partial_{\mu}A_{3}^{+}+i\; q_{2}\; \lambda_{j,\; SU(6)}\; \left(
\begin{array}{cccccc}
\textcolor{green}{{\cal A}_{2}^{P}} \\
\textcolor{green}{{\cal A}_{4}^{P}} \\
\textcolor{green}{{\cal A}_{3}^{P}} \\
{\cal A}_{2}^{P} \\
{\cal A}_{4}^{P} \\
{\cal A}_{3}^{P}
\end{array}
\right)A_{3}^{+}
\label{eq_139}
\end{equation}
\noindent
where we can substitute the red SU(3) sub-quark conjugate antiparticle block with the $\begin{array}{ccc}
\textcolor{red}{{\cal A}_{2}^{P}} \\
\textcolor{red}{{\cal A}_{4}^{P}} \\
\textcolor{red}{{\cal A}_{3}^{P}}
\end{array}$ sub-quark particle block, and/or we can substitute the blue SU(3) sub-quark conjugate antiparticle block with the $\begin{array}{ccc}
\textcolor{blue}{{\cal A}_{2}^{P}} \\
\textcolor{blue}{{\cal A}_{4}^{P}} \\
\textcolor{blue}{{\cal A}_{3}^{P}}
\end{array}$ sub-quark particle block, and/or we can substitute the green SU(3) sub-quark particle block with the $\begin{array}{ccc}
\textcolor{green}{{\cal A}_{1}^{+AP}} \\
\textcolor{green}{{\cal A}_{4}^{+AP}} \\
\textcolor{green}{{\cal A}_{3}^{+AP}}
\end{array}$ sub-quark conjugate antiparticle block, and/or we can substitute the black SU(3) sub-quark particle block with the $\begin{array}{ccc}
{\cal A}_{1}^{+AP} \\
{\cal A}_{4}^{+AP} \\
{\cal A}_{3}^{+AP}
\end{array}$ sub-quark conjugate antiparticle block.

For simplicity, we will denote (in any of the corresponding combinations) the SU(6) column field matrix from (\ref{eq_138}) with ${\cal A}_{SU(6)}$ and the SU(6) column field matrix from (\ref{eq_139}) with ${\cal A}_{SU(6)}^{+}$. The generators for SU(6) will be:

\begin{equation}
\lambda_{i,\; SU(6)}=\left( \begin{array}{cc}
\lambda_{i} & \lambda_{i} \\
\lambda_{i} & \lambda_{i}
\end{array} \right)
\label{eq_140}
\end{equation}
\noindent
for $\lambda_{i,\; SU(6)}$, where the $\lambda_{i}$ are the \cite{deus11} SU(3) generators, and in the same way for $\lambda_{j,\; SU(6)}$.

The interactions between sub-quark particles (and sub-quark conjugate antiparticles) are possible if all the ${\cal Z}_{x}=q_{2}\left\{\lambda_{j,\; SU(6)}\; {\cal A}^{+}_{SU(6)} \right\}_{x}-q_{1}\left\{\lambda_{i,\; SU(6)}\; {\cal A}_{SU(6)} \right\}_{x}$ are equal with the ${\cal Z}_{y}=q_{2}\left\{\lambda_{j,\; SU(6)}\; {\cal A}^{+}_{SU(6)} \right\}_{y}-q_{1}\left\{\lambda_{i,\; SU(6)}\; {\cal A}_{SU(6)} \right\}_{y}$,  where {\it x} and {\it y} are indexing the elements of the resulting 6x6 matrix, $x\in \{1,\; 2,\; 3 \}$ and $y=x+3$. The interaction proves to be possible for all the SU(3) $\lambda_{i}=\lambda_{j}$ generators of the {\bf b)} case and for the ones in the {\bf c)} case, having $J_{\mu}=0$ and $Q=0$;

\item between sub-quark antiparticles (and sub-quark conjugate particles) as:

\begin{equation}
\displaystyle {\cal D}_{\mu}A_{3}=\partial_{\mu}A_{3}+i\; q_{1}\; \lambda_{i,\; SU(6)}\; \left(
\begin{array}{cccccc}
\textcolor{red}{{\cal A}_{2}^{+P}} \\
\textcolor{red}{{\cal A}_{4}^{+P}} \\
\textcolor{red}{{\cal A}_{3}^{+P}} \\
\textcolor{blue}{{\cal A}_{2}^{+P}} \\
\textcolor{blue}{{\cal A}_{4}^{+P}} \\
\textcolor{blue}{{\cal A}_{3}^{+P}}
\end{array}\right)A_{3} \; ,
\label{eq_141}
\end{equation}

\begin{equation}
\displaystyle {\cal D}_{\mu}A_{3}^{+}=\partial_{\mu}A_{3}^{+}+i\; q_{2}\; \lambda_{j,\; SU(6)}\; \left(
\begin{array}{cccccc}
\textcolor{green}{{\cal A}_{1}^{AP}} \\
\textcolor{green}{{\cal A}_{4}^{AP}} \\
\textcolor{green}{{\cal A}_{3}^{AP}} \\
{\cal A}_{1}^{AP} \\
{\cal A}_{4}^{AP} \\
{\cal A}_{3}^{AP}
\end{array}
\right)A_{3}^{+}
\label{eq_142}
\end{equation}
\noindent
where we can substitute the red SU(3) sub-quark conjugate particle block with the $\begin{array}{ccc}
\textcolor{red}{{\cal A}_{1}^{AP}} \\
\textcolor{red}{{\cal A}_{4}^{AP}} \\
\textcolor{red}{{\cal A}_{3}^{AP}}
\end{array}$ sub-quark antiparticle block, and/or we can substitute the blue SU(3) sub-quark conjugate particle block with the $\begin{array}{ccc}
\textcolor{blue}{{\cal A}_{1}^{AP}} \\
\textcolor{blue}{{\cal A}_{4}^{AP}} \\
\textcolor{blue}{{\cal A}_{3}^{AP}}
\end{array}$ sub-quark antiparticle block, and/or we can substitute the green SU(3) sub-quark antiparticle block with the $\begin{array}{ccc}
\textcolor{green}{{\cal A}_{2}^{+P}} \\
\textcolor{green}{{\cal A}_{4}^{+P}} \\
\textcolor{green}{{\cal A}_{3}^{+P}}
\end{array}$ sub-quark conjugate particle block, and/or we can substitute the black SU(3) sub-quark antiparticle block with the $\begin{array}{ccc}
{\cal A}_{2}^{+P} \\
{\cal A}_{4}^{+P} \\
{\cal A}_{3}^{+P}
\end{array}$ sub-quark conjugate particle block.

For simplicity, we will denote (in any of the corresponding combinations) the SU(6) column field matrix from (\ref{eq_141}) with ${\cal A}_{SU(6)}$ and the SU(6) column field matrix from (\ref{eq_142}) with ${\cal A}_{SU(6)}^{+}$. The generators for SU(6) will be:

\begin{equation}
\lambda_{i,\; SU(6)}=\left( \begin{array}{cc}
\lambda_{i} & \lambda_{i} \\
\lambda_{i} & \lambda_{i}
\end{array} \right)
\label{eq_143}
\end{equation}
\noindent
for $\lambda_{i,\; SU(6)}$, where the $\lambda_{i}$ are the \cite{deus11} SU(3) generators, and in the same way for $\lambda_{j,\; SU(6)}$.

The interactions between sub-quarks antiparticles (and sub-quark conjugate particles) are possible if all the ${\cal Z}_{x}=q_{2}\left\{\lambda_{j,\; SU(6)}\; {\cal A}^{+}_{SU(6)} \right\}_{x}-q_{1}\left\{\lambda_{i,\; SU(6)}\; {\cal A}_{SU(6)} \right\}_{x}$ are equal with the ${\cal Z}_{y}=q_{2}\left\{\lambda_{j,\; SU(6)}\; {\cal A}^{+}_{SU(6)} \right\}_{y}-q_{1}\left\{\lambda_{i,\; SU(6)}\; {\cal A}_{SU(6)} \right\}_{y}$,  where {\it x} and {\it y} are indexing the elements of the resulting 6x6 matrix, $x\in \{1,\; 2,\; 3 \}$ and $y=x+3$. The interaction proves to be possible for all the SU(3) $\lambda_{i}=\lambda_{j}$ generators of the {\bf b)} case or for the ones in the {\bf c)} case, having $J_{\mu}=0$ and $Q=0$.
\end{itemize}

It is not possible to have interactions between sub-quark particles with sub-quark antiparticles (or between sub-quark conjugate particles with sub-quark conjugate antiparticles).

\section{Final Conclusions}

In our cycle of Universe ("time") the quarks from the $\displaystyle \left(
\begin{array}{cc}
q \\
q'
\end{array}
\right)$ generation are what in the previous Universe cycle ("before time") were $\displaystyle \left(
\begin{array}{cc}
\overline{q'} \\
\overline{q}
\end{array}
\right)$. For exemplification, the "before time" $\displaystyle \left(
\begin{array}{cc}
\overline{d} \\
\overline{u}
\end{array}\right)$ transforms into "time" $\left(
\begin{array}{cc}
u \\
d
\end{array}\right)$. Equivalently, in "time" Universe view of fourth self-similar level, we have $\displaystyle \left(
\begin{array}{cc}
u \\
d
\end{array}
\right)\equiv $ $\left(
\begin{array}{cc}
u \\
\overline{d'}
\end{array}\right)$.

The same observations work for anti-neutrinos that transform into leptons, or for neutrinos that become anti-leptons (and reciprocal) at the Big Crunch/Big Bang crossing, or for anti-mesons and anti-baryons that transform into mesons and, respectively, baryons (and reciprocal). While the particles are transforming as above, the Universe or the Anti-Universe cross the "before time"/"time" singularity without changing their nature (Universe to Anti-Universe or Anti-Universe to Universe).

In the {\bf b)} case we can construct the baryons of our Universe. For example, a $(q\; q\; q)$ particle can be a $(u\;u\;u)$ particle ($\Delta^{++}$). In the $(\overline{q'}\; \overline{q'}\; \overline{q'})$ general description we can include the $(d\; d\; d)$ (the $\Delta^{-}$ baryon), $(s\;s\;s)$ (the $\Omega^{-}$ baryon), $(d\;d\;s)$ (the $\Sigma^{-}$ baryon), $(d\; s\; s)$ (the $\Xi$ baryon), etc.

In the $(q\; \overline{q'}\; \overline{q'})$ category enter $(u\; d\; d)$ (the neutron and the $\Delta$ baryon), $(u\; s\; s)$ (the $\Xi^{0}$ baryon), $(u\; d\; s)$ (the $\Sigma^{0}$ baryon), etc., while for $(q\; q\; \overline{q'})$ we can have $(u\; u\; d)$ (the proton), $(u\; u\; s)$ (the $\Sigma^{+}$ baryon), $(u\;u\;d)$ (the $\Delta^{+}$ baryon), etc.

In the same we can construct the anti-baryons.

Because in the {\bf b)} SU(2) case, we saw that we cannot have interactions between a quark {\it q} and an anti-quark $\overline{q}$, or between a quark conjugate $q'$ and an anti-quark conjugate $\overline{q'}$, the existence of baryons and mesons containing one of these pairs is not possible. However, in the {\bf c)} case, while between the $N$ and $\overline{N}$ or between the $N'$ and $\overline{N'}$ baryons no strong or weak interaction is allowed, between fermions and anti-fermions or between mesons and anti-mesons the weak interactions are possible. 

For the interactions to be possible we must have also equality between the masses of the DEUS helicoid equivalents (at the corresponding self-similar DEUS level) of the particles (or antiparticles). The mass difference between these interacting particles is contained in the mass of their catenoid. At a smaller particle (or antiparticle) corresponds a more evolved catenoid (closer to cylinder and collapse), emptier of its self-similar structure.

$\\$
While we centered our study on the interactions occurring in the matter Universe, for the Anti-Universe the reasoning is the same.

\ack
Manuscript registered at the Astronomical Institute, Romanian Academy, record No. 261 from 04/19/07.


\title[D.E.U.S.]{XIV: Neutrinos and Quarks in DEUS Atoms}

\begin{abstract}
We exploit the correlations between neutrinos (at the third self-similar DEUS level) and quarks (at the fourth self-similar DEUS level) inside the DEUS atom. These correlations will allow us to establish the neutrino flavor hierarchy (for a timelike or a spacelike observer) and to determine the mass of each neutrino flavor.
\end{abstract}


\section{Introduction}

At the third level of self-similarity, the atoms are non-evaporated DEUS objects previously contained in the catenoid of a higher DEUS object (pre-Big Bang DEUS object or black holes) and released in the Universe. In fact, the atoms are "deeper" than what we can see from the FRW surface. They have a catenoid on which "live" lepton DEUS objects and a BHIG spacetime in which "live" the quark DEUS objects, each of them, seen from the DEUS atom exterior, having a particle interpretation (helicoid) and a wave interpretation (catenoid). Due to the fact that the leptons are on the DEUS atom catenoid, their energy emission toward the atom exterior is partly timelike, partly spacelike: real and imaginary Yamabe energy \cite{deus3}. The energy levels are given by the separation between two consecutive DEUS objects (by their scale) situated on one of the three real (discrete) catenoid hyper-surfaces (the fourth is the cylinder which is a FRW continuum) \cite{deus2}. Each of these catenoids mark a lepton generation.

Also, inside the DEUS atom, the DEUS lepton-objects do not interact with their neighbors (are conserved as objects), the interaction taking place at the DEUS atom collapse, when they meet the anti-leptons situated across the central singularity. Still, they are emitted in the external spacetime from the junction point with the external spacetime (the upper or the lower limit of the DEUS atom) when the catenoid on which they live reaches one of the three real values, different of the pre-collapse one (cylinder).

In the following study we will resume ourseves to neutrinos. We saw in paper \cite{deus10} that, at the pre-collapse cylinder, applying a trivial treatment to neutrinos, $\overline{\Psi}\Psi$ from the mass term of ${\cal L}(\Psi,{\bf A})$ is zero, making impossible any neutrino mass determination. At the pre-collapse cylinder the time becomes a continuous one, $t_{FRW}$, the generatios from the previous catenoid evolution coexisting. Before the collapse, this time is correlated to the atom's catenoid evolution through the $\alpha_{k}$ contact angle between the catenoid and the cylinder \cite{deus5}. The real catenoid hypersurfaces are at $\displaystyle \alpha_{k}=i \left(\frac{\pi}{4}+n\frac{\pi}{12} \right)$ for neutrinos (and, in general, for leptons) and at $\displaystyle \beta_{k}=i \left(\frac{\pi}{4}-n\frac{\pi}{12} \right)$ for anti-neutrinos (and, in general, for anti-leptons), where $n=\overline{1,3}$, the collapse taking place at $\displaystyle \alpha_{k}=i\; \frac{\pi}{4}$. The neutrino and anti-neutrino particles (helicoids) on the DEUS atom catenoids are at the angle $\alpha$ relative to the external spacetime, where $\alpha_{k}=i\; \alpha$.

Now, inside the inner horizon of the DEUS atom, and forming its BHIG spacetime, we have the DEUS objects from the fourth self-similarity DEUS level, seen by the external observer as quarks. In the DEUS atom catenoid evolution toward cylinder, it drags after it the BHIG spacetime. When the catenoid reaches, at $\alpha_{k}$, its real energy values, we are able to see, as external observers, the BHIG's DEUS objects as quarks: three quark generations.

As observed in \cite{deus12}, at self-similarity DEUS level three or four, the Lagrangian is:

\begin{equation}
{\cal L}={\cal L}(\Psi,{\bf A})=i\; \overline{\Psi}\; \gamma^{\mu}\; \partial_{\mu}\Psi-q\; \overline{\Psi}\; A_{\mu}\; \gamma^{\mu}\; \Psi-\overline{\Psi}\; m\; \Psi \; ,
\label{eq_1}
\end{equation}
\noindent
where, for the entire DEUS atom, we must have \cite{deus9}:

\begin{equation}
\Psi=\Psi_{0}\left(
\begin{array}{cccc}
\displaystyle e^{\mp i k_{3}x^{3}} \\
\displaystyle e^{-i\omega x^{5}} \\
\displaystyle e^{\mp i k_{1}x^{1}} \\
\displaystyle e^{\mp i k_{2}x^{2}}
\end{array}
\right)
\label{eq_2}
\end{equation}
\noindent
in the local chart (as function of $\alpha_{k}$):

\begin{equation}
\left\{
\begin{array}{llll}
\displaystyle x^{1}=-\sinh \left[\textnormal{arcctg} \left(\textnormal{tg}\; \alpha_{k} \right) \right]\sin \alpha_{k} \\
\displaystyle x^{2}=\sinh \left[\textnormal{arcctg} \left(\textnormal{tg}\; \alpha_{k} \right) \right]\cos \alpha_{k} \\
\displaystyle x^{3}=\alpha_{k} \\
\displaystyle x^{5}=\mp i\; \sinh \left[\textnormal{arcctg} \left(\textnormal{tg}\; \alpha_{k} \right) \right]
\end{array}
\right. \; .
\label{eq_3}
\end{equation}

The particle (and antiparticle) interaction absence, while on catenoid, implies that the $i\; \overline{\Psi}\; \gamma^{\mu}\; \partial_{\mu}\Psi$ interaction term from (\ref{eq_1}) is zero, where $\mu=\overline{0,3}$, which, when solved, gives:

\begin{equation}
\left\{
\begin{array}{lll}
\displaystyle k_{1}=\pm i \sqrt{2}\; \omega\; \frac{1}{\cosh \alpha_{k}-\sinh \alpha_{k}}\; \frac{\sinh \alpha_{k}\; (2\cosh \alpha_{k}+\sinh \alpha_{k})}{\sinh \alpha_{k}+\alpha_{k}\;(\sinh \alpha_{k}+\cosh \alpha_{k})} \\
\displaystyle k_{2}=- i \sqrt{2}\; \omega\; \frac{\cosh \alpha_{k}}{\sinh \alpha_{k}+\cosh \alpha_{k}} \\
\displaystyle k_{3}=- i\; \omega\; \frac{\sinh \alpha_{k}\; (2\cosh \alpha_{k}+\sinh \alpha_{k})}{\sinh \alpha_{k}+\alpha_{k}\;(\sinh \alpha_{k}+\cosh \alpha_{k})}
\end{array}
\right. \; ,
\label{eq_4}
\end{equation}
\noindent
which makes the wave function (\ref{eq_2}) in the initial state (at one of the real catenoid states) to be:

\begin{equation}
\!\!\!\!\!\!\!\!\!\!\!\!\!\!\!\!\!\!\!\!\!\!\!\!\!\!\!\!\!\!\!\!\!\!\!\!\!\!\!\!\!\!\!\!\!\Psi_{i}=\Psi_{0}\left(
\begin{array}{cccc}
\displaystyle \exp\left\{\displaystyle \mp \omega\; \frac{\sinh \alpha_{k}\; (2\cosh \alpha_{k}+\sinh \alpha_{k})}{\sinh \alpha_{k}+\alpha_{k}\;(\sinh \alpha_{k}+\cosh \alpha_{k})}\; \alpha_{k}\right\} \\
\displaystyle \exp\left\{\mp \omega\; \sinh \left[\textnormal{arcctg} \left(\textnormal{tg}\; \alpha_{k} \right) \right]\right\} \\
\displaystyle \exp\left\{- \sqrt{2}\; \omega\; \frac{1}{\cosh \alpha_{k}-\sinh \alpha_{k}}\; \frac{\sinh \alpha_{k}\; (2\cosh \alpha_{k}+\sinh \alpha_{k})}{\sinh \alpha_{k}+\alpha_{k}\;(\sinh \alpha_{k}+\cosh \alpha_{k})}\; \sin \alpha_{k}\; \sinh \left[\textnormal{arcctg} \left(\textnormal{tg}\; \alpha_{k} \right) \right]\right\} \\
\displaystyle \exp\left\{\mp \sqrt{2}\; \omega\; \frac{\cosh \alpha_{k}}{\sinh \alpha_{k}+\cosh \alpha_{k}}\; \cos \alpha_{k}\; \sinh \left[\textnormal{arcctg} \left(\textnormal{tg}\; \alpha_{k} \right) \right]\right\}
\end{array}
\right)\; .
\label{eq_5}
\end{equation}

The wave function at DEUS object final state (at pre-collapse cylinder) is:

\begin{equation}
\Psi_{f}=\Psi_{0}\left(
\begin{array}{cccc}
\displaystyle \exp\left\{\displaystyle -i\; \frac{\sqrt{2}}{2}\; \omega\right\} \\
\displaystyle \exp\left\{\displaystyle -i\; \frac{\sqrt{2}}{2}\; \omega\right\} \\
\displaystyle 1 \\
\displaystyle 1
\end{array}
\right)\; ,
\label{eq_6}
\end{equation}
\noindent
where we used the fact that, at this surface, $\displaystyle \alpha_{k}=i\;\frac{\pi}{4}$ (or $\displaystyle \alpha=\frac{\pi}{4}$).

From the mass term of ${\cal L}(\Psi,{\bf A})$ we obtain that $\Psi_{i}\overline{\Psi}_{i}=\Psi_{f}\overline{\Psi}_{f}$. The first two elements in $\Psi$ (and, naturally, in $\overline{\Psi}$) will describe the quarks. The last two elements in $\Psi$ (and $\overline{\Psi}$) will describe neutrinos (or, in general, leptons), with a final state (in the helicoidal representation) independent of $\omega$ and, in consequence, of the DEUS neutrino-particle energy or mass.

\section{Neutrinos}

From the third element of the $\Psi_{i}\overline{\Psi}_{i}=\Psi_{f}\overline{\Psi}_{f}$ equality results:

\begin{equation}
\!\!\!\!\!\!\!\!\!\!\!\!\!\!\!\!\!\!\!\!\begin{array}{ccc}
\displaystyle \omega \left\{2\;i\; \sin \alpha\; \cos \alpha\; (\beta +\sin \beta\; \cos \beta)-\sin^{2}\alpha\; (\beta +\sin \beta\; \cos \beta)+2\; \sin \alpha\; \cos \alpha\; \sin^{2}\beta+ \right. \\
\displaystyle \left. +i\; \sin^{2}\alpha\; \sin^{2}\beta \right\}\sin \alpha\; \sin \left[\textnormal{arcctg} \left(\textnormal{tg}\; \alpha \right) \right]-\overline{\omega}\left\{ 2\;i\; \sin \beta\; \cos \beta\; (\alpha +\sin \alpha\; \cos \alpha)- \right. \\
\displaystyle \left. -\sin^{2}\beta\; (\alpha +\sin \alpha\; \cos \alpha)+2\; \sin \beta\; \cos \beta\; \sin^{2}\alpha+i\; \sin^{2}\beta\; \sin^{2}\alpha \right\}\sin \beta\; \sin \left[\textnormal{arcctg} \left(\textnormal{tg}\; \beta \right) \right]=0\; ,
\end{array}
\label{eq_7}
\end{equation}
\noindent
where, for the antiparticle's $\overline{\Psi}$, we used the notation $\beta$ instead of $\alpha$. In (\ref{eq_7}) we have $\displaystyle \alpha=\frac{\pi}{4}+n\frac{\pi}{12}$ for neutrinos and $\displaystyle \beta=\frac{\pi}{4}-n\frac{\pi}{12}$ for anti-neutrinos, $n=\overline{1,3}$.

For the imaginary part of (\ref{eq_7}) we have:

\begin{equation}
\left\{ \begin{array}{ccc}
\displaystyle \omega_{1} \left[ 2 \cos \alpha\; (\beta +\sin \beta\; \cos \beta)+ \sin \alpha\; \sin^{2} \beta\right]\; \sin^{2}\alpha\; \sin \left[\textnormal{arcctg} \left(\textnormal{tg}\; \alpha \right) \right] - \\
\displaystyle  -\overline{\omega_{1}}\left[ 2 \cos \beta\; (\alpha +\sin \alpha\; \cos \alpha)+ \sin^{2}\alpha\; \sin\beta\right]\; \sin^{2}\beta\; \sin \left[\textnormal{arcctg} \left(\textnormal{tg}\; \beta \right) \right]=0 \\
\left[\displaystyle \Im(E_{Yamabe,\; helicoid})\right]^{2}=\omega_{1}\overline{\omega_{1}}
\; ,
\end{array} \right.
\label{eq_8}
\end{equation}
\noindent
with $\omega_{1}=\Im\left\{ \omega \right\}$, $\overline{\omega_{1}}=\Im\left\{ \overline{\omega} \right\}$ and $\Im(E_{Yamabe,\; helicoid})$ from \cite{deus3}, where $E_{Yamabe,\; helicoid}=E_{Yamabe,\; helicoid}(r,a)$.

The real part of (\ref{eq_7}) is:

\begin{equation}
\left\{ \begin{array}{ccc}
\displaystyle \omega_{2} \left[ 2 \cos \alpha\; \sin^{2}\beta -\sin \alpha\; (\beta +\sin \beta\; \cos \beta)\right]\; \sin^{2}\alpha\; \sin \left[\textnormal{arcctg} \left(\textnormal{tg}\; \alpha \right) \right] - \\
\displaystyle  -\overline{\omega_{2}} \left[ 2 \cos \beta\; \sin^{2}\alpha -\sin \beta\; (\alpha +\sin \alpha\; \cos \alpha)\right]\; \sin^{2}\beta\; \sin \left[\textnormal{arcctg} \left(\textnormal{tg}\; \beta \right) \right]=0 \\
\left[\displaystyle \Re(E_{Yamabe,\; helicoid})\right]^{2}=\omega_{2}\overline{\omega_{2}}
\; ,
\end{array} \right.
\label{eq_9}
\end{equation}
\noindent
with $\omega_{2}=\Re\left\{ \omega \right\}$, $\overline{\omega_{2}}=\Re\left\{ \overline{\omega} \right\}$ and $\Re(E_{Yamabe,\; helicoid})$ from \cite{deus3}.

From the fourth element of $\Psi_{i}\overline{\Psi}_{i}=\Psi_{f}\overline{\Psi}_{f}$ results:

\begin{equation}
\begin{array}{cc}
\displaystyle \mp \omega \left(i\; \cos \beta\; -\sin \beta\right)\; \cos^{2} \alpha\; \sin \left[\textnormal{arcctg} \left(\textnormal{tg}\; \alpha \right) \right] \pm \\ 
\displaystyle \pm \overline{\omega}\left(i\; \cos \alpha\; -\sin \alpha\right)\; \cos^{2} \beta\; \sin \left[\textnormal{arcctg} \left(\textnormal{tg}\; \beta \right) \right]=0\; ,
\end{array}
\label{eq_10}
\end{equation}
\noindent
where the imaginary part is ($\omega_{3}=\Im\left\{ \omega \right\}$; $\overline{\omega_{3}}=\Im\left\{ \overline{\omega} \right\}$):

\begin{equation}
\left\{ \begin{array}{cc}
\displaystyle \mp \omega_{3}\; \cos^{2}\alpha\; \cos \beta\; \sin \left[\textnormal{arcctg} \left(\textnormal{tg}\; \alpha \right) \right] \pm \overline{\omega_{3}}\; \cos^{2}\beta\; \cos \alpha\; \sin \left[\textnormal{arcctg} \left(\textnormal{tg}\; \beta \right) \right]=0 \\
\left[\displaystyle \Im(E_{Yamabe,\; helicoid})\right]^{2}=\omega_{3}\overline{\omega_{3}}
\; ,
\end{array} \right.
\label{eq_11}
\end{equation}
\noindent
and the real part ($\omega_{4}=\Re\left\{ \omega \right\}$; $\overline{\omega_{4}}=\Re\left\{ \overline{\omega} \right\}$):

\begin{equation}
\left\{ \begin{array}{ccc}
\displaystyle \pm \omega_{4}\; \cos^{2}\alpha\; \sin \beta\; \sin \left[\textnormal{arcctg} \left(\textnormal{tg}\; \alpha \right) \right] \mp \overline{\omega_{4}}\; \cos^{2}\beta\; \sin \alpha\; \sin \left[\textnormal{arcctg} \left(\textnormal{tg}\; \beta \right) \right]=0 \\
\left[\displaystyle \Re(E_{Yamabe,\; helicoid})\right]^{2}=\omega_{4}\overline{\omega_{4}}
\; .
\end{array} \right.
\label{eq_12}
\end{equation}

Solving the (\ref{eq_8}), (\ref{eq_9}), (\ref{eq_11}) and (\ref{eq_12}) systems in $\omega_{1}$-$\omega_{4}$ and $\overline{\omega_{1}}$-$\overline{\omega_{4}}$, we obtain for the three neutrino flavors the table 1 values. In this table we use for the Yamabe energy the two extreme values of the \cite{deus2} determined Hubble constant $H_{0}$. For the same {\it n} flavor, the change of sign between two consecutive $\omega_{4}$, $\overline{\omega_{4}}$, $\omega_{2}$, $\overline{\omega_{2}}$ or $\omega_{1}$, $\overline{\omega_{1}}$, $\omega_{3}$, $\overline{\omega_{3}}$ sets is a symmetry related to time reversal transformation ("{\bf T}" symmetry). For the same $H_{0}$ values, between the {\it Down Sign} (which, for example, for a $\pm$ in equations, keeps only the "-" sign) and the {\it Up Sign} (which, for example, for a $\pm$ in equations, keeps only the "+" sign) we have a symmetry related to the parity inversion ("{\bf P}" symmetry conservation). Also, between a $\omega_{4}$, $\overline{\omega_{4}}$, $\omega_{2}$, $\overline{\omega_{2}}$ data set and $\omega_{1}$, $\overline{\omega_{1}}$, $\omega_{3}$, $\overline{\omega_{3}}$ data set (for the same {\it n}) with same signs, we have a $\displaystyle \frac{m_{helicoid}}{q}$ conservation, equivalent to a {\it q} conservation, or a conservation under the charge-conjugation transformation ("{\bf C}" symmetry conservation). In a global view of table 1 we have a {\bf CPT} symmetry conservation.

In table 2 we summarize the real and imaginary parts of the Yamabe energy resulting from (\ref{eq_8}), (\ref{eq_9}), (\ref{eq_11}) and (\ref{eq_12}). At the same {\it n} value, the difference of the real and the imaginary parts of the Yamabe energy values for the $H_{0}$ extremes are marked with red and blue. With blue we want to accentuate the existing difference between two real or two imaginary Yamabe energy values (again, at the same {\it n}). In this table a virtual particle is described by imaginary $\omega$ values and a real particle by real $\omega$ values (see table 1).

In order that the virtual particle to be the same particle as the real one (neutrino), but spacelike instead of timelike, for each of the three neutrino flavors we should have equality between the imaginary and the real parts of the Yamabe energy, $\left[\Im(E_{Yamabe,\; helicoid})\right]^{2}=\left[\Re(E_{Yamabe,\; helicoid})\right]^{2}$. From the table 2 values results (in the neutrino ($r,a$) reference system):

\begin{itemize}
\item for $n=1$, with $\omega_{4}\overline{\omega_{4}}=\omega_{1}\overline{\omega_{1}}=\omega_{3}\overline{\omega_{3}}$, we have:

\begin{equation}
\frac{a^{2}}{r^{2}}=\left\{ \begin{array}{ll}
\displaystyle 1035.02841 \;\;\;\; \textnormal{for}\;\; H_{0}=72.35 \\
\displaystyle 1029.07804 \;\;\;\; \textnormal{for}\;\; H_{0}=72.31 \; ;
\end{array} \right.
\label{eq_13}
\end{equation}

\item for $n=2$, with $\omega_{4}\overline{\omega_{4}}=\omega_{1}\overline{\omega_{1}}=\omega_{3}\overline{\omega_{3}}$, we have:

\begin{equation}
\frac{a^{2}}{r^{2}}=\left\{ \begin{array}{ll}
\displaystyle 1035.03254 \;\;\;\; \textnormal{for}\;\; H_{0}=72.35 \\
\displaystyle 1029.07804 \;\;\;\; \textnormal{for}\;\; H_{0}=72.31 \; ;
\end{array} \right.
\label{eq_14}
\end{equation}

\item for $n=3$, with $\omega_{4}\overline{\omega_{4}}=\omega_{3}\overline{\omega_{3}}$, we have:

\begin{equation}
\frac{a^{2}}{r^{2}}=\left\{ \begin{array}{ll}
\displaystyle 1035.031928 \;\;\;\; \textnormal{for}\;\; H_{0}=72.35 \\
\displaystyle 1029.075517 \;\;\;\; \textnormal{for}\;\; H_{0}=72.31 \; ,
\end{array} \right.
\label{eq_15}
\end{equation}
\noindent
or, with $\omega_{2}\overline{\omega_{2}}=\omega_{1}\overline{\omega_{1}}$, we have:

\begin{equation}
\frac{a^{2}}{r^{2}}=\left\{ \begin{array}{ll}
\displaystyle 1035.029506 \;\;\;\; \textnormal{for}\;\; H_{0}=72.35 \\
\displaystyle 1029.07671 \;\;\;\; \textnormal{for}\;\; H_{0}=72.31 \; .
\end{array} \right.
\label{eq_16}
\end{equation}
\end{itemize}

\section{Neutrino Flavor Hierarchy}

The mass difference between two {\it n} neutrino flavors will be:

\begin{equation}
\left(\Delta m^{2}_{ij}\right)^{\nu}=\left| \left. \left[\Re(E_{Yamabe,\; helicoid})\right]^{2}\right|_{n=i}-\left. \left[\Re(E_{Yamabe,\; helicoid})\right]^{2}\right|_{n=j} \right|
\label{eq_17}
\end{equation}
\noindent
in the real particle case, or:

\begin{equation}
\left(\Delta m^{2}_{ij}\right)^{\nu}=\left| \left. \left[\Im(E_{Yamabe,\; helicoid})\right]^{2}\right|_{n=i}-\left. \left[\Im(E_{Yamabe,\; helicoid})\right]^{2}\right|_{n=j} \right|
\label{eq_18}
\end{equation}
\noindent
in the virtual particle case, where $i\in \{1,2,3 \}$, $j\in \{1,2,3 \}$ and $i\neq j$. In consequence:

\begin{equation}
\begin{array}{cccccc}
\displaystyle \left(\Delta m^{2}_{12}\right)^{\nu}_{A}=\left| \left. \left(\omega_{4}\overline{\omega_{4}}\right)\right|_{n=1}-\left. \left(\omega_{4}\overline{\omega_{4}}\right)\right|_{n=2} \right| \\
\displaystyle \left(\Delta m^{2}_{12}\right)^{\nu}_{B}=\left| \left. \left(\omega_{2}\overline{\omega_{2}}\right)\right|_{n=1}-\left. \left(\omega_{2}\overline{\omega_{2}}\right)\right|_{n=2} \right| \\
\displaystyle \left(\Delta m^{2}_{12}\right)^{\nu}_{C}=\left| \left. \left(\omega_{1}\overline{\omega_{1}}\right)\right|_{n=1}-\left. \left(\omega_{1}\overline{\omega_{1}}\right)\right|_{n=2} \right| \\
\displaystyle \left(\Delta m^{2}_{12}\right)^{\nu}_{D}=\left| \left. \left(\omega_{3}\overline{\omega_{3}}\right)\right|_{n=1}-\left. \left(\omega_{3}\overline{\omega_{3}}\right)\right|_{n=2} \right|\equiv \left(\Delta m^{2}_{12}\right)^{\nu}_{A} \\
\displaystyle \left(\Delta m^{2}_{13}\right)^{\nu}_{C}=\left| \left. \left(\omega_{1}\overline{\omega_{1}}\right)\right|_{n=1}-\left. \left(\omega_{1}\overline{\omega_{1}}\right)\right|_{n=3} \right| \\
\displaystyle \left(\Delta m^{2}_{23}\right)^{\nu}_{C}=\left| \left. \left(\omega_{1}\overline{\omega_{1}}\right)\right|_{n=2}-\left. \left(\omega_{1}\overline{\omega_{1}}\right)\right|_{n=3} \right| \; .
\end{array}
\label{eq_19}
\end{equation}

$\displaystyle \left(\Delta m^{2}_{13}\right)^{\nu}_{A}$, $\displaystyle \left(\Delta m^{2}_{13}\right)^{\nu}_{B}$, $\displaystyle \left(\Delta m^{2}_{13}\right)^{\nu}_{D}$, $\displaystyle \left(\Delta m^{2}_{23}\right)^{\nu}_{A}$, $\displaystyle \left(\Delta m^{2}_{23}\right)^{\nu}_{B}$ and $\displaystyle \left(\Delta m^{2}_{23}\right)^{\nu}_{D}$ are not defined because in each substraction one term is for a real particle and the other for a virtual particle. For the virtual particles we will have:

\begin{equation}
\begin{array}{ccc}
\displaystyle \left(\Delta m^{2}_{12}\right)^{\nu}=\left| \left. \left(\omega_{2}\overline{\omega_{2}}\right)\right|_{n=1}-\left. \left(\omega_{2}\overline{\omega_{2}}\right)\right|_{n=2} \right|\equiv \left(\Delta m^{2}_{12}\right)^{\nu}_{B} \\
\displaystyle \left(\Delta m^{2}_{13}\right)^{\nu}=\left| \left. \left(\omega_{2}\overline{\omega_{2}}\right)\right|_{n=1}-\left. \left(\omega_{4}\overline{\omega_{4}}\right)\right|_{n=3} \right| \\
\displaystyle \left(\Delta m^{2}_{23}\right)^{\nu}=\left| \left. \left(\omega_{2}\overline{\omega_{2}}\right)\right|_{n=2}-\left. \left(\omega_{4}\overline{\omega_{4}}\right)\right|_{n=3} \right| \; .
\end{array}
\label{eq_20}
\end{equation}

\begin{figure}[htb]
\centering
\leavevmode
\includegraphics[width=7cm,height=3cm]{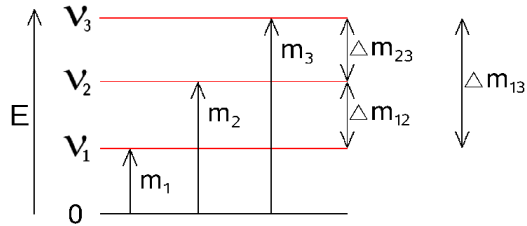}
\caption{Real (timelike) neutrinos "normal" hierarchy.}
\label{ierarh1}
\end{figure}

\begin{figure}[htb]
\centering
\leavevmode
\includegraphics[width=7cm,height=3cm]{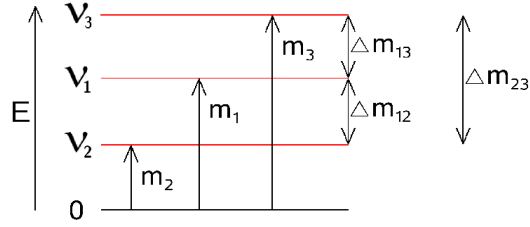}
\caption{Virtual (spacelike) neutrinos "inverted" hierarchy.}
\label{ierarh2}
\end{figure}

With the values for $\displaystyle \left(\Delta m^{2}_{ij}\right)^{\nu}$ (see table 3) we can construct the neutrino flavor hierarchic schemes (Figures \ref{ierarh1} and \ref{ierarh2}). Relative to the hierarchy of the neutrinos as real particles (seen as timelike particles), the neutrinos as virtual particles (seen as spacelike particles) are in an inverted hierarchy. 

\begin{table}[htb]
\begin{center}
\begin{tabular}{|c|c|c|c|c|c|} \hline\hline
\multicolumn{2}{|c|}{\bf Table 1} & \multicolumn{2}{c|}{\bf Down Sign} & \multicolumn{2}{c|}{\bf Up Sign}  \\ \cline{3-6}
\multicolumn{2}{|c|}{}  & $H_{0}=72.35$ & $H_{0}=72.31$ & $H_{0}=72.35$ & $H_{0}=72.31$  \\\hline\hline
 & $\omega_{4}$ & -241178 & -240911 & -241178 & -240911 \\\cline{2-6}
 & $\overline{\omega_{4}}$ & -7273.82 & -7265.78 & -7273.82 & -7265.78 \\\cline{2-6}
 & $\omega_{2}$ & -35267.4 {\it i} & -35228.4 {\it i} & -35267.4 {\it i} & -35228.4 {\it i} \\\cline{2-6}
 & $\overline{\omega_{2}}$ & 49742.4 {\it i} & 49687.4 {\it i} & 49742.4 {\it i} & 49687.4 {\it i} \\[.2ex] \cline{2-6}
 & & & & & \\[-2.3ex] \cline{2-6}
 & $\omega_{4}$ & 241178 & 240911 & 241178 & 240911 \\\cline{2-6}
 & $\overline{\omega_{4}}$ & 7273.82 & 7265.78 & 7273.82 & 7265.78 \\\cline{2-6}
 & $\omega_{2}$ & 35267.4 {\it i} & 35228.4 {\it i} & 35267.4 {\it i} & 35228.4 {\it i} \\\cline{2-6}
$n=1$ & $\overline{\omega_{2}}$ & -49742.4 {\it i} & -49687.4 {\it i} & -49742.4 {\it i} & -49687.4 {\it i} \\[.2ex] \cline{2-6}
 & & & & & \\[-2.3ex] \cline{2-6}
 & $\omega_{1}$ & -1767.03 $\times$ $(a/r)$ & -1770.17 $\times$ $(a/r)$ & -1767.03 $\times$ $(a/r)$ & -1770.17 $\times$ $(a/r)$ \\\cline{2-6}
 & $\overline{\omega_{1}}$ & -959.19 $\times$ $(a/r)$ & -960.896 $\times$ $(a/r)$ & -959.19 $\times$ $(a/r)$ & -960.896 $\times$ $(a/r)$ \\\cline{2-6}
 & $\omega_{3}$ & -3124.05 $\times$ $(a/r)$ & -3129.6 $\times$ $(a/r)$ & -3124.05 $\times$ $(a/r)$ & -3129.6 $\times$ $(a/r)$ \\\cline{2-6}
 & $\overline{\omega_{3}}$ & -542.538 $\times$ $(a/r)$ & -543.503 $\times$ $(a/r)$ & -542.538 $\times$ $(a/r)$ & -543.503 $\times$ $(a/r)$ \\[.2ex] \cline{2-6}
 & & & & & \\[-2.3ex] \cline{2-6}
 & $\omega_{1}$ & 1767.03 $\times$ $(a/r)$ & 1770.17 $\times$ $(a/r)$ & 1767.03 $\times$ $(a/r)$ & 1770.17 $\times$ $(a/r)$ \\\cline{2-6}
 & $\overline{\omega_{1}}$ & 959.19 $\times$ $(a/r)$ & 960.896 $\times$ $(a/r)$ & 959.19 $\times$ $(a/r)$ & 960.896 $\times$ $(a/r)$ \\\cline{2-6}
 & $\omega_{3}$ & 3124.05 $\times$ $(a/r)$ & 3129.6 $\times$ $(a/r)$ & 3124.05 $\times$ $(a/r)$ & 3129.6 $\times$ $(a/r)$ \\\cline{2-6}
 & $\overline{\omega_{3}}$ & 542.538 $\times$ $(a/r)$ & 543.503 $\times$ $(a/r)$ & 542.538 $\times$ $(a/r)$ & 543.503 $\times$ $(a/r)$ \\[.2ex] \hline
 & & & & & \\[-2.3ex] \hline
 & $\omega_{4}$ & -2.33827 $\times$ $10^{6}$ & -2.33568 $\times$ $10^{6}$ & -2.33827 $\times$ $10^{6}$ & -2.33568 $\times$ $10^{6}$ \\\cline{2-6}
 & $\overline{\omega_{4}}$ & -750.25 & -749.42 & -750.25 & -749.42 \\\cline{2-6}
 & $\omega_{2}$ & -39821.3 {\it i} & -39777.3 {\it i} & -39821.3 {\it i} & -39777.3 {\it i} \\\cline{2-6}
 & $\overline{\omega_{2}}$ & 44053.9 {\it i} & 44005.2 {\it i} & 44053.9 {\it i} & 44005.2 {\it i} \\[.2ex] \cline{2-6}
 & & & & & \\[-2.3ex] \cline{2-6}
 & $\omega_{4}$ & 2.33827 $\times$ $10^{6}$ & 2.33568 $\times$ $10^{6}$ & 2.33827 $\times$ $10^{6}$ & 2.33568 $\times$ $10^{6}$ \\\cline{2-6}
 & $\overline{\omega_{4}}$ & 750.25 & 749.42 & 750.25 & 749.42 \\\cline{2-6}
 & $\omega_{2}$ & 39821.3 {\it i} & 39777.3 {\it i} & 39821.3 {\it i} & 39777.3 {\it i} \\\cline{2-6}
$n=2$ & $\overline{\omega_{2}}$ & -44053.9 {\it i} & -44005.2 {\it i} & -44053.9 {\it i} & -44005.2 {\it i} \\[.2ex] \cline{2-6}
 & & & & & \\[-2.3ex] \cline{2-6}
 & $\omega_{1}$ & -2896.64 $\times$ $(a/r)$ & -2901.79 $\times$ $(a/r)$ & -2896.64 $\times$ $(a/r)$ & -2901.79 $\times$ $(a/r)$ \\\cline{2-6}
 & $\overline{\omega_{1}}$ & -585.131 $\times$ $(a/r)$ & -586.172 $\times$ $(a/r)$ & -585.131 $\times$ $(a/r)$ & -586.172 $\times$ $(a/r)$ \\\cline{2-6}
 & $\omega_{3}$ & -9727.38 $\times$ $(a/r)$ & -9744.68 $\times$ $(a/r)$ & -9727.38 $\times$ $(a/r)$ & -9744.68 $\times$ $(a/r)$ \\\cline{2-6}
 & $\overline{\omega_{3}}$ & -174.242 $\times$ $(a/r)$ & -174.551 $\times$ $(a/r)$ & -174.242 $\times$ $(a/r)$ & -174.551 $\times$ $(a/r)$ \\[.2ex] \cline{2-6}
 & & & & & \\[-2.3ex] \cline{2-6}
 & $\omega_{1}$ & 2896.64 $\times$ $(a/r)$ & 2901.79 $\times$ $(a/r)$ & 2896.64 $\times$ $(a/r)$ & 2901.79 $\times$ $(a/r)$ \\\cline{2-6}
 & $\overline{\omega_{1}}$ & 585.131 $\times$ $(a/r)$ & 586.172 $\times$ $(a/r)$ & 585.131 $\times$ $(a/r)$ & 586.172 $\times$ $(a/r)$ \\\cline{2-6}
 & $\omega_{3}$ & 9727.38 $\times$ $(a/r)$ & 9744.68 $\times$ $(a/r)$ & 9727.38 $\times$ $(a/r)$ & 9744.68 $\times$ $(a/r)$ \\\cline{2-6}
 & $\overline{\omega_{3}}$ & 174.242 $\times$ $(a/r)$ & 174.551 $\times$ $(a/r)$ & 174.242 $\times$ $(a/r)$ & 174.551 $\times$ $(a/r)$ \\[.2ex] \hline
 & & & & & \\[-2.3ex] \hline
 & $\omega_{4}$ & -2.49874 $\times$ $10^{6}$ {\it i} & -2.49597 $\times$ $10^{6}$ {\it i} & -2.49874 $\times$ $10^{6}$ {\it i} & -2.49597 $\times$ $10^{6}$ {\it i} \\\cline{2-6}
 & $\overline{\omega_{4}}$ & 702.069 {\it i} & 701.293 {\it i} & 702.069 {\it i} & 701.293 {\it i} \\\cline{2-6}
 & $\omega_{2}$ & -44246.3 & -44197.4 & -44246.3 & -44197.4 \\\cline{2-6}
 & $\overline{\omega_{2}}$ & -39648.2 & -39604.3 & -39648.2 & -39604.3 \\[.2ex] \cline{2-6}
 & & & & & \\[-2.3ex] \cline{2-6}
 & $\omega_{4}$ & 2.49874 $\times$ $10^{6}$ {\it i} & 2.49597 $\times$ $10^{6}$ {\it i} & 2.49874 $\times$ $10^{6}$ {\it i} & 2.49597 $\times$ $10^{6}$ {\it i} \\\cline{2-6}
 & $\overline{\omega_{4}}$ & -702.069 {\it i} & -701.293 {\it i} & -702.069 {\it i} & -701.293 {\it i} \\\cline{2-6}
 & $\omega_{2}$ & 44246.3 & 44197.4 & 44246.3 & 44197.4 \\\cline{2-6}
$n=3$ & $\overline{\omega_{2}}$ & 39648.2 & 39604.3 & 39648.2 & 39604.3 \\[.2ex] \cline{2-6}
 & & & & & \\[-2.3ex] \cline{2-6}
 & $\omega_{1}$ & -2827.78 $\times$ $(a/r)$ & -2832.81 $\times$ $(a/r)$ & -2827.78 $\times$ $(a/r)$ & -2832.81 $\times$ $(a/r)$ \\\cline{2-6}
 & $\overline{\omega_{1}}$ & -599.38 $\times$ $(a/r)$ & -600.446 $\times$ $(a/r)$ & -599.38 $\times$ $(a/r)$ & -600.446 $\times$ $(a/r)$ \\\cline{2-6}
 & $\omega_{3}$ & -10055.6 {\it i} $\times$ $(a/r)$ & -10073.5 {\it i} $\times$ $(a/r)$ &  -10055.6 {\it i} $\times$ $(a/r)$ & -10073.5 {\it i} $\times$ $(a/r)$ \\\cline{2-6}
 & $\overline{\omega_{3}}$ & 168.554 {\it i} $\times$ $(a/r)$ & 168.854 {\it i} $\times$ $(a/r)$ & 168.554 {\it i} $\times$ $(a/r)$ & 168.854 {\it i} $\times$ $(a/r)$ \\[.2ex] \cline{2-6}
 & & & & & \\[-2.3ex] \cline{2-6}
 & $\omega_{1}$ & 2827.78 $\times$ $(a/r)$ & 2832.81 $\times$ $(a/r)$ & 2827.78 $\times$ $(a/r)$ & 2832.81 $\times$ $(a/r)$ \\\cline{2-6}
 & $\overline{\omega_{1}}$ & 599.38 $\times$ $(a/r)$ & 600.446 $\times$ $(a/r)$ & 599.38 $\times$ $(a/r)$ & 600.446 $\times$ $(a/r)$ \\\cline{2-6}
 & $\omega_{3}$ & 10055.6 {\it i} $\times$ $(a/r)$ & 10073.5 {\it i} $\times$ $(a/r)$ &  10055.6 {\it i} $\times$ $(a/r)$ & 10073.5 {\it i} $\times$ $(a/r)$ \\\cline{2-6}
 & $\overline{\omega_{3}}$ & -168.554 {\it i} $\times$ $(a/r)$ & -168.854 {\it i} $\times$ $(a/r)$ & -168.554 {\it i} $\times$ $(a/r)$ & -168.854 {\it i} $\times$ $(a/r)$ \\\hline
\noalign{\smallskip} \hline
\end{tabular}
\end{center}
\label{unu}
\end{table}

\begin{table}[htb]
\begin{center}
\begin{tabular}{|c|c|c|c|c|} \hline\hline
\multicolumn{2}{|c|}{\bf Table 2}  & $H_{0}=72.35$ & $H_{0}=72.31$ & {\bf particle type} \\\hline\hline
 & $\left[\displaystyle \Re(E_{Yamabe,\; helicoid})\right]^{2}=\omega_{4}\overline{\omega_{4}}$ & 1.75\textcolor{red}{4285}\textcolor{blue}{36} $\times$ $10^{9}$ & 1.75\textcolor{red}{040}\textcolor{blue}{6326} $\times$ $10^{9}$ & {\bf real}  \\\cline{2-5}
 & $\left[\displaystyle \Re(E_{Yamabe,\; helicoid})\right]^{2}=\omega_{2}\overline{\omega_{2}}$ & 1.75\textcolor{red}{4285}\textcolor{blue}{118} $\times$ $10^{9}$ & 1.75\textcolor{red}{040}\textcolor{blue}{7602} $\times$ $10^{9}$ & {\bf virtual} \\\cline{2-5}
 $n=1$ & $\left[\displaystyle \Im(E_{Yamabe,\; helicoid})\right]^{2}=\omega_{1}\overline{\omega_{1}}$ & 1.69\textcolor{red}{491}\textcolor{blue}{7506} $\times$ $10^{6}$ $\times$ $\displaystyle \frac{a^{2}}{r^{2}}$ & 1.70\textcolor{red}{094}\textcolor{blue}{9272} $\times$ $10^{6}$ $\times$ $\displaystyle \frac{a^{2}}{r^{2}}$ & {\bf real} \\\cline{2-5}
 & $\left[\displaystyle \Im(E_{Yamabe,\; helicoid})\right]^{2}=\omega_{3}\overline{\omega_{3}}$ & 1.69\textcolor{red}{491}\textcolor{blue}{5839} $\times$ $10^{6}$ $\times$ $\displaystyle \frac{a^{2}}{r^{2}}$ & 1.70\textcolor{red}{094}\textcolor{blue}{6989} $\times$ $10^{6}$ $\times$ $\displaystyle \frac{a^{2}}{r^{2}}$ & {\bf real} \\[.2ex] \hline
 & & & & \\[-2.3ex] \hline
 & $\left[\displaystyle \Re(E_{Yamabe,\; helicoid})\right]^{2}=\omega_{4}\overline{\omega_{4}}$ & 1.75\textcolor{red}{428}\textcolor{blue}{7068} $\times$ $10^{9}$ & 1.75\textcolor{red}{040}\textcolor{blue}{5306} $\times$ $10^{9}$ & {\bf real} \\\cline{2-5}
 & $\left[\displaystyle \Re(E_{Yamabe,\; helicoid})\right]^{2}=\omega_{2}\overline{\omega_{2}}$ & 1.75\textcolor{red}{428}\textcolor{blue}{3568} $\times$ $10^{9}$ & 1.75\textcolor{red}{040}\textcolor{blue}{8042} $\times$ $10^{9}$ & {\bf virtual} \\\cline{2-5}
 $n=2$ & $\left[\displaystyle \Im(E_{Yamabe,\; helicoid})\right]^{2}=\omega_{1}\overline{\omega_{1}}$  & 1.69\textcolor{red}{491}\textcolor{blue}{386} $\times$ $10^{6}$ $\times$ $\displaystyle \frac{a^{2}}{r^{2}}$ & 1.70\textcolor{red}{094}\textcolor{blue}{8048} $\times$ $10^{6}$ $\times$ $\displaystyle \frac{a^{2}}{r^{2}}$ & {\bf real} \\\cline{2-5}
 & $\left[\displaystyle \Im(E_{Yamabe,\; helicoid})\right]^{2}=\omega_{3}\overline{\omega_{3}}$ & 1.69\textcolor{red}{491}\textcolor{blue}{8146} $\times$ $10^{6}$ $\times$ $\displaystyle \frac{a^{2}}{r^{2}}$ & 1.70\textcolor{red}{094}\textcolor{blue}{3639} $\times$ $10^{6}$ $\times$ $\displaystyle \frac{a^{2}}{r^{2}}$ & {\bf real} \\[.2ex] \hline
 & & & & \\[-2.3ex] \hline
 & $\left[\displaystyle \Re(E_{Yamabe,\; helicoid})\right]^{2}=\omega_{4}\overline{\omega_{4}}$ & 1.75\textcolor{red}{428}\textcolor{blue}{7893} $\times$ $10^{9}$ & 1.75\textcolor{red}{040}\textcolor{blue}{6289} $\times$ $10^{9}$ & {\bf virtual} \\\cline{2-5}
 & $\left[\displaystyle \Re(E_{Yamabe,\; helicoid})\right]^{2}=\omega_{2}\overline{\omega_{2}}$ & 1.75\textcolor{red}{428}\textcolor{blue}{6152} $\times$ $10^{9}$ & 1.75\textcolor{red}{040}\textcolor{blue}{7089} $\times$ $10^{9}$ & {\bf real} \\\cline{2-5}
 $n=3$ & $\left[\displaystyle \Im(E_{Yamabe,\; helicoid})\right]^{2}=\omega_{1}\overline{\omega_{1}}$ & 1.69\textcolor{red}{491}\textcolor{blue}{4776} $\times$ $10^{6}$ $\times$ $\displaystyle \frac{a^{2}}{r^{2}}$ & 1.70\textcolor{red}{0949}\textcolor{blue}{433} $\times$ $10^{6}$ $\times$ $\displaystyle \frac{a^{2}}{r^{2}}$ & {\bf real} \\\cline{2-5}
 & $\left[\displaystyle \Im(E_{Yamabe,\; helicoid})\right]^{2}=\omega_{3}\overline{\omega_{3}}$ & 1.69\textcolor{red}{491}\textcolor{blue}{1602} $\times$ $10^{6}$ $\times$ $\displaystyle \frac{a^{2}}{r^{2}}$ & 1.70\textcolor{red}{0950}\textcolor{blue}{769} $\times$ $10^{6}$ $\times$ $\displaystyle \frac{a^{2}}{r^{2}}$ & {\bf virtual} \\\hline
\noalign{\smallskip} \hline
\end{tabular}
\label{doi}
\end{center}
\end{table}

\begin{table}[htb]
\begin{center}
{\renewcommand{\arraystretch}{1.8}
\begin{tabular}{|c|c|c|c|} \hline\hline
{\bf Table 3}  & $H_{0}=72.35$ & $H_{0}=72.31$ & {\bf particle type} \\\hline\hline
$\displaystyle \left(\Delta m^{2}_{12}\right)^{\nu}_{A}$ & 2 $\times$ $10^{3}$ & 1 $\times$ $10^{3}$ & {\bf real}  \\\hline
$\displaystyle \left(\Delta m^{2}_{12}\right)^{\nu}_{B}$ & 2 $\times$ $10^{3}$ & 1 $\times$ $10^{3}$ & {\bf virtual} \\\hline
$\displaystyle \left(\Delta m^{2}_{12}\right)^{\nu}_{C}$ & 2 $\times$ $10^{3}$ & 1 $\times$ $10^{3}$ & {\bf real}  \\\hline
$\displaystyle \left(\Delta m^{2}_{12}\right)^{\nu}_{D}$ & 2 $\times$ $10^{3}$ & 1 $\times$ $10^{3}$ & {\bf real} \\\hline
$\displaystyle \left(\Delta m^{2}_{13}\right)^{\nu}_{A}$ & - & - & -  \\\hline
$\displaystyle \left(\Delta m^{2}_{13}\right)^{\nu}_{B}$ & - & - & - \\\hline
$\displaystyle \left(\Delta m^{2}_{13}\right)^{\nu}_{C}$ & 2.8 $\times$ $10^{3}$ & 2.2 $\times$ $10^{3}$ & {\bf real}  \\\hline
$\displaystyle \left(\Delta m^{2}_{13}\right)^{\nu}_{D}$ & - & - & - \\\hline
$\displaystyle \left(\Delta m^{2}_{23}\right)^{\nu}_{A}$ & - & - & -  \\\hline
$\displaystyle \left(\Delta m^{2}_{23}\right)^{\nu}_{B}$ & - & - & - \\\hline
$\displaystyle \left(\Delta m^{2}_{23}\right)^{\nu}_{C}$ & 1 $\times$ $10^{3}$ & 1.2 $\times$ $10^{3}$ & {\bf real}  \\\hline
$\displaystyle \left(\Delta m^{2}_{23}\right)^{\nu}_{D}$ & - & - & - \\\hline
\noalign{\smallskip} \hline
\end{tabular}}
\label{trei}
\end{center}
\end{table}

\section{Quarks}

From the first element of the $\Psi_{i}\overline{\Psi}_{i}=\Psi_{f}\overline{\Psi}_{f}$ equality results:

\begin{equation}
\!\!\!\!\!\!\!\!\!\!\!\!\!\!\!\!\!\!\!\!\!\!\!\!\!\begin{array}{cccccc}
\displaystyle \pm \omega \left(2\; i\; \alpha\; \sin \alpha\; \cos \alpha\; \sin \beta+2\; i\; \alpha\; \beta\; \sin \alpha\; \cos \alpha\; \cos \beta -2\; \alpha\; \beta\; \sin \alpha\; \cos \alpha\; \sin \beta -\right. \\
\displaystyle \left. -\alpha\; \sin^{2}\alpha\; \sin \beta -\alpha\; \beta\; \sin^{2}\alpha\; \cos \beta-i\; \alpha\; \beta\; \sin^{2}\alpha\; \sin \beta\right)\mp \overline{\omega}\left( 2\; i\; \beta\; \sin \beta\; \cos \beta\; \sin \alpha +\right. \\
\displaystyle +2\; i\; \alpha\; \beta\; \sin \beta\; \cos \beta\; \cos \alpha -2\; \alpha\; \beta\; \sin \beta\; \cos \beta\; \sin \alpha -\beta\; \sin^{2}\beta\; \sin \alpha -\alpha\; \beta\; \sin^{2}\beta\; \cos \alpha - \\
\displaystyle \left. -i\; \alpha\; \beta\; \sin^{2}\beta\; \sin \alpha\right)=i\; \frac{\sqrt{2}}{2}\left(\overline{\omega}-\omega \right)\left( -\sin \alpha\; \sin \beta-\beta\; \sin \alpha\; \cos \beta-i\; \beta\; \sin \alpha\; \sin \beta- \right. \\
\displaystyle -\alpha\; \cos \alpha\; \sin \beta -\alpha\; \beta\; \cos \alpha\; \cos \beta-i\; \alpha\; \beta\; \cos \alpha\; \sin \beta-i\; \alpha\; \sin \alpha\; \sin \beta-i\; \alpha\; \beta\; \sin \alpha\; \cos \beta + \\
\displaystyle \left. +\alpha\; \beta\; \sin \alpha\; \sin \beta\right) \; ,
\end{array}
\label{eq_21}
\end{equation}
\noindent
where the imaginary part is ($\omega_{1}=\Im\{\omega\}$; $\overline{\omega_{1}}=\Im\{\overline{\omega}\}$):

\begin{equation}
\!\!\!\!\!\!\!\!\!\!\!\!\!\!\!\!\!\!\!\!\!\!\!\!\!\!\!\!\!\!\!\!\!\!\!\!\!\!\!\!\!\!\!\left\{\begin{array}{llll}
\displaystyle \pm \omega_{1} \left(2\; \alpha\; \sin \alpha\; \cos \alpha\; \sin \beta+2\; \alpha\; \beta\; \sin \alpha\; \cos \alpha\; \cos \beta - \alpha\; \beta\; \sin^{2} \alpha\; \sin \beta\right)\mp \\
\displaystyle \mp \overline{\omega_{1}} \left(2\; \beta\; \sin \beta\; \cos \beta\; \sin \alpha+2\; \alpha\; \beta\; \sin \beta\; \cos \beta\; \cos \alpha - \alpha\; \beta\; \sin^{2} \beta\; \sin \alpha\right)= \\
\displaystyle =\frac{\sqrt{2}}{2}\left(\overline{\omega_{1}}-\omega_{1} \right) \left(-\sin \alpha\; \sin \beta -\beta\; \sin \alpha\; \cos \beta-\alpha\; \cos \alpha\; \sin \beta-\alpha\; \beta\; \cos \alpha\; \cos \beta+\alpha\; \beta\; \sin \alpha\; \sin \beta\right) \\
\displaystyle \left[\displaystyle \Im(E_{Yamabe,\; helicoid}(r,a))\right]^{2}=\left[\displaystyle \Re(E_{Yamabe,\; helicoid}(r,A))\right]^{2}=\omega_{1}\overline{\omega_{1}}
\end{array} \right.
\label{eq_22}
\end{equation}
\noindent
and the real part is ($\omega_{2}=\Re\{\omega\}$; $\overline{\omega_{2}}=\Re\{\overline{\omega}\}$):

\begin{equation}
\!\!\!\!\!\!\!\!\!\!\!\!\!\!\!\!\!\!\!\!\!\!\!\!\!\!\!\!\!\!\!\!\!\!\!\!\!\!\!\!\!\!\!\left\{\begin{array}{lllll}
\displaystyle \mp \omega_{2} \left(2\; \alpha\; \beta\; \sin \alpha\; \cos \alpha\; \sin \beta +\alpha\; \sin^{2}\alpha\; \sin \beta +\alpha\; \beta\; \sin^{2}\alpha\; \cos \beta \mp \frac{1}{\sqrt{2}}\; \beta\; \sin \alpha\; \sin \beta \mp \right. \\
\displaystyle \left. \mp \frac{1}{\sqrt{2}}\; \alpha\; \beta\; \cos \alpha\; \sin \beta \mp \frac{1}{\sqrt{2}}\; \alpha\; \sin \alpha\; \sin \beta \mp \frac{1}{\sqrt{2}}\; \alpha\; \beta\; \sin \alpha\; \cos \beta\right)\pm \overline{\omega_{2}}\left(2\; \alpha\; \beta\; \sin \beta\; \cos \beta\; \sin \alpha+ \right. \\
\displaystyle +\beta\; \sin^{2}\beta\; \sin \alpha +\alpha\; \beta\; \sin^{2}\beta\; \cos \alpha \mp \frac{1}{\sqrt{2}}\; \beta\; \sin \alpha\; \sin \beta \mp \frac{1}{\sqrt{2}}\; \alpha\; \beta\; \cos \alpha\; \sin \beta \mp \frac{1}{\sqrt{2}}\; \alpha\; \sin \alpha\; \sin \beta \mp \\
\displaystyle \left. \mp \frac{1}{\sqrt{2}}\; \alpha\; \beta\; \sin \alpha\; \cos \beta\right)=0 \\
\displaystyle \left[\displaystyle \Re(E_{Yamabe,\; helicoid}(r,a))\right]^{2}=\left[\displaystyle \Im(E_{Yamabe,\; helicoid}(r,A))\right]^{2}=\omega_{2}\overline{\omega_{2}}\; .
\end{array} \right.
\label{eq_23}
\end{equation}

From the second element of $\Psi_{i}\overline{\Psi}_{i}=\Psi_{f}\overline{\Psi}_{f}$ results:

\begin{equation}
\displaystyle \mp i\; \omega\; \sin \left[\textnormal{arcctg} \left(\textnormal{tg}\; \alpha \right) \right] \pm i\; \overline{\omega}\; \sin \left[\textnormal{arcctg} \left(\textnormal{tg}\; \beta \right) \right]=-i\; \frac{1}{\sqrt{2}}\; \omega+i\; \frac{1}{\sqrt{2}}\; \overline{\omega} \; ,
\label{eq_24}
\end{equation}
\noindent
where the imaginary part is ($\omega_{4}=\Im\{\omega\}$; $\overline{\omega_{4}}=\Im\{\overline{\omega}\}$):

\begin{equation}
\left\{\begin{array}{ll}
\displaystyle \mp \omega_{4} \left\{\sin \left[\textnormal{arcctg} \left(\textnormal{tg}\; \alpha \right) \right]\mp\frac{1}{\sqrt{2}} \right\}\pm \overline{\omega_{4}}\left\{\sin \left[\textnormal{arcctg} \left(\textnormal{tg}\; \beta \right) \right]\mp\frac{1}{\sqrt{2}} \right\}=0 \\
\displaystyle \left[\displaystyle \Im(E_{Yamabe,\; helicoid}(r,a))\right]^{2}=\left[\displaystyle \Re(E_{Yamabe,\; helicoid}(r,A))\right]^{2}=\omega_{4}\overline{\omega_{4}}
\end{array} \right.
\label{eq_25}
\end{equation}
\noindent
and the real part is ($\omega_{3}=\Re\{\omega\}$; $\overline{\omega_{3}}=\Re\{\overline{\omega}\}$):

\begin{equation}
\displaystyle \left[\displaystyle \Re(E_{Yamabe,\; helicoid}(r,a))\right]^{2}=\left[\displaystyle \Im(E_{Yamabe,\; helicoid}(r,A))\right]^{2}=\omega_{3}\overline{\omega_{3}}=0 \; ,
\label{eq_26}
\end{equation}
\noindent
which is possible to be achieved for $r\rightarrow \infty$ in the helicoid's Yamabe energy \cite{deus3}.

Because of the distortion of the embedded DEUS particles, the angles to be used for solving the above systems will be $\displaystyle \alpha=\frac{\pi}{4}+n\; \frac{\pi}{12}+d$ and $\displaystyle \beta=\frac{\pi}{4}-n\; \frac{\pi}{12}-d$, with $\displaystyle d\in \left.\left(0,\frac{\pi}{12}\right]\right.$ and $n=\overline{1,3}$.

Solving the (\ref{eq_22}), (\ref{eq_23})  and (\ref{eq_25}) systems in $\omega_{1}$, $\omega_{2}$, $\omega_{4}$ and $\overline{\omega_{1}}$, $\overline{\omega_{2}}$, $\overline{\omega_{4}}$, we obtain for the three quark generations the table 4 values. In this table we observe the same conservation as for neutrinos (with the same correlations) of the "{\bf T}" and "{\bf C}" symmetries, while the "{\bf P}" symmetry is violated.

The real and imaginary parts of the Yamabe energy (in {\it r} and {\it A}) resulting from (\ref{eq_22}), (\ref{eq_23}), (\ref{eq_25}) and (\ref{eq_26}) are given in table 5. We kept in black the relevant decimals for which we had equality between the two real parts or between the two imaginary parts of the Yamabe energy values (for the same $H_{0}$ and {\it n}) in the neutrino case. Again, a virtual particle is described by imaginary $\omega$ values and a real particle by real $\omega$ values (see table 4).

\begin{table}[htbc]
\begin{center}
\begin{tabular}{|c|c|c|c|c|c|} \hline\hline
\multicolumn{2}{|c|}{\bf Table 4} & \multicolumn{2}{c|}{\bf Up Sign} & \multicolumn{2}{c|}{\bf Down Sign}  \\ \cline{3-6}
\multicolumn{2}{|c|}{}  & $H_{0}=72.35$ & $H_{0}=72.31$ & $H_{0}=72.35$ & $H_{0}=72.31$  \\\hline\hline
 & $\omega_{4}$ & -60562.8 & -60495.9 & -38858.1 & -38815.2 \\\cline{2-6}
 & $\overline{\omega_{4}}$ & -28966.4 & -28934.3 & -45145.9 & -45096 \\\cline{2-6}
 & $\omega_{2}$ & -34276.8 {\it i} & -34239 {\it i} & -51179.9 & -51123.3 \\\cline{2-6}
 & $\overline{\omega_{2}}$ & 51179.9 {\it i} & 51123.3 {\it i} & -34276.8 & -34239 \\[.2ex] \cline{2-6}
 & & & & & \\[-2.3ex] \cline{2-6}
 & $\omega_{4}$ & 60562.8 & 60495.9 & 38858.1 & 38815.2 \\\cline{2-6}
 & $\overline{\omega_{4}}$ & 28966.4 & 28934.3 & 45145.9 & 45096 \\\cline{2-6}
$n=1$  & $\omega_{2}$ & 34276.8 {\it i} & 34239 {\it i} & 51179.9 & 51123.3 \\\cline{2-6}
 & $\overline{\omega_{2}}$ & -51179.9 {\it i} & -51123.3 {\it i} & 34276.8 & 34239 \\[.2ex] \cline{2-6}
 & & & & & \\[-2.3ex] \cline{2-6}
 & $\omega_{1}$ & -1924.2 {\it i} $\times$ $(A/r)$ & -1927.62 {\it i} $\times$ $(A/r)$ & -1083.06 $\times$ $(A/r)$ & -1084.99 $\times$ $(A/r)$ \\\cline{2-6}
 & $\overline{\omega_{1}}$ & 880.84 {\it i} $\times$ $(A/r)$ & 882.407 {\it i} $\times$ $(A/r)$ & -1564.93 $\times$ $(A/r)$ & -1567.71 $\times$ $(A/r)$ \\[.2ex] \cline{2-6}
 & & & & & \\[-2.3ex] \cline{2-6}
 & $\omega_{1}$ & 1924.2 {\it i} $\times$ $(A/r)$ & 1927.62 {\it i} $\times$ $(A/r)$ & 1083.06 $\times$ $(A/r)$ & 1084.99 $\times$ $(A/r)$ \\\cline{2-6}
 & $\overline{\omega_{1}}$ & -880.84 {\it i} $\times$ $(A/r)$ & -882.407 {\it i} $\times$ $(A/r)$ & 1564.93 $\times$ $(A/r)$ & 1567.71 $\times$ $(A/r)$ \\[.2ex] \hline
 & & & & & \\[-2.3ex] \hline
 & $\omega_{4}$ & -47555.4 & -47502.9 & -38995.8 & -38952.8 \\\cline{2-6}
 & $\overline{\omega_{4}}$ & -36889.2 & -36848.5 & -44986.4 & -44936.7 \\\cline{2-6}
 & $\omega_{2}$ & -29452.7 {\it i} & -29420.1 {\it i} & -59562.8 & -59497 \\\cline{2-6}
 & $\overline{\omega_{2}}$ & 59562.8 {\it i} & 59497 {\it i} & -29452.7 & -29420.1 \\[.2ex] \cline{2-6}
 & & & & & \\[-2.3ex] \cline{2-6}
 & $\omega_{4}$ & 47555.4 & 47502.9 & 38995.8 & 38952.8 \\\cline{2-6}
 & $\overline{\omega_{4}}$ & 36889.2 & 36848.5 & 44986.4 & 44936.7 \\\cline{2-6}
$n=2$ & $\omega_{2}$ & 29452.7 {\it i} & 29420.1 {\it i} & 59562.8 & 59497 \\\cline{2-6}
 & $\overline{\omega_{2}}$ & -59562.8 {\it i} & -59497 {\it i} & 29452.7 & 29420.1 \\[.2ex] \cline{2-6}
 & & & & & \\[-2.3ex] \cline{2-6}
 & $\omega_{1}$ & -1978.63 {\it i} $\times$ $(A/r)$ & -1982.15 {\it i} $\times$ $(A/r)$ & -924.911 $\times$ $(A/r)$ & -926.556 $\times$ $(A/r)$ \\\cline{2-6}
 & $\overline{\omega_{1}}$ & 856.608 {\it i} $\times$ $(A/r)$ & 858.131 {\it i} $\times$ $(A/r)$ & -1832.52 $\times$ $(A/r)$ & -1835.77 $\times$ $(A/r)$ \\[.2ex] \cline{2-6}
 & & & & & \\[-2.3ex] \cline{2-6}
 & $\omega_{1}$ & 1978.63 {\it i} $\times$ $(A/r)$ & 1982.15 {\it i} $\times$ $(A/r)$ & 924.911 $\times$ $(A/r)$ & 926.556 $\times$ $(A/r)$ \\\cline{2-6}
 & $\overline{\omega_{1}}$ & -856.608 {\it i} $\times$ $(A/r)$ & -858.131 {\it i} $\times$ $(A/r)$ & 1832.52 $\times$ $(A/r)$ & 1835.77 $\times$ $(A/r)$ \\[.2ex] \hline
 & & & & & \\[-2.3ex] \hline
 & $\omega_{4}$ & -36911.8 & -36871 & -54751.1 & -54690.6 \\\cline{2-6}
 & $\overline{\omega_{4}}$ & -47526.3 & -47473.8 & -32041.1 & -32005.6 \\\cline{2-6}
 & $\omega_{2}$ & -59724.2 & -59658.2 & -29373.1 {\it i} & -29340.6 {\it i} \\\cline{2-6}
 & $\overline{\omega_{2}}$ & -29373.1 & -29340.6 & 59724.2 {\it i} & 59658.2 {\it i} \\[.2ex] \cline{2-6}
 & & & & & \\[-2.3ex] \cline{2-6}
 & $\omega_{4}$ & 36911.8 & 36871 & 54751.1 & 54690.6 \\\cline{2-6}
 & $\overline{\omega_{4}}$ & 47526.3 & 47473.8 & 32041.1 & 32005.6 \\\cline{2-6}
$n=3$ & $\omega_{2}$ & 59724.2 & 59658.2 & 29373.1 {\it i} & 29340.6 {\it i} \\\cline{2-6}
 & $\overline{\omega_{2}}$ & 29373.1 & 29340.6 & -59724.2 {\it i} & -59658.2 {\it i} \\[.2ex] \cline{2-6}
 & & & & & \\[-2.3ex] \cline{2-6}
 & $\omega_{1}$ & -2064.9 {\it i} $\times$ $(A/r)$ & -2068.57 {\it i} $\times$ $(A/r)$ & -749.715 $\times$ $(A/r)$ & -751.048 $\times$ $(A/r)$ \\\cline{2-6}
 & $\overline{\omega_{1}}$ & 820.822 {\it i} $\times$ $(A/r)$ & 822.282 {\it i} $\times$ $(A/r)$ & -2260.74 $\times$ $(A/r)$ & -2264.77 $\times$ $(A/r)$ \\[.2ex] \cline{2-6}
 & & & & & \\[-2.3ex] \cline{2-6}
 & $\omega_{1}$ & 2064.9 {\it i} $\times$ $(A/r)$ & 2068.57 {\it i} $\times$ $(A/r)$ & 749.715 $\times$ $(A/r)$ & 751.048 $\times$ $(A/r)$ \\\cline{2-6}
 & $\overline{\omega_{1}}$ & -820.822 {\it i} $\times$ $(A/r)$ & -822.282 {\it i} $\times$ $(A/r)$ & 2260.74 $\times$ $(A/r)$ & 2264.77 $\times$ $(A/r)$ \\ \hline
\noalign{\smallskip} \hline
\end{tabular}
\label{patru}
\end{center}
\end{table}

\begin{table}[htbc]
\begin{center}
\begin{footnotesize}
\begin{tabular}{|c|c|c|c|c|c|} \hline\hline
\multicolumn{2}{|c|}{\bf Table 5} & \multicolumn{2}{c|}{\bf Up Sign} & \multicolumn{2}{c|}{\bf Down Sign} \\ \cline{3-6}
\multicolumn{2}{|c|}{} & $H_{0}=72.35$ & $H_{0}=72.31$ & $H_{0}=72.35$ & $H_{0}=72.31$ \\\hline\hline
 & $\left[\displaystyle \Re(E_{Yamabe,\; helicoid})\right]^{2}=$ & 1.754286\textcolor{blue}{290} $\times$ $10^{9}$ & 1.750406\textcolor{blue}{519} $\times$ $10^{9}$ & 1.754283\textcolor{blue}{897} $\times$ $10^{9}$ & 1.750410\textcolor{blue}{259} $\times$ $10^{9}$ \\\cline{3-6}
 & $=\omega_{4}\overline{\omega_{4}}$ & \multicolumn{2}{|c|}{\bf real particle} & \multicolumn{2}{|c|}{\bf real particle} \\\cline{2-6}
 & $\left[\displaystyle \Im(E_{Yamabe,\; helicoid})\right]^{2}=$ & 1.754283\textcolor{blue}{196} $\times$ $10^{9}$ & 1.750410\textcolor{blue}{669} $\times$ $10^{9}$ & 1.754283\textcolor{blue}{196} $\times$ $10^{9}$ & 1.750410\textcolor{blue}{669} $\times$ $10^{9}$ \\\cline{3-6}
 & $=\omega_{2}\overline{\omega_{2}}$ & \multicolumn{2}{|c|}{\bf virtual particle} & \multicolumn{2}{|c|}{\bf real particle} \\\cline{2-6}
$n=1$ & $\left[\displaystyle \Re(E_{Yamabe,\; helicoid})\right]^{2}=$ & 1.694912\textcolor{blue}{328} $\times$ $10^{6}$ $\times$ $\displaystyle \frac{A^{2}}{r^{2}}$ & 1.700945\textcolor{blue}{381} $\times$ $10^{6}$ $\times$ $\displaystyle \frac{A^{2}}{r^{2}}$ & 1.694913\textcolor{blue}{086} $\times$ $10^{6}$ $\times$ $\displaystyle \frac{A^{2}}{r^{2}}$ & 1.700949\textcolor{blue}{673} $\times$ $10^{6}$ $\times$ $\displaystyle \frac{A^{2}}{r^{2}}$ \\\cline{3-6}
 & $=\omega_{1}\overline{\omega_{1}}$ & \multicolumn{2}{|c|}{\bf real particle} & \multicolumn{2}{|c|}{\bf virtual particle} \\\cline{2-6}
 & $\left[\displaystyle \Im(E_{Yamabe,\; helicoid})\right]^{2}=$ & 0 & 0 & 0 & 0 \\\cline{3-6}
 & $=\omega_{3}\overline{\omega_{3}}$ & \multicolumn{2}{|c|}{-} & \multicolumn{2}{|c|}{-} \\[.2ex] \hline
 & & & & & \\[-2.3ex] \hline
 & $\left[\displaystyle \Re(E_{Yamabe,\; helicoid})\right]^{2}=$ & 1.754280\textcolor{blue}{662} $\times$ $10^{9}$ & 1.750410\textcolor{blue}{611} $\times$ $10^{9}$ & 1.75428\textcolor{blue}{0657} $\times$ $10^{9}$ & 1.7504\textcolor{blue}{10288} $\times$ $10^{9}$ \\\cline{3-6}
 & $=\omega_{4}\overline{\omega_{4}}$ & \multicolumn{2}{|c|}{\bf real particle} & \multicolumn{2}{|c|}{\bf real particle} \\\cline{2-6}
 & $\left[\displaystyle \Im(E_{Yamabe,\; helicoid})\right]^{2}=$ & 1.754285\textcolor{blue}{280} $\times$ $10^{9}$ & 1.750407\textcolor{blue}{690} $\times$ $10^{9}$ & 1.75428\textcolor{blue}{5280} $\times$ $10^{9}$ & 1.7504\textcolor{blue}{07690} $\times$ $10^{9}$ \\\cline{3-6}
 & $=\omega_{2}\overline{\omega_{2}}$ & \multicolumn{2}{|c|}{\bf virtual particle} & \multicolumn{2}{|c|}{\bf real particle} \\\cline{2-6}
$n=2$ & $\left[\displaystyle \Re(E_{Yamabe,\; helicoid})\right]^{2}=$ & 1.694910\textcolor{blue}{287} $\times$ $10^{6}$ $\times$ $\displaystyle \frac{A^{2}}{r^{2}}$ & 1.700944\textcolor{blue}{362} $\times$ $10^{6}$ $\times$ $\displaystyle \frac{A^{2}}{r^{2}}$ & 1.69491\textcolor{blue}{7906} $\times$ $10^{6}$ $\times$ $\displaystyle \frac{A^{2}}{r^{2}}$ & 1.7009\textcolor{blue}{43708} $\times$ $10^{6}$ $\times$ $\displaystyle \frac{A^{2}}{r^{2}}$ \\\cline{3-6}
 & $=\omega_{1}\overline{\omega_{1}}$ & \multicolumn{2}{|c|}{\bf real particle} & \multicolumn{2}{|c|}{\bf virtual particle} \\\cline{2-6}
 & $\left[\displaystyle \Im(E_{Yamabe,\; helicoid})\right]^{2}=$ & 0 & 0 & 0 & 0 \\\cline{3-6}
 & $=\omega_{3}\overline{\omega_{3}}$ & \multicolumn{2}{|c|}{-} & \multicolumn{2}{|c|}{-} \\[.2ex] \hline
 & & & & & \\[-2.3ex] \hline
 & $\left[\displaystyle \Re(E_{Yamabe,\; helicoid})\right]^{2}=$ & 1.75428\textcolor{blue}{1280} $\times$ $10^{9}$ & 1.75040\textcolor{blue}{6480} $\times$ $10^{9}$ & 1.754285\textcolor{blue}{470} $\times$ $10^{9}$ & 1.750405\textcolor{blue}{467} $\times$ $10^{9}$ \\\cline{3-6}
 & $=\omega_{4}\overline{\omega_{4}}$ & \multicolumn{2}{|c|}{\bf real particle} & \multicolumn{2}{|c|}{\bf real particle} \\\cline{2-6}
 & $\left[\displaystyle \Im(E_{Yamabe,\; helicoid})\right]^{2}=$ & 1.75428\textcolor{blue}{4899} $\times$ $10^{9}$ & 1.75040\textcolor{blue}{7383} $\times$ $10^{9}$ & 1.754284\textcolor{blue}{899} $\times$ $10^{9}$ & 1.750407\textcolor{blue}{383} $\times$ $10^{9}$ \\\cline{3-6}
 & $=\omega_{2}\overline{\omega_{2}}$ & \multicolumn{2}{|c|}{\bf real particle} & \multicolumn{2}{|c|}{\bf virtual particle} \\\cline{2-6}
$n=3$ & $\left[\displaystyle \Re(E_{Yamabe,\; helicoid})\right]^{2}=$ & 1.69491\textcolor{blue}{5348} $\times$ $10^{6}$ $\times$ $\displaystyle \frac{A^{2}}{r^{2}}$ & 1.70094\textcolor{blue}{7877} $\times$ $10^{6}$ $\times$ $\displaystyle \frac{A^{2}}{r^{2}}$ & 1.694910\textcolor{blue}{689} $\times$ $10^{6}$ $\times$ $\displaystyle \frac{A^{2}}{r^{2}}$ & 1.700950\textcolor{blue}{979} $\times$ $10^{6}$ $\times$ $\displaystyle \frac{A^{2}}{r^{2}}$ \\\cline{3-6}
 & $=\omega_{1}\overline{\omega_{1}}$ & \multicolumn{2}{|c|}{\bf real particle} & \multicolumn{2}{|c|}{\bf virtual particle} \\\cline{2-6}
 & $\left[\displaystyle \Im(E_{Yamabe,\; helicoid})\right]^{2}=$ & 0 & 0 & 0 & 0 \\\cline{3-6}
 & $=\omega_{3}\overline{\omega_{3}}$ & \multicolumn{2}{|c|}{-} & \multicolumn{2}{|c|}{-} \\\hline
\noalign{\smallskip} \hline
\end{tabular}
\end{footnotesize}
\label{cinci}
\end{center}
\end{table}

From the table 5 values, where the quark as a virtual particle (spacelike) is the same particle as the quark as a real particle (timelike), for each of the three quark generations we should have equality between the imaginary and the real parts of the Yamabe energy, $\left[\Im(E_{Yamabe,\; helicoid})\right]^{2}=\left[\Re(E_{Yamabe,\; helicoid})\right]^{2}$. From the equality of these energies results (in the quark ($r,A$) reference system):

\begin{equation}
\frac{A^{2}}{r^{2}}\simeq\left\{ \begin{array}{ll}
\displaystyle 1035.03 \;\;\;\; \textnormal{for}\;\; H_{0}=72.35 \\
\displaystyle 1029.08 \;\;\;\; \textnormal{for}\;\; H_{0}=72.31 \; .
\end{array} \right.
\label{eq_27}
\end{equation}

\section{Quark Mass}

The squared mass difference between two {\it n} quark generations will be:

\begin{equation}
\left(\Delta m^{2}_{ij}\right)^{Q}=\left| \left. \left[\Re(E_{Yamabe,\; helicoid})\right]^{2}\right|_{n=i}-\left. \left[\Re(E_{Yamabe,\; helicoid})\right]^{2}\right|_{n=j} \right|
\label{eq_31}
\end{equation}
\noindent
in the real particle case, or:

\begin{equation}
\left(\Delta m^{2}_{ij}\right)^{Q}=\left| \left. \left[\Im(E_{Yamabe,\; helicoid})\right]^{2}\right|_{n=i}-\left. \left[\Im(E_{Yamabe,\; helicoid})\right]^{2}\right|_{n=j} \right|
\label{eq_32}
\end{equation}
\noindent
for the virtual particle case, where $i\in \{1,2,3 \}$, $j\in \{1,2,3 \}$ and $i\neq j$. In consequence:

\begin{equation}
\begin{array}{ccccccccc}
\displaystyle \left(\Delta m^{2}_{12}\right)^{Q}_{A}=\left| \left. \left(\omega_{4}\overline{\omega_{4}}\right)\right|_{n=1}-\left. \left(\omega_{4}\overline{\omega_{4}}\right)\right|_{n=2} \right| \\
\displaystyle \left(\Delta m^{2}_{12}\right)^{Q}_{B}=\left| \left. \left(\omega_{2}\overline{\omega_{2}}\right)\right|_{n=1}-\left. \left(\omega_{2}\overline{\omega_{2}}\right)\right|_{n=2} \right| \\
\displaystyle \left(\Delta m^{2}_{12}\right)^{Q}_{C}=\left| \left. \left(\omega_{1}\overline{\omega_{1}}\right)\right|_{n=1}-\left. \left(\omega_{1}\overline{\omega_{1}}\right)\right|_{n=2} \right|\equiv \left(\Delta m^{2}_{12}\right)^{Q}_{A} \\
\displaystyle \left(\Delta m^{2}_{12}\right)^{Q}_{D}=\left| \left. \left(\omega_{3}\overline{\omega_{3}}\right)\right|_{n=1}-\left. \left(\omega_{3}\overline{\omega_{3}}\right)\right|_{n=2} \right|=0 \\
\displaystyle \left(\Delta m^{2}_{13}\right)^{Q}_{A}=\left| \left. \left(\omega_{4}\overline{\omega_{4}}\right)\right|_{n=1}-\left. \left(\omega_{4}\overline{\omega_{4}}\right)\right|_{n=3} \right| \\
\displaystyle \left(\Delta m^{2}_{13}\right)^{Q}_{C}=\left| \left. \left(\omega_{1}\overline{\omega_{1}}\right)\right|_{n=1}-\left. \left(\omega_{1}\overline{\omega_{1}}\right)\right|_{n=3} \right|\equiv \left(\Delta m^{2}_{13}\right)^{Q}_{A} \\
\displaystyle \left(\Delta m^{2}_{23}\right)^{Q}_{A}=\left| \left. \left(\omega_{4}\overline{\omega_{4}}\right)\right|_{n=2}-\left. \left(\omega_{4}\overline{\omega_{4}}\right)\right|_{n=3} \right| \\
\displaystyle \left(\Delta m^{2}_{23}\right)^{Q}_{C}=\left| \left. \left(\omega_{1}\overline{\omega_{1}}\right)\right|_{n=2}-\left. \left(\omega_{1}\overline{\omega_{1}}\right)\right|_{n=3} \right|\equiv \left(\Delta m^{2}_{23}\right)^{Q}_{A} \\
\displaystyle \left(\Delta m^{2}_{13}\right)^{Q}_{D}=\left(\Delta m^{2}_{23}\right)^{Q}_{D}=0 \; .
\end{array}
\label{eq_33}
\end{equation}

$\displaystyle \left(\Delta m^{2}_{13}\right)^{Q}_{B}$ and $\displaystyle \left(\Delta m^{2}_{23}\right)^{Q}_{B}$ are not defined because in each substraction one term is for a real particle and the other for a virtual particle. In table 5 we observe that the $\left(\Delta m^{2}_{ij}\right)^{Q}_{A}=\left(\Delta m^{2}_{ij}\right)^{Q}_{C}$ "Up Sign" is the proper case for describing the quarks as real (timelike) particles and that the $\left(\Delta m^{2}_{ij}\right)^{Q}_{D}=0$ is the proper case for describing the quarks as imaginary (spacelike) particles. The values are given in table 6.

\begin{table}[htb]
\begin{center}
{\renewcommand{\arraystretch}{1.8}
\begin{tabular}{|c|c|c|c|c|c|c|} \hline\hline
{\bf Table 6} & \multicolumn{3}{c|}{\bf Up Sign} & \multicolumn{3}{c|}{\bf Down Sign} \\ \cline{2-7}
 & $H_{0}=72.35$ & $H_{0}=72.31$ & {\bf particle type} & $H_{0}=72.35$ & $H_{0}=72.31$ & {\bf particle type} \\\hline\hline
$\displaystyle \left(\Delta m^{2}_{12}\right)^{Q}_{A}$ & 6 $\times$ $10^{3}$ & 4 $\times$ $10^{3}$ & {\bf real} & 3 $\times$ $10^{3}$ & $\sim$ 0 & {\bf real} \\\hline
$\displaystyle \left(\Delta m^{2}_{12}\right)^{Q}_{B}$ & 2 $\times$ $10^{3}$ & 3 $\times$ $10^{3}$ & {\bf virtual} & 2 $\times$ $10^{3}$ & 3 $\times$ $10^{3}$ & {\bf real} \\\hline
$\displaystyle \left(\Delta m^{2}_{12}\right)^{Q}_{C}$ & 6 $\times$ $10^{3}$ & 4 $\times$ $10^{3}$ & {\bf real} & 3 $\times$ $10^{3}$ & $\sim$ 0 & {\bf virtual} \\\hline
$\displaystyle \left(\Delta m^{2}_{12}\right)^{Q}_{D}$ & 0 & 0 & {\bf virtual} & 0 & 0 & {\bf virtual} \\\hline
$\displaystyle \left(\Delta m^{2}_{13}\right)^{Q}_{A}$ & 5 $\times$ $10^{3}$ & $\sim$ 0 & {\bf real} & 2 $\times$ $10^{3}$ & 5 $\times$ $10^{3}$ & {\bf real} \\\hline
$\displaystyle \left(\Delta m^{2}_{13}\right)^{Q}_{B}$ & - & - & - & - & - & - \\\hline
$\displaystyle \left(\Delta m^{2}_{13}\right)^{Q}_{C}$ & 5 $\times$ $10^{3}$ & $\sim$ 0 & {\bf real} & 2 $\times$ $10^{3}$ & 5 $\times$ $10^{3}$ & {\bf virtual} \\\hline
$\displaystyle \left(\Delta m^{2}_{13}\right)^{Q}_{D}$ & 0 & 0 & {\bf virtual} & 0 & 0 & {\bf virtual} \\\hline
$\displaystyle \left(\Delta m^{2}_{23}\right)^{Q}_{A}$ & 1 $\times$ $10^{3}$ & 4 $\times$ $10^{3}$ & {\bf real} & 5 $\times$ $10^{3}$ & 5 $\times$ $10^{3}$ & {\bf real} \\\hline
$\displaystyle \left(\Delta m^{2}_{23}\right)^{Q}_{B}$ & - & - & - & - & - & - \\\hline
$\displaystyle \left(\Delta m^{2}_{23}\right)^{Q}_{C}$ & 1 $\times$ $10^{3}$ & 4 $\times$ $10^{3}$ & {\bf real} & 5 $\times$ $10^{3}$ & 5 $\times$ $10^{3}$ & {\bf virtual} \\\hline
$\displaystyle \left(\Delta m^{2}_{23}\right)^{Q}_{D}$ & 0 & 0 & {\bf virtual} & 0 & 0 & {\bf virtual} \\\hline
\noalign{\smallskip} \hline
\end{tabular}}
\label{sase}
\end{center}
\end{table}

\section{Neutrino Masses}

In paper \cite{deus3} we saw that $E_{DEUS}=2\; E_{Yamabe,\; catenoid}$ and $E_{DEUS}=4\; E_{Yamabe,\; helicoid}$, the catenoid and the helicoid being representations of the same particle-wave object. In consequence, we have:

\begin{equation}
E_{Yamabe,\; catenoid}=2\; E_{Yamabe,\; helicoid}=2\left[\Re(E_{Yamabe,\; helicoid})+i\; \Im(E_{Yamabe,\; helicoid}) \right]\; .
\label{eq_34}
\end{equation}

In the same time we know that:

\begin{equation}
E_{Yamabe,\; catenoid}=\Re(E_{Yamabe,\; catenoid})+i\; \Im(E_{Yamabe,\; catenoid})\; .
\label{eq_35}
\end{equation}

Then, in a DEUS atom, for neutrinos we will have $\Re(E_{Yamabe,\; catenoid})=2\; \Re(E_{Yamabe,\; helicoid})$, with $\displaystyle \frac{a}{r}$ from (\ref{eq_13})-(\ref{eq_16}), while for quarks we will have $\Im(E_{Yamabe,\; catenoid})=2\; \Im(E_{Yamabe,\; helicoid})$, with $\displaystyle \frac{A}{r}$ from (\ref{eq_27}). At the same {\it r}, from this system of equations we can compute (in the reference system of the external observer):

\begin{equation}
\frac{a}{A}\simeq 3.7053 \times 10^{4}\; ,
\label{eq_36}
\end{equation}
\noindent
or:

\begin{equation}
R_{neutrino}\simeq 3.7053 \times 10^{4}\; R_{quark}\; ,
\label{eq_37}
\end{equation}
\noindent
where the quark and the neutrino appertain to the same fermion generation.

In natural units we have $\displaystyle m\equiv E=\frac{\hbar\; c}{R}=\frac{1}{R}$. Results that, for the external observer:

\begin{equation}
m_{quark}\simeq 3.7053 \times 10^{4}\; m_{neutrino}\; .
\label{eq_38}
\end{equation}

When we have the mass of the {\it up} quark $m_{u}=0.004$ GeV/$c^{2}$, results a neutrino mass $m_{\nu}^{I}=1.08\times 10^{-7}$ GeV/$c^{2}$, or when we have the mass of the {\it down} quark $m_{d}=0.008$ GeV/$c^{2}$, results a neutrino mass $m_{\nu}^{I}=2.16\times 10^{-7}$ GeV/$c^{2}$. For the second generation quarks, at a {\it charm} quark mass $m_{c}=1.5$ GeV/$c^{2}$ corresponds a neutrino mass $m_{\nu}^{II}=4.05\times 10^{-5}$ GeV/$c^{2}$, while at a {\it strange} quark mass $m_{s}=0.15$ GeV/$c^{2}$ corresponds a neutrino mass $m_{\nu}^{II}=4.05\times 10^{-6}$ GeV/$c^{2}$. For the third generation quarks, at a {\it top} quark mass $m_{t}=176$ GeV/$c^{2}$ corresponds a neutrino mass $m_{\nu}^{III}=4.75\times 10^{-3}$ GeV/$c^{2}$, while at a {\it bottom} quark mass $m_{b}=4.7$ GeV/$c^{2}$ corresponds a neutrino mass $m_{\nu}^{III}=1.27\times 10^{-4}$ GeV/$c^{2}$.

\section{Conclusions}

In this paper we dealt with the existing correlations between neutrinos (at the third self-similar DEUS level) and quarks (at the fourth self-similar DEUS level) in the DEUS atom (the Riemannian extended version of the FRW-Minkowski atom). These correlations allowed us to establish the neutrino flavor hierarchy (for a timelike or a spacelike observer) and to determine the mass of each neutrino flavor. Our results are in a good agreement with the present knowledge of these masses.

With more accurate Hubble constant and quark masses we would be able to determine the neutrino masses with a better precision.

\ack
Manuscript registered at the Astronomical Institute, Romanian Academy, record No. 262 from 04/19/07.


\title[D.E.U.S.]{XV: The Cabibbo and the CKM Matrices}

\begin{abstract}
We present a Cabibbo matrix and a CKM matrix study, applied to level three and level four self-similar DEUS objects, with emphasis on the CP symmetry violation and on the anisotropy matter-antimatter.
\end{abstract}


\section{Introduction}

In the Standard Model of particle physics, with SU(2) $\times$ U(1) as the gauge group of electroweak interactions, both the quarks and leptons are assigned to be left-handed doublets and right-handed singlets. The quark mass eigenstates are not the same as the weak eigenstates, and the matrix relating these bases was constructed for six quarks, and given an explicit parametrization by Kobayashi and Maskawa \cite{koba}. The Cabibbo-Kobayashi-Maskawa (CKM) matrix is described by a single parameter, the Cabibbo angle \cite{cabibbo}.

In the following sections we will study the situations in which, first, the Cabibbo matrix, and after that, the CKM matrix, satisfy the unitarity relation. For this we will use the DEUS object wave function \cite{deus9}, differentiated for matter and antimatter in Universe and Anti-Universe \cite{deus11,deus12}.

The final section will justify the CP violation relevance in our Universe.

\section{Cabibbo Matrix}

In SU(2), for the weak (W) or electroweak interaction (EW), instead of the $\left(\begin{array}{cc}
{\cal A}_{3} \\
{\cal A}_{4}
\end{array}\right)$ or $\left(\begin{array}{cc}
{\cal A}_{4} \\
{\cal A}_{3}
\end{array}\right)$ field matrices \cite{deus12,deus13} we will work with $\Psi^{W\; {\scriptsize \textnormal{or}}\; EW}=\left(\begin{array}{cc}
\Psi_{3} \\
\Psi_{4}
\end{array}\right)$ or  $\Psi^{W\; {\scriptsize \textnormal{or}}\; EW}=\left(\begin{array}{cc}
\Psi_{4} \\
\Psi_{3}
\end{array}\right)$ from the global wave function $\Psi^{{\scriptsize \textnormal{global}}}$, where:

\begin{equation}
\Psi^{{\scriptsize \textnormal{global}}}=\left(
\begin{array}{cccc}
\displaystyle \Psi_{1}^{{\scriptsize \textnormal{global}}} \\
\displaystyle \Psi_{2}^{{\scriptsize \textnormal{global}}} \\
\displaystyle \Psi_{3}^{{\scriptsize \textnormal{global}}} \\
\displaystyle \Psi_{4}^{{\scriptsize \textnormal{global}}}
\end{array}
\right) \; ,
\label{eq_1}
\end{equation}
\noindent
with $\Psi_{3}\stackrel{not}{=}\Psi_{3}^{{\scriptsize \textnormal{global}}}$ and $\Psi_{4}\stackrel{not}{=}\Psi_{4}^{{\scriptsize \textnormal{global}}}$. The components are to be found also in $\Psi^{{\scriptsize \textnormal{local}}}$, where:

\begin{equation}
\Psi^{{\scriptsize \textnormal{local}}}=\left(
\begin{array}{cccc}
\displaystyle \Psi_{3}^{{\scriptsize \textnormal{local}}} \\
\displaystyle \Psi_{4}^{{\scriptsize \textnormal{local}}} \\
\displaystyle \Psi_{1}^{{\scriptsize \textnormal{local}}} \\
\displaystyle \Psi_{2}^{{\scriptsize \textnormal{local}}}
\end{array}
\right) \; ,
\label{eq_2}
\end{equation}
\noindent
or:

\begin{equation}
\Psi^{{\scriptsize \textnormal{local}}}=\left(
\begin{array}{cccc}
\displaystyle \Psi_{4}^{{\scriptsize \textnormal{local}}} \\
\displaystyle \Psi_{3}^{{\scriptsize \textnormal{local}}} \\
\displaystyle \Psi_{1}^{{\scriptsize \textnormal{local}}} \\
\displaystyle \Psi_{2}^{{\scriptsize \textnormal{local}}}
\end{array}
\right) \; .
\label{eq_3}
\end{equation}

Using the Cabibbo matrix $U^{W\; {\scriptsize \textnormal{or}}\; EW}$ we must have:

\begin{equation}
\Psi_{i}^{W\; {\scriptsize \textnormal{or}}\; EW}=U^{W\; {\scriptsize \textnormal{or}}\; EW}\; \Psi_{f}^{W\; {\scriptsize \textnormal{or}}\; EW} \; ,
\label{eq_4}
\end{equation}
\noindent
and:

\begin{equation}
\Psi_{f}^{W\; {\scriptsize \textnormal{or}}\; EW}=U^{W\; {\scriptsize \textnormal{or}}\; EW}\; \Psi_{i}^{W\; {\scriptsize \textnormal{or}}\; EW} \; ,
\label{eq_5}
\end{equation}
\noindent
where $\Psi_{i}^{W\; {\scriptsize \textnormal{or}}\; EW}$ and $\Psi_{f}^{W\; {\scriptsize \textnormal{or}}\; EW}$ are the initial and, respectively, the final states of the $\Psi^{W\; {\scriptsize \textnormal{or}}\; EW}$ wave function.

In the {\bf hyperbolic coordinates} of the DEUS hypersurfaces:

\begin{equation}
U^{W\; {\scriptsize \textnormal{or}}\; EW}=\left(
\begin{array}{cc}
\cosh \theta_{W\; {\scriptsize \textnormal{or}}\; EW} & \sinh \theta_{W\; {\scriptsize \textnormal{or}}\; EW} \\
\sinh \theta_{W\; {\scriptsize \textnormal{or}}\; EW} & \cosh \theta_{W\; {\scriptsize \textnormal{or}}\; EW} \\
\end{array}
\right)
\label{eq_6}
\end{equation}
\noindent
is the Cabibbo rotation matrix and:

\begin{equation}
\Psi_{i}^{W\; {\scriptsize \textnormal{or}}\; EW}=\left(\begin{array}{cc}
\Psi_{3,\; i}^{W\; {\scriptsize \textnormal{or}}\; EW} \\
\Psi_{4,\; i}^{W\; {\scriptsize \textnormal{or}}\; EW}
\end{array}\right) \; ,
\label{eq_7}
\end{equation}
\noindent
or:

\begin{equation}
\Psi_{i}^{W\; {\scriptsize \textnormal{or}}\; EW}=\left(\begin{array}{cc}
\Psi_{4,\; i}^{W\; {\scriptsize \textnormal{or}}\; EW} \\
\Psi_{3,\; i}^{W\; {\scriptsize \textnormal{or}}\; EW}
\end{array}\right) \; ,
\label{eq_8}
\end{equation}
\noindent
is the initial state wave function, and:

\begin{equation}
\Psi_{f}^{W\; {\scriptsize \textnormal{or}}\; EW}=\left(\begin{array}{cc}
\Psi_{3,\; f}^{W\; {\scriptsize \textnormal{or}}\; EW} \\
\Psi_{4,\; f}^{W\; {\scriptsize \textnormal{or}}\; EW}
\end{array}\right) \; ,
\label{eq_9}
\end{equation}
\noindent
or:

\begin{equation}
\Psi_{f}^{W\; {\scriptsize \textnormal{or}}\; EW}=\left(\begin{array}{cc}
\Psi_{4,\; f}^{W\; {\scriptsize \textnormal{or}}\; EW} \\
\Psi_{3,\; f}^{W\; {\scriptsize \textnormal{or}}\; EW}
\end{array}\right) \; ,
\label{eq_10}
\end{equation}
\noindent
is the final state wave function.

In (\ref{eq_7})-(\ref{eq_10}) we have \cite{deus14}:

\begin{equation}
\!\!\!\!\!\!\!\!\!\!\!\!\!\!\!\!\!\!\!\!\!\!\!\!\!\!\!\!\!\!\!\!\!\!\!\!\!\!\!\!\!\!\!\!\!\!\!\!\begin{array}{ccc}
\displaystyle \Psi_{3,\;i}^{W\; {\scriptsize \textnormal{or}}\; EW}=\Psi_{0}\; \exp\left\{- \sqrt{2}\; \omega\; \frac{1}{\cosh \alpha_{k}^{1}-\sinh \alpha_{k}^{1}}\; \frac{\sinh \alpha_{k}^{1}\; (2\cosh \alpha_{k}^{1}+\sinh \alpha_{k}^{1})}{\sinh \alpha_{k}^{1}+\alpha_{k}^{1}\;(\sinh \alpha_{k}^{1}+\cosh \alpha_{k}^{1})}\; \sin \alpha^{1}\; \sinh \left[i\; \textnormal{arcctg} \left(\textnormal{tg}\; \alpha^{1} \right) \right] \right\} \\
\displaystyle \Psi_{4,\;i}^{W\; {\scriptsize \textnormal{or}}\; EW}=\Psi_{0}\; \exp\left\{\mp \sqrt{2}\; \omega\; \frac{\cosh \alpha_{k}^{2}}{\sinh \alpha_{k}^{2}+\cosh \alpha_{k}^{2}}\; \cos \alpha^{2}\; \sinh \left[i\; \textnormal{arcctg} \left(\textnormal{tg}\; \alpha^{2} \right) \right] \right\} \\
\displaystyle \Psi_{3,\;f}^{W\; {\scriptsize \textnormal{or}}\; EW}=\Psi_{4,\;f}^{W\; {\scriptsize \textnormal{or}}\; EW}=\Psi_{0} \; ,
\end{array}
\label{eq_11}
\end{equation}
\noindent
with $\alpha_{k}^{1}$ and $\alpha_{k}^{2}$ (or $\alpha^{1}$ and $\alpha^{2}$) two different angles.

For the above hyperbolic case it proves to be impossible to have satisfied the (\ref{eq_4}) or (\ref{eq_5}) relations (for the catenoid or helicoid representations), evidence that inside the DEUS object (non-evaporated) there is no interaction between the embedded DEUS particles.

In {\bf spherical} coordinates (evaporated DEUS object, or the FRW representation of the helicoidal hyper-surface):

\begin{equation}
U^{W\; {\scriptsize \textnormal{or}}\; EW}=\left(
\begin{array}{cc}
\cos \theta_{W\; {\scriptsize \textnormal{or}}\; EW} & \sin \theta_{W\; {\scriptsize \textnormal{or}}\; EW} \\
-\sin \theta_{W\; {\scriptsize \textnormal{or}}\; EW} & \cos \theta_{W\; {\scriptsize \textnormal{or}}\; EW} \\
\end{array}
\right) \; .
\label{eq_12}
\end{equation}

As function of the $\Psi_{i}^{W\; {\scriptsize \textnormal{or}}\; EW}$ and $\Psi_{f}^{W\; {\scriptsize \textnormal{or}}\; EW}$ choice we distinguish two possible situations:

\subsection{Situation 1}

\begin{equation}
\!\!\!\!\!\!\!\!\!\!\!\!\!\!\!\!\!\!\!\!\!\!\!\!\!\!\!\!\!\!\!\!\!\!\!\!\!\!\!\!\!\!\begin{array}{cc}
\displaystyle \Psi_{i}^{W\; {\scriptsize \textnormal{or}}\; EW}=\left(\begin{array}{cc}
\Psi_{3,\; i}^{W\; {\scriptsize \textnormal{or}}\; EW} \\
\Psi_{4,\; i}^{W\; {\scriptsize \textnormal{or}}\; EW}
\end{array}\right)= \\
\displaystyle = \Psi_{0}\; \left(\begin{array}{cc}
\displaystyle \exp\left\{- \sqrt{2}\; \omega\; \frac{1}{\cos \alpha^{1}-\sin \alpha^{1}}\; \frac{\sin \alpha^{1}\; (2\cos \alpha^{1}+\sin \alpha^{1})}{\sin \alpha^{1}+\alpha^{1}\;(\sin \alpha^{1}+\cos \alpha^{1})}\; \sin \alpha^{1}\; \sinh \left[i\; \textnormal{arcctg} \left(\textnormal{tg}\; \alpha^{1} \right) \right] \right\} \\
\displaystyle \exp\left\{\mp \sqrt{2}\; \omega\; \frac{\cos \alpha^{2}}{\sin \alpha^{2}+\cos \alpha^{2}}\; \cos \alpha^{2}\; \sinh \left[i\; \textnormal{arcctg} \left(\textnormal{tg}\; \alpha^{2} \right) \right] \right\}
\end{array}\right)
\end{array}
\label{eq_13}
\end{equation}
\noindent
and:

\begin{equation}
\Psi_{f}^{W\; {\scriptsize \textnormal{or}}\; EW}=\left(\begin{array}{cc}
\Psi_{3,\; f}^{W\; {\scriptsize \textnormal{or}}\; EW} \\
\Psi_{4,\; f}^{W\; {\scriptsize \textnormal{or}}\; EW}
\end{array}\right)=\Psi_{0}\; \left(\begin{array}{cc}
1 \\
1
\end{array}\right) \; .
\label{eq_14}
\end{equation}

If we want that $\Psi_{i}^{W\; {\scriptsize \textnormal{or}}\; EW}=U^{W\; {\scriptsize \textnormal{or}}\; EW}\; \Psi_{f}^{W\; {\scriptsize \textnormal{or}}\; EW}$ we must have for the DEUS object evaporation $\displaystyle \alpha^{1}=-\frac{\pi}{4}$ and $\displaystyle \alpha^{2}=\frac{\pi}{4}$ angles (at the cylinder):

\begin{equation}
e^{i\theta}-e^{-i\theta}=0 \; ,
\label{eq_15}
\end{equation}
possible only if $\theta=0$ and:

\begin{equation}
\displaystyle \Psi_{4,\;i}^{{\scriptsize \textnormal{global}}}=\Psi_{0}\; \exp\left\{\textcolor{red}{+} \sqrt{2}\; \omega\; \frac{\cos \alpha^{2}}{\sin \alpha^{2}+\cos \alpha^{2}}\; \cos \alpha^{2}\; \sinh \left[i\; \textnormal{arcctg} \left(\textnormal{tg}\; \alpha^{2} \right) \right] \right\}\; ,
\label{eq_16}
\end{equation}
\noindent
with which results that:

\begin{equation}
\displaystyle \theta=-\frac{\sqrt{2}}{4}\; \omega\; .
\label{eq_17}
\end{equation}

Because $\omega \geq 0$, results that $\theta \leq 0$. In consequence, for the above $\Psi_{3,\; i}^{{\scriptsize \textnormal{global}}}$, $\Psi_{4,\; i}^{{\scriptsize \textnormal{global}}}$, $\Psi_{3,\; f}^{{\scriptsize \textnormal{global}}}$ and $\Psi_{4,\; f}^{{\scriptsize \textnormal{global}}}$, the $\left(\begin{array}{cc}
\Psi_{3} \\
\Psi_{4}
\end{array}\right)$ form describes an {\bf Anti-Universe FRW bubble}. When we take into account also that $e^{i\theta}-e^{-i\theta}=0$ is solvable only if $\theta =0$, then $\omega=0$. This implies that, for $\displaystyle \alpha^{1}=-\frac{\pi}{4}$ and $\displaystyle \alpha^{2}=\frac{\pi}{4}$, the Cabibbo matrix is unitary, having only the particle representation (helicoid) of an empty space or of a space filled with particles that do not emit or absorb radiation (impossible to be seen from the FRW matter Universe).

If we want that $\Psi_{f}^{W\; {\scriptsize \textnormal{or}}\; EW}=U^{W\; {\scriptsize \textnormal{or}}\; EW}\; \Psi_{i}^{W\; {\scriptsize \textnormal{or}}\; EW}$, ($\forall$) $ \alpha^{1}\in \mathbb{R}$ and $\alpha^{2}\in \mathbb{R}$, the only possible solution is $\theta=\omega=0$.

$\\$
\noindent
{\bf Conclusion}: For $\Psi_{i}=U\; \Psi_{f}$ and $\Psi_{f}=U\; \Psi_{i}$ simultaneous satisfied ({\it U} unitary), we should have $\omega=0$ and $\theta=0$, with  $\displaystyle \alpha^{1}=-\frac{\pi}{4}$ (antimatter) and $\displaystyle \alpha^{2}=\frac{\pi}{4}$ (matter). In this situation, $\Psi_{3}^{{\scriptsize \textnormal{global}}}$ as function of $\alpha^{1}$ describes the antimatter and $\Psi_{4}^{{\scriptsize \textnormal{global}}}$ as function of $\alpha^{2}$ describes the matter in the Anti-Universe. The matter and the antimatter components are equal ($\Psi_{3,\;f}=\Psi_{4,\;f}=\Psi_{0}$) before their annihilation (on the pre-collapse cylinder of a DEUS level three or four of self-similarity), where $\Psi_{i}=U\; \Psi_{f}$.

If (as obtained experimentally) $\theta \neq 0$:

\begin{enumerate}
\item $\Psi_{f}=U\; \Psi_{i}$ is impossible, resulting that {\it U} is not unitary;

\item $\Psi_{i}=U\; \Psi_{f}$ is valid only if $\theta$ depends on the particle associated pulsation (frequency); $\theta$ is obtained from $\omega$. For the case when $\Psi_{i}=U\; \Psi_{f}$ is satisfied we have the wave representation (catenoid), while for $\Psi_{f}=U\; \Psi_{i}$ we have the particle representation (helicoid; $\omega=\theta=0$). If we look for duality in our observations at $\theta=0$ (unitary {\it U}), we must develop in the future new experiments through which to observe the particle perspective of the implied processes.
\end{enumerate}

\subsection{Situation 2}

\begin{equation}
\!\!\!\!\!\!\!\!\!\!\!\!\!\!\!\!\!\!\!\!\!\!\!\!\!\!\!\!\!\!\!\!\!\!\!\!\!\!\!\!\!\!\begin{array}{cc}
\displaystyle \Psi_{i}^{W\; {\scriptsize \textnormal{or}}\; EW}=\left(\begin{array}{cc}
\Psi_{4,\; i}^{W\; {\scriptsize \textnormal{or}}\; EW} \\
\Psi_{3,\; i}^{W\; {\scriptsize \textnormal{or}}\; EW}
\end{array}\right)= \\
\displaystyle = \Psi_{0}\; \left(\begin{array}{cc}
\displaystyle \exp\left\{\mp \sqrt{2}\; \omega\; \frac{\cos \alpha^{2}}{\sin \alpha^{2}+\cos \alpha^{2}}\; \cos \alpha^{2}\; \sinh \left[i\; \textnormal{arcctg} \left(\textnormal{tg}\; \alpha^{2} \right) \right] \right\} \\
\displaystyle \exp\left\{- \sqrt{2}\; \omega\; \frac{1}{\cos \alpha^{1}-\sin \alpha^{1}}\; \frac{\sin \alpha^{1}\; (2\cos \alpha^{1}+\sin \alpha^{1})}{\sin \alpha^{1}+\alpha^{1}\;(\sin \alpha^{1}+\cos \alpha^{1})}\; \sin \alpha^{1}\; \sinh \left[i\; \textnormal{arcctg} \left(\textnormal{tg}\; \alpha^{1} \right) \right] \right\}
\end{array}\right)
\end{array}
\label{eq_18}
\end{equation}
\noindent
and:

\begin{equation}
\Psi_{f}^{W\; {\scriptsize \textnormal{or}}\; EW}=\left(\begin{array}{cc}
\Psi_{4,\; f}^{W\; {\scriptsize \textnormal{or}}\; EW} \\
\Psi_{3,\; f}^{W\; {\scriptsize \textnormal{or}}\; EW}
\end{array}\right)=\Psi_{0}\; \left(\begin{array}{cc}
1 \\
1
\end{array}\right) \; .
\label{eq_19}
\end{equation}

If we want that $\Psi_{i}^{W\; {\scriptsize \textnormal{or}}\; EW}=U^{W\; {\scriptsize \textnormal{or}}\; EW}\; \Psi_{f}^{W\; {\scriptsize \textnormal{or}}\; EW}$ we must have for the DEUS object evaporation $\displaystyle \alpha^{1}=-\frac{\pi}{4}$ and $\displaystyle \alpha^{2}=\frac{\pi}{4}$ angles (at the cylinder) the (\ref{eq_15}) relation, possible for $\theta =0$, and (\ref{eq_16}). Results:

\begin{equation}
\displaystyle \theta=\frac{\sqrt{2}}{4}\; \omega\; .
\label{eq_20}
\end{equation}

Because $\omega \geq 0$ in (\ref{eq_20}), then $\theta \geq 0$. In consequence, for the above $\Psi_{3,\; i}^{{\scriptsize \textnormal{global}}}$, $\Psi_{4,\; i}^{{\scriptsize \textnormal{global}}}$, $\Psi_{3,\; f}^{{\scriptsize \textnormal{global}}}$ and $\Psi_{4,\; f}^{{\scriptsize \textnormal{global}}}$, the $\left(\begin{array}{cc}
\Psi_{4} \\
\Psi_{3}
\end{array}\right)$ form describes an {\bf Universe FRW bubble}. When we take into account also that $e^{i\theta}-e^{-i\theta}=0$ is solvable only if $\theta =0$, then $\omega=0$. This implies that, for $\displaystyle \alpha^{1}=-\frac{\pi}{4}$ (antimatter) and $\displaystyle \alpha^{2}=\frac{\pi}{4}$ (matter), the Cabibbo matrix {\it U} is unitary, having only the particle representation (helicoid) of particles that do not interact with the radiation field (the particles are "invisible" to the FRW matter Universe observer).

For $\Psi_{f}^{W\; {\scriptsize \textnormal{or}}\; EW}=U^{W\; {\scriptsize \textnormal{or}}\; EW}\; \Psi_{i}^{W\; {\scriptsize \textnormal{or}}\; EW}$, ($\forall$) $ \alpha^{1}\in \mathbb{R}$ and $\alpha^{2}\in \mathbb{R}$, the only possible solution is $\theta=\omega=0$.

$\\$
\noindent
{\bf Conclusion}: The conclusions from {\it Situation 1} are valid also here, but for a FRW Universe instead of a FRW Anti-Universe. Here, $\Psi_{3}^{{\scriptsize \textnormal{global}}}$ as function of $\alpha^{1}$ describes the antimatter and $\Psi_{4}^{{\scriptsize \textnormal{global}}}$ as function of $\alpha^{2}$ describes the matter in the Universe.

\section{CKM Matrix}

In SU(2) $\times$ U(1), for the electroweak interaction, instead of the $\left(\begin{array}{ccc}
{\cal A}_{1} \\
{\cal A}_{4} \\
{\cal A}_{3}
\end{array}\right)$ and $\left(\begin{array}{ccc}
{\cal A}_{2} \\
{\cal A}_{4} \\
{\cal A}_{3}
\end{array}\right)$ or $\left(\begin{array}{ccc}
{\cal A}_{1} \\
{\cal A}_{3} \\
{\cal A}_{4}
\end{array}\right)$ and $\left(\begin{array}{ccc}
{\cal A}_{2} \\
{\cal A}_{3} \\
{\cal A}_{4}
\end{array}\right)$ field matrices \cite{deus12,deus13}, we will have in $\Psi^{{\scriptsize \textnormal{global}}}$ (see (\ref{eq_1})) $\left(\begin{array}{ccc}
\Psi_{1} \\
\Psi_{4} \\
\Psi_{3}
\end{array}\right)$ and $\left(\begin{array}{ccc}
\Psi_{2} \\
\Psi_{4} \\
\Psi_{3}
\end{array}\right)$ or $\left(\begin{array}{ccc}
\Psi_{1} \\
\Psi_{3} \\
\Psi_{4}
\end{array}\right)$ and $\left(\begin{array}{ccc}
\Psi_{2} \\
\Psi_{3} \\
\Psi_{4}
\end{array}\right)$, with $\Psi_{1}\stackrel{not}{=}\Psi_{1}^{{\scriptsize \textnormal{global}}}$, $\Psi_{2}\stackrel{not}{=}\Psi_{2}^{{\scriptsize \textnormal{global}}}$, $\Psi_{3}\stackrel{not}{=}\Psi_{3}^{{\scriptsize \textnormal{global}}}$ and $\Psi_{4}\stackrel{not}{=}\Psi_{4}^{{\scriptsize \textnormal{global}}}$.

The CKM matrix is obtained by the product of three (complex) rotation matrices, where the rotations are characterized by the Euler angles $\theta_{12}$, $\theta_{13}$ and $\theta_{23}$, which are the mixing angles between the generations, and one overall phase $\delta$:

\begin{equation}
V=\left(
\begin{array}{ccc}
c_{12}c_{13} & s_{12}c_{13} & s_{13}e^{-i\delta} \\
-s_{12}c_{23}-c_{12}s_{23}s_{13}e^{i\delta} & c_{12}c_{23}-s_{12}s_{23}s_{13}e^{i\delta} & s_{23}c_{13} \\
s_{12}s_{23}-c_{12}c_{23}s_{13}e^{i\delta} & -c_{12}s_{23}-s_{12}c_{23}s_{13}e^{i\delta} & c_{23}c_{13} \\
\end{array}
\right) \; ,
\label{eq_21}
\end{equation}
\noindent
where $c_{ij}=\cos \theta_{ij}$ and $s_{ij}=\sin \theta_{ij}$, for $i<j=1,2,3$. This parametrization must satisfy strictly the unitarity relation $VV^{+}=I$.

Between the "correct" and physical states of quarks we will have: $\Psi_{i}=V\;\Psi_{f}$ and/or $\Psi_{f}=V\;\Psi_{i}$.

When two of the Euler angles are zero, the {\it V} matrix reduces to the Cabibbo matrix in the remaining $\theta_{ij}=\theta$, for which we saw, in the previous SU(2) study, that $\displaystyle \theta=\frac{\sqrt{2}}{4}\; \omega$ (for Universe) or $\displaystyle \theta=-\frac{\sqrt{2}}{4}\; \omega$ (for Anti-Universe) must verify $\Psi_{i}=V\;\Psi_{f}$ or $\Psi_{f}=V\;\Psi_{i}$ , where $V=U$. This is equivalent with having the fourth self-similar DEUS objects collapsed on the catenoid of the third self-similar DEUS level objects.

When $\theta_{12}=\theta_{13}=\theta_{23}=0$, it is evident that $V=I$. Also, because we saw that is impossible to describe the SU(2) Cabibbo matrix in the hyperbolic spacetime, this fact will reflect also on the CKM case, the only possible situation being for the spherical spacetime.

We distinguish four distinct situations:

\begin{enumerate}
\item Matter in Universe:

\begin{equation}
\Psi^{{\scriptsize \textnormal{global}}}=\left(\begin{array}{ccc}
\Psi_{2} \\
\Psi_{4} \\
\Psi_{3}
\end{array}\right) \; ,
\label{eq_22}
\end{equation}
\noindent
with the particularizations $\Psi_{i}^{{\scriptsize \textnormal{global}}}$ for the initial state and $\Psi_{f}^{{\scriptsize \textnormal{global}}}$ for the final state. The block $\begin{array}{cc}
\Psi_{4} \\
\Psi_{3}
\end{array}$ describes the Universe (as seen in the Cabibbo matrix study);

\item Antimatter in Universe:

\begin{equation}
\Psi^{{\scriptsize \textnormal{global}}}=\left(\begin{array}{ccc}
\Psi_{1} \\
\Psi_{4} \\
\Psi_{3}
\end{array}\right) \; ,
\label{eq_23}
\end{equation}
\noindent
with the particularizations $\Psi_{i}^{{\scriptsize \textnormal{global}}}$ for the initial state and $\Psi_{f}^{{\scriptsize \textnormal{global}}}$ for the final state. The block $\begin{array}{cc}
\Psi_{4} \\
\Psi_{3}
\end{array}$ describes the Universe (as seen in the Cabibbo matrix study);

\item Matter in Anti-Universe:

\begin{equation}
\Psi^{{\scriptsize \textnormal{global}}}=\left(\begin{array}{ccc}
\Psi_{2} \\
\Psi_{3} \\
\Psi_{4}
\end{array}\right) \; ,
\label{eq_24}
\end{equation}
\noindent
with the particularizations $\Psi_{i}^{{\scriptsize \textnormal{global}}}$ for the initial state and $\Psi_{f}^{{\scriptsize \textnormal{global}}}$ for the final state. The block $\begin{array}{cc}
\Psi_{3} \\
\Psi_{4}
\end{array}$ describes the Anti-Universe (as seen in the Cabibbo matrix study);

\item Antimatter in Anti-Universe:

\begin{equation}
\Psi^{{\scriptsize \textnormal{global}}}=\left(\begin{array}{ccc}
\Psi_{1} \\
\Psi_{3} \\
\Psi_{4}
\end{array}\right) \; ,
\label{eq_25}
\end{equation}
\noindent
with the particularizations $\Psi_{i}^{{\scriptsize \textnormal{global}}}$ for the initial state and $\Psi_{f}^{{\scriptsize \textnormal{global}}}$ for the final state. The block $\begin{array}{cc}
\Psi_{3} \\
\Psi_{4}
\end{array}$ describes the Anti-Universe (as seen in the Cabibbo matrix study);
\end{enumerate}

Let us remember that in the SU(2) case:

\begin{itemize}
\item for $\begin{array}{cc}
\Psi_{3} \\
\Psi_{4}
\end{array}$ we had:

\noindent
1) $\displaystyle \theta_{ij}=-\frac{\sqrt{2}}{4}\; \omega_{ij}$ from $\Psi_{i}=U\;\Psi_{f}$, with $\displaystyle \alpha^{1}=-\frac{\pi}{4}$ and $\displaystyle \alpha^{2}=\frac{\pi}{4}$ in $\Psi_{3,\;i}^{{\scriptsize \textnormal{global}}}(\alpha^{1})$ and $\Psi_{4,\;i}^{{\scriptsize \textnormal{global}}}(\alpha^{2})$;

\noindent
2) $\displaystyle \theta_{ij}=\omega_{ij}=0$ from $\Psi_{f}=U\;\Psi_{i}$, ($\forall$) $ \alpha^{1}\in \mathbb{R}$ and $\alpha^{2}\in \mathbb{R}$;

\item for $\begin{array}{cc}
\Psi_{4} \\
\Psi_{3}
\end{array}$ we had:

\noindent
1) $\displaystyle \theta_{ij}=\frac{\sqrt{2}}{4}\; \omega_{ij}$ from $\Psi_{i}=U\;\Psi_{f}$, with $\displaystyle \alpha^{1}=-\frac{\pi}{4}$ and $\displaystyle \alpha^{2}=\frac{\pi}{4}$ in $\Psi_{3,\;i}^{{\scriptsize \textnormal{global}}}(\alpha^{1})$ and $\Psi_{4,\;i}^{{\scriptsize \textnormal{global}}}(\alpha^{2})$;

\noindent
2) $\displaystyle \theta_{ij}=\omega_{ij}=0$ from $\Psi_{f}=U\;\Psi_{i}$, ($\forall$) $ \alpha^{1}\in \mathbb{R}$ and $\alpha^{2}\in \mathbb{R}$;

\item
\begin{equation}
\displaystyle \Psi_{4,\;i}^{{\scriptsize \textnormal{global}}}=\Psi_{0}\; \exp\left\{\sqrt{2}\; \omega\; \frac{\cos \alpha^{2}}{\sin \alpha^{2}+\cos \alpha^{2}}\; \cos \alpha^{2}\; \sinh \left[i\; \textnormal{arcctg} \left(\textnormal{tg}\; \alpha^{2} \right) \right] \right\}\; .
\label{eq_26}
\end{equation}
\end{itemize}

\noindent
For $\displaystyle \Psi_{1,\;i}^{{\scriptsize \textnormal{global}}}(\alpha^{3})=\Psi_{0}\; \exp\left\{\displaystyle \mp \omega\; \frac{\sin \alpha^{3}\; (2\cos \alpha^{3}+\sin \alpha^{3})}{\sin \alpha^{3}+\alpha^{3}\;(\sin \alpha^{3}+\cos \alpha^{3})}\; \alpha^{3}_{k}\right\}$ and $\displaystyle \Psi_{2,\;i}^{{\scriptsize \textnormal{global}}}(\alpha^{4})=\Psi_{0}\; \exp\left\{\mp \omega\; \sinh \left[i\; \textnormal{arcctg} \left(\textnormal{tg}\; \alpha^{4} \right) \right]\right\}$ we check the previous $\alpha$ values: 0, $\displaystyle -\frac{\pi}{4}$ and $\displaystyle \frac{\pi}{4}$. We observe that the only possible values are $\alpha^{3}=\displaystyle -\frac{\pi}{4}$ (antimatter) and $\alpha^{4}=\displaystyle \frac{\pi}{4}$ (matter). In conclusion:

\begin{equation}
\begin{array}{cccc}
\displaystyle \Psi_{1,\; i}^{{\scriptsize \textnormal{global}}}\left(\displaystyle -\frac{\pi}{4}\right)=\Psi_{0}\; \exp\left\{\pm i\; \frac{\sqrt{2}}{2}\; \frac{\pi}{4}\; \omega\right\} \\
\displaystyle \Psi_{2,\; i}^{{\scriptsize \textnormal{global}}}\left(\displaystyle \frac{\pi}{4}\right)=\Psi_{0}\; \exp\left\{\mp i\; \frac{\sqrt{2}}{2}\; \omega\right\} \\
\displaystyle \Psi_{3,\; i}^{{\scriptsize \textnormal{global}}}\left(\displaystyle -\frac{\pi}{4}\right)=\Psi_{0}\; \exp\left\{- i\; \frac{\sqrt{2}}{4}\; \omega\right\} \\
\displaystyle \Psi_{4,\; i}^{{\scriptsize \textnormal{global}}}\left(\displaystyle \frac{\pi}{4}\right)=\Psi_{0}\; \exp\left\{i\; \frac{\sqrt{2}}{4}\; \omega\right\} \; .
\end{array}
\label{eq_27}
\end{equation}

So, $\displaystyle \Psi_{1,\; i}^{{\scriptsize \textnormal{global}}}$ is for antimatter and $\displaystyle \Psi_{2,\; i}^{{\scriptsize \textnormal{global}}}$ is for matter.

\subsection{Matter in Universe:}

\begin{equation}
\Psi_{i}^{{\scriptsize \textnormal{global}}}=\left(\begin{array}{ccc}
\Psi_{2,\; i}^{{\scriptsize \textnormal{global}}} \\
\Psi_{4,\; i}^{{\scriptsize \textnormal{global}}} \\
\Psi_{3,\; i}^{{\scriptsize \textnormal{global}}}
\end{array}\right)=\Psi_{0}\left(\begin{array}{ccc}
\displaystyle \exp\left\{\mp i\; \frac{\sqrt{2}}{2}\; \omega\right\} \\
\displaystyle \exp\left\{i\; \frac{\sqrt{2}}{4}\; \omega\right\} \\
\displaystyle \exp\left\{-i\; \frac{\sqrt{2}}{4}\; \omega\right\}
\end{array}\right) \; ,
\label{eq_28}
\end{equation}

\begin{equation}
\Psi_{f}^{{\scriptsize \textnormal{global}}}=\left(\begin{array}{ccc}
\Psi_{2,\; f}^{{\scriptsize \textnormal{global}}} \\
\Psi_{4,\; f}^{{\scriptsize \textnormal{global}}} \\
\Psi_{3,\; f}^{{\scriptsize \textnormal{global}}}
\end{array}\right)=\Psi_{0}\left(\begin{array}{ccc}
\displaystyle \exp\left\{- i\; \frac{\sqrt{2}}{2}\; \omega\right\} \\
\displaystyle 1 \\
\displaystyle 1
\end{array}\right) \; .
\label{eq_29}
\end{equation}

Because $\Psi_{1}^{{\scriptsize \textnormal{global}}}$ does not appear in $\Psi^{{\scriptsize \textnormal{global}}}$, in (\ref{eq_21}) we have $\theta_{12}=\theta_{13}=0$, and then:

\begin{equation}
V=\left(
\begin{array}{ccc}
1 & 0 & 0 \\
0 & \cos \theta_{23} & \sin \theta_{23} \\
0 & -\sin \theta_{23} & \cos \theta_{23} \\
\end{array}
\right) \; .
\label{eq_30}
\end{equation}

In this case there are three conditions that must be fulfilled in order that $\Psi_{i}^{{\scriptsize \textnormal{global}}}=V\; \Psi_{f}^{{\scriptsize \textnormal{global}}}$ to be satisfied:

\begin{equation}
\begin{array}{ccc}
\Psi_{2,\; i}^{{\scriptsize \textnormal{global}}}=\Psi_{0}\; \displaystyle \exp\left\{- \omega\; \sinh \left[i\; \textnormal{arcctg} \left(\textnormal{tg}\; \alpha^{4} \right) \right]\right\} \\
\theta_{12}=\theta_{13}=\omega_{12}=\omega_{13}=0 \\
\displaystyle \theta_{23}=\frac{\sqrt{2}}{4}\; \omega_{23}
\end{array} \; .
\label{eq_31}
\end{equation}

For $\Psi_{f}^{{\scriptsize \textnormal{global}}}=V\; \Psi_{i}^{{\scriptsize \textnormal{global}}}$ we must have (as for the Cabibbo case) $\theta_{12}=\omega_{12}=0$, $\theta_{13}=\omega_{13}=0$ and $\theta_{23}=\omega_{23}=0$.

\subsection{Antimatter in Universe:}

\begin{equation}
\Psi_{i}^{{\scriptsize \textnormal{global}}}=\left(\begin{array}{ccc}
\Psi_{1,\; i}^{{\scriptsize \textnormal{global}}} \\
\Psi_{4,\; i}^{{\scriptsize \textnormal{global}}} \\
\Psi_{3,\; i}^{{\scriptsize \textnormal{global}}}
\end{array}\right)=\Psi_{0}\left(\begin{array}{ccc}
\displaystyle  \exp\left\{\pm i\; \frac{\sqrt{2}}{2}\; \frac{\pi}{4}\; \omega\right\} \\
\displaystyle \exp\left\{i\; \frac{\sqrt{2}}{4}\; \omega\right\} \\
\displaystyle \exp\left\{-i\; \frac{\sqrt{2}}{4}\; \omega\right\}
\end{array}\right) \; ,
\label{eq_32}
\end{equation}

\begin{equation}
\Psi_{f}^{{\scriptsize \textnormal{global}}}=\left(\begin{array}{ccc}
\Psi_{1,\; f}^{{\scriptsize \textnormal{global}}} \\
\Psi_{4,\; f}^{{\scriptsize \textnormal{global}}} \\
\Psi_{3,\; f}^{{\scriptsize \textnormal{global}}}
\end{array}\right)=\Psi_{0}\left(\begin{array}{ccc}
\displaystyle \exp\left\{- i\; \frac{\sqrt{2}}{2}\; \omega\right\} \\
\displaystyle 1 \\
\displaystyle 1
\end{array}\right) \; .
\label{eq_33}
\end{equation}

Because $\Psi_{2}^{{\scriptsize \textnormal{global}}}$ does not appear in $\Psi^{{\scriptsize \textnormal{global}}}$, in (\ref{eq_21}) we have $\theta_{12}=\theta_{23}=0$, and then:

\begin{equation}
V=\left(
\begin{array}{ccc}
\cos \theta_{13} & 0 & \sin \theta_{13}\; e^{-i\delta} \\
0 & 1 & 0 \\
-\sin \theta_{13}\; e^{i\delta} & 0 & \cos \theta_{13} \\
\end{array}
\right) \; .
\label{eq_34}
\end{equation}

$\Psi_{i}^{{\scriptsize \textnormal{global}}}=V\; \Psi_{f}^{{\scriptsize \textnormal{global}}}$ is satisfied when $\theta_{13}=0$ and $\omega\equiv \omega_{13}=0$, ($\forall$) $ \delta\in \mathbb{R}$.

If $\theta_{13}\neq 0$, is not possible to have $\Psi_{i}^{{\scriptsize \textnormal{global}}}=V\; \Psi_{f}^{{\scriptsize \textnormal{global}}}$ and, in the same time, it appears an anisotropy between matter and antimatter (see the {\it Matter in Universe} situation) that translates into the {\bf CP} violation. All this is related to the $\delta$ phase existence.

For $\Psi_{f}^{{\scriptsize \textnormal{global}}}=V\; \Psi_{i}^{{\scriptsize \textnormal{global}}}$, when $\theta_{12}=\theta_{23}=0$, we should have, as for $\Psi_{i}^{{\scriptsize \textnormal{global}}}=V\; \Psi_{f}^{{\scriptsize \textnormal{global}}}$, $\theta_{13}=\omega_{13}=0$, ($\forall$) $ \delta\in \mathbb{R}$. If it is observed that $\theta_{13}\neq 0$ (as obtained from experiments), $\Psi_{f}^{{\scriptsize \textnormal{global}}}=V\; \Psi_{i}^{{\scriptsize \textnormal{global}}}$ is not satisfied. This means that, when we observe only the wave associated to a phenomenon, we observe an anisotropy between matter and antimatter (see the {\it Matter in Universe} situation). This translates in the {\bf CP} violation, being related to the existence of $\delta$.

\subsection{Matter in Anti-Universe:}

\begin{equation}
\Psi_{i}^{{\scriptsize \textnormal{global}}}=\left(\begin{array}{ccc}
\Psi_{2,\; i}^{{\scriptsize \textnormal{global}}} \\
\Psi_{3,\; i}^{{\scriptsize \textnormal{global}}} \\
\Psi_{4,\; i}^{{\scriptsize \textnormal{global}}}
\end{array}\right)=\Psi_{0}\left(\begin{array}{ccc}
\displaystyle \exp\left\{\mp i\; \frac{\sqrt{2}}{2}\; \omega\right\} \\
\displaystyle \exp\left\{-i\; \frac{\sqrt{2}}{4}\; \omega\right\} \\
\displaystyle \exp\left\{i\; \frac{\sqrt{2}}{4}\; \omega\right\}
\end{array}\right) \; ,
\label{eq_35}
\end{equation}

\begin{equation}
\Psi_{f}^{{\scriptsize \textnormal{global}}}=\left(\begin{array}{ccc}
\Psi_{2,\; f}^{{\scriptsize \textnormal{global}}} \\
\Psi_{3,\; f}^{{\scriptsize \textnormal{global}}} \\
\Psi_{4,\; f}^{{\scriptsize \textnormal{global}}}
\end{array}\right)=\Psi_{0}\left(\begin{array}{ccc}
\displaystyle \exp\left\{- i\; \frac{\sqrt{2}}{2}\; \omega\right\} \\
\displaystyle 1 \\
\displaystyle 1
\end{array}\right) \; .
\label{eq_36}
\end{equation}

Because $\Psi_{1}^{{\scriptsize \textnormal{global}}}$ does not appear in $\Psi^{{\scriptsize \textnormal{global}}}$, in (\ref{eq_21}) we have $\theta_{12}=\theta_{13}=0$, and then {\it V} takes the (\ref{eq_30}) form. In $\Psi^{{\scriptsize \textnormal{global}}}$ the pulsation is $\omega\equiv \omega_{23}$.

$\Psi_{i}^{{\scriptsize \textnormal{global}}}=V\; \Psi_{f}^{{\scriptsize \textnormal{global}}}$ is verified if there are fulfilled three conditions:

\begin{equation}
\begin{array}{ccc}
\Psi_{2,\; i}^{{\scriptsize \textnormal{global}}}=\Psi_{0}\; \displaystyle \exp\left\{- \omega\; \sinh \left[i\; \textnormal{arcctg} \left(\textnormal{tg}\; \alpha^{4} \right) \right]\right\} \\
\theta_{12}=\theta_{13}=\omega_{12}=\omega_{13}=0 \\
\displaystyle \theta_{23}=-\frac{\sqrt{2}}{4}\; \omega_{23}
\end{array} \; ,
\label{eq_37}
\end{equation}
\noindent
the last of them being the result of the Cabibbo case for Anti-Universe.

In order that $\Psi_{f}^{{\scriptsize \textnormal{global}}}=V\; \Psi_{i}^{{\scriptsize \textnormal{global}}}$ it is required that $\theta_{12}=\omega_{12}=0$, $\theta_{13}=\omega_{13}=0$ and $\theta_{23}=\omega_{23}=0$ (see also the Cabibbo case).

\subsection{Antimatter in Anti-Universe:}

\begin{equation}
\Psi_{i}^{{\scriptsize \textnormal{global}}}=\left(\begin{array}{ccc}
\Psi_{1,\; i}^{{\scriptsize \textnormal{global}}} \\
\Psi_{3,\; i}^{{\scriptsize \textnormal{global}}} \\
\Psi_{4,\; i}^{{\scriptsize \textnormal{global}}}
\end{array}\right)=\Psi_{0}\left(\begin{array}{ccc}
\displaystyle  \exp\left\{\pm i\; \frac{\sqrt{2}}{2}\; \frac{\pi}{4}\; \omega\right\} \\
\displaystyle \exp\left\{-i\; \frac{\sqrt{2}}{4}\; \omega\right\} \\
\displaystyle \exp\left\{i\; \frac{\sqrt{2}}{4}\; \omega\right\}
\end{array}\right) \; ,
\label{eq_38}
\end{equation}

\begin{equation}
\Psi_{f}^{{\scriptsize \textnormal{global}}}=\left(\begin{array}{ccc}
\Psi_{1,\; f}^{{\scriptsize \textnormal{global}}} \\
\Psi_{3,\; f}^{{\scriptsize \textnormal{global}}} \\
\Psi_{4,\; f}^{{\scriptsize \textnormal{global}}}
\end{array}\right)=\Psi_{0}\left(\begin{array}{ccc}
\displaystyle \exp\left\{- i\; \frac{\sqrt{2}}{2}\; \omega\right\} \\
\displaystyle 1 \\
\displaystyle 1
\end{array}\right) \; .
\label{eq_39}
\end{equation}

Because $\Psi_{2}^{{\scriptsize \textnormal{global}}}$ does not appear in $\Psi^{{\scriptsize \textnormal{global}}}$, in (\ref{eq_21}) we have $\theta_{12}=\theta_{23}=0$, and then {\it V} takes the (\ref{eq_34}) form. In $\Psi^{{\scriptsize \textnormal{global}}}$ the pulsation is $\omega\equiv \omega_{13}$.

$\Psi_{i}^{{\scriptsize \textnormal{global}}}=V\; \Psi_{f}^{{\scriptsize \textnormal{global}}}$ is possible if $\theta_{13}=0$ and $\omega_{13}=0$, ($\forall$) $\delta\in \mathbb{R}$. When we consider (as obtained from the particle experiments) $\theta_{13}\neq 0$, is not possible to have $\Psi_{i}^{{\scriptsize \textnormal{global}}}=V\; \Psi_{f}^{{\scriptsize \textnormal{global}}}$, appearing an anisotropy between matter and antimatter (see the {\it Matter in Anti-Universe} situation) that translates into the {\bf CP} violation. This is related to the $\delta$ phase in {\it V}, or physically speaking, to the observation of a wave as non-correlated to its associated particle (no particle representation for the observed event).

For $\Psi_{f}^{{\scriptsize \textnormal{global}}}=V\; \Psi_{i}^{{\scriptsize \textnormal{global}}}$, when $\theta_{12}=\theta_{23}=0$, we should have, as for $\Psi_{i}^{{\scriptsize \textnormal{global}}}=V\; \Psi_{f}^{{\scriptsize \textnormal{global}}}$, $\theta_{13}=\omega_{13}=0$, ($\forall$) $ \delta\in \mathbb{R}$. If, again, as is observed, $\theta_{13}\neq 0$, $\Psi_{f}^{{\scriptsize \textnormal{global}}}=V\; \Psi_{i}^{{\scriptsize \textnormal{global}}}$ is not satisfied. Results that we observe only the wave associated to a phenomenon, this translating in an anisotropy between matter and antimatter (see the {\it Matter in Anti-Universe} situation): {\bf CP} violation.

$\\$
In conclusion, for a better picture of how is related the $\theta$ angle with the particle and wave perception, we can exemplify with the figure 6 from paper \cite{deus2}, where the component parallel with the cylinder surface is lost as a spacelike effect, the FRW external observer being able to see only the component perpendicular to the pre-collapse DEUS cylinder as a timelike effect. For objects situated on the pre-collapse cylinder, the $\theta$ angle is zero, and the observer sees the complete effect, both from the catenoid and the helicoid of the embedded DEUS-particle objects.

\section{$\theta\neq 0$}

From observations we know that for the weak interactions we have the angle $\theta_{W}\simeq 13.1^{\circ}$, while for the electroweak interactions the angle is $\theta_{EW}\simeq 28.7^{\circ}$.

\subsection{In Anti-Universe}

From the Cabibbo and the CKM matrices we obtained $\displaystyle \theta=-\frac{\sqrt{2}}{4}\; \omega$, case in which the pulsation is:

\begin{equation}
\omega =-\frac{4}{\sqrt{2}}\; \frac{13.1}{180}\; \pi
\label{eq_40}
\end{equation}
\noindent
in the weak interaction case, or:

\begin{equation}
\omega =-\frac{4}{\sqrt{2}}\; \frac{28.7}{180}\; \pi
\label{eq_41}
\end{equation}
\noindent
in the electroweak interaction case.

This means that the frequency (and, in natural units, the energy) is:

\begin{equation}
\nu\equiv E =-\sqrt{2}\; \frac{13.1}{180}
\label{eq_42}
\end{equation}
\noindent
in the weak interaction case, or:

\begin{equation}
\nu\equiv E =-\sqrt{2}\; \frac{28.7}{180}
\label{eq_43}
\end{equation}
\noindent
in the electroweak interaction case.

From the self-similarity of the DEUS objects we have \cite{deus3}:

\begin{equation}
\begin{array}{cccc}
\Im(E_{Yamabe,\; helicoid})=0 \\
\Im(E_{Yamabe,\; catenoid})\equiv \textnormal{gravitational\; wave} \\
\Re(E_{Yamabe,\; helicoid})\equiv \textnormal{particle}=E \\
\Re(E_{Yamabe,\; catenoid})\equiv \textnormal{wave\; associate\; to\; the\; particle}
\end{array}
\label{eq_44}
\end{equation}

When we have $\Re(E_{Yamabe,\; helicoid})=0$ (requiring $\theta=0$ and the embedding DEUS object's catenoid angle $\alpha=0$ \cite{deus4}) we will be able to observe only the wave behavior given by $\Re(E_{Yamabe,\; catenoid})$. If both $\Re(E_{Yamabe,\; helicoid})\neq 0$ and $\Re(E_{Yamabe,\; catenoid})\neq 0$, our observations will contain a particle and a wave component of the same phenomenon, the total energy of the DEUS-particle object being distributed between them. $\Re(E_{Yamabe,\; helicoid})$ can be zero only on the pre-collapse cylinder, where, for the four-dimensional FRW observer, the radial coordinate of the collapsing DEUS object is $r\rightarrow \infty$. However, the (\ref{eq_42}) and (\ref{eq_43}) energy {\it E} is negative, making impossible for us to observe the events occurring into the Anti-Universe.

\subsection{In Universe}

From the Cabibbo and the CKM matrices we obtained $\displaystyle \theta=\frac{\sqrt{2}}{4}\; \omega$, case in which the pulsation is:

\begin{equation}
\omega =\frac{4}{\sqrt{2}}\; \frac{13.1}{180}\; \pi
\label{eq_45}
\end{equation}
\noindent
in the weak interaction case, or:

\begin{equation}
\omega =\frac{4}{\sqrt{2}}\; \frac{28.7}{180}\; \pi
\label{eq_46}
\end{equation}
\noindent
in the electroweak interaction case.

This means that the frequency (and, in natural units, the energy) is:

\begin{equation}
\nu\equiv E =\sqrt{2}\; \frac{13.1}{180}
\label{eq_47}
\end{equation}
\noindent
in the weak interaction case, or:

\begin{equation}
\nu\equiv E =\sqrt{2}\; \frac{28.7}{180}
\label{eq_48}
\end{equation}
\noindent
in the electroweak interaction case.

The remarks for the {\it Anti-Universe} are valid also for the {\it Universe}, with the only difference that here the energy {\it E} is positive (observable).

\section{Conclusions}

Because, in practice, $\theta\neq 0$, we can not count on the validity of $\Psi_{f}^{{\scriptsize \textnormal{global}}}=V\; \Psi_{i}^{{\scriptsize \textnormal{global}}}$, being impossible to convert the DEUS topology to FRW coordinates (spherical representation). However, this conversion can be done when the catenoid is evolved to the pre-collapse cylinder, where $\theta =0$. So, the observation takes place before the DEUS object collapse, the greater the "distance" of the catenoid from cylinder, the bigger the $\theta$ angle (equivalent with the $\alpha$ catenoid angle \cite{deus5}). At $\theta_{EW}\simeq 28.7^{\circ}$, the gravitational waves and the electromagnetic perturbations are not yet emitted, while at $\theta_{W}\simeq 13.1^{\circ}$ these perturbations were already emitted toward the external observer (see the \cite{deus5} discussions related to the $\alpha$ angle).

Also, at a greater "distance" of the catenoid from the cylinder correspond bigger DEUS-particle objects (on the catenoid), with more self-similar embeddings and, so, to heavier particles. The experiments used to determine the CKM matrix elements use more energetic (or more massive) particles than the ones (nuclear $\beta$-decay, neutron $\beta$-decay, pion $\beta$-decay) used for the determination of the Cabibbo angle.

Another observation is that each of the CKM matrix elements is determined through a different experiment, the value of $\theta$ being slightly different from one to another. This difference is due to the same reason: different experiments involving different particles and their corresponding total energy; so, different evolution states of their associate catenoid or different position on the catenoid in which they are embedded. From this comes naturally the observed deviation from $\theta=0$, whose value will be bigger for heavier or more energetic particles.

When the matter from level four of DEUS self-similarity meets at level three DEUS object creation/annihilation the antimatter from level four of DEUS self-similarity, $\Psi_{2}^{{\scriptsize \textnormal{global}}}$ and $\overline{\Psi_{2}}^{{\scriptsize \textnormal{global}}}$ "disappear" (and, analogous, for $\Psi_{1}^{{\scriptsize \textnormal{global}}}$ and $\overline{\Psi_{1}}^{{\scriptsize \textnormal{global}}}$) from $\Psi^{{\scriptsize \textnormal{global}}}$, and the case will be described by a Cabibbo situation.

\ack
Manuscript registered at the Astronomical Institute, Romanian Academy, record No. 263 from 04/19/07.


\title[D.E.U.S.]{XVI: Cross Sections}

\begin{abstract}
In this paper we compute the cross sections for the DEUS weak and electroweak interactions and we explain how other interactions from our local spacetime originate in the DEUS geometry. 
\end{abstract}


\section{Introduction}

Into the Cabibbo-GIM scheme, instead of the physical quarks {\it d} and {\it s}, are used the "correct" states in the weak interactions {\it d'} and {\it s'}, given by:

\begin{equation}
\begin{array}{ll}
\displaystyle d'=d\; \cos \theta_{w}+s\; \sin \theta_{w} \\
\displaystyle s'=-d\; \sin \theta_{w}+s\; \cos \theta_{w}
\end{array} \; ,
\label{eq_1}
\end{equation}
\noindent
or, in matrix form:

\begin{equation}
\left(
\begin{array}{cc}
\displaystyle d' \\
\displaystyle s'
\end{array}
\right)=\left(\begin{array}{cc}
\cos \theta_{W} & \sin \theta_{W} \\
-\sin \theta_{W} & \cos \theta_{W} \\
\end{array}
\right) \left(
\begin{array}{cc}
\displaystyle d \\
\displaystyle s
\end{array}
\right) \; .
\label{eq_2}
\end{equation}

In this way we can couple the "Cabibbo-rotated" states $\left(
\begin{array}{cc}
\displaystyle u \\
\displaystyle d'
\end{array}
\right)$ and $\left(
\begin{array}{cc}
\displaystyle c \\
\displaystyle s'
\end{array}
\right)$ in the same way that we can couple to the lepton pairs $\left(
\begin{array}{cc}
\displaystyle \nu_{e} \\
\displaystyle e
\end{array}
\right)$ and $\left(
\begin{array}{cc}
\displaystyle \nu_{\mu} \\
\displaystyle \mu
\end{array}
\right)$ through {\it W} particles. Their couplings to the physical particles (states of specific flavor) are then given by:

\begin{equation}
\left(
\begin{array}{cc}
\displaystyle u \\
\displaystyle d'
\end{array}
\right)=\left(\begin{array}{cc}
u \\
d\; \cos \theta_{w}+s\; \sin \theta_{w}
\end{array}
\right) \; ,
\label{eq_3}
\end{equation}

\begin{equation}
\left(
\begin{array}{cc}
\displaystyle c \\
\displaystyle s'
\end{array}
\right)=\left(\begin{array}{cc}
c \\
-d\; \sin \theta_{w}+s\; \cos \theta_{w}
\end{array}
\right) \; .
\label{eq_4}
\end{equation}

That is, $d\rightarrow u + W^{-}$ carries a factor $\cos \theta_{w}$, and $s\rightarrow u + W^{-}$ a factor $\sin \theta_{w}$.

It is purely conventional that we "rotate" {\it d} and {\it s}, rather than {\it u} and {\it c}. We could accomplish the same by introducing $u'=u\; \cos \theta_{w}-c\; \sin \theta_{w}$ and $c'=u\; \sin \theta_{w}+c\; \cos \theta_{w}$.

Kobayashi and Maskawa \cite{koba} had generalized the Cabibbo-GIM scheme to handle three generations of quarks, the "weak interaction quark generations": $\left( \begin{array}{cc}
\displaystyle u \\
\displaystyle d'
\end{array}
\right)$, $\left( \begin{array}{cc}
\displaystyle c \\
\displaystyle s'
\end{array}
\right)$ and $\left( \begin{array}{cc}
\displaystyle t \\
\displaystyle b'
\end{array}
\right)$, which are related to the physical quark states by the Kobayashi-Maskawa matrix:

\begin{equation}
\left(
\begin{array}{ccc}
\displaystyle d' \\
\displaystyle s' \\
\displaystyle b'
\end{array}
\right)=\left(\begin{array}{ccc}
U_{ud} & U_{us} & U_{ub} \\
U_{cd} & U_{cs} & U_{cb} \\
U_{td} & U_{ts} & U_{tb} \\
\end{array}
\right) \left(
\begin{array}{ccc}
\displaystyle d \\
\displaystyle s \\
\displaystyle b
\end{array}
\right) \; ,
\label{eq_5}
\end{equation}
\noindent
where, for example, $U_{ud}$ specifies the coupling of {\it u} to {\it d} ($d\rightarrow u+W^{-}$). {\it U} can be reduced to a "canonical form", in which there remain just three "generalized Cabibbo angles" and one phase factor: the CKM matrix.

In 1958, Bludman \cite{blud} suggested that there might exist neutral weak interactions (between any lepton or any quark), mediated by uncharged partner of {\it W}'s: the $Z^{0}$. However, in the "classical" picture, the same fermion comes out as went in (as in QED or QCD), not allowing couplings of the form $\mu^{-}\rightarrow e^{-}+Z^{0}$, for example (this would violate conservation of muon and electron number), nor of the form $s\rightarrow d +Z^{0}$ (such a strangeness-changing neutral process would lead to $K^{0}\rightarrow \mu^{+}+\mu^{-}$ which is strongly suppressed). The reaction $\overline{\nu_{\mu}}+e\rightarrow \overline{\nu_{\mu}}+e$ is also mediated by a $Z^{0}$ particle. The corresponding neutrino-quark process, in the form of inclusive neutrino-nucleon scattering, are $\overline{\nu_{\mu}}+N\rightarrow \overline{\nu_{\mu}}+X$ and $\nu_{\mu}+N\rightarrow \nu_{\mu}+X$.

\section{Neutral DEUS Weak and Electroweak Interactions}

In the case of neutral processes, it doesn't matter whether you use the physical states ({\it d}, {\it s}, {\it b}) or the "Cabibbo-rotated" states ({\it d'}, {\it s'}, {\it b'}):

\begin{equation}
{\cal M}\sim \overline{d'}\; d'=\overline{d}\; d\; \cos^{2}\theta_{w}+\overline{s}\; s\; \sin^{2}\theta_{w}+\left(\overline{d}\; s+\overline{s}\; d \right)\sin\theta_{w}\; \cos\theta_{w} \; ,
\label{eq_6}
\end{equation}
\noindent
for the interaction between two {\it d'} quarks mediated by a $Z^{0}$ boson, or:

\begin{equation}
{\cal M}\sim \overline{s'}\; s'=\overline{d}\; d\; \sin^{2}\theta_{w}+\overline{s}\; s\; \cos^{2}\theta_{w}-\left(\overline{d}\; s+\overline{s}\; d \right)\sin\theta_{w}\; \cos\theta_{w} \; ,
\label{eq_7}
\end{equation}
\noindent
for the interaction between two {\it s'} quarks mediated by a $Z^{0}$ boson. ${\cal M}$ is the scattering amplitude of the process.

The sum of the two is:

\begin{equation}
{\cal M}_{tot}\sim \overline{d'}\; d' + \overline{s'}\; s'=\overline{d}\; d + \overline{s}\; s \; .
\label{eq_8}
\end{equation}

Thus, the net amplitude is the same whichever states we use. The same argument generalizes to three generations, as long the CKM matrix is unitary.

The differential cross section will be:

\begin{equation}
\frac{d\sigma}{d\Omega}=\left< \left|\; {\cal M}\; \right|^{2} \right> \; .
\label{eq_9}
\end{equation}

In what follows we will limit ourselves to the computation of the differential cross section for DEUS objects interactions mediated by neutral bosons ($Q=0$ \cite{deus13}), with states given by the \cite{deus15} global wave functions. In both SU(2) and SU(2) $\times$ U(1), $\Psi_{0}$ is the one determined in \cite{deus9}.

In SU(2), we obtain \cite{deus15}:

\begin{equation}
{\cal M}= \Psi_{3}^{{\scriptsize \textnormal{global}}}\; \Psi_{4}^{{\scriptsize \textnormal{global}}} \; .
\label{eq_10}
\end{equation}

For both Universe and Anti-Universe:

\begin{itemize}
\item when $\theta_{w}=0$ (on the pre-collapse cylinder),

\begin{equation}
{\cal M}= \Psi_{3,\; f}^{{\scriptsize \textnormal{global}}}\; \Psi_{4,\; f}^{{\scriptsize \textnormal{global}}}=\Psi_{0}\; \Psi_{0}=\Psi_{0}^{2} \; ,
\label{eq_11}
\end{equation}
\noindent
with which, the (\ref{eq_9}) differential cross section:

\begin{equation}
\frac{d\sigma}{d\Omega}=\Psi_{0}^{4} \; ;
\label{eq_12}
\end{equation}

\item when $\theta_{w}\neq 0$ (on the DEUS catenoid),

\begin{equation}
{\cal M}= \Psi_{3,\; i}^{{\scriptsize \textnormal{global}}}\; \Psi_{4,\; i}^{{\scriptsize \textnormal{global}}}=\left[ \Psi_{0}\; \exp\left(-i\;\frac{\sqrt{2}}{4}\; \omega \right) \right]\left[ \Psi_{0}\; \exp\left(i\;\frac{\sqrt{2}}{4}\; \omega \right) \right]=\Psi_{0}^{2} \; ,
\label{eq_13}
\end{equation}
\noindent
where $\Psi_{3,\; i}^{{\scriptsize \textnormal{global}}}$ is seen by the external observer at $\displaystyle \alpha=-\frac{\pi}{4}$ and $\Psi_{4,\; i}^{{\scriptsize \textnormal{global}}}$ at $\displaystyle \alpha=\frac{\pi}{4}$. Again, the differential cross section is:

\begin{equation}
\frac{d\sigma}{d\Omega}=\Psi_{0}^{4} \; .
\label{eq_14}
\end{equation}
\end{itemize}

For SU(2) $\times$ U(1), for both matter and antimatter in Universe \cite{deus15}:

\begin{itemize}
\item when $\theta_{12}=\theta_{13}=\theta_{23}=0$ (on the pre-collapse cylinder),

\begin{equation}
{\cal M}= \Psi_{3,\; f}^{{\scriptsize \textnormal{global}}}\; \Psi_{4,\; f}^{{\scriptsize \textnormal{global}}}=\Psi_{0}^{2} \; ,
\label{eq_15}
\end{equation}
\noindent
with which, the (\ref{eq_9}) differential cross section:

\begin{equation}
\frac{d\sigma}{d\Omega}=\Psi_{0}^{4} \; ;
\label{eq_16}
\end{equation}

\item when $\displaystyle \theta=\frac{\sqrt{2}}{4}\; \omega$ (on the DEUS catenoid; for matter in Universe $\displaystyle \theta\equiv \theta_{23}=\frac{\sqrt{2}}{4}\; \omega_{23}$, while for antimatter in Universe $\displaystyle \theta\equiv \theta_{13}=\frac{\sqrt{2}}{4}\; \omega_{13}$; the other two CKM angles for matter or antimatter are equal to zero),

\begin{equation}
{\cal M}= \Psi_{3,\; i}^{{\scriptsize \textnormal{global}}}\; \Psi_{4,\; i}^{{\scriptsize \textnormal{global}}}=\left[ \Psi_{0}\; \exp\left(-i\;\frac{\sqrt{2}}{4}\; \omega \right) \right]\left[ \Psi_{0}\; \exp\left(i\;\frac{\sqrt{2}}{4}\; \omega \right) \right]=\Psi_{0}^{2} \; ,
\label{eq_17}
\end{equation}
\noindent
where $\Psi_{3,\; i}^{{\scriptsize \textnormal{global}}}$ is seen by the external observer at $\displaystyle \alpha=-\frac{\pi}{4}$ and $\Psi_{4,\; i}^{{\scriptsize \textnormal{global}}}$ at $\displaystyle \alpha=\frac{\pi}{4}$. The differential cross section is given by (\ref{eq_16}).
\end{itemize}

The same scattering amplitude and differential cross section is obtained for matter and antimatter in Anti-Universe, when $\theta_{12}=\theta_{13}=\theta_{23}=0$ (on the DEUS pre-collapse cylinder), or when $\displaystyle \theta=-\frac{\sqrt{2}}{4}\; \omega$ (on the DEUS catenoid; for matter in Anti-Universe $\displaystyle \theta\equiv \theta_{23}=-\frac{\sqrt{2}}{4}\; \omega_{23}$, while for antimatter in Anti-Universe $\displaystyle \theta\equiv \theta_{13}=-\frac{\sqrt{2}}{4}\; \omega_{13}$; the other two CKM angles for matter or antimatter are equal to zero) \cite{deus15}.

In the scattering amplitude we see that, passing from matter to antimatter (or from Universe to Anti-Universe) through the initial four-dimensional singularity, the field $\Psi_{3,\; i}^{{\scriptsize \textnormal{global}}}$ of antimatter "correct" state rotates to a new matter $\Psi_{4,\; i}^{{\scriptsize \textnormal{global}}}$ physical state field.

\section{DEUS Objects Interaction Cross Sections}

In the previous chapter we concluded that the differential cross section for DEUS object interactions is given by the (\ref{eq_16}) formula, where \cite{deus9}:

\begin{equation}
\Psi_{0}^{2}=\frac{1}{8\sqrt{3}\; \pi}\; (2\mp 1)\; \frac{m}{q^{2}}\; \frac{a^{2}}{r^{3}}  \; .
\label{eq_18}
\end{equation}

The mass {\it m} is the global contribution of all the self-similar DEUS objects embedded in the catenoid (equivalent with the contribution to the interaction of all the nucleons, for example, or, at lower level, of quarks) \cite{deus3}:

\begin{equation}
\begin{array}{cc}
m=\sum{m'_{helicoid}}=m_{catenoid}\equiv 2\; \Re(E_{Yamabe,\; catenoid})= \\
\displaystyle = 2\times 3\; \frac{\sqrt{2}}{a} \sqrt{H^{2}-H\sqrt{H^{2}+1}+1}\left[1+\frac{\sin^{2}\left(\displaystyle \frac{r}{a} \right)}{\cos^{4}\left(\displaystyle \frac{r}{a} \right)} \right]^{-1} \frac{\sin\left(\displaystyle \frac{r}{a} \right)}{\cos^{3}\left(\displaystyle \frac{r}{a} \right)} \; ,
\end{array}
\label{eq_19}
\end{equation}
\noindent
where $a\equiv t_{FRW}$, $t_{FRW}\in (0,1)$, and $H\equiv H_{0}=72.35$.

In the center of mass, the energy of the interaction of a particle with its antiparticle (the interaction of a FRW matter bubble with a FRW antimatter bubble) will be the total helicoid energy \cite{deus3}:

\begin{equation}
E=\Re(E_{Yamabe,\; helicoid})\equiv E_{Yamabe,\; helicoid} \; .
\label{eq_20}
\end{equation}

In figures 1 and 2 we represented the way in which $\displaystyle \frac{d\sigma}{d\Omega}$ varies with {\it E}, considering the present value of time $t_{FRW}=0.524$ \cite{deus6}. We took the radial coordinate as $r\in [0.01,3]$, where $\Delta r=1$ is the size of one DEUS self-similar level. In figure 1 we have the differential cross section for a charge $|q|=1$, while in the figure 2, for $|q|=2$. In both these figures there are three peaks, each peak representing the resonance of one of the visible FRW self-similar levels. From left to right we have the formation of a DEUS object at the fourth self-similarity level, followed by the resonance of the third self-similar DEUS objects and, finally, the resonance for the second DEUS self-similar level. The final ascendent trend (and going to infinity) is related to the embedding of the fourth, the third and the second self-similar DEUS levels in the first self-similar DEUS level (the Universe observed without its initial and final singularities). 
Even when we increase {\it r} to a higher value, the number of peaks remains, for the present $t_{FRW}$ time, unchanged, as a statement of our impossibility to see smaller than level four of self-similarity DEUS objects (quarks). Decreasing the value of $t_{FRW}$, many other picks appear into picture, meaning that an observer of our FRW Universe's past was able to see more self-similar DEUS levels (sub-quark and beyond). Increasing the $t_{FRW}$ value (our Universe future) makes the resonances disappear one by one, from left to right (from quarks to black holes), finally remaining only the increasing trend of the first self-similar DEUS level, the observer being unable to see any structure in his Universe.

To not be missleaded to think that it can appear a contradiction between the above interpretation of the cross sections and paper's \cite{deus6} final remarque, we should specify that in the present case we have the image of a pre-Big Bang evaporating object, the evaporated objects appertaining to the other three self-similar levels being seen just at the "surface" of the FRW Universe, without their implicit self-similar structure. In paper \cite{deus6} we immersed ourseves into the structure of the self-similar level two (black holes; third peak from 1 and 2 figures) observing only how this level evaporates in time. Taking this into account, the future's FRW Universe observer's dimmining perception of the fine structure of the Universe can be delayed by the evaporation of the black holes composing "particles".

\begin{figure}
\begin{center}
\includegraphics[scale=0.8,angle=90]{sectiuni_q1}
\caption{Differential cross section of non-evaporated level one of self-similarity DEUS object containing three observable embedded DEUS levels. q=1}
\end{center}
\label{cross1}
\end{figure}

\begin{figure}
\begin{center}
\includegraphics[scale=0.8,angle=90]{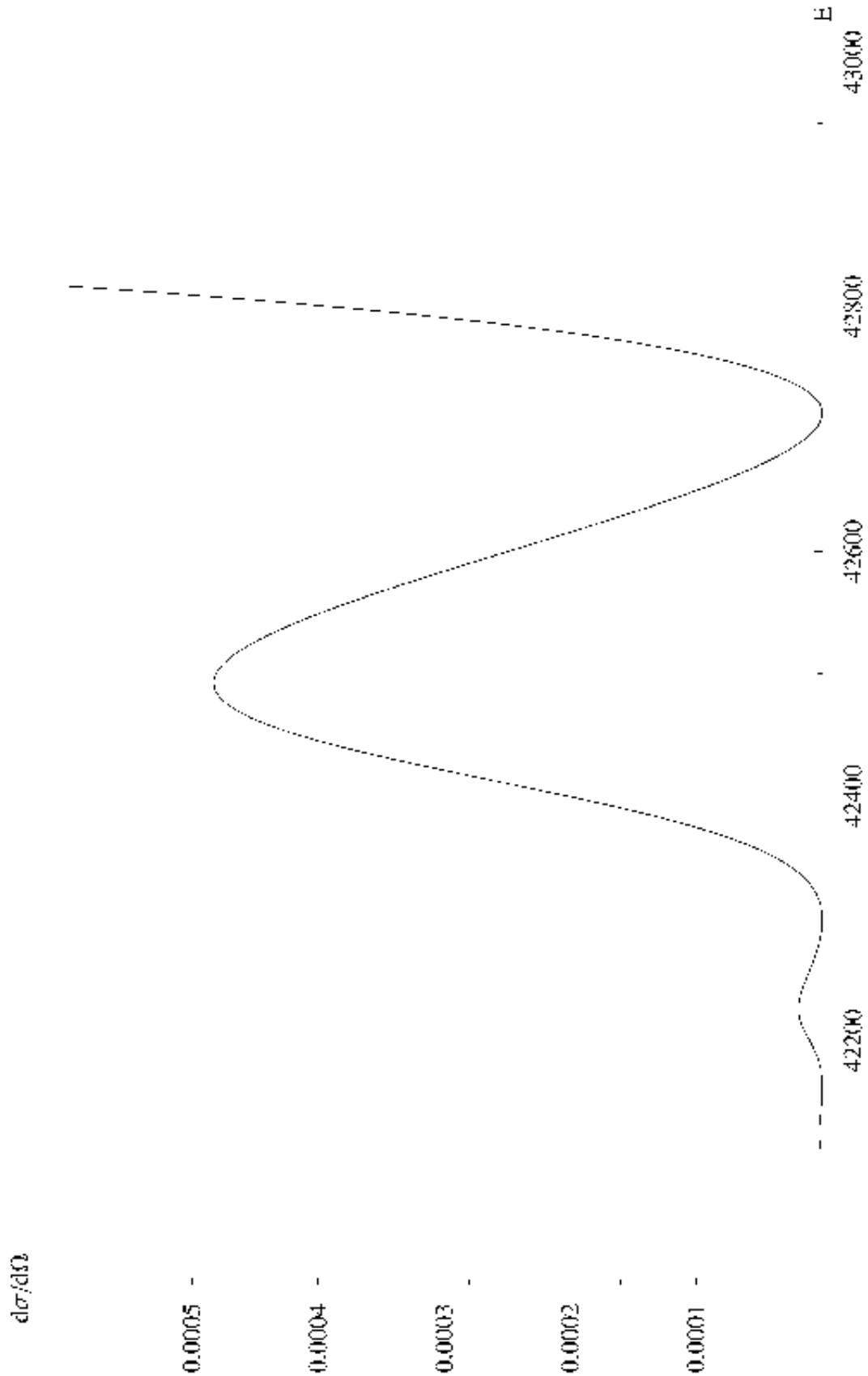}
\caption{Differential cross section of non-evaporated level one of self-similarity DEUS object containing three observable embedded DEUS levels. q=2}
\end{center}
\label{cross2}
\end{figure}

When two DEUS objects interact (the only possibility being the interaction at the upper and lower contact rings) in our local Minkowski spacetime, their catenoidal hyper-surfaces unify in a bigger catenoid in one of its three possible real states ($n=\{1,2,3\}$) \cite{deus2}. If the resulting catenoid has an energy corresponding to a $n>3$ hyper-surface, the excess is emitted in the external spacetime as gravitational and electromagnetic radiation (as, for example, for interacting black holes), the catenoidal hyper-surface becoming a $n=3$ hyper-surface. If the resulting hyper-surface is in the proximity of $n=1$ or $n=2$, the excess energy is emitted from it in the form of embedded DEUS objects, the catenoid relaxing to $n=1$ or $n=2$ state. For the third self-similar DEUS atom-objects this translates in an emission of nucleons: $X+Y\rightarrow Z$ + neutron (or proton). If the energy difference to $n=1$ or $n=2$ catenoidal hyper-surface is smaller than the energy of its embedded objects, the excess will be emitted toward the external space from the catenoids of these embedded DEUS objects. For even lower energy differences the emission will take place from even deeper self-similar levels.

$\\$
Let us describe now, in the DEUS perspective, few of the most important interactions of our Minkowski spacetime:

An annihilation of a matter with an antimatter FRW particle-bubble occurs at the pre-collapse cylinder.

The photoelectric effect is an emission from the catenoid, the difference in the photons energy originating in the difference of energy (from the catenoid neck, where it is maximal, to its cylinder contact rings, where it is minimal) between catenoid and cylinder. The quantification comes from the discrete self-similar inner structure of the catenoid. The continuum is the ergosphere equivalent: the spacetime between the catenoid with $n=3$ and the cylinder.

The Compton effect is the interaction of the DEUS atom-object with the electromagnetic radiation contained in our spacetime, while the catenoid evolves from $n=3$ to cylinder. The emission takes place at an embedded self-similar level closer to the DEUS atoms' catenoid "neck", evolved to be in contact with the pre-collapse cylinder and seen as if it took place at the un-evolved catenoid (process "frozen" on ergosphere).

Last but not least, the excitation and the de-excitation or the ionization and recombination, occur when the DEUS atom-object absorbs and re-emits electromagnetic radiation or particles for remaining a DEUS object whose evolution time (and lifetime) is correlated with the lifetime of its superior DEUS object in which is embedded ($t_{FRW}''\propto t_{FRW}$).

\ack
Manuscript registered at the Astronomical Institute, Romanian Academy, record No. 264 from 04/19/07.


\title[D.E.U.S.]{XVII: Multiverse}

\begin{abstract}
We interpret the DEUS objects embedded in the catenoid of a higher DEUS object as Stephani Universes. In this assumption, we check for the validity of the Gibbs-Duhem equation, one of the four conditions needed in order that the fluid to evolve in local thermal equilibrium. The second subject approached is the global action of the DEUS object, as a continuous action describing the transition from one self-similar DEUS level to another.
\end{abstract}


\section{Introduction}

A special class of Stephani Universes \cite{steph} can be interpreted as an ideal gas evolving in local thermal equilibrium.

The local thermal equilibrium evolution of a fluid is determined by the following four conditions:

\begin{itemize}
\item Energy-momentum conservation: 

\begin{equation}
\nabla_{\cdot}\textnormal{\bf T}=0 \; .
\label{eq_1}
\end{equation}

\item The energy density $\rho$ is decomposed in terms of matter density ${\cal R}$ and the specific internal energy $\epsilon$ :

\begin{equation}
\rho={\cal R}\; (1+\epsilon) \; .
\label{eq_2}
\end{equation}

\item The equation of conservation of matter:

\begin{equation}
\nabla_{\cdot}({\cal R}\; u)=0 \; ,
\label{eq_3}
\end{equation}
\noindent
with {\it u} the fluid velocity.

\item The thermodynamic quantities temperature {\it T} and specific entropy {\it s} are related by equations of state compatible with the thermodynamic Gibbs-Duhem relation:

\begin{equation}
T\; ds=d\epsilon + p\; d(1/{\cal R}) \; .
\label{eq_4}
\end{equation}
\end{itemize}

\section{DEUS Objects as Stephani Universes}

In \cite{coll} it is considered the following thermodynamic characterization:

$\\$
\noindent
{\bf Proposition 1}  The necessary and sufficient condition for a non barotropic and non isoenergetic divergent-free perfect energy tensor $\textnormal{\bf T}=\left(\rho +p\right) u\otimes u-p\; g$ to represent the local thermal equilibrium evolution of an ideal gas is that its indicatrix spacetime function $\displaystyle \chi\equiv \frac{u(p)}{u(\rho)}$ be a non identity function of the variable $\displaystyle \pi\equiv \frac{p}{\rho}$:

\begin{equation}
d\chi \wedge d\pi =0 \; ,
\label{eq_5}
\end{equation}
\noindent
with $\chi\neq \pi$.

For the DEUS object catenoid we saw that, from the external Lorenzian frame, we have \cite{deus3}:

\begin{equation}
\begin{array}{ll}
\displaystyle u(\rho)=U^{0}_{catenoid}\equiv \frac{dt_{FRW}}{dt}=-\frac{1}{r}\; \frac{\sqrt{\textnormal{tg}^{2}t_{FRW}+1}}{\textnormal{tg}\; t_{FRW}} \\
\displaystyle u(p)=U^{1}_{catenoid}\equiv \frac{dt_{FRW}}{d\phi}=-\frac{1}{r}\; \sqrt{\textnormal{tg}^{2}t_{FRW}+1}
\end{array} \; ,
\label{eq_6}
\end{equation}
\noindent
with $\textnormal{tg}\; t_{FRW}=\sqrt{H^{2}+1}-H$ and {\it H} the Hubble parameter. From (\ref{eq_6}) results:

\begin{equation}
\chi\equiv \frac{u(p)}{u(\rho)}=\textnormal{tg}\; t_{FRW} \; .
\label{eq_7}
\end{equation}

Also, because the catenoid is associated to a wave representation (radiation):

\begin{equation}
\pi\equiv \frac{p}{\rho}=1 \; .
\label{eq_8}
\end{equation}

At $\textnormal{tg}\; t_{FRW}\neq 1$, or $\displaystyle t_{FRW}\neq \frac{\pi}{4}$ (the equality being the signature of the DEUS object collapse, or, from the FRW external observer's perspective, to the event horizon of the DEUS black hole), we have $\chi\neq \pi$ and $d\chi\wedge d\pi=0$. In consequence, inside the catenoid the fluid is in local thermal equilibrium and is described by a perfect energy tensor.

$\\$
\noindent
{\bf Proposition 2}  A non barotropic and non isoenergetic divergence-free perfect fluid tensor verifying (\ref{eq_5}) represents the local thermal equilibrium evolution of the ideal fluid with specific internal energy $\epsilon$, temperature {\it T}, matter density ${\cal R}$, and specific entropy {\it s} given by:

\begin{equation}
\begin{array}{llll}
\displaystyle \epsilon(\rho,p)=\epsilon(\pi)\equiv e(\pi)-1  \\
\displaystyle T(\rho,p)=T(\pi)\equiv \frac{\pi}{k}\; e(\pi) \\
\displaystyle {\cal R}(\rho,p)=\frac{\rho}{e(\pi)} \\
\displaystyle s(\rho,p)=k\; \ln\frac{f(\pi)}{\rho} \; ,
\end{array}
\label{eq_9}
\end{equation}
\noindent
$e(\pi)$ and $f(\pi)$ being, respectively,

\begin{equation}
\begin{array}{lllll}
\displaystyle e(\pi)=e_{0}\; \exp\left\{\int{\psi(\pi)\; d\pi} \right\} \\ \\
\displaystyle \psi(\pi)\equiv \frac{\pi}{\left(\chi(\pi)-\pi \right)(\pi+1)} \\
\displaystyle f(\pi)=f_{0}\; \exp\left\{\int{\phi(\pi)\; d\pi} \right\} \\
\displaystyle \phi(\pi)\equiv \frac{1}{\chi(\pi)-\pi} \; .
\end{array}
\label{eq_10}
\end{equation}

In the case of our DEUS catenoid, because $\pi=1$ and $\chi(\pi)=\chi=\textnormal{tg}\; t_{FRW}$:

\begin{equation}
\begin{array}{llll}
\displaystyle \psi(\pi)=\frac{1}{2\left(\textnormal{tg}\; t_{FRW}-1 \right)} \\
\displaystyle \phi(\pi)=\frac{1}{\textnormal{tg}\; t_{FRW}-1} \\
\displaystyle e(\pi)=e_{0}\; \exp\{\psi(\pi) \}=e_{0}\; \exp\left\{\frac{1}{2\left(\textnormal{tg}\; t_{FRW}-1 \right)} \right\} \\
\displaystyle f(\pi)=f_{0}\; \exp\{\phi(\pi) \}=f_{0}\; \exp\left\{\frac{1}{\textnormal{tg}\; t_{FRW}-1} \right\} \; ,
\end{array}
\label{eq_11}
\end{equation}
\noindent
with which:

\begin{equation}
\epsilon(\rho,p)=e_{0}\; \exp\left\{\frac{1}{2\left(\textnormal{tg}\; t_{FRW}-1 \right)} \right\}-1 \; ,
\label{eq_12}
\end{equation}

\begin{equation}
T(\rho,p)=\frac{e_{0}}{k}\; \exp\left\{\frac{1}{2\left(\textnormal{tg}\; t_{FRW}-1 \right)} \right\} \; ,
\label{eq_13}
\end{equation}
\noindent
and:

\begin{equation}
s(\rho,p)=k\; \ln\left\{\frac{f_{0}}{\rho}\; \exp\left[\frac{1}{\textnormal{tg}\; t_{FRW}-1} \right] \right\}=k\; \ln\left(\frac{f_{0}}{\rho} \right)+\frac{k}{\textnormal{tg}\; t_{FRW}-1} \; ,
\label{eq_14}
\end{equation}
\noindent
where \cite{deus4}:

\begin{equation}
\rho=p= \frac{15}{256\; \pi}\; \frac{m^{2}}{q^{2}}\; \frac{a^{4}}{r^{4}} \; .
\label{eq_15}
\end{equation}

Results that the specific entropy is:

\begin{equation}
s(\rho,p)= k\; \ln\left\{f_{0}\; \frac{256\; \pi}{15}\; \frac{q^{2}}{m^{2}}\; \frac{r^{4}}{a^{4}} \right\}+\frac{k}{\textnormal{tg}\; t_{FRW}-1} \; .
\label{eq_16}
\end{equation}

The derivatives of (\ref{eq_16}) with respect to the radial coordinate {\it r} and to the $t_{FRW}$ time of the FRW spacetime are:

\begin{equation}
\begin{array}{cc}
\displaystyle \frac{\partial s}{\partial r}=4\; \frac{k}{r} \\
\displaystyle \frac{\partial s}{\partial t_{FRW}}=-\frac{k}{1-\sin (2\; t_{FRW})} \; .
\end{array}
\label{eq_17}
\end{equation}

Then:

\begin{equation}
ds=\frac{\partial s}{\partial r}\; dr+\frac{\partial s}{\partial t_{FRW}}\; dt_{FRW}=4\; \frac{k}{r}\; dr-\frac{k}{1-\sin (2\; t_{FRW})}\; dt_{FRW} \; .
\label{eq_18}
\end{equation}

Now, in (\ref{eq_4}) we have (from (\ref{eq_12})):

\begin{equation}
d\epsilon=\frac{\partial \epsilon}{\partial r}\; dr+\frac{\partial \epsilon}{\partial t_{FRW}}\; dt_{FRW}=-\frac{e_{0}}{2}\; \frac{1}{1-\sin (2\; t_{FRW})}\; \exp\left\{\frac{1}{2\left(\textnormal{tg}\; t_{FRW}-1 \right)} \right\}\; dt_{FRW}
\label{eq_19}
\end{equation}
\noindent
and (from (\ref{eq_9}), (\ref{eq_11}) and (\ref{eq_15})):

\begin{equation}
\!\!\!\!\!\!\!\!\!d(1/{\cal R})=e_{0}\; \frac{256\; \pi}{15}\; \frac{q^{2}}{m^{2}}\; \exp\left\{\frac{1}{2\left(\textnormal{tg}\; t_{FRW}-1 \right)} \right\}\; \frac{r^{3}}{a^{4}}\left[4\; dr-\frac{1}{2}\; \frac{r}{1-\sin (2\; t_{FRW})}\; dt_{FRW} \right]\; .
\label{eq_20}
\end{equation}

In conclusion, the result of (\ref{eq_13}) multiplied with (\ref{eq_18}) proves to be equal with $d\epsilon+p\; d(1/{\cal R})$, with {\it p} from (\ref{eq_15}). In other words, the Gibbs-Duhem equation (\ref{eq_4}) is satisfied for the catenoid fluid.

When the catenoid reaches the DEUS pre-collapse cylinder (or, from the perception of a FRW external observer, at the DEUS object event horizon) we have $\displaystyle t_{FRW}\rightarrow \frac{\pi}{4}$, case in which:

\begin{equation}
s(\rho,p)\rightarrow \pm \infty \; ,
\label{eq_21}
\end{equation}

\begin{equation}
T(\rho,p)=\frac{e_{0}}{k}\; \exp\left\{-\frac{1}{2\left(1-\textnormal{tg}\; t_{FRW}\right)} \right\}\rightarrow 0 \; ,
\label{eq_22}
\end{equation}
\noindent
representing the "classical" result for the external observer perception of the black hole interior, and also:

\begin{equation}
{\cal R}(\rho,p)=\frac{\rho}{e_{0}}\; \exp\left\{\frac{1}{2\left(1-\textnormal{tg}\; t_{FRW}\right)} \right\} \rightarrow \infty \; ,
\label{eq_23}
\end{equation}
\noindent
the mattter density going to infinity. In the same conditions, the external observer sees that the black hole is having negative specific internal energy:

\begin{equation}
\epsilon(\rho,p)=e_{0}\; \exp\left\{-\frac{1}{2\left(1-\textnormal{tg}\; t_{FRW}\right)} \right\}-1\rightarrow -1 \; .
\label{eq_24}
\end{equation}

\section{DEUS Model Global Action}

Considering an arbitrary self-similar DEUS object indexed as {\it level 1} (and without self-similar embeddings) we have:

\begin{equation}
S=\frac{K_{t}}{K_{\phi}}\int dS_{1} \; ,
\label{eq_25}
\end{equation}
\noindent
where, in the catenoid representation, $K_{t}$ and $K_{\phi}$ are the Gaussian curvatures of the catenoidal hyper-surface \cite{deus2} and the action $S_{1}$ has as Lagrangian the \cite{deus9} ${\cal L}_{F}$, while in the helicoid representation the curvatures are for the helicoidal hyper-surface (with the metric tensor components of the helicoid) and the corresponding ${\cal L}_{B}$ Lagrangian \cite{deus9}.

At the pre-collapse cylinder, in both representations (catenoid or helicoid), $K_{t}=K_{\phi}$, resulting:

\begin{equation}
S=\int dS_{1} \; .
\label{eq_26}
\end{equation}

At the following, lower self-similar level of DEUS objects ({\it level 2}):

\begin{equation}
S_{1}=\int dS_{2}-\frac{K'_{t}}{K'_{\phi}}\int dS_{2} \; ,
\label{eq_27}
\end{equation}
\noindent
with the same interpretation of $K'_{t}$ and $K'_{\phi}$, but this time for a second level DEUS self-similar object. Here, {\it level 1} is seen as having one embedded level of DEUS objects. When $K'_{t}=K'_{\phi}$ (at the pre-collapse cylinder of {\it level 2}), $S_{1}=0$.

For {\it level 3}:

\begin{equation}
S_{2}=\int dS_{3}-\frac{K''_{t}}{K''_{\phi}}\int dS_{3} \; ,
\label{eq_28}
\end{equation}
\noindent
and so on.

This algorithm repeats to an infinite number of embeddings, where:

\begin{equation}
S_{\infty-1}=\int dS_{\infty}-\frac{K^{\infty}_{t}}{K^{\infty}_{\phi}}\int dS_{\infty} \; .
\label{eq_29}
\end{equation}

\ack
Manuscript registered at the Astronomical Institute, Romanian Academy, record No. 265 from 04/19/07.


\title[D.E.U.S.]{XVIII: DEUS vs. Loop Quantum Gravity}

\begin{abstract}
In this short paper we do a comparative study of loop quantum gravity and DEUS model.
\end{abstract}

$\\$
In loop quantum gravity \cite{ash1,ash2} the entropy is defined as being a statistical entropy:

\begin{equation}
S=-\textnormal{Tr}\left(\rho\; \ln\rho\right) \; ,
\label{eq_1}
\end{equation}
\noindent
where the density matrix $\rho$ can be computed by counting the number ${\cal N}$ of states existent on the black hole horizon:

\begin{equation}
S=\ln {\cal N} \; .
\label{eq_2}
\end{equation}

For DEUS model, the number of states on the horizon is zero, all the internal structure of the DEUS black hole horizon being, at the observation moment, already "ejected" in the external observer's spacetime. In (\ref{eq_2}) we have $S\rightarrow \infty$, result obtained also in \cite{deus17} for the $s(\rho,p)$ specific entropy at $\displaystyle t_{FRW}\rightarrow \frac{\pi}{4}$, where, from its global FRW and local Minkowski spacetime, the external observer sees the horizon as spherical.

Second, at least in \cite{ash2}, the necessity of the supplementary $\displaystyle \frac{1}{16 \pi\; G}\; f(\phi)\; R$ term in the action:

\begin{equation}
{\cal S}\left[g_{ab},\phi \right]=\int{d^{4}x\; \sqrt{-g}\left[\frac{1}{16\pi\; G}\; f(\phi)\; R-\frac{1}{2}\; g^{ab}\; \nabla_{a}\phi\; \nabla_{b}\phi-V(\phi) \right] } \; ,
\label{eq_3}
\end{equation}
\noindent
is not justified by the authors. In the DEUS model we saw \cite{deus12} that is enough having the classical action of the nonlinear, relativistic scalar field theory, with better choices of the covariant derivatives. If we still insist to consider this term, then the function $f(\phi)$, defined as a nowhere vanishing function (the minimal coupling being for $f(\phi)=1$), {\bf has to be zero}. If $f(\phi)=0$, then the "suggested" entropy:

\begin{equation}
S_{\Delta}=\frac{1}{4\; l^{\;2}_{planck}}\; \oint_{S}{f(\phi)\; d^{2}V}
\label{eq_4}
\end{equation}
\noindent
is zero, which disagrees with the $S\rightarrow \infty$ result.

We should mention that, in the DEUS model, the Planck length is a fundamental length {\bf only} for one level of self-similar DEUS objects at a time and it materializes from the impossibility of an observer to see (from his own DEUS level) into the DEUS embedded structure "deeper" than the fourth self-similarity level.

The "loop thinking" has its basis in the {\bf considered} and {\bf semi-classical} Bekenstein-Hawking entropy formula:

\begin{equation}
S=\frac{1}{4\; l^{\;2}_{planck}}\; a_{0} \; ,
\label{eq_5}
\end{equation}
\noindent
where $a_{0}$ is the horizon area, and which has no real possibility to be checked and validated experimentally. Apart from that, the Bekenstein-Hawking formula is constructed for classical black holes and not for quantum ones.

The loop quantum gravity starting point is a semi-classical effect - the Hawking effect, when it seemed more appropriate to consider a quantum effect, more general than the Hawking effect: the Unruh effect. Unfortunately, also for the Unruh effect the temperature is derived from the $\displaystyle T=\frac{\kappa}{2\pi}$ Hawking temperature, where the Killing field surface gravity (Killing black hole horizon) is $\displaystyle \kappa=\sqrt{\frac{\Lambda}{3}}$, with $\Lambda$ the cosmological constant. Results that, in the considered deSitter space \cite{gib}:

\begin{equation}
T=\frac{1}{2\pi}\sqrt{\frac{\Lambda}{3}} \; .
\label{eq_6}
\end{equation}

By using the cosmological constant, one black hole theory problem is hidden behind a cosmological one: the $\Lambda$ meaning.

Throughout the development of the DEUS model \cite{deus1}-\cite{deus17} we gave a clear interpretation of the $\Lambda$ constant (or dark energy).

\section{Conclusions}

Between the Loop Theory and the DEUS model there is a principial incompatibility.

\ack
Manuscript registered at the Astronomical Institute, Romanian Academy, record No. 266 from 04/19/07.


\title{XIX: Topological invariants. Constructing the Universe from Particles to Large Scale Structure}

\section{Introduction \cite{labastida1}}

Though connections between quantum physics and topology can be traced back to the fifties, it is in the eighties when a new and 
unprecedented kind of relation between the two takes place. In 1982 Witten \cite{witten} considered {\it N} = 2 supersymmetric sigma 
models in two dimensions and rewrote Morse theory in the language of quantum field theory. Furthermore, he constructed out of those 
models a refined version of Morse theory known nowadays as Morse-Witten theory. Witten's arguments in \cite{witten} made use of functional 
integrals and therefore can be regarded as non-rigorous. Nevertheless, some years later, Floer reformulated Morse-Witten theory providing 
a rigorous mathematical structure \cite{floer}.

The influence of Atiyah \cite{atiyah} on Witten in the fall of 1987 culminated with the construction by the latter of the first 
topological quantum field theory (TQFT) in January 1988 \cite{witten1}. The quantum theory turned out to be a ``twisted'' version of 
{\it N} = 2 supersymmetric Yang-Mills. This theory, whose existence was conjectured by Atiyah, is related to Donaldson invariants for 
four-manifolds \cite{donaldson}, and it is known nowadays as Donaldson-Witten theory.

In 1988 Witten formulated also two models which have been of fundamental importance in two and three dimensions: topological sigma 
models \cite{witten2} and Chern-Simons gauge theory \cite{witten3}. The first one can be understood as a twist of the {\it N} = 2 
supersymmetric sigma model considered by Witten on his work on Morse theory \cite{witten}, and is related to Gromov invariants 
\cite{gromov}. The second one is not the result of a twist but a model whose action is the integral of the Chern-Simons form. In this 
casethe corresponding topological invariants are knot and link invariants as the Jones polynomial \cite{jones} and its generalizations.

TQFT provided a new point of view to study the topological invariants which were discovered only a few years before the formulation of 
this type of quantum theory. One of the important aspects of this new approach is that they could be generalized in a variety of 
directions. Since 1988 there are two main lines of work: on one hand, the rigorous constructions (without using functional integration) 
of the generalizations predicted by TQFT; on the other hand, the use of quantum field theory techniques to analyze and compute the 
generalized invariants.

TQFT's have been studied from both, perturbative and non-perturbative approaches. In physical theories it is well known that both 
approaches provide very valuable information on the features on the features of the model under consideration. In general, the 
non-perturbative methods are less developed than perturbative ones. However, precisely in {\it N} = 2 supersymmetric theories, the ones 
intimately related to TQFT's, important progress has been done recently \cite{seiberg}. It is also important to notice that TQFT's are 
in general much simplier than their physical counterparts and one expects that the use of these methods is much more tractable.

In three dimensions, non-perturbative methods have been applied to Chern-Simons gauge theory to obtain properties of knot and link 
invariants as well as general procedures for their computation. On the other hand, perturbative methods have provided an integral 
representation for Vassiliev invariants \cite{vassiliev} which, among other things, allows to extend the formulation of these invariants 
to arbitrary smooth three-manifolds. Vassiliev invariants are strong candidates to classify knots and links.

In four dimensions, perturbative methods show that Donaldson-Witten theory is related to Donaldson invariants. On the other hand, 
non-perturbative methods indicate that those invariants are related to other rather different topological invariants which are called 
Seiberg-Witten invariants. In sharp contrast to Donaldson invariants, which are defined on the moduli spaces of instantons, Seiberg-Witten 
invariants are associated to moduli spaces of abelian monopoles \cite{witten4,hyun}. Recently, Donaldson-Witten theory has been 
generalized to a theory involving non-abelian monopoles \cite{labastida2}, which provides a rich set of new topological invariants for 
the four-manifolds study. Nevertheless, there are indications that these new invariants can also be written, at least in some situations, 
in terms of Seiberg-Witten invariants. Therefore, it might happen that no new topological information is gained.

The above theories seem to share a common structure. Their topological invariants can be labeled with group theoretical data: Wilson 
lines for different representations and gauge groups (Jones polynomial and its generalizations) and non-abelian monopoles for different 
representations and gauge groups (generalized Donaldson polynomials). However, these invariants can all be written in terms of topological 
invariants which are independent of the group and representations chosen: Vassiliev invariants and Seiberg-Witten invariants, 
respectively. Both depend strictly on the topology. The group-theoretical data labeling generalized Jones and Donaldson polynomials enter 
in the coefficients of the expressions of these polynomials as a power series in Vassiliev and Seiberg-Witten invariants, respectively.

The resemblance between the two pictures is very appealing. Nevertheless, there are important differences which rise some important 
questions. In the case of knot theory, Vassiliev invariants constitute an infinite set. However, in Donaldson theory, for the cases 
studied so far, only a finite set of invariants seems to play a relevant role. One would like to know if this is general or if this fact 
is just a peculiarity of the only two cases (gauge group {\it SU(2)} without matter and with one multiplet of matter in the fundamental 
representation) which have been studied so far. The general picture of non-perturbative {\it N} = 2 supersymmetric Yang-Mills theories 
seems to sugest that the set of invariants entering the expressions for the generalized Donaldson polynomials is going to be finite. 
However, one might find unexpected results by studying different kinds of matter.

\section{Schwarz Type Theories}

In the case of Schwarz type theories one must first construct an action which is independent of the metric $g_{\mu\nu}$. The method is 
best illustrated by considering, for example, the Chern-Simons gauge theory. This is a three-dimensional theory whose action is the 
integral of the Chern-Simons form associated to a gauge connection {\it A} corresponding to a group {\it G}:

\begin{equation}
S_{CS}(A)=\int_{M}\textnormal{Tr}(A\wedge dA+\frac{2}{3}A\wedge A\wedge A) \; .
\label{eq_1}
\end{equation}

Observables must be constructed out of operators which do not contain the metric $g_{\mu\nu}$. In gauge invariant theories, as it is the 
case, one must also demand invariance under gauge transformations for these operators. An important set of observables in Chern-Simons 
gauge theory is constructed by the trace of the holonomy of the gauge connection {\it A} in some representation {\it R} along a 1-cycle 
$\gamma$, the Wilson line:

\begin{equation}
\textnormal{Tr}_{R}(\textnormal{Hol}_{\gamma}(A))=\textnormal{Tr}_{R}\textnormal{P}\; \textnormal{exp}\int_{\gamma}A \; .
\label{eq_2}
\end{equation}

The non-perturbative analysis of the theory shows that the invariants associated to the observables are knot and link invariants with 
the same properties as Jones polynomial and its generalizations. If one considers $M=S^{3}$, {\it G = SU(2)}, and takes all the Wilson 
lines entering in the fundamental representation, the non-perturbative analysis proves that the vevs associated to three links whose 
only difference is in an overcrossing, in an undercrossing or in no-crossing, satisfy the following relation:

\begin{equation}
q^{-1}\; \overcrossing - q\; \undercrossing = (q^{1/2}-q^{-\;1/2})\uparrow\;\uparrow
\label{eq_3}
\end{equation}
\noindent
where $q=\textnormal{exp}(2\pi i/(2yk+g))$, {\it g} being the the dual Coxeter number of the group {\it G}. These are precisely the skein 
rules which define the Jones polynomial. The great advantage of Chern-Simons gauge thaory is that it allows to generalize very simply 
these invariants to other groups and other representations. The HOMFLY \cite{freyd} and the Kauffman \cite{kauffman} polynomials are 
obtained after considering the fundamental representation of the groups {\it SU(N)} and {\it SO(N)}, respectively. The Akutsu-Wadati 
\cite{akutsu} or colored Jones polynomial is obtained considering the group {\it SU(2)} with the Wilson lines in different 
representations. Other non-perturbative methods have allowed to obtain these invariants for classes of knots and links as, for example, 
torus knot and links \cite{labastida3}. Methods for general computations of these invariants have been proposed in \cite{martin} and 
\cite{kaul}.

From the point of view of perturbation theory, Chern-Simons gauge theory has been studied in both, the Hamiltonian (non-covariant) and 
the Lagrangian (covariant) approaches providing a variety of interesting results.

\section{Topology}

One of the central problems of topology is to classify manifolds. Two manifolds are said to be the same if there is a diffeomorphism 
between them. To illustrate the problem, consider the classification of compact connected surfaces without boundary. This problem was 
solved by Poincar\'{e} in the early twentieth century, and it goes like this: some surfaces are orientable, and some are not. Here is 
a list of the compact connected surfaces without boundary that are orientable: the two sphere $S^{2}$, the two-torus $T^{2}$, the 
double-torus (like a torus but with two handles), the triple-torus, and so on. A good way to think these is as a connected sum of tori. 
The {\it connected sum} of two connected surfaces {\it X} and {\it Y} is what you get when you remove a disk from {\it X} and a disk from 
{\it Y}, then glue the result along the boundary.

The Euler characteristic $\chi$ is a number that is easy to assign to each surface. Once you know whether or not a surface is orientable, 
the Euler characteristic uniquely determines the surface.

More generally, we would like to classify manifolds, the {\it n}-dimensional version of surfaces. Whether or not we insist on 
connectedness is not very important, since any disconnected manifold is just a union of connected manifolds. The criterion of compactness 
is more worthwhile, since any open subset of a manifold is also a manifold, but for the moment, we will not get bogged in the 
classification of open subsets.

If we were to pattern the project of classification of manifolds after the above classification for surfaces, then one way to describe 
the problem would be to say that we wish to assign some mathematical object (such as a number, a group, or anything just as easy to 
understand) to each manifold (hopefully in a way that is easy to compute) so that if two manifolds are diffeomorphic, they have the same 
mathematical object (in which case the object is called a topological invariant), and so that if two manifolds are not diffeomorphic, 
then they are not assigned the same mathematical object (in which case the topological invariant is called a complete topological 
invariant).

In the case of surfaces, we had two important topological invariants: the Euler characteristic and the orientability. Neither alone is 
a complete topological invariant of compact connected surfaces without boundary, but the ordered pair is.

In general, we do not hope to come up with a single object that is our complete topological invariant right away. We expect to come 
up with many topological invariants which together classify manifolds completely.

Algebraic topology defined many kinds of topological invariants for the {\it n}-dimensional manifolds. In fact they were usually defined 
for arbitrary topological spaces. For instance, if {\it X} is a connected space, its fundamental group $\pi_{1}(X)$ is a group, and if 
two manifolds are diffeomorphic, then they have the same fundamental group. Therefore, $\pi_{1}$ is a topological invariant.

There are generalizations $\pi_{2}(X)$, $\pi_{3}(X)$, ..., that are also topological invariants, which are actually abelian groups. 
There are other sequences of topological invariants that are groups: the homology of a manifold {\it X} is a sequence of abelian groups 
$H_{0}(X)$, $H_{1}(X)$, ..., and the cohomology $H^{0}(X)$, $H^{1}(X)$, ..., and there are others. A brief account for physicists is 
found in \cite{nash} and a more complete text on the subject is \cite{munkres}.

For compact manifolds, these groups are all finitely generated, and the point is that inasmuch as finitely generated groups or abelian 
groups are understood, these invariants should make it easier to understand the problem of classification of manifolds. The problem is 
that it is not clear whether or not these form a set of complete invariants and, furthermore, which values of the invariants are 
possible. Actually, it is possible to prove that in dimension 4 and higher, any group with finitely many generators and relations can 
be $\pi_{1}(X)$ for some manifold {\it X}. This can be done explicitly enough that the classification of manifolds would also produce 
a classification of groups with finitely many generators and relations. The bad news is that the classification of groups with finitely 
many generators and relations has been proven to be impossible \cite{boone1,boone2}, and therefore, the classification of manifolds 
must be impossible too.

This would seem to answer the main problem in a spectacularly negative fashion: if $n\geq 4$, then the classification of compact 
{\it n}-manifolds without boundary is algorithmically impossible.

The above statement means that if we try to charaterize some manifolds that are having four or five dimensions (let say a catenoid 
hypersurface or a helicoid hypersurface) our observables that uniquely determine (classify) the properties of the manifolds are infinite 
in number. This is equivalent with seeing (simultaneously) an infinite number of self-similar levels of embedded manifolds. The 
problem is solved either by considering a $n\leq 3$ projection of the hypersurfaces which brings with it a finite number of observables 
(or levels) and a 3-dimensional FRW bubble evolving in time ($t_{FRW}$) between two singularities, either by decomposing (as we will see 
in a moment) the group associated to our considered DEUS geometry into groups (which are for one observable 3-dimensional DEUS lower level 
effects {\it SU(2)} or {\it SO(2)} or for two 3-dimensional DEUS lower levels effects {\it SU(3)} or {\it SO(3)}) simply connected (by the 
direct product) as subgroups of the DEUS group. For the $n\leq 3$ observer the subgroups are not seen as connected, partly because it can 
not perceive the totality of the effects living into the supplementary dimension and partly because he can not see across the initial and 
final singularities in other than its DEUS object spacetime projection. This is the same as saying that it is not possible to find the 
``Holy Grail of Unification'' excepting the situation in which of the observer is five dimensional or in the case in which it can find a 
method of observing the totality (not only the indirect effects) of the 5-dimensional spacetime effects acting in that spacetime. The 
first conclusion that follows is that for a (3 + 1) spacetime the representations are simple (the groups does not contain any non-trivial 
invariant subgroup - it cannot be written as a direct-product group) and disconnected, while the five-dimensional spacetime effects can be 
described by simply connected manifolds of other than simple group representation. The five-dimensional group representations cannot be 
simple because each of them can be decomposed in subgroup representations of the first lower level embedded DEUS objects (visible for a 
$n=5$ observer because of the ``lack'' of singularities) and this decomposition being possible to be done at infinity for lower and lower 
self-similar DEUS levels. So, the second conclusion that follows is that even that the observer ``sees'' in five-dimensions it cannot 
construct a theory of ``Unification'' based on simple groups and, worst, he cannot construct it because of the necesity of a mathematical 
description of an infinit level of self-similarity. He would have to limit himself to a finite number of levels, the best choice beeing 
to construct a theory for the number of levels visible from the projection of his $n$-dimensional spacetime in a $(n-1)$-dimensional 
spacetime.

But this is not the end of the story. We could restrict our attention to simply connected manifolds (those for which $\pi_{1}(X)$ is 
the trivial group), or manifolds with $\pi_{1}(X)$ some group that is easy to understand (finite groups, cyclic groups, etc.). And it 
is precisely for dimensions four and higher that we know of many manifolds that are simply connected.

In dimensions five and higher, remarkably, the problem of classifying simply connected compact manifolds without boundary is solved, 
whereas the analogous classification in dimensions three and four is still unsolved today. This strange circumstance, suggesting that 
dimensions five and higher are easier than dimensions three and four, comes about because there are certain techniques that are very 
powerful \cite{kosinski,ranicki}. This classification also extends to the classification of manifolds whose $\pi_{1}(X)$ is understood 
sufficiently well.

In dimension three it is not known if there are other simply connected compact three-dimensional manifolds without boundary other than 
the three-sphere $S^{3}$.

Before 1980s, there was not much known about simply connected four-dimensional manifolds. It was possible to compute homology and 
cohomology groups, but invariants like these, from algebraic topology, gave limited information and it was not clear whether or not there 
was more to the classification story.

The homology groups look like $H_{0}(X^{4})\simeq Z$, $H_{1}(X^{4})\simeq 0$, $H_{2}(X^{4})\simeq Z^{b_{2}}$, $H_{3}(X^{4})\simeq 0$, 
$H_{4}(X^{4})\simeq Z$, the higher homology groups being all trivial. The vanishing of $H_{1}$ occurs because $X^{4}$ is simply 
connected (using the Hurewicz theorem) and the vanishing $H_{3}$ occurs because of Poincar\'{e} duality. So, if you were to use only 
homology, the only topological invariant we could get was one number: the second Betti number $b_{2}$.

The cohomology groups can be calculated using the universal coefficient theorem and, in this case, the table of cohomology groups is 
identical to the one for the above homology groups. But the cohomology groups have some extra information, because cohomology classes 
can be multiplied via the wedge product. In our case, the only situation to consider is multiplying two elements of $H^{2}(X^{4})$, which 
gives rise to an element of $H^{4}(X^{4})$. We can view this as a bilinear form on $H^{2}(X^{4})$, taking two cohomology classes and 
returning a number. By Poincar\'{e} duality, we can interpret it in terms of homology instead of cohomology, and this is what happens: 
an element of $H_{2}(X^{4})$ can be viewed as a surface embedded in $X^{4}$, and if $\Sigma_{1}$ and $\Sigma_{2}$ are two such, they 
will generically intersect in a finite set of points. If these are counted with appropriate signs, the number of points in the 
intersection will be an integer.

Whichever way we wish to think of it, there is a bilinear form on $H_{2}(X^{4})$ or equivalently on $H^{2}(X^{4})$ called the intersection 
form, and it is symmetric, integer-valued and non-degenerate. If we choose a basis for $H_{2}(X^{4})$, this intersection form can be 
viewed as a square $b_{2}\times b_{2}$ matrix of integers. This matrix is symmetric and its determinant is $\pm 1$.

This intersection form is a topological invariant. Every simply connected compact four-dimensional manifold without boundary gives rise 
to an integer-valued $b_{2}\times b_{2}$ symmetric matrix with determinant $\pm 1$. But identifying it as a matrix requires choosing a 
basis.If we were to allow any real change of basis, the classification of these bilinear forms is just a matter of counting the number 
of positive and negative eigenvalues (since the determinant is $\pm 1$, there are no zero eigenvalues). Let $b_{2}^{+}$ be the number of 
positive eigenvalues and $b_{2}^{-}$ the number of negative eigenvalues. In each case $b_{2}=b_{2}^{+}+b_{2}^{-}$ and 
$\sigma(X^{4})=b_{2}^{+}-b_{2}^{-}$ is called {\it signature}. If the orientation of the manifold is reversed, the matrix is replaced by 
its negative, and therefore $b_{2}^{+}$ and $b_{2}^{-}$ reverse roles. So, $b_{2}^{+}$ and $b_{2}^{-}$ are not really topological 
invariants, but $|\sigma|$ is. Alternately, we can try to classify manifolds together with their orientation, and then we have $b_{2}^{+}$ 
and $b_{2}^{-}$ as invariants of manifolds with orientation.

There were two breakthroughs in the 1980s that added remarkable clarity to what was going on for simply connected four-dimensional 
manifolds, and they happened at roughly the same time. On the one hand was the work of Freedman that was completely topological and, 
on the other hand, was the work of Donaldson that used instantons. These two breakthroughs were complementary in the sense that they 
addressed two disjoint sides of the question.

Freedman's work \cite{freedman} classified topological manifolds (where the coordinate charts need not patch together smoothly) up to 
homeomorphism (for two topological manifolds to be homeomorphic, all that is necessary is the existence of a continous map from one to the 
other with a continous inverse) as opposed to Donaldson's work which described what happens to smooth manifolds (where the coordinate 
charts patch together differentiably) up to a diffeomorphism (so that the map relating the two and its inverse must be differentiable).

The idea behind Freedman's work is to show that a more sophisticated version of what works for dimensions five and higher actually works 
for dimension four. In dimensions five and higher it is often necessary to ``simplify'' a description of a manifold by finding a 
complicated subset and showing it is really a ball. The same idea works in dimension four, except that sometimes the necessary subset 
is infinitely complicated, and Freedman was able to show that such a subset is homeomorphic (though perhaps not diffeomorphic) to a ball.
The same behavior is reproduced also in the DEUS model where we fist have to introduce a fifth dimension in order to describe the 
four-dimensional spacetime projections of our manifolds and get across the singularities of these projections.

On the other hand, by considering a Yang-Mills {\it SU(2)} gauge field on the four-dimensional manifold and studying instantons, 
Donaldson \cite{donaldson1} was able to prove that the intersection form must be either indefinite or plus or minus the identity. In 
other words, the situation where we didn't know how to classify intersection forms, the case where it was definite, is the situation where 
this classification is unnecessary, since smooth manifolds can't have them as intersection forms anyway, with the exception of the 
identity and minus the identity.

\section{Instantons}

Consider a pure {\it SU(2)} gauge field theory on flat $R^{4}$ as described in standard textbooks like \cite{peskin}. Let $i\sigma^{a}$ be 
the standard Pauli basis for the Lie algebra of {\it SU(2)}, where {\it a} = 1,..., 3. Let $A_{\mu}=A_{\mu}^{a}\sigma_{a}$ be an {\it 
SU(2)} connection, with $\mu$ = 1,..., 4 a spatial index, and $F_{\mu\nu}^{a}=\partial_{\mu}A_{\nu}^{a}-\partial_{\nu}A_{\mu}^{a}+
\epsilon^{abc}A_{\mu}^{b}A_{\nu}^{c}$ is its curvature tensor, so that:

\begin{equation}
F_{\mu\nu}=A_{[\mu ,\nu]}+[A_{\mu},A_{\nu}]\; .
\label{eq_4}
\end{equation}

Consider the action:

\begin{equation}
S=\int_{R^{4}}||F||^{2}d^{4}x=\int_{R^{4}}F_{\mu\nu}^{a}F_{a}^{\mu\nu}d^{4}x\; .
\label{eq_5}
\end{equation}

If we replace the Lorentzian (-+++) metric with the Euclidian (++++) metric, we can obtain classical minima of the action above. These 
are called {\it instantons}, and are useful in calculating tunneling amplitudes \cite{coleman} (the rotation from time to imaginary time 
is what is involved in the WKB approximation).

We care not about $R^{4}$ but about arbitrary compact manifolds ($R^{4}$ is not compact). The question of finding instantons is basically 
unchanged, except when $R^{4}$ is replaced by a non-trivial manifold, we need to consider some topological considerations. Namely, the 
gauge field corresponds to a vector bundle {\it E} (in this case, a two dimensional complex vector bundle) on the manifold. The 
connection is locally defined on coordinate patches and transforms as we go from one patch to another by gauge transformations.

For each such vector bundle {\it E} over our manifold $X^{4}$ we can associate the second Chern class:

\begin{equation}
c_{2}(E)=-\frac{1}{8\pi^{2}}\int_{X^{4}}F_{\mu\nu}^{a}\textnormal{{\it \~{F}}}_{a}^{\mu\nu}
\label{eq_6}
\end{equation}
\noindent
which is an integer. $\textnormal{{\it \~{F}}}_{a}^{\mu\nu}$ is the dual of $F_{a}^{\mu\nu}$ and $\textnormal{{\it \~{F}}}_{a}^{\mu\nu}=
\epsilon^{\mu\nu\lambda\rho}F_{a}^{\lambda\rho}$. More precisely, $c_{2}(E)$ is the four form in the integrand. It is an element of 
$H^{4}(X^{4})\simeq Z$, the isomorphism being realized by taking the integral. The second Chern class is defined above in terms of the 
connection {\it A}, through its curvature {\it F}, but in fact it is independent of the connection and only depends on the vector bundle 
{\it E}. The first Chern class $c_{1}(E)$, incidentally, is zero because the group is {\it SU(2)}. In the {\it U(1)} gauge theory, the 
first Chern class is generally non-zero and measures the monopole charge for a Dirac monopole. There are higher Chern classes but they are 
all zero for {\it SU(2)}.

It turns out that the second Chern class completely classifies the vector bundle topologically, so that there is a unique vector bundle 
up to topological vector bundle isomorphism for every integer value of $c_{2}$. The trivial bundle has $c_{2}=0$.

For each vector bundle {\it E} we can look for connections {\it A} that minimize the Yang-Mills action. One choice might be the trivial 
connection {\it A} = 0, which gives rise to the action being equal to zero. This is clearly an absolute minimum, because the action in our 
case can not be negative. But this trivial connection only exists in the trivial bundle. More generally, any flat connection is a 
minimum, but also exists only in the trivial bundle with $c_{2}=0$.

For other vector bundles the minima are not obvious. The trick to understand these minima is to split the curvature {\it F} into the +1 
and -1 eigenvalues of the duality operator *, where *{\it F}={\it \~{F}}. We define $F^{+}=\displaystyle \frac{F+\textnormal{{\it\~{F}}}}
{2}$ and $F^{-}=\displaystyle \frac{F-\textnormal{{\it\~{F}}}}{2}$. Then $F=F^{+}+F^{-}$, where $*F^{+}=F^{+}$ and $*F^{-}=-F^{-}$.
Furthermore, $F^{+}$ and $F^{-}$ are orthogonal. The formula for $c_{2}$ gives:

\begin{equation}
\begin{array}{llll}
c_{2}(E)&=-\displaystyle \frac{1}{8\pi^{2}}\int_{X^{4}}(F^{+}+F^{-})_{\mu\nu} (*(F^{+}+F^{-}))^{\mu\nu} \\
&= -\displaystyle \frac{1}{8\pi^{2}}\int_{X^{4}}F_{\mu\nu}^{+}(*F^{+})^{\mu\nu}+F_{\mu\nu}^{-}(*F^{-})^{\mu\nu} \\
&= -\displaystyle \frac{1}{8\pi^{2}}\int_{X^{4}}F_{\mu\nu}^{+}F^{+\mu\nu}-F_{\mu\nu}^{-}F^{-\mu\nu} \\
&= \displaystyle \frac{1}{8\pi^{2}}\int_{X^{4}}-||F^{+}||^{2}+||F^{-}||^{2}
\end{array}
\label{eq_7}
\end{equation}
\noindent
while the formula for the action is:

\begin{equation}
S=\int_{X^{4}}(F^{+}+F^{-})_{\mu\nu} (F^{+}+F^{-})^{\mu\nu}=\int_{X^{4}}||F^{+}||^{2}+||F^{-}||^{2} \; .
\label{eq_8}
\end{equation}

Thus we see that when $c_{2}(E)<0$, the action is minimized when $F^{-}=0$, so that for instantons, $*F=F$ (in which case we call {\it 
F self-dual}) and when $c_{2}(E)>0$, the action is minimized when $F^{+}=0$, so that instantons have $*F=-F$ (in which case we call 
{\it F anti-self-dual} and, sometimes, call such solutions {\it anti-instantons}). When $c_{2}=0$, the action is minimized when $F=0$, 
which we observed before.

Supose that we have an instanton with $c_{2}(E)=1$. We view this as a minimum of the action. When we ask the question why this is the 
minimum when the connection $A=0$ clearly gives a lower value for the action, the answer is that $A=0$ does not exist in our bundle. 
To ``decay'' from our instanton to zero would require that we ``tear'' our bundle (to untwist it first). This is what we mean when we say 
that the instanton can not decay for topological reasons. The number $c_{2}(E)$ (more conventionally, $-c_{2}(E)$) is called the 
{\it instanton number} of the solution, and we imagine that instantons with $c_{2}(E)=2$ are in some sense ``non-linear'' combinations 
of two instantons with $c_{2}(E)=1$. When we combine a solution with $c_{2}=-1$ (an instanton) with a solution with $c_{2}(E)=1$ (an 
anti-instanton), they can cancel and flow down to a flat connection.

We now consider instantons on $S^{4}$. Readers who are familiar with instantons on $R^{4}$ will see many similarities. The reason is 
that the above Yang-Mills action has a conformal symmetry, and there is a conformal map from $R^{4}$ to $S^{4}$ that covers everything 
except for one point. We will describe here the results from \cite{atiyah1} for these instantons having $c_{2}(E)=-1$.

In the case of $c_{2}(E)=-1$, we are looking for self-dual connections on {\it E}, which involves solving the differential equation 
$*F=F$ for {\it A}. It turns out that the set of instantons on a bundle with $c_{2}(E)=-1$ on $S^{4}$, modulo gauge symmetry, is 
naturally a {\bf a five-dimensional non-compact manifold}. More specifically, it is a five-dimensional open ball (as it is, for example a 
FRW Universe bubble), and this happens {\bf only} because we are working in $S^{4}$. We call this set the {\it moduli space}. It turns out 
we can identify $S^{4}$ with the missing boundary of the ball in a sense I will describe in a moment.

But, before doing that, we should first consider how it came to be that the set of minima is not unique. Usually, a function has a unique 
absolute minimum. It is possible to have functions that have many absolute minima by arranging it so that many points take on the same 
minimum value of the function. But, we usually regard this as an unusual phenomenon and, in the world of physics, where the formulas are 
given to us by nature rather than specifically dreamed up to have multiple minima, we should expect there to be only one absolute 
minimum. If we see more than one absolute minimum, this is a phenomenon to be explained.

There are, indeed, circumstances in physics that give multiple absolute minima, and even continuous families of absolute minima, but 
these are usually explained by the existence of a group of symmetries. Take, for example, the Higgs mechanism in a $\phi^{4}$ theory. 
The theory has a spherical symmetry (and so the set of minima might be a sphere) and small perturbations that preserve this symmetry will 
still have a spherical set of minima.

In the case of instantons on $S^{4}$, the existence of many minima can also be explained by symmetry. There is the gauge symmetry, but 
recall that we have already quotiented out by this symmetry. But there are also conformal symmetries of $S^{4}$ and, since the action 
is conformally invariant, these conformal symmetries will take instantons to other instantons. In fact, the conformal symmetries of 
$S^{4}$ are enough to explain the entire set of solutions in this case. Therefore, from one solution, we can use the conformal symmetries 
to explain the entire moduli space.

Taking the idea of using the conformal symmetry, we can take a conformal symmetry that flows all of $S^{4}$ concentrating more and more 
of it closer to any given point of $S^{4}$. The effect of this is to concentrate the instanton near a given point of $S^{4}$. This 
explains why the boundary of the set of solutions is $S^{4}$. The conformal symmetry that concentrates most of $S^{4}$ near a point 
$p\in S^{4}$ will also move instantons in the moduli space (recall it is a five-dimensional ball) near a corresponding point on its 
boundary. The limiting connection is degenerate and, in a sense that is reminescent of a Dirac delta function, is flat everywhere on 
$S^{4}$ except at {\it p}, where it has infinite curvature.Thus we can add to our moduli space these extra limiting connections, thereby 
turning our non-compact ball to a compact ball with boudary. These limiting degenerate configurations are sometimes called {\it small 
instantons} and, while physicists are used to view them as instantons of a special kind, mathematicians tend not to view them as 
instantons, since $A_{\mu}$ is not well-defined at the point {\it p}. But it is possible to define a ``small instanton'' and add these 
small instantons to the moduli space in a natural way. The result makes the moduli space compact and this process is called compactifying 
the moduli space. From these results that if we see the FRW Universe as a sphere in $S^{4}$, all the matter contained into it will be 
distributed on the inner surface of this sphere, while the outer boundary of it, as part of a DEUS object (the time disappears 
at the ``edge'' of the Universe), will contain one instanton (each of four bubbles composing the DEUS helicoid is contained in a 
instanton shell; two instantons for the matter bubbles and two for the antimatter bubble). In this situation the observer on the sphere 
will see matter (galaxies and cluster of galaxies) of an open Universe to a distance limited by its perception horizon given by the 
curvature of the sphere.

More generally, the moduli space of instantons on a four-dimensional manifold $X^{4}$, with $c_{2}(E)<0$ given, is a manifold of dimension:

\begin{equation}
d=-8c_{2}(E)-3(1-b_{1}(X^{4})+b_{2}^{-}(X^{4})) \; .
\label{eq_9}
\end{equation}

This formula is obtained by the Atiyah-Singer index theorem, by viewing the self-dual equations as zeros of a differential operator, 
together with a suitable gauge-fixing condition like $d*(A-A_{0})=0$ once a fixed reference connection $A_{0}$ is identified.

Similarly, when $c_{2}(E)>0$, we are solving the anti-self-dual equation $*F=-F$, with the same gauge-fixing condition, and the Atiyah-
Singer index theorem gives the dimension as:

\begin{equation}
d=8c_{2}(E)-3(1-b_{1}(X^{4})+b_{2}^{+}(X^{4})) \; .
\label{eq_10}
\end{equation}

The dimension may be zero, in which case the moduli space would be a set of points, or the dimension may be negative, in which case the 
moduli space will be empty (so that there would be generically no instantons with that value of $c_{2}$).

If the dimension {\it d} is positive, we should in general have many absolute minima, the explanation of this fact can be found (no 
longer having the conformal symmetry of $S^{4}$) in \cite{iga}.

There is no group that guarantees a non-zero-dimensional family of solutions. The ``correct'' dimension of the set of minima is simply 
given by the Atiyah-Singer index theorem. If we plug in $S^{4}$ and $c_{2}(E)=-1$ in the dimension formula (\ref{eq_9}) (note that for 
$S^{4}$ we have $b_{1}=0$ and $b_{2}^{-}=0$) we get {\it d} = 5, which says that the five-dimensionality of the moduli space is not really 
a consequence of the conformal symmetry group after all, in the sense that the moduli space would continue to be five-dimensional even if 
we were to slightly perturb the metric on $S^{4}$ so that it no longer has conformal symmetry.

\section{Donaldson' Theorem}

{\bf Theorem:} Let $X^{4}$ be a simply connected compact four-dimensional manifold (no boudary) with definite intersection form. The its 
intersection form, in some basis, is plus or minus the identity matrix.

A rough proof goes as follows: By changing the orientation on $X^{4}$ we can assume that the intersection form is positive-definite. 
Then $b_{2}^{-}=0$. For simply connected manifolds, we saw above that $b_{1}=0$. Then, if we are interested in the bundle {\it E} over 
$X^{4}$ with $c_{2}(E)=-1$, we see that the formula for the dimension of the moduli space (\ref{eq_9}) gives us that the moduli space of 
instantons will be a five dimensional manifold.

Analogously to the case for $S^{4}$, where $S^{4}$ could be viewed as the boundary of the moduli space, we can similarly ``compactify'' 
the moduli space by including small instantons (the set of small instantons looks like a copy of $X^{4}$) so that the resulting moduli 
space is a five-dimensional manifold with boundary $X^{4}$.

We have to mention that the moduli space may not be quite a manifold, because it may have singularities. It turns out that, in the 
situation we are describing, there are finitely many singularities, each isolated and locally isomorphic to a cone on $\overline{CP^{2}}$ 
(the complex projective plane). They can be counted in the following way: let {\it m} be the number of elements $v\in H_{2}(X^{4})$ so 
that $v^{T}Iv=1$, where {\it I} is the intersection form of $X^{4}$. Then there will be {\it m}/2 singularities.

These singularities come about from the fact that the gauge group does not always act freely. When the complex two-dimensional bundle 
{\it E} can be split into a direct sum of two one-dimensional bundles $L_{1}$ and $L_{2}$, in such way that the connection {\it A} turns 
out to be the product of connections on each of the one-dimensional bundle factors, so that the connection is actually aproduct of {\it 
U(1)} connections, then a part of the gauge group will fix {\it A}. In particular, a constant {\it U(1)} gauge transformation will leave 
this reducible connection {\it A} invariant. Such connections are called {\it reducible}, and if this does not occur, we call it {\it 
irreducible}.

The result is that, when {\it A} is reducible and we quotient by the global gauge group, there will be the kind of singularity mentioned 
above: a cone on $\overline{CP^{2}}$ (or, for example, a light cone). Studying the self-dual equations for connections of this special 
type shows that at each splitting 
of {\it E} into two factors contributes a unique reducible connection, and for this splitting to happen $c_{1}(L_{1})+c_{1}(L_{2})=0$ and 
$c_{1}(L_{1})^{T}Ic_{1}(L_{2})=c_{2}(E)$. So these correspond to elements $v=c_{1}(L_{1})\in H_{2}(X^{4})$ so that $v^{T}Iv=1$, and this 
is a one-to-one correspondence up to swapping the roles of $L_{1}$ and $L_{2}$. This explains the number of singularities.

These singularities are isolated and do not occur on the glued-in $X^{4}$. Therefore, we can take our moduli space of instantons and 
modify it as follows: first, glue in the $X^{4}$ so that the moduli space becomes a compact manifold with boundary and with 
singularities. Then excise a small open ball around each of the {\it m}/2 singularities. What we now have is a five-dimensional manifold 
with $X^{4}$ as one boundary component, and {\it m}/2 other boundary components, each of which is a $\overline{CP^{2}}$.

The question of discerning different manifolds that have the same intersection form comes down to finding new invariants. Using 
instantons, Donaldson defined what are now known as {\it Donaldson invariants}, or {\it Donaldson polynomials} \cite{donaldson}.

To get some idea of how these might be defined, consider a simply connected four-dimensional manifold $X^{4}$. Suppose some choice of 
$c_{2}(E)>0$ makes the dimension of the moduli space (\ref{eq_10}) equal to zero. For instance, if $b_{1}=0$ (as is required for $X^{4}$ 
to be simply connected) and $b_{2}^{+}=7$ (as is the case with a connected sum of seven $CP^{2}$'s), then for the bundle {\it E} over 
$X^{4}$ with $c_{2}(E)=3$ (three anti-instantons), the moduli space of instantons would have dimension zero, and so would be a collection 
of points. These points come with multiplicity and sign. The Donaldson invariant of the $X^{4}$ manifold would be the count of how many 
points there are in the moduli space, counted with appropriate multiplicity and sign.

What makes this invariant a topological invariant is that this count is independent of the metric. The reason is that if $g_{0}$ and 
$g_{1}$ are two metrics on $X^{4}$, then, since the set of metrics is connected, we can consider a path of metrics $g_{t}$, $0\leq 
t\leq 1$ on $X^{4}$. Then, over $X^{4}\times [0,1]$, with the metric $g_{t}$ on the slice $X^{4}\times {t}$, {\bf the moduli space over 
each slice joins together to form a one-dimensional manifold} (or, as in the DEUS model, a after-collapse string).

Note that {\it t} does not really mean time. It is the parameter through which we are changing our metric. Because we are looking for 
instantons, time has already been made spacelike.

Positive solutions and negative solutions may cancel, or pairs of positive and negative solutions may appear. So, if we count these 
solutions with appropriate multiplicity and sign, the number does not change. When two points ``annihilate'', two 3-anti-instantons 
solutions become more and more similar as the metric is varied and, at a certain metric, they become identical and then the solution 
disappears completely. Fans of catastrophe theory may recognize this phenomenon.

\section{Knots and Polynomial Invariants}

In knot theory we study knots and knot types (isotopy classes of knots) as mathematical objects. We often deal with knots by depicting 
them in a plane. The depicted pictures of knots are called knot diagrams. Further, we describe isotopy of knots by some moves among knot 
diagrams, which are called the Reidemeister moves. Then, the set of isotopy classes of knots can be identified with the quotient set of 
knot diagrams modulo the Reidemeister moves. A remarkable advantage of such identification is that although knots are topological objects 
in a (at least) three-dimensional space they can be treated as combinatorial objects, like graphs embedded in a plane. In this way we can 
reduce studies of knots to studies of the combinatorics of knot diagrams.

A main topic in knot theory is the study of invariants of knots. An invariant is a map from a set of knots to a well-known set, such as 
a polynomial ring, such that two isotopic knots have the same image by the map. By using invariants of knots we can distinguish isotopy 
classes of knots concretely. A typical way of constructing an invariant is to construct a function of the set of knot diagrams, in a 
combinatorial way, such that it is unchanged under Reidemeister moves.

The historic discovery of the Jones polynomial took place in the middle of 1980s, relatively recent in the history of knot theory. In the 
1980s many invariants of knots, what are called quantum invariants, were discovered in active interaction between low dimensional 
topology and mathematical physics. The Jones polynomial can be regarded as the simplest quantum invariant. On the other hand a most 
classical invariant of knots is the Alexander polynomial, which was discovered in the 1920s. It is defined in a classical way, using the 
homology of the infinite cyclic covering space of a knot complement.

\begin{figure}
\centering
\leavevmode
\includegraphics[width=8.0cm,height=6.0cm]{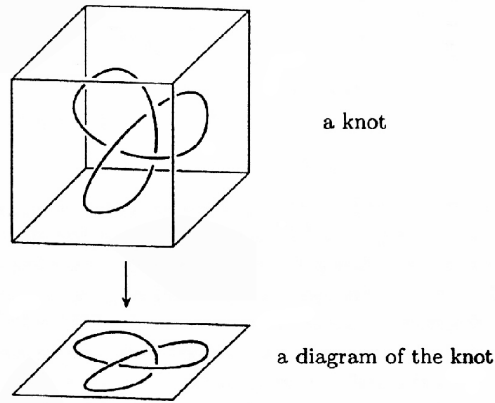}
\caption{A knot and a diagram of it}
\label{diagram}
\end{figure}

\subsection{Three-dimensional knots and links and their diagrams}

Intuitively, a {\it knot} is a circle embedded in the ambient space up to elastic deformation. A {\it link} is a family of disjoint 
knots.

In mathematical language, a link is a compact one-dimensional smooth submanifold of $R^{3}$. A connected link is called a knot. A link may 
be oriented or not. Every knot is the image of a embedding {\it f} from the circle $S^{1}$ into $R^{3}$. For a link, the situation is 
similar but the embedding is defined on a disjoint union of finitely many copies of the circle.

{\bf Definitions:} A link {\it L} is called banded if {\it L} is equipped with a vector field {\it V} from {\it L} to $R^{3}$ such that 
{\it V(x)} is transverse to {\it L} for every point $x\in L$.

A link is called framed if it is oriented and banded.

An isotopy of $R^{3}$ is a $C^{\infty}$ map $h:R^{3}\times I\rightarrow R^{3}$ such that $h_{t}=h(.,t)$ is a diffeomorphism for all $t\in 
[0,1]$. Two links $L_{0}$ and $L_{1}$ are isotopic if there exists an isotopy $h_{t}$ of the ambient space $R^{3}$ such that $h_{0}$ is 
the identity and $L_{1}$ is the link $h_{1}(L_{0})$. If the links are oriented we suppose also that $L_{1}$ has the same orientation as 
$h_{1}(L_{0})$. If ($L_{0},V_{0}$) and ($L_{1},V_{1}$) are banded links, there are called isotopic if $L_{0}$ and $L_{1}$ are isotopic 
via an isotopy $h_{t}$ in such a way that $V_{1}$ is homotopic to the vector field $h_{1}(V_{0})$ by a homotopy which is allways 
transverse to $L_{1}$. So we have four isotopy relations corresponding to the four classes of links: non-oriented, oriented, banded and 
framed.

An invariant of knots (or links) is a function from the set of knots (or links) to some module which is invariant under isotopy. It is 
also possible to define an invariant of oriented knots (or links), or an invariant of banded knots (or links) or an invariant of framed 
knots (or links).

A map can be defined as function of diagrams that is invariant under Reidemeister moves.

Every link can be described by its projection on the plane if it is generic. Such a projection is called a diagram of a link. A diagram 
of a link is a finite graph {\it D} contained in the plane such every vertex is of order four. Moreover near every vertex {\it x} two 
edges arriving at {\it x} correspond to the over branch and the two other ones correspond to the under branch. The edges corresponding 
to the over branch are represented by a connected path.

A {\it positive crossing} in a diagram is a crossing that looks like $\overcrossing$ (up to rotation of the plane). (The ``shortest'' arc 
that goes from the arrow of the top strand to the arrow of the bottom strand turns counterclockwise.) A {\it negative crossing} in a 
diagram is a crossing that looks like $\undercrossing$.

The {\it linking number} for two-component links is half the number of the positive crossings that involve the two components minus half 
the number of negative crossings that involve the two components.

{\bf Reidemeister theorem \cite{reide}:} Up to orientation-preserving diffeomorphism of the plane, two diagrams of a link can be related 
by a finite sequence of Reidemeister moves that are local changes of the following type:

\begin{figure}[htp]
\centering
\leavevmode
\includegraphics[width=7.0cm,height=3.0cm]{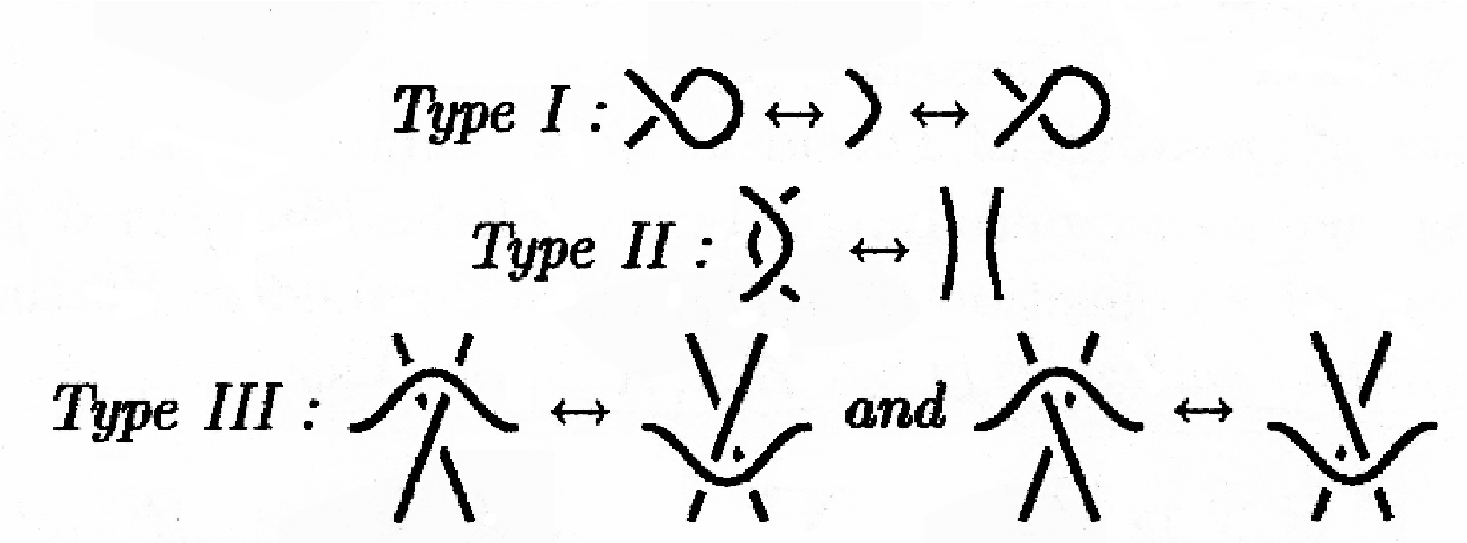}
\end{figure}

Let {\it L} be a link represented by a diagram {\it D}. If {\it L} is oriented, the orientation is represented by an orientation of {\it 
D}. If it is banded it is possible to choose the diagram {\it D} in such a way that the transverse vector field is normal to the plane 
$R^{2}$ with positive last coordonate.

Using this convention every diagram defines a banded link and every oriented diagram defines a framed link and these links are well 
defined up to isotopy.

Suppose that $L_{0}$ and $L_{1}$ are two links related by a family $L_{t}$, $0\leq t\leq 1$, of geometric objects. If every $L_{t}$ is 
a link which depends smoothly on {\it t} (that is the union $L_{[0,1]}$ of all $L_{t}\times {t}$ is a submanifold of $R^{3}\times [0,1]$) 
the links $L_{0}$ and $L_{1}$ are isotopic. But it is possible to consider singular deformation when $L_{t}$ becomes singular, for some 
particular values of {\it t}. The simplest example of such singular deformation is when a branch of {\it L} crosses another one. When 
this crossing happens the link becomes a singular link in the following sense:

{\bf Definition:} A singular link {\it L} is the image of an immersion {\it f} from a one-dimensional compact manifold $\Gamma$ to $R^{3}$ 
such that {\it f} has only finitely many multiple points and every multiple point is double and transverse, together with local 
orientations in $\Gamma$ near each singular point of {\it f}.

A singular link {\it L} is oriented if the source $\Gamma$ of the immersion {\it f} is oriented and the local orientations are induced 
by the orientation of $\Gamma$. It is banded if {\it L} is equipped with a transverse vector field {\it V} such that, for every double 
point {\it x} of {\it L}, {\it V(x)} is transverse to the plan which is tangent at {\it x} to the two branches of {\it L} containing 
{\it x}.

If {\it D} is a diagram of a link and {\it P} a subset of the set of vertices of {\it D}, one can associate to ({\it D,P}) a singular 
link {\it L} where the double points correspond to the points in {\it P}. With the same way as before, the diagram induces a well 
defined banded structure on {\it L}. If {\it D} is oriented, {\it L} is naturally framed.

A crossing change is a local modification of the type $\backoverslash \leftrightarrow \slashoverback$.

{\bf Proposition:} Any link can be unknotted by a finite number of crossing changes.

{\bf Proof:} At a philosophical level, it comes from the fact that $R^{3}$ is simply connected, and that a homotopy $h:S^{1}\times 
[0,1]\rightarrow R^{3}$ that transforms a link into a trivial one can be replaced by a homotopy that is an isotopy except at a finite 
number of times where it is a crossing change. (Consider $h\times [0,1]:(x,t)\mapsto (h(x,t),t)\in R^{3}\times [0,1]$. The homotopy {\it 
h} can be perturbed so that $h\times [0,1]$ is an immersion with a finite number of multiple points that are transverse double points 
\cite{hirsch}).

Let {\it L} be a singular link and {\it x} be a double point in {\it L}. One can modify {\it L} a little bit near {\it x} and obtain a 
new singular link {\it L'} with one double point less. But it is possible to do that in two different ways, and one gets two new links 
$L_{+}$ and $L_{-}$.

\begin{figure}[htp]
\centering
\leavevmode
\includegraphics[width=9.0cm,height=1.5cm]{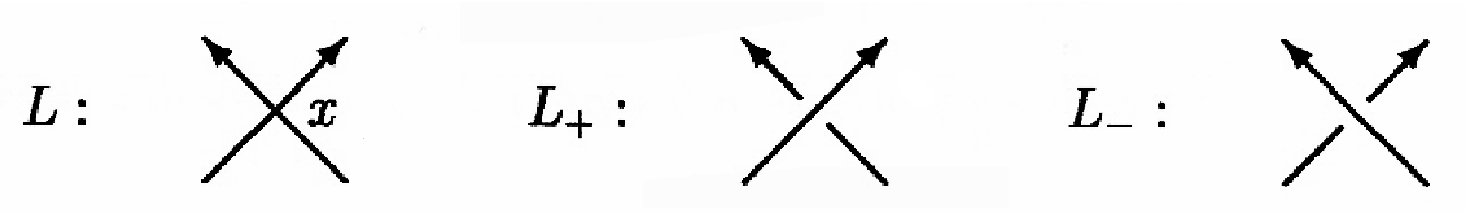}
\end{figure}

Since {\it L} is supposed to be oriented near {\it x}, there is no ambiguity between $L_{+}$ and $L_{-}$. If {\it L} is banded, the two 
desingularized links $L_{+}$ and $L_{-}$ are still banded.

{\bf Lemma:} Let {\it I} be an invariant of oriented knots. Then {\it I} extends uniquely to an invariant defined on the set of all 
singular oriented knots and satisfying the following property:

If {\it K} is a singular oriented knot and $K_{+}$ and $K_{-}$ are two knots obtained by desingularization near a double point in {\it K}, 
one has:

\[
I(K)=I(K_{+})-I(K_{-})
\]

The extension of {\it I} may be defined in the following way:

Let {\it K} be a singular oriented knot. Denote by {\it X} the set of double points in {\it K} and by $\cal{F}$ the set of functions from 
{\it X} to ${\pm 1}$. If $\alpha$ is a function in $\cal{F}$ one can desingularize {\it K} near every double point in {\it K} by using 
the positive or the negative move near a point {\it x} if $\alpha (x)=1$ or -1. So for every $\alpha\in \cal{F}$ one gets a knot 
$K_{\alpha}$. Then one sets:

\[
I(K)=\sum_{\alpha\in \cal{F}}\epsilon (\alpha)I(K_{\alpha})
\]
\noindent
where $\epsilon (\alpha)$ is the product of all numbers $\alpha(x)$, $x\in X$.

{\bf Definition:} Let {\it I} be an invariant of knots. One said that {\it I} is a Vassiliev invariant of degree $\leq n$ if {\it I} 
vanishes on every oriented singular knot with at least $n+1$ double points.

{\bf Remark:} If {\it I} is an invariant of oriented links, or an invariant of knots (or links) or banded knots (or links) or framed 
knots (or links), it is possible to extend {\it I} to the corresponding set of singular knots or links and one can define a Vassiliev 
invariant of knots (or links), or banded knots (or links) or framed knots (or links).

{\bf Example:} Let {\it L} be a singular oriented link with only one double point {\it x}. One can modify {\it L} near {\it x} in three 
different ways:

\begin{figure}[htp]
\centering
\leavevmode
\includegraphics[width=11.0cm,height=1.5cm]{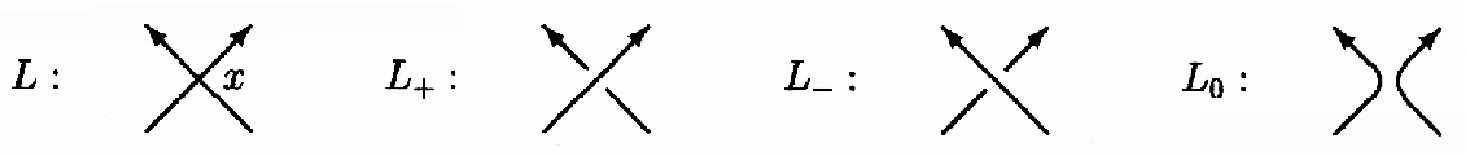}
\end{figure} 

These three links have no double point. The Conway polynomial $\nabla$ is the only polynomial invariant of oriented links which is equal 
to 1 for the trivial knot and satisfies the following skein relation:

\begin{equation}
\nabla (L_{+})-\nabla (L_{-})=t\; \nabla(L_{0}) \; .
\label{eq_11}
\end{equation}

For every oriented link {\it L}, $\nabla (L)$ is a polynomial in the ring {\bf Z}[{\it t}].

{\bf Proposition:} The $n^{th}$ coefficient of the polynomial $\nabla$ is a Vassiliev invariant of degree {\it n}.

{\bf Proof:} The skein relation shows that $\nabla (L)$ is divisible by $t^{n}$ if {\it L} is a singular link with at least {\it n} 
double points and the $n^{th}$ coefficient $a_{n}$ of the polynomial $\nabla$ is an integral invariant of oriented links which vanishes 
on every singular link with at least $n+1$ double points. The result follows.

{\bf Application:} Let us consider a trefoil knot (the simplest knot possible to be defined) with a diagram as in Figure \ref{diagram}. 
Then, for this knot we will have three double points, each of them being modified in such a way that to be the solution of a polynomial 
invariant (Jones, Kauffman, HOMFLY, etc.). In our particular case (trefoil), the diagram vertices will be disposed in a triangle. Each 
edge length (the distance between two solution of a specific polynomial) will be later described as spanning trees of the trefoil graph, 
while the angles of the triangle will be fixed (and will represent also a first checking method for the correctitude of the model, in an 
Euclidian or non-Euclidian plane) by the curvature of the plane (spacetime) in which the diagram lives. Because for each set of 
interactions between quarks, leptons or nucleons we can construct a triangle graph (and also for the interaction between the ``time'' 
matter Universe with the ``time'' anti-matter Anti-Universe and with the ``before-time'' matter Universe), each DEUS particle being a 
vertex and each interaction being an edge, one DEUS object will be represented as a $n$-dimensional ``knot in a box'' of the trefoil type 
whose projection in the $(n-1)$-dimensional diagram will be a triangle in the $(n-1)$-dimensional diagram, connected to its neighbors 
through edges of spanning trees (the second checking method). This ``triangulation'' method will construct a box structure (domains) in a 
trefoil knot and will fail at the boundary of this knot if the polynomial will fail to have solution or if the spanning trees rules will fail to 
apply. Example: Let us consider that at quark level we have a knot representation formed by string links (uni-dimensional; see Figure \ref
{diagram}). Then the diagram of this knot will be immersed in a square box. The totality of these square boxes will define a trefoil knot 
band (nucleon DEUS level) constrained by the above checking rules (locally) and by a polynomial of a further type (Jones, Kauffman, 
HOMFLY, etc.) for its projection diagram. At this level, the diagrams must satisfy the same ``triangulation'' rules and define (as boxes) 
a 3-dimensional trefoil-shaped knot (black hole self-similarity level) which, again, in projection must give invariant solutions of a
polynomial. This final projection where this invariance has to be checked is to be done on a sphere (FRW bubble level) and the result 
compared with the observed Large Scale Structure of the Universe, or on a cylinder (the pre-collapse catenoid seen as the Last Scattering 
Surface) with the observer placed either on a point of the bottom either in a point of the top circle of the cylinder. By 
transforming the representation (and also the observer position) from cylindrical coordinates into spherical coordinates we will have to 
compare the simulation result with the CMB observational map. The number of points (nuclei) will be normed to an average number of nuclei 
forming a galaxy at a particular age of the Universe.

In what follows we will define the knot polynomials and their invariants as it is done in the literature, with emphasis on the level of 
self-similarity where they fit in our simulation.

\section{The Jones Polynomial \cite{lescop,ohtsuki}}

In this section we introduce the Jones polynomial of links. To introduce it in an elementary way we use the Kauffman bracked of link 
diagrams, through, historically speaking, it was introduced by Jones using the theory of operator algebra.

Let {\it D} be an unoriented link diagram. A crossing $\slashoverback$ of {\it D} can be removed in two different ways:

the left-handed one where $\slashoverback$ becomes $\smoothing$ (someone walking on the upper strand towards the crossing turns left 
just before reaching the crossing),

and the right-handed one where $\slashoverback$ becomes $\hsmoothing$.

Let $C(D)$ denote the set of crossings of {\it D} and let {\it f} be a map from $C(D)$ to $\{L,R\}$, then $D_{f}$ will denote the diagram 
obtained by removing every crossing {\it x} in the left-handed way if $f(x)=L$ and in the right-handed way otherwise. $D_{f}$ is 
nothing but a collection of $n(D_{f})$ circles embedded in the plane.

We define the Kauffman bracket $<D>\in \textnormal{{\bf Z}}[A,A^{-1}]$ of {\it D} as:

\begin{equation}
 <D>=\sum_{f:C(D)\rightarrow \{L,R\}}A^{(\#f^{-1}(L)-\#f^{-1}(R))}\delta^{(n(D_{f})-1)}\; ,
\label{eq_12}
\end{equation}
\noindent
with $\delta=-A^{2}-A^{-2}$.

The Kauffman bracket satisfies the following properties:

\[
\textnormal{1.  }<\textnormal{{\it n} disjoint circles}>=\delta^{n-1}
\]
\noindent
or $\left< \bigcirc D\right>=(-A^{2}-A^{-2})<D>$ for any diagram {\it D} without crossings, $<$the empty diagram$>$ = 1; and we have the 
following equalities that relate brackets of diagrams that are identical anywhere except where they are drawn.

\begin{figure}[htp]
\centering
\leavevmode
\includegraphics[width=10.0cm,height=2.7cm]{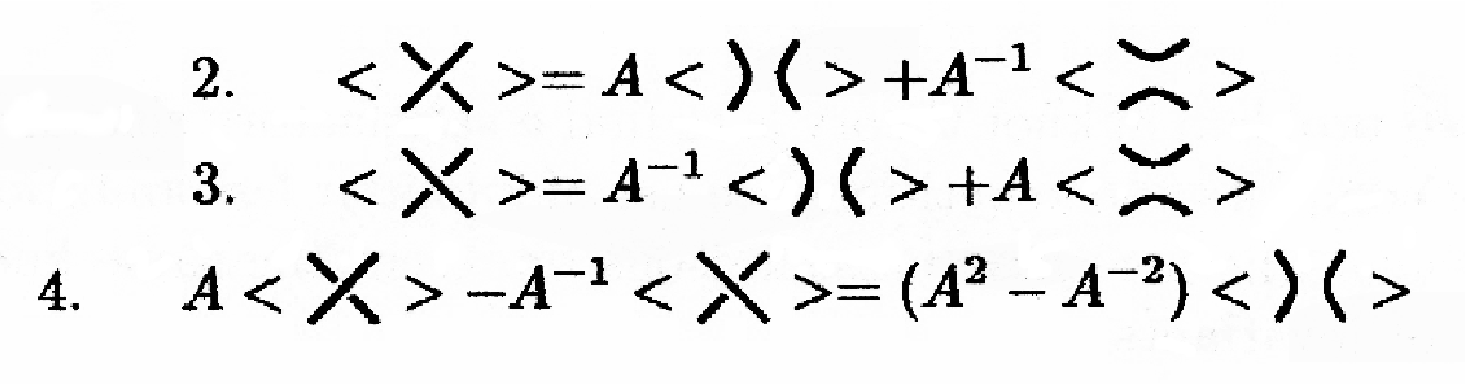}
\end{figure}
\begin{figure}[htp]
\centering
\leavevmode
\includegraphics[width=11.0cm,height=6.0cm]{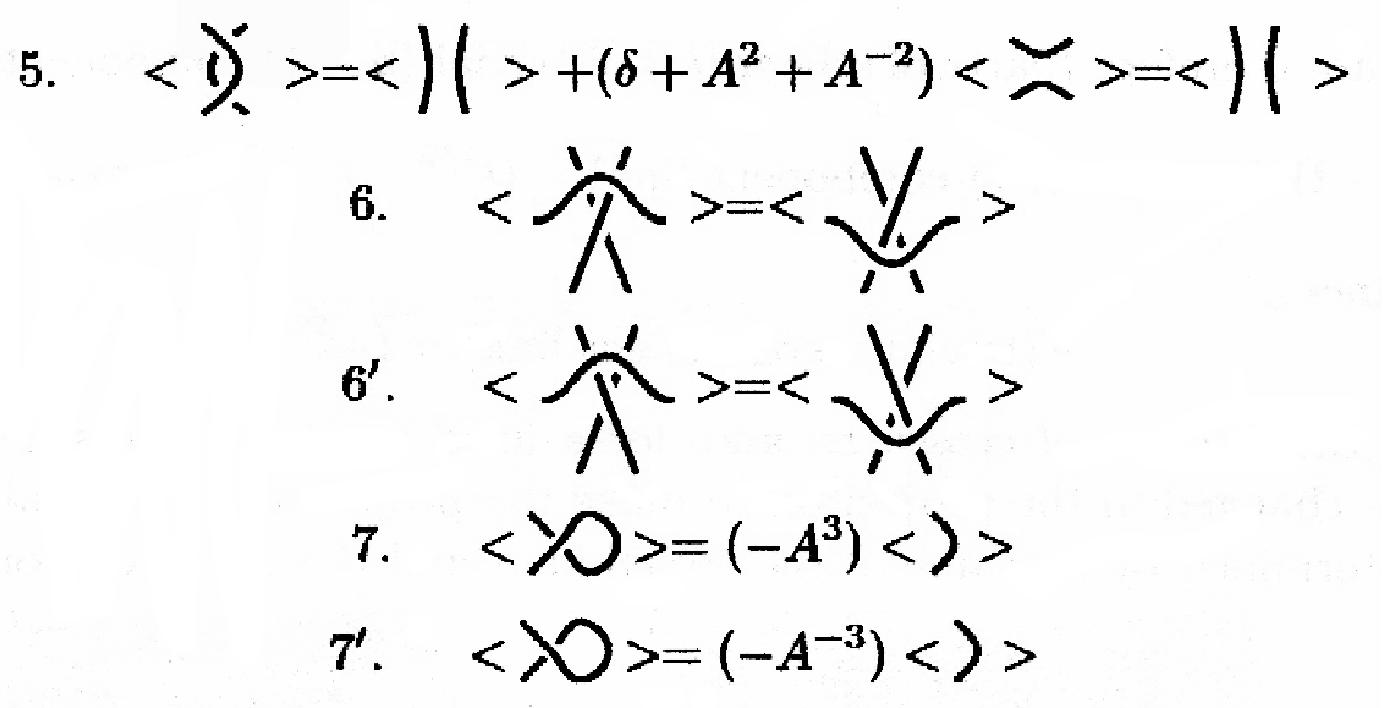}
\end{figure}
8. The Kauffman bracket of the mirror image of a diagram {\it D} is obtained from $<D>$ by exchanging {\it A} and $A^{-1}$.

For example, for a trefoil diagram we have:

\begin{figure}[htp]
\centering
\leavevmode
\includegraphics[width=8cm]{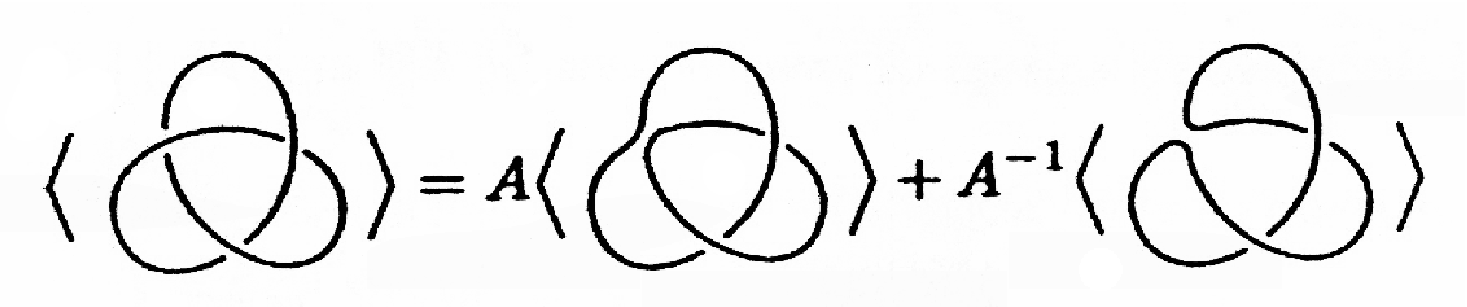}
\end{figure}
\newpage
\begin{figure}[htp]
\centering
\leavevmode
\includegraphics[width=12cm]{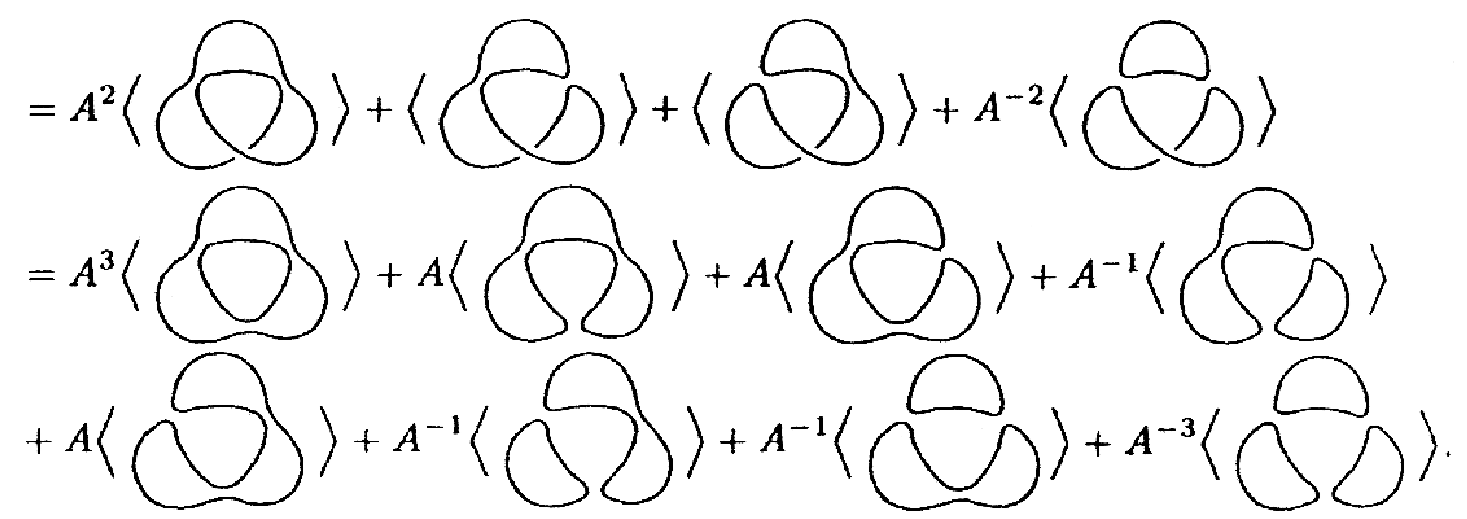}
\end{figure}
In general, for a link diagram {\it D} with {\it k} crossings, we obtain a linear sum of $2^{k}$ diagrams without crossings. Note that 
the linear sum is obtained independently of the order of expansion of the {\it k} crossings. Further, since a diagram without crossings 
is the disjoint union of loops, we obtain the value of its bracket recursively with the above {\bf 1.} property. The bracket of a diagram 
consisting of {\it l} disjoint loops has the value $(-A^{2}-A^{-2})^{l}$. Hence, we obtain $<D>$ from the above linear sum of $2^{k}$ 
diagrams by replacing each diagram, say, {\it l} loops, by $(-A^{2}-A^{-2})^{l}$. For example, for the trefoil knot diagram, we have 
that:

\begin{equation}
\begin{array}{lll}
 <\righttrefoil>=A^{3}(-A^{2}-A^{-2})^{2}+A(-A^{2}-A^{-2})+A(-A^{2}-A^{-2})+A^{-1}(-A^{2}-A^{-2})^{2}\\
+A(-A^{2}-A^{-2})+A^{-1}(-A^{2}-A^{-2})^{2}+A^{-1}(-A^{2}-A^{-2})^{2}+A^{-3}(-A^{2}-A^{-2})^{3} =\\
=(-A^{2}-A^{-2})(-A^{5}-A^{-3}+A^{-7})\; .
\end{array}
\label{eq_13}
\end{equation}

For an oriented diagram {\it D} we define the {\it writhe} of {\it D} by $w(D)$ = (the number of positive crossings of {\it D}) - (the 
number of negative crossings of {\it D}).

Modifying the Kauffman bracket with the writhe we obtain an isotopy invariant of oriented links as follows.

{\bf Theorem:} Let {\it L} be an oriented link, and {\it D} an oriented diagram of {\it L}. Then, $-A^{-3w(D)}<D>$ is invariant under the 
Reidemeister moves, where $<D>$ is the Kauffman bracket of {\it D} with its orientations forgotten. In particular, it is an isotopy 
invariant of {\it L}.

{\bf Theorem:} The {\it Jones polynomial} $V(L)$ of an oriented link {\it L} is the Laurent polynomial of {\bf Z}[$t^{1/2},t^{-1/2}$] 
defined from an oriented diagram {\it D} of {\it L} by $$V(L)=(-A)^{-3w(D)}<D>_{A^{-2}=t^{1/2}}\;.$$ {\it V} is an invariant of oriented 
links. It is the unique invariant of oriented links that satisfies:

1. {\it V}(trivial knot) = 1

2. the skein relation: $$t^{-1}V(\slashoverback)-t\;V(\backoverslash)=(t^{1/2}-t^{-1/2})V(\smoothing)\; ,$$
\noindent
when forgeting the orientation. The proof of this theorem can be found in \cite{lescop}.

For the right-hand trefoil knot ($\righttrefoil$) we can obtain from above:

\begin{equation}
V(t)=(-A)^{-9}(-A^{5}-A^{-3}+A^{-7})_{A^{2}=t^{-1/2}}=t+t^{3}-t^{4}=t(1+t^{2}-t^{3})\; .
\label{eq_14}
\end{equation}
\noindent
where $w(D)=3$, while for the left-hand trefoil ($\lefttrefoil$): $V(t)=t^{-1}+t^{-3}-t^{-4}=t^{-1}(1+t^{-2}-t^{-3})$.

{\bf Application:} In our simulation we will use the Jones polynomial for the 3-dimensional trefoil knot (black hole self-similarity level)
 seen as a tube collapsed to a string. We will have the liberty to work with the right-hand or with the left-hand trefoil where, in the 
formula (\ref{eq_14}) for the right-hand trefoil or in the one for the left-hand trefoil, $t=\textnormal{arctg}
\left( \displaystyle \frac{t_{k}}{\phi_{k}}\right)$ as for the matter (string; helicoid) representation. The three solutions of $V(t_{k},
\phi_{k})=0$ will give the spacetime position $(t_{k},\phi_{k})$ of the three double points on the knot diagram, seen as before-time or 
time matter bubbles. The fourth solution, $t=0$ will be for the empty spacetime in which these bubbles manifest themselves.

The results obtained on Jones polynomial basis (particle interpretation) must be the same as the one using a modified Kauffman 
polynomial in catenoidal (wave) representation, as we will see in another subsection.

\section{The HOMFLY Polynomial \cite{vogel}}

The HOMFLY polynomial was found by several authors just after the discovery of the Jones polynomial \cite{freyd,prz}.

Consider now the lie algebra $L=sl_{n}$ of $n\times n$ matrices with zero trace. This Lie algebra is quadratic by taking the trace of 
the direct product as bilinear form. The standard representation {\it E} is {\it n}-dimensional. The module {\it E} is a {\it L}-module. 
So we get an invariant of framed links.

{\bf Theorem:} Let $K\mapsto P(K)$ be the invariant of framed links induced by the quadratic Lie algebra $sl_{n}=sl(E)$ equipped with 
the standard representation {\it E}. Set $\alpha=\textnormal{exp}[t/(2n)]$, $\beta=\textnormal{exp}[nt/2]$ and $z=\textnormal{exp}[t/2] 
-\textnormal{exp}[-t/2]$. Then this invariant satisfies the following properties:
\begin{itemize}
\item for every framed link {\it K}, $P(K)$ belongs to $k[[t]]$
\item {\it P} is multiplicative with respect to the disjoint union
\item $P(\bigcirc)=1$ and the invariant of the trivial banded knot $\delta$ is:
\[
F(\delta)=\frac{\beta-\beta^{-1}}{z}
\]
\item if {\it K'} is obtained from a banded link {\it K} by a positive twist, one has $P(K')=\beta \alpha^{-1}P(K)$
\item if $K_{+}$, $K_{-}$ and $K_{0}$ are obtained from a singular framed link by the three standard modifications, one has:
\[
\alpha^{-1} P(K_{+})-\alpha\; P(K_{-})=z\; P(K_{0})
\]
\end{itemize}

We deduce from the HOMFLY skein relation the following property:

\[
P(L\cup \bigcirc)=\frac{\alpha^{-1}-\alpha}{z}\; P(L)
\]

Hence the value of {\it P} on the {\it m} components trivial link must be $\left(\displaystyle \frac{\alpha^{-1}-\alpha}{z}\right)^{m-1}$. 
Note that $P(L)$ belongs to the subring $\Lambda=\textnormal{{\bf Z}}[\alpha^{\pm 1}]\left[\displaystyle \frac{\alpha^{-1}-\alpha}{z} 
\right]$.

{\bf Proposition:} For any integer $n\geq 2$, there exists a unique map $L\mapsto F_{n}(L)$ from isotopy classes of oriented links to 
{\bf Z}[$s^{\pm 1}$] which satisfies the relations in the above theorem, with $z=s-s^{-1}$ and $\alpha=s^{-n}$.

If {\it s} from HOMFLY skein relation is equal with $t^{1/2}$ from the Jones skew relation, then the Jones polynomial is just the $n=2$ 
specialization of the HOMFLY polynomial.

Considering now trivalent plane graphs (Feynman diagrams; loops are allowed). If {\it G} is such a graph, then we denote by 
$\mathcal{V}_{G}$ the set of trivalent vertices in {\it G}, and by $\mathcal{E}_{G}$ the set of connected components of {\it 
G} $\backslash\mathcal{V}_{G}$ (the edges). A 2-flow for such a graph is an orientation of the edges, together with a map $f:\mathcal
{E}_{G} \rightarrow \{1,2\}$ which defines the integral cycle.

\subsection{The SU(n) specialization of HOMFLY polynomial}

We consider a link diagram {\it D} and we denote by $\mathcal{C}_{D}$ the set of its crossings. We define $<D>_{n}$ by:

\begin{equation}
<D>_{n}=\sum_{X\subset \mathcal{C}_{D}}s^{\epsilon(X)}(-1)^{\sharp(\mathcal{C}_{D}\backslash X)}<D_{X}>_{n}\; .
\label{eq_15}
\end{equation}

Here, $\epsilon(X)=\sum_{c\in X}\epsilon(c)$, with $\epsilon(\overcrossing)=1$ and $\epsilon(\undercrossing)=-1$.

The graph with 2-flow $D_{X}$ is obtained by replacing each crossing {\it c} by two parallel edges if {\it c} is in {\it X}, and by a 
double edge otherwise.

{\bf Theorem:} $<D>_{n}$ defines an invariant of oriented framed links.

We give a sketchy proof.

Let {\it V} be the free $\Lambda$-module with basis ($v_{i}$)$_{i\in \mathcal{N}}$. We define an endomorphism {\it E} of $V\otimes V$ by 
the following matrix:

\[
E_{ij}^{kl}=\left\{ \begin{array}{llll}
s\;\;\;\;\;\; \textnormal{if}\; i<j\;,\; k=i\; \textnormal{and}\; l=j\; ; \\
s^{-1} \;\;\; \textnormal{if}\; i>j\;,\; k=i\; \textnormal{and}\; l=j\; ; \\
1\;\;\;\;\;\; \textnormal{if}\; i\neq j\;,\; k=j\; \textnormal{and}\; l=i\; ; \\
0\;\;\;\;\;\; \textnormal{else}\; .
\end{array} \right.
\]

{\bf Proposition A:} a) $E^{2}=(s+s^{-1})E$.

b) $(E\otimes Id_{V})\circ (Id_{V}\otimes E)\circ (E\otimes Id_{V})-(E\otimes Id_{V})= (Id_{V}\otimes E)\circ (E\otimes Id_{V})\circ 
(Id_{V}\otimes E)-(Id_{V}\otimes E)$.

Let {\it R} be defined by $R=s\;Id_{V\otimes V}-E$.

{\bf Proposition B:} a) $R^{-1}=s^{-1}Id_{V\otimes V}-E$,

b) $R-R^{-1}=(s-s^{-1})Id_{V\otimes V}$.

The key point is that {\it R} is the Yang-Baxter operator.

{\bf Theorem (Yang-Baxter equation):} $(R\otimes Id_{V})\circ (Id_{V}\otimes R)\circ (R\otimes Id_{V})=(Id_{V}\otimes R)\circ 
(R\otimes Id_{V})\circ (Id_{V}\otimes R)$.

The statement above is the Yang-Baxter equation and prove that $<D>_{n}$ satisfies the Reidemeister relation III (with coherent 
orientations). For the Reidemeister relation II, we have to consider two cases depending on the orientations of the strings. The first 
one follows from the proposition B. The second one uses the graphical lemmas below. Here the quantum integer [{\it m}] is defined by 
\cite{kassel,kirillov,turaev2,lusztig}:

\begin{equation}
[m]=\frac{s^{m}-s^{-m}}{s-s^{-1}}\; .
\label{eq_16}
\end{equation}

{\bf Lemma A:}

\begin{figure}[h]
\leavevmode
\centering
\includegraphics[width=8.0cm,height=3cm]{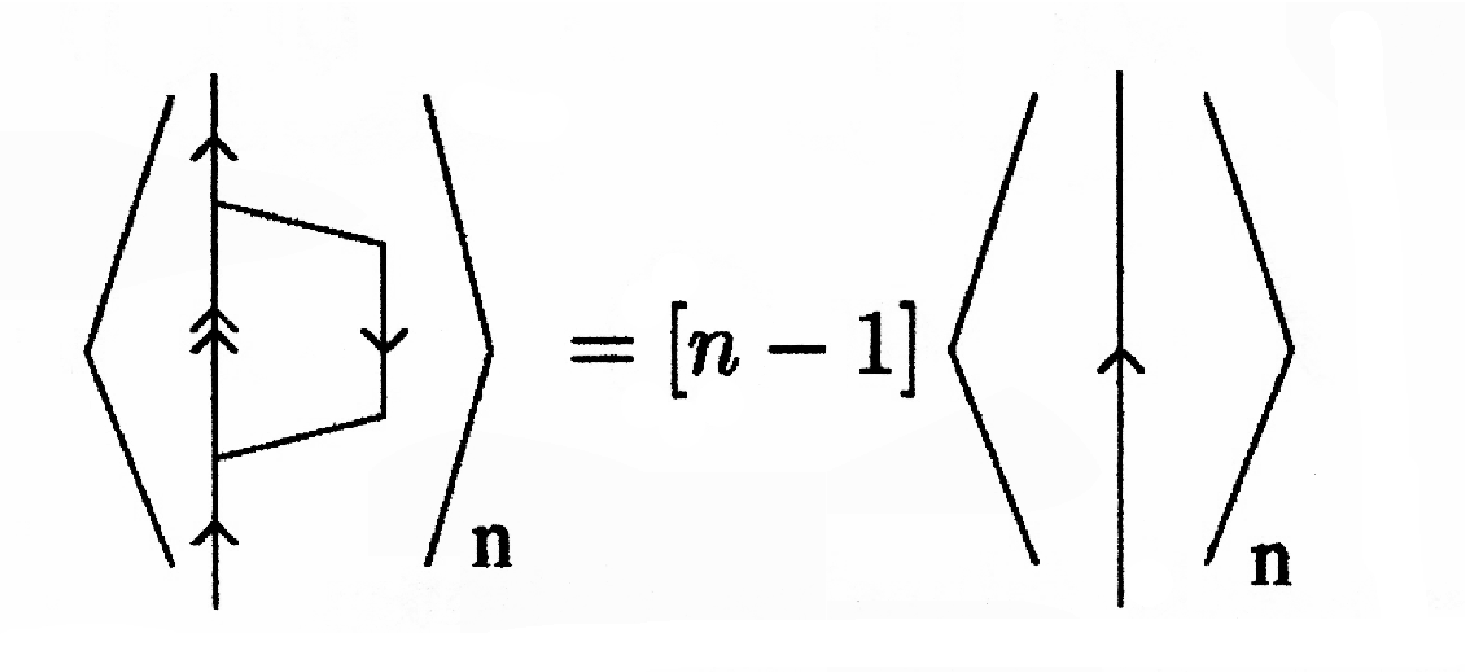}
\end{figure} 

{\bf Lemma B:}

\begin{figure}[h]
\leavevmode
\centering
\includegraphics[width=14.0cm,height=3cm]{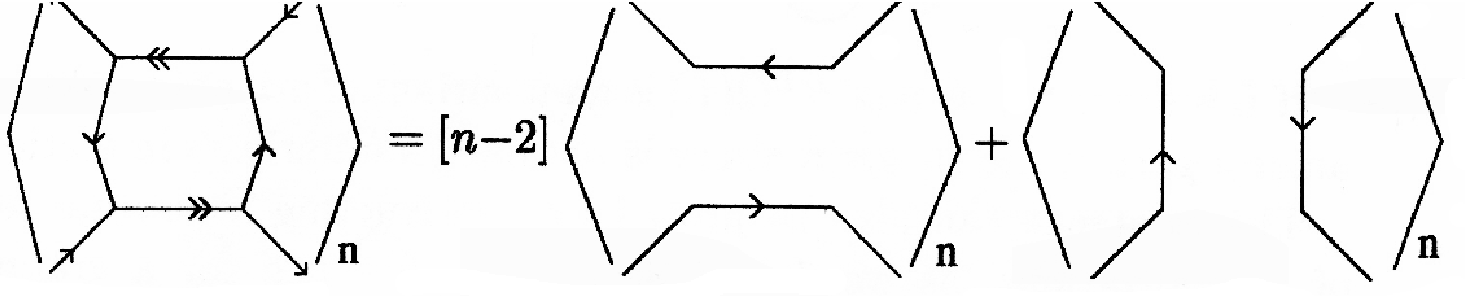}
\end{figure}

Hence we can get an invariant $<L>_{n}$ of oriented framed links {\it L}. Moreover \cite{blanchet}:

\begin{equation}
F_{n}(L)=\frac{s-s^{-1}}{s^{n}-s^{-n}}s^{-n\; w}<L>_{n}\; .
\label{eq_17}
\end{equation}

Here {\it w} is the {\it writhe} of the framed link {\it L} (the sum of the signs of the crossings in a diagram).

{\bf The Framed HOMFLY polynomial}. Here {\it k} is an integral domain containing the invertible elements {\it a}, $\alpha$, {\it s}; we 
suppose moreover that $s-s^{-1}$ is invertible in {\it k}. Let $z=s-s^{-1}$. Then we can show that, for a framed link {\it L}, 
the following is an invariant:

\begin{equation}
<L>=\frac{\alpha^{-1}-\alpha}{s-s^{-1}}(a\;\alpha^{-1})^{w}P(L)\; ,
\label{eq_18}
\end{equation}

{\bf Application:} With the HOMFLY polynomial as a generalization of the Jones polynomial we can construct banded trefoil knots which we 
will use for the nucleon DEUS self-similarity level having the $V(t_{k},\phi_{k})=0$ from (\ref{eq_14}). An important 
check will be to verify if (\ref{eq_16}) can be solved for an integer value of {\it m}, where $s=t^{1/2}=\left[ \textnormal{arctg}
\left( \displaystyle \frac{t_{k}}{\phi_{k}}\right)\right]^{1/2}$.

\section{The Kauffman Polynomial \cite{vogel}}

Consider a {\it n}-dimensional vector space {\it E} equipped with a non singular symmetric form {\it b}. Let $L=o(E)$ be the Lie algebra 
of antisymmetric endomorphism of {\it E}. The trace of the product induces a form $<\;,\;>$ on {\it L} and {\it L} is a quadratic Lie 
algebra. The module {\it E} is a {\it L}-module. So we get an invariant of banded links.

{\bf Theorem:} Let $K\mapsto F(K)$ be the invariant of banded links induced by the quadratic Lie algebra $o(E)$ equipped with the standard 
representation {\it E}. Set $\alpha = \textnormal{exp}[(n-1)t/4]$ and $z=2\textnormal{sinh}(t/4)$. This map is from isotopy classes of 
unoriented framed links in the 3-sphere to {\bf Z}[$\alpha^{\pm 1},z^{\pm1}$]. Then this invariant satisfies the following properties:

\begin{itemize}
\item for every banded link {\it K}, $F(K)$ belongs to $k[[t]]$
\item {\it F} is multiplicative with respect to the disjoint union operation
\item $F(\bigcirc)=1$ and the invariant of the trivial banded knot $\delta$ is:
\[
F(\delta)=1+\frac{\alpha-\alpha^{-1}}{z}
\]
\item if {\it K'} is obtained from a banded link {\it K} by a positive twist, one has $F(K')=\alpha F(K)$
\item the invariant {\it F} satisfies the following skein relation:

\begin{figure}[htp]
\centering
\leavevmode
\includegraphics[width=12.0cm,height=2cm]{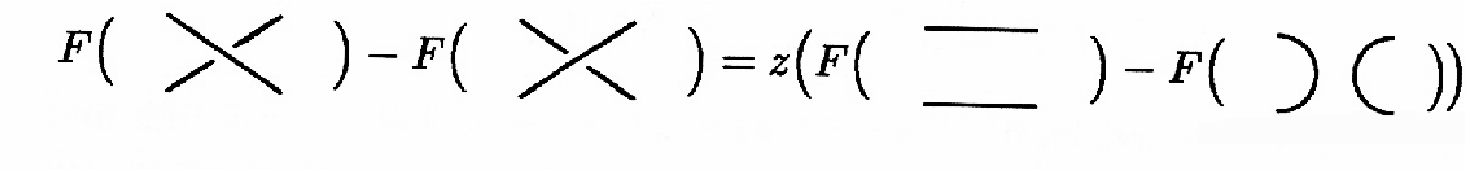}
\end{figure} 
\end{itemize}

We can deduce the following relation:

\[
F(L\cup \bigcirc)=\left(1+\frac{\alpha-\alpha^{-1}}{z} \right)\; F(L)\; .
\]

If {\it L} is oriented, then $\alpha^{-w}F(L)$ is an invariant of oriented links. One can construct three discrete series of 
specializations, using the Yang-Baxter operators associated with the deformations of the fundamental representations of the classical 
simple Lie algebras $so(2n+1)$, $sp(2n)$ and $so(2n)$ \cite{turaev,turaev1}.

{\bf Application:} If in the skein for this polynomial is $F(\backoverslash)-F(\slashoverback)=z_{1}$, with $z_{1}=2\textnormal{sinh}\left( 
\displaystyle\frac{\sqrt{t^{2}+\phi^{2}}}{A}\right)$ and $F(\backoverslash)+F(\slashoverback)=z_{2}$ with $z_{2}=2\textnormal{cosh}\left( 
\displaystyle\frac{\sqrt{t^{2}+\phi^{2}}}{A}\right)$, then $F(\backoverslash)=\textnormal{exp}\left( \displaystyle\frac{\sqrt{t^{2}+
\phi^{2}}}{A}\right)$ and $F(\slashoverback)=\textnormal{exp}\left( -\;\displaystyle\frac{\sqrt{t^{2}+\phi^{2}}}{A}\right)$. The 
representation of a link or knot having these skein relations will be a tube knot. For a trefoil tube knot the representation must be 
equivalent with the image given by the Jones polynomial (catenoid evolved to a cylinder as part of a tube that collapses to a string). We 
will use the above polynomial to complete and check (we know that $t/\phi=-\phi_{k}/t_{k}$) the description through the Jones polynomial 
of the black hole self-similar level.

\section{Hopf Algebras}

We suppose here that {\it k} is a field and {\it A} is an algebra over {\it k}, with unit $1_{A}$.

{\bf Definition:} a) A bialgebra ($A,\Delta,\eta$) is an algebra {\it A} equipped with two algebra morphisms:

\[
\Delta:A\rightarrow A\otimes A
\]
\[
\eta:A\rightarrow k
\]
\noindent
such that $\Delta$ (the comultiplication) is coassociative, and $\eta$ (the augmentation) is a counit.

b) A Hopf algebra ($A,\Delta,\eta,S$) is a bialgebra with an antipode {\it S} (an endomorphism {\it S} of {\it A}), such that
\[
\forall x\in A\;\; (m\circ (S\otimes Id_{A})\circ \Delta)(x)=(m\circ (Id_{A}\otimes S)\circ \Delta)(x)=\eta(x)1_{A}\; .
\]

As a fundamental example, we have the universal enveloping algebra $U_{\mathfrak{g}}$ of a Lie algebra $\mathfrak{g}$.

Consider the category of finite dimensional representations of a Hopf algebra, denoted by {\it Rep}({\it A}). This category has a tensor 
product, defined by using the coproduct on A, making {\it Rep}({\it A}) into a linear monoidal category. Moreover, by using the antipode, 
we can provide {\it Rep}({\it A}) with a duality. In order to construct a functor from the tangle category to the category 
{\it Rep}({\it A}), we need, on the category {\it Rep}({\it A}) of finite dimensional representations of {\it A}, a braiding and a twist, 
compatible with the duality, we will then obtain a {\it ribbon category}.

\bigskip

If the above $\mathfrak{g}=sl_{2}$ we can define for a knot the colored Jones polynomial \cite{thang}. But since in this study we intend 
to analize the behavior of knots under {\it SU(n)} specialization, we will skip the description of this case, but we will return to 
another representation of the colored Jones polynomial later.

\section{The Alexander-Conway polynomial \cite{masbaum3}}

Let {\it L} be an oriented link in $S^{3}$ with {\it m} (numbered) components. Its Alexander-Conway:

\begin{equation}
\nabla_{L}(z)=\sum_{i\geq 0}c_{i}(L)z^{i}\in \textnormal{{\bf Z}}[z]
\label{eq_19}
\end{equation}
\noindent
is one of the most studied classical isotopy invariants of links. It can be defined in various ways. For example, if {\it V} is a Seifert 
matrix for {\it L}, then:

\begin{equation}
\nabla_{L}(z)=\textnormal{det}(tV-t^{-1}V^{T})
\label{eq_20}
\end{equation}
\noindent
where $z=t-t^{-1}$. Another definition is via the skein relation:

\begin{equation}
\nabla_{L_{+}}-\nabla_{L_{-}}=z\nabla_{L_{0}}\; ,
\label{eq_21}
\end{equation}
\noindent
with $L_{+}$, $L_{-}$ and $L_{0}$ previously defined.

The Alexander-Conway polynomial is uniquely determined by the skein relation (\ref{eq_21}) and the initial conditions:

\begin{equation}
\nabla_{U_{m}}=\left\{ \begin{array}{ll}
1\;\;\;\;\;\; \textnormal{if}\; m=1 \\
0\;\;\;\;\;\; \textnormal{if}\; m\geq 2
\end{array} \right.
\label{eq_22}
\end{equation}
\noindent
where $U_{m}$ is the trivial link with {\it m} components.

Hosokawa \cite{hoso}, Hartley \cite{hartley} and Hoste \cite{hoste} showed that the coefficients $c_{i}(L)$ of $\nabla_{L}$ for an {\it 
m}-component link {\it L} vanish when $i\leq m-2$ and that the coefficient $c_{m-1}(L)$ depends only on the linking numbers $\ell
_{ij}(L)$ between the {\it i}th and the {\it j}th components of {\it L}. Namely,

\begin{equation}
c_{m-1}(L)=\textnormal{det}\; \Lambda^{(p)}\; ,
\label{eq_23}
\end{equation}
\noindent
where $\Lambda=(\lambda_{ij})$ is the matrix formed by linking numbers

\begin{equation}
\lambda_{ij}=\left\{ \begin{array}{ll}
-\ell_{ij}(L)\;\;\;\;\;\;\;\;\;\;\; \textnormal{if}\; i\neq j \\
\sum_{k\neq i}\ell_{ik}(L)\;\;\;\;\;\; \textnormal{if}\; i=j
\end{array} \right.
\label{eq_24}
\end{equation}
\noindent
and $\Lambda^{(p)}$ denotes the matrix obtained by removing from $\Lambda$ the {\it p}th row and column (it is easy to see that $\Lambda
^{(p)}$ does not depend on {\it p}).

Formula (\ref{eq_23}) can be proved using the Seifert matrix definition (\ref{eq_20}) of $\nabla_{L}$. We will not give the proof here, 
but let us indicate how linking numbers come in from this point of view. Let $\Sigma$ be a Seifert surface for {\it L}. The key point is 
that the Seifert form restricted to $H_{1}(\partial\Sigma;\textnormal{{\bf Z}})\subset H_{1}(\Sigma;\textnormal{{\bf Z}})$ is just given 
by the linking numbers $\ell_{ij}$. In particular, for an appropriate choice of basis for $ H_{1}(\Sigma;\textnormal{{\bf Z}})$, the 
Seifert matrix {\it V} contains the matrix $\Lambda^{(p)}$ as a submatrix, which then leads to formula (\ref{eq_23}).

Hartley and Hoste also gave a second expression for $c_{m-1}(L)$ as a sum over trees:

\begin{equation}
c_{m-1}(L)=\sum_{T}\prod_{\{i,j\}\in edges(T)}\; \ell_{ij}(L)\; ,
\label{eq_25}
\end{equation}
\noindent
where {\it T} runs through the spanning trees in the complete graph $K_{m}$. (The complete graph $K_{m}$ has vertices {1,2,...,m}, and 
one and only one edge for every unordered pair {$i,j$} of distinct vertices.)

For example, if $m=2$ then $c_{1}(L)=\ell_{12}(L)$, corresponding to the only spanning tree in:

\begin{figure}[h]
\centering
\includegraphics[width=2cm]{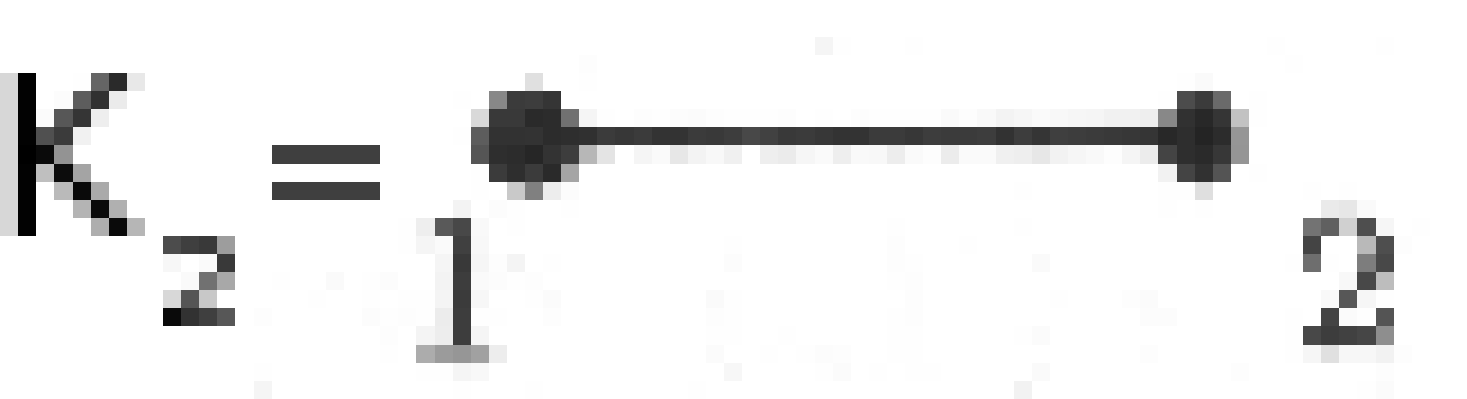}
\end{figure}

If $m=3$, then:

\begin{equation}
c_{2}(L)=\ell_{12}(L)\ell_{23}(L)+\ell_{23}(L)\ell_{13}(L)+\ell_{13}(L)\ell_{12}(L)\; ,
\label{eq_26}
\end{equation}
\noindent
corresponding to the three spanning trees of $K_{3}$ from the following figure:

\begin{figure}[h]
\centering
\leavevmode
\includegraphics[width=9cm]{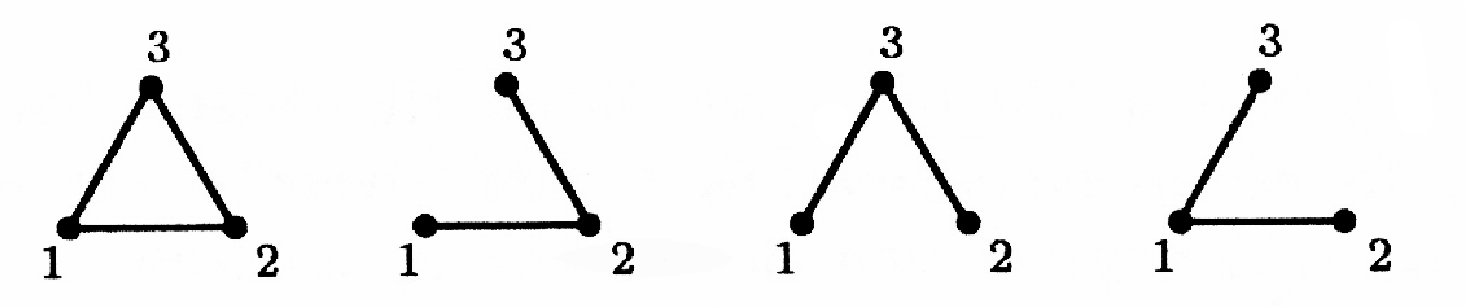}
\end{figure}

{\bf Application:} We will use the Alexander-Conway polynomial in our simulation for checking the validity of the three double point 
solutions of the other polynomials used and applied for the trefoil knots. The distance between two solutions will be the linking 
number $\ell_{ij}$ up to a constant. We must have satisfied the expression (\ref{eq_25}) where we have to have in the case of one edge 
($m=2$) $c_{1}(L)=\ell_{12}$ and for all three edges the (\ref{eq_26}) formulation. The check of the curvature of the plane in which the 
diagram lives will be done with the help of the angles which must be consistent with each other for the computed $\ell_{ij}$ values. Then 
we will be able to compute the (\ref{eq_19}) topological invariant for our particular case:

\begin{equation}
\nabla_{L}(z)=c_{1}(L)\; z^{(1)}+c_{2}(L)\; z^{(2)}\; ,
\label{eq_27a}
\end{equation}
\noindent
with $z^{(1)}=\textnormal{cosh}^{\beta}\left( \displaystyle\frac{\sqrt{t^{2}+\phi^{2}}}{A}\right)$ (where $\beta=\pm 1$), as for a 
catenoidal bridge evolved to cylinder and then collapsed, and $z^{(2)}=\textnormal{arctg}\left( \displaystyle \frac{t_{k}}{\phi_{k}}
\right)$ for the helicoid matter representation of the knot. Then, for each triangle diagram box (composing the 2-dimensional trefoil knot) 
we must have the same $\nabla_{L}(z)$ invariant, keeping in mind that two neighbor boxes have one common edge which will enter as a fixed 
$\ell$ in the neighbor box invariant computation.

Unfortunately, because the trivial solution ($t=0$; or the center of the ``Y''-shaped graph) of the Jones polynomial invariant may not 
be in the same plane as the other three solutions (vertices), and because we are not able to set an unique relation between the Milnor 
numbers and $\ell_{ijk}$ (also, $\mu_{ijk}=y_{ijk}$ being too particular in order to be useful) the following two subsections, related to 
algebraically split links, will not be used in our application.

\subsection{Algebraically split links and Levine's formula}

If the link is {\it algebraically split} (all linking numbers $\ell_{ij}$ vanish) then not only $c_{m-1}(L)=0$, but, as was proved by 
Traldi \cite{traldi1,traldi2} and \cite{levine}, the next $m-2$ coefficients of $\nabla_{L}$ also vanish:

\[
c_{m-1}(L)=c_{m}(L)=...=c_{2m-3}(L)=0
\]

For algebraically split oriented links, there exist well-defined integer-valued isotopy invariants $\mu_{ijk}(L)$ called the {\it Milnor 
triple linking numbers}. These invariants generalize ordinary linking numbers, but unlike $\ell_{ij}$, the triple linking numbers are 
antisymmetric with respect to their indices, $\mu_{ijk}(L)=-\mu_{jik}(L)=\mu_{jki}(L)$. Thus, for an algebraically split link {\it L} 
with {\it m} components, we have 
$\left(\begin{array}{ll}
m \\
3
\end{array}\right)$ triple linking numbers $\mu_{ijk}(L)$ corresponding to the different 3-component sublinks 
of {\it L}.

Levine \cite{levine} found an expression for the coefficient $c_{2m-2}$ of $\nabla_{L}$ for an algebraically split {\it m}-component 
link in terms of triple Milnor numbers:

\begin{equation}
c_{2m-2}(L)=\textnormal{det}\; \Lambda^{(p)}\; ,
\label{eq_27}
\end{equation}
\noindent
where $\Lambda^{(p)}=(\lambda_{ij})$ is an $m\times m$ skew-symmetric matrix with entries:

\begin{equation}
\lambda_{ij}=\sum_{k}\mu_{ijk}(L)\; ,
\label{eq_28}
\end{equation}
\noindent
and $\Lambda^{(p)}$, as before, is the result of removing the {\it p}th row and column.

For example, if $m=3$, we have:

\begin{equation}
\Lambda=\left(\begin{array}{lll}
0 & \mu_{123}(L) & \mu_{132}(L)\\
\mu_{213}(L) & 0 & \mu_{231}(L)\\
\mu_{312}(L) & \mu_{321}(L) & 0
\end{array}\right)
\label{eq_29}
\end{equation}
\noindent
and:

\begin{equation}
c_{4}(L)=\textnormal{det}\; \Lambda^{(3)}=-\mu_{123}(L)\mu_{213}(L)=\mu_{123}(L)^{2}\; ,
\label{eq_30}
\end{equation}

\subsection{The Pfaffian-tree polynomial $\mathcal{P}_m$}

Formula (\ref{eq_27}) is similar to the first determinantal expression (\ref{eq_23}). One of the main results of \cite{masbaum1,masbaum2} 
is that there is an analog of the tree sum formula (\ref{eq_25}) for algebraically split links. To state this result, we need to 
introduce another tree-generating polynomial analogous to the Kirchhoff polynomial.

Namely, instead of usual graphs whose edges can be thought of as segments joining pairs of points, we consider 3-{\it graphs} whose 
edges have three (distinct) vertices and can be visualized as triangles or ``Y''-shaped objects with the three vertices at their endpoints.

\begin{figure}[h]
\centering
\leavevmode
\includegraphics[width=7cm]{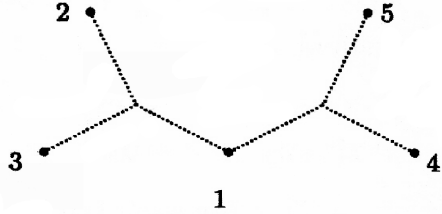}
\caption{A spanning tree in the complete 3-graph $\Gamma_{5}$. It has two edges \{1,2,3\} and \{1,4,5\}, and contributes the term $y_{123}
y_{145}$ to $\mathcal{P}_{5}$.}
\label{tree}
\end{figure}

The notion of spanning trees on a 3-graph is defined in the natural way. A sub-3-graph {\it T} of a 3-graph {\it G} is {\it spanning} if 
its vertex set equals that of {\it G}, and it is a {\it tree} if its topological realization (the 1-complex obtained by gluing together 
Y shaped objects corresponding to the edges of {\it T}) is a tree (it is connected and simply connected). See Figure \ref{tree} for an 
example.

Similarly to the variables $x_{ij}$ of $\mathcal{D}_{m}$, for each triple of distinct numbers $i,j,k\in \{1,2,...,m\}$ we introduce 
variables $y_{ijk}$ antisymmetric in $i,j,k$: $y_{ijk}=-y_{jik}=y_{jki}$ and $y_{iij}=0$.

These variables correspond to edges $\{i,j,k\}$ of the {\it complete 3-graph} $\Gamma_{m}$ with vertex set $\{1,2,...,m\}$. 

As in the case of ordinary graphs, the correspondence:

\[
\textnormal{variable}\;\; y_{ijk}\;\; \mapsto\;\; \textnormal{edge}\;\; \{i,j,k\}\; \textnormal{of}\; \Gamma_{m}
\]

assigns to each monomial in $y_{ijk}$ a sub-3-graph of $\Gamma_{m}$.

The generating function of spanning trees in the complete 3-graph $\Gamma_{m}$ is called the {\it Pfaffian-tree polynomial} 
$\mathcal{P}_{m}$ in \cite{masbaum1,masbaum2}. It is:

\[
\mathcal{P}_{m}=\sum_{T}y_{T}
\]
\noindent
where the sum is over all spanning trees {\it T} of $\Gamma_{m}$, and $y_{T}$ is, up to sign, just the product of the variables $y_{ijk}$ 
over the edges of {\it T}. Because of the antisymmetry of the $y_{ijk}$'s signs cannot be avoided here. In fact, the correspondence 
between monomials and sub-3-graphs of $\Gamma_{m}$ is not one-to-one and a sub-3-graph determines a monomial only up to sign. But these 
signs can be fixed unambiguously, although we won't explain this here.

If {\it m} is even, then one has $\mathcal{P}_{m}=0$, because there are no spanning trees in 3-graphs with even number of vertices. If 
{\it m} is odd, then $\mathcal{P}_{m}$ is a homogeneous polynomial of degree $\displaystyle\frac{m-1}{2}$ in the $y_{ijk}$'s. For example, 
one has $\mathcal{P}_{3}=y_{123}$ (the 3-graph $\Gamma_{3}$ with three vertices and one edge is itself a tree). If $m=5$, we have:

\begin{equation}
\mathcal{P}_{5}=y_{123}y_{145}-y_{124}y_{135}+y_{125}y_{134}\pm\; ...\; ,
\label{eq_31}
\end{equation}
\noindent
where the right-hand side is a sum of 15 similar terms corresponding to the 15 spanning trees of $\Gamma_{5}$. If we visualize the edges 
of $\Gamma_{m}$ as ``Y''-shaped objects, then the spanning tree corresponding to the first term of (\ref{eq_31}) will look like on Figure 
\ref{tree}.

We can now state one of the main results of \cite{masbaum1,masbaum2}:

{\bf Theorem:} Let {\it L} be an algebraically split oriented link with {\it m} components. Then:

\begin{equation}
c_{2m-2}(L)=(\mathcal{P}_{m}(\mu_{ijk}(L)))^{2}\; ,
\label{eq_32}
\end{equation}
\noindent
where $\mathcal{P}_{m}(\mu_{ijk}(L))$ means the result of evaluating the polynomial $\mathcal{P}_{m}$ at $y_{ijk}=\mu_{ijk}(L)$.

For $m=3$, we find the Cochran's formula (\ref{eq_30}), and for $m\geq 5$ the formula is from \cite{masbaum3}, which obtains that the 
first non-vanishing coefficient of $\nabla_{L}(z)$ for algebraically split links with 5 components is equal to:

\begin{equation}
c_{8}(L)=\mathcal{P}_{5}(\mu_{ijk}(L))^{2}=(\mu_{123}(L)\mu_{145}(L)-\mu_{124}(L)\mu_{135}(L)+\mu_{125}(L)\mu_{134}(L)\pm\; ...)^{2}\; ,
\label{eq_33}
\end{equation}
\noindent
where $\mathcal{P}_{5}(\mu_{ijk}(L))$ consists of 15 terms corresponding to the spanning trees of $\Gamma_{5}$.

{\bf Theorem:} The generating function of spanning trees on the complete 3-graph $\Gamma_{m}$ is given by:

\begin{equation}
\mathcal{P}_{m}=(-1)^{p-1}\textnormal{Pf}(\Lambda(\Gamma_{m})^{(p)})\; ,
\label{eq_34}
\end{equation}
\noindent
where $\Lambda(\Gamma_{m})$ is the $m\times m$ skew-symmetric matrix with entries $\Lambda(\Gamma_{m})_{ij}=\sum_{k}y_{ijk}$, and Pf 
denotes the Pfaffian.

Recall that the Pfaffian of a skew-symmetric matrix {\it A} is a polynomial in the coefficients of {\it A} such that:

\begin{equation}
(\textnormal{Pf}\; A)^{2}=\textnormal{det}\; A\; .
\label{eq_35}
\end{equation}

\section{Cobordisms \cite{dror1}}

Cobordisms are movies ``starring'' knots and links \cite{carter}. The main difficulty in showing that cobordisms induce maps of homology 
groups is to show that trivial movies induce trivial maps on homology.

{\bf Definition:} $Cob^{3}(0)$ is the category whose objects are smoothings (simple curves in the plane) and whose morphisms are 
cobordisms between such smoothings regarded up to boundary-preserving isotopies. Likewise, if {\it B} is a finite set of points on the 
circle (such as the boundary $\partial T$ of a tangle {\it T}, where the tangles are defined as knot pieces), then $Cob^{3}(B)$ is the 
category whose objects are smoothings with the boundary {\it B} and whose morphisms are cobordisms between such smoothings, regarded up to 
boundary-preserving isotopies. In either case, the composition of morphisms is given by placing one cobordism atop the other. We will use 
the notation $Cob^{3}$ as a generic reference either to $Cob^{3}(0)$ or to $Cob^{3}(B)$ for some {\it B}.

\subsection{The quotient $Cob^{3}_{/l}$ of $Cob^{3}$}

We mod out the morphisms of the category $Cob^{3}$ by the relations {\bf S}, {\bf T} and {\bf 4Tu} defined below and call the resulting 
quotient $Cob^{3}_{/l}$ (the $/l$ stands for ``modulo local relations'').

The {\bf S} relation says that whenever a cobordism contains a connected component which is a closed sphere (with no boundary), it is set 
equal to zero (we make all categories pre-additive, so that 0 always makes sense).

\begin{figure}[h]
\centering
\leavevmode
\includegraphics[width=2cm]{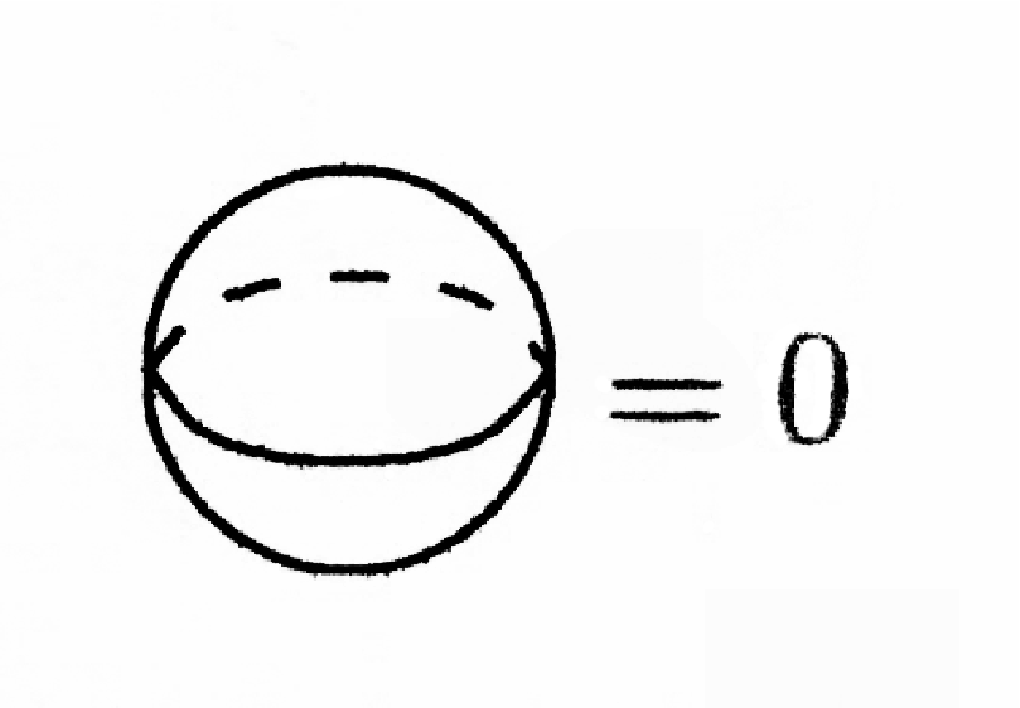}
\end{figure}

The {\bf T} relation says that whenever a cobordism contains a connected component which is a closed torus (with no boundary), that 
component may be dropped and replaced by a numerical factor of 2 (again, we make all categories pre-additive, so that multiplying a 
cobordism by a numerical factor makes sense).

\begin{figure}[h]
\centering
\leavevmode
\includegraphics[width=3cm]{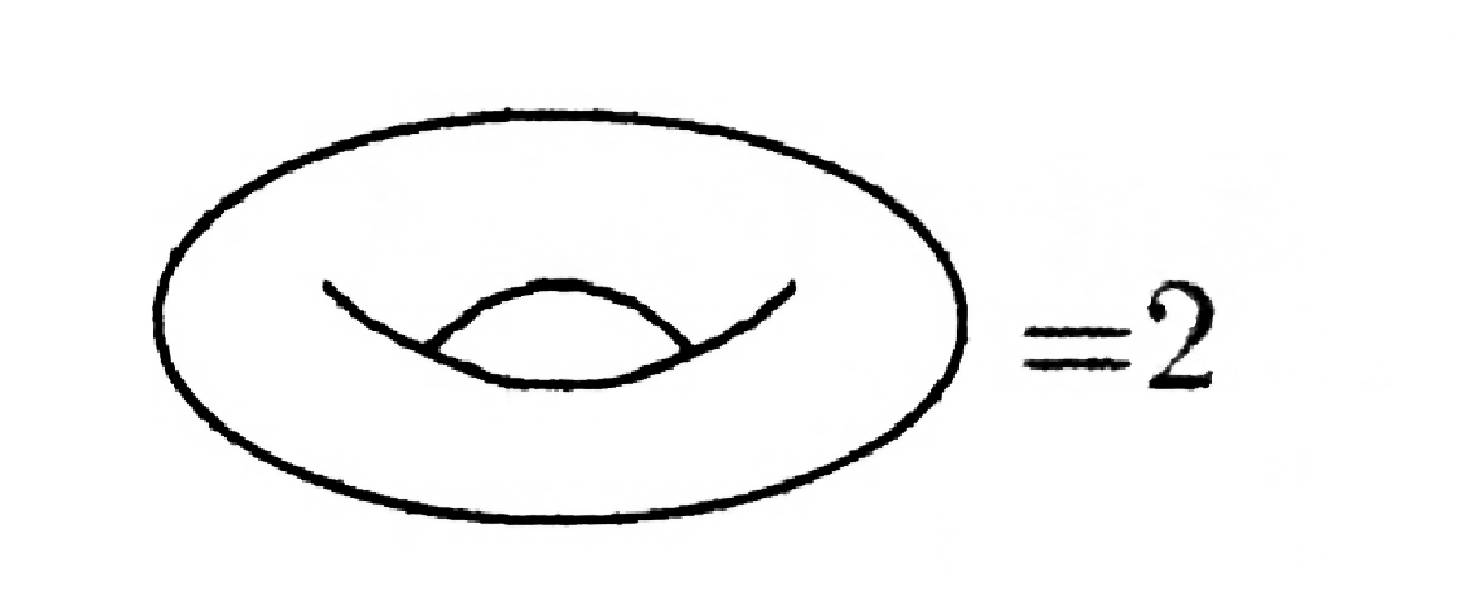}
\end{figure}

To understand {\bf 4Tu}, start from some given cobordism {\it C} and assume its intersection with a certain ball is the union of four 
disks $D_{1}$ through $D_{4}$ (these disks may well be on different connected components of {\it C}). Let $C_{ij}$ denote the result of 
removing $D_{i}$ and $D_{j}$ from {\it C} and replacing them by a tube that has the same boundary. The ``four tube'' relation {\bf 4Tu} 
assets that $C_{12}+C_{34}=C_{13}+C_{24}$.

\begin{figure}[h]
\centering
\leavevmode
\includegraphics[width=9cm]{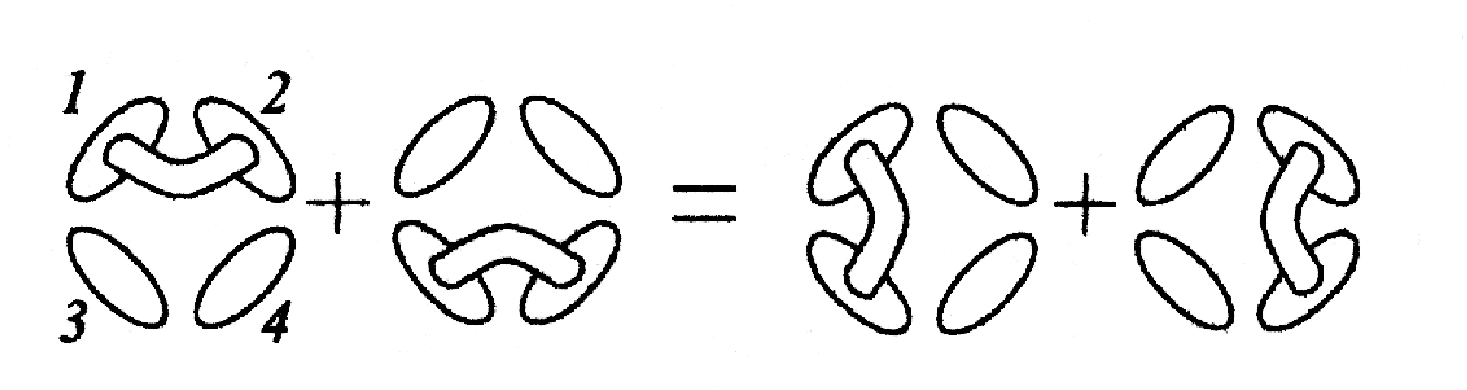}
\end{figure}

The local nature of the {\bf S}, {\bf T} and {\bf 4Tu} relations implies that the composition operations remain well defined in 
$Cob^{3}_{/l}$ and, hence, it is also a pre-additive category.

\subsection{Dotted cobordisms}

In this section we briefly describe a weaker variant of the theory which on links is equivalent to the original Khovanov theory 
\cite{khovanov}, but still rich enough.

\begin{figure}[h]
\centering
\leavevmode
\includegraphics[width=12cm]{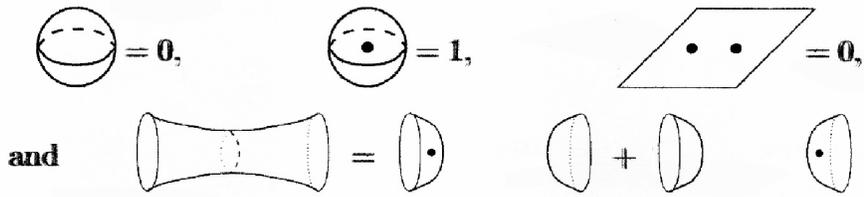}
\caption{Dotted cobordism relations.}
\label{punct}
\end{figure}

Extend the category $Cob^{3}$ to a new category $Cob^{3}_{\bullet}$ that has the same objects as $Cob^{3}$ and nearly the same morphisms - 
the only difference is that we now allow ``dots'' (of degree -2) that can be marked on cobordisms and moved freely within each connected 
component of a given cobordism. We then form the quotient category $Cob^{3}_{\bullet/l}$ by reducing $Cob^{3}_{\bullet}$ modulo the local 
relations from Figure \ref{punct}.

The {\bf S}, {\bf T} and {\bf 4Tu} relations follow from the above relations. Now we have a neck cutting relation.

{\bf Application:} The dotted cobordism relations are useful in the representation of spacetimes containing instantons, where the dots 
(the singularities) are for instantons while the normal matter lives in the rest of the manifold. So, in Figure \ref{punct} first 
sphere there is no instanton (=0), the second sphere contains one instanton and, as we saw previosly, the point can be transported to the 
sphere boundary as a small instanton (instanton envelope of the FRW spacetime). From the same figure we see that in a plane (Euclidian) 
containing an instanton and an anti-instanton the instanton number is 0 and, finally, that a catenoid can be written as one instanton, its 
spacetime and an anti-spacetime plus one anti-instanton, its anti-spacetime and a spacetime. 

\section{Lie algebras and the four color theorem \cite{dror4}}

Let us recall a well-known construction that associates to any finite dimensional metrized Lie algebra {\it L} a numerical-valued 
functional $W_{L}$ defined on the set of all oriented trivalent graphs {\it G} (that is, trivalent graphs in which every vertex is endowed 
with a cyclic ordering of the edges emaning from it). This construction underlies the gauge-group dependence of gauge theories in general 
and of the Chern-Simons topological field theory in particular and plays a prominent role in the theory of finite type (Vassiliev) 
invariants of knots and most likely also in the theory of finite type invariants of 3-manifolds.

Fix a finite dimensional metrized Lie algebra {\it L} (that is, a finite dimensional Lie algebra with an {\it ad}-invariant symmetric non-
degenerate bilinear form), choose some basis $\{L_{a}\}_{a=1}^{dim\; L}$ of {\it L}, let $t_{ab}=\left<L_{a},L_{b}\right>$ be the metric 
tensor, let $t^{ab}$ be the inverse matrix of $t_{ab}$, and let $f_{abc}$ be the structure constants of {\it L} relative to $\{L_{a}\}$:

\begin{equation}
f_{abc}=\left<L_{a},\left[L_{b},L_{c}\right] \right>\; .
\label{eq_36}
\end{equation}

\begin{figure}[h]
\centering
\leavevmode
\includegraphics[width=15cm]{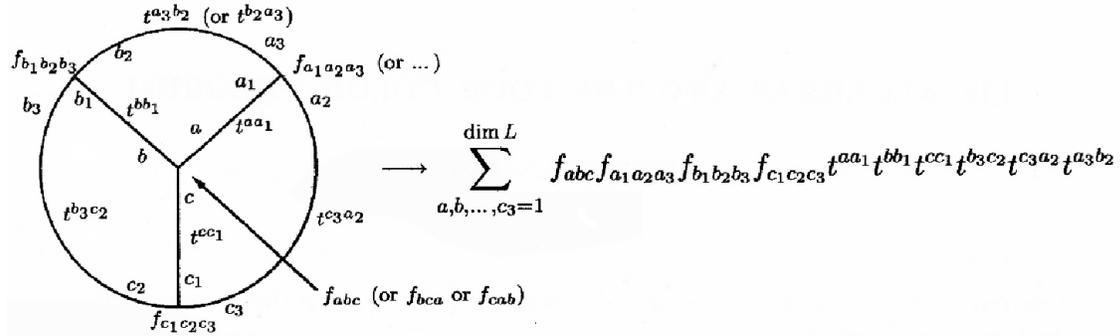}
\caption{An example illustrating the construction of $W_{L}(G)$. Notice that when {\it G} is drawn in the plane, we assume 
counterclockwise orientation for all vertices (unless noted otherwise), and that the cyclic symmetry $f_{abc}=f_{bca}=f_{cab}$ of the 
structure constants and the symmetry $t^{ab}=t^{ba}$ of the inverse metric ensures that $W_{L}(G)$ is well defined.}
\label{culori}
\end{figure}

Let {\it G} be some oriented trivalent graph. To define $W_{L}$, label all half-edges of {\it G} by symbols from the list $a,b,c,...,
a_{1},b_{1},...,$ and sum over $a,b,...,a_{1},...\in \left\{ 1,...,\textnormal{dim}\; L \right\}$ the product over the vertices of {\it G} 
of the structure constants ``seen'' around each vertex times the product over the edges of the $t^{..}$'s seen on each edge. This 
definition is better explained by an example, as in Figure \ref{culori}.

By introducing an explicit change-of-basis matrix as in \cite{dror2} or by re-interpreting $W_{L}(G)$ in terms of abstract tensor calculus 
as in \cite{dror3}, one can verify that $W_{L}(G)$ does not depend on the choice of the basis $\{L_{a}\}$. Typically one chooses a 
``nice'' orthonormal (or almost orthonormal) basis $\{L_{a}\}$, so that most of the constants $t^{ab}$ and $f_{abc}$ vanish, thus greatly 
reducing the number of summands in the definition of $W_{L}(G)$.

Unless otherwise stated, whenever dealing with a Lie algebra of matrices, we will take the metric to be the matrix trace in the defining 
representation:

\begin{equation}
\left< L_{a},L_{b} \right> =\textnormal{tr}(L_{a}L_{b})\; .
\label{eq_37}
\end{equation}

{\bf Lemma-Definition:} If a connected {\it G} has {\it v} vertices, then $W_{sl(N)}(G)$ is a polynomial in {\it N} of degree at most 
$\displaystyle \frac{v}{2}+2$ in {\it N}. Thus we can set $W^{top}_{sl(N)}(G)$ to be the coefficient $N^{\frac{v}{2}+2}$ in $W_{sl(N)}(G)$.

The following statement sounds rather resonable; it just says that if {\it G} is ``{\it sl(2)}-trivial'', then it is at least ``{\it sl(N)}
-degenerate''. For us who grew up thinking that all that there is to learn about {\it sl(N)} is already in {\it sl(2)}, this is not a 
big surprise:

{\bf Statement:} For a connected oriented trivalent graph {\it G}, $W_{sl(2)}(G)=0$ implies $W_{sl(N)}^{top}(G)=0$.

Lie-theoretically, there is much to say about {\it sl(2)} and {\it sl(N)}. There are representations of {\it sl(2)} into {\it sl(N)}, 
there is an ``almost decomposition'' of {\it sl(N)} into a product of {\it sl(2)}'s, and there are many other similarities. A-priori, the 
above statement sounds within reach, being equivalent to the Four Color Theorem, conjectured by Francis Guthrie in 1852 and proven by 
K.I. Appel and W. Haken \cite{appel} in 1976.

This equivalence follows from the following two propositions:

{\bf Proposition 1:} Let {\it G} be a connected oriented trivalent graph. If {\it G} is 2-connected, $\left| W^{top}_{sl(N)}(G) \right|$ 
is equal to the number of embeddings of {\it G} in an oriented sphere. Otherwise, $W^{top}_{sl(N)}(G)=0$.

{\bf Proposition 2 (\cite{penrose,kauff1,kauff2,kauff3}):} If {\it G} is planar with {\it v} vertices and $G^{c}$ is the map defined by 
its complement, than $\left| W_{sl(2)}(G) \right|$ is $2^{\frac{v}{2}-2}$ times the total number of colorings of $G^{c}$ with four 
colors so that adjacent states are colored with different colors.

Indeed, the above statement is clearly equivalent to:

\[
\left| W_{sl(N)}^{top}(G)\right| \neq 0\;\; \Rightarrow\;\;  \left| W_{sl(2)}(G)\right| \neq 0\; ,
\]
\noindent
which by propositions 1 and 2 is the same as saying: {\it G has a planar embedding with} $G^{c}$ {\it a map} $\Rightarrow G^{c}$ 
{\it has a 4-coloring}.

Notice that if {\it G} is connected, $G^{c}$ is a map (does not have states that border themselves) if {\it G} is 2-connected.

{\it Remark:} We have choosen the formulation of the above statement that we felt was the most appealing. With no change to the end 
result, one can replace $sl(N)=A_{N-1}$ by $B_{N}$, $C_{N}$, $D_{N}$, or $gl(N)$ and {\it sl(2)} by {\it so(3)} in the formulation of the
above statement.

\begin{figure}[h]
\centering
\caption{}
\leavevmode
\includegraphics[width=3.7cm]{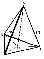}
\label{piramida}
\end{figure}

{\bf Application:} If we consider that $t^{ij}_{SU(3)}=\left< \lambda_{i},\lambda_{j} \right> = tr(\mathcal{B})$, where $\lambda_{i}$, 
$\lambda_{j}$ are our {\it SU(3)} matrices for the DEUS object, and $\mathcal{B}$ is obtained from $\lambda_{i}\times \lambda_{j}$ by 
interchanging the second and the third row in the product result, we have that the only non zero $t^{ij}$'s are $t^{2\;7}_{SU(3)}=-1$, 
$t^{2\;4}_{SU(3)}=\displaystyle \frac{i}{2}$, $t^{3\;5}_{SU(3)}=-\displaystyle \frac{i}{2}$, $t^{7\;4}_{SU(3)}=i$. In {\it SU(2)}, 
$t^{ij}_{SU(2)}=\left< c_{i},c_{j} \right> = tr(\mathcal{C})$, with $\mathcal{C}$ obtained from $c_{i}\times c_{j}$ by interchanging again 
the second and the third rows in the product result. So, $t^{1\;2}_{SU(2)}=t^{2\;3}_{SU(2)}=0$ and $t^{1\;3}_{SU(2)}=i\equiv t^{7\;4}_
{SU(3)}$. These can be represented as in Figure \ref{piramida} in which the fourth solution ($t=0$) of the Jones polynomial is illustrated 
by the point {\it A} which is not in same plane with the other three solutions, $\overline{AO}=i$ being perpendicular on the {\it BCD} 
plane.

From the {\it AOB} triangle we obtain that $\overline{BO}=\pm \sqrt{2}$. In the same way, from the {\it AOC} triangle results that 
$\overline{CO}=\pm \displaystyle \frac{\sqrt{3}}{2}$ and, from the {\it AOD} triangle, that $\overline{DO}=\pm \displaystyle 
\frac{\sqrt{3}}{2}$.

In the Figure \ref{piramida} we see that:

\begin{equation}
\begin{array}{llll}
\overline{BD}^{2}=\overline{BO}^{2}+\overline{DO}^{2}-2\; \overline{BO}\; \overline{DO}\; \textnormal{cos}\widehat{BOD}\\
\overline{BC}^{2}=\overline{BO}^{2}+\overline{CO}^{2}-2\; \overline{BO}\; \overline{CO}\; \textnormal{cos}\widehat{BOC}\\
\overline{CD}^{2}=\overline{CO}^{2}+\overline{DO}^{2}-2\; \overline{CO}\; \overline{DO}\; \textnormal{cos}\widehat{COD}\\
\widehat{BOD}+\widehat{BOC}+\widehat{COD}=2\pi
\end{array}
\label{eq_38}
\end{equation}
\noindent

With the (\ref{eq_38}) relations and $\overline{BC}=\overline{CD}=\overline{BD}$, as we must have if we intend to consider that the points 
(identical particles) interact with each other in the same way (same interaction type), we obtain that:

\begin{equation}
\ell=\overline{BC}=\overline{CD}=\overline{BD}=\frac{1}{2}\sqrt{7\pm 2\sqrt{6}}\simeq \left\{ \begin{array}{ll}
1.724745 \\
0.724745
\end{array} \right. \; ,
\label{eq_39}
\end{equation}
\noindent
or as multiples of one of these values.

So, between two solutions of the Jones polynomial we must have:

\begin{equation}
\left| V(t_{1})-V(t_{2}) \right| = \left| V(t_{1})-V(t_{3}) \right|=\left| V(t_{2})-V(t_{3}) \right|\propto \ell\; .
\label{eq_40}
\end{equation}
\noindent
and is intended to be applied in the case quarks as zero level of similarity visible at level one of self-similarity (nucleons) only in a 
wave representation. We won't be able to observe the complete DEUS object of this level (with a particle and a wave representation) but 
only the catenoid. The Jones polynomial will be ``colored'' and the solutions (quarks) are at $V(t,\phi)=0$, where the variable in the 
Jones polynomial is $\textnormal{cosh}^{\alpha}\left( \displaystyle\frac{\sqrt{t^{2}+\phi^{2}}}{A}\right)$ with $\alpha=\pm 1$.

Because $[\lambda_{i},\lambda_{j}]=\lambda_{i}\times \lambda_{j}-\lambda_{j}\times \lambda_{i}=0$, the structure constants from 
(\ref{eq_36}) are all zero.

If $t^{ij}_{SU(3)}=\left< \lambda_{i},\lambda_{j} \right> = tr(\lambda_{i}\times \lambda_{j})$ and  $t^{ij}_{SU(2)}=\left< c_{i},c_{j} 
\right> = tr(c_{i}\times c_{j})$ (without any rotation in the $\lambda_{i}\times \lambda_{j}$, respectively in the $c_{i}\times c_{j}$ 
matrix), then only $t^{2\;7}_{SU(3)}=-1$, $t^{2\;4}_{SU(3)}=i$, $t^{7\;4}_{SU(3)}=i$ are not zero. This means that the representation is 
not three-dimensional ($t^{3\;5}_{SU(3)}=0$), one of the $t\neq 0$ Jones polynomial solutions being not defined. For {\it SU(2)}, 
all the metric tensor components $t^{ij}_{SU(2)}=0$, this situation being impossible.

\section{Scalar Potential of a DEUS Object}

The right-hand trefoil knot is described by the (\ref{eq_14}) Jones polynomial.

We define the uni-dimensional potential ${\cal V}(t)$ of a self-similar DEUS level object's scalar field {\it t}, in the FRW representation (for example, the potential of the FRW Universe), \cite{deus12} as the Jones polynomial (let say, of the right-hand trefoil knot) from which we eliminate the $t=0$ solution, corresponding to a pointlike field of the DEUS object:

\begin{equation}
{\cal V}(t)=t^{2}-t^{3}+1\; .
\label{eq_41}
\end{equation}

\begin{figure}[htb]
\centering
\caption{}
\leavevmode
\includegraphics[scale=0.6]{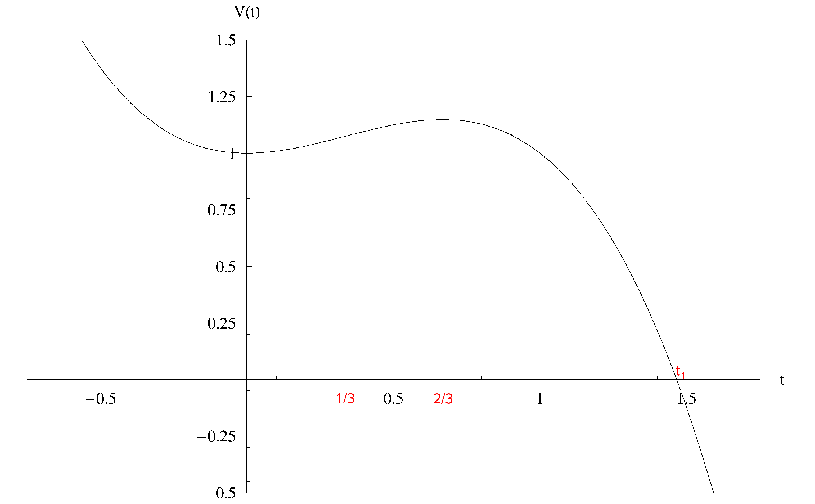}
\label{sp1}
\end{figure}

Figure \ref{sp1} shows how the potential ${\cal V}$ varies with {\it t}. When we eliminate the $t=0$ value, in the Jones polynomial remains the solution $t_{1}\in (1.46,1.47)$. Another observation is that at the zero value of the {\it t} field the value of the potential is not zero.

Taking the diagram of the right-hand trefoil knot as the global representation of the Universe field (self-similarity DEUS level one), then the left-hand trefoil knot described by:

\begin{equation}
V(p)=p^{-1}(p^{-2}-p^{-3}+1)\; ,
\label{eq_42}
\end{equation}
\noindent
represents a lower DEUS level (self-similarity DEUS level two), where:

\begin{equation}
{\cal V}(p)=p^{-2}-p^{-3}+1\; ,
\label{eq_43}
\end{equation}
\noindent
is the potential of the field {\it p}.

\begin{figure}[htb]
\centering
\caption{}
\leavevmode
\includegraphics[scale=0.39,angle=90]{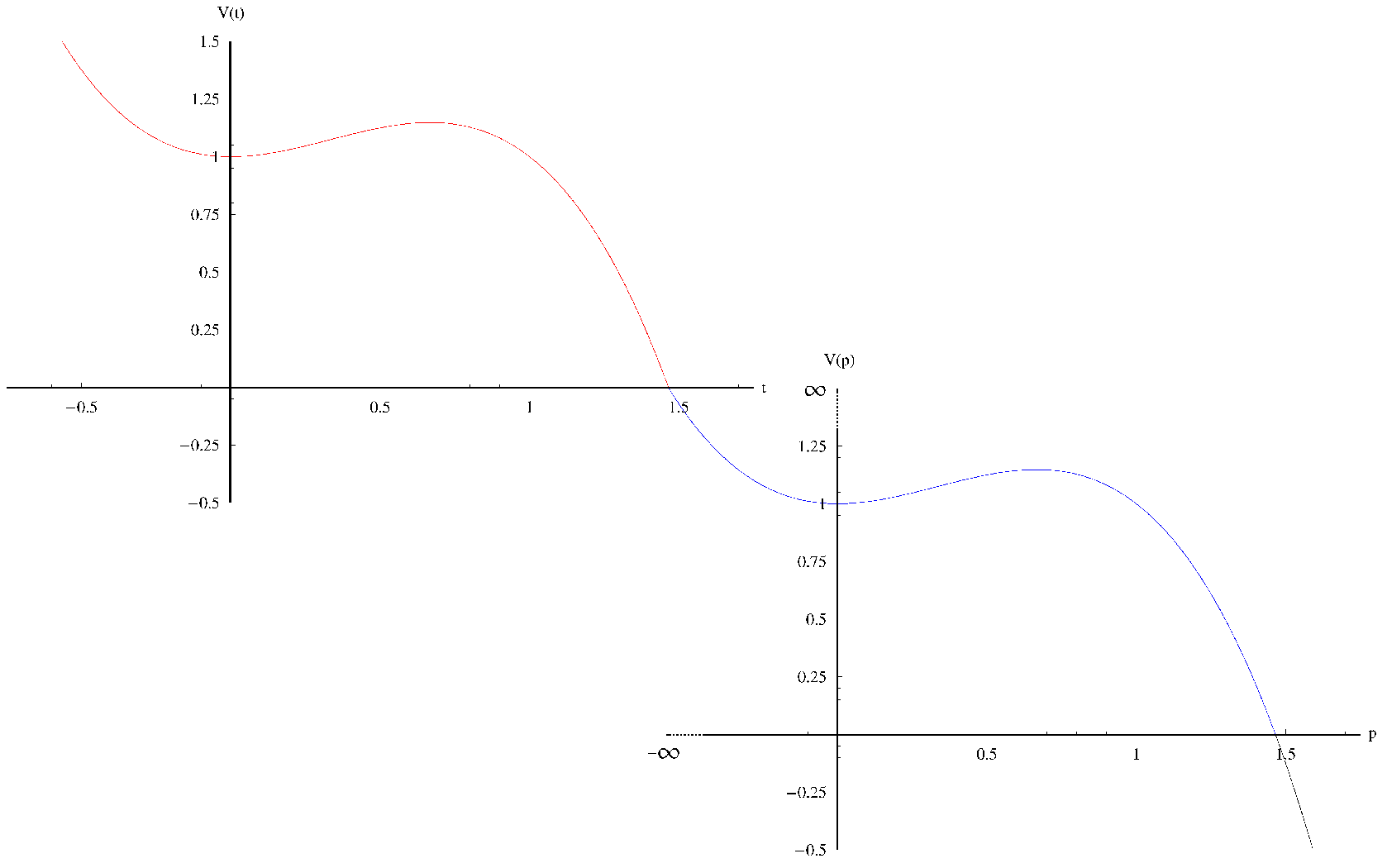}
\label{sp2}
\end{figure}

\begin{figure}[htb]
\centering
\caption{}
\leavevmode
\includegraphics[scale=0.5]{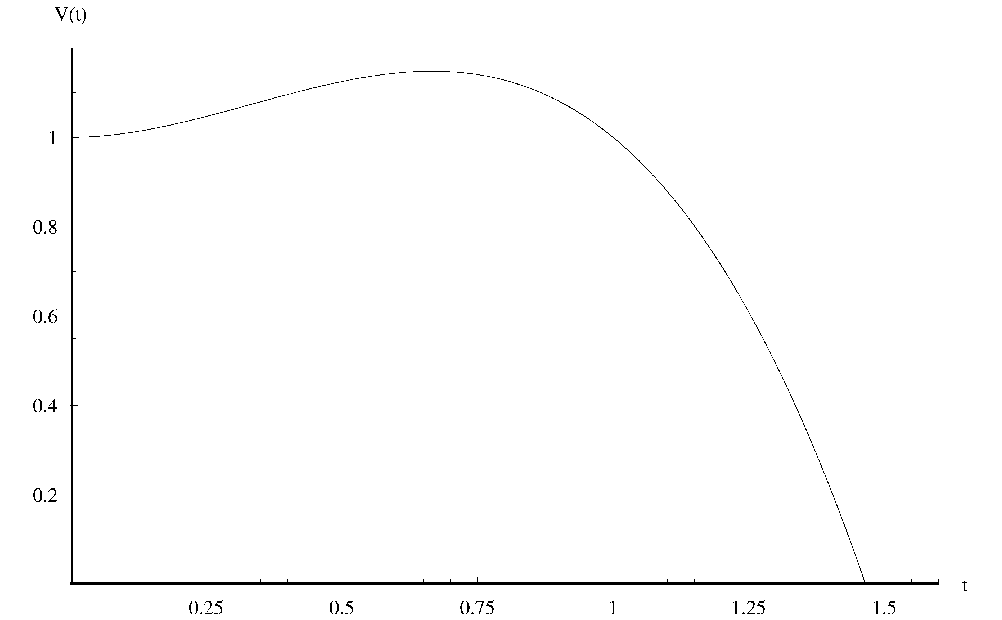}
\label{sp3}
\end{figure}

If $\displaystyle p=\frac{1}{t}$, when the field $t\rightarrow 0$, a lower level of self-similarity "opens" to the observer, in $p\rightarrow -\infty$ and ${\cal V}(p)\rightarrow \infty$ (see figure \ref{sp2}), and at $p=0$ an even lower DEUS level (right-hand trefoil knot in {\it t}). So, the potential of an empty space (without any others self-similar DEUS level objects) of an arbitrary DEUS level has a representation as the one in figure \ref{sp3}.

For figure \ref{sp3} representing the potential of the Universe, we have for $\displaystyle t\in\left(0,\frac{1}{3}\right)$ a radiation dominated era, for $\displaystyle t\in\left(\frac{1}{3},\frac{2}{3}\right)$ a matter dominated era, and for $\displaystyle t\in\left(\frac{2}{3},1.47\right)$ a dark energy dominated era. The present Universe just crossed the $\displaystyle t=\frac{2}{3}$ border between matter and dark energy.

\ack
Manuscript registered at the Astronomical Institute, Romanian Academy, record No. 267 from 04/19/07.